# Direct Measurements of Giant Star Effective Temperatures and Linear Radii: Calibration Against Spectral Types and V-K Color

Gerard T. van Belle,[1] Kaspar von Braun,[1] David R. Ciardi,[2] Genady Pilyavsky,[3] Ryan S. Buckingham,[1] Andrew F. Boden,[4] Catherine A. Clark,[1,5] Zachary Hartman,[1,6] Gerald van Belle,[7] William Bucknew,[1] and Gary Cole[8,*]

[1]*Lowell Observatory*
*1400 West Mars Hill Road*
*Flagstaff, AZ 86001, USA*
[2]*California Institute of Technology, NASA Exoplanet Science Institute*
*Mail Code 100-22*
*1200 East California Blvd.*
*Pasadena, CA 91125, USA*
[3]*Systems & Technology Research*
*600 West Cummings Park*
*Woburn, MA 01801, USA*
[4]*California Institute of Technology*
*Mail Code 11-17*
*1200 East California Blvd.*
*Pasadena, CA 91125, USA*
[5]*Northern Arizona University*
*Department of Astronomy and Planetary Science*
*NAU Box 6010*
*Flagstaff, Arizona 86011, USA*
[6]*Georgia State University*
*Department of Physics and Astronomy*
*P.O. Box 5060*
*Atlanta, GA 30302, USA*
[7]*University of Washington*
*Department of Biostatistics*
*Box 357232*
*Seattle, WA 98195-7232, USA*
[8]*Starphysics Observatory*
*14280 W. Windriver Lane*
*Reno, NV 89511, USA*

## ABSTRACT

We calculate directly determined values for effective temperature ($T_{\rm EFF}$) and radius ($R$) for 191 giant stars based upon high resolution angular size measurements from optical interferometry at the Palomar Testbed Interferometer. Narrow- to wide-band photometry data for the giants are used to establish bolometric fluxes and luminosities through spectral energy distribution fitting, which allow for homogeneously establishing an assessment of spectral type and dereddened $V_0 - K_0$ color; these two parameters are used as calibration indices for establishing trends in $T_{\rm EFF}$ and $R$. Spectral types range from G0III to M7.75III, $V_0 - K_0$ from 1.9 to 8.5. For the $V_0 - K_0 = \{1.9, 6.5\}$ range, median

Corresponding author: Gerard T. van Belle
gerard@lowell.edu

$T_{\rm EFF}$ uncertainties in the fit of effective temperature versus color are found to be less than 50K; over this range, $T_{\rm EFF}$ drops from 5050K to 3225K. Linear sizes are found to be largely constant at 11 $R_\odot$ from G0III to K0III, increasing linearly with subtype to 50 $R_\odot$ at K5III, and then further increasing linearly to 150 $R_\odot$ by M8III. Three examples of the utility of this data set are presented: first, a fully empirical Hertzsprung-Russell Diagram is constructed and examined against stellar evolution models; second, values for stellar mass are inferred based on measures of $R$ and literature values for $\log g$. Finally, an improved calibration of an angular size prediction tool, based upon $V$ and $K$ values for a star, is presented.

## 1. INTRODUCTION

One of the key contributions of long-baseline optical interferometry to date has been the empirical determination of fundamental stellar parameters. The range of values for parameters such as radius and temperature can be simply inferred from blackbody assumptions about stars, or with greater precision through using stellar models such as PHOENIX (Husser et al. 2013), but ultimately are most accurate when directly measured. Their utility is consequently extended by calibrating stellar models.

Calibration of 'canonical' values for temperature and radius has wide-ranging utility in astrophysics. Evaluation of X-ray transient host stars (Skopal 2015; McCollum et al. 2018), interpretation of protoplanetary environments (Arulanantham et al. 2017), establishing the tempeature of a galactic runaway giant (Massey et al. 2018), and modeling of galactic long-period variables (Barthès & Luri 2001) are a few examples of the use of our earlier calibration of effective temperature (vB99; van Belle et al. 1999). Improved temperature calibrations can have wide-ranging effects, particularly when they refine calibrations used in large surveys (e.g., the symbotic star survey of Akras et al. 2019). General values for such calibrations appear in references such as Cox (2000) and facilitate a myriad of back-of-the-envelope calculations.

The Palomar Testbed Interferometer (PTI; Colavita et al. 1999), whose technical details are discussed in §3, was a particularly productive long-baseline optical interferometer (LBOI) that operated from 1996 until 2008. Its highly efficient, semi-robotic operations enabled the collection of large amounts of stellar fringe visibility data on any given night; these visibility data can be used to establish a direct measure of source angular size for resolved stellar sources greater in size than ∼0.75 milliarcseconds (mas). Given the sensitivity limit of the facility ($m_K$ <5) and its angular resolving power, PTI was particularly well-suited for large surveys of evolved stars; main sequence stars on the faint end of this sensitivity range tended to be too small to resolve.

Our first investigation on the calibration of surfaces temperatures of giant stars, in vB99, was published early in the operation of the instrument. Subsequent to that investigation, notable improvements were made that motivate this larger follow-up study. Considerable effort was invested in increasing our understanding the operations of PTI and the implications for its data products (Colavita 1999), the night-to-night repeatability of the data (Boden et al. 1998b, 1999), the atmospheric conditions of the site (Linfield et al. 2001), the nature of absolute calibration of the fringe visibility data (van Belle & van Belle 2005), and a strictly vetted set of on-sky calibration sources (van Belle et al. 2008). In addition to those improvements, the instrument during its decade of operation was used to conduct a wide range of scientific investigations, including a broad survey to measure the angular sizes of giant stars.

The development of the `sedFit` code subsequent to our initial vB99 investigation has provided a superior means of calculating bolometric fluxes through spectral energy distribution fitting, which are necessary for robust determination of temperature. Utilization of `sedFit` is furthered by the availability of modern stellar spectra templates such as PHOENIX (Husser et al. 2013), as well as empirical spectral templates such as the INGS library[1], a substantial improvement over the earlier Pickles Flux Library (Pickles 1998). Additionally, given the availability of `sedFit`, considerable effort was invested in collection of ancillary photometric data, both through observation and examination of archival sources, and improving our zero-point calibrations of those data (Bohlin et al. 2014; Mann & von Braun

---

2015). Finally, for each individual science target reported upon herein, data from a larger number of observing nights and baseline configurations were typically collected, allowing for better control of occasional spurious data points. Our intent with this investigation was to establish the definitive effective temperature scale for giant stars, and as such great care was taken in having each of these steps be as empirical as possible, with the greatest accuracy and precision available.

The previous surveys of PTI and other notable LBOI facilities are presented in §2. Details on the PTI facility are given in §3, as well as particulars of the target selection (§3.1). Bolometric flux determination using `sedFit` is detailed in §4; derived effective temperatures are given in §5; distances and their determinations are discussed in §6. With the establishment of these fundamental parameters, relationships between $T_{\mathrm{EFF}}$ and $R$ and indicator indices, $V_0 - K_0$ (dereddened) color and spectral type, are explored in §7. An intriguing gap in the otherwise smooth continuum of points in the $T_{\mathrm{EFF}}$ versus $V_0 - K_0$ is examined statistically for significance in §7.1. A serendipitous result from the steps taken in this investigation, the calibration of spectral type versus $V_0 - K_0$, is presented in §7.2.3. We then take a broader look at some of the possible applications of this data with examples in §8. First, a comparison of our results to stellar evolutionary tracks (§8.1) ; second, we demonstrate that these measures of $R$ , when combined with $\log g$, can be used to infer evolved star masses (§8.2). Finally, a new calibration of a predictive tool for stellar angular diameters is presented in §9.

## 2. PREVIOUS LARGE SURVEYS

Measures of stellar angular diameters are particularly useful when conducted in surveys covering multiple targets. A summary of the surveys by LBOI facilities is presented in Table 1. Given the intersection of sensitivity and angular resolution of earlier facilities that typically had baselines only up to ∼100 m, a focus on evolved stars is seen in those surveys from roughly before 2005. More modern facilities have baselines in excess of ≳100 m (VLTI), ≳300 m (CHARA), and ≳400 m (NPOI), enabling studies of smaller (<≲1.0 mas) objects such as main sequence stars. However, the most highly automated facilities – PTI and the Mark III, which could hop star-to-star in times of ≲5 minutes – are now in the past, meaning the largest surveys are more difficult and time-consuming observationally.

Although use of the lunar occultation (LO) technique is not a focus of this investigation, it is worth noting that the very earliest surveys at milliarcsecond scale, from the 1970's onwards, were carried out by the LO technique. These include the extensive work of the Kitt Peak group (Ridgway et al. 1977, 1979, 1980a, 1982; Schmidtke et al. 1986) as well as the UT-Austin group (see papers I - XVI of the series that concludes with Evans et al. 1986). A summary of 348 measures on 124 stars is presented in White & Feierman (1987). There continue to be LO measures carried out with more modern equipment (Tej & Chandrasekhar 2000; Mondal & Chandrasekhar 2005; Baug & Chandrasekhar 2013; Ertel et al. 2014), though the principal body of high-resolution work in the last 15 years has been with LBOI facilities.

In relation to these previous surveys, the intent of this particular investigation is two-fold. First, the homogenous dataset presented herein is twice as large as the next-largest survey, and benefits from advances in understanding of atmospheric and instrumental effects. Second, the ancillary data sets and supporting modeling have improved substantially. Together, these advances build on the experience from those surveys, but make for substantially improved characterization of stellar fundamental parameters for giant stars.

## 3. ANGULAR SIZE MEASUREMENTS WITH PTI

PTI was an 85 to 110 m baseline H- and K-band (1.6 μm and 2.2 μm) interferometer located at Palomar Observatory in San Diego County, California, and is described in detail in Colavita et al. (1999). It had three 40-cm apertures used in pairwise combination for measurement of stellar fringe visibility on sources that range in angular size from 0.05 to 5.0 milliarcseconds, being able to resolve individual sources with angular sizes $\theta > 0.75$ mas in size. PTI was in nightly operation between 1997 and 2008, with minimum downtime throughout the intervening years. The data from PTI considered herein covers the range from the beginning of 1998 (when the standardized data collection and pipeline reduction went into place) until the beginning of 2008 (when operations at PTI concluded). In addition to the giant stars discussed herein, appropriate calibration sources were observed as well and can be found in van Belle et al. (2008).

The calibration of the giant star visibility ($V^2$) data was performed by estimating the interferometer system visibility ($V_{\mathrm{SYS}}^2$) using the calibration sources with model angular diameters and then normalizing the raw giant visibility by $V_{\mathrm{SYS}}^2$ to estimate the $V^2$ measured by an ideal interferometer at that epoch (Mozurkewich et al. 1991; Boden et al. 1998a;

| Facility | Survey size | Reference |
|---|---|---|
| **Giants** | | |
| Mark III | 24 giants | Hutter et al. (1989) |
| Mark III | 12 giants | Mozurkewich et al. (1991) |
| IOTA | 37 giants | Dyck et al. (1996a) |
| IOTA | 74 giants | Dyck et al. (1998) |
| PTI | 69 giants/supergiants | van Belle et al. (1999) |
| NPOI | 50 giants | Nordgren et al. (1999) |
| NPOI | 41 giants | Nordgren et al. (2001) |
| Mark III | 85 giants | Mozurkewich et al. (2003) |
| CHARA | 25 K giants | Baines et al. (2010) |
| NPOI | 69 giants, 18 additional stars | Baines et al. (2018) |
| PTI | 191 giants | This work |
| **Other Evolved Stars** | | |
| IOTA | 15 carbon stars | Dyck et al. (1996b) |
| IOTA | 18 O-rich Miras | van Belle et al. (1996) |
| | 9 carbon/S-type Miras | |
| IOTA | 4 non-Mira S-type | van Belle et al. (1997) |
| IOTA | 22 O-rich Miras | van Belle et al. (2002) |
| VLTI-VINCI | 14 Miras | Richichi & Wittkowski (2003) |
| VLTI-VINCI | 7 Cepheids | Kervella et al. (2004) |
| PTI | 74 supergiants | van Belle et al. (2009) |
| PTI | 5 carbon stars | Paladini et al. (2011) |
| PTI | 41 carbon stars | van Belle et al. (2013) |
| VLTI-PIONIER | 9 Cepheids | Breitfelder et al. (2016) |
| VLTI-PIONIER | 23 post-AGB disks | Kluska et al. (2019) |
| **Main Sequence Stars** | | |
| PTI | 5 Main-sequence stars | Lane et al. (2001) |
| VLTI-VINCI | 5 Vega-like stars | Di Folco et al. (2004) |
| VLTI-VINCI/AMBER | 7 low mass stars | Demory et al. (2009) |
| PTI | 40 stars (incl. 12 exoplanet hosts) | van Belle & von Braun (2009) |
| CHARA | 44 AFG MS stars | Boyajian et al. (2012a) |
| CHARA | 22 KM stars | Boyajian et al. (2012b) |
| CHARA | 23 A-K stars, 5 exoplanet hosts | Boyajian et al. (2013) |
| CHARA | 11 exoplanet hosts | von Braun et al. (2014) |
| CHARA | 7 A-type stars | Jones et al. (2015) |
| **Young Stellar Objects** | | |
| VLTI-PIONIER | 92 Debris disk stars | Ertel et al. (2014) |
| VLTI-PIONIER | 21 T-Tauri stars | Anthonioz et al. (2015) |
| VLTI-PIONIER | 7 Debris disk stars | Ertel et al. (2016) |
| VLTI-PIONIER | 51 Herbig AeBe disks | Lazareff et al. (2017) |
| VLTI-GRAVITY | 27 Herbig AeBe disks | GRAVITY Collaboration et al. (2019) |

**Table 1.** Long-baseline optical interferometry diameter surveys of 5 or more stars. See Table 3.1 in von Braun & Boyajian (2017) for a detailed list of stars with interferometrically determined radii. See Section 2 for details.

van Belle & van Belle 2005). Uncertainties in the system visibility and the calibrated target visibility are inferred from internal scatter among the data in an observation using standard error-propagation calculations (Colavita 1999). Calibrating our point-like calibration objects against each other produced no evidence of systematics, with all objects delivering reduced $V^2 = 1$.

PTI's limiting night-to-night measurement error is $\sigma_{V^2_{SYS}} \approx 1.5\%$, the source of which is most likely a combination of effects: uncharacterized atmospheric seeing (in particular, scintillation), detector noise, and other instrumental effects. This measurement error limit is an empirically established floor from the previous study of Boden et al. (1999).

From the relationship between visibility and uniform disk angular size ($\theta_{UD}$), $V^2 = [2J_1(x)/x]^2$ (Airy 1835; Born & Wolf 1980), where $J_1$ is the first Bessel function and spatial frequency $x = \pi B \theta_{UD} \lambda^{-1}$, we established uniform disk angular sizes ($\theta$) for the giants observed by PTI since the accompanying parameters (projected telescope-to-telescope separation, or baseline, $B$ and wavelength of observation $\lambda$) are well-characterized during the observation. This uniform disk angular size will be connected to a more physical limb darkened angular size ($\theta_{LD}$) in §5.1; these uniform disk angular sizes are presented in Table 16.

### 3.1. Target Selection for PTI

Given the highly efficient queue-scheduled nature of observing with PTI, a large program of observing as many evolved stars as possible was undertaken at the facility. These were targets of opportunity, available for observing when the facility wasn't tasked with other observing. The scope of this manuscript will focus on the field giant stars. This leaves objects such as S-type stars and Miras for forthcoming articles, or objects such as supergiants (van Belle et al. 2009) and carbon stars (van Belle et al. 2013) already having been published. This broad sweep meant target selection was largely limited by the 'sweet spot' of PTI sensitivity, for both target brightness and angular size.

*Angular size range.* Angular sizes at PTI for robust size determination, noting the night-to-night precision cited above, were typically intended to be in the range of 1.5 to 4.0 mas. This range is consistent with $K$-band operation of a 109 meter baseline, resulting in $V^2$ contrast below roughly 90% (to ensure target resolution) and above roughly 10% (to avoid degenerate diameter solutions). *A priori* estimators, such as the $V - K$ technique found in van Belle (1999), allowed reasonable expectations of the gross size of a given target prior to observing. The number of resulting targets over the range of angular sizes is found in Table 2 and plotted in Figure 1. Some angular sizes in excess of 4.0 mas are seen from large objects observed with PTI's shorter 85 meter baseline. Based on the minimum expected night-to-night $V^2$ repeatability discussed above, the limiting fractional error expected for various angular sizes is presented in Table 3.

*Brightness range.* PTI's limiting magnitude of $K < 5$ primarily limited the available point-like calibrator observations interleaved with resolved target star observations. Generally speaking, targets that satisfied the angular size range constraints for resolvability above were quite bright, with $K < 3$. The bigger impact on operations were those targets that were extremely red (corresponding to the targets with the lowest effective temperature); a lack of visible light photons made it difficult for the facility's tip-tilt tracker to follow atmospheric turbulence. However, this was more of a concern for studies that targeted even redder targets such as carbon stars (van Belle et al. 2013) than this particular investigation.

*Spectral types.* Spectral types were initially taken from various literature references (e.g., see Skiff 2014b, and references therein); care was taken to select those that had been previously typed as luminosity class III, or were otherwise indicated to be off the main sequence (e.g. based on parallax values). As will be presented below in §4, these types were taken as starting points for our own solutions for fitting spectral types. These initial spectral types are presented in Table 17, as well as our best fit values.

*Completeness.* A cursory examination of the 2MASS and Skiff catalogs (Cutri et al. 2003a; Skiff 2014b) can give an indication of the completeness of our giant star sample. First, the 2MASS catalog was sampled for all objects brighter than $K < 3.5$, resulting in approximately 13,000 objects. For stars in this brightness range, 2MASS photometry is saturated and increasingly unreliable, although for this assessment it is more than sufficient. Second, from the $B$ and $K$ magnitudes in 2MASS, all such objects were selected for angular sizes in the range of $\theta = \{1.0, 4.75\}$ mas, using the $B - K$ size estimator in van Belle (1999). Finally, those ∼11,600 objects were compared against the Skiff catalog for objects previously typed to be luminosity class III objects, resulting in ∼4,500 targets, of which 1,723 are northern hemisphere objects. As such, our sample herein constitute a sample of roughly ∼10% all possible giant stars that PTI could have observed.

This is an *ex post facto* assessment and could not have guided target selection at the time. The Skiff catalog was not available at the time of the PTI target selection, and 2MASS available only for the latter portion of PTI operations. During PTI operations, our guideposts were the Bright Star Catalog (Hoffleit & Jaschek 1982), the recently released (for that era) Hipparcos catalog (Perryman et al. 1997), and the *Catalog of Infrared Observations* (CIO; Gezari et al. 1993). During that era, luminosity classification was somewhat less clearly defined during PTI operations, even if in the course of this investigation our targets will be diligently sorted for luminosity class III objects.

**Table 2**. Number of targets per angular size bin.

| Angular Size Bin | Number of Stars |
|---|---|
| 0.75 | 1 |
| 1.00 | 2 |
| 1.25 | 9 |
| 1.50 | 17 |
| 1.75 | 29 |
| 2.00 | 36 |
| 2.25 | 23 |
| 2.50 | 29 |
| 2.75 | 16 |
| 3.00 | 10 |
| 3.25 | 15 |
| 3.50 | 7 |
| 3.75 | 5 |
| 4.00 | 3 |
| 4.25 | 3 |
| 4.50 | 1 |
| 4.75 | 2 |

Note—For more information, see 3.1 and Figure 1.

**Table 3**. Minimum possible fractional error for measure angular sizes, based upon PTI's night-to-night limiting $V^2$ repeatability of 1.5%, as discussed in §3.

| Angular Size Bin (mas) | Minimum Fractional Error |
|---|---|
| 0.75 | 10% |
| 1.00 | 6.0% |
| 1.25 | 4.1% |
| 1.50 | 3.1% |
| 1.75 | 2.4% |
| 2.00 | 2.1% |
| 2.25 | 1.9% |
| 2.50 | 1.8% |
| 2.75 | 1.7% |
| 3.00 | 1.8% |
| 3.25 | 2.0% |
| 3.50 | 2.2% |
| 3.75 | 2.6% |
| 4.00 | 3.1% |
| 4.25 | 4.1% |
| 4.50 | 5.5% |
| 4.75 | 8.0% |

Note—For more information, see 3.1 and Figure 1.

## 4. BOLOMETRIC FLUXES

For the program stars, measurements of the stellar bolometric fluxes ($F_{\rm BOL}$) were needed in order to compute the stellar effective temperature, $T_{\rm EFF}$. A direct determination of $T_{\rm EFF}$ can be made from a stellar angular diameter ($\theta$), in conjunction with a measurement of $F_{\rm BOL}$ (Equation 1). $F_{\rm BOL}$ is calculated by integrating a stellar spectrum (stellar model, spectral template, or spectrophotometric data product) over frequency or wavelength after shifting this spectrum in flux units to fit calibrated photometry or spectrophotometry. For this study, these photometry data are a combination of literature photometry and data we obtained ourselves in support of this particular survey.

The stellar bolometric flux is the total integrated flux from a star, as if there were no intervening obscuration between us and the object of interest. Any attempts of measuring $F_{\rm BOL}$ need to be careful to account for two problems that can limit our attempts to characterize the flux from the star. First, phenomena that attenuate the apparent flux of a star at the location of the observer need to be accounted for. This includes interstellar extinction, extinction local to the source (such as circumstellar material), and effects local to the observer such as atmospheric extinction. Second, photometric measures of stellar flux are rarely (if ever) done in a comprehensively multi-wavelength fashion - hence, gaps in the wavelength coverage of flux observations need to be interpolated or extrapolated in a plausible manner, using appropriate spectral type templates (§4.1). We discuss literature photometry products in §4.2 and Table 4, as well as photometry obtained ourselves in support of this project. An approach to determination of $F_{\rm BOL}$ in a robust manner that accommodates these complications and deliveries ∼2% levels of accuracy for $F_{\rm BOL}$ seen in §4.3 follows these steps:

- Photometry is collected for the target of interest. Ground-based photometry must be extinction-corrected for atmospheric extinction (Johnson & Morgan 1953a; Johnson 1965a; Hardie 1964; Linnell 1982; Cousins 1985).

- The instrument response, and in particular, the filter function, must be known in detail; the zero-point calibration needs to be known to ∼1-2% or better. Ideally, zero-point calibrations are directly traceable to National Institute of Standards and Technology (NIST) standards or similar absolute standards. See, for example, the ACCESS experiment (Kaiser et al. 2008, 2010) or the HST STIS Next Generation Spectral Library (NGSL) (Gregg et al. 2004; Heap & Lindler 2010).

- Generalized broad-band photometry is the easiest to collect, due to its high photon throughput and ready filter availability. However, narrow-band or specialized broad-band photometry is of great utility in disentangling degeneracies between spectral type, luminosity class, and interstellar or circumstellar extinction – such photometry has been used for stellar classification since the advent of high-precision phototubes (e.g. Canterna 1976; Johnson & Mitchell 1995). Our fitting approach can utilize both broad- and narrow-band photometry from multiple sources, which provides an element of cross-checking data points from multiple heterogeneous sources.

- Photometry ideally are on both the long- and short-wavelength sides of the 'blackbody' peak. These data help to break the degeneracy between gross flux levels and interstellar or circumstellar extinction.

- Once the photometry is in hand, a spectral template is fit to these data. The necessity of this step is to interpolate or extrapolate over the unsampled wavelength regimes; a secondary purpose of this step is to fit for interstellar/circumstellar extinction. In this step, a grid of templates is typically searched to determine the most appropriate one for use with the star being examined.

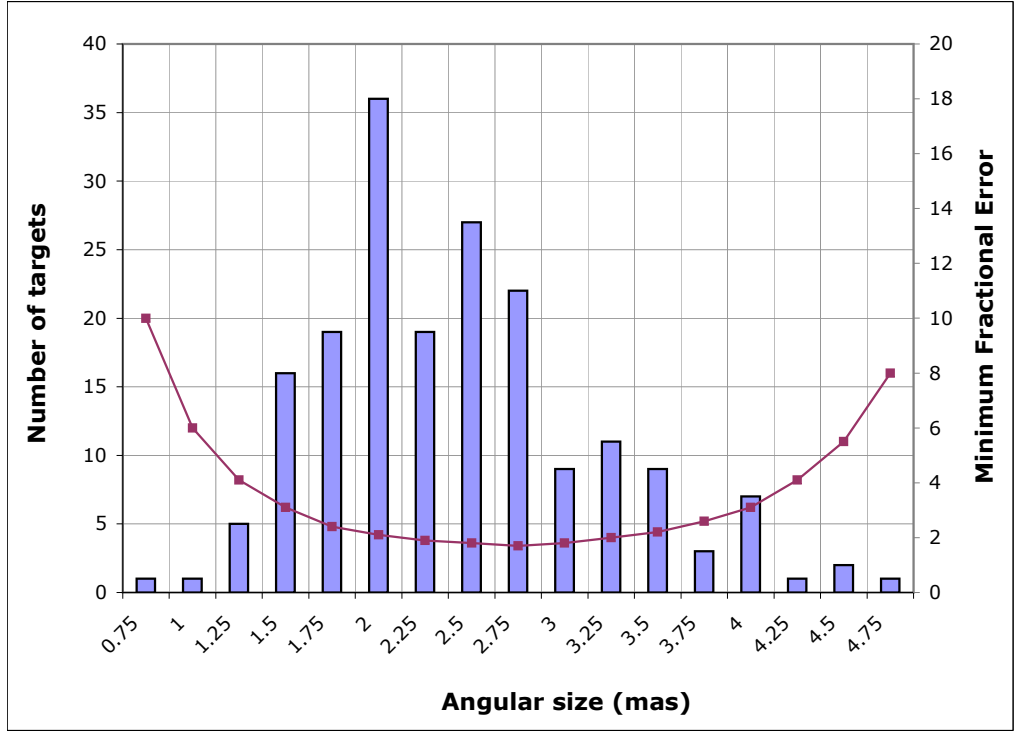

**Figure 1.** Histogram of the measured angular sizes for bins centered at intervals of 0.25 mas (blue columns, left vertical axis), along with minimum fractional error for each of those bins (red line, right vertical axis). For more information, see Section 3.1 and Tables 2 and 3.

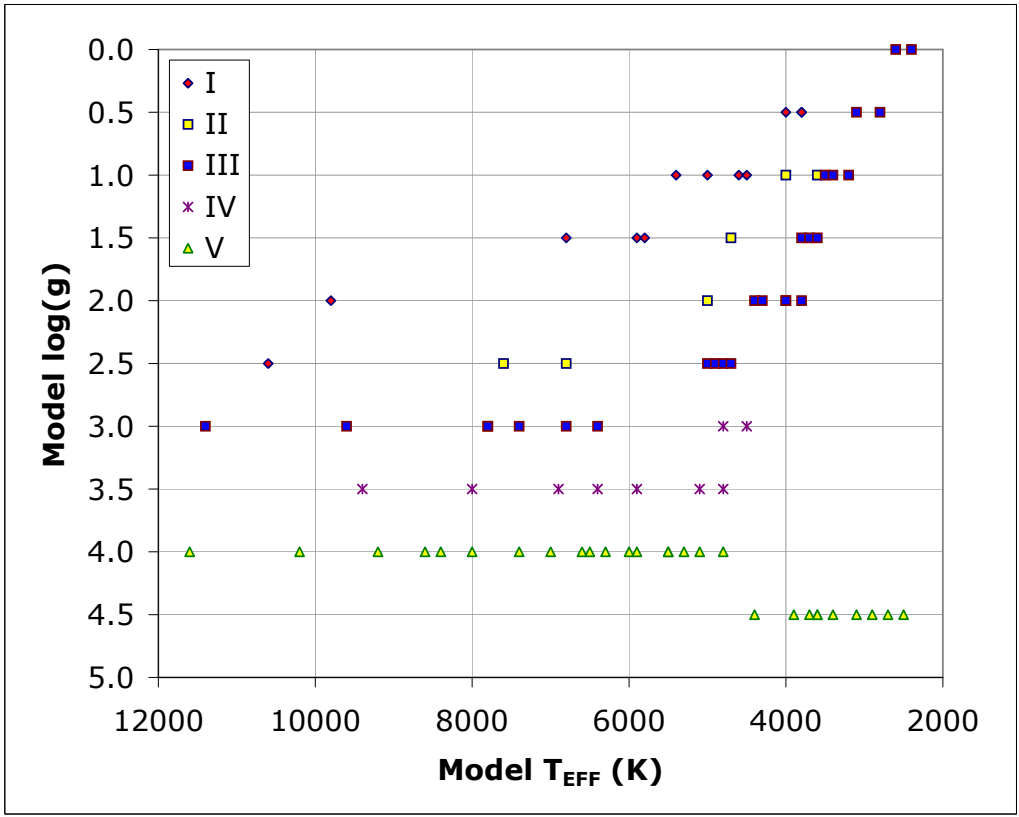

**Figure 2.** PHOENIX fitting of INGS spectra, by luminosity class, as a Kiel diagram. For more detail, see Section 4.1.

### 4.1. *Spectral templates and spectral types*

A significant question that we frequently encountered during this study was whether to fit our photometric data with empirical spectral templates such as the INGS library, or to work with model spectral templates based on models such as the PHOENIX. The INGS (IUE/NGSL/SpeX) library[2] is a refinement of the empirical templates originally found in Pickles (1998), by the same author; it has the virtue of being a model-independent approach to providing a stellar spectral template. Unfortunately, the drawback of the INGS templates is that they are, relative to model grids such as the PHOENIX models, less densely available in a grid of stellar spectral type versus luminosity class.

In contrast to the INGS empirical templates, a dense grid of PHOENIX model templates is available (Husser et al. 2013)[3] which offers a significantly denser sampling in $T_{EFF}$ (i.e. spectral type) versus surface gravity (i.e. luminosity class), and also sidesteps any potential concerns about uncorrected interstellar extinction in the INGS empirical templates.

Our solution to get the best of these complementary approaches to stellar spectra template was to utilize a set of PHOENIX models that had been calibrated against the INGS spectra data. Using the INGS data as input spectrophotometry into our sedFit code (described in §4.3, with the reddening option for sedFit disabled), the best-fit PHOENIX model spectra from a grid of PHOENIX models were matched to each INGS empirical template. The results in log(g)-$T_{EFF}$ space of fitting of INGS spectra against PHOENIX models can be seen as a Kiel diagram (cf. Fuhrmann 1998) in Figure 2. Comparing the PHOENIX $T_{EFF}$ model values against the Pickles $T_{EFF}$ values, we find an average $T_{EFF}$ difference of -12K (-0.2%) with no particular evident trend against spectral type of those differences. Overall, this approach allows us the density of the PHOENIX grid, while still being based on the empirical INGS grid. Table 15 outlines the INGS-to-PHOENIX mapping, and indicates where spectral types not present in the INGS are mapped onto PHOENIX.

---

[2] https://lco.global/user/apickles/dev/INGS/

[3] http://phoenix.astro.physik.uni-goettingen.de/.

### 4.2. *Photometry*

Execution of the `sedFit` code, and its ability to provide meaningful results on our program stars, was wholly dependent upon providing the code with meaningful photometry. A large number of archival sources were researched for narrow- to broad-band photometry. Meta-references such as SIMBAD[4] (Set of Identifications, Measurements and Bibliography for Astronomical Data; Wenger et al. 2000), the *General Catalogue of Photometric Data*[5] (Mermilliod et al. 1997), and the *Catalog of Infrared Observations* (CIO; Gezari et al. 1993) were useful in locating primary data sources; the sources that provided the largest number of points are noted in Table 4, with the individual data points called out explicitly in Table 21. The meta-references are uniquely empowering in collecting a large number of archival data points; thus, for the sake of traceability and proper attribution credit we were scrupulous in Table 21 to provide the original source reference.

In addition to the archival sources, we engaged in an extensive program of photometric observing at the Lowell 31-inch telescope located at Anderson Mesa. These visible data included broad-band Johnson $UBVRI$, as well as narrow-band photometry taken in the Hale-Bopp ('HB') system of Farnham et al. (2000). The telescope and CCD characteristics for this observing are described in Schleicher & Bair (2011) and Knight & Schleicher (2015), with HB system filters that isolate emission bands in OH, NH, CN, $C_3$ and $C_2$, as well as narrow-band continuum points in the UV, blue, and green.

For best results with the `sedFit` code, we found 3 elements were needed in the photometric input data. First, visible data were necessary to broadly characterize the Wien short-wavelength side of the SED. Second, near-infrared data were essential for characterizing the Rayleigh-Jeans long wavelength side of the SED. Since most of our targets' spectral peaks are located between $3,000 - 6,000$ K, data bracketing the peak at $\lambda \sim 1 \mu m$ were necessary. Third, for resolution of degeneracies between interstellar extinction, spectral types and luminosity classes, narrow-band ($\Delta \lambda \sim$ 60-100 Å) photometric data were particularly helpful and motivated our observing program at the 31-inch.

*A note on V and K-band magnitudes.* Given the extensive use of $V$ and $K$ magnitudes (and their dereddened counterparts, $V_0$ and $K_0$) in the analyses of §7 and 9, it is worthwhile to note in detail the nature of these parameters. Both symbols are intended to denote the standard Johnson bandpasses and zeropoints; for the visible $V$-band (Arp 1958; Johnson & Morgan 1951; Bessell 1990) this is to be expected, but for the near-infrared $K$-band (Mendoza v. 1963; Bessell & Brett 1988) a likely alternative would be $K_s$ found in 2MASS (Cutri et al. 2003a). Unfortunately, for the majority of our stars – which, particularly in the near-infrared were rather bright – we found that the $K_s$ data was saturated, and that use of the $K_s$ data in our `sedFit` process led to poor fits. Since most of our program stars had Johnson $K$-band data present in the CIO, which did not make for problematic `sedFit` fits, we elected to use this variant of $K$-band in our study.

**Table 4**. Sources of photometry for our program stars with $N \geq 20$ data points.

| Reference | System | $N_{\mathrm{POINTS}}$ |
|---|---|---|
| Smith et al. (2004) | COBE DIRBE | 1038 |
| This Work | Farnham-HB | 822 |
| Johnson & Mitchell (1995) | Johnson 13-color | 748 |
| Kornilov et al. (1991) | WBVR | 707 |
| Ducati (2002) | Johnson-IR | 509 |
| McClure & Forrester (1981) | DDO | 507 |
| Häggkvist & Oja (1970) | Oja | 436 |
| Johnson et al. (1966) | Johnson-IR | 396 |
| Golay (1972) | Geneva | 378 |

**Table 4** *continued on next page*

**Table 4** *(continued)*

| Reference | System | $N_{\mathrm{POINTS}}$ |
|---|---|---|
| Mermilliod (1986) | Johnson-visible | 325 |
| Neugebauer & Leighton (1969) | Johnson-IR | 241 |
| Kazlauskas et al. (2005) | Vilnius | 157 |
| Häggkvist & Oja (1966) | Johnson-visible | 150 |
| Argue (1966) | Johnson-visible | 149 |
| Argue (1963) | Johnson-visible | 144 |
| Jennens & Helfer (1975) | Johnson-visible | 105 |
| Hauck & Mermilliod (1998) | Stromgren | 104 |
| Zdanavicius et al. (1969) | Vilnius | 102 |
| Zdanavicius et al. (1972) | Vilnius | 88 |
| Shenavrin et al. (2011) | Johnson-IR | 80 |
| Johnson (1964) | Johnson-IR | 79 |
| Haggkvist & Oja (1970) | Johnson-visible | 73 |
| Mermilliod & Nitschelm (1989) | DDO | 67 |
| Olsen (1993) | Stromgren | 60 |
| Alonso et al. (1998) | Johnson-IR | 55 |
| Jasevicius et al. (1990) | Vilnius | 50 |
| Straizys et al. (1989a) | Vilnius | 49 |
| Gutierrez-Moreno & et al. (1966) | Johnson-visible | 40 |
| Crawford & Barnes (1970) | Stromgren | 40 |
| Nicolet (1978) | Johnson-visible | 35 |
| Johnson & Morgan (1953b) | Johnson-visible | 34 |
| Bartkevicius et al. (1973) | Vilnius | 34 |
| Roman (1955) | Johnson-visible | 32 |
| Fernie (1983) | Johnson-visible | 32 |
| Glass (1974) | Johnson-IR | 29 |
| Sudzius et al. (1970) | Vilnius | 28 |
| Ljunggren & Oja (1965) | Johnson-visible | 26 |
| Lee (1970) | Johnson-IR | 26 |
| Voelcker (1975) | Johnson-IR | 24 |
| McWilliam & Lambert (1984) | Johnson-IR | 24 |
| Gray & Olsen (1991) | Stromgren | 24 |
| Neckel (1974) | Johnson-visible | 22 |
| Straizys & Meistas (1989) | Vilnius | 22 |
| Oja (1991) | Johnson-visible | 22 |
| McClure (1970) | Johnson-visible | 21 |
| Oja (1986) | Johnson-visible | 21 |
| Selby et al. (1988) | Johnson-IR | 21 |
| Forbes et al. (1993) | Vilnius | 21 |
| Laney et al. (2012) | Johnson-IR | 21 |
| Moffett & Barnes (1979) | Johnson-visible | 20 |

**Table 4** *(continued)*

| Reference | System | $N_{\rm POINTS}$ |
|-----------|--------|----------|
| Olsen (1983) | Stromgren | 20 |
| Oja (1984) | Johnson-visible | 20 |

Note—In the *system* column, 'Johnson-visible' is used to denote any passbands from $UBVRI$, and 'Johnson-IR' denotes passbands from $JHKLM$. Additional sources are listed on a star-by-star basis in Table 21. For more details, see Section 4.2.

### 4.3. `sedFit` *Process*

We have previously used the `sedFit` code in similar investigations (van Belle et al. 2007, 2009, 2013, 2016; van Belle & von Braun 2009) for determinations of $F_{\rm BOL}$ and $A_{\rm V}$. The process is rather straightforward: input photometry is matched against an input spectral template via amplitude scaling of that template, optimized through a Marquardt-Levenberg (M-L) least-squares technique. As an option, the overall spectral template can also be optimized for reddening $A_V$. Reddening corrections are based upon the empirical determination by Cardelli et al. (1989), which is only marginally different from van de Hulst's theoretical reddening curve No. 15 (Johnson 1968; Dyck et al. 1996a).

If the input spectral template has an assigned effective temperature estimate, then an angular size prediction can also be produced[6]. A multi-threaded wrapper script enables many ($\sim$ dozens) of input spectral templates to be tested against the same input photometry for establishing which template is most appropriate for that photometry. An important recent improvement to the `sedFit` code has been the ability to incorporate detailed spectral profiles and optimized zeropoints for individual filters of the photometric systems; a number of specific filters have significantly skewed and/or side-weighted profiles (e.g. Johnson $R$-band) which benefit significantly from this improvement. As a result, the detailed empirical characterization of those filters based upon HST NGSL data computed by Mann & von Braun (2015) were included as part of the SED fitting.

The `sedFit` code works as follows. Input photometry (consisting of values and their uncertainties) is read in, matched to its photometric system, and converted if necessary from astronomical magnitudes to flux values using standard values for its photometric system. These points are then scaled in amplitude for comparison against a reference spectral template (e.g. an INGS-linked PHOENIX model). That spectral template can be reddened as noted above, and the whole process converges to an optimum solution using a M-L least-squares technique. Once optimized, the code explores the relevant reduced chi-squared space to establish uncertainties on template amplitude and reddening. Finally, the code computes the indicated bolometric flux via a sum across wavelength of the spectral template. This process can be automatically repeated to explore a grid of templates for the best fit template against a set of photometry, as indicated by the reduced chi-squared value for each template.

A significant element of the `sedFit` process is establishing the correct input spectral template for SED fitting. Our technique for arriving at the optimal template began with literature search for previously published spectral type for a given star; ideally this was reported from at least two sources. These spectral types and original references are given in Table 17.

Following the mapping of that spectral type to PHOENIX models from our analysis in §4.1, a grid of spectral templates in $T_{\rm EFF}$ and $\log g$ about that location was examined for the optimum spectral template. The best fit in $\{T_{\rm EFF}, \log g\}$, and the corresponding spectral type, is also given in Table 17, along with the resultant estimates for $F_{\rm BOL}$ and $A_{\rm V}$. While the range of $A_{\rm V}$ estimates are small – typically a few tenths of a magnitude – corrections at this level are important in attempting to achieve $F_{\rm BOL}$ measurements, and derived $T_{\rm EFF}$ values, at our optimum levels of accuracy and precision. Estimation of appropriate reddening $A_{\rm V}$ for a given object also allowed us to establish dereddened photometric values for $V$ and $K$, as well as the dereddened color $V_0 - K_0$.

The resulting values for $F_{\rm BOL}$ are evenly distributed as a function of source angular size, as seen in Figures 3 and 4; we have used Equation 1 to show 'standard' $F_{\rm BOL}$ versus $\theta$ curves for given values of $T_{\rm EFF}$.

---

[6] Although this is a fine way to produce an angular size *estimate*, such a process does not replace the significant value of actually making an angular size measurement if possible.

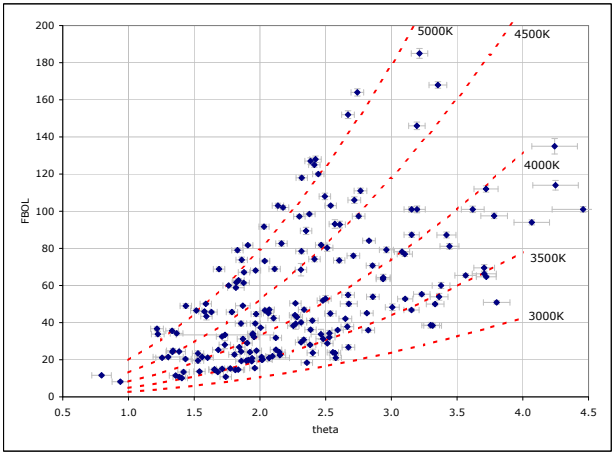

**Figure 3.** Bolometric flux $F_{\mathrm{BOL}}$ (in units of $10^{-8}$ erg s$^{-1}$ cm$^{-2}$) versus angular size $\theta$ (milliarcseconds) for our program stars. Canonical fit lines for fixed values of $T_{\mathrm{EFF}}$ from Equation 1 are shown as dotted lines.

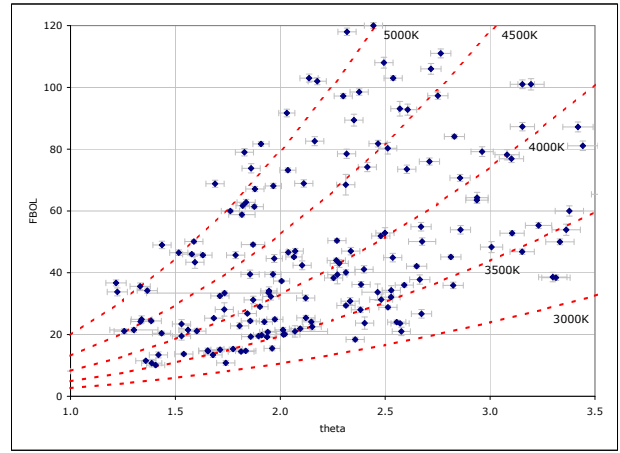

**Figure 4.** As Figure 3, with a zoom on the central region. For more details, see Section 4.3.

The $F_{\mathrm{BOL}}$ estimation process has the following sources of error. First, the observables - the photometry - have inherent measurement error, as well as error in the corrections applied for Earth's atmospheric extinction. Next, we interpret those observables radiometrically, using zeropoint values, which themselves have uncertainty. Finally, we model those radiometric points with a model SED - the PHOENIX models - while applying interstellar extinction corrections. Once the `sedFit` process has determined the optimum template for a given star, the median one-sigma uncertainty reported by the code after exploration of reduced chi-squared space for that optimum template in our $F_{\mathrm{BOL}}$ values is 2.1%. This level of uncertainty is consistent with (a) a limiting photometric (observable) accuracy of 1-2% as seen in Table 21; (b) zeropoint errors, which as seen in Mann & von Braun (2015) have a median error of 0.5%; and (c) the SED modeling with its interstellar extinction corrections. On that latter point, the overall flux levels are set by the agreement between the SED model, with the notable caveat that application of interstellar extinction could potentially lead to spurious results when a mismatched spectral template is overly reddened. To break this degeneracy, we found that narrow-band photometry was particularly useful in ensuring that the correct template was utilized for a given star. Specifically, when using narrow-band photometry, we found that the reduced chi-squared values for spectral templates offset from the optimum template increased on average by a value of $\Delta\chi^2_{red} = 1.4$; for wide-band photometry alone, this value was less than a half (ie. not statistically significant).

Taking these elements into account, the SED fitting process established the estimates of uncertainty for $F_{\mathrm{BOL}}$ and $A_{\mathrm{V}}$ simultaneously through the appropriate exploration of $\chi^2$ space; the median uncertainty in $A_{\mathrm{V}}$ of 2% is consistent with our median $F_{\mathrm{BOL}}$ uncertainty. Finally, one minor potential source of systematic error could result from our spectral type mapping exercise of §4.1, with disagreement between the Pickles templates and PHOENIX models. However, the matched Pickles-PHOENIX pairs showed disagreement in overall bolometric flux at only the ∼0.1% level, well below the general level of uncertainties in the $F_{\mathrm{BOL}}$ estimates.

## 5. EFFECTIVE TEMPERATURES

### 5.1. *Limb Darkening Corrections*

To convert the wavelength-specific uniform disk (UD) angular size measurements from PTI into more general limb-darkened (LD) Rosseland mean angular diameters, we applied the standard procedure of estimating a multiplicative factor for converting the $K$-band UD diameters to LD diameters. One advantage of the $K$-band PTI data is this correction is small (on the order 2-6%) relative to corrections for observations at shorter wavelengths; systematic uncertainties, if any, that exist in these corrections will therefore have a smaller impact on the results. For example, the visible-light study of Mozurkewich et al. (2003) had UD-to-LD factors that averaged ∼9% and went as high as ∼20%.

Previous estimates of such conversion factors by Davis et al. (2000, DTB00) are now superseded by those in Neilson & Lester (2013, NL13), the latter of which we used in our investigation. The principal improvement between these two investigations is a transition from plane-parallel to spherically-symmetric model atmospheres. NL13 also report results from a plane-parallel evaluation, which is similar to DTB00, with a slight decrease in the magnitude of the correction. The spherically-symmetric results for the stars hotter than >4000K remain similar to the two plane-parallel results; but for stars cooler than 4000K, the UD-to-LD correction trends upwards from ~1.03 to ~1.06. This is consistent with the expectation that, for these cooler extended atmospheres, the spherically-symmetric model atmospheres start to deviate from plane-parallel models and improve predictions of the wavelength-dependence of the structure (Scholz & Takeda 1987; Hofmann & Scholz 1998). Values taken from NL13 were for 2.5 $M_\odot$ stars (out of a choice of 1.0, 2.5, and 5.0 $M_\odot$ stars) and were interpolated between NL13 temperature data point spacing of 100K for use in our study; see the final column of Table 5. As demonstrated in §8.2, this mass range is a reasonable choice for our program stars.

To quantify the numerical impact of these corrections, if we were not to utilize them: imposing a uniform UD-to-LD conversion value of 1.026 (the average of the DTB00 values in Table 5 for our typical program star $T_{EFF}$ range of 3000K to 5200K) results in a shift downwards in $T_{EFF}$ for the hottest stars in the range by -4K, and a shift downwards for the cooler stars by -60K, relative to our final adopted answer using the NL13-fit values. Quantifying the impact of uncertainties in this correction factor: assuming the correction factors presented in Table 5 have an uncertainty of ±0.01, the final $T_{EFF}$ values presented in Table 18 would be impacted by ±20K, well below the errors in those values from other sources.

**Table 5.** Multiplicative correction for converting 2.2$\mu$m uniform disk angular sizes to Roseland limb darkened angular sizes, based upon spherically-symmetric model atmospheres analysis in Neilson & Lester (2013) (NL13); comparable values for a plane-parallel atmospheres analyses in Davis et al. (2000) (DTB00) and NL13 are also given.

| $T_{EFF}$ | DTB00 | NL13 | NL13 | NL13-fit |
| (K) | (plane) | (plane) | (spherical) | (spherical) |
| --- | --- | --- | --- | --- |
| 3000 | ⋯ | 1.025 | 1.061 | 1.065 |
| 3100 | ⋯ | 1.022 | 1.066 | 1.060 |
| 3200 | ⋯ | 1.024 | 1.050 | 1.057 |
| 3400 | ⋯ | 1.022 | 1.055 | 1.050 |
| 3500 | 1.030 | 1.021 | 1.056 | 1.047 |
| 3600 | 1.030 | 1.021 | 1.043 | 1.045 |
| 3700 | 1.029 | 1.020 | 1.043 | 1.042 |
| 3800 | 1.029 | 1.021 | 1.036 | 1.040 |
| 3900 | 1.029 | 1.019 | 1.035 | 1.038 |
| 4000 | 1.028 | 1.019 | 1.035 | 1.037 |
| 4100 | 1.027 | 1.019 | 1.035 | 1.035 |
| 4200 | 1.027 | 1.018 | 1.034 | 1.034 |
| 4300 | 1.026 | 1.018 | 1.034 | 1.033 |
| 4400 | 1.025 | 1.018 | 1.033 | 1.032 |
| 4500 | 1.025 | 1.017 | 1.033 | 1.031 |
| 4600 | 1.024 | 1.017 | 1.028 | 1.030 |
| 4700 | 1.024 | 1.017 | 1.028 | 1.029 |
| 4800 | 1.023 | 1.017 | 1.028 | 1.029 |
| 4900 | 1.023 | 1.016 | 1.027 | 1.028 |
| 5000 | 1.022 | 1.016 | 1.027 | 1.027 |
| 5100 | 1.021 | 1.015 | 1.027 | 1.027 |

**Table 5** *(continued)*

| $T_{\rm EFF}$ | DTB00 | NL13 | NL13 | NL13-fit |
|---|---|---|---|---|
| (K) | (plane) | (plane) | (spherical) | (spherical) |
| 5200 | 1.020 | 1.015 | 1.027 | 1.026 |
| 5300 | 1.020 | 1.015 | 1.027 | 1.026 |
| 5400 | 1.019 | 1.015 | 1.027 | 1.026 |
| 5500 | 1.019 | 1.014 | 1.027 | 1.025 |
| 5600 | 1.019 | 1.014 | 1.027 | 1.025 |
| 5700 | 1.019 | 1.013 | 1.027 | 1.024 |
| 5800 | 1.019 | 1.013 | 1.023 | 1.024 |
| 5900 | 1.019 | 1.013 | 1.023 | 1.024 |
| 6000 | 1.019 | 1.013 | 1.022 | 1.023 |
| 6100 | $\cdots$ | 1.012 | 1.022 | 1.023 |
| 6200 | $\cdots$ | 1.012 | 1.022 | 1.022 |
| 6300 | $\cdots$ | 1.012 | 1.021 | 1.022 |
| 6400 | $\cdots$ | 1.012 | 1.021 | 1.022 |
| 6800 | $\cdots$ | 1.010 | 1.020 | 1.020 |
| 7400 | $\cdots$ | 1.009 | 1.019 | 1.019 |
| 7500 | $\cdots$ | 1.009 | 1.020 | 1.019 |
| 7600 | $\cdots$ | 1.009 | 1.020 | 1.019 |
| 7700 | $\cdots$ | 1.009 | 1.020 | 1.019 |
| 7800 | $\cdots$ | 1.009 | 1.020 | 1.019 |
| 7900 | $\cdots$ | 1.009 | 1.020 | 1.020 |
| 8000 | $\cdots$ | 1.009 | 1.020 | 1.020 |

Note—For more detail, see Section 5.1.

## 5.2. *Effective Temperature Computations*

Incorporating the limb darkening corrections of the previous subsection, the final values for effective temperature can be computed, as seen in Table 18. The uniform disk angular sizes as measured by PTI are given, with their formal error and systematic errors (as noted in Table 3). An initial estimate of $T_{\rm EFF}$ is then computed using angular sizes and the computed bolometric fluxes (Table 17) in the standard relationship:

$$T_{\rm EFF} = 2341 \times \left[\frac{F_{\rm BOL}}{\theta^2}\right]^{1/4} \tag{1}$$

where the units of $F_{\rm BOL}$ are $10^{-8}$ erg s$^{-1}$ cm$^{-2}$, and $\theta$ is in milliarcseconds. Equation 1 is used with $\theta_{\rm UD}$ for an initial estimate of $T_{\rm EFF}$ to select the UD-to-LD correction, which is consequently applied to $\theta_{\rm UD}$ to obtain limb-darkened Rosseland mean diameter $\theta_{\rm R}$. Additional iterations on the $T_{\rm EFF}$-to-correction step are not necessary, since this would have the effect of revising $T_{\rm EFF}$ further by $10^{-4}$, which is far below our measurement error. With $\theta_{\rm R}$ and $F_{\rm BOL}$ in Equation 1, a final value for $T_{\rm EFF}$ is then computed, and presented in the final column of Table 18.

The median uncertainty of our $T_{\rm EFF}$ values is 49K (1.25%). Of this error, the principal contribution is error in $\theta_{\rm UD}$; if there were no uncertainty in $F_{\rm BOL}$, the total random error in $T_{\rm EFF}$ values is still 42K (1.00%). Conversely, if $\theta_{\rm UD}$ uncertainty were eliminated, the total random error in $T_{\rm EFF}$ values would be 22K (0.55%).

## 6. DISTANCES & LINEAR RADII

The combination of measured angular diameter with measured trigonometric parallax value trivially yields an empirical estimate of linear stellar radius $R$:

$$R = \frac{\theta}{0.009292 \times \pi} R_{\odot} \tag{2}$$

where $\pi$ is parallax and $\theta$ is angular size, both in in milliarcseconds; $R$ is given in units of solar radius, calibrated to $R_\odot = 6.957 \times 10^8$ m (Brown & Christensen-Dalsgaard 1998; Haberreiter et al. 2008). A significant number of our targets have *Gaia* magnitudes of brighter than 6, for which calibration issues exist in the astrometry in Data Release 2 in April 2018 (Lindegren et al. 2018). To calculate distances and consequently linear radii of our targets, we thus preferentially utilize *Hipparcos* parallax values (van Leeuwen 2007) instead, resorting to *Gaia* DR2 for dimmer objects, and those not present in the *Hipparcos* catalog. The parallax values we employed, source, and resulting $R$ values are in Table 19.

Regardless of source of distance data, a significant fraction of the targets, however, are sufficiently far away for distance estimates to carry very large associated uncertainties. By and large, the influence of distance uncertainties on the error budget for linear diameter estimates is greater than that of the measured angular diameters for the stars in our sample, quite unlike for effective temperatures (§5.2). We list target distances and calculated linear radii in Table 19 along with associated uncertainties. The parallax zero-point effects reported for *Gaia* at $\overline{\omega}_0 = -48 \pm 1\mu$as (Chan & Bovy 2020) DR2 are below the noise threshold for our stars. (Applying this offset to the parallaxes for our program stars that utilize DR2 has the net effect of reducing their sizes by an average of ∼1.1%.)

It is also worth noting that, while these stars are not as severely affected with spotting phenomenology as Mira variables, they will suffer similarly in ways that affect distance measures, particularly on the red end. These reddest, coolest stars in our sample are expected to have increasingly large convection cells (Chiavassa 2018), up to > 10 − 20% the radius of the star. Coincidentally, for a given angular size, the characteristic distance of such stars in our instrumentation is such that the magnitude of the parallactic shift can be on order the size of the stellar disk (e.g. compare the values in Tables 18 and 19). Given that the stellar photocenter – the principal observable for parallax measures – is shifting by such spots by a significant fraction of disk, on timescales similar to parallax measurement epochs, the impact here is in an unfortunate confluence of factors that operate in a correlated fashion to degrade parallax measures. This effect is more pronounced in the visible (where *Hipparcos* and *Gaia* work) than in the near-infrared. (See the detailed discussion in §3.5 of van Belle et al. 2002).

Overall the median uncertainty in linear radius for our program stars is 3.7%, of which the greater contribution is parallax error. If the uncertainty from the parallaxes for our program stars were zero, the average uncertainty in linear radius for our program stars would drop to 2.0%; likewise, if angular size uncertainty were zero, linear radius uncertainty would drop to 2.5%.

## 7. TEMPERATURE AND RADIUS VERSUS $V_0 - K_0$ AND SPECTRAL TYPE

We shall first present the *directly* measured fundamental stellar parameters of stellar effective temperature ($T_{\rm EFF}$) and linear radius ($R$), as functions of $V_0 - K_0$ and spectral type indices. Examples of expanded analyses enabled by these measurements follow in the next section, §8.

### 7.1. $T_{\rm EFF}$ versus $V_0 - K_0$

Our $T_{\rm EFF}$ results as indexed by $V_0 - K_0$ color are seen in Figure 5 with a zoom in on the central results in the range of $V_0 - K_0 = \{2, 6\}$ in Figure 6. $T_{\rm EFF}$, unsurprisingly, is tightly correlated with this color. The general progression of $T_{\rm EFF}$ with $V_0 - K_0$ is summarized in Table 6, wherein we step through our sample in steps of $\Delta V_0 - K_0 = 0.1$, increasing the bin sizes on the red end to account of increasing sparsity of data. For this table we estimate the unbiased sample variance taking into account the varying weights $\{\sigma_{V_0-K_0}, \sigma_{T_{\rm EFF}}\}$ for the parameters $\{V_0 - K_0, T_{\rm EFF}\}$:

$$s^2 = \frac{\sum_{i=1}^{N} w_i (x_i - \mu^*)^2}{V_1 - (V_2/V_1)} \tag{3}$$

where $\mu^*$ is the weighted mean for each of these parameters, $w_i = 1/\sigma^2$, $V_1 = \sum_{i=1}^{N} w_i$, and $V_2 = \sum_{i=1}^{N} w_i^2$. Given the occasional large spread in measurement error of $T_{\rm EFF}$ for a given star as seen in Table 18, an approach such as this was necessary to give proper consideration to the varying weights of each pair of data points in determining the bin weight (Kendall et al. 1987). For the data in Table 6, the median $T_{\rm EFF}$ uncertainty is 66K per bin.

#### 7.1.1. Gap Analysis: $T_{\rm EFF}$ versus $V_0 - K_0$

A striking characteristic of the $T_{\rm EFF}$ versus $V_0 - K_0$ plots (Figures 5 and 6) that merited further investigation were the apparent gaps in the continuum of data points, in the region of densest sampling of $V_0 - K_0$ color space, between $V_0 - K_0 = \{1.9, 6.5\}$. Two such gaps are apparent to visual inspection, located at at $\{T_{\rm EFF}, V_0 - K_0\} =$

**Table 6.** $T_{\rm EFF}$ versus $V_0 - K_0$, binned.

| Bin Center | Bin Width | Bin Bounds | $N$ | $V_0 - K_0$ (mag) | $T_{\rm EFF}$ (K) |
|---|---|---|---|---|---|
| 2.0 | 0.1 | {1.95, 2.05} | 4 | $2.013 \pm 0.032$ | $5043 \pm 73$ |
| 2.1 | 0.1 | {2.05, 2.15} | 12 | $2.105 \pm 0.026$ | $5007 \pm 57$ |
| 2.2 | 0.1 | {2.15, 2.25} | 10 | $2.210 \pm 0.033$ | $4896 \pm 115$ |
| 2.3 | 0.1 | {2.25, 2.35} | 4 | $2.281 \pm 0.022$ | $4762 \pm 62$ |
| 2.4 | 0.1 | {2.35, 2.45} | 7 | $2.400 \pm 0.031$ | $4776 \pm 88$ |
| 2.5 | 0.1 | {2.45, 2.55} | 5 | $2.490 \pm 0.010$ | $4695 \pm 81$ |
| 2.6 | 0.1 | {2.55, 2.65} | 5 | $2.575 \pm 0.016$ | $4580 \pm 61$ |
| 2.7 | 0.1 | {2.65, 2.75} | 3 | $2.686 \pm 0.023$ | $4490 \pm 17$ |
| 2.8 | 0.1 | {2.75, 2.85} | 4 | $2.770 \pm 0.010$ | $4492 \pm 44$ |
| 2.9 | 0.1 | {2.85, 2.95} | 7 | $2.882 \pm 0.023$ | $4408 \pm 64$ |
| 3.0 | 0.1 | {2.95, 3.05} | 2 | $3.025 \pm 0.020$ | $4438 \pm 37$ |
| 3.1 | 0.1 | {3.05, 3.15} | 3 | $3.125 \pm 0.019$ | $4183 \pm 40$ |
| 3.2 | 0.1 | {3.15, 3.25} | 5 | $3.199 \pm 0.036$ | $4197 \pm 115$ |
| 3.3 | 0.1 | {3.25, 3.35} | 6 | $3.304 \pm 0.030$ | $4162 \pm 64$ |
| 3.4 | 0.1 | {3.35, 3.45} | 5 | $3.433 \pm 0.009$ | $4040 \pm 68$ |
| 3.5 | 0.1 | {3.45, 3.55} | 7 | $3.505 \pm 0.025$ | $3959 \pm 56$ |
| 3.6 | 0.1 | {3.55, 3.65} | 4 | $3.623 \pm 0.030$ | $3951 \pm 57$ |
| 3.7 | 0.1 | {3.65, 3.75} | 3 | $3.704 \pm 0.030$ | $3887 \pm 45$ |
| 3.8 | 0.1 | {3.75, 3.85} | 5 | $3.804 \pm 0.025$ | $3941 \pm 102$ |
| 3.9 | 0.1 | {3.85, 3.95} | 4 | $3.876 \pm 0.014$ | $3852 \pm 52$ |
| 4.0 | 0.1 | {3.95, 4.05} | 3 | $3.984 \pm 0.038$ | $3840 \pm 49$ |
| 4.1 | 0.1 | {4.05, 4.15} | 3 | $4.091 \pm 0.044$ | $3751 \pm 131$ |
| 4.2 | 0.1 | {4.15, 4.25} | 2 | $4.183 \pm 0.019$ | $3809 \pm 129$ |
| 4.3 | 0.1 | {4.25, 4.35} | 6 | $4.279 \pm 0.019$ | $3728 \pm 69$ |
| 4.4 | 0.1 | {4.35, 4.45} | 7 | $4.399 \pm 0.024$ | $3707 \pm 98$ |
| 4.5 | 0.1 | {4.45, 4.55} | 2 | $4.462 \pm 0.012$ | $3685 \pm 233$ |
| 4.6 | 0.1 | {4.55, 4.65} | 0 | | |
| 4.7 | 0.1 | {4.65, 4.75} | 4 | $4.699 \pm 0.037$ | $3547 \pm 68$ |
| 4.8 | 0.1 | {4.75, 4.85} | 3 | $4.818 \pm 0.027$ | $3540 \pm 140$ |
| 4.9 | 0.1 | {4.85, 4.95} | 6 | $4.911 \pm 0.039$ | $3565 \pm 113$ |
| 5.0 | 0.1 | {4.95, 5.05} | 6 | $4.988 \pm 0.034$ | $3486 \pm 35$ |
| 5.1 | 0.1 | {5.05, 5.15} | 4 | $5.111 \pm 0.045$ | $3493 \pm 20$ |
| 5.2 | 0.1 | {5.15, 5.25} | 5 | $5.202 \pm 0.029$ | $3447 \pm 69$ |
| 5.3 | 0.1 | {5.25, 5.35} | 4 | $5.289 \pm 0.035$ | $3411 \pm 45$ |
| 5.4 | 0.1 | {5.35, 5.45} | 4 | $5.406 \pm 0.034$ | $3439 \pm 68$ |
| 5.5 | 0.15 | {5.45, 5.60} | 4 | $5.490 \pm 0.033$ | $3469 \pm 18$ |
| 5.7 | 0.2 | {5.60, 5.80} | 3 | $5.650 \pm 0.025$ | $3414 \pm 53$ |
| 5.9 | 0.2 | {5.80, 6.00} | 4 | $5.858 \pm 0.024$ | $3241 \pm 117$ |
| 6.1 | 0.2 | {6.00, 6.20} | 4 | $6.113 \pm 0.042$ | $3363 \pm 46$ |
| 6.3 | 0.2 | {6.20, 6.40} | 2 | $6.213 \pm 0.005$ | $3296 \pm 67$ |
| 6.6 | 0.35 | {6.40, 6.75} | 2 | $6.460 \pm 0.051$ | $3225 \pm 162$ |
| 7.0 | 0.5 | {6.75, 7.25} | 1 | | |
| 7.5 | 0.5 | {7.25, 7.75} | 2 | $7.656 \pm 0.030$ | $3117 \pm 4$ |
| 8.0 | 0.5 | {7.75, 8.25} | 2 | $8.062 \pm 0.219$ | $3110 \pm 70$ |
| 8.5 | 0.5 | {8.25, 8.75} | 2 | $8.565 \pm 0.013$ | $3090 \pm 66$ |

Note—For more detail, see Section 7.1 and Figures 5 and 6.

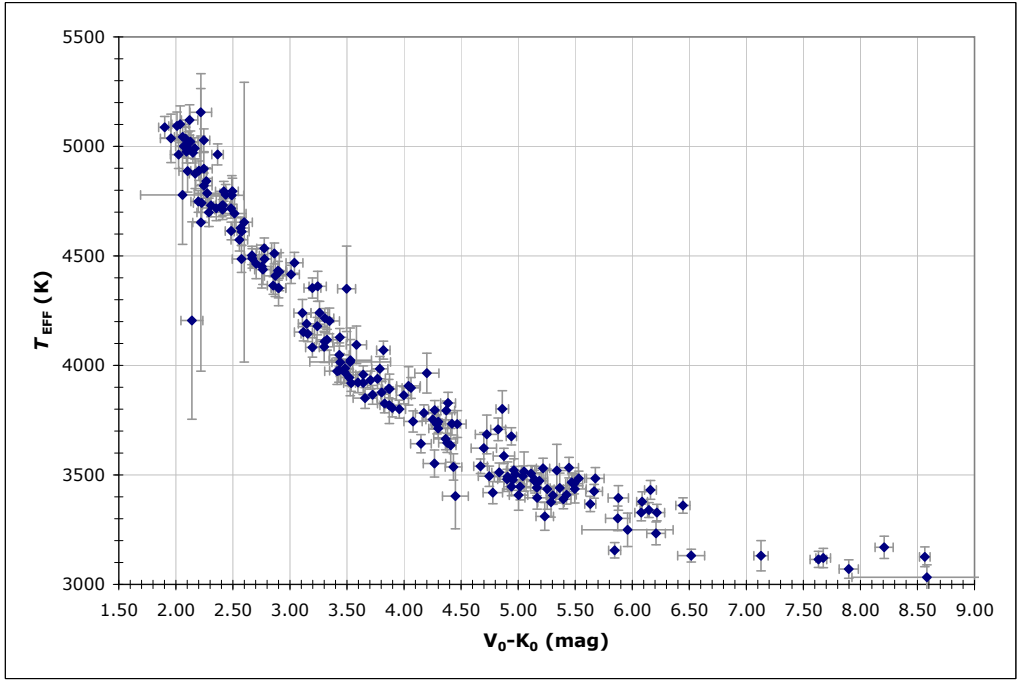

**Figure 5.** Effective temperature $T_{\mathrm{EFF}}$ versus dereddened V-K color, $V_0 - K_0$. The corresponding data are shown in Table 6.

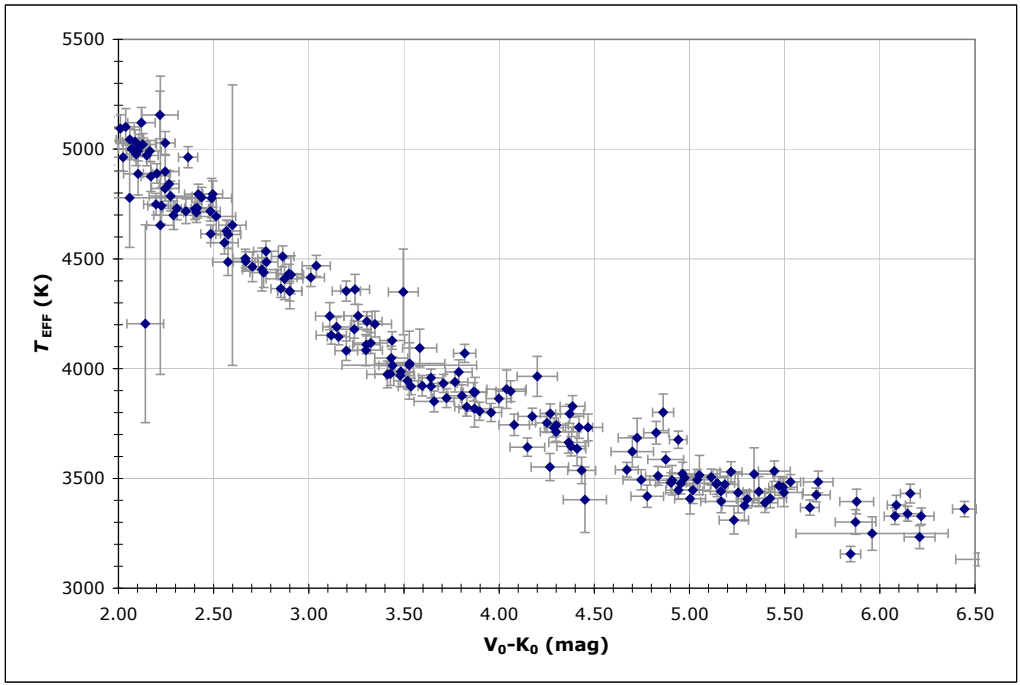

**Figure 6.** Effective temperature versus dereddened V-K color, $V_0 - K_0$, zoomed into the principal range of our colors, $V_0 - K_0$ = {2.0, 6.5}. The corresponding data are shown in Table 6. Notable gaps are seen at $V_0 - K_0 = 3.0$ and 4.6 and are discussed in §TBD.

{4350K, 3.10mag} and = {3600K, 4.55mag}. However, to test our intuition in this regard, we developed a Monte Carlo test to examine the likelihood that these gaps in our data were simply the product of random fluctuation.

The initial step in this analysis was to collect the observational data in the region of interest, $V_0 - K_0 = \{1.9, 6.5\}$, which encompasses of a subsample of $N = 183$ of our stars. A polynomial fit of $T_{EFF}$ versus $V_0 - K_0$ gave the following result:

$$T_{EFF} = -9.301 \times (V_0 - K_0)^3 + 212.4 \times (V_0 - K_0)^2 - 1648 \times (V_0 - K_0) + 7602 \qquad (4)$$

For our subsample in this range of interest, the median individual measurement errors of our observational data are $\sigma_{T_{EFF}} = 49$ K and $\sigma_{V_0 - K_0} = 0.07$ mag; the median absolute difference between the individual $T_{EFF}$ measurements and the fit in Equation 4 is 50K, which agrees well with our measurement error.

Given these data, we were able to construct synthetic populations of observational data points in $\{T_{EFF}, V_0 - K_0\}$ space. Each population consisted of the same number of observations ($N$=183), and randomly generated values for $V_0 - K_0$ were generated in the range of $\{1.9, 6.5\}$ from a uniform distribution, with a corresponding $T_{EFF}$ value determined from Eqn. 4. Each synthetic data points was then scattered by a typical measurement error, by random generation of a $\sigma_{T_{EFF}}$ from a normal distribution of width equal to the median error of the true observational data (49K), which was then added to the synthetic data point. An identical step added a measurement error to the synthetic data point's $V_0 - K_0$ value. Each of the synthetic data points was then assigned values for their measurement error equal to the median errors in our true observational data.

Once constructed, these synthetic populations could be 'gap tested' by scanning a test particle along the Eqn. 4 fit line in steps of $\Delta V_0 - K_0 = 0.01$ mag, in the range of interest, $\{1.9, 6.5\}$. At each step, a region around the test particle would be examined for the presence of data points from the test population; if at any point along the scan region, a test particle's examination region was found to be free of data points, that population would be considered to have a gap. For sufficiently small scan regions, nearly all populations would be found to have gaps; for sufficiently large regions, all populations would fail that test. (It is worthwhile to note that this test broadly seeks for gaps *anywhere* along the population's $V_0 - K_0$ sequence, not just at specific locations.) The statistics of gap likelihood for various region sizes can then be built by running a large number of synthetic populations; for our investigation, we found $N_{POP} = 1,000$ was a reasonable size for each test run.

*Gap 1 testing.* Examination of the first gap that we noted in our data at $\{T_{EFF}, V_0 - K_0\} = \{4350K, 3.10mag\}$ shows a gap of size $\Delta T_{EFF} = \pm 88$ K and $\Delta V_0 - K_0 = \pm 0.15$ mag. Taking that as our test scan region envelope, we find that a run of $N_{POP} = 1,000$ synthetic populations turns up only 53 as having gaps of this size anywhere along the range of $V_0 - K_0$ values.

*Gap 2 testing.* Examination of the second gap that we noted in our data at $\{T_{EFF}, V_0 - K_0\} = \{3500K, 4.55mag\}$ shows a gap of size $\Delta T_{EFF} = \pm 400$ K and $\Delta V_0 - K_0 = \pm 0.10$ mag. For this sample, we find that a $N_{POP} = 1,000$ synthetic population run indicates only 70 as having gaps of this size.

*'Control' run.* Finally, as a check on the overall likelihood of being able to conceal a data point of 'typical' size in the sample, we tested our process against a a gap matched in size to the median errors in both $T_{EFF}$ and $V_0 - K_0$ - namely, $\Delta T_{EFF} = \pm 49$ K and $\Delta V_0 - K_0 = \pm 0.07$ mag. Interestingly, this only slightly smaller scan region has only 1 of 1,000 synthetic populations as *not* having gaps.

Overall, we took these results from our first and second gap analyses as motivating evidence that one or more astrophysical phenomenon is quite likely causing these overall discontinuities in what would otherwise be a continuum of points in $\{T_{EFF}, V_0 - K_0\}$ space.

Gaps being noted in photometric color sequences are not a new phenomenon. Böhm-Vitense (1970b,a, 1981, 1982) noted that the onset of surface convection in hotter stars ($T_{EFF} \sim 7250$ K) would result in a discontinuity in $M_V$ versus $B - V$ data, given how the temperature differences between convective and radiative layers were expected to affect $B$-band filters. Evidence for detection of these gaps was initially sparse (Böhm-Vitense & Canterna 1974; Rachford & Canterna 2000), and their existence disputed (Mazzei & Pigatto 1988; Newberg & Yanny 1998) until gaps in Hipparcos $M_V$ versus $B - V$ data were noted by de Bruijne et al. (2000, 2001). Böhm-Vitense (1995a,b) further predicted a second gap at $T_{EFF} \sim 6400$ K, associated with a "a sudden increase in convection zone depths", for which evidence was found in the de Bruijne et al. (2001) investigation.

Recently the detection of a possible gap in *Gaia* data for $M_G$ versus $G_{BP} - G_{RP}$ for cooler stars was noted by Jao et al. (2018), who attribute it to the luminosity-temperature regime where M-dwarf stars transition from partially to fully convective (roughly M3.0V). This detection is particularly relevant to our finding here, in that the second of our gaps – at $\{T_{EFF}, V_0 - K_0\} = \{3500K, 4.55mag\}$ – correspond well with the gap detected by Jao et al.

**Table 7.** Spectral type as a function of effective temperature for our program stars

| Sp. Type | Sp. Type Num | $N$ | $T_{\mathrm{EFF}}$ | $\sigma_T$ | $T_{\mathrm{FIT}}$ | $\chi^2$/DOF |
|----------|--------------|-----|--------|------------|--------|-----------|
| G0III    | 50    | 1  | 4963 | 0   | 0    | 0.00 |
| G1III    | 51    | 1  | 5120 | 0   | 5166 | ⋯    |
| G3III    | 53    | 2  | 5062 | 24  | 5061 | ⋯    |
| G4III    | 54    | 6  | 5023 | 46  | 5008 | 0.11 |
| G5III    | 55    | 13 | 4963 | 71  | 4955 | 0.01 |
| G8III    | 58    | 11 | 4781 | 122 | 4797 | 0.02 |
| K0III    | 60    | 8  | 4715 | 43  | 4692 | 0.30 |
| K1III    | 61    | 4  | 4634 | 107 | 4639 | ⋯    |
| K1.25III | 61.25 | 5  | 4528 | 47  | 4537 | 0.03 |
| K1.5III  | 61.5  | 9  | 4424 | 80  | 4487 | 0.62 |
| K2III    | 62    | 4  | 4387 | 32  | 4387 | ⋯    |
| K3III    | 63    | 1  | 4452 | 0   | 4188 | ⋯    |
| K3.25III | 63.25 | 10 | 4183 | 59  | 4138 | 0.61 |
| K3.5III  | 63.5  | 8  | 4063 | 98  | 4088 | 0.06 |
| K4III    | 64    | 8  | 3951 | 63  | 3988 | 0.35 |
| K5III    | 65    | 18 | 3911 | 54  | 3902 | 0.03 |
| M0III    | 66    |    |      |     | *3816* |    |
| M1III    | 67    | 13 | 3777 | 102 | 3730 | 0.21 |
| M2III    | 68    | 10 | 3617 | 92  | 3644 | 0.09 |
| M3III    | 69    | 8  | 3559 | 120 | 3558 | ⋯    |
| M4III    | 70    | 27 | 3466 | 114 | 3472 | ⋯    |
| M5III    | 71    | 14 | 3381 | 69  | 3386 | 0.01 |
| M5.5III  | 71.5  | 4  | 3301 | 147 | 3343 | 0.08 |
| M6III    | 72    | 2  | 3112 | 29  | 3134 | 0.60 |
| M7III    | 73    | 1  | 3114 | 0   | 3134 | ⋯    |
| M7.5III  | 73.5  | 1  | 3121 | 0   | 3134 | ⋯    |
| M7.75III | 73.75 | 2  | 3145 | 22  | 3134 | 0.27 |

NOTE—Spectral types of our program stars as determined by our SED fitting approach as a function of effective temperatures for our program stars, with italicized values for where interpolation provided a $T_{\mathrm{EFF}}$ value. For more detail, see Section 7.2.

## 7.2. $T_{\mathrm{EFF}}$ versus Spectral Type

For examining the relationship between $T_{\mathrm{EFF}}$ and spectral type, we took two approaches to arranging the data. First, sorting by straight spectral type, and second, by sorting by $V_0 - K_0$ color, with an analysis of the resulting relationship seen between $T_{\mathrm{EFF}}$ and spectral type in each approach.

### 7.2.1. Sorted into spectral type bins

Although stellar spectral typing may at times be considered to be subjective (e.g. different practitioners can arrive a different spectral types of individual stars, as discussed in §4.3 and seen in the range of results for individual stars in Skiff 2014a), it remains a useful shorthand for quickly referencing stellar properties. We have collected the general $T_{\mathrm{EFF}}$ properties using the spectral type values assigned during the sedFit process in §4.3. The raw $T_{\mathrm{EFF}}$ versus spectral type values, and general trends, can be seen in Figure 7.

For each spectral subtype, the weighted average $T_{\mathrm{EFF}}$ and weighted variance $\sigma_{\mathrm{T}}$ (as described in §7.1) was computed; these values are seen in Table 7 and Figure 8. A simple linear fit ( $T_{\mathrm{EFF}} = a \times ST + b$ ) was applied to the data

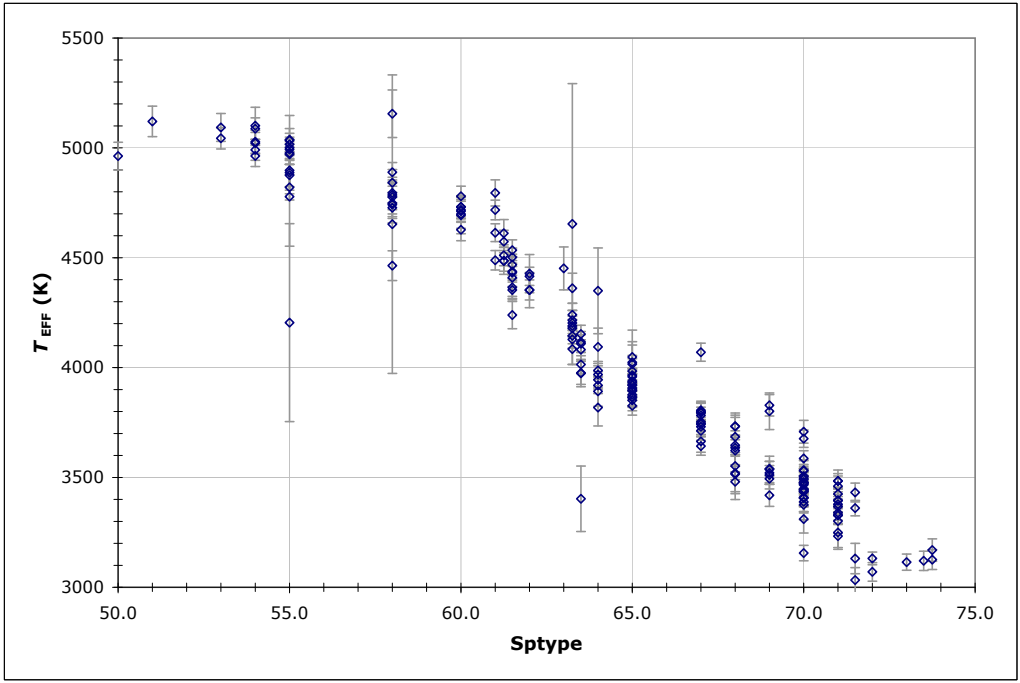

**Figure 7.** Effective temperature $T_{\rm EFF}$ versus spectral type. For more details, see Section 7.2.

**Table 8.** Linear fits of $T_{\rm EFF}$ versus spectral type for specific spectral type ranges.

| Spectral Type range | Slope $a$ | Intercept $b$ | $\chi^2$ | Median $\Delta T$ (K) |
|---|---|---|---|---|
| 51 – 61 | -52.74 | 7856 | 0.44 | 52 |
| 61 – 64 | -199.41 | 16751 | 1.68 | 53 |
| 64 – 71.5 | -85.98 | 9491 | 0.77 | 52 |
| 72 – 74 | 0.00 | 3134 | 0.87 | 28 |

NOTE—These fits are plotted in Figure 8. Spectral type indices range from G0 = 50, K0 = 60, M0 = 66. Median $\Delta T$ is the median average difference between the linear fit and the individual data points in the spectral type range. For more details, see Section 7.2.

seen in Figure 7 and can be seen as well in Figure 8; the fit values are presented in Table 8. Minimization of the $\chi^2$ metric was used not only to optimize the line fits, but also to correctly select when to bracket spectral type ranges. A continuity requirement was enforced at spectral type values that joined two linear fit regions.

### 7.2.2. Sorted by $V_0 - K_0$ color into bins of 5

As a secondary check against the possible subjectivity of spectral typing in exploring the $T_{\rm EFF}$ versus spectral type trends seen in Figures 7 and 8, we explored arranging our spectral type data in the following alternative way. First, we sorted the stars by $V_0 - K_0$ color, and then binned the data into bins of 5. Once done, we then determined the average spectral type and weighted average $T_{\rm EFF}$ values for each bin of 5 stars. These data can be seen in Table 9 and Figure 9, and qualitatively duplicate the results seen in Figures 7 and 8. The principal difference between the latter Figure and the first two is a measure of spectral type 'smearing' in each of the $V_0 - K_0$ bins (e.g. the bin widths in spectral type are 1-2 subtypes), illustrating that spectral type, $T_{EFF}$, and $V_0 - K_0$ color do not uniquely map in 3D space for these stars.

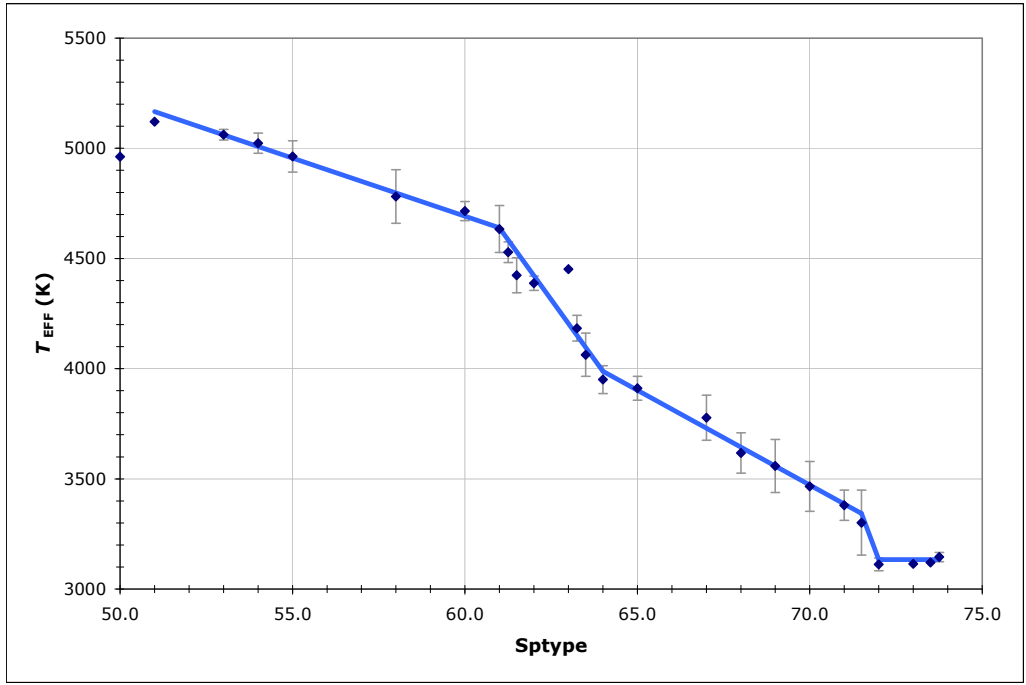

**Figure 8.** Effective temperature $T_{EFF}$ versus spectral type, binned by spectral type with weighted averages, and weighted variance. For more details, see Section 7.2.

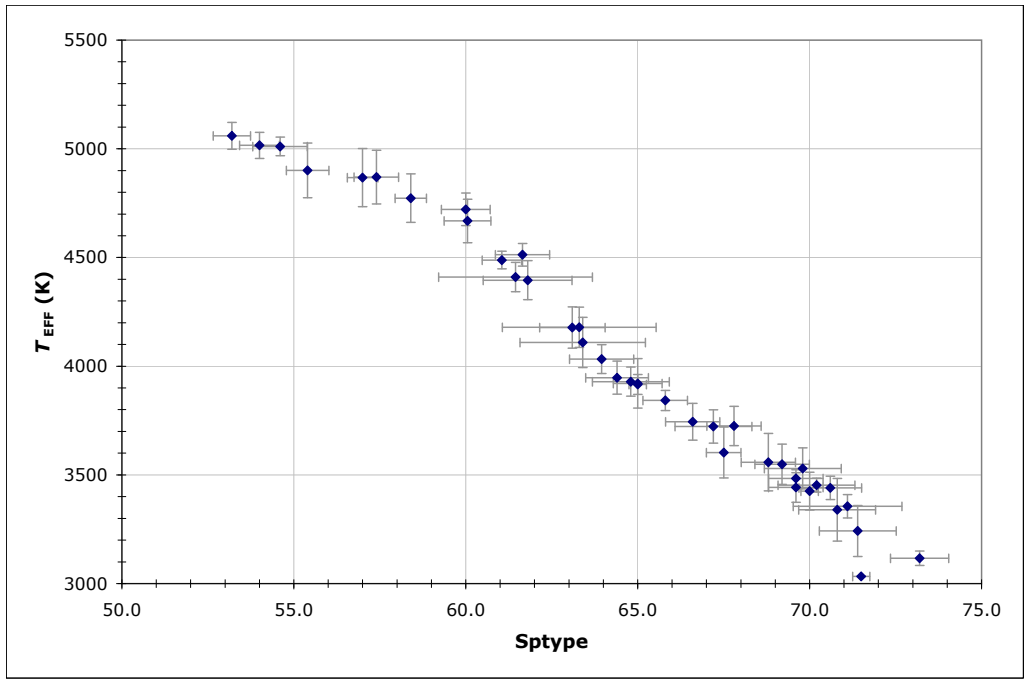

**Figure 9.** Effective temperature $T_{EFF}$ versus spectral type, with our $T_{EFF}$ results sorted by $V_0 - K_0$ and then collected into bins of $N = 5$. For more details, see Section 7.2 and Table 9.

7.2.3. *Calibration of Spectral Type versus $V_0 - K_0$*

**Table 9.** $T_{\mathrm{EFF}}$ versus spectral type

| $V_0 - K_0$ | Sp. Type Num | $T_{\mathrm{EFF}}$ |
|---|---|---|
| $1.974 \pm 0.031$ | $53.20 \pm 0.54$ | $5060 \pm 62$ |
| $2.081 \pm 0.032$ | $54.60 \pm 0.79$ | $5011 \pm 43$ |
| $2.111 \pm 0.029$ | $54.00 \pm 0.58$ | $5016 \pm 60$ |
| $2.167 \pm 0.028$ | $55.40 \pm 0.62$ | $4901 \pm 126$ |
| $2.223 \pm 0.033$ | $57.40 \pm 0.65$ | $4870 \pm 124$ |
| $2.258 \pm 0.030$ | $57.00 \pm 0.45$ | $4867 \pm 134$ |
| $2.369 \pm 0.029$ | $58.40 \pm 0.46$ | $4773 \pm 111$ |
| $2.454 \pm 0.025$ | $60.00 \pm 0.71$ | $4721 \pm 75$ |
| $2.524 \pm 0.033$ | $60.05 \pm 0.68$ | $4668 \pm 100$ |
| $2.622 \pm 0.031$ | $61.65 \pm 0.79$ | $4513 \pm 52$ |
| $2.745 \pm 0.028$ | $61.05 \pm 0.57$ | $4488 \pm 41$ |
| $2.874 \pm 0.028$ | $61.45 \pm 2.24$ | $4410 \pm 67$ |
| $2.983 \pm 0.031$ | $61.80 \pm 1.29$ | $4396 \pm 90$ |
| $3.166 \pm 0.030$ | $63.10 \pm 0.95$ | $4178 \pm 95$ |
| $3.269 \pm 0.035$ | $63.30 \pm 2.24$ | $4179 \pm 93$ |
| $3.360 \pm 0.039$ | $63.40 \pm 1.83$ | $4110 \pm 115$ |
| $3.456 \pm 0.032$ | $63.95 \pm 0.93$ | $4033 \pm 66$ |
| $3.520 \pm 0.040$ | $64.40 \pm 0.91$ | $3947 \pm 76$ |
| $3.627 \pm 0.035$ | $64.80 \pm 1.12$ | $3929 \pm 67$ |
| $3.759 \pm 0.032$ | $65.00 \pm 0.25$ | $3916 \pm 46$ |
| $3.849 \pm 0.029$ | $65.00 \pm 0.71$ | $3921 \pm 114$ |
| $3.981 \pm 0.032$ | $65.80 \pm 0.65$ | $3842 \pm 47$ |
| $4.181 \pm 0.035$ | $66.60 \pm 0.79$ | $3744 \pm 85$ |
| $4.285 \pm 0.028$ | $67.20 \pm 1.12$ | $3723 \pm 77$ |
| $4.390 \pm 0.029$ | $67.80 \pm 0.79$ | $3725 \pm 91$ |
| $4.510 \pm 0.033$ | $67.50 \pm 0.50$ | $3603 \pm 117$ |
| $4.772 \pm 0.039$ | $68.80 \pm 0.79$ | $3558 \pm 132$ |
| $4.867 \pm 0.033$ | $69.20 \pm 0.79$ | $3549 \pm 93$ |
| $4.952 \pm 0.028$ | $69.80 \pm 1.12$ | $3530 \pm 94$ |
| $5.045 \pm 0.036$ | $69.60 \pm 0.79$ | $3484 \pm 40$ |
| $5.159 \pm 0.036$ | $70.20 \pm 1.12$ | $3453 \pm 32$ |
| $5.257 \pm 0.030$ | $70.00 \pm 0.25$ | $3426 \pm 87$ |
| $5.397 \pm 0.037$ | $69.60 \pm 0.79$ | $3442 \pm 68$ |
| $5.523 \pm 0.024$ | $70.60 \pm 0.91$ | $3440 \pm 54$ |
| $5.786 \pm 0.033$ | $70.80 \pm 1.12$ | $3340 \pm 144$ |
| $6.113 \pm 0.028$ | $71.10 \pm 1.58$ | $3356 \pm 54$ |
| $6.542 \pm 0.032$ | $71.40 \pm 1.12$ | $3242 \pm 117$ |
| $8.109 \pm 0.029$ | $73.20 \pm 0.85$ | $3116 \pm 33$ |

NOTE—$T_{\mathrm{EFF}}$ results are sorted by $V_0 - K_0$ and then collected into bins of $N = 5$. Spectral type indices range from G0 = 50, K0 = 60, M0 = 66. For more details, see Section 7.2 and Figure 9.

**Table 10.** $V_0 - K_0$ by spectral type from our SED fitting.

| Spectral Type | Sp. Type Num | $N$ | $V_0 - K_0$ (BB88) | $V_0 - K_0$ (this work) |
|---|---|---|---|---|
| G0III | 50 | 1 | 1.75 | $2.024 \pm 0.073$ |
| G1III | 51 | 1 | 1.83 | $2.121 \pm 0.072$ |
| G2III | 52 | | 1.90 | *$2.078 \pm 0.088$* |
| G3III | 53 | 2 | 1.98 | $2.035 \pm 0.051$ |
| G4III | 54 | 6 | 2.05 | $2.156 \pm 0.024$ |
| G5III | 55 | 13 | 2.10 | $2.126 \pm 0.019$ |
| G6III | 56 | | 2.15 | *$2.207 \pm 0.028$* |
| G7III | 57 | | 2.16 | *$2.288 \pm 0.028$* |
| G8III | 58 | 11 | 2.16 | $2.369 \pm 0.021$ |
| G9III | 59 | | 2.24 | *$2.393 \pm 0.032$* |
| K0III | 60 | 8 | 2.31 | $2.416 \pm 0.025$ |
| K1III | 61 | 4 | 2.50 | $2.519 \pm 0.027$ |
| K1.25III | 61.25 | 5 | 2.55 | $2.694 \pm 0.030$ |
| K1.5III | 61.5 | 9 | 2.60 | $2.865 \pm 0.022$ |
| K2III | 62 | 4 | 2.70 | $2.990 \pm 0.034$ |
| K3III | 63 | 1 | 3.00 | $2.753 \pm 0.101$ |
| K3.25III | 63.25 | 10 | 3.07 | $3.194 \pm 0.023$ |
| K3.5III | 63.5 | 8 | 3.13 | $3.373 \pm 0.028$ |
| K4III | 64 | 8 | 3.26 | $3.612 \pm 0.026$ |
| K5III | 65 | 18 | 3.60 | $3.785 \pm 0.019$ |
| M0III | 66 | | 3.85 | *$3.974 \pm 0.026$* |
| M1III | 67 | 13 | 4.05 | $4.163 \pm 0.018$ |
| M2III | 68 | 10 | 4.30 | $4.603 \pm 0.024$ |
| M3III | 69 | 8 | 4.64 | $4.718 \pm 0.024$ |
| M4III | 70 | 27 | 5.10 | $5.222 \pm 0.013$ |
| M5III | 71 | 14 | 5.96 | $5.842 \pm 0.019$ |
| M5.5III | 71.5 | 4 | 6.40 | $6.527 \pm 0.034$ |
| M6III | 72 | 2 | 6.84 | $7.435 \pm 0.068$ |
| M7III | 73 | 1 | 7.80 | $7.632 \pm 0.073$ |
| M7.5III | 73.5 | 1 | . . . | $7.674 \pm 0.064$ |
| M7.75III | 73.75 | 2 | . . . | $8.473 \pm 0.040$ |

NOTE—For comparison, the values from Bessell & Brett ('BB88'; 1988) are also presented. Our values are in the final column, with italicized values for where interpolation provided $V_0 - K_0$ values. These data are plotted in Figure 10. For more details, see Section 7.2.3.

One useful tangential result from the analyses performed in support of this investigation is that a calibration of $V_0 - K_0$ versus spectral type can be explored. 'Intrinsic' values for giant stars of this color were presented in Bessell & Brett ('BB88'; 1988), which provide a comparison basis for our own values, presented in Table 10 and Figure 10. Interestingly, our calibration of this relationship shows a consistently redward trend; the difference between our $V_0 - K_0$ colors for a given spectral subclass and BB88 is that our values are redder by a median value of $\Delta V_0 - K_0 = +0.11$.

### 7.3. $R$ versus $V_0 - K_0$

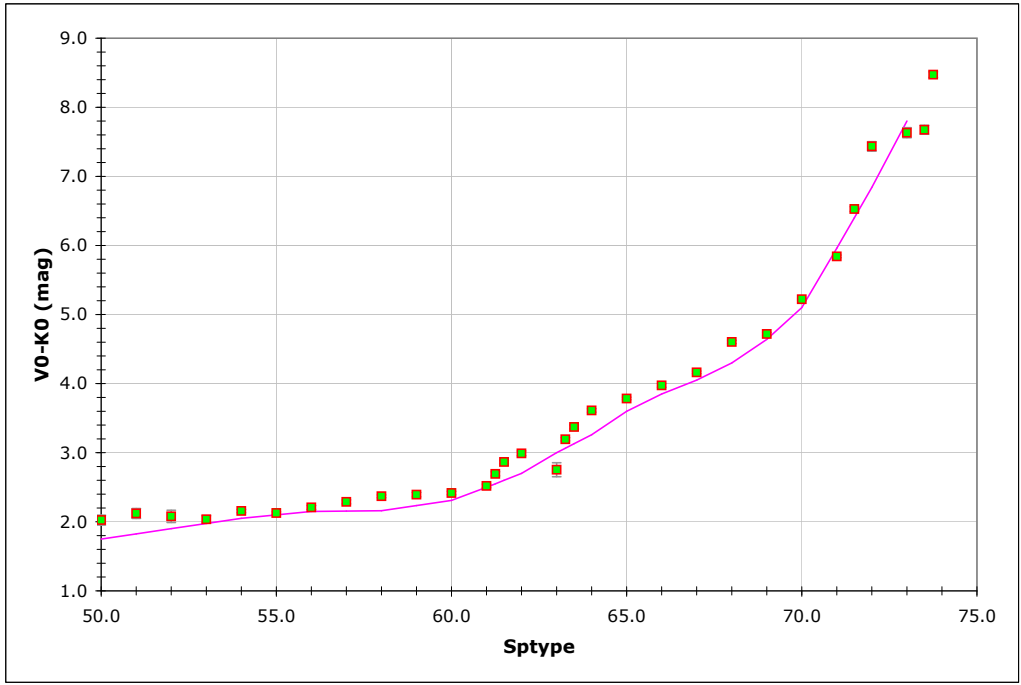

**Figure 10.** $V_0 - K_0$ versus spectral type. Solid line is the values from Bessell & Brett (1988); data points are from this study. For more details, see Section 7.2.3 and Table 10.

In contrast to the tight correlations seen in $T_{\mathrm{EFF}}$ versus $V_0 - K_0$ in §7.1, we find that values for stellar radius $R$ show considerably more scatter. As discussed in §6, there are uncertainties in our targets' distance determinations and consequently significant intrinsic scatter for $R$ as a function of $V_0 - K_0$; the raw data are seen in Figure 11.

In order to assess any possible general trend in the relation between stellar radius and dereddened color, we binned the data into sections of width $\Delta V_0 - K_0 = 0.5$. The first and final bins ($\Delta V_0 - K_0 = \{1.5, 2.5\}$ and $\Delta V_0 - K_0 = \{6.0, 9.0\}$) are wider to capture the color outliers. For each one of the bins, a weighted average radius for all the stars in that given bin was computed. To assess the impact of significant outliers, a second step was taken; the data for a given bin were assessed with a linear fit ($R = a \times (V_0 - K_0) + b$), and the 10-$\sigma$ outliers were excluded from a second weighted average computation of mean $R$ for that bin. In most bins this did not result in a significant change in the expected $R$ value for that bin. These results are presented in Table 11.

The general trend shows for a relatively flat size distribution for the bluest $\Delta V_0 - K_0 = \{1.5, 2.5\}$ stars at $\sim 11 R_\odot$, and then a steady increase in linear size as $V_0 - K_0$ color becomes redder. This becomes more gradual as stellar linear sizes exceed $100 R_\odot$, but still trends upwards with increasing $V_0 - K_0$ as the expectation for stellar linear size hits $\sim 150 R_\odot$ for the reddest stars. Overall, $V_0 - K_0$ is indicative of stellar linear size at only the $\sim 30\%$ level.

## 7.4. R versus Spectral Type

Similar to §7.2, we examined $R$ versus spectral type with arranging the data by spectral type, and then by $V_0 - K_0$ color.

### 7.4.1. Sorted into spectral type bins

As in Sections 7.2 and 7.3, linear radius $R$ shows a considerable amount of scatter versus spectral type, although there are some interesting trends. Specifically, the linear size of the giants appears to be roughly constant between G0III and K0III, at $\sim 12 R_\odot$; this is consistent with the flat linear size Section 7.3 for the bluer stars.

### 7.4.2. Sorted by $V_0 - K_0$ color into bins of 5

Similar to the second part of §7.2, to check against the possible subjectivity of spectral typing in exploring the $R$ versus spectral type trends seen in Figure 12, we arranged our spectral type data again into bins of 5. We then determined the average spectral type and weighted average $R$ values; these data are presented in Table 9 and Figure 13.

**Table 11.** $R$ versus binned $V_0 - K_0$

| $V_0 - K_0$ | | All Data | | | 10$\sigma$ removed | |
|---|---|---|---|---|---|---|
| Bin | $N$ | $R$ | $a$ | $b$ | $N$ | $R$ |
| (mag) | | ($R_\odot$) | | | | ($R_\odot$) |
| $1.5 - 2.5$ | 42 | $12.1 \pm 2.7$ | 1.3 | 7.3 | 32 | $10.7 \pm 1.1$ |
| $2.5 - 3.0$ | 20 | $24.3 \pm 9.3$ | 26.5 | -56.2 | 14 | $25.7 \pm 9.2$ |
| $3.0 - 3.5$ | 24 | $42.6 \pm 15.9$ | 20.4 | -37.6 | 20 | $39.9 \pm 12.8$ |
| $3.5 - 4.0$ | 22 | $61.5 \pm 20.8$ | 53.5 | -156.0 | 20 | $61.3 \pm 20.2$ |
| $4.0 - 4.5$ | 21 | $71.6 \pm 12.7$ | 44.4 | -125.0 | 19 | $74.4 \pm 11.4$ |
| $4.5 - 5.0$ | 16 | $93.6 \pm 25.4$ | 17.9 | -13.0 | 14 | $92.5 \pm 22.6$ |
| $5.0 - 5.5$ | 23 | $94.3 \pm 16.3$ | 29.2 | -69.0 | 22 | $95.9 \pm 16.0$ |
| $5.5 - 6.0$ | 8 | $112.7 \pm 22.0$ | 7.0 | 57.9 | 7 | $104.3 \pm 14.6$ |
| $6.0 - 9.0$ | 15 | $170.4 \pm 37.9$ | 20.1 | -0.1 | 15 | $170.4 \pm 37.9$ |

NOTE—Details of the fitting are presented in §7.3 and shown in Figure 11.

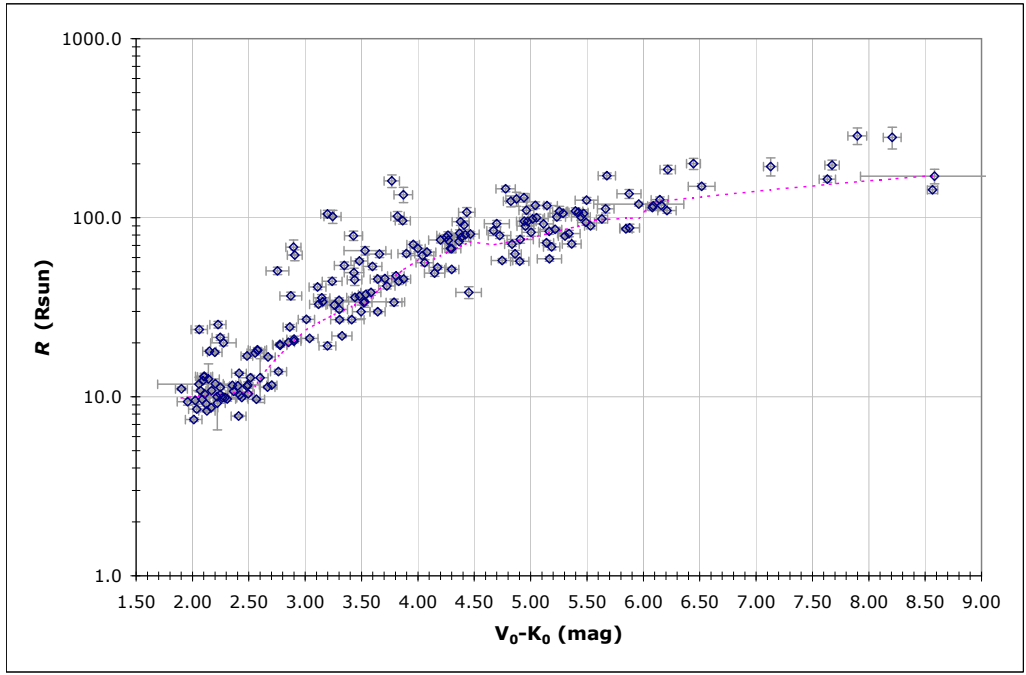

**Figure 11.** Radius $R$ versus $V_0 - K_0$. The dotted red line is for the linear fits noted in Table 11. For more details, see Section 7.3.

It is apparent in the figure that arranging our data in this fashion results in a qualitatively 'cleaner' data, exhibiting clear trends in $R$ versus the binned spectral type values. For each of the GKM spectral types in this data set, we performed a linear fit to assess the $R$ trend as a function of binned spectral type; the linear fit parameters are presented in Table 14. A continuity constraint was enforced for the fitting procedure so that there were no 'jumps' in $R$ at the boundaries of the linear fit regimes. The results in this table are consistent with the previous sections – the G-type giants show an almost flat linear radius across all subtypes at $\sim 11R_\odot$; the K-types present a trend of increasing size with subtype, hitting a maximum of $\sim 50R_\odot$ at K5III; and the M-types have an increased slope in their size-subtype

**Table 12.** Weighted average $R$ versus weighed average spectral type

| Spectral Type | Sp. Type Num | $N$ | $R$ ($R_\odot$) |
|---|---|---|---|
| G0III | 50 | 1 | $9.78 \pm \ldots$ |
| G1III | 51 | 1 | $9.37 \pm \ldots$ |
| G3III | 53 | 2 | $8.18 \pm 2.90$ |
| G4III | 54 | 6 | $9.70 \pm 1.14$ |
| G5III | 55 | 13 | $11.76 \pm 2.40$ |
| G8III | 58 | 11 | $11.97 \pm 2.69$ |
| K0III | 60 | 8 | $9.96 \pm 1.72$ |
| K1III | 61 | 4 | $12.87 \pm 2.62$ |
| K1.25III | 61.25 | 5 | $19.41 \pm 1.90$ |
| K1.5III | 61.5 | 9 | $16.08 \pm 5.53$ |
| K2III | 62 | 4 | $23.50 \pm 5.49$ |
| K3III | 63 | 1 | $52.06 \pm \ldots$ |
| K3.25III | 63.25 | 10 | $33.82 \pm 7.51$ |
| K3.5III | 63.5 | 8 | $28.22 \pm 8.37$ |
| K4III | 64 | 8 | $41.28 \pm 8.54$ |
| K5III | 65 | 18 | $43.60 \pm 14.16$ |
| M0III | 66 | | *53.6* |
| M1III | 67 | 13 | $63.76 \pm 11.09$ |
| M2III | 68 | 10 | $79.30 \pm 14.15$ |
| M3III | 69 | 8 | $75.87 \pm 20.28$ |
| M4III | 70 | 27 | $92.42 \pm 15.58$ |
| M5III | 71 | 14 | $97.83 \pm 30.41$ |
| M5.5III | 71.5 | 4 | $145.59 \pm 35.54$ |
| M6III | 72 | 2 | $162.74 \pm 24.84$ |
| M7III | 73 | 1 | $173.31 \pm \ldots$ |
| M7.5III | 73.5 | 1 | $208.23 \pm \ldots$ |
| M7.75III | 73.75 | 2 | $155.09 \pm 23.67$ |

NOTE—Data are shown in Figure 12, with italicized values for where interpolation provided a $R$ value. For more details, see §7.4.

relationship, hitting a size of $\sim 150 R_\odot$ for the reddest in this group. The average deviation of the data with respect to these linear fits across the GKM range is 11%.

## 8. DISCUSSION: LUMINOSITIES, MASSES AND EVOLUTIONARY TRACKS

The principal intent of this investigation is to provide reference values for giant star angular sizes, effective temperatures, and, to the extent possible by current distance measurements, radii, all calibrated against the indices of $V_0 - K_0$ and spectral type. These data present a launchpad for further investigations of interest. We illustrate below how the stellar fundamental parameters presented in this investigation can be used to establish giant star luminosities, build an Hertzsprung-Russell diagram (Hertzsprung 1909, 1911; Russell 1914), and explore the determination of stellar masses.

### 8.1. *Luminosities and Evolutionary Tracks*

**Table 13.** $R$ versus spectral type sorted by $V_0 - K_0$ and collected into bins of $N = 5$.

| Sp. Type Num | $V_0 - K_0$ | $R$ | $R_{\text{fit}}$ |
|---|---|---|---|
| $53.2 \pm 0.5$ | $1.97 \pm 0.03$ | $9.2 \pm 1.8$ | 9.8 |
| $54.6 \pm 0.8$ | $2.08 \pm 0.03$ | $11.1 \pm 2.6$ | 10.1 |
| $54.0 \pm 0.6$ | $2.11 \pm 0.03$ | $10.0 \pm 2.0$ | 10.0 |
| $55.4 \pm 0.6$ | $2.17 \pm 0.03$ | $11.4 \pm 5.1$ | 10.2 |
| $57.4 \pm 0.6$ | $2.22 \pm 0.03$ | $12.0 \pm 3.0$ | 10.5 |
| $57.0 \pm 0.4$ | $2.26 \pm 0.03$ | $11.1 \pm 3.4$ | 10.4 |
| $58.4 \pm 0.5$ | $2.37 \pm 0.03$ | $9.8 \pm 1.9$ | 10.7 |
| $60.0 \pm 0.7$ | $2.45 \pm 0.03$ | $11.5 \pm 2.1$ | 10.7 |
| $60.1 \pm 0.7$ | $2.52 \pm 0.03$ | $11.5 \pm 3.0$ | 11.0 |
| $61.7 \pm 0.8$ | $2.62 \pm 0.03$ | $14.2 \pm 4.1$ | 22.9 |
| $61.1 \pm 0.6$ | $2.74 \pm 0.03$ | $16.5 \pm 5.5$ | 18.4 |
| $61.5 \pm 2.2$ | $2.87 \pm 0.03$ | $22.2 \pm 4.8$ | 21.4 |
| $61.8 \pm 1.3$ | $2.98 \pm 0.03$ | $24.4 \pm 7.7$ | 24.0 |
| $63.1 \pm 1.0$ | $3.17 \pm 0.03$ | $30.2 \pm 12.5$ | 33.6 |
| $63.3 \pm 2.2$ | $3.27 \pm 0.04$ | $34.6 \pm 7.0$ | 35.1 |
| $63.4 \pm 1.8$ | $3.36 \pm 0.04$ | $26.0 \pm 8.5$ | 35.8 |
| $64.0 \pm 0.9$ | $3.46 \pm 0.03$ | $43.3 \pm 10.7$ | 39.9 |
| $64.4 \pm 0.9$ | $3.52 \pm 0.04$ | $37.6 \pm 8.4$ | 43.2 |
| $64.8 \pm 1.1$ | $3.63 \pm 0.04$ | $37.2 \pm 12.6$ | 46.2 |
| $65.0 \pm 0.3$ | $3.76 \pm 0.03$ | $43.2 \pm 10.9$ | 47.6 |
| $65.0 \pm 0.7$ | $3.85 \pm 0.03$ | $52.3 \pm 23.0$ | 47.6 |
| $65.8 \pm 0.6$ | $3.98 \pm 0.03$ | $65.5 \pm 6.2$ | 53.6 |
| $66.6 \pm 0.8$ | $4.18 \pm 0.03$ | $61.2 \pm 12.0$ | 61.6 |
| $67.2 \pm 1.1$ | $4.28 \pm 0.03$ | $62.5 \pm 13.6$ | 68.3 |
| $67.8 \pm 0.8$ | $4.39 \pm 0.03$ | $82.3 \pm 8.6$ | 75.0 |
| $67.5 \pm 0.5$ | $4.51 \pm 0.03$ | $76.9 \pm 22.8$ | 71.6 |
| $68.8 \pm 0.8$ | $4.77 \pm 0.04$ | $75.4 \pm 33.5$ | 86.1 |
| $69.2 \pm 0.8$ | $4.87 \pm 0.03$ | $67.4 \pm 12.1$ | 90.5 |
| $69.8 \pm 1.1$ | $4.95 \pm 0.03$ | $103.2 \pm 13.5$ | 97.2 |
| $69.6 \pm 0.8$ | $5.04 \pm 0.04$ | $102.0 \pm 12.2$ | 95.0 |
| $70.2 \pm 1.1$ | $5.16 \pm 0.04$ | $75.1 \pm 19.7$ | 101.6 |
| $70.0 \pm 0.3$ | $5.26 \pm 0.03$ | $90.1 \pm 11.8$ | 99.4 |
| $69.6 \pm 0.8$ | $5.40 \pm 0.04$ | $90.8 \pm 18.7$ | 95.0 |
| $70.6 \pm 0.9$ | $5.52 \pm 0.02$ | $104.2 \pm 13.9$ | 106.1 |
| $70.8 \pm 1.1$ | $5.79 \pm 0.03$ | $100.6 \pm 33.3$ | 108.3 |
| $71.1 \pm 1.6$ | $6.11 \pm 0.03$ | $124.8 \pm 4.6$ | 111.6 |
| $71.4 \pm 1.1$ | $6.54 \pm 0.03$ | $138.8 \pm 40.7$ | 115.0 |
| $73.2 \pm 0.8$ | $8.11 \pm 0.03$ | $172.0 \pm 37.4$ | 135.0 |

NOTE—The fit values in the final column are from the linear fits noted in Table 14. Spectral type indices range from G0 = 50, K0 = 60, M0 = 66. Data are shown in Figure 13. For more details, see 7.4

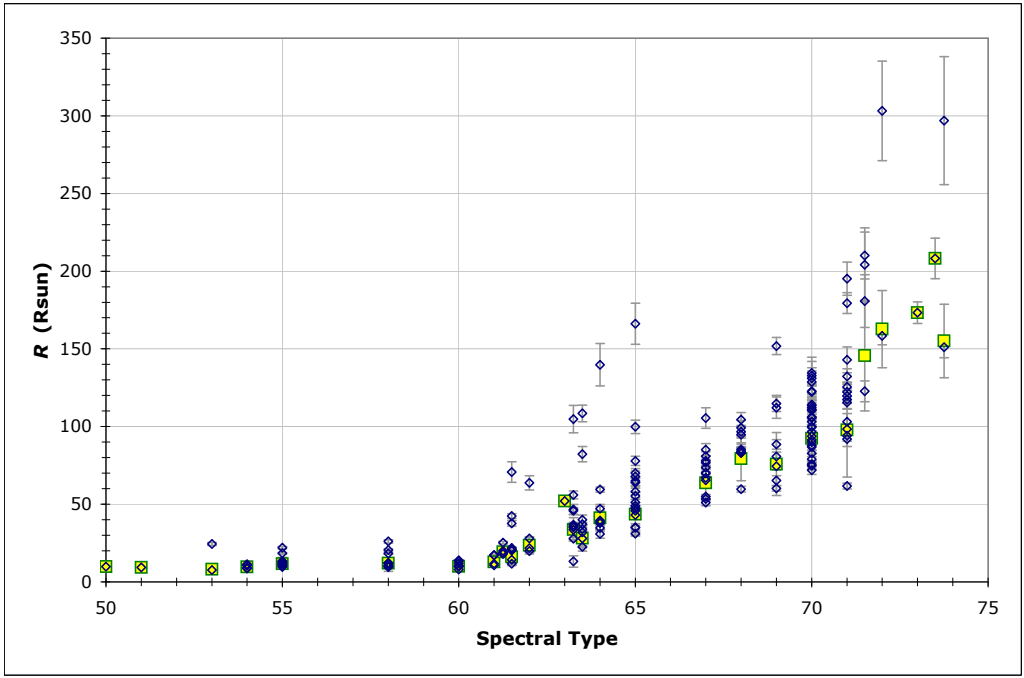

**Figure 12.** Radius $R$ versus spectral type. The yellow boxes are the weighted averages for each spectral type, as presented in Table 12. For more details, see §7.4.

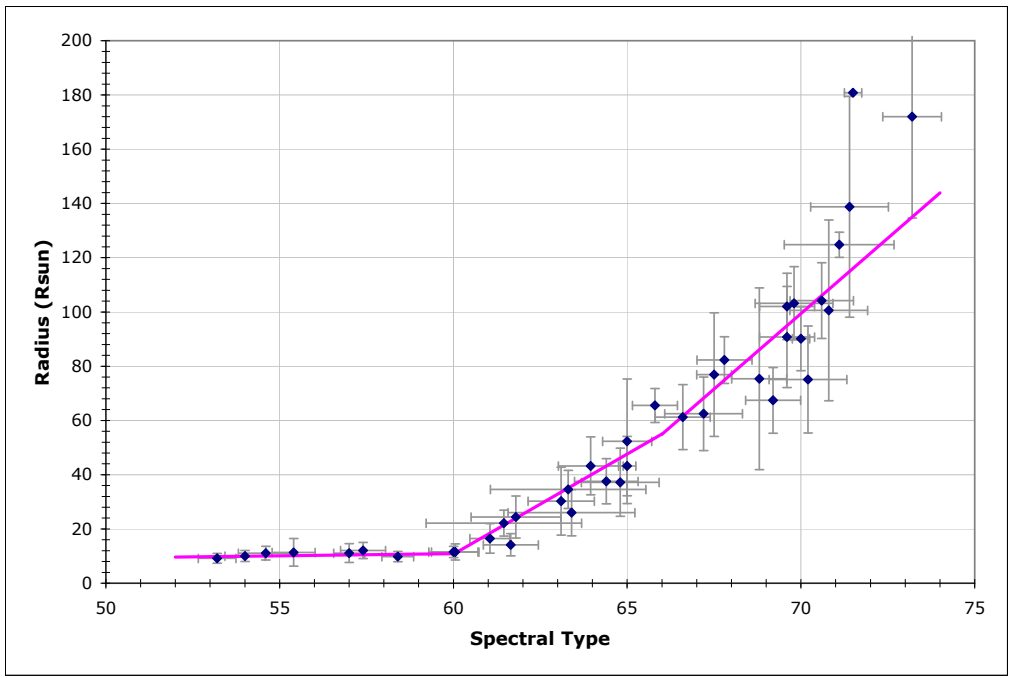

**Figure 13.** Effective temperature $R$ versus spectral type, with our $R$ results sorted by $V_0 - K_0$ and then collected into bins of $N = 5$. The linear fit parameters can be found in Table 14 and are discussed in §7.4.

**Table 14.** Linear fits for $R$ versus the spectral type ranges in Table 9 shown Figure 13.

| Sp. type range | $a$ | $b$ | $\chi^2$ |
|:---:|:---:|:---:|:---:|
| G | 0.16 | 1.41 | 0.85 |
| K | 7.39 | -432.90 | 5.41 |
| M | 11.11 | -678.55 | 7.05 |

Note—For more details, see §7.4.

Given the values of stellar $F_{\rm BOL}$ and distance, stellar luminosity $L$ can be calculated. Note that, as discussed in detail in Appendix A, $L$ is *not* dependent upon our determinations of $R$ and $T_{\rm EFF}$, but rather $F_{\rm BOL}$ and $\pi$ – see Equation A3. Our target luminosity values are seen in Table 19 and Figure 14.

For further illustration, we took stellar evolutionary tracks in $L$ versus $T_{\rm EFF}$ from the BaSTI models[7] in Pietrinferni et al. (2006), using the solar abundance tracks ([M/H]=0.058, Z=0.0198, Y=0.273) for two representative mass tracks, $M = 1.2$ and $2.4 M_{\odot}$. Our choice of solar abundance was rather perfunctory and motivated only by there being no significant impetus to choose any other track. The mass values were chosen to represent values expected to be characteristic of the lowest mass giants at $M = 1.2 M_{\odot}$, with twice that to explore the impact of mass on the track path. Additionally, we cross-referenced our target list against those stars called out in Gontcharov (2008) as red giant clump stars. Clump giants are post-helium flash giants, undergoing core helium burning, with an inert hydrogen envelope surrounding the core (Gontcharov 2017). These tracks, along with our program stars, are shown in Figure 14. Along the evolutionary tracks, time intervals of $\Delta t$=10Myr are indicated, and there is a pile-up of those interval ticks in the lower left of the plot, characteristic of red giant clump stars.

Interestingly, when identified as clump giants, it is qualitatively apparent that there is an overdensity of our program stars in that region of our HR diagram, which is consistent with increased loiter time of the evolutionary tracks in that area of the plot. Overall, the BaSTI evolutionary tracks bracket our program stars well, with most outliers coming to the upper left of the $M = 2.4 M_{\odot}$ stars, indicative of higher mass objects.

### 8.2. *Masses*

A precision measurements of stellar linear radius enables the calculation of stellar mass, provided a value for a given star's surface gravity $\log g$ is available. For that purpose, we utilized the most recent version of the PASTEL catalog of Soubiran et al. (2016), using the 2020-01-30 version, which features an inventory of such values from the literature. 91 of our program stars can be found in this catalog; the values we used from PASTEL are seen in Table 20, along with their respective primary references.

For stars that have $R$ and $\log g$, $(g = GM/R^2)$ can trivially be rearranged to solve for $M$, minding the usual conversion factors and error propagation, so long as errors were provided for $\log g$. For the PASTEL catalog measurements that had no value for the uncertainty in $\log g$, a value of 0.25 was assigned as a placeholder.

Overall, this approach is quite crude, and limited as well. On the former point, our average mass error was typically 50%; this improved only slightly, by about 10%, if the measures were restricted to those that reported formal errors. For this approach's limitations, t is worth noting that few of our program stars have $\log g$ measures if they are below $T_{\rm EFF} < 4000$K, which is indicative of the difficulty currently found in determining $\log g$ for these lower temperature stars.

Nevertheless, this approach is at least qualitatively illustrative of mass effects on our results. If we replot our Figure 14 with the corresponding mass information, in Figure 15, we can see that the points that lie above and to the left of the BaSTI evolutionary tracks are, as expected, the stars that indicate higher masses. Quantitatively, the values in Table 20 are consistent with expectations: the median value of our program stars for which we can calculate a value for mass is $M = 1.9 M_\odot$; these program stars are bracketed well by the $M = \{1.2, 2.4\} M_\odot$ evolutionary tracks in Figure 14.

---

[7] 'Bag of Stellar Tracks and Isochrones', http://basti.oa-teramo.inaf.it/index.html

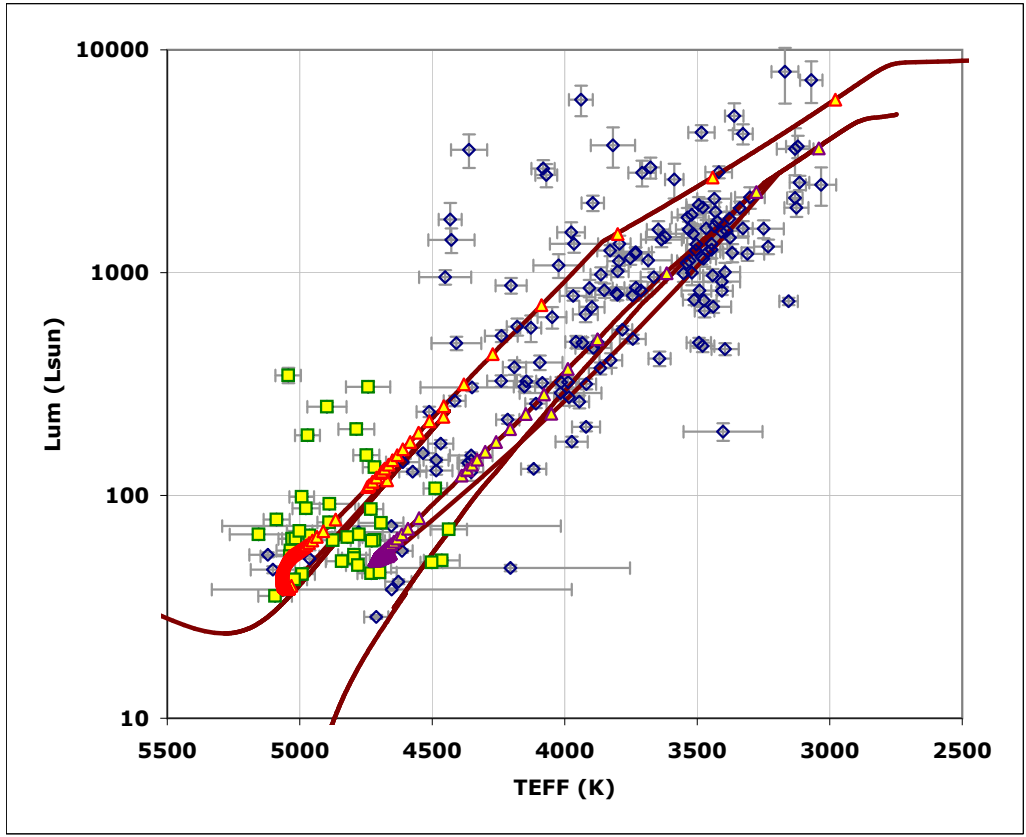

**Figure 14.** $L$ versus $T_{EFF}$ for our program stars. The upper and lower lines are $M = 1.2$ and $2.4 M_\odot$ stellar evolutionary tracks from Pietrinferni et al. (2006), with red-yellow triangles at 10Myr intervals. Our program stars are blue diamonds and yellow squares; the latter are those stars that are also identified in Gontcharov (2008) as red giant clump stars. See discussion in 8.1.

If stellar $\log g$ measures can be (a) improved to better than 0.05 dex, and (b) extended to the lower temperature ranges, the community will have a useful tool for ~10% or better mass determinations of stars in this region of the HR diagram. In so doing, the $\log g$ measures would approach the current noise floor on the uncertainty in $M$ that is contributed by the linear radius measures, which is approximately 8.5%.

## 9. PREDICTION OF ANGULAR DIAMETERS

Stellar angular diameter prediction is a useful tool for various observational endeavors, particularly small solar system body occultation events and microlensing events. Previous calibrations of evolved star angular sizes can be found in van Belle (1999) and Di Benedetto (2005); main sequence predictions have followed more recently with significant data sets on such stars from The CHARA Array (Boyajian et al. 2014; Adams et al. 2018).

Implementing an analysis similar to van Belle (1999), we can use our results to produce a robust predictive tool for stellar angular size, based upon $V_0 - K_0$ color, improving substantially upon that prior investigation. The utility of this tool lies in that it is strictly empirical and sidesteps considerations of spectral type, distance, and atmospheric modeling. To begin, we scale all of our angular sizes to a common $V = 0$ zero magnitude basis ($\theta_{V=0} = \theta \times 10^{V/5}$). Minimization of the two-axis $\chi^2$ (e.g. see Eqn. 15.3.2 in Bevington & Robinson 1992) gives a fit of:

$$\log(\theta_{V=0}) = 0.2292 \pm 0.0016 \times (V_0 - K_0) + 0.6164 \pm 0.0067 \tag{5}$$

with the data and the fit plotted in Figure 16; the reduced $\chi^2_\nu$ is 0.87. The resulting median absolute scatter in predicted versus measured $\theta_{V=0}$ values is 2.9%, indicating a factor of 4 improvement over the calibration in van Belle (1999).

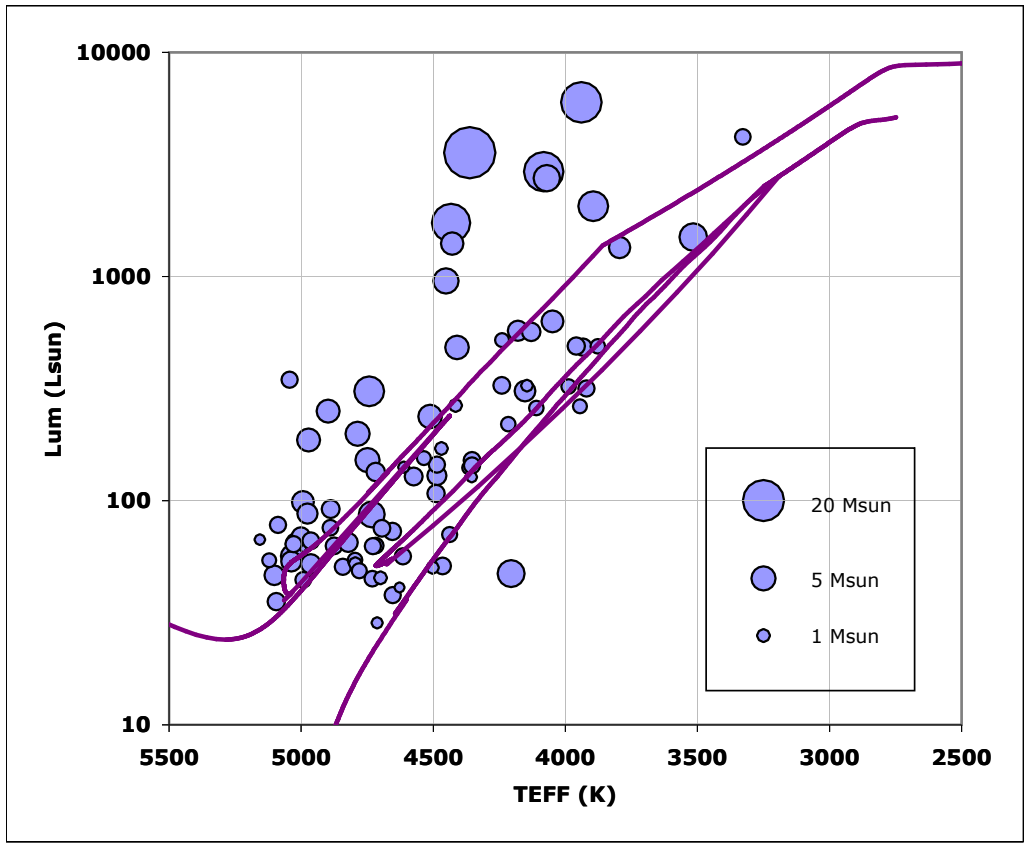

**Figure 15.** $L$ versus $T_{EFF}$ versus $M$ for our program stars. Evolutionary tracks are those seen in Figure 14. See §8.2 for more information.

## 10. CONCLUSIONS

This study has been the intersection of multiple, significant lines of investigation. First, it is the product of considerable effort in data collection, both with optical interferometry and photometry. Second, literature data and meta-databases, such as distances from Hipparcos and Gaia, photometry in the GCPD (Mermilliod et al. 1997), as well as spectral types from Skiff (Skiff 2014b) and the INGS Library of A. Pickles, have all played major supporting roles. Third, advances in modeling, including the PHOENIX models (Husser et al. 2013) and our sedFit package have allowed us to take the data to build a large, precise set of fundamental parameters for giant stars. Our intent has been to provide a meticulously calibrated reference dataset of giant star $T_{EFF}$ and $R$. Our main findings are as follows:

1. When indexed against $V_0 - K_0$, $T_{EFF}$ can be predicted to a precision of 50K (Equation 4) over the range $V_0 - K_0 = \{1.9, 6.5\}$. Within that range there appear to be two statistically significant gaps, as discussed in §7.1.1.

2. When indexed against spectral type, $T_{EFF}$ can be linearly fit against spectral subtype with a median absolute difference of fit versus our data points of $52 - 53$K for most stars (§7.2 and Table 8).

3. For $R$, the color index $V_0 - K_0$ is indicative of linear radius at only the ~30% level (§7.3).

4. Using spectral type as a guide, linear radius can be predicted to ~11% (§7.4 and Table 14).

5. Luminosities are dependent upon our bolometric fluxes and published parallax measures, and, with our values for $T_{EFF}$, can be used to directly build an empirical HR diagram, which qualitatively agrees well with theoretical evolutionary tracks. Masses can be inferred from our values for $R$ and literature $\log g$ values, and are also qualitatively in agreement with those tracks (§8).

6. An improved calibration of $V_0 - K_0$ as a predictor of stellar angular size is presented, with a median absolute scatter of 2.9% (§9).

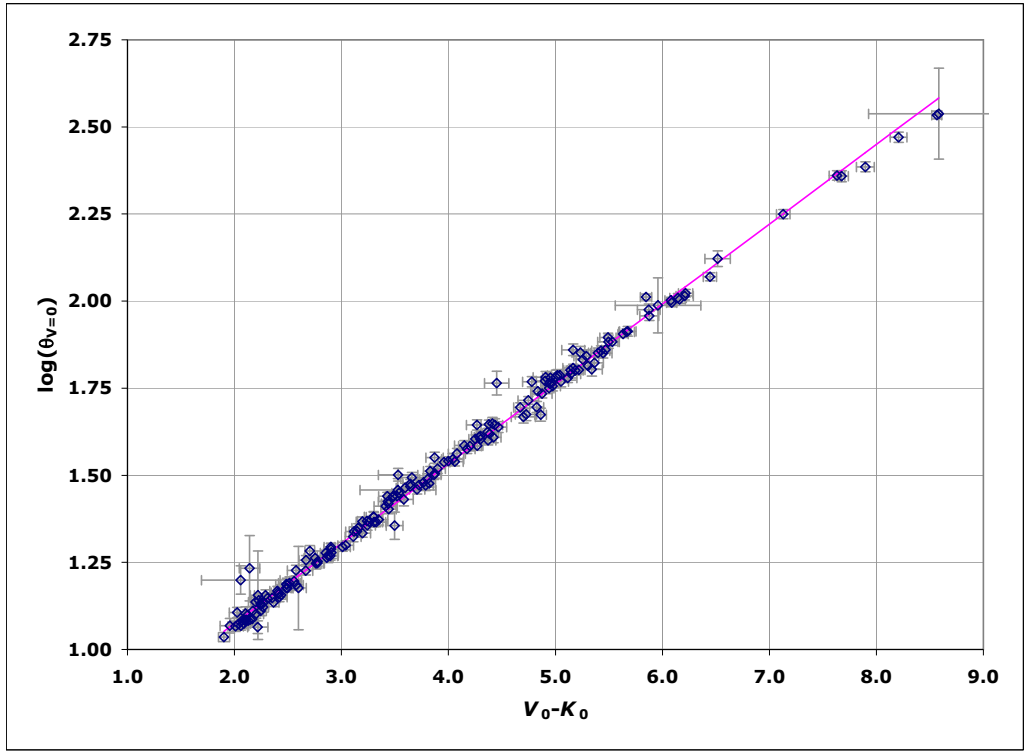

**Figure 16.** Zero-magnitude angular size $\theta_{V=0}$ versus $V_0 - K_0$ color.

Additionally, we present a serendipitous calibration of $V_0 - K_0$ versus spectral type, exhibiting colors slightly redder at the 0.11 mag level for a given spectral subtype than Bessell & Brett (1988) (§7.2.3).

For future work, improvements in parallax measurements is the most obvious avenue for extending the data herein, for both linear radii and luminosities. More subtly, sub-percent absolute photometry would also improve luminosities, but angular size precision would have to be improved by a factor of 2 to 3 to realize a substantive improvement in effective temperatures. As noted in §8, improvements in measurement of giant star $\log g$ could mean significant insights into the nature of giant star masses. As an increasing number of asteroseismic results from TESS are released, possibly on our very program stars, the combination on constraints from those studies and ours could provide even sharper insights into the nature of post-main sequence stellar evolution.

## 10.1. *Data Release*

All data produced for this investigation will be made available at the NASA Exoplanet Science Institute (NExScI) PTI archive[8]. This includes all tables in this paper, `sedFit` figures for each of our program stars.

---

[8] https://nexsci.caltech.edu/software/PTISupport/

ACKNOWLEDGMENTS

*Acknowledgments.* We would like to graciously acknowledge productive discussions and helpful suggestions from Chengjie Xiong (Washington University); we also acknowledge and appreciate the helpful feedback from an anonymous referee. We have made extensive use of the SIMBAD database and the VizieR catalogue access tool, operated by the CDS in Strasbourg, France (Ochsenbein et al. 2000). This research has made use of the AFOEV database, operated at CDS, France, and the GCPD database at the University of Lausanne, Switzerland (Mermilliod et al. 1997). This research has made use of NASA's Astrophysics Data System. Portions of this work were performed at the Jet Propulsion Laboratory, California Institute of Technology under contract with the National Aeronautics and Space Administration. As always, we caution users of the Palomar site to watch out for giant bumblebees in the sky. This work has made use of data from the European Space Agency (ESA) mission *Gaia* (https://www.cosmos.esa.int/gaia), processed by the *Gaia* Data Processing and Analysis Consortium (DPAC, https://www.cosmos.esa.int/web/gaia/dpac/consortium). Funding for the DPAC has been provided by national institutions, in particular the institutions participating in the *Gaia* Multilateral Agreement.

Funding for this research has been generously provided in part by Lowell Observatory. This material is based upon work supported by the National Science Foundation (NSF) under Grant No. AST-1212203, and NASA Grant No. NNX13AF01G. This work was supported in part through a NASA grant awarded to the Arizona/NASA Space Grant Consortium, and by the NSF Research Experiences for Undergraduates (REU) program.

*Facilities:* PO:PTI,LO:0.8m

APPENDIX

A. THE RELATIONSHIP BETWEEN $L$, $R$, $T$ – AND $\theta$, $\pi$, $F_{\mathrm{BOL}}$

A common misconception that we have encountered frequently during the last 30 years of carrying out these sorts of angular size measurements is the perceived value for measurements of stellar luminosity $L$. This is understandable, given that $L$ is *defined* as

$$L = \text{stellar surface area} \times \text{emitted flux at stellar surface} = 4\pi R^2 \sigma_{\mathrm{SB}} T^4 \tag{A1}$$

where $\pi$ is the usual $3.14159\ldots$ and $\sigma_{\mathrm{SB}}$ is the Stefan-Boltzman constant, $5.6704\ldots \times 10^5$ erg cm$^{-2}$ s$^{-1}$ K$^{-4}$. On the face of it, since our measurements of angular size provide new information on stellar $R$ and $T$, the expectation is that additional information has also been derived for $L$. This unfortunately is not true, and can be revealed if the underlying measurements that establish $L$ from $R$ and $T$ are examined further. Expanding Equation A1 with Equations 1 and 2, it can be shown how $\theta$ cancels out:

$$L = 4\pi \left[ \frac{\theta}{0.009292 \times \pi} \right]^2 \sigma_{\mathrm{SB}} \left[ 2341 \times \left[ \frac{F_{\mathrm{BOL}}}{\theta^2} \right]^{1/4} \right]^4 \tag{A2}$$

$$= 312.1 \times \frac{F_{\mathrm{BOL}}}{\pi^2} L_\odot \tag{A3}$$

(carrying only the first 4 significant figures for the various constants; noting some symbol collision between $\pi$ for the constant in Eqn. A1 and for parallax in Eqn. 2; as before, units for $F_{\mathrm{BOL}}$ are $10^{-8}$ erg s$^{-1}$ cm$^{-2}$, milliseconds for $\pi$, and $L$ is in units of solar luminosity $L_\odot$, based on 2015 IAU Resolution B3's $L_\odot \equiv 3.828 \times 10^{33}$ erg s$^{-1}$). From this we can see

$$\sigma_L = L \left[ \left( \frac{\sigma_F}{F_{\mathrm{BOL}}} \right)^2 + 4 \times \left( \frac{\sigma_\pi}{\pi} \right)^2 \right]^{1/2} \tag{A4}$$

It *is* true that our efforts in determining new values for stellar $F_{\mathrm{BOL}}$ in §4 constitutes 'new' information on $L$, and as such we have included discussion of this parameter in §8. Why is this important?[9] The most significant element of realizing the underlying nature of Eqn. A3 in the determination of $L$ is for error propagation. If one were to simply propagate errors from Eqn. A1, calculations of $\sigma_L$ would unnecessarily be carrying additional uncertainty from $\sigma_\theta$. In practice we find that with the dataset presented herein, this makes for uncertainties in $L$ that are roughly 30% greater than necessary.

---

[9] And will the editor allow a rhetorical question in our manuscript?

## REFERENCES

Abt, H. A. 1981, ApJS, 45, 437

—. 1985, ApJS, 59, 95

—. 2008, ApJS, 176, 216

Abt, H. A. & Morrell, N. I. 1995, ApJS, 99, 135

Adamczak, J. & Lambert, D. L. 2014, ApJ, 791, 58

Adams, A. D., Boyajian, T. S., & von Braun, K. 2018, MNRAS, 473, 3608

Adams, W. S., Joy, A. H., & Humason, M. L. 1926, ApJ, 64

Adams, W. S., Joy, A. H., Humason, M. L., & Brayton, A. M. 1935, ApJ, 81, 187

Airy, G. B. 1835, Transactions of the Cambridge Philosophical Society, 5, 283

Akras, S., Guzman-Ramirez, L., Leal-Ferreira, M. L., & Ramos-Larios, G. 2019, ApJS, 240, 21

Alexander, J. B., Jones, D. H. P., & Sinclair, J. E. 1983, Royal Greenwich Observatory Bulletins, 191

Alonso, A., Arribas, S., & Martinez-Roger, C. 1994, A&A, 282, 684

—. 1998, A&AS, 131, 209

Anthonioz, F., Ménard, F., Pinte, C., Le Bouquin, J. B., Benisty, M., Thi, W. F., Absil, O., Duchêne, G., Augereau, J. C., Berger, J. P., Casassus, S., Duvert, G., Lazareff, B., Malbet, F., Millan-Gabet, R., Schreiber, M. R., Traub, W., & Zins, G. 2015, A&A, 574, A41

Appenzeller, I. 1966, ZA, 64, 269

—. 1967, PASP, 79, 102

Arellano Ferro, A., Parrao, L., Schuster, W., Gonzalez-Bedolla, S., Peniche, R., & Pena, J. H. 1990, A&AS, 83, 225

Argue, A. N. 1963, MNRAS, 125, 557

—. 1966, MNRAS, 133, 475

Arp, H. C. 1958, AJ, 63, 118

Arribas, S. & Martinez Roger, C. 1987, A&AS, 70, 303

Arulanantham, N. A., Herbst, W., Gilmore, M. S., Cauley, P. W., & Leggett, S. K. 2017, ApJ, 834, 119

Baines, E. K., Armstrong, J. T., Schmitt, H. R., Zavala, R. T., Benson, J. A., Hutter, D. J., Tycner, C., & van Belle, G. T. 2018, AJ, 155, 30

Baines, E. K., Döllinger, M. P., Cusano, F., Guenther, E. W., Hatzes, A. P., McAlister, H. A., ten Brummelaar, T. A., Turner, N. H., Sturmann, J., Sturmann, L., Goldfinger, P. J., Farrington, C. D., & Ridgway, S. T. 2010, ApJ, 710, 1365

Bakos, G. A. 1968, AJ, 73, 187

—. 1974, AJ, 79, 866

Barbier, M. 1971, A&A, 14, 396

Barbier, M. & Maiocchi, R. 1966, Journal des Observateurs, 49, 290

Barnes, T. G., Evans, D. S., & Moffett, T. J. 1978, MNRAS, 183, 285

Bartasiute, S. 1981, Vilnius Astronomijos Observatorijos Biuletenis, 58, 41

Barthès, D. & Luri, X. 2001, A&A, 365, 519

Bartkevicius, A., Gurklyte, A., Kavaliauskaite, G., Kazlauskas, A., Kalytis, R., Sviderskiene, Z., Sperauskas, J., Sudzius, J., & Jasevicius, V. 1973, Vilnius Astronomijos Observatorijos Biuletenis, 36, 17

Bartkevicius, A. & Metik, L. P. 1969, Vilnius Astronomijos Observatorijos Biuletenis, 26, 13

Bartkevicius, A. & Sperauskas, I. 1974, Vilnius Astronomijos Observatorijos Biuletenis, 40, 3

Baug, T. & Chandrasekhar, T. 2013, Research in Astronomy and Astrophysics, 13, 1363

Beichman, C. A., Neugebauer, G., Habing, H. J., Clegg, P. E., & Chester, T. J., eds. 1988, Infrared astronomical satellite (IRAS) catalogs and atlases. Volume 1: Explanatory supplement, Vol. 1

Bergeat, J. & Lunel, M. 1980, A&A, 87, 139

Bergeat, J., Lunel, M., Garnier, R., Sibille, F., Roux, S., & vant Veer, F. 1981, A&A, 94, 350

Bergner, Y. K., Miroshnichenko, A. S., Yudin, R. V., Kuratov, K. S., Mukanov, D. B., & Shejkina, T. A. 1995, A&AS, 112, 221

Bessell, M. S. 1990, PASP, 102, 1181

Bessell, M. S. & Brett, J. M. 1988, PASP, 100, 1134

Bevington, P. R. & Robinson, D. K. 1992, Data reduction and error analysis for the physical sciences (McGraw-Hill College)

Bidelman, W. P. 1957, PASP, 69, 326

—. 1985, ApJS, 59, 197

Blackwell, D. E., Petford, A. D., Arribas, S., Haddock, D. J., & Selby, M. J. 1990, A&A, 232, 396

Blackwell, D. E., Shallis, M. J., & Selby, M. J. 1979, MNRAS, 188, 847

Boden, A. F., Colavita, M. M., van Belle, G. T., & Shao, M. 1998a, in Society of Photo-Optical Instrumentation Engineers (SPIE) Conference Series, Vol. 3350, Society of Photo-Optical Instrumentation Engineers (SPIE) Conference Series, ed. R. D. Reasenberg, 872–880

Boden, A. F., Koresko, C. D., van Belle, G. T., Colavita, M. M., Dumont, P. J., Gubler, J., Kulkarni, S. R., Lane, B. F., Mobley, D., Shao, M., Wallace, J. K., Henry, G. W., & PII Collaboration. 1999, ApJ, 515, 356

Boden, A. F., van Belle, G. T., Colavita, M. M., Dumont, P. J., Gubler, J., Koresko, C. D., Kulkarni, S. R., Lane, B. F., Mobley, D. W., Shao, M., Wallace, J. K., & The PTI Collaboration. 1998b, ApJL, 504, L39+

Bohlin, R. C., Gordon, K. D., & Tremblay, P. E. 2014, PASP, 126, 711

Böhm-Vitense, E. 1970a, A&A, 8, 299

—. 1970b, A&A, 8, 283

—. 1981, ARA&A, 19, 295

—. 1982, ApJ, 255, 191

—. 1995a, A&A, 297, L25

—. 1995b, AJ, 110, 228

Böhm-Vitense, E. & Canterna, R. 1974, ApJ, 194, 629

Bopp, B. W. 1984, PASP, 96, 894

Born, M. & Wolf, E. 1980, Principles of Optics Electromagnetic Theory of Propagation, Interference and Diffraction of Light, ed. Born, M. & Wolf, E.

Bouigue, R. 1959, Annales de l'Observatoire Astron. et Meteo. de Toulouse, 27, 47

Bouigue, R., Boulon, J., & Pedoussaut, A. 1961, Annales de l'Observatoire Astron. et Meteo. de Toulouse, 28, 33

Boyajian, T. S., McAlister, H. A., van Belle, G., Gies, D. R., ten Brummelaar, T. A., von Braun, K., Farrington, C., Goldfinger, P. J., O'Brien, D., Parks, J. R., Richardson, N. D., Ridgway, S., Schaefer, G., Sturmann, L., Sturmann, J., Touhami, Y., Turner, N. H., & White, R. 2012a, ApJ, 746, 101

Boyajian, T. S., van Belle, G., & von Braun, K. 2014, AJ, 147, 47

Boyajian, T. S., von Braun, K., van Belle, G., Farrington, C., Schaefer, G., Jones, J., White, R., McAlister, H. A., ten Brummelaar, T. A., Ridgway, S., Gies, D., Sturmann, L., Sturmann, J., Turner, N. H., Goldfinger, P. J., & Vargas, N. 2013, ApJ, 771, 40

Boyajian, T. S., von Braun, K., van Belle, G., McAlister, H. A., ten Brummelaar, T. A., Kane, S. R., Muirhead, P. S., Jones, J., White, R., Schaefer, G., Ciardi, D., Henry, T., López-Morales, M., Ridgway, S., Gies, D., Jao, W.-C., Rojas-Ayala, B., Parks, J. R., Sturmann, L., Sturmann, J., Turner, N. H., Farrington, C., Goldfinger, P. J., & Berger, D. H. 2012b, ApJ, 757, 112

Breitfelder, J., Mérand, A., Kervella, P., Gallenne, A., Szabados, L., Anderson, R. I., & Le Bouquin, J.-B. 2016, A&A, 587, A117

Brown, T. M. & Christensen-Dalsgaard, J. 1998, ApJL, 500, L195

Cameron, R. C. 1966, Georgetown Obs. Monogram, Vol. 21, p. 0 (1966), 21, 0

Campins, H., Rieke, G. H., & Lebofsky, M. J. 1985, AJ, 90, 896

Cannon, A. J. & Mayall, M. W. 1949, Annals of Harvard College Observatory, 112, 1

Cannon, A. J. & Pickering, E. C. 1993, VizieR Online Data Catalog, 3135

Canterna, R. 1976, AJ, 81, 228

Cardelli, J. A., Clayton, G. C., & Mathis, J. S. 1989, ApJ, 345, 245

Carney, B. W. 1983, AJ, 88, 610

Carter, B. S. 1990, MNRAS, 242, 1

Castor, J. I. & Simon, T. 1983, ApJ, 265, 304

Cernis, K., Meistas, E., Straizys, V., & Jasevicius, V. 1989, Vilnius Astronomijos Observatorijos Biuletenis, 84, 9

Chan, V. C. & Bovy, J. 2020, MNRAS, 493, 4367

Chen, P.-s., Gao, H., Hao, Y.-x., Chu, Q.-r., & Zhou, K.-p. 1982, ChA&A, 6, 153

Chen, P. S., Wang, X. H., & Xiong, G. Z. 1998, A&A, 333, 613

Chiavassa, A. 2018, arXiv e-prints, arXiv:1812.11002

Clariá, J. J., Piatti, A. E., Mermilliod, J.-C., & Palma, T. 2008, Astronomische Nachrichten, 329, 609

Clark, J. P. A. & McClure, R. D. 1979, PASP, 91, 507

Colavita, M. M. 1999, PASP, 111, 111

Colavita, M. M., Wallace, J. K., Hines, B. E., Gursel, Y., Malbet, F., Palmer, D. L., Pan, X. P., Shao, M., Yu, J. W., Boden, A. F., Dumont, P. J., Gubler, J., Koresko, C. D., Kulkarni, S. R., Lane, B. F., Mobley, D. W., & van Belle, G. T. 1999, ApJ, 510, 505

Coleman, L. A. 1982, AJ, 87, 369

Cousins, A. W. J. 1962a, Monthly Notes of the Astronomical Society of South Africa, 21, 20

—. 1962b, Monthly Notes of the Astronomical Society of South Africa, 21, 61

—. 1963a, Monthly Notes of the Astronomical Society of South Africa, 22, 130

—. 1963b, Monthly Notes of the Astronomical Society of South Africa, 22, 58

—. 1963c, Monthly Notes of the Astronomical Society of South Africa, 22, 12

—. 1964a, Monthly Notes of the Astronomical Society of South Africa, 23, 175

—. 1964b, Monthly Notes of the Astronomical Society of South Africa, 23, 10

—. 1965, Monthly Notes of the Astronomical Society of South Africa, 24, 120

—. 1984, South African Astronomical Observatory Circular, 8, 59

—. 1985, Monthly Notes of the Astronomical Society of South Africa, 44, 10

Cousins, A. W. J. & Caldwell, J. A. R. 1996, MNRAS, 281, 522

Cousins, A. W. J. & Stoy, R. H. 1962, Royal Greenwich Observatory Bulletins, 64

Cowley, A. P. & Bidelman, W. P. 1979, PASP, 91, 83

Cowley, A. P., Hiltner, W. A., & Witt, A. N. 1967, AJ, 72, 1334

Cox, A. N. 2000, Allen's astrophysical quantities (Allen's astrophysical quantities, 4th ed. Publisher: New York: AIP Press; Springer, 2000. Editedy by Arthur N. Cox. ISBN: 0387987460)

Crawford, D. L. & Barnes, J. V. 1970, AJ, 75, 978

Crawford, D. L., Barnes, J. V., Faure, B. Q., Golson, J. C., & Perry, C. L. 1966, AJ, 71, 709

Crawford, D. L. & Perry, C. L. 1966, AJ, 71, 206

Cuffey, J. 1973, AJ, 78, 408

Cutri, R. M., Skrutskie, M. F., van Dyk, S., Beichman, C. A., Carpenter, J. M., Chester, T., Cambresy, L., Evans, T., Fowler, J., Gizis, J., Howard, E., Huchra, J., Jarrett, T., Kopan, E. L., Kirkpatrick, J. D., Light, R. M., Marsh, K. A., McCallon, H., Schneider, S., Stiening, R., Sykes, M., Weinberg, M., Wheaton, W. A., Wheelock, S., & Zacarias, N. 2003a, 2MASS All Sky Catalog of point sources.

—. 2003b, VizieR Online Data Catalog, 2246

da Silva, R., Milone, A. d. C., & Rocha-Pinto, H. J. 2015, A&A, 580, A24

Davis, J., Tango, W. J., & Booth, A. J. 2000, MNRAS, 318, 387

de Bruijne, J. H. J., Hoogerwerf, R., & de Zeeuw, P. T. 2000, ApJL, 544, L65

—. 2001, A&A, 367, 111

de Vaucouleurs, G. 1958, ApJ, 128, 465

—. 1959, Planet. Space Sci., 2, 26

Dean, J. F. 1981, South African Astronomical Observatory Circular, 6, 10

Deka-Szymankiewicz, B., Niedzielski, A., Adamczyk, M., Adamów, M., Nowak, G., & Wolszczan, A. 2018, A&A, 615, A31

Demory, B.-O., Ségransan, D., Forveille, T., Queloz, D., Beuzit, J.-L., Delfosse, X., di Folco, E., Kervella, P., Le Bouquin, J.-B., Perrier, C., Benisty, M., Duvert, G., Hofmann, K.-H., Lopez, B., & Petrov, R. 2009, A&A, 505, 205

Di Benedetto, G. P. 2005, MNRAS, 357, 174

Di Folco, E., Thévenin, F., Kervella, P., Domiciano de Souza, A., Coudé du Foresto, V., Ségransan, D., & Morel, P. 2004, A&A, 426, 601

Ducati, J. R. 2002, VizieR Online Data Catalog, 2237

Dyck, H. M., Benson, J. A., van Belle, G. T., & Ridgway, S. T. 1996a, AJ, 111, 1705

Dyck, H. M., van Belle, G. T., & Benson, J. A. 1996b, AJ, 112, 294

Dyck, H. M., van Belle, G. T., & Thompson, R. R. 1998, AJ, 116, 981

Dzervitis, U. & Paupers, O. 1981, Issledovaniya Solntsa i Krasnykh Zvezd, 13, 22

Eggen, O. J. 1963, AJ, 68, 483

—. 1965, AJ, 70, 19

—. 1989, PASP, 101, 54

Engels, D., Sherwood, W. A., Wamsteker, W., & Schultz, G. V. 1981, A&AS, 45, 5

Ertel, S., Absil, O., Defrère, D., Le Bouquin, J.-B., Augereau, J.-C., Marion, L., Blind, N., Bonsor, A., Bryden, G., Lebreton, J., & Milli, J. 2014, A&A, 570, A128

Ertel, S., Defrère, D., Absil, O., Le Bouquin, J. B., Augereau, J. C., Berger, J. P., Blind, N., Bonsor, A., Lagrange, A. M., Lebreton, J., Marion, L., Milli, J., & Olofsson, J. 2016, A&A, 595, A44

Evans, D. S., McWilliam, A., Sandmann, W. H., & Frueh, M. 1986, AJ, 92, 1210

Fabregat, J. & Reglero, V. 1990, A&AS, 82, 531

Farnham, T. L., Schleicher, D. G., & A'Hearn, M. F. 2000, Icarus, 147, 180

Feast, M. W., Whitelock, P. A., & Carter, B. S. 1990, MNRAS, 247, 227

Fehrenbach, C. 1966, Publications of the Observatoire Haute-Provence, 8, 25

Fehrenbach, C., Petit, M., Cruvellier, G., & Peyrin, Y. 1961, Journal des Observateurs, 44, 233

Fernie, J. D. 1969, JRASC, 63, 133

—. 1983, ApJS, 52, 7

Forbes, M. C., Dodd, R. J., & Sullivan, D. J. 1993, Baltic Astronomy, 2, 246

Frogel, J. A., Persson, S. E., Matthews, K., & Aaronson, M. 1978, ApJ, 220, 75

Fuhrmann, K. 1998, A&A, 338, 161

Garrison, R. F. & Gray, R. O. 1994, AJ, 107, 1556

Garrison, R. F. & Kormendy, J. 1976, PASP, 88, 865

Gaze, V. F. & Shajn, G. A. 1952, Izvestiya Ordena Trudovogo Krasnogo Znameni Krymskoj Astrofizicheskoj Observatorii, 9, 52

Gehrels, T., Coffeen, T., & Owings, D. 1964, AJ, 69

Gezari, D. Y., Schmitz, M., Pitts, P. S., & Mead, J. M. 1993, Far infrared supplement: Catalog of infrared observations (lambda >= 4.6 micrometer) (NASA)

Ghosh, S. K., Iyengar, K. V. K., Tandon, S. N., Verma, R. P., Daniel, R. R., & Rengarajan, T. N. 1984, MNRAS, 206, 611

Giclas, H. L. 1954, AJ, 59, 128

Ginestet, N. & Carquillat, J. M. 2002, ApJS, 143, 513

Ginestet, N., Carquillat, J. M., Jaschek, M., & Jaschek, C. 1994, A&AS, 108, 359

Glass, I. S. 1974, Monthly Notes of the Astronomical Society of South Africa, 33, 53

—. 1975, MNRAS, 171, 19P

Gnedin, I. N., Khozov, G. V., & Larionov, V. M. 1982, Soviet Astronomy Letters, 8, 689

Gnedin, Y. N., Khozov, G. V., & Larionov, V. M. 1981, Soviet Astronomy Letters, 7, 466

Golay, M. 1972, Vistas in Astronomy, 14, 13

Gontcharov, G. A. 2008, Astronomy Letters, 34, 785

—. 2017, Astronomy Letters, 43, 545

Grant, G. 1959, ApJ, 129, 62

Grasdalen, G. L. 1974, AJ, 79, 1047

GRAVITY Collaboration, Perraut, K., Labadie, L., Lazareff, B., Klarmann, L., Segura-Cox, D., Benisty, M., Bouvier, J., Brandner, W., Caratti O Garatti, A., Caselli, P., Dougados, C., Garcia, P., Garcia-Lopez, R., Kendrew, S., Koutoulaki, M., Kervella, P., Lin, C. C., Pineda, J., Sanchez-Bermudez, J., van Dishoeck, E., Aabuter, R., Amorim, A., Berger, J. P., Bonnet, H., Buron, A., Cantalloube, F., Clénet, Y., Coudé Du Foresto, V., Dexter, J., de Zeeuw, P. T., Duvert, G., Eckart, A., Eisenhauer, F., Eupen, F., Gao, F., Gendron, E., Genzel, R., Gillessen, S., Gordo, P., Grellmann, R., Haubois, X., Haussmann, F., Henning, T., Hippler, S., Horrobin, M., Hubert, Z., Jocou, L., Lacour, S., Le Bouquin, J. B., Léna, P., Mérand, A., Ott, T., Paumard, T., Perrin, G., Pfuhl, O., Rabien, S., Ray, T., Rau, C., Rousset, G., Scheithauer, S., Straub, O., Straubmeier, C., Sturm, E., Vincent, F., Waisberg, I., Wank, I., Widmann, F., Wieprecht, E., Wiest, M., Wiezorrek, E., Woillez, J., & Yazici, S. 2019, A&A, 632, A53

Gray, R. O., Corbally, C. J., Garrison, R. F., McFadden, M. T., & Robinson, P. E. 2003, AJ, 126, 2048

Gray, R. O. & Olsen, E. H. 1991, A&AS, 87, 541

Gregg, M. D., Silva, D., Rayner, J., Valdes, F., Worthey, G., Pickles, A., Rose, J. A., Vacca, W., & Carney, B. 2004, in American Astronomical Society Meeting Abstracts, Vol. 205, American Astronomical Society Meeting Abstracts, 94.06

Grønbech, B., Olsen, E. H., & Strömgren, B. 1976, A&AS, 26, 155

Guetter, H. H. 1980, PASP, 92, 215

Guetter, H. H. & Hewitt, A. V. 1984, PASP, 96, 441

Gutierrez-Moreno, A. & et al. 1966, Publications of the Department of Astronomy University of Chile, 1

Gyldenkerne, K. 1955, ApJ, 121, 38

Haberreiter, M., Schmutz, W., & Kosovichev, A. G. 2008, ApJL, 675, L53

Häggkvist, L. 1966, Arkiv for Astronomi, 4, 165

Häggkvist, L. & Oja, T. 1966, Arkiv for Astronomi, 4, 137

—. 1969a, Arkiv for Astronomi, 5, 303

—. 1969b, Arkiv for Astronomi, 5, 125

—. 1970, A&AS, 1, 199

Haggkvist, L. & Oja, T. 1970, Private Communication

—. 1973, A&AS, 12, 381

Halliday, I. 1955, ApJ, 122, 222

Hardie, R. H. Photoelectric Reductions, ed. W. A. Hiltner, 178

Harlan, E. A. 1969, AJ, 74, 916

Harmanec, P., Horn, J., Koubsky, P., Zdarsky, F., Kriz, S., & Pavlovski, K. 1980, Bulletin of the Astronomical Institutes of Czechoslovakia, 31, 144

Hauck, B. & Mermilliod, M. 1998, A&AS, 129, 431

Heap, S. R. & Lindler, D. 2010, in American Astronomical Society Meeting Abstracts, Vol. 215, American Astronomical Society Meeting Abstracts #215, 463.02

Heard, J. F. 1956, Publications of the David Dunlap Observatory, 2, 107

Heiter, U., Jofré, P., Gustafsson, B., Korn, A. J., Soubiran, C., & Thévenin, F. 2015, A&A, 582, A49

Hekker, S. & Meléndez, J. 2007, A&A, 475, 1003

Helfer, H. L. & Sturch, C. 1970, AJ, 75, 971

Henry, G. W., Fekel, F. C., Henry, S. M., & Hall, D. S. 2000, ApJS, 130, 201

Herbig, G. H. 1960, ApJ, 131, 632

Hertzsprung, E. 1909, Astronomische Nachrichten, 179, 373

—. 1911, Publikationen des Astrophysikalischen Observatoriums zu Potsdam, 63

Hoffleit, D. 1942, Harvard College Observatory Circular, 448, 1

Hoffleit, D. & Jaschek, C. 1982, The Bright Star Catalogue

Hoffleit, D. & Shapley, H. 1937, Annals of Harvard College Observatory, 105, 45

Hofmann, K.-H. & Scholz, M. 1998, A&A, 335, 637

Hogg, A. R. 1958, Mount Stromlo Observatory Mimeographs, 2

Hossack, W. R. 1954, ApJ, 119, 613

Houk, N. 1982, Michigan Catalogue of Two-dimensional Spectral Types for the HD stars. Volume_3. Declinations -40 to -26 (Dept. of Astronomy, University of Michigan)

Husser, T.-O., Wende-von Berg, S., Dreizler, S., Homeier, D., Reiners, A., Barman, T., & Hauschildt, P. H. 2013, A&A, 553, A6

Hutter, D. J., Johnston, K. J., Mozurkewich, D., Simon, R. S., Colavita, M. M., Pan, X. P., Shao, M., Hines, B. E., Staelin, D. H., Hershey, J. L., Hughes, J. A., & Kaplan, G. H. 1989, ApJ, 340, 1103

Hyland, A. R., Becklin, E. E., Frogel, J. A., & Neugebauer, G. 1972, A&A, 16, 204

Hyland, A. R., Becklin, E. E., Neugebauer, G., & Wallerstein, G. 1969, ApJ, 158, 619

Iijima, T. & Ishida, K. 1978, PASJ, 30, 657

Iriarte, B. 1970, Boletin de los Observatorios Tonantzintla y Tacubaya, 5, 317

Irwin, J. B. 1961, ApJS, 6, 253

Ito, M., Kasaba, Y., Ueno, M., Sato, S., & Kimata, M. 1995, PASP, 107, 691

Jameson, R. F. & Akinci, R. 1979, MNRAS, 188, 421

Janes, K. A. 1979, ApJS, 39, 135

Janulis, R. 1986, Vilnius Astronomijos Observatorijos Biuletenis, 75, 8

Jao, W.-C., Henry, T. J., Gies, D. R., & Hambly, N. C. 2018, ApJL, 861, L11

Jasevicius, V., Kuriliene, G., Strazdaite, V., Kazlauskas, A., Sleivyte, J., & Cernis, K. 1990, Vilnius Astronomijos Observatorijos Biuletenis, 85, 50

Jennens, P. A. & Helfer, H. L. 1975, MNRAS, 172, 667

Jofré, E., Petrucci, R., Saffe, C., Saker, L., Artur de la Villarmois, E., Chavero, C., Gómez, M., & Mauas, P. J. D. 2015, A&A, 574, A50

Johansen, K. T. & Gyldenkerne, K. 1970, A&AS, 1, 165

Johnson, H. L. 1953, ApJ, 117, 361

—. 1964, Boletin de los Observatorios Tonantzintla y Tacubaya, 3, 305

—. 1965a, Communications of the Lunar and Planetary Laboratory, 3, 67

—. 1965b, ApJ, 141, 923

—. 1965c, Communications of the Lunar and Planetary Laboratory, 3, 79

—. 1967, ApJ, 149, 345

—. 1968, in Nebulae and Interstellar Matter, ed. B. M. Middlehurst & L. H. Aller (University of Chicago Press), 167

Johnson, H. L. & Harris, D. L. 1954, ApJ, 120, 196

Johnson, H. L. & Knuckles, C. F. 1955, ApJ, 122, 209

—. 1957, ApJ, 126, 113

Johnson, H. L., Low, F. J., & Steinmetz, D. 1965a, ApJ, 142, 808

—. 1965b, Communications of the Lunar and Planetary Laboratory, 3, 95

Johnson, H. L., Mendoza V., E. E., & Wisniewski, W. Z. 1965c, ApJ, 142, 1249

Johnson, H. L., Mendoza v., E. E., & Wisniewski, W. Z. 1965d, Communications of the Lunar and Planetary Laboratory, 3, 97

Johnson, H. L. & Mitchell, R. I. 1995, VizieR Online Data Catalog, 2084

Johnson, H. L., Mitchell, R. I., Iriarte, B., & Wisniewski, W. Z. 1966, Communications of the Lunar and Planetary Laboratory, 4, 99

Johnson, H. L. & Morgan, W. W. 1951, ApJ, 114, 522

—. 1953a, ApJ, 117, 313

—. 1953b, ApJ, 117, 313

Jones, D. H. P. 1972, ApJ, 178, 467

Jones, J., White, R. J., Boyajian, T., Schaefer, G., Baines, E., Ireland, M., Patience, J., ten Brummelaar, T., McAlister, H., Ridgway, S. T., Sturmann, J., Sturmann, L., Turner, N., Farrington, C., & Goldfinger, P. J. 2015, ApJ, 813, 58

Joy, A. H. 1942, ApJ, 96, 344

Kaiser, M. E., Kruk, J. W., McCand liss, S. R., Sahnow, D. J., Barkhouser, R. H., Van Dixon, W., Feldman, P. D., Moos, H. W., Orndorff, J., Pelton, R., Riess, A. G., Rauscher, B. J., Kimble, R. A., Benford, D. J., Gardner, J. P., Hill, R. J., Woodgate, B. E., Bohlin, R. C., Deustua, S. E., Kurucz, R., Lampton, M., Perlmutter, S., & Wright, E. L. 2010, arXiv e-prints, arXiv:1001.3925

Kaiser, M. E., Kruk, J. W., McCand liss, S. R., Sahnow, D. J., Rauscher, B. J., Benford, D. J., Bohlin, R. C., Deustua, S. E., Dixon, W. V., Feldman, P. D., Gardner, J. P., Kimble, R. A., Kurucz, R., Lampton, M., Moos, H. W., Perlmutter, S., Riess, A. G., Woodgate, B. E., & Wright, E. L. 2008, in Society of Photo-Optical Instrumentation Engineers (SPIE) Conference Series, Vol. 7014, Ground-based and Airborne Instrumentation for Astronomy II, 70145Y

Kakaras, G., Straizys, V., Sudzius, J., & Zdanavicius, K. 1968, Vilnius Astronomijos Observatorijos Biuletenis, 22, 3

Kazlauskas, A., Boyle, R. P., Philip, A. G. D., Straižys, V., Laugalys, V., Černis, K., Bartašiūtė, S., & Sperauskas, J. 2005, Baltic Astronomy, 14, 465

Keenan, P. C. 1942, ApJ, 95, 461

Keenan, P. C. & Keller, G. 1953, ApJ, 117, 241

Keenan, P. C. & McNeil, R. C. 1989, ApJS, 71, 245

Keenan, P. C. & Wilson, O. C. 1977, ApJ, 214, 399

Kendall, M. G., Stuart, A., & Ord, J. K. 1987, Kendall's Advanced Theory of Statistics (USA: Oxford University Press, Inc.)

Kenyon, S. J. 1988, AJ, 96, 337

Kenyon, S. J. & Gallagher, J. S. 1983, AJ, 88, 666

Kerschbaum, F. 1995, A&AS, 113, 441

Kerschbaum, F. & Hron, J. 1994, A&AS, 106, 397

Kerschbaum, F., Lazaro, C., & Habison, P. 1996, A&AS, 118, 397

Kervella, P., Nardetto, N., Bersier, D., Mourard, D., & Coudé du Foresto, V. 2004, A&A, 416, 941

Kluska, J., Van Winckel, H., Hillen, M., Berger, J. P., Kamath, D., Le Bouquin, J. B., & Min, M. 2019, A&A, 631, A108

Knight, M. M. & Schleicher, D. G. 2015, AJ, 149, 19

Kodaira, K. & Lenzen, R. 1983, A&A, 126, 440

Kornilov, V. G., Volkov, I. M., Zakharov, A. I., Kozyreva, V. S., Kornilova, L. N., Krutyakov, A. N., Krylov, A. V., Kusakin, A. V., Leont'ev, S. E., Mironov, A. V., Moshkalev, V. G., Pogrosheva, T. M., Sementsov, V. N., & Khaliullin, K. F. 1991, Trudy Gosudarstvennogo Astronomicheskogo Instituta, 63, 1

Kraft, R. P. & Hiltner, W. A. 1961, ApJ, 134, 850

Kyrolainen, J., Tuominen, I., Vilhu, O., & Virtanen, H. 1986, A&AS, 65, 11

Landolt, A. U. 1967, AJ, 72, 1012

—. 1968, PASP, 80, 749

—. 1975, PASP, 87, 379

Lane, B. F., Boden, A. F., & Kulkarni, S. R. 2001, ApJL, 551, L81

Laney, C. D., Joner, M. D., & Pietrzyński, G. 2012, MNRAS, 419, 1637

Lazareff, B., Berger, J. P., Kluska, J., Le Bouquin, J. B., Benisty, M., Malbet, F., Koen, C., Pinte, C., Thi, W. F., Absil, O., Baron, F., Delboulbé, A., Duvert, G., Isella, A., Jocou, L., Juhasz, A., Kraus, S., Lachaume, R., Ménard, F., Millan-Gabet, R., Monnier, J. D., Moulin, T., Perraut, K., Rochat, S., Soulez, F., Tallon, M., Thiébaut, E., Traub, W., & Zins, G. 2017, A&A, 599, A85

Lee, O. J., Baldwin, R. J., Hamlin, D. W., Bartlett, T. J., Gore, G. D., & Baldwin, T. J. 1943, Annals of the Dearborn Observatory, 5

Lee, T. A. 1970, ApJ, 162, 217

Leitherer, C. & Wolf, B. 1984, A&A, 132, 151

Levato, H. & Abt, H. A. 1978, PASP, 90, 429

Lindegren, L., Hernández, J., Bombrun, A., Klioner, S., Bastian, U., Ramos-Lerate, M., de Torres, A., Steidelmüller, H., Stephenson, C., Hobbs, D., Lammers, U., Biermann, M., Geyer, R., Hilger, T., Michalik, D., Stampa, U., McMillan, P. J., Castañeda, J., Clotet, M., Comoretto, G., Davidson, M., Fabricius, C., Gracia, G., Hambly, N. C., Hutton, A., Mora, A., Portell, J., van Leeuwen, F., Abbas, U., Abreu, A., Altmann, M., Andrei, A., Anglada, E., Balaguer-Núñez, L., Barache, C., Becciani, U., Bertone, S., Bianchi, L., Bouquillon, S., Bourda, G., Brüsemeister, T., Bucciarelli, B., Busonero, D., Buzzi, R., Cancelliere, R., Carlucci, T., Charlot, P., Cheek, N., Crosta, M., Crowley, C., de Bruijne, J., de Felice, F., Drimmel, R., Esquej, P., Fienga, A., Fraile, E., Gai, M., Garralda, N., González-Vidal, J. J., Guerra, R., Hauser, M., Hofmann, W., Holl, B., Jordan, S., Lattanzi, M. G., Lenhardt, H., Liao, S., Licata, E., Lister, T., Löffler, W., Marchant, J., Martin-Fleitas, J. M., Messineo, R., Mignard, F., Morbidelli, R., Poggio, E., Riva, A., Rowell, N., Salguero, E., Sarasso, M., Sciacca, E., Siddiqui, H., Smart, R. L., Spagna, A., Steele, I., Taris, F., Torra, J., van Elteren, A., van Reeven, W., & Vecchiato, A. 2018, A&A, 616, A2

Linfield, R. P., Colavita, M. M., & Lane, B. F. 2001, ApJ, 554, 505

Linnell, A. P. 1982, PASP, 94, 374

Ljunggren, B. 1966, Arkiv for Astronomi, 3, 535

Ljunggren, B. & Oja, T. 1965, Arkiv for Astronomi, 3, 439

—. 1966, Arkiv for Astronomi, 3, 501

Lomaeva, M., Jönsson, H., Ryde, N., Schultheis, M., & Thorsbro, B. 2019, A&A, 625, A141

Low, F. J. & Johnson, H. L. 1964, ApJ, 139, 1130

Low, F. J., Johnson, H. L., Kleinmann, D. E., Latham, A. S., & Geisel, S. L. 1970, ApJ, 160, 531

Lu, P. K. 1991, AJ, 101, 2229

Lu, P. K., Dawson, D. W., Upgren, A. R., & Weis, E. W. 1983, ApJS, 52, 169

Luck, R. E. 2015, AJ, 150, 88

Lutz, T. E. & Lutz, J. H. 1977, AJ, 82, 431

Maldonado, J. & Villaver, E. 2016, A&A, 588, A98

Mallik, S. V. 1998, A&A, 338, 623

Malmquist, K. G. 1960, Uppsala Astronomical Observatory Annals, 4

Mann, A. W. & von Braun, K. 2015, PASP, 127, 102

Massey, P., Levine, S. E., Neugent, K. F., Levesque, E., Morrell, N., & Skiff, B. 2018, AJ, 156, 265

Matrozis, E., Ryde, N., & Dupree, A. K. 2013, A&A, 559, A115

Mazzei, P. & Pigatto, L. 1988, A&A, 193, 148

McClure, R. D. 1970, AJ, 75, 41

McClure, R. D. & Forrester, W. T. 1981, Publications of the Dominion Astrophysical Observatory Victoria, 15, 439

McCollum, B., Laine, S., & McCollum, M. 2018, Research Notes of the American Astronomical Society, 2, 193

McCuskey, S. W. 1955, ApJS, 2, 75

—. 1967, AJ, 72, 1199

McWilliam, A. 1990, ApJS, 74, 1075

McWilliam, A. & Lambert, D. L. 1984, PASP, 96, 882

Medhi, B. J., Messina, S., Parihar, P. S., Pagano, I., Muneer, S., & Duorah, K. 2007, A&A, 469, 713

Meistas, E. & Zitkevicius, V. 1976, Vilnius Astronomijos Observatorijos Biuletenis, 45, 13

Mendoza, E. E. 1967, Boletin de los Observatorios Tonantzintla y Tacubaya, 4, 114

Mendoza, E. E., Gomez, V. T., & Gonzalez, S. 1978, AJ, 83, 606

Mendoza v., E. E. 1963, Boletin de los Observatorios Tonantzintla y Tacubaya, 3, 137

Mermilliod, J.-C. 1986, Catalogue of Eggen's UBV data., 0 (1986)

Mermilliod, J.-C., Mermilliod, M., & Hauck, B. 1997, A&AS, 124, 349

Mermilliod, J. C. & Nitschelm, C. 1989, A&AS, 81, 401

Miczaika, G. R. 1954, AJ, 59, 233

Moffett, T. J. & Barnes, III, T. G. 1979, PASP, 91, 180

—. 1980, ApJS, 44, 427

Mondal, S. & Chandrasekhar, T. 2005, AJ, 130, 842

Moore, J. H. 1932, Publications of Lick Observatory, 18

Moore, J. H. & Paddock, G. F. 1950, ApJ, 112, 48

Moreno, H. 1971, A&A, 12, 442

Morgan, W. W. & Hiltner, W. A. 1965, ApJ, 141, 177

Morgan, W. W. & Keenan, P. C. 1973, ARA&A, 11, 29

Mortier, A., Santos, N. C., Sousa, S. G., Adibekyan, V. Z., Delgado Mena, E., Tsantaki, M., Israelian, G., & Mayor, M. 2013, A&A, 557, A70

Mozurkewich, D., Armstrong, J. T., Hindsley, R. B., Quirrenbach, A., Hummel, C. A., Hutter, D. J., Johnston, K. J., Hajian, A. R., Elias, II, N. M., Buscher, D. F., & Simon, R. S. 2003, AJ, 126, 2502

Mozurkewich, D., Johnston, K. J., Simon, R. S., Bowers, P. F., Gaume, R., Hutter, D. J., Colavita, M. M., Shao, M., & Pan, X. P. 1991, AJ, 101, 2207

Nassau, J. J. & Blanco, V. M. 1954, ApJ, 120, 118

Neckel, H. 1958, ApJ, 128, 510

—. 1974, A&AS, 18, 169

Neilson, H. R. & Lester, J. B. 2013, A&A, 554, A98

Neugebauer, G. & Leighton, R. B. 1969, Two-micron sky survey. A preliminary catalogue (NASA)

Newberg, H. J. & Yanny, B. 1998, ApJL, 499, L57

Ney, E. P. & Merrill, K. M. 1980, Study of Sources in AFGL Rocket Infrared Study, Final Scientific Report 1 July 1986-30 Sept 1979

Nicolet, B. 1978, A&AS, 34, 1

Noguchi, K., Kawara, K., Kobayashi, Y., Okuda, H., Sato, S., & Oishi, M. 1981, PASJ, 33, 373

Nordgren, T. E., Germain, M. E., Benson, J. A., Mozurkewich, D., Sudol, J. J., Elias, II, N. M., Hajian, A. R., White, N. M., Hutter, D. J., Johnston, K. J., Gauss, F. S., Armstrong, J. T., Pauls, T. A., & Rickard, L. J. 1999, AJ, 118, 3032

Nordgren, T. E., Sudol, J. J., & Mozurkewich, D. 2001, AJ, 122, 2707

Ochsenbein, F., Bauer, P., & Marcout, J. 2000, A&AS, 143, 23

O'Connell, D. J. K. 1964, Ricerche Astronomiche, 6

Oja, T. 1963, Arkiv for Astronomi, 3, 273

—. 1970, Private Communication

—. 1983, A&AS, 52, 131

—. 1984, A&AS, 57, 357

—. 1985a, A&AS, 61, 331

—. 1985b, A&AS, 59, 461

—. 1986, A&AS, 65, 405

—. 1987, A&AS, 71, 561

—. 1991, A&AS, 89, 415

—. 1993, A&AS, 100, 591

Olsen, E. H. 1982, A&AS, 48, 165

—. 1983, A&AS, 54, 55

—. 1993, A&AS, 102, 89

Olson, E. C. 1974, AJ, 79, 1424

Oosterhoff, P. T. 1960, BAN, 15, 199

Paladini, C., van Belle, G. T., Aringer, B., Hron, J., Reegen, P., Davis, C. J., & Lebzelter, T. 2011, A&A, 533, A27

Parkhurst, J. A. 1912, ApJ, 36

Parsons, S. B. & Ake, T. B. 1998, ApJS, 119, 83

Peña, J. H., Peniche, R., Mújica, R., Ibanoglu, C., Ertan, A. Y., Tumer, O., Tunca, Z., & Evren, S. 1993, RMxAA, 25, 129

Pedoussaut, A. 1962, Journal des Observateurs, 45, 257

Perraud, H. 1961, Journal des Observateurs, 44, 247

Perryman, M. A. C., Lindegren, L., Kovalevsky, J., Hoeg, E., Bastian, U., Bernacca, P. L., Crézé, M., Donati, F., Grenon, M., van Leeuwen, F., van der Marel, H., Mignard, F., Murray, C. A., Le Poole, R. S., Schrijver, H., Turon, C., Arenou, F., Froeschlé, M., & Petersen, C. S. 1997, A&A, 323, L49

Persi, P., Ferrari-Toniolo, M., Spada, G., Conti, G., di Benedetto, P., Tanzi, E. G., & Tarenghi, M. 1979, MNRAS, 187, 293

Philip, A. G. D. & Philip, K. D. 1973, ApJ, 179, 855

Phillips, J. P., Selby, M. J., Wade, R., & Sanchez Magro, C. 1980, MNRAS, 190, 337

Pickles, A. J. 1998, PASP, 110, 863

Pietrinferni, A., Cassisi, S., Salaris, M., & Castelli, F. 2006, ApJ, 642, 797

Piirola, V. 1976, Observatory and Astrophysics Laboratory University of Helsinki Report, 1, 0

Pilachowski, C. A. 1978, PASP, 90, 683

Popper, D. M. 1959, ApJ, 129, 647

Rachford, B. L. & Canterna, R. 2000, AJ, 119, 1296

Ramírez, I., Allende Prieto, C., & Lambert, D. L. 2013, ApJ, 764, 78

Reglero, V., Gimenez, A., de Castro, E., & Fernandez-Figueroa, M. J. 1987, A&AS, 71, 421

Richichi, A. & Wittkowski, M. 2003, Ap&SS, 286, 219

Ridgway, S. T., Jacoby, G. H., Joyce, R. R., Siegel, M. J., & Wells, D. C. 1982, AJ, 87, 808

Ridgway, S. T., Jacoby, G. H., Joyce, R. R., & Wells, D. C. 1980a, AJ, 85, 1496

Ridgway, S. T., Joyce, R. R., White, N. M., & Wing, R. F. 1980b, ApJ, 235, 126

Ridgway, S. T., Wells, D. C., & Joyce, R. R. 1977, AJ, 82, 414

Ridgway, S. T., Wells, D. C., Joyce, R. R., & Allen, R. G. 1979, AJ, 84, 247

Roman, N. G. 1952, ApJ, 116, 122

—. 1955, ApJS, 2, 195

Russell, H. N. 1914, Popular Astronomy, 22, 275

Rybka, E. 1969, AcA, 19, 229

Rydgren, A. E. & Vrba, F. J. 1983, AJ, 88, 1017

S., X. 1932, Meddelanden fran Astronomiska Observatorium Uppsala, 54

Sanwal, N. B., Parthasarathy, M., & Abhyankar, K. D. 1973, The Observatory, 93, 30

Sato, K. & Kuji, S. 1990, A&AS, 85, 1069

Schild, R. E. 1973, AJ, 78, 37

Schleicher, D. G. & Bair, A. N. 2011, AJ, 141, 177

Schmidtke, P. C., Africano, J. L., Jacoby, G. H., Joyce, R. R., & Ridgway, S. T. 1986, AJ, 91, 961

Schmitt, J. L. 1971, ApJ, 163, 75

Scholz, M. & Takeda, Y. 1987, A&A, 186, 200

Selby, M. J., Hepburn, I., Blackwell, D. E., Booth, A. J., Haddock, D. J., Arribas, S., Leggett, S. K., & Mountain, C. M. 1988, A&AS, 74, 127

Serkowski, K. 1961, Lowell Observatory Bulletin, 5, 157

Shao, C. Y. 1964, AJ, 69, 858

Sharpless, S. 1952, ApJ, 116, 251

Shaw, J. S. & Guinan, E. F. 1989, AJ, 97, 836

Shenavrin, V. I., Taranova, O. G., & Nadzhip, A. E. 2011, Astronomy Reports, 55, 31

Skiff, B. A. 2014a, VizieR Online Data Catalog, B/mk

—. 2014b, VizieR Online Data Catalog, 1

Skopal, A. 2015, NewA, 36, 128

Sleivyte, J. 1985, Vilnius Astronomijos Observatorijos Biuletenis, 69, 9

—. 1987, Vilnius Astronomijos Observatorijos Biuletenis, 77, 32

Slutskij, V. E., Stalbovskij, O. I., & Shevchenko, V. S. 1980, Soviet Astronomy Letters, 6, 750

Smak, J. 1964, ApJS, 9, 141

Smith, B. J., Price, S. D., & Baker, R. I. 2004, ApJS, 154, 673

Smith, V. V. & Lambert, D. L. 1986, ApJ, 311, 843

—. 1990, ApJS, 72, 387

Soubiran, C., Le Campion, J.-F., Brouillet, N., & Chemin, L. 2016, A&A, 591, A118

Sperauskas, J., Bartkevicius, A., & Zdanavicius, K. 1981, Vilnius Astronomijos Observatorijos Biuletenis, 58, 3

Stephenson, C. B. 1960, AJ, 65, 60

Stephenson, C. B. & Sanwal, N. B. 1969, AJ, 74, 689

Straizys, V., Kazlauskas, A., Jodinskiene, E., & Bartkevicius, A. 1989a, Bulletin d'Information du Centre de Donnees Stellaires, 37, 179

Straizys, V., Kazlauskas, A., Vansevicius, V., & Cernis, K. 1993, Baltic Astronomy, 2, 171

Straizys, V. & Meistas, E. 1989, Vilnius Astronomijos Observatorijos Biuletenis, 84, 26

Straizys, V., Meistas, E., Vansevicius, V., & Goldberg, E. P. 1989b, Vilnius Astronomijos Observatorijos Biuletenis, 83, 3

Strassmeier, K. G. & Fekel, F. C. 1990, A&A, 230, 389

Strecker, D. W., Erickson, E. F., & Witteborn, F. C. 1979, ApJS, 41, 501

Sturch, C. R. & Helfer, H. L. 1972, AJ, 77, 726

Sudzius, J., Zdanavicius, K., Sviderskiene, Z., Straizys, V., Bartkevicius, A., Zitkevicius, V., Kavaliauskaite, G., & Kakaras, G. 1970, Vilnius Astronomijos Observatorijos Biuletenis, 29, 3

Swings, J. P. & Allen, D. A. 1972, PASP, 84, 523

Swings, P., McKellar, A., & Rao, K. N. 1953, MNRAS, 113, 571

Szabados, L. 1977, Mitt. Sternw. Ungarisch. Akad. Wiss., Budapest-Szabadsághegy, No. 70, 123 p., 70

Tej, A. & Chandrasekhar, T. 2000, MNRAS, 317, 687

Thygesen, A. O., Frandsen, S., Bruntt, H., Kallinger, T., Andersen, M. F., Elsworth, Y. P., Hekker, S., Karoff, C., Stello, D., Brogaard, K., Burke, C., Caldwell, D. A., & Christiansen, J. L. 2012, A&A, 543, A160

Tolbert, C. R. 1964, ApJ, 139, 1105

Turnshek, D. A., Turnshek, D. E., & Craine, E. R. 1980, AJ, 85, 1638

Uranova, T. A. 1977, Astronomicheskij Tsirkulyar, 946, 4

van Belle, G. T. 1999, PASP, 111, 1515

van Belle, G. T., Ciardi, D. R., & Boden, A. F. 2007, ApJ, 657, 1058

van Belle, G. T., Creech-Eakman, M. J., & Hart, A. 2009, MNRAS, 394, 1925

van Belle, G. T., Creech-Eakman, M. J., & Ruiz-Velasco, A. E. 2016, AJ, 152, 16

van Belle, G. T., Dyck, H. M., Benson, J. A., & Lacasse, M. G. 1996, AJ, 112, 2147

van Belle, G. T., Dyck, H. M., Thompson, R. R., Benson, J. A., & Kannappan, S. J. 1997, AJ, 114, 2150

van Belle, G. T., Lane, B. F., Thompson, R. R., Boden, A. F., Colavita, M. M., Dumont, P. J., Mobley, D. W., Palmer, D., Shao, M., Vasisht, G. X., Wallace, J. K., Creech-Eakman, M. J., Koresko, C. D., Kulkarni, S. R., Pan, X. P., & Gubler, J. 1999, AJ, 117, 521

van Belle, G. T., Paladini, C., Aringer, B., Hron, J., & Ciardi, D. 2013, ApJ, 775, 45

van Belle, G. T., Thompson, R. R., & Creech-Eakman, M. J. 2002, AJ, 124, 1706

van Belle, G. T. & van Belle, G. 2005, PASP, 117, 1263

van Belle, G. T., van Belle, G., Creech-Eakman, M. J., Coyne, J., Boden, A. F., Akeson, R. L., Ciardi, D. R., Rykoski, K. M., Thompson, R. R., Lane, B. F., & PTI Collaboration. 2008, ApJS, 176, 276

van Belle, G. T. & von Braun, K. 2009, ApJ, 694, 1085

van de Kamp, P. 1958, Publications of the Leander McCormick Observatory, 13

van Leeuwen, F. 2007, A&A, 474, 653

Voelcker, K. 1975, A&AS, 22, 1

von Braun, K. & Boyajian, T. 2017, ArXiv e-prints

von Braun, K., Boyajian, T. S., van Belle, G. T., Kane, S. R., Jones, J., Farrington, C., Schaefer, G., Vargas, N., Scott, N., ten Brummelaar, T. A., Kephart, M., Gies, D. R., Ciardi, D. R., López-Morales, M., Mazingue, C., McAlister, H. A., Ridgway, S., Goldfinger, P. J., Turner, N. H., & Sturmann, L. 2014, MNRAS, 438, 2413

Walker, Jr., R. L. 1971, PASP, 83, 177

Warren, Jr., W. H. 1973, AJ, 78, 192

Wenger, M., Ochsenbein, F., Egret, D., Dubois, P., Bonnarel, F., Borde, S., Genova, F., Jasniewicz, G., Laloë, S., Lesteven, S., & Monier, R. 2000, A&AS, 143, 9

Westerlund, B. 1962, Arkiv for Astronomi, 3, 21

Westerlund, B. E. 1953, Uppsala Astronomical Observatory Annals, 15

White, N. M. & Feierman, B. H. 1987, AJ, 94, 751

Williams, J. A. 1966, AJ, 71, 615

Wilson, R. E. & Joy, A. H. 1950, ApJ, 111, 221

—. 1952, ApJ, 115, 157

Wing, R. F. 1967, PhD thesis, UNIVERSITY OF CALIFORNIA, BERKELEY.

Wisniewski, W. Z., Wing, R. F., Spinrad, H., & Johnson, H. L. 1967, ApJL, 148, L29

Wu, F. & Wang, Z.-z. 1985, ChA&A, 9, 44

Yamashita, Y. 1967, Publications of the Dominion Astrophysical Observatory Victoria, 13, 47

Yamashita, Y. & Norimoto, Y. 1981, Annals of the Tokyo Astronomical Observatory, 18, 125

Yoss, K. M. 1961, ApJ, 134, 809

—. 1977, AJ, 82, 832

Yoss, K. M. & Griffin, R. F. 1997, Journal of Astrophysics and Astronomy, 18, 161

Zdanavicius, K. & Cerniene, E. 1985, Vilnius Astronomijos Observatorijos Biuletenis, 69, 3

Zdanavicius, K., Nikonov, V. B., Sudzius, J., Straizys, V., Sviderskiene, Z., Kalytis, R., Jodinskiene, E., Meistas, E., Kavaliauskaite, G., Jasevicius, V., Kakaras, G., Bartkevicius, A., Gurklyte, A., Bartkus, R., Azusienis, A., Sperauskas, J., Kazlauskas, A., & Zitkevicius, V. 1972, Vilnius Astronomijos Observatorijos Biuletenis, 34, 3

Zdanavicius, K., Sudzius, J., Sviderskiene, Z., Straizys, V., Burnasov, V., Drazdys, R., Bartkevicius, A., Kakaras, G., Kavaliauskaite, G., & Jasevicius, V. 1969, Vilnius Astronomijos Observatorijos Biuletenis, 26, 3

Zhao, G., Qiu, H. M., & Mao, S. 2001, ApJL, 551, L85

Zorec, J., Cidale, L., Arias, M. L., Frémat, Y., Muratore, M. F., Torres, A. F., & Martayan, C. 2009, A&A, 501, 297

**Table 15**. PHOENIX model parameters fit to INGS spectral templates

| Spectral Type | $T_{\mathrm{EFF}}$ | $\log(g)$ | $\chi^2_{\mathrm{DOF}}$ |
|---|---|---|---|
|  | 12000 | 3.0 |  |
|  | 11800 | 3.0 |  |
|  | 11600 | 3.0 |  |
| B9III | 11400 | 3.0 | 0.34 |
|  | 11200 | 3.0 |  |
|  | 11000 | 3.0 |  |
|  | 10800 | 3.0 |  |
|  | 10600 | 3.0 |  |
|  | 10400 | 3.0 |  |
|  | 10200 | 3.0 |  |
|  | 10000 | 3.0 |  |
|  | 9800 | 3.0 |  |
| A0III | 9600 | 3.0 | 0.48 |
|  | 9400 | 3.0 |  |
|  | 9200 | 3.0 |  |
| {A1III} | 9000 | 3.0 |  |
|  | 8800 | 3.0 |  |
|  | 8600 | 3.0 |  |
| {A2III} | 8400 | 3.0 |  |
|  | 8200 | 3.0 |  |
|  | 8000 | 3.0 |  |
| A3III | 7800 | 3.0 | 0.46 |
| A5III | 7600 | 3.0 | 0.16 |
| A7III | 7400 | 3.0 | 0.56 |
| {A8III} | 7200 | 3.0 |  |
| {A9III} | 7000 | 3.0 |  |
|  | 6900 | 3.0 |  |
| F0III | 6800 | 3.0 | 0.51 |
|  | 6700 | 3.0 |  |
| {F1III} | 6600 | 3.0 |  |
|  | 6500 | 3.0 |  |
| F2III | 6400 | 3.0 | 0.55 |
|  | 6300 | 3.0 |  |
| {F3III} | 6200 | 3.0 |  |
| {F4III} | 6100 | 3.0 |  |
| {F5III} | 6000 | 3.0 |  |
| {F6III} | 5900 | 3.0 |  |
| {F7III} | 5800 | 3.0 |  |

**Table 15** *(continued)*

| Spectral Type | $T_{\mathrm{EFF}}$ | $\log(g)$ | $\chi^2_{\mathrm{DOF}}$ |
|---|---|---|---|
| {F8III} | 5700 | 2.5 | |
| {F9III} | 5600 | 2.5 | |
| {G0III} | 5500 | 2.5 | |
| {G1III} | 5400 | 2.5 | |
| {G2III} | 5300 | 2.5 | |
| {G3III} | 5200 | 2.5 | |
| {G4III} | 5100 | 2.5 | |
| G5III | 5000 | 2.5 | 0.96 |
| {G6III} | 5000 | 2.5 | |
| {G7III} | 4900 | 2.5 | |
| G8III | 4900 | 2.5 | 0.85 |
| {G9III} | 4800 | 2.5 | |
| K0III | 4800 | 2.5 | 1.21 |
| {K0.5III} | 4700 | 2.5 | |
| K1III | 4700 | 2.5 | 0.71 |
| | 4600 | 2.0 | |
| {K1.5III} | 4500 | 2.0 | |
| K2III | 4400 | 2.0 | 0.93 |
| {K2.5III} | 4400 | 2.0 | |
| K3III | 4300 | 2.0 | 1.16 |
| | 4200 | 2.0 | |
| {K3.5III} | 4100 | 2.0 | |
| K4III | 4000 | 2.0 | 3.13 |
| {K4.5III} | 3900 | 1.5 | |
| K5III | 3900 | 1.5 | 2.21 |
| {K6III} | 3900 | 1.5 | |
| {K7III} | 3900 | 1.5 | |
| M0III | 3800 | 1.5 | 2.49 |
| {M0.5III} | 3800 | 1.5 | |
| M1III | 3800 | 1.5 | 3.19 |
| {M1.5III} | 3800 | 1.5 | |
| M2III | 3700 | 1.5 | 2.43 |
| {M2.5III} | 3700 | 1.5 | |
| M3III | 3600 | 1.5 | 2.77 |
| {M3.5III} | 3500 | 1.0 | |
| M4III | 3500 | 1.0 | 5.54 |
| {M4.5III} | 3400 | 1.0 | |
| M5III | 3400 | 1.0 | 4.06 |
| {M5.5III} | 3300 | 1.0 | |
| M6III | 3200 | 1.0 | 8.44 |
| {M6.5III} | 3200 | 1.0 | |

**Table 15** *(continued)*

| Spectral Type | $T_{\rm EFF}$ | $\log(g)$ | $\chi^2_{\rm DOF}$ |
|:---:|:---:|:---:|:---:|
| M7III | 3100 | 0.5 | 11.28 |
| {M7.5III} | 3000 | 0.5 | |
| | 2900 | 0.5 | |
| M8III | 2800 | 0.5 | 16.57 |
| {M8.5III} | 2700 | 0.5 | |
| M9III | 2600 | 0.0 | 26.35 |
| {M9.5III} | 2500 | 0.0 | |
| M10III | 2400 | 0.0 | 229.16 |
| | 2300 | 0.0 | |

NOTE—Spectral types in brackets do not have INGS spectral templates and are linearly inferred interpolations associated with the corresponding PHOENIX models. For more detail, see §4.1.

**Table 16**. Broadband $H-$ and $K-$band angular sizes for the program stars. Errors cited are (in order) the formal error, and the systematic error based upon a limiting $\sigma_{V^2}$ of 1.5%.

| Star ID | $N_{\mathrm{pts,H}}$ | $\theta_{\mathrm{H}}$ | $N_{\mathrm{pts,K}}$ | $\theta_{\mathrm{K}}$ | Avg. Res. | $\chi^2$/DOF |
|---|---|---|---|---|---|---|
| HD598 | – | – | 30 | $2.591 \pm 0.021 \pm 0.046$ | 0.033 | 0.28 |
| HD1632 | 10 | $2.251 \pm 0.009 \pm 0.042$ | 22 | $2.281 \pm 0.014 \pm 0.042$ | 0.029 | 1.51 |
| HD1795 | – | – | 15 | $2.024 \pm 0.049 \pm 0.042$ | 0.120 | 1.80 |
| HD3346 | – | – | 11 | $3.102 \pm 0.007 \pm 0.055$ | 0.005 | 4.43 |
| HD3546 | – | – | 75 | $1.689 \pm 0.011 \pm 0.042$ | 0.029 | 0.69 |
| HD3574 | – | – | 20 | $2.270 \pm 0.040 \pm 0.042$ | 0.037 | 0.55 |
| HD3627 | – | – | 82 | $3.968 \pm 0.005 \pm 0.120$ | 0.011 | 1.13 |
| HD5006 | – | – | 18 | $2.152 \pm 0.031 \pm 0.042$ | 0.054 | 1.49 |
| HD5575 | – | – | 25 | $3.019 \pm 0.012 \pm 0.055$ | 0.035 | 3.03 |
| HD6186 | – | – | 29 | $1.879 \pm 0.041 \pm 0.042$ | 0.074 | 0.40 |
| HD6409 | – | – | 75 | $1.937 \pm 0.014 \pm 0.042$ | 0.031 | 0.44 |
| HD7000 | – | – | 5 | $1.835 \pm 0.049 \pm 0.042$ | 0.037 | 0.57 |
| HD7087 | – | – | 35 | $1.578 \pm 0.030 \pm 0.042$ | 0.037 | 0.43 |
| HD8126 | – | – | 3 | $2.038 \pm 0.027 \pm 0.042$ | 0.009 | 0.15 |
| HD9500 | – | – | 16 | $2.313 \pm 0.012 \pm 0.042$ | 0.010 | 0.80 |
| HD9927 | – | – | 121 | $3.354 \pm 0.008 \pm 0.069$ | 0.027 | 0.69 |
| HD10380 | – | – | 130 | $2.830 \pm 0.008 \pm 0.051$ | 0.032 | 0.75 |
| HD12929 | – | – | 19 | $4.740 \pm 0.021 \pm 0.375$ | 0.007 | 0.37 |
| HD14146 | – | – | 20 | $1.974 \pm 0.027 \pm 0.042$ | 0.026 | 0.21 |
| HD14512 | – | – | 5 | $2.824 \pm 0.026 \pm 0.051$ | 0.019 | 1.87 |
| HD14872 | 12 | $3.116 \pm 0.020 \pm 0.134$ | 562 | $3.154 \pm 0.006 \pm 0.060$ | 0.055 | 0.61 |
| HD14901 | 3 | $3.330 \pm 0.004 \pm 0.208$ | 9 | $3.322 \pm 0.005 \pm 0.065$ | 0.003 | 3.54 |
| HD15656 | – | – | 64 | $2.479 \pm 0.015 \pm 0.042$ | 0.023 | 0.70 |
| HD16396 | 1 | $2.511 \pm 1.892 \pm 0.051$ | 4 | $2.213 \pm 0.097 \pm 0.042$ | 0.144 | 0.87 |
| HD17361 | – | – | 120 | $1.837 \pm 0.009 \pm 0.042$ | 0.052 | 0.96 |
| HD17709 | – | – | 185 | $3.720 \pm 0.003 \pm 0.093$ | 0.016 | 1.65 |
| HD18322 | 16 | $2.390 \pm 0.047 \pm 0.046$ | 10 | $2.538 \pm 0.037 \pm 0.046$ | 0.044 | 2.75 |
| HD19787 | – | – | 45 | $1.762 \pm 0.026 \pm 0.042$ | 0.038 | 0.59 |
| HD20844 | 8 | $2.462 \pm 0.008 \pm 0.051$ | 16 | $2.528 \pm 0.009 \pm 0.046$ | 0.008 | 0.22 |
| HD21465 | – | – | 23 | $2.095 \pm 0.052 \pm 0.042$ | 0.113 | 0.90 |
| HD25555 | – | – | 15 | $1.545 \pm 0.037 \pm 0.042$ | 0.035 | 0.77 |
| HD25604 | – | – | 20 | $1.876 \pm 0.034 \pm 0.042$ | 0.024 | 0.23 |
| HD26605 | – | – | 25 | $0.795 \pm 0.074 \pm 0.074$ | 0.038 | 0.23 |
| HD27308 | 5 | $2.535 \pm 0.011 \pm 0.056$ | 11 | $2.513 \pm 0.012 \pm 0.046$ | 0.036 | 2.57 |
| HD27348 | – | – | 5 | $1.937 \pm 0.315 \pm 0.042$ | 0.163 | 0.66 |
| HD27697 | – | – | 361 | $2.176 \pm 0.003 \pm 0.042$ | 0.023 | 0.77 |
| HD28100 | – | – | 15 | $1.516 \pm 0.027 \pm 0.046$ | 0.032 | 1.16 |
| HD28292 | – | – | 10 | $1.870 \pm 0.056 \pm 0.042$ | 0.027 | 0.38 |

**Table 16** *(continued)*

| Star ID | $N_{\mathrm{pts,H}}$ | $\theta_{\mathrm{H}}$ | $N_{\mathrm{pts,K}}$ | $\theta_{\mathrm{K}}$ | Avg. Res. | $\chi^2/\mathrm{DOF}$ |
|---|---|---|---|---|---|---|
| HD28305 | – | – | 175 | $2.444 \pm 0.005 \pm 0.042$ | 0.051 | 1.08 |
| HD28307 | – | – | 20 | $2.031 \pm 0.033 \pm 0.042$ | 0.021 | 0.27 |
| HD28581 | – | – | 5 | $2.083 \pm 0.090 \pm 0.042$ | 0.032 | 0.26 |
| HD28595 | – | – | 12 | $3.007 \pm 0.024 \pm 0.056$ | 0.023 | 0.72 |
| HD29094 | – | – | 80 | $2.607 \pm 0.009 \pm 0.046$ | 0.052 | 1.08 |
| HD30504 | – | – | 96 | $2.711 \pm 0.009 \pm 0.046$ | 0.049 | 1.02 |
| HD30605 | – | – | 85 | $1.843 \pm 0.013 \pm 0.042$ | 0.037 | 0.55 |
| HD30834 | – | – | 114 | $2.603 \pm 0.006 \pm 0.046$ | 0.027 | 0.54 |
| HD31139 | – | – | 15 | $3.442 \pm 0.018 \pm 0.074$ | 0.017 | 0.87 |
| HD31421 | – | – | 15 | $2.718 \pm 0.031 \pm 0.046$ | 0.032 | 0.95 |
| HD33463 | 12 | $2.572 \pm 0.008 \pm 0.055$ | 24 | $2.650 \pm 0.006 \pm 0.046$ | 0.011 | 1.72 |
| HD34334 | – | – | 180 | $2.512 \pm 0.010 \pm 0.046$ | 0.029 | 0.66 |
| HD34559 | – | – | 25 | $1.223 \pm 0.043 \pm 0.051$ | 0.099 | 0.47 |
| HD34577 | 4 | $2.218 \pm 0.378 \pm 0.042$ | 8 | $2.333 \pm 0.011 \pm 0.042$ | 0.013 | 1.91 |
| HD35620 | – | – | 222 | $2.065 \pm 0.013 \pm 0.042$ | 0.038 | 0.50 |
| HD38656 | – | – | 97 | $1.588 \pm 0.020 \pm 0.042$ | 0.057 | 0.90 |
| HD39003 | – | – | 60 | $2.376 \pm 0.015 \pm 0.042$ | 0.064 | 1.28 |
| HD39045 | 2 | $1.744 \pm 0.406 \pm 0.032$ | 13 | $3.105 \pm 0.010 \pm 0.055$ | 0.011 | 0.67 |
| HD39225 | – | – | 15 | $2.385 \pm 0.012 \pm 0.042$ | 0.015 | 0.79 |
| HD39732 | 16 | $2.208 \pm 0.041 \pm 0.042$ | 5 | $2.481 \pm 0.055 \pm 0.042$ | 0.013 | 0.25 |
| HD40441 | – | – | 148 | $1.529 \pm 0.010 \pm 0.042$ | 0.055 | 1.45 |
| HD41116 | – | – | 319 | $1.861 \pm 0.005 \pm 0.042$ | 0.029 | 0.83 |
| HD42049 | – | – | 10 | $2.272 \pm 0.046 \pm 0.042$ | 0.042 | 0.58 |
| HD43039 | – | – | 15 | $1.967 \pm 0.037 \pm 0.042$ | 0.038 | 4.52 |
| HD46709 | – | – | 370 | $1.683 \pm 0.008 \pm 0.042$ | 0.051 | 0.83 |
| HD48450 | 48 | $1.985 \pm 0.015 \pm 0.032$ | 136 | $2.006 \pm 0.008 \pm 0.042$ | 0.063 | 0.73 |
| HD49738 | – | – | 15 | $1.339 \pm 0.050 \pm 0.046$ | 0.024 | 0.32 |
| HD49968 | 8 | $1.869 \pm 0.074 \pm 0.032$ | 112 | $1.871 \pm 0.013 \pm 0.042$ | 0.042 | 0.80 |
| HD54719 | 4 | $2.336 \pm 0.023 \pm 0.046$ | 52 | $2.316 \pm 0.009 \pm 0.042$ | 0.039 | 0.89 |
| HD60136 | – | – | 5 | $2.464 \pm 0.075 \pm 0.042$ | 0.010 | 0.07 |
| HD61913 | – | – | 14 | $3.782 \pm 0.028 \pm 0.097$ | 0.026 | 1.77 |
| HD62044 | – | – | 568 | $2.351 \pm 0.003 \pm 0.042$ | 0.037 | 1.08 |
| HD62285 | – | – | 45 | $2.499 \pm 0.015 \pm 0.042$ | 0.030 | 0.68 |
| HD62345 | – | – | 30 | $2.318 \pm 0.016 \pm 0.042$ | 0.017 | 0.41 |
| HD62509 | – | – | 2 | $4.671 \pm 0.031 \pm 0.333$ | 0.000 | inf |
| HD62721 | – | – | 216 | $2.857 \pm 0.005 \pm 0.051$ | 0.030 | 0.94 |
| HD66216 | – | – | 40 | $1.593 \pm 0.023 \pm 0.042$ | 0.036 | 0.74 |
| HD74442 | – | – | 111 | $2.300 \pm 0.014 \pm 0.042$ | 0.056 | 0.69 |
| HD76294 | – | – | 5 | $3.214 \pm 0.062 \pm 0.060$ | 0.013 | 0.25 |
| HD76830 | – | – | 450 | $3.231 \pm 0.003 \pm 0.060$ | 0.034 | 1.43 |
| HD81817 | – | – | 30 | $3.195 \pm 0.028 \pm 0.060$ | 0.025 | 0.46 |

**Table 16** *continued on next page*

**Table 16** (continued)

| Star ID | $N_{\mathrm{pts,H}}$ | $\theta_{\mathrm{H}}$ | $N_{\mathrm{pts,K}}$ | $\theta_{\mathrm{K}}$ | Avg. Res. | $\chi^2/\mathrm{DOF}$ |
|---------|------|------|------|------|------|------|
| HD82198 | – | – | 24 | $2.671 \pm 0.015 \pm 0.046$ | 0.036 | 0.38 |
| HD82381 | – | – | 52 | $2.071 \pm 0.018 \pm 0.042$ | 0.037 | 1.12 |
| HD82635 | – | – | 44 | $1.436 \pm 0.020 \pm 0.046$ | 0.072 | 0.47 |
| HD85503 | – | – | 225 | $2.765 \pm 0.004 \pm 0.046$ | 0.016 | 0.87 |
| HD87046 | – | – | 25 | $1.896 \pm 0.011 \pm 0.042$ | 0.013 | 0.28 |
| HD87837 | – | – | 443 | $3.154 \pm 0.003 \pm 0.060$ | 0.030 | 1.30 |
| HD90254 | 16 | $2.977 \pm 0.012 \pm 0.102$ | 156 | $2.938 \pm 0.009 \pm 0.051$ | 0.035 | 0.87 |
| HD92620 | – | – | 83 | $3.565 \pm 0.005 \pm 0.079$ | 0.018 | 1.81 |
| HD94264 | 12 | $2.347 \pm 0.014 \pm 0.046$ | 84 | $2.494 \pm 0.004 \pm 0.042$ | 0.019 | 1.15 |
| HD94336 | – | – | 10 | $1.940 \pm 0.032 \pm 0.042$ | 0.035 | 0.43 |
| HD95212 | – | – | 29 | $2.104 \pm 0.029 \pm 0.042$ | 0.092 | 2.15 |
| HD95345 | – | – | 174 | $1.787 \pm 0.016 \pm 0.042$ | 0.050 | 0.82 |
| HD96274 | – | – | 15 | $1.911 \pm 0.023 \pm 0.042$ | 0.017 | 0.50 |
| HD96833 | – | – | 83 | $3.797 \pm 0.005 \pm 0.102$ | 0.028 | 2.02 |
| HD100236 | – | – | 35 | $1.679 \pm 0.022 \pm 0.042$ | 0.071 | 1.11 |
| HD102224 | – | – | 5 | $3.194 \pm 0.019 \pm 0.060$ | 0.006 | 0.65 |
| HD104207 | – | – | 404 | $3.153 \pm 0.002 \pm 0.060$ | 0.019 | 1.42 |
| HD104575 | – | – | 40 | $1.654 \pm 0.019 \pm 0.042$ | 0.052 | 0.90 |
| HD104831 | – | – | 20 | $1.540 \pm 0.042 \pm 0.042$ | 0.049 | 0.29 |
| HD104979 | – | – | 10 | $1.908 \pm 0.022 \pm 0.042$ | 0.013 | 0.25 |
| HD106714 | – | – | 20 | $1.217 \pm 0.037 \pm 0.051$ | 0.039 | 1.45 |
| HD107256 | – | – | 213 | $1.775 \pm 0.008 \pm 0.042$ | 0.048 | 0.78 |
| HD109282 | – | – | 64 | $1.859 \pm 0.013 \pm 0.042$ | 0.030 | 0.61 |
| HD113226 | – | – | 80 | $3.066 \pm 0.004 \pm 0.055$ | 0.012 | 1.45 |
| HD114780 | 16 | $2.175 \pm 0.017 \pm 0.037$ | 146 | $2.254 \pm 0.005 \pm 0.042$ | 0.031 | 1.21 |
| HD115898 | – | – | 223 | $4.315 \pm 0.003 \pm 0.190$ | 0.009 | 2.89 |
| HD116207 | 16 | $2.054 \pm 0.012 \pm 0.037$ | 132 | $2.148 \pm 0.005 \pm 0.042$ | 0.029 | 1.29 |
| HD118669 | – | – | 29 | $1.713 \pm 0.021 \pm 0.042$ | 0.052 | 0.72 |
| HD119584 | – | – | 25 | $1.280 \pm 0.060 \pm 0.051$ | 0.054 | 0.40 |
| HD120819 | – | – | 10 | $2.536 \pm 0.035 \pm 0.046$ | 0.012 | 0.18 |
| HD121860 | – | – | 100 | $2.013 \pm 0.010 \pm 0.042$ | 0.025 | 0.61 |
| HD122316 | – | – | 18 | $3.801 \pm 0.007 \pm 0.102$ | 0.005 | 1.71 |
| HD127093 | – | – | 90 | $2.665 \pm 0.005 \pm 0.046$ | 0.016 | 1.61 |
| HD127762 | – | – | 701 | $1.066 \pm 0.006 \pm 0.056$ | 0.037 | 0.73 |
| HD128902 | – | – | 15 | $1.944 \pm 0.026 \pm 0.042$ | 0.018 | 0.42 |
| HD130084 | – | – | 20 | $2.121 \pm 0.024 \pm 0.042$ | 0.019 | 0.29 |
| HD133124 | – | – | 9 | $2.962 \pm 0.004 \pm 0.051$ | 0.005 | 2.37 |
| HD133208 | – | – | 285 | $2.414 \pm 0.004 \pm 0.042$ | 0.023 | 1.16 |
| HD133582 | – | – | 5 | $2.415 \pm 0.009 \pm 0.042$ | 0.007 | 1.13 |
| HD135722 | – | – | 245 | $2.670 \pm 0.003 \pm 0.046$ | 0.019 | 1.21 |
| HD136404 | – | – | 35 | $1.419 \pm 0.017 \pm 0.046$ | 0.021 | 0.75 |

**Table 16** continued on next page

**Table 16** *(continued)*

| Star ID | $N_{\text{pts,H}}$ | $\theta_{\text{H}}$ | $N_{\text{pts,K}}$ | $\theta_{\text{K}}$ | Avg. Res. | $\chi^2$/DOF |
|---|---|---|---|---|---|---|
| HD137071 | – | – | 79 | $2.336 \pm 0.008 \pm 0.042$ | 0.090 | 0.97 |
| HD137853 | – | – | 119 | $2.399 \pm 0.004 \pm 0.042$ | 0.015 | 1.21 |
| HD138481 | – | – | 58 | $3.081 \pm 0.008 \pm 0.056$ | 0.024 | 1.21 |
| HD139153 | – | – | 133 | $3.419 \pm 0.004 \pm 0.069$ | 0.028 | 2.23 |
| HD139374 | – | – | 74 | $1.406 \pm 0.016 \pm 0.046$ | 0.055 | 0.87 |
| HD139971 | – | – | 112 | $1.655 \pm 0.012 \pm 0.042$ | 0.102 | 1.54 |
| HD142176 | 84 | $0.946 \pm 0.014 \pm 0.032$ | 399 | $0.938 \pm 0.010 \pm 0.060$ | 0.119 | 0.96 |
| HD143107 | – | – | 10 | $2.751 \pm 0.006 \pm 0.046$ | 0.008 | 1.77 |
| HD144065 | – | – | 20 | $1.387 \pm 0.028 \pm 0.046$ | 0.015 | 0.27 |
| HD144578 | – | – | 126 | $1.662 \pm 0.023 \pm 0.042$ | 0.102 | 0.58 |
| HD147749 | – | – | 51 | $3.619 \pm 0.005 \pm 0.083$ | 0.009 | 1.47 |
| HD148897 | – | – | 510 | $1.856 \pm 0.007 \pm 0.042$ | 0.062 | 0.95 |
| HD150047 | – | – | 35 | $3.363 \pm 0.010 \pm 0.069$ | 0.018 | 1.55 |
| HD150450 | – | – | 25 | $4.242 \pm 0.011 \pm 0.171$ | 0.011 | 1.17 |
| HD150997 | – | – | 80 | $2.423 \pm 0.006 \pm 0.042$ | 0.024 | 1.05 |
| HD151732 | – | – | 25 | $4.459 \pm 0.012 \pm 0.236$ | 0.014 | 1.95 |
| HD152173 | – | – | 35 | $2.313 \pm 0.006 \pm 0.042$ | 0.026 | 2.60 |
| HD153698 | 24 | $2.177 \pm 0.029 \pm 0.037$ | 217 | $2.122 \pm 0.005 \pm 0.042$ | 0.032 | 1.11 |
| HD153834 | – | – | 70 | $1.334 \pm 0.010 \pm 0.046$ | 0.032 | 0.59 |
| HD154301 | – | – | 50 | $1.602 \pm 0.024 \pm 0.042$ | 0.045 | 0.57 |
| HD156966 | – | – | 15 | $1.806 \pm 0.020 \pm 0.042$ | 0.020 | 0.46 |
| HD157617 | – | – | 157 | $1.256 \pm 0.009 \pm 0.051$ | 0.036 | 1.07 |
| HD157881 | 9 | $0.619 \pm 0.049 \pm 0.051$ | – | – | 0.041 | 1.28 |
| HD163547 | – | – | 20 | $1.385 \pm 0.045 \pm 0.046$ | 0.061 | 1.12 |
| HD163947 | – | – | 225 | $1.961 \pm 0.003 \pm 0.042$ | 0.017 | 0.85 |
| HD163993 | 12 | $2.029 \pm 0.020 \pm 0.037$ | 84 | $2.137 \pm 0.009 \pm 0.042$ | 0.015 | 0.75 |
| HD164064 | – | – | 15 | $1.946 \pm 0.030 \pm 0.042$ | 0.021 | 0.57 |
| HD166013 | – | – | 94 | $2.021 \pm 0.004 \pm 0.042$ | 0.017 | 0.89 |
| HD168775 | 4 | $2.119 \pm 0.027 \pm 0.037$ | 85 | $2.112 \pm 0.006 \pm 0.042$ | 0.014 | 0.46 |
| HD169305 | – | – | 30 | $4.249 \pm 0.008 \pm 0.176$ | 0.012 | 2.41 |
| HD170137 | – | – | 61 | $1.806 \pm 0.013 \pm 0.042$ | 0.029 | 0.80 |
| HD170970 | – | – | 195 | $1.861 \pm 0.007 \pm 0.042$ | 0.106 | 1.21 |
| HD171779 | – | – | 30 | $3.621 \pm 0.112 \pm 0.083$ | 0.149 | 0.65 |
| HD173213 | – | – | 19 | $2.402 \pm 0.016 \pm 0.042$ | 0.067 | 1.72 |
| HD173954 | – | – | 15 | $1.529 \pm 0.014 \pm 0.042$ | 0.010 | 0.49 |
| HD176411 | – | – | 183 | $2.164 \pm 0.006 \pm 0.042$ | 0.025 | 0.95 |
| HD176670 | – | – | 114 | $2.312 \pm 0.008 \pm 0.042$ | 0.020 | 0.59 |
| HD176844 | – | – | 25 | $3.378 \pm 0.014 \pm 0.069$ | 0.015 | 0.97 |
| HD176981 | – | – | 5 | $1.733 \pm 0.039 \pm 0.042$ | 0.020 | 0.36 |
| HD178690 | – | – | 2 | $2.230 \pm 0.039 \pm 0.042$ | 0.034 | 0.81 |
| HD180450 | 57 | $2.676 \pm 0.005 \pm 0.065$ | – | – | 0.012 | 1.33 |

**Table 16** (continued)

| Star ID | $N_{\text{pts,H}}$ | $\theta_{\text{H}}$ | $N_{\text{pts,K}}$ | $\theta_{\text{K}}$ | Avg. Res. | $\chi^2/\text{DOF}$ |
|---------|---------|---------|---------|---------|---------|---------|
| HD185958 | 52 | $1.706 \pm 0.010 \pm 0.032$ | 20 | $1.817 \pm 0.063 \pm 0.042$ | 0.095 | 1.11 |
| HD186675 | – | – | 460 | $1.332 \pm 0.007 \pm 0.046$ | 0.049 | 1.03 |
| HD186776 | – | – | 125 | $2.813 \pm 0.011 \pm 0.051$ | 0.032 | 0.77 |
| HD187849 | – | – | 34 | $4.068 \pm 0.008 \pm 0.139$ | 0.012 | 1.44 |
| HD188310 | 109 | $1.588 \pm 0.012 \pm 0.028$ | 386 | $1.631 \pm 0.004 \pm 0.042$ | 0.042 | 0.88 |
| HD189695 | – | – | 220 | $1.903 \pm 0.004 \pm 0.042$ | 0.017 | 0.65 |
| HD191178 | – | – | 158 | $2.859 \pm 0.003 \pm 0.051$ | 0.013 | 1.25 |
| HD193347 | – | – | 245 | $1.924 \pm 0.008 \pm 0.042$ | 0.063 | 1.01 |
| HD193579 | 4 | $1.868 \pm 0.015 \pm 0.032$ | 5 | $1.955 \pm 0.018 \pm 0.042$ | 0.009 | 0.36 |
| HD194097 | – | – | 20 | $1.434 \pm 0.028 \pm 0.046$ | 0.028 | 1.11 |
| HD194317 | 4 | $2.478 \pm 0.009 \pm 0.051$ | 8 | $2.570 \pm 0.016 \pm 0.046$ | 0.011 | 0.48 |
| HD197912 | 670 | $1.960 \pm 0.002 \pm 0.032$ | 110 | $2.036 \pm 0.007 \pm 0.042$ | 0.023 | 0.86 |
| HD198237 | – | – | 186 | $1.857 \pm 0.008 \pm 0.042$ | 0.042 | 0.78 |
| HD199101 | 328 | $2.197 \pm 0.004 \pm 0.037$ | 8 | $2.267 \pm 0.026 \pm 0.042$ | 0.030 | 0.80 |
| HD199697 | – | – | 159 | $1.971 \pm 0.006 \pm 0.042$ | 0.035 | 0.89 |
| HD199799 | – | – | 23 | $3.333 \pm 0.011 \pm 0.065$ | 0.017 | 1.04 |
| HD202109 | – | – | 30 | $2.741 \pm 0.008 \pm 0.046$ | 0.014 | 1.08 |
| HD205435 | – | – | 245 | $1.829 \pm 0.020 \pm 0.042$ | 0.125 | 0.60 |
| HD205733 | 2 | $2.399 \pm 0.020 \pm 0.046$ | 4 | $2.571 \pm 0.025 \pm 0.046$ | 0.037 | 4.25 |
| HD206330 | – | – | 199 | $3.297 \pm 0.010 \pm 0.065$ | 0.045 | 1.12 |
| HD206445 | – | – | 10 | $1.712 \pm 0.040 \pm 0.042$ | 0.022 | 0.22 |
| HD206749 | – | – | 171 | $2.937 \pm 0.012 \pm 0.051$ | 0.046 | 0.78 |
| HD207001 | – | – | 15 | $1.359 \pm 0.015 \pm 0.046$ | 0.009 | 0.32 |
| HD207328 | – | – | 15 | $2.529 \pm 0.020 \pm 0.046$ | 0.007 | 0.17 |
| HD209857 | – | – | 176 | $3.707 \pm 0.003 \pm 0.092$ | 0.011 | 1.59 |
| HD210514 | 8 | $2.312 \pm 0.009 \pm 0.042$ | 16 | $2.382 \pm 0.020 \pm 0.042$ | 0.051 | 1.27 |
| HD211800 | – | – | 25 | $1.560 \pm 0.018 \pm 0.042$ | 0.014 | 0.59 |
| HD212470 | 5 | $3.432 \pm 0.007 \pm 0.259$ | 20 | $3.407 \pm 0.010 \pm 0.069$ | 0.001 | 0.19 |
| HD212496 | 104 | $1.821 \pm 0.014 \pm 0.032$ | – | – | 0.040 | 0.65 |
| HD214868 | – | – | 39 | $2.466 \pm 0.010 \pm 0.042$ | 0.034 | 1.71 |
| HD215182 | – | – | 91 | $3.033 \pm 0.007 \pm 0.055$ | 0.018 | 0.89 |
| HD215547 | – | – | 10 | $2.555 \pm 0.037 \pm 0.046$ | 0.075 | 4.77 |
| HD216131 | – | – | 89 | $2.385 \pm 0.006 \pm 0.042$ | 0.014 | 0.61 |
| HD216174 | 499 | $1.425 \pm 0.005 \pm 0.028$ | 20 | $1.432 \pm 0.054 \pm 0.046$ | 0.044 | 0.51 |
| HD216930 | – | – | 10 | $1.813 \pm 0.062 \pm 0.042$ | 0.045 | 0.18 |
| HD218452 | – | – | 35 | $1.965 \pm 0.026 \pm 0.042$ | 0.107 | 1.13 |
| HD220074 | 8 | $1.832 \pm 0.043 \pm 0.032$ | 20 | $1.949 \pm 0.031 \pm 0.042$ | 0.037 | 0.86 |
| HD220211 | – | – | 5 | $2.015 \pm 0.079 \pm 0.042$ | 0.049 | 0.30 |
| HD221345 | – | – | 23 | $1.366 \pm 0.028 \pm 0.046$ | 0.036 | 0.82 |
| HD222107 | 24 | $2.565 \pm 0.018 \pm 0.055$ | 479 | $2.651 \pm 0.003 \pm 0.046$ | 0.019 | 0.78 |
| HIP8682 | – | – | 114 | $2.357 \pm 0.004 \pm 0.042$ | 0.033 | 1.56 |

**Table 16** *(continued)*

| Star ID | $N_{\mathrm{pts,H}}$ | $\theta_{\mathrm{H}}$ | $N_{\mathrm{pts,K}}$ | $\theta_{\mathrm{K}}$ | Avg. Res. | $\chi^2/\mathrm{DOF}$ |
|---------|---------|---------|---------|---------|---------|---------|
| HIP35915 | – | – | 100 | $1.740 \pm 0.018 \pm 0.042$ | 0.063 | 0.73 |
| HIP68357 | – | – | 57 | $3.314 \pm 0.006 \pm 0.065$ | 0.011 | 1.59 |
| HIP86677 | – | – | 14 | $2.185 \pm 0.045 \pm 0.042$ | 0.039 | 1.08 |
| HI95024 | – | – | 38 | $3.931 \pm 0.019 \pm 0.116$ | 0.002 | 0.05 |
| HIP113390 | – | – | 130 | $2.577 \pm 0.012 \pm 0.046$ | 0.030 | 0.83 |
| IRC+30095 | – | – | 38 | $2.673 \pm 0.020 \pm 0.046$ | 0.057 | 2.16 |
| IRC+40533 | – | – | 85 | $3.299 \pm 0.007 \pm 0.065$ | 0.045 | 1.64 |
| IRC+50261 | – | – | 41 | $1.953 \pm 0.017 \pm 0.042$ | 0.037 | 0.91 |
| IRC+60005 | – | – | 140 | $3.223 \pm 0.013 \pm 0.060$ | 0.074 | 1.57 |

**Table 17.** Computed bolometric fluxes for the program stars.

| Star | Primary | Primary | Secondary | Secondary | Model | Model | Fit | Flux[a] | | Reddening | Fit | Fit |
|---|---|---|---|---|---|---|---|---|---|---|---|---|
| ID | ST | Reference | ST | Reference | TEFF | logg | ST | $F_{\rm BOL}$ | Err | $A_V$ | $\chi^2$/DOF | DOF |
| HD178 | M5/6III | Honk (1982) | M6III | Jones (1972) | 3200 | 1.0 | M6III | 30.4±0.9 | 3.0% | 0.43±0.03 | 4.9 | 6 |
| HD598 | M4III | Wilson & Joy (1952) | M7 | Lee et al. (1943) | 3400 | 1.0 | M5III | 36.0±0.7 | 1.8% | 0.00±0.03 | 21.5 | 8 |
| HD672 | M5III | Jones (1972) | | | 3300 | 1.0 | M5.5III | 25.6±0.8 | 3.2% | 0.18±0.03 | 5.1 | 5 |
| HD787 | K5III | Roman (1952) | K5III | Adams et al. (1935) | 3900 | 1.5 | K5III | 51.9±1.1 | 2.2% | 0.00±0.03 | 3.2 | 16 |
| HD1522 | K1+ III | Keenan & McNeil (1989) | K1.5III | Morgan & Keenan (1973) | 4600 | 2.0 | K1.3III | 141.0±1.5 | 1.0% | 0.01±0.01 | 0.9 | 54 |
| HD1632 | K5III | S. (1932) | | | 3900 | 1.5 | K5III | 43.0±1.3 | 3.0% | 0.31±0.03 | 2.3 | 14 |
| HD2436 | K5III | Adams et al. (1935) | | | 3900 | 1.5 | K5III | 29.8±0.8 | 2.6% | 0.24±0.02 | 4.3 | 26 |
| HD3346 | K6IIIa | Keenan & McNeil (1989) | M0III | Henry et al. (2000) | 3900 | 1.5 | K5III | 76.9±1.2 | 1.6% | 0.29±0.02 | 3.2 | 24 |
| HD3546 | G7III | Keenan & McNeil (1989) | G8III | Roman (1955) | 5400 | 2.5 | G1III | 68.8±1.2 | 1.7% | 0.31±0.01 | 0.4 | 87 |
| HD3574 | K4III | Abt (1985) | K7III | Bakos (1974) | 3800 | 1.5 | M0III | 50.4±0.6 | 1.2% | 0.09±0.01 | 13.0 | 20 |
| HD3627 | K3III-IIIb | Keenan & McNeil (1989) | K3III | Gray et al. (2003) | 4500 | 2.0 | K1.5III | 216.0±2.1 | 1.0% | 0.14±0.01 | 1.6 | 62 |
| HD5006 | K8III | Gaze & Shajn (1952) | | | 3400 | 1.0 | M5III | 22.5±1.0 | 4.4% | 0.01±0.04 | 12.8 | 12 |
| HD5575 | G6III | Adams et al. (1935) | | | 4800 | 2.5 | G9III | 23.9±0.3 | 1.4% | 0.07±0.01 | 0.7 | 24 |
| HD6186 | G9IIIb | Keenan & McNeil (1989) | G9 | Hossack (1954) | 5000 | 2.5 | G5III | 67.1±0.8 | 1.2% | 0.07±0.01 | 0.3 | 84 |
| HD6262 | M3III | Moore & Paddock (1950) | M5 | Lee et al. (1943) | 3600 | 1.5 | M3III | 25.3±0.8 | 3.1% | 0.31±0.03 | 7.3 | 9 |
| HD6409 | M5 | Lee et al. (1943) | | | 3500 | 1.0 | M4III | 19.2±0.3 | 1.4% | 0.01±0.02 | 8.9 | 10 |
| HD7000 | M4 | Lee et al. (1943) | | | 3400 | 1.0 | M5III | 14.7±0.6 | 4.2% | 0.19±0.04 | 8.4 | 8 |
| HD7087 | G8.5III | Keenan & McNeil (1989) | K0III | Sato & Kuji (1990) | 4900 | 2.5 | G7III | 46.0±0.6 | 1.2% | 0.06±0.01 | 0.8 | 48 |
| HD7318 | G8III | Abt (1985) | G8III-IV | Yoss (1961) | 4900 | 2.5 | G7III | 46.7±0.8 | 1.6% | 0.08±0.02 | 1.4 | 57 |
| HD8126 | K5III | Adams et al. (1935) | | | 4200 | 2.0 | K3.III | 46.6±0.8 | 1.7% | 0.20±0.02 | 2.1 | 20 |
| HD9500 | M3III | Wilson & Joy (1952) | M4III | Moore & Paddock (1950) | 3500 | 1.0 | M4III | 29.4±0.5 | 1.7% | 0.03±0.02 | 4.5 | 17 |
| HD9927 | K3- III | Keenan & McNeil (1989) | K3 | Hossack (1954) | 4500 | 2.0 | K1.5III | 168.0±1.8 | 1.0% | 0.16±0.01 | 1.2 | 53 |
| HD10380 | K3III | Keenan & McNeil (1989) | K3III | Lu (1991) | 4200 | 2.0 | K3.III | 84.1±0.7 | 0.8% | 0.05±0.01 | 1.1 | 91 |
| HD11928 | M2III | Wilson & Joy (1950) | M4 | Lee et al. (1943) | 3700 | 1.5 | M2III | 54.9±0.6 | 1.2% | 0.20±0.02 | 6.2 | 20 |
| HD12274 | M0III | Keenan & McNeil (1989) | K6III | Hoffleit & Shapley (1937) | 3800 | 1.5 | M0III | 193.0±2.5 | 1.3% | 0.00±0.02 | 1.5 | 44 |
| HD12929 | K2- IIIb | Keenan & McNeil (1989) | K1IIIb | Gray et al. (2003) | 4200 | 2.0 | K3.III | 574.0±3.5 | 0.6% | 0.00±0.01 | 45.4 | 91 |

**Table 17** *continued on next page*

**Table 17** (continued)

| Star ID | Primary ST | Primary Reference | Secondary ST | Secondary Reference | Model TEFF | Model logg | Fit ST | Flux[a] $F_{BOL}$ | Err | Reddening $A_V$ | Fit $\chi^2$/DOF | Fit DOF |
|---|---|---|---|---|---|---|---|---|---|---|---|---|
| HD14146 | M0III | Heard (1956) | M1III | Wilson & Joy (1952) | 3700 | 1.5 | M2III | 24.9 ± 0.7 | 2.7% | 0.20 ± 0.03 | 2.2 | 12 |
| HD14512 | M5 D | Color Fitting | | | 3400 | 1.0 | M5III | 35.9 ± 1.0 | 2.9% | 0.34 ± 0.02 | 12.1 | 9 |
| HD14770 | G8III | Roman (1952) | G5III | Adams et al. (1935) | 5000 | 2.5 | G5III | 28.8 ± 0.5 | 1.7% | 0.08 ± 0.02 | 3.2 | 33 |
| HD14872 | K4.5III | Keenan & McNeil (1989) | K4III | Roman (1952) | 3900 | 1.5 | K5III | 87.3 ± 1.4 | 1.6% | 0.03 ± 0.02 | 2.6 | 40 |
| HD14901 | K5III | S. (1932) | | | 3300 | 1.0 | M5.5III | 64.7 ± 0.6 | 1.0% | 0.19 ± 0.02 | 33.0 | 11 |
| HD15656 | K5III | Roman (1952) | M1 | Lee et al. (1943) | 4000 | 2.0 | K4III | 51.9 ± 0.8 | 1.5% | 0.02 ± 0.02 | 2.7 | 19 |
| HD16396 | K2III | Stephenson (1960) | K2III | Adams et al. (1935) | 4700 | 2.5 | K1III | 7.0 ± 0.2 | 2.8% | 0.12 ± 0.02 | 3.7 | 14 |
| HD17228 | G8III | Halliday (1955) | G8III | | 5100 | 2.5 | G4III | 11.9 ± 0.4 | 3.2% | 0.22 ± 0.02 | 2.6 | 28 |
| HD17361 | K0.5IIIb | Keenan & McNeil (1989) | K1III | Hossack (1954) | 4700 | 2.5 | K1III | 62.8 ± 0.9 | 1.5% | 0.15 ± 0.01 | 2.1 | 78 |
| HD17709 | K5.5III | Keenan & McNeil (1989) | K5+ IIIab | Yamashita (1967) | 3900 | 1.5 | K5III | 112.0 ± 1.7 | 1.5% | 0.11 ± 0.02 | 1.2 | 34 |
| HD18322 | K1+ IIIb | Keenan & McNeil (1989) | K1IIIb | Lu (1991) | 4700 | 2.5 | K1III | 103.0 ± 0.7 | 0.7% | 0.05 ± 0.01 | 12.4 | 84 |
| HD18449 | K2III | Roman (1952) | K2III | Adams et al. (1935) | 4500 | 2.0 | K1.5III | 49.6 ± 1.1 | 2.2% | 0.17 ± 0.02 | 0.5 | 63 |
| HD19613 | M5III | Honk (1982) | | | 3300 | 1.0 | M5.5III | 24.0 ± 0.9 | 3.9% | 0.20 ± 0.03 | 17.8 | 7 |
| HD19787 | G9.5IIIb | Keenan & McNeil (1989) | K2III | Roman (1952) | 4900 | 2.5 | G7III | 60.0 ± 0.5 | 0.9% | 0.01 ± 0.01 | 0.6 | 49 |
| HD20644 | K3IIIa | Keenan & McNeil (1989) | K4III | Gyldenkerne (1955) | 4000 | 2.0 | K4III | 116.0 ± 1.6 | 1.3% | 0.25 ± 0.01 | 5.4 | 38 |
| HD20844 | M3III | Turnshek et al. (1980) | M0 | Cannon & Mayall (1949) | 3400 | 1.0 | M5III | 32.1 ± 0.4 | 1.3% | 0.02 ± 0.02 | 28.1 | 8 |
| HD21465 | K5 | Lee et al. (1943) | K7 | Moore (1932) | 3600 | 1.5 | M3III | 21.9 ± 0.7 | 3.1% | 0.19 ± 0.04 | 22.0 | 7 |
| HD25604 | K0III-IIIb | Keenan & McNeil (1989) | K0III | Roman (1952) | 4800 | 2.5 | G9III | 61.4 ± 0.7 | 1.1% | 0.00 ± 0.01 | 0.7 | 57 |
| HD26605 | G9III | Halliday (1955) | G5III | | 5000 | 2.5 | G5III | 11.6 ± 0.3 | 2.6% | 0.37 ± 0.02 | 3.6 | 23 |
| HD27308 | M2 | Cannon & Mayall (1949) | M5 | Lee et al. (1943) | 3400 | 1.0 | M5III | 28.8 ± 0.6 | 2.1% | 0.40 ± 0.03 | 12.7 | 8 |
| HD27348 | G8+ IIIb | Keenan & McNeil (1989) | G8III | Schmitt (1971) | 5000 | 2.5 | G5III | 33.4 ± 0.7 | 2.0% | 0.00 ± 0.02 | 0.7 | 32 |
| HD27697 | G9.5III | Keenan & McNeil (1989) | K1III | Morgan & Hiltner (1965) | 5000 | 2.5 | G5III | 102.0 ± 1.1 | 1.1% | 0.03 ± 0.01 | 4.6 | 103 |
| HD28100 | G7III | Keenan & McNeil (1989) | G9III-III | Yoss (1961) | 5000 | 2.5 | G5III | 46.5 ± 0.9 | 1.9% | 0.11 ± 0.02 | 0.5 | 50 |
| HD28292 | K1IIIb | Keenan & McNeil (1989) | K2III | Harlan (1969) | 4900 | 2.5 | G7III | 49.1 ± 0.6 | 1.1% | 0.47 ± 0.01 | 2.8 | 22 |
| HD28305 | G9.5III | Keenan & McNeil (1989) | K1III | Morgan & Hiltner (1965) | 4900 | 2.5 | G7III | 120.0 ± 0.9 | 0.8% | 0.00 ± 0.01 | 1.8 | 95 |
| HD28307 | G9III | Keenan & McNeil (1989) | G9III | Morgan & Hiltner (1965) | 5000 | 2.5 | G5III | 91.7 ± 1.2 | 1.4% | 0.00 ± 0.01 | 0.9 | 71 |
| HD28581 | K2III D | Color Fitting | | | 4100 | 2.0 | K3.5III | 21.0 ± 1.8 | 8.8% | 1.08 ± 0.03 | 2.6 | 16 |
| HD28595 | M2 | Perraud (1961) | M3III | Roman (1955) | 3700 | 1.5 | M2III | 48.3 ± 1.8 | 3.7% | 0.53 ± 0.03 | 3.4 | 8 |

**Table 17** continued on next page

**Table 17** (continued)

| Star ID | Primary ST | Primary Reference | Secondary ST | Secondary Reference | Model TEFF | Model logg | Fit ST | Flux $F_{\mathrm{BOL}}$[a] | Err | Reddening $A_V$ | Fit $\chi^2$/DOF | Fit DOF |
|---|---|---|---|---|---|---|---|---|---|---|---|---|
| HD29094 | G7Ib | Ginestet & Carquillat (2002) | B7.5 | Parsons & Ake (1998) | 4500 | 2.0 | K1.5III | 92.8 ± 1.8 | 1.9% | 0.15 ± 0.02 | 0.7 | 33 |
| HD30504 | K3+ III | Keenan & McNeil (1989) | K3III | Lu (1991) | 4200 | 2.0 | K3III | 76.0 ± 1.1 | 1.5% | 0.36 ± 0.01 | 1.2 | 24 |
| HD30605 | K2II | Abt (2008) | K3III | Adams et al. (1935) | 3900 | 1.5 | K5III | 26.8 ± 0.4 | 1.5% | 0.09 ± 0.02 | 8.0 | 11 |
| HD30834 | K3- III | Keenan & McNeil (1989) | K2IIIb | Lu (1991) | 4200 | 2.0 | K3III | 73.5 ± 1.3 | 1.8% | 0.23 ± 0.02 | 1.5 | 39 |
| HD31139 | M0 | Lee et al. (1943) | K6III | Hoffleit (1942) | 3800 | 1.5 | M0III | 81.1 ± 1.5 | 1.9% | 0.32 ± 0.03 | 1.8 | 12 |
| HD31421 | K2- IIIb | Keenan & McNeil (1989) | K1III | Abt (2008) | 4700 | 2.5 | K1III | 106.0 ± 1.8 | 1.7% | 0.30 ± 0.01 | 1.4 | 49 |
| HD33463 | M2 | Neckel (1958) | M2III | Heard (1956) | 3500 | 1.0 | M4III | 42.1 ± 0.6 | 1.4% | 0.00 ± 0.03 | 14.1 | 8 |
| HD34334 | K2.5IIIb | Keenan & McNeil (1989) | K3.5III | Abt (1985) | 4400 | 2.0 | K2III | 80.3 ± 1.9 | 2.3% | 0.23 ± 0.02 | 0.4 | 43 |
| HD34559 | G8III | Roman (1952) | G5III | Adams et al. (1935) | 5000 | 2.5 | G5III | 33.8 ± 1.3 | 3.8% | 0.00 ± 0.03 | 1.3 | 22 |
| HD34577 | M3 | Neckel (1958) | M1 | Cannon & Mayall (1949) | 3500 | 1.0 | M4III | 30.8 ± 1.1 | 3.7% | 0.51 ± 0.03 | 6.5 | 9 |
| HD35497 | B5.5IV | Zorec et al. (2009) | B7III | Garrison & Gray (1994) | 12000 | 3.0 | B8III | 515.0 ± 4.0 | 0.8% | 0.00 ± 0.01 | 3.6 | 110 |
| HD35620 | K3III Fe1 | Keenan & McNeil (1989) | K4IIIp | Schmitt (1971) | 4100 | 2.0 | K3.5III | 45.1 ± 0.5 | 1.1% | 0.00 ± 0.01 | 4.2 | 48 |
| HD37329 | G9III | Halliday (1955) | G9III | | 4900 | 2.5 | G7III | 9.0 ± 0.2 | 2.2% | 0.00 ± 0.02 | 1.4 | 27 |
| HD38152 | M4 | Neckel (1958) | M1 | Lee et al. (1943) | 3400 | 1.0 | M5III | 14.8 ± 0.4 | 2.4% | 0.32 ± 0.03 | 4.1 | 9 |
| HD38656 | G8IIIb | Keenan & McNeil (1989) | G9III | Abt (2008) | 5000 | 2.5 | G5III | 50.1 ± 0.5 | 1.1% | 0.00 ± 0.01 | 0.4 | 101 |
| HD39003 | K0.5III | Keenan & McNeil (1989) | K1III | Abt (2008) | 4700 | 2.5 | K1III | 98.5 ± 1.0 | 1.0% | 0.11 ± 0.01 | 2.8 | 95 |
| HD39045 | M3III | Keenan & McNeil (1989) | M4- IIab | Yamashita (1967) | 3500 | 1.0 | M4III | 52.8 ± 0.8 | 1.5% | 0.10 ± 0.02 | 6.9 | 35 |
| HD39225 | M1+ III | Keenan & McNeil (1989) | M2+ IV | Yamashita (1967) | 3800 | 1.5 | M0III | 36.2 ± 1.0 | 2.7% | 0.15 ± 0.03 | 2.1 | 15 |
| HD39732 | M0 | Cannon & Mayall (1949) | M4 | Lee et al. (1943) | 3500 | 1.0 | M4III | 31.3 ± 1.8 | 5.8% | 0.37 ± 0.05 | 3.9 | 7 |
| HD40441 | K5III | Fehrenbach (1966) | K5III | | 3900 | 1.5 | K5III | 19.5 ± 1.3 | 6.7% | 0.36 ± 0.05 | 5.0 | 11 |
| HD41116 | K1III | Strassmeier & Fekel (1990) | G5III-IV | Cowley & Bidelman (1979) | 5500 | 2.5 | G0III | 73.8 ± 1.5 | 2.0% | 0.28 ± 0.02 | 1.8 | 38 |
| HD42049 | K5Ib | McCuskey (1967) | K4III | Wilson & Joy (1952) | 4000 | 2.0 | K4III | 39.4 ± 3.0 | 7.6% | 0.57 ± 0.04 | 5.0 | 5 |
| HD43039 | G8.5IIIb | Keenan & McNeil (1989) | G8III | Roman (1955) | 4900 | 2.5 | G7III | 68.1 ± 0.8 | 1.2% | 0.12 ± 0.01 | 0.3 | 93 |
| HD46709 | K4 | Iijima & Ishida (1978) | K4III | Schmitt (1971) | 4100 | 2.0 | K3.5III | 25.3 ± 0.3 | 1.2% | 0.16 ± 0.01 | 13.0 | 33 |
| HD48450 | K3III | Adams et al. (1935) | K4III | | 4100 | 2.0 | K3.5III | 37.3 ± 0.4 | 1.1% | 0.12 ± 0.01 | 3.6 | 24 |
| HD49738 | K5 | Lee et al. (1943) | K3III | Adams et al. (1935) | 4300 | 2.0 | K3III | 24.9 ± 1.2 | 4.6% | 0.09 ± 0.04 | 2.6 | 18 |
| HD49968 | K5III | Adams et al. (1935) | K5III | | 4100 | 2.0 | K3.5III | 31.2 ± 1.2 | 3.8% | 0.12 ± 0.04 | 1.3 | 16 |
| HD52960 | K3III | Roman (1952) | K5III | Adams et al. (1935) | 4200 | 2.0 | K3.III | 50.8 ± 1.5 | 3.0% | 0.19 ± 0.02 | 2.2 | 34 |

**Table 17** continued on next page

**Table 17** *(continued)*

| Star ID | Primary ST | Primary Reference | Secondary ST | Secondary Reference | Model TEFF | Model logg | Fit ST | Flux[a] $F_{\rm BOL}$ | Err | Reddening $A_{\rm V}$ | Fit $\chi^2$/DOF | Fit DOF |
|---|---|---|---|---|---|---|---|---|---|---|---|---|
| HD54719 | K2III | Keenan & McNeil (1989) | K2II | Ginestet et al. (1994) | 4600 | 2.0 | K1.3III | 78.5 ± 1.5 | 1.9% | 0.22 ± 0.02 | 0.7 | 47 |
| HD57669 | K0.5IIIa | Keenan & McNeil (1989) | K0III | Schmitt (1971) | 4500 | 2.0 | K1.5III | 32.8 ± 1.2 | 3.6% | 0.04 ± 0.03 | 1.3 | 19 |
| HD60136 | M5 | Lee et al. (1943) | | | 3700 | 1.5 | M2III | 33.7 ± 2.7 | 7.9% | 0.63 ± 0.04 | 4.1 | 10 |
| HD61294 | M0III | Wilson & Joy (1950) | M4 | Lee et al. (1943) | 3800 | 1.5 | M0III | 45.8 ± 1.4 | 3.0% | 0.18 ± 0.03 | 4.9 | 24 |
| HD61913 | M3+ I1-III | Keenan & McNeil (1989) | M3+SIIb | Yamashita (1967) | 3500 | 1.0 | M4III | 97.5 ± 1.5 | 1.5% | 0.00 ± 0.02 | 10.0 | 50 |
| HD62044 | K1IIIe | Keenan & McNeil (1989) | K1III | Roman (1952) | 4800 | 2.5 | G9III | 89.4 ± 1.9 | 2.1% | 0.30 ± 0.02 | 1.3 | 53 |
| HD62285 | K4.5III | Keenan & McNeil (1989) | K5IIIa | Yamashita (1967) | 3900 | 1.5 | K5III | 52.9 ± 1.7 | 3.2% | 0.04 ± 0.03 | 3.3 | 23 |
| HD62345 | G8III-IIIb | Keenan & McNeil (1989) | G9III | Abt (2008) | 5000 | 2.5 | G5III | 118.0 ± 1.2 | 1.0% | 0.00 ± 0.01 | 1.2 | 120 |
| HD62509 | K0IIIb | Keenan & McNeil (1989) | G9III | Abt (2008) | 4900 | 2.5 | G7III | 1130 ± 11.4 | 1.0% | 0.00 ± 0.01 | 1.3 | 104 |
| HD62721 | K4III | Keenan & McNeil (1989) | K5III | Roman (1955) | 4000 | 2.0 | K4III | 70.7 ± 1.1 | 1.6% | 0.06 ± 0.02 | 1.3 | 53 |
| HD65345 | G8IIIb | Keenan & McNeil (1989) | K0III | Roman (1955) | 5000 | 2.5 | G5III | 24.5 ± 1.2 | 5.1% | 0.00 ± 0.04 | 0.3 | 30 |
| HD65759 | K3III | Adams et al. (1935) | | | 4400 | 2.0 | K2III | 27.5 ± 1.2 | 4.5% | 0.16 ± 0.04 | 2.4 | 19 |
| HD66216 | K1- IIIb | Keenan & McNeil (1989) | K2II | Roman (1952) | 4800 | 2.5 | G9III | 43.4 ± 2.0 | 4.5% | 0.20 ± 0.04 | 0.6 | 21 |
| HD67743 | M2II | Yoss (1977) | M2III | Wilson & Joy (1952) | 3500 | 1.5 | M4III | 19.3 ± 0.9 | 4.5% | 0.00 ± 0.05 | 6.5 | 8 |
| HD74442 | K0+ IIIb | Keenan & McNeil (1989) | G9III | Abt (2008) | 4800 | 2.5 | G9III | 97.2 ± 0.8 | 0.9% | 0.12 ± 0.01 | 1.5 | 68 |
| HD75506 | K0III | Roman (1952) | G6III | Adams et al. (1935) | 5000 | 2.5 | G5III | 29.2 ± 1.1 | 3.8% | 0.06 ± 0.04 | 0.7 | 36 |
| HD76294 | G8.5III | Keenan & McNeil (1989) | G9II-III | Morgan & Keenan (1973) | 4900 | 2.5 | G7III | 185.0 ± 2.6 | 1.4% | 0.00 ± 0.01 | 0.8 | 67 |
| HD76830 | M4.5III | Keenan & McNeil (1989) | M4III | Roman (1955) | 3500 | 1.0 | M4III | 55.3 ± 1.0 | 1.8% | 0.04 ± 0.02 | 3.4 | 26 |
| HD79554 | K1III | Adams et al. (1935) | | | 4500 | 2.0 | K1.5III | 40.5 ± 2.0 | 4.9% | 0.40 ± 0.03 | 1.1 | 21 |
| HD81817 | K3II-III | Keenan & McNeil (1989) | K3III | Roman (1952) | 4100 | 2.0 | K3.5III | 101.0 ± 1.8 | 1.8% | 0.00 ± 0.02 | 4.0 | 23 |
| HD82198 | M1IIIab | Yamashita (1967) | K6III | Hoffleit (1942) | 3900 | 1.5 | K5III | 54.9 ± 1.4 | 2.5% | 0.14 ± 0.02 | 1.0 | 23 |
| HD82381 | K2.5III | Keenan & McNeil (1989) | K3III | Abt (1985) | 4200 | 2.0 | K3.III | 47.0 ± 0.9 | 1.9% | 0.03 ± 0.02 | 2.9 | 47 |
| HD82635 | G7.5IIIb | Keenan & McNeil (1989) | G8.5III | Morgan & Keenan (1973) | 5100 | 2.5 | G4III | 49.0 ± 1.2 | 2.4% | 0.07 ± 0.02 | 0.7 | 74 |
| HD82741 | G9.5IIIb | Keenan & McNeil (1989) | K0III | Roman (1952) | 5000 | 2.5 | G5III | 44.7 ± 1.4 | 3.2% | 0.18 ± 0.02 | 0.7 | 56 |
| HD84181 | M... | Color Fitting | | | 3500 | 1.0 | M4III | 15.8 ± 0.9 | 5.7% | 0.18 ± 0.05 | 11.5 | 6 |
| HD84914 | K4III | Sato & Kuji (1990) | | | 3900 | 1.5 | K5III | 19.2 ± 1.0 | 5.2% | 0.24 ± 0.04 | 2.0 | 13 |
| HD85503 | K2IIIb | Keenan & McNeil (1989) | K2IIIb | Lu (1991) | 4500 | 2.0 | K1.5III | 111.0 ± 1.5 | 1.3% | 0.07 ± 0.01 | 3.0 | 94 |
| HD87046 | M... | Color Fitting | | | 3500 | 1.0 | M4III | 19.5 ± 0.7 | 3.3% | 0.00 ± 0.03 | 13.4 | 8 |

**Table 17** *continued on next page*

**Table 17** *(continued)*

| Star ID | Primary ST | Primary Reference | Secondary ST | Secondary Reference | Model $T_{EFF}$ | Model $\log g$ | Fit ST | Flux[a] $F_{BOL}$ | Err | Reddening $A_V$ | Fit $\chi^2/\mathrm{DOF}$ | Fit DOF |
|---|---|---|---|---|---|---|---|---|---|---|---|---|
| HD87837 | K3.5IIb | Keenan & McNeil (1989) | K5III | Abt (1981) | 4100 | 2.0 | K3.5III | 101.0±1.4 | 1.4% | 0.08±0.02 | 1.5 | 54 |
| HD90254 | M2+SIIIab | Yamashita (1967) | M4 | Lee et al. (1943) | 3800 | 1.5 | M0III | 64.3±1.7 | 2.7% | 0.37±0.02 | 1.6 | 21 |
| HD92620 | M4III | Keenan & McNeil (1989) | M4IIIa | Yamashita (1967) | 3500 | 1.0 | M4III | 65.4±1.0 | 1.6% | 0.00±0.02 | 4.3 | 26 |
| HD93287 | M... | Color Fitting | | | 3500 | 1.0 | M4III | 11.5±0.5 | 4.3% | 0.18±0.03 | 14.5 | 8 |
| HD94252 | M4 | Lee et al. (1943) | | | 3600 | 1.5 | M3III | 13.9±0.5 | 3.6% | 0.03±0.03 | 12.4 | 8 |
| HD94264 | K0+ III-IV | Keenan & McNeil (1989) | G9 | Bidelman (1985) | 4800 | 2.5 | G9III | 108.0±1.8 | 1.6% | 0.06±0.02 | 0.4 | 61 |
| HD94336 | M3III | Moore & Paddock (1950) | M3III | Wilson & Joy (1950) | 3600 | 1.5 | M3III | 20.8±0.6 | 2.8% | 0.12±0.02 | 15.6 | 15 |
| HD94600 | K1III | Roman (1952) | K1III | Adams et al. (1935) | 4700 | 2.5 | K1III | 39.0±2.0 | 5.1% | 0.12±0.04 | 0.3 | 17 |
| HD95212 | K5III | Adams et al. (1935) | | | 3900 | 1.5 | K5III | 42.4±1.4 | 3.4% | 0.00±0.04 | 4.5 | 12 |
| HD95345 | K0.5III | Keenan & McNeil (1989) | K1III | Yamashita & Norimoto (1981) | 4600 | 2.0 | K1.3III | 45.7±1.1 | 2.4% | 0.06±0.02 | 0.3 | 66 |
| HD96274 | M3 | Lee et al. (1943) | | | 3600 | 1.5 | M3III | 19.8±0.7 | 3.3% | 0.02±0.03 | 8.6 | 12 |
| HD96833 | K1III | Keenan & McNeil (1989) | K2II | Ljunggren & Oja (1966) | 4600 | 2.0 | K1.3III | 230.0±2.4 | 1.1% | 0.00±0.01 | 0.7 | 67 |
| HD100236 | M... | Color Fitting | | | 3500 | 1.0 | M4III | 13.4±0.8 | 6.1% | 0.06±0.06 | 5.0 | 8 |
| HD102224 | K0.5IIIb: | Keenan & McNeil (1989) | K0III | Schild (1973) | 4600 | 2.0 | K1.3III | 146.0±2.0 | 1.4% | 0.19±0.01 | 0.9 | 54 |
| HD104207 | M4III | Wilson & Joy (1950) | M5 | Lee et al. (1943) | 3400 | 1.0 | M5III | 46.8±0.9 | 1.8% | 0.02±0.02 | 10.9 | 14 |
| HD104575 | M4 | Lee et al. (1943) | | | 3500 | 1.0 | M4III | 14.8±0.8 | 5.4% | 0.00±0.06 | 8.6 | 7 |
| HD104831 | M3 | Lee et al. (1943) | | | 3700 | 1.5 | M2III | 13.7±0.5 | 4.0% | 0.04±0.04 | 11.2 | 8 |
| HD104979 | G8III | Keenan & McNeil (1989) | G8III | Keenan & Wilson (1977) | 5100 | 2.5 | G4III | 81.7±0.7 | 0.9% | 0.21±0.01 | 0.6 | 100 |
| HD105943 | K5III | Adams et al. (1935) | K2 | Parkhurst (1912) | 3900 | 1.5 | K5III | 27.7±1.4 | 5.0% | 0.11±0.05 | 3.0 | 6 |
| HD106714 | G8III-IIIb | Keenan & McNeil (1989) | G8III | Westerlund (1953) | 4900 | 2.5 | G7III | 36.7±1.0 | 2.7% | 0.00±0.02 | 5.5 | 26 |
| HD107256 | M5 | Lee et al. (1943) | | | 3400 | 1.0 | M5III | 15.3±0.7 | 4.3% | 0.17±0.04 | 10.1 | 10 |
| HD107274 | M1+ IIab | Yamashita (1967) | M0III | Sato & Kuji (1990) | 3800 | 1.5 | M0III | 74.6±1.4 | 1.8% | 0.27±0.02 | 3.5 | 28 |
| HD109282 | M0III | Westerlund (1953) | M3III | Wilson & Joy (1952) | 3600 | 1.5 | M3III | 19.3±0.7 | 3.6% | 0.29±0.03 | 10.5 | 22 |
| HD109317 | K0III | Schild (1973) | K0III | Ljunggren (1966) | 4900 | 2.5 | G7III | 28.3±1.7 | 5.9% | 0.21±0.05 | 1.2 | 23 |
| HD110296 | K5III | Ljunggren (1966) | K6III | Malmquist (1960) | 3900 | 1.5 | K5III | 6.4±0.4 | 6.8% | 0.20±0.05 | 3.2 | 15 |
| HD111067 | K3III | Roman (1952) | K5III | Wilson & Joy (1950) | 4300 | 2.0 | K3III | 46.7±1.2 | 2.6% | 0.18±0.02 | 5.5 | 31 |
| HD113226 | G8III-IIIb | Keenan & McNeil (1989) | G8III | Gray et al. (2003) | 5100 | 2.5 | G4III | 221.0±1.7 | 0.8% | 0.00±0.01 | 1.5 | 63 |
| HD114780 | M0III | Appenzeller (1967) | M0III | | 3800 | 1.5 | M0III | 38.3±0.7 | 1.9% | 0.00±0.02 | 2.8 | 19 |

**Table 17** (continued)

| Star ID | Primary ST | Primary Reference | Secondary ST | Secondary Reference | Model $T_{EFF}$ | Model $\log g$ | Fit ST | Flux[a] $F_{BOL}$ | Err | Reddening $A_V$ | Fit $\chi^2/DOF$ | Fit DOF |
|---|---|---|---|---|---|---|---|---|---|---|---|---|
| HD115723 | K4/5III | Appenzeller (1967) | K5III | Wilson & Joy (1950) | 4300 | 2.0 | K3III | 25.0 ± 1.0 | 3.9% | 0.19 ± 0.03 | 2.1 | 19 |
| HD116207 | M5 | Lee et al. (1943) | | | 3400 | 1.0 | M5III | 24.1 ± 0.8 | 3.5% | 0.00 ± 0.04 | 11.6 | 8 |
| HD118669 | M... | Color Fitting | | | 3500 | 1.5 | M4III | 15.1 ± 0.7 | 4.8% | 0.00 ± 0.05 | 4.7 | 8 |
| HD119584 | K4III | Appenzeller (1967) | K5 | Lee et al. (1943) | 4000 | 2.0 | K4III | 21.5 ± 0.7 | 3.4% | 0.01 ± 0.03 | 3.4 | 17 |
| HD120819 | M1IIIa | Yamashita (1967) | M2III | Appenzeller (1967) | 3800 | 1.5 | M0III | 44.9 ± 1.4 | 3.1% | 0.29 ± 0.03 | 4.4 | 22 |
| HD121860 | M5 | Lee et al. (1943) | | | 3500 | 1.0 | M4III | 21.5 ± 0.8 | 3.7% | 0.06 ± 0.04 | 10.5 | 8 |
| HD122316 | M8 D | Color Fitting | | | 3000 | 0.5 | M7.5III | 50.9 ± 0.9 | 1.8% | 0.00 ± 0.03 | 22.4 | 10 |
| HD126307 | K4III | Heard (1956) | M3 | Lee et al. (1943) | 4000 | 2.0 | K4III | 22.9 ± 1.2 | 5.2% | 0.38 ± 0.03 | 8.7 | 11 |
| HD127093 | M4III | Wilson & Joy (1950) | M4 | Lee et al. (1943) | 3500 | 1.0 | M4III | 37.8 ± 1.5 | 4.0% | 0.00 ± 0.04 | 2.6 | 13 |
| HD128902 | K4III | Roman (1955) | K2III | Keenan & Keller (1953) | 4000 | 2.0 | K4III | 33.5 ± 1.1 | 3.1% | 0.13 ± 0.02 | 3.0 | 30 |
| HD129312 | G7IIIa | Keenan & McNeil (1989) | G8III | Roman (1952) | 4900 | 2.5 | G7III | 37.2 ± 1.0 | 2.6% | 0.01 ± 0.02 | 0.5 | 61 |
| HD130084 | M1+ IIIb | Yamashita (1967) | M1III | Wilson & Joy (1950) | 3800 | 1.5 | M0III | 31.8 ± 1.0 | 3.1% | 0.27 ± 0.03 | 3.0 | 20 |
| HD131507 | K4III | Roman (1955) | M1 | Lee et al. (1943) | 4200 | 2.0 | K3.III | 35.5 ± 1.4 | 3.9% | 0.16 ± 0.03 | 1.7 | 20 |
| HD133124 | K4III | Keenan & McNeil (1989) | K4III | Roman (1952) | 4000 | 2.0 | K4III | 79.2 ± 1.6 | 2.0% | 0.13 ± 0.02 | 2.0 | 30 |
| HD133208 | G8IIIa | Keenan & McNeil (1989) | G8IIIa | Lu (1991) | 5000 | 2.5 | G5III | 125.0 ± 1.3 | 1.0% | 0.00 ± 0.01 | 0.6 | 79 |
| HD133485 | G8III-IV | Halliday (1955) | | | 4900 | 2.5 | G7III | 8.7 ± 0.4 | 4.6% | 0.14 ± 0.04 | 1.5 | 22 |
| HD133582 | K2III | Keenan & McNeil (1989) | K2III | Roman (1952) | 4500 | 2.0 | K1.5III | 74.2 ± 1.5 | 2.0% | 0.19 ± 0.02 | 3.6 | 54 |
| HD135722 | G8III | Keenan & McNeil (1989) | G8IV | Gray et al. (2003) | 5100 | 2.5 | G4III | 152.0 ± 2.1 | 1.4% | 0.21 ± 0.01 | 3.3 | 98 |
| HD136404 | M... | Color Fitting | | | 3700 | 1.5 | M2III | 13.4 ± 1.0 | 7.2% | 0.37 ± 0.05 | 3.3 | 15 |
| HD137071 | K4III | Adams et al. (1935) | | | 3900 | 1.5 | K5III | 47.0 ± 1.2 | 2.6% | 0.19 ± 0.02 | 6.3 | 20 |
| HD137853 | M2- IIIab | Yamashita (1967) | M2 | Lee et al. (1943) | 3800 | 1.5 | M0III | 41.1 ± 1.2 | 2.9% | 0.32 ± 0.02 | 2.3 | 18 |
| HD138481 | K4.5IIIb | Keenan & McNeil (1989) | M0- IIIab | Yamashita (1967) | 3900 | 1.5 | K5III | 78.2 ± 1.7 | 2.1% | 0.21 ± 0.02 | 2.2 | 37 |
| HD139153 | M1.5IIIb | Keenan & McNeil (1989) | M2- IIIab | Yamashita (1967) | 3800 | 1.5 | M0III | 87.2 ± 1.6 | 1.8% | 0.26 ± 0.02 | 2.8 | 19 |
| HD139374 | Mb E | Cannon & Pickering (1993) | | | 3500 | 1.0 | M4III | 10.1 ± 0.5 | 4.9% | 0.09 ± 0.05 | 7.6 | 8 |
| HD139971 | Ma E | Cannon & Pickering (1993) | | | 3500 | 1.0 | M4III | 14.6 ± 0.7 | 4.6% | 0.01 ± 0.05 | 1.9 | 8 |
| HD142176 | K5III | Moore & Paddock (1950) | M2 | Lee et al. (1943) | 3900 | 1.5 | K5III | 8.2 ± 0.6 | 7.4% | 0.07 ± 0.07 | 1.9 | 8 |
| HD143107 | K2IIab | Keenan & McNeil (1989) | K3III | Moore & Paddock (1950) | 4500 | 2.0 | K1.5III | 97.3 ± 1.1 | 1.2% | 0.11 ± 0.01 | 0.6 | 155 |
| HD144065 | M4 | Lee et al. (1943) | | | 3700 | 1.5 | M2III | 10.7 ± 1.2 | 11.4% | 0.77 ± 0.05 | 15.1 | 8 |

**Table 17** continued on next page

**Table 17** *(continued)*

| Star ID | Primary ST | Primary Reference | Secondary ST | Secondary Reference | Model TEFF | Model logg | Fit ST | Flux[a] $F_{\rm BOL}$ | Err | Reddening $A_{\rm V}$ | Fit $\chi^2$/DOF | Fit DOF |
|---|---|---|---|---|---|---|---|---|---|---|---|---|
| HD147749 | M2+ IIIab | Yamashita (1967) | M2 | Pedoussaut (1962) | 3600 | 1.5 | M3III | 101.0 ± 1.7 | 1.7% | 0.04 ± 0.02 | 12.6 | 21 |
| HD148897 | G8.5III | Keenan & McNeil (1989) | G8III | Lu (1991) | 4500 | 2.0 | K1.5III | 39.5 ± 1.3 | 3.4% | 0.21 ± 0.02 | 1.9 | 54 |
| HD150047 | M6 D | Color Fitting | M2III | | 3400 | 1.0 | M5III | 53.9 ± 1.8 | 3.4% | 0.32 ± 0.03 | 12.6 | 11 |
| HD150450 | M3- IIIab | Yamashita (1967) | M2III | Sato & Kuji (1990) | 3600 | 1.5 | M3III | 135.0 ± 4.2 | 3.1% | 0.02 ± 0.00 | 8.3 | 19 |
| HD150997 | G7III | Keenan & McNeil (1989) | G5III | Ljunggren & Oja (1966) | 5100 | 2.5 | G4III | 128.0 ± 1.8 | 1.4% | 0.07 ± 0.01 | 0.7 | 68 |
| HD151732 | M4.5III | Keenan & McNeil (1989) | M4+ IIIab | Yamashita (1967) | 3500 | 1.0 | M4III | 101.0 ± 1.5 | 1.5% | 0.20 ± 0.02 | 8.8 | 22 |
| HD152173 | M1+ IIIa | Yamashita (1967) | M4 | Lee et al. (1943) | 3800 | 1.5 | M0III | 40.1 ± 0.8 | 2.1% | 0.04 ± 0.02 | 5.2 | 21 |
| HD152863 | G8III | Abt (1985) | G8IIIb | Levato & Abt (1978) | 5100 | 2.5 | G4III | 13.4 ± 0.7 | 5.3% | 0.16 ± 0.04 | 0.1 | 33 |
| HD153287 | G5III | Adams et al. (1935) | G3III | | 5200 | 2.5 | G3III | 11.0 ± 0.3 | 2.8% | 0.20 ± 0.02 | 1.5 | 26 |
| HD153698 | M4III | Moore & Paddock (1950) | M3.5III | Wilson & Joy (1950) | 3500 | 1.0 | M4III | 25.4 ± 0.8 | 3.1% | 0.12 ± 0.03 | 13.9 | 19 |
| HD153834 | K3III | Adams et al. (1935) | K3III | | 4400 | 2.0 | K2III | 24.2 ± 0.7 | 3.0% | 0.09 ± 0.03 | 3.9 | 23 |
| HD154301 | K4III | Abt (1985) | K4III | Stephenson (1960) | 4000 | 2.0 | K4III | 21.1 ± 0.9 | 4.3% | 0.26 ± 0.03 | 4.3 | 15 |
| HD155816 | K2 D | Color Fitting | | | 3800 | 1.5 | M0III | 19.1 ± 0.7 | 3.5% | 0.06 ± 0.03 | 9.0 | 10 |
| HD156966 | M2III | Heard (1956) | M1III | Roman (1955) | 3700 | 1.5 | M2III | 22.8 ± 1.0 | 4.3% | 0.22 ± 0.04 | 4.0 | 15 |
| HD157617 | K1III | Wilson & Joy (1950) | K1III | | 4500 | 2.0 | K1.5III | 21.1 ± 0.8 | 3.7% | 0.07 ± 0.03 | 8.3 | 35 |
| HD163547 | K3III | Adams et al. (1935) | K3III | | 4400 | 2.0 | K2III | 24.4 ± 0.6 | 2.3% | 0.05 ± 0.02 | 9.9 | 25 |
| HD163947 | M5III | Wilson & Joy (1950) | M6 | Lee et al. (1943) | 3400 | 1.0 | M5III | 15.5 ± 0.8 | 5.0% | 0.24 ± 0.04 | 21.3 | 8 |
| HD163993 | G8.5III | Keenan & McNeil (1989) | G8III | Morgan & Keenan (1973) | 5000 | 2.5 | G5III | 103.0 ± 1.3 | 1.3% | 0.00 ± 0.01 | 1.2 | 62 |
| HD164064 | K5III | Wilson & Joy (1950) | K5III | | 3900 | 1.5 | K5III | 34.1 ± 1.3 | 3.9% | 0.15 ± 0.03 | 2.0 | 16 |
| HD165760 | G8III | Keenan & McNeil (1989) | G8III-IV | Roman (1952) | 5000 | 2.5 | G5III | 46.1 ± 1.0 | 2.2% | 0.06 ± 0.02 | 0.5 | 64 |
| HD166013 | M... | Color Fitting | | | 3500 | 1.0 | M4III | 20.1 ± 0.6 | 2.8% | 0.00 ± 0.03 | 14.9 | 11 |
| HD168775 | K2- IIIa | Keenan & McNeil (1989) | K2III | Morgan & Keenan (1973) | 4600 | 2.0 | K1.3III | 68.9 ± 1.2 | 1.7% | 0.03 ± 0.02 | 1.1 | 53 |
| HD169191 | K3III | Roman (1952) | K2III | Adams et al. (1935) | 4500 | 2.0 | K1.5III | 41.3 ± 2.0 | 4.8% | 0.27 ± 0.03 | 0.4 | 31 |
| HD169305 | M2IIIab | Yamashita (1967) | M2III | Sato & Kuji (1990) | 3700 | 1.5 | M2III | 114.0 ± 2.6 | 2.3% | 0.18 ± 0.02 | 6.8 | 24 |
| HD171779 | G9III | Abt (1981) | K0II-III | Stephenson & Sanwal (1969) | 4700 | 2.5 | K1III | 25.7 ± 1.1 | 4.2% | 0.00 ± 0.04 | 0.7 | 26 |
| HD173213 | M... | Color Fitting | | | 3400 | 1.0 | M5III | 23.7 ± 2.0 | 8.4% | 0.30 ± 0.39 | 1.3 | 1 |
| HD173215 | M2 | Lee et al. (1943) | | | 3700 | 1.5 | M2III | 26.1 ± 1.7 | 6.6% | 0.24 ± 0.06 | 7.0 | 2 |
| HD173780 | K2III | Keenan & McNeil (1989) | K2III | Yoss (1961) | 4500 | 2.0 | K1.5III | 47.2 ± 0.9 | 1.9% | 0.01 ± 0.02 | 0.7 | 61 |

**Table 17** *continued on next page*

**Table 17** *(continued)*

VAN BELLE, VON BRAUN, CIARDI ET AL.

| Star ID | Primary ST | Secondary ST | Primary Reference | Secondary Reference | Model TEFF | Model logg | Fit ST | Flux[a] $F_{\mathrm{BOL}}$ | Err | Reddening $A_{\mathrm{V}}$ | Fit $\chi^2$/DOF | Fit DOF |
|---|---|---|---|---|---|---|---|---|---|---|---|---|
| HD173954 | K5 D | | Color Fitting | | 4000 | 2.0 | K4III | 23.4 ± 1.4 | 6.0% | 0.23 ± 0.04 | 2.0 | 6 |
| HD176411 | K1- III | K3III | Keenan & McNeil (1989) | Lu (1991) | 4800 | 2.5 | G9III | 82.6 ± 1.5 | 1.8% | 0.00 ± 0.02 | 0.6 | 32 |
| HD176670 | K2.5III: | K3II | Keenan & McNeil (1989) | Roman (1952) | 4200 | 2.0 | K3.III | 68.5 ± 3.3 | 4.9% | 0.36 ± 0.03 | 3.0 | 30 |
| HD176844 | M4- IIIa | M5 | Yamashita (1967) | Lee et al. (1943) | 3500 | 1.0 | M4III | 60.0 ± 1.7 | 2.8% | 0.35 ± 0.02 | 3.7 | 10 |
| HD176081 | K3III | | Medhi et al. (2007) | | 3900 | 1.5 | K5III | 28.1 ± 2.2 | 7.7% | 0.38 ± 0.05 | 2.6 | 18 |
| HD180450 | M1IIIab | M1 | Yamashita (1967) | Lee et al. (1943) | 3700 | 1.5 | M2III | 50.1 ± 1.2 | 2.4% | 0.16 ± 0.03 | 1.6 | 12 |
| HD184293 | K1III | K2III | Wilson & Joy (1950) | Adams et al. (1935) | 4500 | 2.0 | K1.5III | 30.4 ± 1.7 | 5.7% | 0.26 ± 0.04 | 0.5 | 18 |
| HD185958 | G8IIIa | K0III | Keenan & McNeil (1989) | Sato & Kuji (1990) | 4900 | 2.5 | G7III | 58.8 ± 0.7 | 1.3% | 0.07 ± 0.01 | 1.4 | 65 |
| HD186675 | G8III | G8III | Keenan & McNeil (1989) | Roman (1952) | 5000 | 2.5 | G5III | 35.6 ± 1.1 | 3.0% | 0.00 ± 0.03 | 0.6 | 32 |
| HD186776 | M3.5III | M4- IIIab | Keenan & McNeil (1989) | Yamashita (1967) | 3600 | 1.5 | M3III | 45.1 ± 0.7 | 1.5% | 0.23 ± 0.02 | 3.1 | 42 |
| HD187649 | M2+ IIIa | M2III | Yamashita (1967) | Abt (1985) | 3600 | 1.5 | M3III | 94.0 ± 1.4 | 1.4% | 0.01 ± 0.02 | 11.7 | 18 |
| HD188310 | G9.5IIIb | K0III | Keenan & McNeil (1989) | Roman (1952) | 4800 | 2.5 | G9III | 45.7 ± 0.7 | 1.4% | 0.03 ± 0.01 | 0.4 | 46 |
| HD189695 | K5III | K5III | Wilson & Joy (1950) | | 3900 | 1.5 | K5III | 29.0 ± 0.5 | 1.8% | 0.02 ± 0.02 | 2.6 | 25 |
| HD189849 | kA8hA9mF3 | | Abt & Morrell (1995) | | 7600 | 3.0 | A5III | 31.0 ± 0.7 | 2.2% | 0.02 ± 0.02 | 0.5 | 67 |
| HD191178 | M3III | M1 | Wilson & Joy (1950) | Cannon & Mayall (1949) | 3500 | 1.0 | M4III | 53.9 ± 1.4 | 2.5% | 0.19 ± 0.02 | 26.1 | 15 |
| HD193347 | M2III | M1.5III | Heard (1956) | Wilson & Joy (1950) | 3800 | 1.5 | M0III | 24.1 ± 0.8 | 3.5% | 0.31 ± 0.03 | 6.6 | 20 |
| HD193579 | K5III | M1 | Wilson & Joy (1950) | Lee et al. (1943) | 3900 | 1.5 | K5III | 32.3 ± 1.2 | 3.6% | 0.00 ± 0.04 | 3.4 | 15 |
| HD194097 | K2 D | | Color Fitting | | 4200 | 2.0 | K3.III | 20.4 ± 0.7 | 3.3% | 0.10 ± 0.03 | 2.2 | 17 |
| HD194193 | K7III | K2.5V | Keenan & McNeil (1989) | Uranova (1977) | 3900 | 1.5 | K5III | 36.1 ± 0.8 | 2.1% | 0.24 ± 0.02 | 13.7 | 43 |
| HD194317 | K2.5III | K3III | Keenan & McNeil (1989) | Morgan & Keenan (1973) | 4500 | 2.0 | K1.5III | 93.1 ± 2.4 | 2.5% | 0.37 ± 0.02 | 1.2 | 32 |
| HD194526 | K5III+ | M1 | Keenan & Keller (1953) | Lee et al. (1943) | 3900 | 1.5 | K5III | 24.2 ± 1.3 | 5.2% | 0.23 ± 0.03 | 1.6 | 23 |
| HD196036 | M3 | M3 | Cannon & Mayall (1949) | Lee et al. (1943) | 3500 | 1.0 | M4III | 28.2 ± 1.4 | 5.0% | 0.39 ± 0.03 | 15.0 | 9 |
| HD196360 | G8III | K0III | Yoss (1961) | Bidelman (1957) | 5000 | 2.5 | G5III | 7.2 ± 0.2 | 3.1% | 0.00 ± 0.03 | 1.0 | 22 |
| HD197912 | K0IIIa | G9III | Keenan & McNeil (1989) | Abt (1985) | 4800 | 2.5 | G9III | 73.2 ± 0.7 | 1.0% | 0.04 ± 0.01 | 0.7 | 52 |
| HD198237 | M2III | K3III | Abt (1985) | Fehrenbach et al. (1961) | 3800 | 1.5 | M0III | 24.4 ± 0.5 | 2.2% | 0.16 ± 0.02 | 4.3 | 28 |
| HD199101 | K5III | K5III | Sato & Kuji (1990) | Adams et al. (1935) | 3900 | 1.5 | K5III | 44.0 ± 0.7 | 1.6% | 0.04 ± 0.02 | 2.1 | 24 |
| HD199697 | K3.5III | K5 | Keenan & McNeil (1989) | Lee et al. (1943) | 4200 | 2.0 | K3.III | 44.6 ± 1.3 | 2.9% | 0.27 ± 0.02 | 2.8 | 19 |
| HD199799 | M4 | M6 | Neckel (1958) | Nassau & Blanco (1954) | 3400 | 1.0 | M5III | 50.0 ± 0.9 | 1.9% | 0.49 ± 0.02 | 14.5 | 14 |

**Table 17** *(continued)*

| Star ID | Primary ST | Primary Reference | Secondary ST | Secondary Reference | Model TEFF | Model logg | Fit ST | Flux[a] $F_{BOL}$ | Err | Reddening $A_V$ | Fit $\chi^2$/DOF | Fit DOF |
|---|---|---|---|---|---|---|---|---|---|---|---|---|
| HD202109 | G8+ III- | Keenan & McNeil (1989) | G8.5II-IIIa | Keenan & Wilson (1977) | 5000 | 2.5 | G5III | 164.0 ± 2.0 | 1.2% | 0.04 ± 0.01 | 1.1 | 115 |
| HD205435 | G8III | Keenan & McNeil (1989) | G8III | Levato & Abt (1978) | 5200 | 2.5 | G3III | 79.0 ± 1.2 | 1.5% | 0.06 ± 0.02 | 1.0 | 56 |
| HD205733 | M8 | Lee et al. (1943) | M4:III | Keenan (1942) | 3200 | 1.0 | M6III | 23.6 ± 0.4 | 1.5% | 0.06 ± 0.04 | 12.4 | 9 |
| HD206645 | K4II-III | Adams et al. (1935) | K3:III | | 4200 | 2.0 | K3.III | 32.5 ± 0.7 | 2.3% | 0.30 ± 0.02 | 3.0 | 30 |
| HD206749 | M2IIIb | Keenan & McNeil (1989) | M2- IIIab | Yamashita (1967) | 3800 | 1.5 | M0III | 63.5 ± 1.0 | 1.6% | 0.26 ± 0.02 | 2.1 | 9 |
| HD207001 | K5 D | Color Fitting | | | 3700 | 1.5 | M2III | 11.5 ± 0.4 | 3.3% | 0.33 ± 0.03 | 20.1 | 7 |
| HD207328 | M3IIIa | Keenan & McNeil (1989) | M3Ia | Garrison & Kormendy (1976) | 3400 | 1.0 | M5III | 34.3 ± 1.5 | 4.4% | 0.11 ± 0.05 | 18.9 | 8 |
| HD209857 | M4+ IIIab | Yamashita (1967) | M4 | Wing (1967) | 3500 | 1.0 | M4III | 66.3 ± 1.0 | 1.6% | 0.06 ± 0.02 | 6.6 | 44 |
| HD210514 | M4 | Barbier & Maiocchi (1966) | M4III | Moore & Paddock (1950) | 3500 | 1.0 | M4III | 28.0 ± 0.9 | 3.2% | 0.20 ± 0.03 | 5.2 | 12 |
| HD211800 | M4 | Lee et al. (1943) | M1III | Adams et al. (1926) | 3900 | 1.5 | K5III | 21.5 ± 1.5 | 7.1% | 0.67 ± 0.03 | 4.8 | 11 |
| HD212470 | M4.5III | Wilson & Joy (1950) | M5 | Lee et al. (1943) | 3300 | 1.0 | M5.5III | 69.5 ± 1.8 | 2.5% | 0.00 ± 0.03 | 39.4 | 8 |
| HD212496 | G9IIIb | Keenan & McNeil (1989) | G5III | McCuskey (1955) | 4900 | 2.5 | G7III | 61.7 ± 0.7 | 1.2% | 0.10 ± 0.01 | 0.4 | 73 |
| HD214868 | K2.5III | Keenan & McNeil (1989) | K3III | Roman (1952) | 4400 | 2.0 | K2III | 81.8 ± 1.1 | 1.3% | 0.21 ± 0.01 | 0.9 | 53 |
| HD215182 | A2 | Parsons & Ake (1998) | G2III | Abt (1985) | 5200 | 2.5 | G3III | 209.0 ± 3.3 | 1.6% | 0.05 ± 0.02 | 0.5 | 49 |
| HD215547 | M... | Color Fitting | | | 3500 | 1.0 | M4III | 24.0 ± 0.7 | 2.8% | 0.65 ± 0.02 | 4.7 | 13 |
| HD216131 | G8+ III | Keenan & McNeil (1989) | K1III | Skiff (2014a) | 5100 | 2.5 | G4III | 127.0 ± 1.7 | 1.4% | 0.09 ± 0.01 | 1.4 | 74 |
| HD216930 | M E | Color Fitting | K5III | | 3500 | 1.0 | M4III | 14.5 ± 0.5 | 3.4% | 0.08 ± 0.03 | 8.9 | 11 |
| HD218452 | K4III | Abt (1985) | K5III | Levato & Abt (1978) | 4100 | 2.0 | K3.5III | 39.5 ± 0.9 | 2.3% | 0.05 ± 0.02 | 2.1 | 19 |
| HD219945 | G7III | Gyldenkerne (1955) | K0III | Roman (1952) | 4900 | 2.5 | G7III | 24.9 ± 1.0 | 3.9% | 0.13 ± 0.03 | 0.2 | 22 |
| HD220211 | M... | Color Fitting | | | 3500 | 1.5 | M4III | 20.0 ± 0.4 | 2.2% | 0.11 ± 0.02 | 30.0 | 12 |
| HD221115 | G8IIa | Keenan & McNeil (1989) | G9III | Halliday (1955) | 5000 | 2.5 | G5III | 46.6 ± 0.4 | 0.8% | 0.00 ± 0.01 | 1.4 | 62 |
| HD221345 | G8III | Roman (1955) | K0III+ | Keenan & Keller (1953) | 4900 | 2.5 | G7III | 34.2 ± 0.8 | 2.5% | 0.25 ± 0.02 | 0.3 | 46 |
| HD221673 | K4IIIb | Abt (1981) | K3III | Stephenson & Sanwal (1969) | 4200 | 2.0 | K3.III | 52.9 ± 0.9 | 1.7% | 0.11 ± 0.02 | 2.0 | 22 |
| HD221905 | M0III | Sato & Kuji (1990) | M3III | | 3600 | 1.5 | M3III | 35.0 ± 0.7 | 1.9% | 0.20 ± 0.02 | 28.6 | 18 |
| HD222842 | G8IIIb | Abt (1985) | K0II-III | Yoss (1961) | 5100 | 2.5 | G4III | 41.0 ± 1.5 | 3.7% | 0.23 ± 0.03 | 0.3 | 40 |
| HD224303 | M2.5III | Wilson & Joy (1950) | M4 | Lee et al. (1943) | 3800 | 1.5 | M0III | 34.6 ± 0.8 | 2.3% | 0.31 ± 0.02 | 4.4 | 16 |
| HIP8682 | M2 | Lee et al. (1943) | M6III | | 3200 | 1.0 | M6III | 18.4 ± 0.7 | 3.9% | 0.69 ± 0.03 | 12.6 | 6 |
| HIP35915 | Meq | Bopp (1984) | M6 | Swings et al. (1953) | 3300 | 1.0 | M5.5III | 10.8 ± 0.8 | 7.1% | 0.65 ± 0.04 | 21.4 | 11 |

**Table 17** *continued on next page*

**Table 17** (continued)

| Star ID | Primary ST | Primary Reference | Secondary ST | Secondary Reference | Model $T_{EFF}$ | Model logg | Fit ST | Flux[a] $F_{BOL}$ | Err | Reddening $A_V$ | Fit $\chi^2/D$ |
|---|---|---|---|---|---|---|---|---|---|---|---|
| HIP68357 | M7III: | Herbig (1960) | M7 | Lee et al. (1943) | 3100 | 0.5 | M7III | $38.4 \pm 0.9$ | 2.3% | $0.00 \pm 0.03$ | 20. |
| HIP86677 | M2III | Wilson & Joy (1952) | | | 3400 | 1.0 | M5III | $42.1 \pm 1.8$ | 4.2% | $0.00 \pm 0.09$ | 14. |
| HIP113390 | M6 | Lee et al. (1943) | M6IIIe | Joy (1942) | 3300 | 1.0 | M5.5III | $21.0 \pm 1.4$ | 6.5% | $1.13 \pm 0.04$ | 5.4 |
| IRC+30095 | M8 D | Color Fitting | | | 2900 | 0.5 | M7.7III | $26.7 \pm 1.4$ | 5.2% | $0.24 \pm 0.04$ | 52. |
| IRC+40533 | M8 | van de Kamp (1958) | M5 | Lee et al. (1943) | 2900 | 0.5 | M7.7III | $38.6 \pm 1.5$ | 4.0% | $0.32 \pm 0.03$ | 22. |
| IRC+60005 | B2 | Shaw & Guinan (1989) | M3e + B | Barbier (1971) | 3300 | 1.0 | M5.5III | $25.2 \pm 0.6$ | 2.3% | $0.00 \pm 0.07$ | 1.9 |

[a]Flux units are $10^{-8}$ erg s$^{-1}$ cm$^{-2}$.

Note.—For more detail, see §4.3.

**Table 18**. Angular sizes, UD sizes, limb darkening corrections, unreddened $V_0 - K_0$ color, and computed effective temperatures for the program stars.

| Star ID | UD size (mas) | Initial $T_{\rm EFF}$ (K) | UD-to-LD | LD size (mas) | $V_0 - K_0$ (mag) | $T_{\rm EFF}$ (K) |
|---------|---------------|---------------------------|----------|---------------|-------------------|-------------------|
| HD598   | $2.591 \pm 0.021 \pm 0.043$ | 3562 | 1.046 | $2.709 \pm 0.045$ | $5.53 \pm 0.05$ | $3484 \pm 34$ |
| HD1632  | $2.281 \pm 0.014 \pm 0.043$ | 3969 | 1.037 | $2.366 \pm 0.045$ | $4.06 \pm 0.08$ | $3897 \pm 47$ |
| HD3346  | $3.102 \pm 0.007 \pm 0.058$ | 3936 | 1.038 | $3.219 \pm 0.060$ | $4.00 \pm 0.07$ | $3864 \pm 39$ |
| HD3546  | $1.689 \pm 0.011 \pm 0.043$ | 5188 | 1.026 | $1.734 \pm 0.045$ | $2.12 \pm 0.07$ | $5120 \pm 69$ |
| HD3574  | $2.270 \pm 0.040 \pm 0.043$ | 4140 | 1.035 | $2.349 \pm 0.045$ | $3.82 \pm 0.06$ | $4070 \pm 41$ |
| HD3627  | $3.968 \pm 0.005 \pm 0.120$ | 4505 | 1.031 | $4.090 \pm 0.124$ | $2.76 \pm 0.07$ | $4438 \pm 68$ |
| HD5006  | $2.152 \pm 0.031 \pm 0.043$ | 3476 | 1.048 | $2.255 \pm 0.045$ | $5.17 \pm 0.11$ | $3395 \pm 51$ |
| HD6186  | $1.879 \pm 0.041 \pm 0.043$ | 4888 | 1.028 | $1.932 \pm 0.045$ | $2.25 \pm 0.07$ | $4821 \pm 58$ |
| HD6409  | $1.937 \pm 0.014 \pm 0.039$ | 3521 | 1.047 | $2.027 \pm 0.040$ | $5.16 \pm 0.10$ | $3442 \pm 36$ |
| HD7000  | $1.835 \pm 0.049 \pm 0.043$ | 3384 | 1.051 | $1.928 \pm 0.051$ | $5.87 \pm 0.11$ | $3301 \pm 56$ |
| HD7087  | $1.578 \pm 0.030 \pm 0.043$ | 4853 | 1.028 | $1.623 \pm 0.045$ | $2.27 \pm 0.11$ | $4786 \pm 68$ |
| HD8126  | $2.038 \pm 0.027 \pm 0.039$ | 4284 | 1.033 | $2.105 \pm 0.040$ | $3.31 \pm 0.08$ | $4216 \pm 44$ |
| HD9500  | $2.313 \pm 0.012 \pm 0.043$ | 3584 | 1.045 | $2.417 \pm 0.045$ | $5.11 \pm 0.09$ | $3506 \pm 36$ |
| HD9927  | $3.354 \pm 0.008 \pm 0.067$ | 4602 | 1.030 | $3.454 \pm 0.069$ | $2.77 \pm 0.07$ | $4535 \pm 47$ |
| HD10380 | $2.830 \pm 0.008 \pm 0.048$ | 4214 | 1.034 | $2.926 \pm 0.050$ | $3.16 \pm 0.06$ | $4145 \pm 36$ |
| HD12929 | $5.891 \pm 0.044 \pm 1.617$ | 4721 | 1.029 | $6.062 \pm 1.664$ | $2.60 \pm 0.07$ | $4654 \pm 639$ |
| HD14146 | $1.974 \pm 0.027 \pm 0.039$ | 3722 | 1.042 | $2.057 \pm 0.040$ | $4.38 \pm 0.07$ | $3646 \pm 44$ |
| HD14512 | $2.824 \pm 0.026 \pm 0.048$ | 3410 | 1.050 | $2.964 \pm 0.051$ | $6.08 \pm 0.06$ | $3328 \pm 38$ |
| HD14872 | $3.154 \pm 0.006 \pm 0.058$ | 4029 | 1.036 | $3.269 \pm 0.060$ | $3.64 \pm 0.07$ | $3958 \pm 40$ |
| HD14901 | $3.722 \pm 0.004 \pm 0.076$ | 3441 | 1.049 | $3.904 \pm 0.079$ | $6.44 \pm 0.06$ | $3360 \pm 35$ |
| HD15656 | $2.479 \pm 0.015 \pm 0.043$ | 3991 | 1.037 | $2.571 \pm 0.045$ | $3.54 \pm 0.07$ | $3919 \pm 37$ |
| HD17361 | $1.837 \pm 0.009 \pm 0.043$ | 4862 | 1.028 | $1.889 \pm 0.045$ | $2.50 \pm 0.05$ | $4795 \pm 60$ |
| HD17709 | $3.720 \pm 0.003 \pm 0.091$ | 3949 | 1.038 | $3.860 \pm 0.095$ | $3.80 \pm 0.07$ | $3876 \pm 50$ |
| HD18322 | $2.538 \pm 0.037 \pm 0.043$ | 4681 | 1.029 | $2.613 \pm 0.045$ | $2.48 \pm 0.05$ | $4614 \pm 40$ |
| HD19787 | $1.762 \pm 0.026 \pm 0.043$ | 4908 | 1.028 | $1.811 \pm 0.045$ | $2.27 \pm 0.05$ | $4841 \pm 60$ |
| HD20844 | $2.528 \pm 0.009 \pm 0.043$ | 3505 | 1.047 | $2.647 \pm 0.045$ | $5.67 \pm 0.07$ | $3425 \pm 31$ |
| HD21465 | $2.095 \pm 0.052 \pm 0.039$ | 3499 | 1.047 | $2.194 \pm 0.054$ | $4.78 \pm 0.09$ | $3419 \pm 50$ |
| HD25604 | $1.876 \pm 0.034 \pm 0.043$ | 4784 | 1.029 | $1.930 \pm 0.045$ | $2.35 \pm 0.07$ | $4717 \pm 56$ |
| HD26605 | $0.795 \pm 0.074 \pm 0.072$ | 4845 | 1.028 | $0.817 \pm 0.077$ | $2.06 \pm 0.37$ | $4778 \pm 226$ |
| HD27308 | $2.513 \pm 0.012 \pm 0.043$ | 3421 | 1.049 | $2.637 \pm 0.045$ | $6.15 \pm 0.08$ | $3339 \pm 34$ |
| HD27348 | $1.734 \pm 0.371 \pm 0.035$ | 4274 | 1.033 | $1.791 \pm 0.383$ | $2.14 \pm 0.10$ | $4205 \pm 451$ |
| HD27697 | $2.176 \pm 0.003 \pm 0.043$ | 5043 | 1.027 | $2.235 \pm 0.045$ | $2.09 \pm 0.05$ | $4976 \pm 52$ |
| HD28100 | $1.516 \pm 0.027 \pm 0.043$ | 4965 | 1.028 | $1.558 \pm 0.045$ | $2.25 \pm 0.07$ | $4898 \pm 74$ |
| HD28292 | $1.870 \pm 0.056 \pm 0.043$ | 4532 | 1.031 | $1.927 \pm 0.057$ | $2.70 \pm 0.05$ | $4464 \pm 68$ |
| HD28305 | $2.444 \pm 0.005 \pm 0.043$ | 4956 | 1.028 | $2.512 \pm 0.045$ | $2.20 \pm 0.07$ | $4889 \pm 45$ |
| HD28307 | $2.031 \pm 0.033 \pm 0.039$ | 5083 | 1.027 | $2.086 \pm 0.040$ | $2.11 \pm 0.05$ | $5016 \pm 51$ |
| HD28581 | $2.070 \pm 0.155 \pm 0.038$ | 3483 | 1.048 | $2.169 \pm 0.163$ | $4.45 \pm 0.11$ | $3403 \pm 149$ |

**Table 18** *continued on next page*

**Table 18** (continued)

| Star ID | UD size | Initial | UD-to-LD | LD size | $V_0 - K_0$ | $T_{EFF}$ |
|---------|---------|---------|----------|---------|-------------|-----------|
|         | (mas)   | $T_{EFF}$ (K) |    | (mas)   | (mag)       | (K)       |
| HD28595 | $3.007 \pm 0.024 \pm 0.053$ | 3559 | 1.046 | $3.144 \pm 0.055$ | $4.90 \pm 0.08$ | $3480 \pm 45$ |
| HD29094 | $2.607 \pm 0.009 \pm 0.043$ | 4500 | 1.031 | $2.687 \pm 0.045$ | $2.90 \pm 0.07$ | $4432 \pm 43$ |
| HD30504 | $2.711 \pm 0.009 \pm 0.048$ | 4198 | 1.034 | $2.803 \pm 0.050$ | $3.44 \pm 0.07$ | $4128 \pm 40$ |
| HD30605 | $1.843 \pm 0.013 \pm 0.043$ | 3923 | 1.038 | $1.913 \pm 0.045$ | $3.66 \pm 0.11$ | $3851 \pm 48$ |
| HD30834 | $2.603 \pm 0.006 \pm 0.043$ | 4249 | 1.033 | $2.690 \pm 0.045$ | $3.24 \pm 0.09$ | $4179 \pm 40$ |
| HD31139 | $3.442 \pm 0.018 \pm 0.072$ | 3787 | 1.041 | $3.582 \pm 0.075$ | $4.30 \pm 0.08$ | $3712 \pm 43$ |
| HD31421 | $2.718 \pm 0.031 \pm 0.048$ | 4556 | 1.030 | $2.800 \pm 0.050$ | $2.67 \pm 0.07$ | $4489 \pm 44$ |
| HD33463 | $2.650 \pm 0.006 \pm 0.048$ | 3663 | 1.043 | $2.764 \pm 0.050$ | $4.88 \pm 0.09$ | $3586 \pm 35$ |
| HD34334 | $2.512 \pm 0.010 \pm 0.043$ | 4422 | 1.032 | $2.591 \pm 0.045$ | $3.20 \pm 0.07$ | $4353 \pm 46$ |
| HD34559 | $1.223 \pm 0.043 \pm 0.048$ | 5104 | 1.027 | $1.256 \pm 0.049$ | $1.96 \pm 0.09$ | $5037 \pm 111$ |
| HD34577 | $2.333 \pm 0.011 \pm 0.043$ | 3611 | 1.044 | $2.437 \pm 0.045$ | $5.45 \pm 0.08$ | $3533 \pm 47$ |
| HD35620 | $2.065 \pm 0.013 \pm 0.039$ | 4222 | 1.034 | $2.135 \pm 0.040$ | $3.12 \pm 0.08$ | $4152 \pm 40$ |
| HD38656 | $1.588 \pm 0.020 \pm 0.043$ | 4942 | 1.028 | $1.632 \pm 0.045$ | $2.17 \pm 0.05$ | $4875 \pm 68$ |
| HD39003 | $2.376 \pm 0.015 \pm 0.043$ | 4785 | 1.029 | $2.444 \pm 0.045$ | $2.48 \pm 0.05$ | $4717 \pm 45$ |
| HD39045 | $3.105 \pm 0.010 \pm 0.058$ | 3581 | 1.045 | $3.245 \pm 0.060$ | $4.97 \pm 0.07$ | $3503 \pm 35$ |
| HD39225 | $2.385 \pm 0.012 \pm 0.043$ | 3718 | 1.042 | $2.485 \pm 0.045$ | $4.15 \pm 0.09$ | $3643 \pm 42$ |
| HD39732 | $2.481 \pm 0.055 \pm 0.043$ | 3515 | 1.047 | $2.597 \pm 0.057$ | $5.50 \pm 0.10$ | $3436 \pm 64$ |
| HD40441 | $1.529 \pm 0.010 \pm 0.043$ | 3978 | 1.037 | $1.586 \pm 0.045$ | $4.04 \pm 0.10$ | $3907 \pm 88$ |
| HD41116 | $1.861 \pm 0.005 \pm 0.043$ | 5030 | 1.027 | $1.912 \pm 0.045$ | $2.02 \pm 0.07$ | $4963 \pm 63$ |
| HD42049 | $2.272 \pm 0.046 \pm 0.043$ | 3891 | 1.039 | $2.360 \pm 0.047$ | $3.87 \pm 0.08$ | $3818 \pm 84$ |
| HD43039 | $1.967 \pm 0.037 \pm 0.039$ | 4795 | 1.029 | $2.023 \pm 0.040$ | $2.40 \pm 0.07$ | $4728 \pm 49$ |
| HD46709 | $1.683 \pm 0.008 \pm 0.043$ | 4047 | 1.036 | $1.744 \pm 0.045$ | $3.43 \pm 0.08$ | $3976 \pm 53$ |
| HD48450 | $2.006 \pm 0.008 \pm 0.039$ | 4085 | 1.036 | $2.077 \pm 0.040$ | $3.44 \pm 0.07$ | $4014 \pm 40$ |
| HD49738 | $1.339 \pm 0.050 \pm 0.048$ | 4519 | 1.031 | $1.380 \pm 0.051$ | $2.75 \pm 0.10$ | $4452 \pm 98$ |
| HD49968 | $1.871 \pm 0.013 \pm 0.043$ | 4045 | 1.036 | $1.939 \pm 0.045$ | $3.41 \pm 0.11$ | $3974 \pm 61$ |
| HD54719 | $2.316 \pm 0.009 \pm 0.043$ | 4579 | 1.030 | $2.386 \pm 0.045$ | $2.86 \pm 0.06$ | $4511 \pm 48$ |
| HD60136 | $2.464 \pm 0.075 \pm 0.043$ | 3593 | 1.045 | $2.574 \pm 0.078$ | $5.05 \pm 0.10$ | $3515 \pm 90$ |
| HD61913 | $3.782 \pm 0.028 \pm 0.101$ | 3783 | 1.041 | $3.936 \pm 0.105$ | $4.83 \pm 0.07$ | $3708 \pm 52$ |
| HD62044 | $2.351 \pm 0.003 \pm 0.043$ | 4695 | 1.029 | $2.420 \pm 0.045$ | $2.57 \pm 0.07$ | $4627 \pm 50$ |
| HD62285 | $2.499 \pm 0.015 \pm 0.043$ | 3994 | 1.037 | $2.591 \pm 0.045$ | $3.60 \pm 0.08$ | $3922 \pm 47$ |
| HD62345 | $2.318 \pm 0.016 \pm 0.043$ | 5068 | 1.027 | $2.381 \pm 0.045$ | $2.07 \pm 0.07$ | $5001 \pm 49$ |
| HD62509 | $8.269 \pm 0.088 \pm 2.414$ | 4720 | 1.029 | $8.510 \pm 2.484$ | $2.22 \pm 0.07$ | $4653 \pm 679$ |
| HD62721 | $2.857 \pm 0.005 \pm 0.048$ | 4016 | 1.037 | $2.961 \pm 0.050$ | $3.52 \pm 0.07$ | $3945 \pm 37$ |
| HD66216 | $1.593 \pm 0.023 \pm 0.043$ | 4761 | 1.029 | $1.639 \pm 0.045$ | $2.51 \pm 0.10$ | $4694 \pm 84$ |
| HD74442 | $2.300 \pm 0.014 \pm 0.043$ | 4847 | 1.028 | $2.365 \pm 0.045$ | $2.44 \pm 0.05$ | $4780 \pm 46$ |
| HD76294 | $3.214 \pm 0.062 \pm 0.063$ | 4816 | 1.028 | $3.305 \pm 0.064$ | $2.20 \pm 0.07$ | $4749 \pm 49$ |
| HD76830 | $3.231 \pm 0.003 \pm 0.063$ | 3552 | 1.046 | $3.379 \pm 0.065$ | $5.19 \pm 0.07$ | $3473 \pm 37$ |
| HD81817 | $3.195 \pm 0.028 \pm 0.063$ | 4152 | 1.035 | $3.305 \pm 0.065$ | $3.20 \pm 0.06$ | $4082 \pm 44$ |
| HD82198 | $2.671 \pm 0.015 \pm 0.048$ | 3899 | 1.038 | $2.774 \pm 0.050$ | $3.83 \pm 0.06$ | $3826 \pm 43$ |
| HD82381 | $2.071 \pm 0.018 \pm 0.039$ | 4259 | 1.033 | $2.140 \pm 0.040$ | $3.15 \pm 0.07$ | $4190 \pm 44$ |

**Table 18** continued on next page

**Table 18** (continued)

| Star ID | UD size (mas) | Initial $T_{\rm EFF}$ (K) | UD-to-LD | LD size (mas) | $V_0 - K_0$ (mag) | $T_{\rm EFF}$ (K) |
|---|---|---|---|---|---|---|
| HD82635 | $1.436 \pm 0.020 \pm 0.043$ | 5169 | 1.027 | $1.474 \pm 0.045$ | $2.04 \pm 0.07$ | $5101 \pm 84$ |
| HD85503 | $2.765 \pm 0.004 \pm 0.048$ | 4570 | 1.030 | $2.849 \pm 0.050$ | $2.67 \pm 0.06$ | $4502 \pm 42$ |
| HD87046 | $1.896 \pm 0.011 \pm 0.038$ | 3573 | 1.045 | $1.982 \pm 0.040$ | $5.04 \pm 0.08$ | $3494 \pm 47$ |
| HD87837 | $3.154 \pm 0.003 \pm 0.058$ | 4179 | 1.034 | $3.262 \pm 0.060$ | $3.30 \pm 0.07$ | $4109 \pm 40$ |
| HD90254 | $2.938 \pm 0.009 \pm 0.053$ | 3867 | 1.039 | $3.053 \pm 0.055$ | $4.37 \pm 0.07$ | $3794 \pm 43$ |
| HD92620 | $3.565 \pm 0.005 \pm 0.082$ | 3526 | 1.047 | $3.731 \pm 0.086$ | $4.94 \pm 0.07$ | $3447 \pm 42$ |
| HD94264 | $2.494 \pm 0.004 \pm 0.043$ | 4779 | 1.029 | $2.566 \pm 0.045$ | $2.41 \pm 0.07$ | $4712 \pm 45$ |
| HD94336 | $1.940 \pm 0.032 \pm 0.039$ | 3589 | 1.045 | $2.027 \pm 0.040$ | $4.83 \pm 0.06$ | $3511 \pm 43$ |
| HD95212 | $2.104 \pm 0.029 \pm 0.043$ | 4118 | 1.035 | $2.178 \pm 0.045$ | $3.43 \pm 0.09$ | $4048 \pm 55$ |
| HD95345 | $1.787 \pm 0.016 \pm 0.043$ | 4553 | 1.030 | $1.841 \pm 0.045$ | $2.58 \pm 0.08$ | $4486 \pm 61$ |
| HD96274 | $1.911 \pm 0.023 \pm 0.039$ | 3572 | 1.045 | $1.998 \pm 0.040$ | $4.75 \pm 0.10$ | $3494 \pm 46$ |
| HD96833 | $3.797 \pm 0.005 \pm 0.101$ | 4679 | 1.029 | $3.909 \pm 0.104$ | $2.58 \pm 0.07$ | $4611 \pm 63$ |
| HD100236 | $1.679 \pm 0.022 \pm 0.043$ | 3457 | 1.048 | $1.760 \pm 0.046$ | $5.29 \pm 0.12$ | $3376 \pm 69$ |
| HD102224 | $3.194 \pm 0.019 \pm 0.063$ | 4553 | 1.030 | $3.291 \pm 0.065$ | $2.78 \pm 0.06$ | $4486 \pm 47$ |
| HD104207 | $3.153 \pm 0.002 \pm 0.058$ | 3448 | 1.049 | $3.306 \pm 0.061$ | $5.63 \pm 0.05$ | $3367 \pm 35$ |
| HD104575 | $1.654 \pm 0.019 \pm 0.043$ | 3570 | 1.045 | $1.729 \pm 0.045$ | $4.91 \pm 0.12$ | $3492 \pm 67$ |
| HD104831 | $1.540 \pm 0.042 \pm 0.043$ | 3629 | 1.044 | $1.608 \pm 0.045$ | $4.27 \pm 0.10$ | $3552 \pm 62$ |
| HD104979 | $1.908 \pm 0.022 \pm 0.039$ | 5095 | 1.027 | $1.959 \pm 0.040$ | $2.25 \pm 0.05$ | $5028 \pm 52$ |
| HD106714 | $1.217 \pm 0.037 \pm 0.048$ | 5223 | 1.026 | $1.249 \pm 0.050$ | $2.22 \pm 0.09$ | $5156 \pm 108$ |
| HD107256 | $1.775 \pm 0.008 \pm 0.043$ | 3475 | 1.048 | $1.860 \pm 0.045$ | $5.88 \pm 0.09$ | $3395 \pm 56$ |
| HD109282 | $1.859 \pm 0.013 \pm 0.043$ | 3599 | 1.045 | $1.942 \pm 0.045$ | $4.96 \pm 0.07$ | $3521 \pm 53$ |
| HD113226 | $3.066 \pm 0.004 \pm 0.058$ | 5155 | 1.027 | $3.148 \pm 0.059$ | $1.90 \pm 0.05$ | $5088 \pm 49$ |
| HD114780 | $2.254 \pm 0.005 \pm 0.043$ | 3879 | 1.039 | $2.341 \pm 0.045$ | $3.90 \pm 0.07$ | $3806 \pm 41$ |
| HD116207 | $2.148 \pm 0.005 \pm 0.043$ | 3539 | 1.046 | $2.247 \pm 0.045$ | $5.49 \pm 0.07$ | $3460 \pm 47$ |
| HD118669 | $1.713 \pm 0.021 \pm 0.043$ | 3526 | 1.047 | $1.793 \pm 0.045$ | $5.02 \pm 0.09$ | $3447 \pm 61$ |
| HD119584 | $1.302 \pm 0.115 \pm 0.049$ | 4418 | 1.032 | $1.343 \pm 0.118$ | $3.50 \pm 0.08$ | $4350 \pm 195$ |
| HD120819 | $2.536 \pm 0.035 \pm 0.043$ | 3805 | 1.040 | $2.638 \pm 0.045$ | $4.29 \pm 0.05$ | $3731 \pm 43$ |
| HD121860 | $2.013 \pm 0.010 \pm 0.039$ | 3553 | 1.046 | $2.105 \pm 0.040$ | $5.14 \pm 0.06$ | $3474 \pm 47$ |
| HD122316 | $3.801 \pm 0.007 \pm 0.101$ | 3207 | 1.056 | $4.015 \pm 0.107$ | $7.67 \pm 0.06$ | $3121 \pm 44$ |
| HD127093 | $2.665 \pm 0.005 \pm 0.048$ | 3556 | 1.046 | $2.787 \pm 0.050$ | $4.95 \pm 0.07$ | $3477 \pm 48$ |
| HD128902 | $1.944 \pm 0.026 \pm 0.039$ | 4039 | 1.036 | $2.014 \pm 0.040$ | $3.48 \pm 0.07$ | $3968 \pm 51$ |
| HD130084 | $2.121 \pm 0.024 \pm 0.043$ | 3817 | 1.040 | $2.206 \pm 0.045$ | $4.30 \pm 0.06$ | $3743 \pm 48$ |
| HD133124 | $2.962 \pm 0.004 \pm 0.053$ | 4058 | 1.036 | $3.068 \pm 0.055$ | $3.48 \pm 0.07$ | $3987 \pm 41$ |
| HD133208 | $2.414 \pm 0.004 \pm 0.043$ | 5038 | 1.027 | $2.480 \pm 0.045$ | $2.15 \pm 0.06$ | $4971 \pm 47$ |
| HD133582 | $2.415 \pm 0.009 \pm 0.043$ | 4421 | 1.032 | $2.491 \pm 0.045$ | $2.90 \pm 0.06$ | $4353 \pm 45$ |
| HD135722 | $2.670 \pm 0.003 \pm 0.048$ | 5030 | 1.027 | $2.743 \pm 0.050$ | $2.37 \pm 0.05$ | $4963 \pm 48$ |
| HD136404 | $1.419 \pm 0.017 \pm 0.043$ | 3760 | 1.041 | $1.477 \pm 0.045$ | $4.73 \pm 0.10$ | $3685 \pm 88$ |
| HD137071 | $2.336 \pm 0.008 \pm 0.043$ | 4010 | 1.037 | $2.422 \pm 0.045$ | $3.77 \pm 0.07$ | $3939 \pm 45$ |
| HD137853 | $2.399 \pm 0.004 \pm 0.043$ | 3827 | 1.040 | $2.494 \pm 0.045$ | $4.25 \pm 0.06$ | $3753 \pm 44$ |
| HD138481 | $3.081 \pm 0.008 \pm 0.058$ | 3966 | 1.037 | $3.196 \pm 0.060$ | $3.87 \pm 0.07$ | $3894 \pm 42$ |

**Table 18** *(continued)*

| Star ID | UD size | Initial | UD-to-LD | LD size | $V_0 - K_0$ | $T_{\mathrm{EFF}}$ |
|---|---|---|---|---|---|---|
|  | (mas) | $T_{\mathrm{EFF}}$ (K) |  | (mas) | (mag) | (K) |
| HD139153 | $3.419 \pm 0.004 \pm 0.072$ | 3869 | 1.039 | $3.552 \pm 0.075$ | $4.27 \pm 0.05$ | $3796 \pm 44$ |
| HD139374 | $1.406 \pm 0.016 \pm 0.043$ | 3520 | 1.047 | $1.472 \pm 0.045$ | $5.36 \pm 0.08$ | $3440 \pm 68$ |
| HD139971 | $1.655 \pm 0.012 \pm 0.043$ | 3557 | 1.046 | $1.731 \pm 0.045$ | $5.15 \pm 0.09$ | $3478 \pm 61$ |
| HD142176 | $0.938 \pm 0.010 \pm 0.063$ | 4087 | 1.035 | $0.971 \pm 0.065$ | $3.53 \pm 0.35$ | $4016 \pm 155$ |
| HD143107 | $2.751 \pm 0.006 \pm 0.048$ | 4433 | 1.031 | $2.837 \pm 0.050$ | $2.85 \pm 0.05$ | $4365 \pm 40$ |
| HD144065 | $1.387 \pm 0.028 \pm 0.048$ | 3598 | 1.045 | $1.449 \pm 0.050$ | $5.34 \pm 0.10$ | $3520 \pm 120$ |
| HD147749 | $3.619 \pm 0.005 \pm 0.087$ | 3901 | 1.038 | $3.758 \pm 0.090$ | $4.39 \pm 0.07$ | $3828 \pm 49$ |
| HD148897 | $1.856 \pm 0.007 \pm 0.043$ | 4308 | 1.033 | $1.917 \pm 0.045$ | $3.11 \pm 0.07$ | $4239 \pm 62$ |
| HD150047 | $3.363 \pm 0.010 \pm 0.067$ | 3459 | 1.048 | $3.526 \pm 0.071$ | $6.09 \pm 0.05$ | $3378 \pm 45$ |
| HD150450 | $4.242 \pm 0.011 \pm 0.173$ | 3874 | 1.039 | $4.407 \pm 0.180$ | $4.86 \pm 0.06$ | $3801 \pm 83$ |
| HD150997 | $2.423 \pm 0.006 \pm 0.043$ | 5059 | 1.027 | $2.489 \pm 0.045$ | $2.17 \pm 0.06$ | $4991 \pm 48$ |
| HD151732 | $4.459 \pm 0.012 \pm 0.236$ | 3514 | 1.047 | $4.668 \pm 0.247$ | $5.26 \pm 0.06$ | $3435 \pm 92$ |
| HD152173 | $2.313 \pm 0.006 \pm 0.043$ | 3873 | 1.039 | $2.403 \pm 0.045$ | $3.96 \pm 0.06$ | $3800 \pm 41$ |
| HD153698 | $2.122 \pm 0.005 \pm 0.043$ | 3608 | 1.044 | $2.216 \pm 0.045$ | $5.22 \pm 0.06$ | $3530 \pm 46$ |
| HD153834 | $1.334 \pm 0.010 \pm 0.048$ | 4495 | 1.031 | $1.375 \pm 0.050$ | $2.91 \pm 0.07$ | $4428 \pm 87$ |
| HD154301 | $1.602 \pm 0.024 \pm 0.043$ | 3964 | 1.037 | $1.662 \pm 0.045$ | $3.87 \pm 0.06$ | $3892 \pm 68$ |
| HD156966 | $1.806 \pm 0.020 \pm 0.043$ | 3807 | 1.040 | $1.879 \pm 0.045$ | $4.47 \pm 0.08$ | $3732 \pm 61$ |
| HD157617 | $1.256 \pm 0.009 \pm 0.048$ | 4477 | 1.031 | $1.295 \pm 0.050$ | $2.87 \pm 0.10$ | $4409 \pm 94$ |
| HD163547 | $1.385 \pm 0.045 \pm 0.048$ | 4421 | 1.032 | $1.429 \pm 0.050$ | $2.90 \pm 0.06$ | $4353 \pm 80$ |
| HD163947 | $1.961 \pm 0.003 \pm 0.039$ | 3317 | 1.053 | $2.064 \pm 0.041$ | $6.21 \pm 0.08$ | $3233 \pm 52$ |
| HD163993 | $2.137 \pm 0.009 \pm 0.043$ | 5102 | 1.027 | $2.194 \pm 0.044$ | $2.09 \pm 0.07$ | $5034 \pm 54$ |
| HD164064 | $1.946 \pm 0.030 \pm 0.039$ | 4055 | 1.036 | $2.016 \pm 0.040$ | $3.79 \pm 0.07$ | $3984 \pm 56$ |
| HD166013 | $2.021 \pm 0.004 \pm 0.039$ | 3487 | 1.048 | $2.117 \pm 0.040$ | $5.30 \pm 0.06$ | $3407 \pm 41$ |
| HD168775 | $2.112 \pm 0.006 \pm 0.043$ | 4641 | 1.030 | $2.175 \pm 0.045$ | $2.56 \pm 0.07$ | $4574 \pm 51$ |
| HD169305 | $4.249 \pm 0.008 \pm 0.173$ | 3711 | 1.042 | $4.428 \pm 0.181$ | $4.41 \pm 0.05$ | $3635 \pm 77$ |
| HD173213 | $2.402 \pm 0.016 \pm 0.043$ | 3333 | 1.052 | $2.527 \pm 0.046$ | $5.96 \pm 0.40$ | $3249 \pm 76$ |
| HD173954 | $1.529 \pm 0.014 \pm 0.043$ | 4164 | 1.034 | $1.582 \pm 0.045$ | $3.58 \pm 0.09$ | $4094 \pm 86$ |
| HD176411 | $2.164 \pm 0.006 \pm 0.043$ | 4798 | 1.029 | $2.226 \pm 0.045$ | $2.31 \pm 0.07$ | $4730 \pm 52$ |
| HD176670 | $2.312 \pm 0.008 \pm 0.043$ | 4429 | 1.031 | $2.385 \pm 0.045$ | $3.24 \pm 0.08$ | $4361 \pm 68$ |
| HD176844 | $3.378 \pm 0.014 \pm 0.067$ | 3545 | 1.046 | $3.533 \pm 0.071$ | $5.47 \pm 0.04$ | $3466 \pm 42$ |
| HD176981 | $1.733 \pm 0.039 \pm 0.043$ | 4094 | 1.035 | $1.794 \pm 0.045$ | $3.53 \pm 0.18$ | $4024 \pm 94$ |
| HD180450 | $2.676 \pm 0.005 \pm 0.065$ | 3807 | 1.040 | $2.783 \pm 0.067$ | $4.42 \pm 0.07$ | $3733 \pm 51$ |
| HD185958 | $1.817 \pm 0.063 \pm 0.043$ | 4809 | 1.028 | $1.869 \pm 0.065$ | $2.23 \pm 0.07$ | $4742 \pm 83$ |
| HD186675 | $1.332 \pm 0.007 \pm 0.048$ | 4955 | 1.028 | $1.369 \pm 0.050$ | $2.10 \pm 0.08$ | $4888 \pm 96$ |
| HD186776 | $2.813 \pm 0.011 \pm 0.048$ | 3617 | 1.044 | $2.937 \pm 0.050$ | $4.67 \pm 0.06$ | $3540 \pm 33$ |
| HD187849 | $4.068 \pm 0.008 \pm 0.135$ | 3614 | 1.044 | $4.248 \pm 0.141$ | $4.43 \pm 0.07$ | $3537 \pm 60$ |
| HD188310 | $1.631 \pm 0.004 \pm 0.043$ | 4766 | 1.029 | $1.678 \pm 0.045$ | $2.29 \pm 0.10$ | $4699 \pm 65$ |
| HD189695 | $1.903 \pm 0.004 \pm 0.039$ | 3938 | 1.038 | $1.975 \pm 0.040$ | $3.72 \pm 0.07$ | $3866 \pm 43$ |
| HD191178 | $2.859 \pm 0.003 \pm 0.048$ | 3751 | 1.041 | $2.977 \pm 0.050$ | $4.94 \pm 0.05$ | $3676 \pm 39$ |
| HD193347 | $1.924 \pm 0.008 \pm 0.039$ | 3739 | 1.042 | $2.004 \pm 0.040$ | $4.36 \pm 0.10$ | $3664 \pm 49$ |

**Table 18** *continued on next page*

**Table 18** *(continued)*

| Star ID | UD size (mas) | Initial $T_{\rm EFF}$ (K) | UD-to-LD | LD size (mas) | $V_0 - K_0$ (mag) | $T_{\rm EFF}$ (K) |
|---|---|---|---|---|---|---|
| HD193579 | $1.955 \pm 0.018 \pm 0.039$ | 3991 | 1.037 | $2.027 \pm 0.040$ | $3.64 \pm 0.07$ | $3920 \pm 53$ |
| HD194097 | $1.434 \pm 0.028 \pm 0.043$ | 4155 | 1.035 | $1.484 \pm 0.045$ | $3.30 \pm 0.09$ | $4085 \pm 71$ |
| HD194317 | $2.570 \pm 0.016 \pm 0.043$ | 4536 | 1.030 | $2.648 \pm 0.045$ | $3.04 \pm 0.07$ | $4468 \pm 48$ |
| HD197912 | $2.036 \pm 0.007 \pm 0.039$ | 4799 | 1.029 | $2.094 \pm 0.040$ | $2.41 \pm 0.06$ | $4732 \pm 46$ |
| HD198237 | $1.857 \pm 0.008 \pm 0.043$ | 3818 | 1.040 | $1.931 \pm 0.045$ | $4.08 \pm 0.08$ | $3744 \pm 48$ |
| HD199101 | $2.267 \pm 0.026 \pm 0.043$ | 4004 | 1.037 | $2.350 \pm 0.045$ | $3.71 \pm 0.07$ | $3933 \pm 41$ |
| HD199697 | $1.971 \pm 0.006 \pm 0.039$ | 4309 | 1.033 | $2.035 \pm 0.040$ | $3.26 \pm 0.07$ | $4240 \pm 52$ |
| HD199799 | $3.333 \pm 0.011 \pm 0.067$ | 3410 | 1.050 | $3.499 \pm 0.071$ | $6.22 \pm 0.07$ | $3328 \pm 37$ |
| HD202109 | $2.741 \pm 0.008 \pm 0.048$ | 5060 | 1.027 | $2.815 \pm 0.049$ | $2.10 \pm 0.06$ | $4993 \pm 47$ |
| HD205435 | $1.829 \pm 0.020 \pm 0.043$ | 5161 | 1.027 | $1.878 \pm 0.044$ | $2.01 \pm 0.07$ | $5093 \pm 63$ |
| HD205733 | $2.571 \pm 0.025 \pm 0.043$ | 3218 | 1.056 | $2.715 \pm 0.046$ | $6.52 \pm 0.12$ | $3131 \pm 29$ |
| HD206445 | $1.712 \pm 0.040 \pm 0.043$ | 4272 | 1.033 | $1.769 \pm 0.045$ | $3.35 \pm 0.09$ | $4203 \pm 59$ |
| HD206749 | $2.937 \pm 0.012 \pm 0.053$ | 3856 | 1.039 | $3.052 \pm 0.055$ | $4.17 \pm 0.07$ | $3783 \pm 37$ |
| HD207001 | $1.359 \pm 0.015 \pm 0.048$ | 3698 | 1.042 | $1.417 \pm 0.050$ | $4.70 \pm 0.11$ | $3622 \pm 71$ |
| HD207328 | $2.529 \pm 0.020 \pm 0.043$ | 3562 | 1.046 | $2.644 \pm 0.045$ | $5.68 \pm 0.08$ | $3484 \pm 50$ |
| HD209857 | $3.707 \pm 0.003 \pm 0.091$ | 3470 | 1.048 | $3.885 \pm 0.096$ | $5.40 \pm 0.07$ | $3389 \pm 44$ |
| HD210514 | $2.382 \pm 0.020 \pm 0.043$ | 3489 | 1.048 | $2.495 \pm 0.045$ | $5.42 \pm 0.08$ | $3409 \pm 42$ |
| HD211800 | $1.560 \pm 0.018 \pm 0.043$ | 4036 | 1.036 | $1.616 \pm 0.045$ | $4.20 \pm 0.11$ | $3965 \pm 91$ |
| HD212470 | $3.706 \pm 0.007 \pm 0.079$ | 3511 | 1.047 | $3.880 \pm 0.082$ | $6.16 \pm 0.05$ | $3432 \pm 43$ |
| HD212496 | $1.821 \pm 0.014 \pm 0.032$ | 4862 | 1.028 | $1.872 \pm 0.033$ | $2.42 \pm 0.07$ | $4795 \pm 45$ |
| HD214868 | $2.466 \pm 0.010 \pm 0.043$ | 4483 | 1.031 | $2.542 \pm 0.045$ | $3.01 \pm 0.07$ | $4415 \pm 42$ |
| HD215182 | $3.033 \pm 0.007 \pm 0.053$ | 5111 | 1.027 | $3.114 \pm 0.054$ | $2.06 \pm 0.07$ | $5044 \pm 48$ |
| HD215547 | $2.555 \pm 0.037 \pm 0.043$ | 3242 | 1.055 | $2.696 \pm 0.046$ | $5.85 \pm 0.05$ | $3156 \pm 35$ |
| HD216131 | $2.385 \pm 0.006 \pm 0.043$ | 5089 | 1.027 | $2.449 \pm 0.045$ | $2.13 \pm 0.07$ | $5022 \pm 49$ |
| HD216930 | $1.813 \pm 0.062 \pm 0.043$ | 3393 | 1.050 | $1.904 \pm 0.065$ | $5.23 \pm 0.08$ | $3311 \pm 63$ |
| HD218452 | $1.965 \pm 0.026 \pm 0.039$ | 4187 | 1.034 | $2.032 \pm 0.040$ | $3.33 \pm 0.09$ | $4117 \pm 47$ |
| HD220211 | $2.015 \pm 0.079 \pm 0.039$ | 3488 | 1.048 | $2.111 \pm 0.082$ | $5.00 \pm 0.08$ | $3407 \pm 69$ |
| HD221345 | $1.366 \pm 0.028 \pm 0.048$ | 4844 | 1.028 | $1.405 \pm 0.050$ | $2.49 \pm 0.10$ | $4777 \pm 90$ |
| HIP8682 | $2.357 \pm 0.004 \pm 0.043$ | 3158 | 1.058 | $2.494 \pm 0.046$ | $7.90 \pm 0.08$ | $3070 \pm 42$ |
| HIP35915 | $1.740 \pm 0.018 \pm 0.043$ | 3217 | 1.056 | $1.837 \pm 0.046$ | $7.13 \pm 0.06$ | $3131 \pm 69$ |
| HIP68357 | $3.314 \pm 0.006 \pm 0.067$ | 3201 | 1.057 | $3.502 \pm 0.071$ | $7.63 \pm 0.07$ | $3114 \pm 37$ |
| HIP113390 | $2.577 \pm 0.012 \pm 0.043$ | 3122 | 1.060 | $2.731 \pm 0.046$ | $8.58 \pm 0.66$ | $3033 \pm 57$ |
| IRC+30095 | $2.673 \pm 0.020 \pm 0.048$ | 3255 | 1.055 | $2.819 \pm 0.051$ | $8.21 \pm 0.08$ | $3169 \pm 51$ |
| IRC+40533 | $3.299 \pm 0.007 \pm 0.067$ | 3213 | 1.056 | $3.484 \pm 0.071$ | $8.56 \pm 0.05$ | $3126 \pm 45$ |

Note—For more information, see §5.2.

**Table 19**. Luminosities $L$ and radii $R$ for the program stars.

| Star ID | $V_0 - K_0$ | Sp. Type | $\pi$ | Source | $R$ | $L$ |
|---|---|---|---|---|---|---|
| | (mag) | | (mas) | | ($R_\odot$) | ($L_\odot$) |
| HD598 | $5.53 \pm 0.05$ | M5III | $3.10 \pm 0.11$ | GaiaDR2 | $94.16 \pm 3.66$ | $1171.91 \pm 85.08$ |
| HD1632 | $4.06 \pm 0.08$ | K5III | $4.37 \pm 0.14$ | GaiaDR2 | $58.20 \pm 2.15$ | $701.15 \pm 48.93$ |
| HD3346 | $4.00 \pm 0.07$ | K5III | $4.95 \pm 0.19$ | GaiaDR2 | $69.98 \pm 2.98$ | $979.20 \pm 76.65$ |
| HD3546 | $2.12 \pm 0.07$ | G1III | $19.91 \pm 0.19$ | HIP | $9.37 \pm 0.26$ | $54.17 \pm 1.37$ |
| HD3574 | $3.82 \pm 0.06$ | M1III | $2.40 \pm 0.14$ | GaiaDR2 | $105.52 \pm 6.59$ | $2740.58 \pm 327.59$ |
| HD3627 | $2.76 \pm 0.07$ | K1.5III | $30.91 \pm 0.15$ | HIP | $14.24 \pm 0.44$ | $70.56 \pm 0.96$ |
| HD5006 | $5.17 \pm 0.11$ | M5III | $3.93 \pm 0.09$ | GaiaDR2 | $61.71 \pm 1.84$ | $454.02 \pm 28.13$ |
| HD6186 | $2.25 \pm 0.07$ | G5III | $17.94 \pm 0.21$ | HIP | $11.59 \pm 0.30$ | $65.07 \pm 1.72$ |
| HD6409 | $5.16 \pm 0.10$ | M4III | $2.48 \pm 0.08$ | GaiaDR2 | $87.85 \pm 3.41$ | $971.52 \pm 66.10$ |
| HD7000 | $5.87 \pm 0.11$ | M5III | $1.45 \pm 0.08$ | GaiaDR2 | $142.90 \pm 8.29$ | $2176.61 \pm 242.46$ |
| HD7087 | $2.27 \pm 0.11$ | G8III | $8.50 \pm 0.21$ | HIP | $20.54 \pm 0.76$ | $198.70 \pm 10.12$ |
| HD8126 | $3.31 \pm 0.08$ | K3.25III | $8.14 \pm 0.16$ | GaiaDR2 | $27.82 \pm 0.77$ | $219.27 \pm 9.62$ |
| HD9500 | $5.11 \pm 0.09$ | M4III | $2.70 \pm 0.18$ | GaiaDR2 | $96.47 \pm 6.78$ | $1261.80 \pm 172.36$ |
| HD9927 | $2.77 \pm 0.07$ | K1.5III | $18.41 \pm 0.18$ | HIP | $20.19 \pm 0.45$ | $154.70 \pm 3.43$ |
| HD10380 | $3.16 \pm 0.06$ | K3.25III | $8.98 \pm 0.23$ | HIP | $35.06 \pm 1.08$ | $325.48 \pm 16.89$ |
| HD12929 | $2.60 \pm 0.07$ | K3.25III | $49.56 \pm 0.25$ | HIP | $13.16 \pm 3.61$ | $72.93 \pm 0.86$ |
| HD14146 | $4.38 \pm 0.07$ | M2III | $2.23 \pm 0.10$ | GaiaDR2 | $99.17 \pm 4.79$ | $1560.01 \pm 144.31$ |
| HD14512 | $6.08 \pm 0.06$ | M5III | $2.66 \pm 0.10$ | GaiaDR2 | $119.74 \pm 4.79$ | $1578.24 \pm 122.91$ |
| HD14872 | $3.64 \pm 0.07$ | K5III | $7.46 \pm 0.23$ | GaiaDR2 | $47.15 \pm 1.68$ | $489.53 \pm 30.94$ |
| HD14901 | $6.44 \pm 0.06$ | M5.5III | $2.00 \pm 0.14$ | GaiaDR2 | $210.18 \pm 15.09$ | $5053.58 \pm 698.03$ |
| HD15656 | $3.54 \pm 0.07$ | K4III | $7.16 \pm 0.18$ | GaiaDR2 | $38.66 \pm 1.18$ | $316.28 \pm 16.55$ |
| HD17361 | $2.50 \pm 0.05$ | K1III | $19.01 \pm 0.21$ | HIP | $10.69 \pm 0.28$ | $54.23 \pm 1.45$ |
| HD17709 | $3.80 \pm 0.07$ | K5III | $8.47 \pm 0.28$ | GaiaDR2 | $49.07 \pm 2.04$ | $487.72 \pm 33.66$ |
| HD18322 | $2.48 \pm 0.05$ | K1III | $23.89 \pm 0.19$ | HIP | $11.77 \pm 0.22$ | $56.32 \pm 0.98$ |
| HD19787 | $2.27 \pm 0.05$ | G8III | $19.22 \pm 0.19$ | HIP | $10.14 \pm 0.27$ | $50.69 \pm 1.09$ |
| HD20844 | $5.67 \pm 0.07$ | M5III | $2.43 \pm 0.12$ | GaiaDR2 | $117.30 \pm 6.01$ | $1698.66 \pm 165.38$ |
| HD21465 | $4.78 \pm 0.09$ | M3III | $1.56 \pm 0.04$ | GaiaDR2 | $151.81 \pm 5.50$ | $2825.48 \pm 173.33$ |
| HD25604 | $2.35 \pm 0.07$ | K0III | $17.43 \pm 0.21$ | HIP | $11.92 \pm 0.31$ | $63.07 \pm 1.67$ |
| HD26605 | $2.06 \pm 0.37$ | G5III | $7.27 \pm 0.04$ | GaiaDR2 | $12.10 \pm 1.13$ | $68.43 \pm 1.95$ |
| HD27308 | $6.15 \pm 0.08$ | M5III | $2.15 \pm 0.07$ | GaiaDR2 | $132.31 \pm 4.89$ | $1953.51 \pm 134.10$ |
| HD27348 | $2.14 \pm 0.10$ | G5III | $14.85 \pm 0.20$ | GaiaDR2 | $12.98 \pm 2.78$ | $47.28 \pm 1.59$ |
| HD27697 | $2.09 \pm 0.05$ | G5III | $19.06 \pm 0.37$ | GaiaDR2 | $12.62 \pm 0.35$ | $87.60 \pm 3.54$ |
| HD28100 | $2.25 \pm 0.07$ | G5III | $7.61 \pm 0.15$ | GaiaDR2 | $22.02 \pm 0.77$ | $250.29 \pm 11.22$ |
| HD28292 | $2.70 \pm 0.05$ | G8III | $17.31 \pm 0.21$ | GaiaDR2 | $11.98 \pm 0.39$ | $51.11 \pm 1.37$ |
| HD28305 | $2.20 \pm 0.07$ | G8III | $22.24 \pm 0.25$ | HIP | $12.15 \pm 0.26$ | $75.72 \pm 1.80$ |
| HD28307 | $2.11 \pm 0.05$ | G5III | $21.13 \pm 0.27$ | HIP | $10.62 \pm 0.24$ | $64.10 \pm 1.85$ |
| HD28581 | $4.45 \pm 0.11$ | K3.5III | $5.82 \pm 0.06$ | GaiaDR2 | $40.09 \pm 3.04$ | $193.34 \pm 17.46$ |
| HD28595 | $4.90 \pm 0.08$ | M2III | $5.67 \pm 0.14$ | GaiaDR2 | $59.68 \pm 1.79$ | $468.86 \pm 28.70$ |

**Table 19** *continued on next page*

**Table 19** *(continued)*

| Star ID | $V_0 - K_0$ | Sp. Type | $\pi$ | Source | $R$ | $L$ |
|---------|-------------|----------|-------|--------|-----|-----|
| | (mag) | | (mas) | | ($R_\odot$) | ($L_\odot$) |
| HD29094 | $2.90 \pm 0.07$ | K1.5III | $4.09 \pm 0.38$ | HIP | $70.71 \pm 6.67$ | $1731.32 \pm 323.44$ |
| HD30504 | $3.44 \pm 0.07$ | K3.25III | $6.48 \pm 0.44$ | HIP | $46.55 \pm 3.27$ | $564.86 \pm 77.18$ |
| HD30605 | $3.66 \pm 0.11$ | K5III | $3.17 \pm 0.12$ | GaiaDR2 | $64.95 \pm 2.81$ | $832.48 \pm 61.77$ |
| HD30834 | $3.24 \pm 0.09$ | K3.25III | $6.33 \pm 0.27$ | GaiaDR2 | $45.70 \pm 2.09$ | $571.75 \pm 49.76$ |
| HD31139 | $4.30 \pm 0.08$ | M1III | $5.52 \pm 0.26$ | HIP | $69.83 \pm 3.60$ | $830.65 \pm 79.80$ |
| HD31421 | $2.67 \pm 0.07$ | K1III | $17.54 \pm 0.21$ | HIP | $17.18 \pm 0.37$ | $107.53 \pm 3.13$ |
| HD33463 | $4.88 \pm 0.09$ | M4III | $2.24 \pm 0.20$ | GaiaDR2 | $132.77 \pm 11.95$ | $2616.69 \pm 462.90$ |
| HD34334 | $3.20 \pm 0.07$ | K2III | $14.04 \pm 0.58$ | HIP | $19.86 \pm 0.89$ | $127.13 \pm 10.92$ |
| HD34559 | $1.96 \pm 0.09$ | G5III | $14.05 \pm 0.17$ | GaiaDR2 | $9.62 \pm 0.40$ | $53.46 \pm 2.41$ |
| HD34577 | $5.45 \pm 0.08$ | M4III | $2.48 \pm 0.09$ | GaiaDR2 | $105.73 \pm 4.49$ | $1562.62 \pm 132.50$ |
| HD35620 | $3.12 \pm 0.08$ | K3.5III | $6.76 \pm 0.17$ | GaiaDR2 | $33.97 \pm 1.05$ | $307.74 \pm 15.46$ |
| HD38656 | $2.17 \pm 0.05$ | G5III | $15.77 \pm 0.20$ | HIP | $11.14 \pm 0.33$ | $62.87 \pm 1.73$ |
| HD39003 | $2.48 \pm 0.05$ | K1III | $15.14 \pm 0.33$ | GaiaDR2 | $17.37 \pm 0.49$ | $134.07 \pm 6.01$ |
| HD39045 | $4.97 \pm 0.07$ | M4III | $3.51 \pm 0.15$ | GaiaDR2 | $99.55 \pm 4.74$ | $1338.96 \pm 118.95$ |
| HD39225 | $4.15 \pm 0.09$ | M1III | $5.24 \pm 0.18$ | GaiaDR2 | $51.03 \pm 2.02$ | $411.27 \pm 30.93$ |
| HD39732 | $5.50 \pm 0.10$ | M4III | $2.13 \pm 0.06$ | GaiaDR2 | $131.01 \pm 4.80$ | $2146.03 \pm 177.77$ |
| HD40441 | $4.04 \pm 0.10$ | K5III | $2.67 \pm 0.08$ | GaiaDR2 | $63.85 \pm 2.64$ | $851.94 \pm 76.74$ |
| HD41116 | $2.02 \pm 0.07$ | G0III | $21.03 \pm 0.90$ | HIP | $9.78 \pm 0.48$ | $52.08 \pm 4.58$ |
| HD42049 | $3.87 \pm 0.08$ | K4III | $1.82 \pm 0.17$ | GaiaDR2 | $139.78 \pm 13.63$ | $3725.28 \pm 764.51$ |
| HD43039 | $2.40 \pm 0.07$ | G8III | $18.43 \pm 0.23$ | HIP | $11.81 \pm 0.27$ | $62.57 \pm 1.74$ |
| HD46709 | $3.43 \pm 0.08$ | K3.5III | $2.28 \pm 0.12$ | GaiaDR2 | $82.24 \pm 4.96$ | $1516.63 \pm 166.44$ |
| HD48450 | $3.44 \pm 0.07$ | K3.5III | $6.02 \pm 0.17$ | GaiaDR2 | $37.14 \pm 1.26$ | $321.39 \pm 18.27$ |
| HD49738 | $2.75 \pm 0.10$ | K3III | $2.85 \pm 0.09$ | GaiaDR2 | $52.06 \pm 2.48$ | $954.85 \pm 72.57$ |
| HD49968 | $3.41 \pm 0.11$ | K3.5III | $7.48 \pm 0.14$ | GaiaDR2 | $27.91 \pm 0.83$ | $174.24 \pm 9.35$ |
| HD54719 | $2.86 \pm 0.06$ | K1.25III | $10.16 \pm 0.25$ | HIP | $25.27 \pm 0.78$ | $237.33 \pm 12.52$ |
| HD60136 | $5.05 \pm 0.07$ | M2III | $2.65 \pm 0.08$ | GaiaDR2 | $104.49 \pm 4.58$ | $1495.86 \pm 151.68$ |
| HD61913 | $4.83 \pm 0.07$ | M4III | $3.29 \pm 0.22$ | GaiaDR2 | $128.72 \pm 9.11$ | $2810.50 \pm 371.01$ |
| HD62044 | $2.57 \pm 0.07$ | K0III | $26.08 \pm 0.19$ | HIP | $9.99 \pm 0.20$ | $41.02 \pm 1.06$ |
| HD62285 | $3.60 \pm 0.08$ | K5III | $5.04 \pm 0.18$ | GaiaDR2 | $55.36 \pm 2.25$ | $650.56 \pm 52.08$ |
| HD62345 | $2.07 \pm 0.07$ | G5III | $23.07 \pm 0.22$ | HIP | $11.11 \pm 0.23$ | $69.19 \pm 1.49$ |
| HD62509 | $2.22 \pm 0.07$ | G8III | $96.54 \pm 0.27$ | HIP | $9.49 \pm 2.77$ | $37.84 \pm 0.44$ |
| HD62721 | $3.52 \pm 0.07$ | K4III | $9.15 \pm 0.30$ | GaiaDR2 | $34.83 \pm 1.27$ | $263.54 \pm 17.57$ |
| HD66216 | $2.51 \pm 0.10$ | K0III | $13.42 \pm 0.23$ | GaiaDR2 | $13.15 \pm 0.42$ | $75.26 \pm 4.29$ |
| HD74442 | $2.44 \pm 0.05$ | K0III | $24.98 \pm 0.24$ | HIP | $10.19 \pm 0.22$ | $48.61 \pm 1.02$ |
| HD76294 | $2.20 \pm 0.07$ | G8III | $19.51 \pm 0.18$ | HIP | $18.23 \pm 0.39$ | $151.68 \pm 3.54$ |
| HD76830 | $5.19 \pm 0.07$ | M4III | $5.05 \pm 0.16$ | GaiaDR2 | $71.94 \pm 2.70$ | $675.45 \pm 45.07$ |
| HD81817 | $3.20 \pm 0.06$ | K3.5III | $3.28 \pm 0.15$ | HIP | $108.46 \pm 5.40$ | $2929.89 \pm 273.18$ |
| HD82198 | $3.83 \pm 0.06$ | K5III | $6.51 \pm 0.23$ | GaiaDR2 | $45.85 \pm 1.82$ | $404.27 \pm 30.30$ |
| HD82381 | $3.15 \pm 0.07$ | K3.25III | $6.25 \pm 0.23$ | GaiaDR2 | $36.88 \pm 1.52$ | $376.09 \pm 28.63$ |
| HD82635 | $2.04 \pm 0.07$ | G4III | $18.15 \pm 0.23$ | GaiaDR2 | $8.74 \pm 0.29$ | $46.44 \pm 1.65$ |

**Table 19** *(continued)*

| Star ID | $V_0 - K_0$ | Sp. Type | $\pi$ | Source | $R$ | $L$ |
|---|---|---|---|---|---|---|
| | (mag) | | (mas) | | $(R_\odot)$ | $(L_\odot)$ |
| HD85503 | $2.67 \pm 0.06$ | K1.5III | $26.28 \pm 0.16$ | HIP | $11.67 \pm 0.22$ | $50.16 \pm 0.90$ |
| HD87046 | $5.04 \pm 0.08$ | M4III | $1.74 \pm 0.07$ | GaiaDR2 | $122.72 \pm 5.26$ | $2014.71 \pm 166.38$ |
| HD87837 | $3.30 \pm 0.07$ | K3.5III | $11.04 \pm 0.24$ | HIP | $31.80 \pm 0.90$ | $258.62 \pm 11.81$ |
| HD90254 | $4.37 \pm 0.07$ | M1III | $3.86 \pm 0.16$ | GaiaDR2 | $85.08 \pm 3.87$ | $1346.00 \pm 117.88$ |
| HD92620 | $4.94 \pm 0.07$ | M4III | $4.01 \pm 0.17$ | GaiaDR2 | $100.07 \pm 4.85$ | $1267.91 \pm 109.97$ |
| HD94264 | $2.41 \pm 0.07$ | K0III | $34.38 \pm 0.21$ | HIP | $8.03 \pm 0.15$ | $28.52 \pm 0.58$ |
| HD94336 | $4.83 \pm 0.06$ | M3III | $2.93 \pm 0.07$ | GaiaDR2 | $74.46 \pm 2.32$ | $756.30 \pm 41.88$ |
| HD95212 | $3.43 \pm 0.09$ | K5III | $4.58 \pm 0.24$ | GaiaDR2 | $51.17 \pm 2.87$ | $630.75 \pm 69.23$ |
| HD95345 | $2.58 \pm 0.08$ | K1.25III | $10.51 \pm 0.18$ | GaiaDR2 | $18.86 \pm 0.57$ | $129.22 \pm 5.52$ |
| HD96274 | $4.75 \pm 0.10$ | M3III | $3.57 \pm 0.07$ | GaiaDR2 | $60.23 \pm 1.72$ | $484.96 \pm 25.33$ |
| HD96833 | $2.58 \pm 0.07$ | K1.25III | $22.57 \pm 0.14$ | HIP | $18.64 \pm 0.51$ | $140.91 \pm 2.30$ |
| HD100236 | $5.29 \pm 0.12$ | M4III | $1.71 \pm 0.08$ | GaiaDR2 | $111.00 \pm 5.79$ | $1435.71 \pm 156.85$ |
| HD102224 | $2.78 \pm 0.06$ | K1.25III | $17.76 \pm 0.16$ | HIP | $19.94 \pm 0.43$ | $144.46 \pm 3.28$ |
| HD104207 | $5.63 \pm 0.05$ | M5III | $3.45 \pm 0.16$ | GaiaDR2 | $103.15 \pm 5.12$ | $1227.40 \pm 115.55$ |
| HD104575 | $4.91 \pm 0.12$ | M4III | $2.36 \pm 0.09$ | GaiaDR2 | $78.97 \pm 3.65$ | $831.91 \pm 77.82$ |
| HD104831 | $4.27 \pm 0.10$ | M2III | $2.08 \pm 0.08$ | GaiaDR2 | $83.27 \pm 3.81$ | $990.26 \pm 81.63$ |
| HD104979 | $2.25 \pm 0.05$ | G4III | $19.98 \pm 0.22$ | HIP | $10.55 \pm 0.24$ | $63.87 \pm 1.52$ |
| HD106714 | $2.22 \pm 0.09$ | G8III | $13.08 \pm 0.30$ | HIP | $10.28 \pm 0.47$ | $66.95 \pm 3.55$ |
| HD107256 | $5.88 \pm 0.09$ | M5III | $2.18 \pm 0.10$ | GaiaDR2 | $91.88 \pm 4.66$ | $1006.04 \pm 99.46$ |
| HD109282 | $4.96 \pm 0.07$ | M3III | $1.82 \pm 0.07$ | GaiaDR2 | $114.86 \pm 5.17$ | $1819.01 \pm 154.78$ |
| HD113226 | $1.90 \pm 0.05$ | G4III | $29.76 \pm 0.14$ | HIP | $11.38 \pm 0.22$ | $77.88 \pm 0.95$ |
| HD114780 | $3.90 \pm 0.07$ | M1III | $3.85 \pm 0.12$ | GaiaDR2 | $65.47 \pm 2.40$ | $806.99 \pm 52.78$ |
| HD116207 | $5.49 \pm 0.07$ | M5III | $2.46 \pm 0.06$ | GaiaDR2 | $98.34 \pm 3.11$ | $1243.57 \pm 74.46$ |
| HD118669 | $5.02 \pm 0.09$ | M4III | $1.87 \pm 0.03$ | GaiaDR2 | $103.11 \pm 3.13$ | $1346.05 \pm 78.46$ |
| HD119584 | $3.50 \pm 0.08$ | K4III | $4.69 \pm 0.05$ | GaiaDR2 | $30.84 \pm 2.73$ | $305.49 \pm 12.12$ |
| HD120819 | $4.29 \pm 0.05$ | M1III | $4.05 \pm 0.13$ | GaiaDR2 | $70.18 \pm 2.48$ | $856.42 \pm 59.15$ |
| HD121860 | $5.14 \pm 0.06$ | M4III | $2.99 \pm 0.08$ | GaiaDR2 | $75.74 \pm 2.41$ | $749.94 \pm 46.99$ |
| HD122316 | $7.67 \pm 0.06$ | M7.5III | $2.08 \pm 0.12$ | GaiaDR2 | $208.23 \pm 13.07$ | $3688.71 \pm 424.93$ |
| HD127093 | $4.95 \pm 0.07$ | M4III | $3.20 \pm 0.11$ | GaiaDR2 | $93.83 \pm 3.63$ | $1154.50 \pm 91.22$ |
| HD128902 | $3.48 \pm 0.07$ | K4III | $3.65 \pm 0.06$ | GaiaDR2 | $59.45 \pm 1.57$ | $786.27 \pm 36.90$ |
| HD130084 | $4.30 \pm 0.06$ | M1III | $4.44 \pm 0.07$ | GaiaDR2 | $53.47 \pm 1.41$ | $503.57 \pm 22.82$ |
| HD133124 | $3.48 \pm 0.07$ | K4III | $8.75 \pm 0.22$ | GaiaDR2 | $37.75 \pm 1.18$ | $322.96 \pm 17.74$ |
| HD133208 | $2.15 \pm 0.06$ | G5III | $14.48 \pm 0.14$ | HIP | $18.43 \pm 0.38$ | $186.06 \pm 4.09$ |
| HD133582 | $2.90 \pm 0.06$ | K1.5III | $12.71 \pm 0.19$ | GaiaDR2 | $21.10 \pm 0.49$ | $143.37 \pm 5.14$ |
| HD135722 | $2.37 \pm 0.05$ | G4III | $26.78 \pm 0.16$ | HIP | $11.02 \pm 0.21$ | $66.15 \pm 1.20$ |
| HD136404 | $4.73 \pm 0.10$ | M2III | $1.92 \pm 0.05$ | GaiaDR2 | $82.84 \pm 3.23$ | $1135.50 \pm 98.08$ |
| HD137071 | $3.77 \pm 0.07$ | K5III | $1.57 \pm 0.12$ | GaiaDR2 | $166.25 \pm 13.22$ | $5969.80 \pm 935.67$ |
| HD137853 | $4.25 \pm 0.06$ | M1III | $3.32 \pm 0.10$ | GaiaDR2 | $80.84 \pm 2.88$ | $1163.21 \pm 78.88$ |
| HD138481 | $3.87 \pm 0.07$ | K5III | $3.45 \pm 0.13$ | GaiaDR2 | $99.80 \pm 4.26$ | $2054.96 \pm 163.47$ |
| HD139153 | $4.27 \pm 0.05$ | M1III | $4.92 \pm 0.20$ | GaiaDR2 | $77.76 \pm 3.50$ | $1125.81 \pm 91.75$ |

**Table 19** (continued)

| Star ID | $V_0 - K_0$ | Sp. Type | $\pi$ | Source | $R$ | $L$ |
|---|---|---|---|---|---|---|
| | (mag) | | (mas) | | ($R_\odot$) | ($L_\odot$) |
| HD139374 | 5.36 ± 0.08 | M4III | 2.12 ± 0.04 | GaiaDR2 | 74.75 ± 2.65 | 702.20 ± 42.13 |
| HD139971 | 5.15 ± 0.09 | M4III | 1.52 ± 0.03 | GaiaDR2 | 122.26 ± 3.85 | 1963.37 ± 112.79 |
| HD142176 | 3.53 ± 0.35 | K5III | 2.97 ± 0.03 | GaiaDR2 | 35.16 ± 2.38 | 288.50 ± 22.00 |
| HD143107 | 2.85 ± 0.05 | K1.5III | 14.73 ± 0.21 | HIP | 20.73 ± 0.47 | 139.95 ± 4.31 |
| HD144065 | 5.34 ± 0.10 | M2III | 1.83 ± 0.04 | GaiaDR2 | 85.25 ± 3.48 | 1000.71 ± 121.67 |
| HD147749 | 4.39 ± 0.07 | M3III | 5.01 ± 0.13 | GaiaDR2 | 80.78 ± 2.88 | 1257.36 ± 69.81 |
| HD148897 | 3.11 ± 0.07 | K1.5III | 4.87 ± 0.13 | GaiaDR2 | 42.35 ± 1.48 | 519.73 ± 32.26 |
| HD150047 | 6.09 ± 0.05 | M5III | 3.10 ± 0.11 | GaiaDR2 | 122.30 ± 4.85 | 1747.60 ± 133.30 |
| HD150450 | 4.86 ± 0.06 | M3III | 7.27 ± 0.18 | GaiaDR2 | 65.20 ± 3.10 | 796.34 ± 45.76 |
| HD150997 | 2.17 ± 0.06 | G4III | 30.02 ± 0.11 | HIP | 8.92 ± 0.16 | 44.33 ± 0.69 |
| HD151732 | 5.26 ± 0.06 | M4III | 4.42 ± 0.17 | GaiaDR2 | 113.68 ± 7.37 | 1614.17 ± 123.47 |
| HD152173 | 3.96 ± 0.06 | M1III | 3.51 ± 0.10 | GaiaDR2 | 73.60 ± 2.45 | 1013.60 ± 59.76 |
| HD153698 | 5.22 ± 0.06 | M4III | 2.65 ± 0.04 | GaiaDR2 | 90.04 ± 2.28 | 1129.57 ± 48.96 |
| HD153834 | 2.91 ± 0.07 | K2III | 2.32 ± 0.14 | GaiaDR2 | 63.76 ± 4.51 | 1401.74 ± 175.80 |
| HD154301 | 3.87 ± 0.06 | K4III | 3.79 ± 0.06 | GaiaDR2 | 47.21 ± 1.48 | 458.87 ± 24.42 |
| HD156966 | 4.47 ± 0.08 | M2III | 2.40 ± 0.04 | GaiaDR2 | 84.28 ± 2.38 | 1236.79 ± 64.86 |
| HD157617 | 2.87 ± 0.10 | K1.5III | 3.69 ± 0.11 | GaiaDR2 | 37.73 ± 1.82 | 482.65 ± 33.33 |
| HD163547 | 2.90 ± 0.06 | K2III | 7.10 ± 0.07 | GaiaDR2 | 21.66 ± 0.79 | 151.18 ± 4.75 |
| HD163947 | 6.21 ± 0.08 | M5III | 1.92 ± 0.05 | GaiaDR2 | 115.53 ± 3.90 | 1308.40 ± 96.91 |
| HD163993 | 2.09 ± 0.07 | G5III | 23.84 ± 0.15 | HIP | 9.91 ± 0.21 | 56.56 ± 1.01 |
| HD164064 | 3.79 ± 0.07 | K5III | 6.21 ± 0.23 | GaiaDR2 | 34.92 ± 1.49 | 275.68 ± 23.45 |
| HD166013 | 5.30 ± 0.06 | M4III | 2.75 ± 0.04 | GaiaDR2 | 82.81 ± 2.06 | 828.70 ± 35.28 |
| HD168775 | 2.56 ± 0.07 | K1.25III | 12.96 ± 0.14 | HIP | 18.06 ± 0.42 | 128.02 ± 3.54 |
| HD169305 | 4.41 ± 0.05 | M2III | 5.03 ± 0.18 | HIP | 94.74 ± 5.14 | 1406.20 ± 105.59 |
| HD173213 | 5.96 ± 0.40 | M5III | 2.17 ± 0.04 | GaiaDR2 | 125.41 ± 3.33 | 1572.63 ± 145.60 |
| HD173954 | 3.58 ± 0.09 | K4III | 4.30 ± 0.10 | GaiaDR2 | 39.58 ± 1.45 | 394.93 ± 29.96 |
| HD176411 | 2.31 ± 0.07 | K0III | 23.99 ± 0.40 | GaiaDR2 | 9.98 ± 0.26 | 44.78 ± 1.69 |
| HD176670 | 3.24 ± 0.08 | K3.25III | 2.45 ± 0.20 | GaiaDR2 | 104.74 ± 8.84 | 3560.94 ± 611.02 |
| HD176844 | 5.47 ± 0.04 | M4III | 3.44 ± 0.15 | GaiaDR2 | 110.45 ± 5.25 | 1579.81 ± 143.13 |
| HD176981 | 3.53 ± 0.18 | K5III | 2.85 ± 0.14 | GaiaDR2 | 67.75 ± 3.64 | 1079.37 ± 131.91 |
| HD180450 | 4.42 ± 0.07 | M2III | 3.58 ± 0.10 | GaiaDR2 | 83.59 ± 3.12 | 1217.52 ± 75.19 |
| HD185958 | 2.23 ± 0.07 | G8III | 7.72 ± 0.19 | GaiaDR2 | 26.04 ± 1.11 | 307.61 ± 15.95 |
| HD186675 | 2.10 ± 0.08 | G5III | 11.01 ± 0.12 | GaiaDR2 | 13.38 ± 0.51 | 91.72 ± 3.41 |
| HD186776 | 4.67 ± 0.06 | M3III | 3.57 ± 0.11 | GaiaDR2 | 88.52 ± 3.05 | 1103.63 ± 67.90 |
| HD187849 | 4.43 ± 0.07 | M3III | 4.08 ± 0.21 | GaiaDR2 | 112.17 ± 6.91 | 1765.87 ± 185.29 |
| HD188310 | 2.29 ± 0.10 | K0III | 17.77 ± 0.18 | GaiaDR2 | 10.16 ± 0.29 | 45.15 ± 1.10 |
| HD189695 | 3.72 ± 0.07 | K5III | 4.93 ± 0.15 | GaiaDR2 | 43.11 ± 1.55 | 372.38 ± 23.09 |
| HD191178 | 4.94 ± 0.05 | M4III | 2.38 ± 0.13 | GaiaDR2 | 134.54 ± 7.44 | 2966.45 ± 321.64 |
| HD193347 | 4.36 ± 0.10 | M1III | 2.81 ± 0.05 | GaiaDR2 | 76.85 ± 2.11 | 955.05 ± 48.92 |
| HD193579 | 3.64 ± 0.07 | K5III | 7.05 ± 0.09 | GaiaDR2 | 30.96 ± 0.73 | 203.03 ± 8.95 |

**Table 19** (continued)

| Star ID | $V_0 - K_0$ | Sp. Type | $\pi$ | Source | $R$ | $L$ |
|---------|-------------|----------|-------|--------|-----|-----|
| | (mag) | | (mas) | | $(R_\odot)$ | $(L_\odot)$ |
| HD194097 | $3.30 \pm 0.09$ | K3.25III | $4.46 \pm 0.06$ | GaiaDR2 | $35.78 \pm 1.19$ | $319.73 \pm 13.69$ |
| HD194317 | $3.04 \pm 0.07$ | K1.5III | $13.05 \pm 0.20$ | HIP | $21.84 \pm 0.50$ | $170.61 \pm 6.79$ |
| HD197912 | $2.41 \pm 0.06$ | K0III | $16.22 \pm 0.19$ | HIP | $13.89 \pm 0.31$ | $86.83 \pm 2.20$ |
| HD198237 | $4.08 \pm 0.08$ | M1III | $3.11 \pm 0.05$ | GaiaDR2 | $66.90 \pm 1.95$ | $788.98 \pm 32.22$ |
| HD199101 | $3.71 \pm 0.07$ | K5III | $5.33 \pm 0.15$ | GaiaDR2 | $47.47 \pm 1.60$ | $483.80 \pm 28.09$ |
| HD199697 | $3.26 \pm 0.07$ | K3.25III | $6.53 \pm 0.17$ | GaiaDR2 | $33.57 \pm 1.09$ | $326.86 \pm 19.38$ |
| HD199799 | $6.22 \pm 0.07$ | M5III | $1.93 \pm 0.10$ | GaiaDR2 | $195.23 \pm 10.71$ | $4194.86 \pm 434.95$ |
| HD202109 | $2.10 \pm 0.06$ | G5III | $22.79 \pm 0.35$ | HIP | $13.29 \pm 0.31$ | $98.54 \pm 3.25$ |
| HD205435 | $2.01 \pm 0.07$ | G3III | $26.39 \pm 0.15$ | HIP | $7.66 \pm 0.19$ | $35.40 \pm 0.67$ |
| HD205733 | $6.52 \pm 0.12$ | M6III | $1.85 \pm 0.06$ | GaiaDR2 | $158.35 \pm 5.67$ | $2163.23 \pm 140.56$ |
| HD206445 | $3.35 \pm 0.09$ | K3.25III | $3.40 \pm 0.13$ | GaiaDR2 | $55.91 \pm 2.58$ | $875.14 \pm 70.21$ |
| HD206749 | $4.17 \pm 0.07$ | M1III | $5.98 \pm 0.11$ | GaiaDR2 | $54.92 \pm 1.40$ | $553.99 \pm 21.65$ |
| HD207001 | $4.70 \pm 0.11$ | M2III | $1.58 \pm 0.04$ | GaiaDR2 | $96.64 \pm 4.12$ | $1442.24 \pm 83.50$ |
| HD207328 | $5.68 \pm 0.08$ | M5III | $1.59 \pm 0.05$ | GaiaDR2 | $179.49 \pm 6.69$ | $4258.32 \pm 338.63$ |
| HD209857 | $5.40 \pm 0.07$ | M4III | $3.67 \pm 0.13$ | GaiaDR2 | $114.07 \pm 4.89$ | $1540.18 \pm 110.69$ |
| HD210514 | $5.42 \pm 0.08$ | M4III | $2.39 \pm 0.09$ | GaiaDR2 | $112.22 \pm 4.61$ | $1526.24 \pm 122.67$ |
| HD211800 | $4.20 \pm 0.11$ | K5III | $2.23 \pm 0.06$ | GaiaDR2 | $77.90 \pm 2.98$ | $1345.43 \pm 118.61$ |
| HD212470 | $6.16 \pm 0.05$ | M5.5III | $3.40 \pm 0.17$ | GaiaDR2 | $122.75 \pm 6.69$ | $1874.54 \pm 194.25$ |
| HD212496 | $2.42 \pm 0.07$ | G8III | $19.19 \pm 0.16$ | HIP | $10.50 \pm 0.21$ | $52.29 \pm 1.07$ |
| HD214868 | $3.01 \pm 0.07$ | K2III | $9.80 \pm 0.26$ | HIP | $27.92 \pm 0.89$ | $265.81 \pm 14.53$ |
| HD215182 | $2.06 \pm 0.07$ | G3III | $13.72 \pm 0.54$ | GaiaDR2 | $24.42 \pm 1.04$ | $346.33 \pm 27.55$ |
| HD215547 | $5.85 \pm 0.05$ | M4III | $3.17 \pm 0.04$ | GaiaDR2 | $91.41 \pm 1.94$ | $743.49 \pm 27.99$ |
| HD216131 | $2.13 \pm 0.07$ | G4III | $30.74 \pm 0.27$ | HIP | $8.57 \pm 0.17$ | $41.94 \pm 0.93$ |
| HD216930 | $5.23 \pm 0.08$ | M4III | $1.93 \pm 0.06$ | GaiaDR2 | $106.10 \pm 4.82$ | $1213.11 \pm 84.30$ |
| HD218452 | $3.33 \pm 0.09$ | K3.5III | $9.68 \pm 0.11$ | GaiaDR2 | $22.60 \pm 0.52$ | $131.64 \pm 4.38$ |
| HD220211 | $5.00 \pm 0.08$ | M4III | $2.62 \pm 0.08$ | GaiaDR2 | $86.85 \pm 4.21$ | $912.36 \pm 56.11$ |
| HD221345 | $2.49 \pm 0.10$ | G8III | $12.63 \pm 0.27$ | HIP | $11.97 \pm 0.49$ | $66.91 \pm 3.30$ |
| HIP8682 | $7.90 \pm 0.08$ | M6III | $0.89 \pm 0.09$ | GaiaDR2 | $303.12 \pm 32.05$ | $7323.48 \pm 1552.06$ |
| HIP35915 | $7.13 \pm 0.06$ | M5.5III | $0.97 \pm 0.11$ | GaiaDR2 | $204.16 \pm 23.68$ | $3592.63 \pm 852.54$ |
| HIP68357 | $7.63 \pm 0.07$ | M7III | $2.17 \pm 0.08$ | GaiaDR2 | $173.31 \pm 7.00$ | $2534.95 \pm 186.44$ |
| HIP113390 | $8.58 \pm 0.66$ | M5.5III | $1.63 \pm 0.15$ | GaiaDR2 | $180.78 \pm 17.00$ | $2480.40 \pm 486.30$ |
| IRC+30095 | $8.21 \pm 0.08$ | M7.75III | $1.02 \pm 0.14$ | GaiaDR2 | $296.93 \pm 41.26$ | $7980.99 \pm 2237.84$ |
| IRC+40533 | $8.56 \pm 0.05$ | M7.75III | $2.48 \pm 0.10$ | GaiaDR2 | $151.14 \pm 6.88$ | $1957.09 \pm 177.22$ |

Note—We discuss dereddened colors in §4.3, spectral typing in §4.1 and 4.3, distances and radii in §6, and luminosities in §8.1.

**Table 20**. Reported $\log g$ values for the program stars found in the PASTEL catalog of Soubiran et al. (2016) along with resulting mass determinations $M$.

| Star ID | $\log g$ | Reference | $R$ | $M$ |
|---------|----------|-----------|-----|-----|
| | | | $(R_\odot)$ | $(M_\odot)$ |
| HD3546 | $2.83 \pm 0.19$ | da Silva et al. (2015) | $9.37 \pm 0.26$ | $1.20 \pm 0.53$ |
| HD3574 | 1.44 | McWilliam (1990) | $105.52 \pm 6.59$ | $6.21 \pm 3.66$ |
| HD3627 | 2.56 | Zhao et al. (2001) | $14.24 \pm 0.44$ | $1.49 \pm 0.86$ |
| HD6186 | 2.99 | McWilliam (1990) | $11.59 \pm 0.30$ | $2.66 \pm 1.54$ |
| HD7087 | 2.77 | McWilliam (1990) | $20.54 \pm 0.76$ | $5.03 \pm 2.92$ |
| HD8126 | 1.93 | McWilliam (1990) | $27.82 \pm 0.77$ | $1.33 \pm 0.77$ |
| HD9927 | $2.17 \pm 0.14$ | Maldonado & Villaver (2016) | $20.19 \pm 0.45$ | $1.22 \pm 0.40$ |
| HD10380 | $1.48 \pm 0.08$ | Matrozis et al. (2013) | $35.06 \pm 1.08$ | $0.75 \pm 0.15$ |
| HD12929 | $2.78 \pm 0.39$ | da Silva et al. (2015) | $13.16 \pm 3.61$ | $2.11 \pm 2.23$ |
| HD14872 | 1.65 | McWilliam (1990) | $47.15 \pm 1.68$ | $2.01 \pm 1.17$ |
| HD15656 | 1.76 | McWilliam (1990) | $38.66 \pm 1.18$ | $1.74 \pm 1.01$ |
| HD17361 | $2.77 \pm 0.28$ | da Silva et al. (2015) | $10.69 \pm 0.28$ | $1.36 \pm 0.88$ |
| HD17709 | 1.42 | McWilliam (1990) | $49.07 \pm 2.04$ | $1.28 \pm 0.75$ |
| HD18322 | $2.79 \pm 0.06$ | Jofré et al. (2015) | $11.77 \pm 0.22$ | $1.73 \pm 0.25$ |
| HD19787 | $2.92 \pm 0.26$ | da Silva et al. (2015) | $10.14 \pm 0.27$ | $1.73 \pm 1.04$ |
| HD25604 | $2.74 \pm 0.09$ | Maldonado & Villaver (2016) | $11.92 \pm 0.31$ | $1.58 \pm 0.34$ |
| HD26605 | 2.85 | Luck (2015) | $12.10 \pm 1.13$ | $2.10 \pm 1.27$ |
| HD27348 | $3.29 \pm 0.20$ | da Silva et al. (2015) | $12.98 \pm 2.78$ | $6.65 \pm 4.19$ |
| HD27697 | $2.97 \pm 0.29$ | da Silva et al. (2015) | $12.62 \pm 0.35$ | $3.01 \pm 2.02$ |
| HD28100 | 2.65 | McWilliam (1990) | $22.02 \pm 0.77$ | $4.38 \pm 2.54$ |
| HD28292 | $2.81 \pm 0.28$ | da Silva et al. (2015) | $11.98 \pm 0.39$ | $1.88 \pm 1.22$ |
| HD28305 | $2.73 \pm 0.09$ | Maldonado & Villaver (2016) | $12.15 \pm 0.26$ | $1.61 \pm 0.34$ |
| HD28307 | $3.21 \pm 0.28$ | da Silva et al. (2015) | $10.62 \pm 0.24$ | $3.71 \pm 2.40$ |
| HD29094 | 2.21 | McWilliam (1990) | $70.71 \pm 6.67$ | $16.42 \pm 9.95$ |
| HD30504 | 1.75 | McWilliam (1990) | $46.55 \pm 3.27$ | $2.47 \pm 1.46$ |
| HD30834 | 1.86 | McWilliam (1990) | $45.70 \pm 2.09$ | $3.06 \pm 1.79$ |
| HD31421 | 2.56 | Mallik (1998) | $17.18 \pm 0.37$ | $2.17 \pm 1.25$ |
| HD34334 | $1.87 \pm 0.06$ | Matrozis et al. (2013) | $19.86 \pm 0.89$ | $0.59 \pm 0.10$ |
| HD34559 | $3.21 \pm 0.20$ | da Silva et al. (2015) | $9.62 \pm 0.40$ | $3.04 \pm 1.42$ |
| HD35620 | 2.15 | McWilliam (1990) | $33.97 \pm 1.05$ | $3.30 \pm 1.91$ |
| HD38656 | 2.90 | Hekker & Meléndez (2007) | $11.14 \pm 0.33$ | $1.99 \pm 1.15$ |
| HD39003 | 2.61 | McWilliam (1990) | $17.37 \pm 0.49$ | $2.49 \pm 1.44$ |
| HD41116 | 3.18 | Mallik (1998) | $9.78 \pm 0.48$ | $2.93 \pm 1.71$ |
| HD43039 | 2.81 | McWilliam (1990) | $11.81 \pm 0.27$ | $1.82 \pm 1.05$ |
| HD49738 | 2.00 | Hekker & Meléndez (2007) | $52.06 \pm 2.48$ | $5.49 \pm 3.20$ |
| HD54719 | 2.55 | Hekker & Meléndez (2007) | $25.27 \pm 0.78$ | $4.59 \pm 2.66$ |
| HD60136 | $1.49 \pm 0.42$ | Adamczak & Lambert (2014) | $104.49 \pm 4.58$ | $6.83 \pm 6.63$ |

<div align="center"><strong>Table 20</strong> <em>continued on next page</em></div>

**Table 20** *(continued)*

| Star ID | $\log g$ | Reference | $R$ | $M$ |
|---------|----------|-----------|-----|-----|
| | | | $(R_\odot)$ | $(M_\odot)$ |
| HD62044 | 2.40 | Mallik (1998) | $9.99 \pm 0.20$ | $0.51 \pm 0.29$ |
| HD62345 | $2.98 \pm 0.16$ | da Silva et al. (2015) | $11.11 \pm 0.23$ | $2.38 \pm 0.88$ |
| HD62509 | $3.00 \pm 0.06$ | Maldonado & Villaver (2016) | $9.49 \pm 2.77$ | $1.82 \pm 1.09$ |
| HD62721 | 1.67 | McWilliam (1990) | $34.83 \pm 1.27$ | $1.15 \pm 0.67$ |
| HD66216 | 2.73 | McWilliam (1990) | $13.15 \pm 0.42$ | $1.88 \pm 1.09$ |
| HD74442 | 2.84 | McWilliam (1990) | $10.19 \pm 0.22$ | $1.45 \pm 0.84$ |
| HD76294 | $2.88 \pm 0.30$ | da Silva et al. (2015) | $18.23 \pm 0.39$ | $5.11 \pm 3.53$ |
| HD81817 | 1.87 | McWilliam (1990) | $108.46 \pm 5.40$ | $17.65 \pm 10.31$ |
| HD82635 | $3.26 \pm 0.26$ | da Silva et al. (2015) | $8.74 \pm 0.29$ | $2.82 \pm 1.70$ |
| HD85503 | $2.51 \pm 0.11$ | Heiter et al. (2015) | $11.67 \pm 0.22$ | $0.89 \pm 0.23$ |
| HD87837 | 1.81 | McWilliam (1990) | $31.80 \pm 0.90$ | $1.32 \pm 0.76$ |
| HD90254 | 1.40 | Smith & Lambert (1986) | $85.08 \pm 3.87$ | $3.68 \pm 2.15$ |
| HD94264 | $2.73 \pm 0.16$ | Ramírez et al. (2013) | $8.03 \pm 0.15$ | $0.70 \pm 0.26$ |
| HD95212 | 1.85 | Hekker & Meléndez (2007) | $51.17 \pm 2.87$ | $3.75 \pm 2.20$ |
| HD95345 | 2.58 | McWilliam (1990) | $18.86 \pm 0.57$ | $2.74 \pm 1.58$ |
| HD96833 | $1.95 \pm 0.08$ | Thygesen et al. (2012) | $18.64 \pm 0.51$ | $0.63 \pm 0.12$ |
| HD102224 | 2.35 | McWilliam (1990) | $19.94 \pm 0.43$ | $1.80 \pm 1.04$ |
| HD104979 | $2.88 \pm 0.20$ | da Silva et al. (2015) | $10.55 \pm 0.24$ | $1.71 \pm 0.79$ |
| HD106714 | $2.46 \pm 0.08$ | Maldonado & Villaver (2016) | $10.28 \pm 0.47$ | $0.62 \pm 0.13$ |
| HD113226 | $2.82 \pm 0.06$ | Deka-Szymankiewicz et al. (2018) | $11.38 \pm 0.22$ | $1.73 \pm 0.25$ |
| HD128902 | 1.09 | Lomaeva et al. (2019) | $59.45 \pm 1.57$ | $0.88 \pm 0.51$ |
| HD133124 | 1.68 | McWilliam (1990) | $37.75 \pm 1.18$ | $1.38 \pm 0.80$ |
| HD133208 | $2.79 \pm 0.21$ | da Silva et al. (2015) | $18.43 \pm 0.38$ | $4.24 \pm 2.06$ |
| HD133582 | 2.30 | McWilliam (1990) | $21.10 \pm 0.49$ | $1.80 \pm 1.04$ |
| HD135722 | $2.89 \pm 0.19$ | da Silva et al. (2015) | $11.02 \pm 0.21$ | $1.91 \pm 0.84$ |
| HD137071 | 1.54 | McWilliam (1990) | $166.25 \pm 13.22$ | $19.40 \pm 11.59$ |
| HD138481 | 1.64 | McWilliam (1990) | $99.80 \pm 4.26$ | $8.80 \pm 5.12$ |
| HD143107 | $2.10 \pm 0.14$ | Mortier et al. (2013) | $20.73 \pm 0.47$ | $1.10 \pm 0.63$ |
| HD148897 | 1.50 | Kyrolainen et al. (1986) | $42.35 \pm 1.48$ | $1.15 \pm 0.67$ |
| HD150997 | $2.98 \pm 0.17$ | da Silva et al. (2015) | $8.92 \pm 0.16$ | $1.54 \pm 0.61$ |
| HD153834 | 1.70 | Hekker & Meléndez (2007) | $63.76 \pm 4.51$ | $4.12 \pm 2.45$ |
| HD157617 | 2.20 | Hekker & Meléndez (2007) | $37.73 \pm 1.82$ | $4.57 \pm 2.67$ |
| HD163547 | 2.32 | McWilliam (1990) | $21.66 \pm 0.79$ | $1.99 \pm 1.15$ |
| HD163993 | $3.28 \pm 0.26$ | da Silva et al. (2015) | $9.91 \pm 0.21$ | $3.79 \pm 2.27$ |
| HD168775 | $2.52 \pm 0.10$ | Maldonado & Villaver (2016) | $18.06 \pm 0.42$ | $2.19 \pm 0.51$ |
| HD176411 | 2.91 | McWilliam (1990) | $9.98 \pm 0.26$ | $1.64 \pm 0.95$ |
| HD176670 | 2.21 | McWilliam (1990) | $104.74 \pm 8.84$ | $36.02 \pm 21.61$ |
| HD176981 | 1.61 | Lomaeva et al. (2019) | $67.75 \pm 3.64$ | $3.79 \pm 2.22$ |
| HD185958 | 2.79 | McWilliam (1990) | $26.04 \pm 1.11$ | $8.46 \pm 4.93$ |
| HD186675 | $2.80 \pm 0.09$ | Thygesen et al. (2012) | $13.38 \pm 0.51$ | $2.29 \pm 0.50$ |

**Table 20** *continued on next page*

**Table 20** *(continued)*

| Star ID | $\log g$ | Reference | $R$ | $M$ |
|---------|----------|-----------|-----|-----|
| | | | $(R_\odot)$ | $(M_\odot)$ |
| HD188310 | $2.66 \pm 0.10$ | Maldonado & Villaver (2016) | $10.16 \pm 0.29$ | $0.96 \pm 0.23$ |
| HD194317 | $2.02 \pm 0.14$ | Maldonado & Villaver (2016) | $21.84 \pm 0.50$ | $1.01 \pm 0.33$ |
| HD197912 | $3.17$ | Hekker & Meléndez (2007) | $13.89 \pm 0.31$ | $5.78 \pm 3.34$ |
| HD199101 | $1.65$ | McWilliam (1990) | $47.47 \pm 1.60$ | $2.04 \pm 1.18$ |
| HD199697 | $1.91$ | McWilliam (1990) | $33.57 \pm 1.09$ | $1.85 \pm 1.07$ |
| HD199799 | $0.30$ | Smith & Lambert (1990) | $195.23 \pm 10.71$ | $1.54 \pm 0.90$ |
| HD202109 | $3.02 \pm 0.26$ | da Silva et al. (2015) | $13.29 \pm 0.31$ | $3.75 \pm 2.25$ |
| HD205435 | $3.24 \pm 0.19$ | da Silva et al. (2015) | $7.66 \pm 0.19$ | $2.06 \pm 0.91$ |
| HD212496 | $2.64 \pm 0.05$ | Maldonado & Villaver (2016) | $10.50 \pm 0.21$ | $0.97 \pm 0.12$ |
| HD214868 | $1.76 \pm 0.05$ | Deka-Szymankiewicz et al. (2018) | $27.92 \pm 0.89$ | $0.91 \pm 0.12$ |
| HD215182 | $2.16 \pm 0.06$ | Maldonado & Villaver (2016) | $24.42 \pm 1.04$ | $1.75 \pm 0.28$ |
| HD216131 | $2.99 \pm 0.05$ | Deka-Szymankiewicz et al. (2018) | $8.57 \pm 0.17$ | $1.45 \pm 0.18$ |
| HD218452 | $1.97$ | Lomaeva et al. (2019) | $22.60 \pm 0.52$ | $0.96 \pm 0.56$ |
| HD221345 | $2.68 \pm 0.04$ | Maldonado & Villaver (2016) | $11.97 \pm 0.49$ | $1.39 \pm 0.17$ |

NOTE—For references that did not report a formal error in $\log g$, a value of 0.25 dex was assumed (and the resulting error in $M$ is italicized). For more detail, see §8.2.

**Table 21**. Photometry for the program stars.

| Star ID | System/Wvlen | Band/Bandpass | Value | Reference |
|---------|--------------|---------------|-------|-----------|
| HD178 | Johnson | V | $9.26 \pm 0.02$ | This work |
| HD178 | 1250 | 310 | $138.80 \pm 11.20$ | Smith et al. (2004) |
| HD178 | Johnson | H | $1.95 \pm 0.05$ | Ducati (2002) |
| HD178 | Johnson | H | $1.96 \pm 0.05$ | Feast et al. (1990) |
| HD178 | Johnson | H | $2.09 \pm 0.05$ | Kerschbaum & Hron (1994) |
| HD178 | 2200 | 361 | $158.30 \pm 8.60$ | Smith et al. (2004) |
| HD178 | Johnson | K | $1.63 \pm 0.04$ | Neugebauer & Leighton (1969) |
| HD178 | Johnson | K | $1.66 \pm 0.05$ | Ducati (2002) |
| HD178 | Johnson | L | $1.43 \pm 0.05$ | Ducati (2002) |
| HD178 | 3500 | 898 | $84.10 \pm 6.30$ | Smith et al. (2004) |
| HD178 | 4900 | 712 | $39.20 \pm 6.00$ | Smith et al. (2004) |
| HD178 | 12000 | 6384 | $10.20 \pm 18.20$ | Smith et al. (2004) |
| HD598 | KronComet | C2 | $7.42 \pm 0.01$ | This work |
| HD598 | KronComet | Gc | $7.32 \pm 0.01$ | This work |
| HD598 | WBVR | V | $7.16 \pm 0.05$ | Kornilov et al. (1991) |
| HD598 | Johnson | V | $7.28 \pm 0.02$ | This work |
| HD598 | WBVR | R | $5.31 \pm 0.05$ | Kornilov et al. (1991) |
| HD598 | KronComet | Rc | $5.76 \pm 0.02$ | This work |
| HD598 | 1250 | 310 | $126.00 \pm 6.10$ | Smith et al. (2004) |
| HD598 | 2200 | 361 | $131.10 \pm 10.40$ | Smith et al. (2004) |
| HD598 | Johnson | K | $1.72 \pm 0.04$ | Neugebauer & Leighton (1969) |
| HD598 | 3500 | 898 | $65.40 \pm 8.00$ | Smith et al. (2004) |
| HD598 | 4900 | 712 | $24.60 \pm 6.70$ | Smith et al. (2004) |
| HD598 | 12000 | 6384 | $3.70 \pm 18.40$ | Smith et al. (2004) |
| HD672 | Johnson | U | $10.92 \pm 0.16$ | This work |
| HD672 | Johnson | B | $9.87 \pm 0.02$ | This work |
| HD672 | Johnson | V | $8.50 \pm 0.02$ | This work |
| HD672 | Johnson | J | $3.16 \pm 0.05$ | Feast et al. (1990) |
| HD672 | Johnson | H | $2.22 \pm 0.05$ | Feast et al. (1990) |
| HD672 | 2200 | 361 | $115.70 \pm 5.10$ | Smith et al. (2004) |
| HD672 | Johnson | K | $1.95 \pm 0.06$ | Neugebauer & Leighton (1969) |
| HD672 | 3500 | 898 | $60.70 \pm 13.30$ | Smith et al. (2004) |
| HD672 | 4900 | 712 | $26.40 \pm 6.60$ | Smith et al. (2004) |
| HD672 | 12000 | 6384 | $12.40 \pm 22.00$ | Smith et al. (2004) |
| HD787 | DDO | m42 | $8.52 \pm 0.05$ | McClure & Forrester (1981) |
| HD787 | Oja | m42 | $8.00 \pm 0.05$ | Häggkvist & Oja (1970) |
| HD787 | Johnson | B | $6.72 \pm 0.05$ | Cousins (1964b) |
| HD787 | Johnson | B | $6.73 \pm 0.05$ | Johnson et al. (1966) |
| HD787 | Johnson | B | $6.74 \pm 0.05$ | Irwin (1961) |

**Table 21** (continued)

| Star ID | System/Wvlen | Band/Bandpass | Value | Reference |
|---------|--------------|---------------|-------|-----------|
| HD787 | Johnson | B | $6.74 \pm 0.05$ | Mermilliod (1986) |
| HD787 | DDO | m45 | $7.25 \pm 0.05$ | McClure & Forrester (1981) |
| HD787 | Oja | m45 | $6.41 \pm 0.05$ | Häggkvist & Oja (1970) |
| HD787 | Vilnius | Y | $6.36 \pm 0.05$ | Bartasiute (1981) |
| HD787 | DDO | m48 | $5.88 \pm 0.05$ | McClure & Forrester (1981) |
| HD787 | Vilnius | Z | $5.78 \pm 0.05$ | Bartasiute (1981) |
| HD787 | Vilnius | V | $5.29 \pm 0.05$ | Bartasiute (1981) |
| HD787 | Johnson | V | $5.24 \pm 0.05$ | Cousins (1964b) |
| HD787 | Johnson | V | $5.25 \pm 0.05$ | Johnson et al. (1966) |
| HD787 | Johnson | V | $5.27 \pm 0.05$ | Mermilliod (1986) |
| HD787 | Johnson | V | $5.28 \pm 0.05$ | Irwin (1961) |
| HD787 | Vilnius | S | $4.26 \pm 0.05$ | Bartasiute (1981) |
| HD787 | 1250 | 310 | $146.80 \pm 3.10$ | Smith et al. (2004) |
| HD787 | 2200 | 361 | $128.90 \pm 6.00$ | Smith et al. (2004) |
| HD787 | Johnson | K | $1.71 \pm 0.08$ | Neugebauer & Leighton (1969) |
| HD787 | 3500 | 898 | $64.40 \pm 25.30$ | Smith et al. (2004) |
| HD787 | 4900 | 712 | $27.80 \pm 5.30$ | Smith et al. (2004) |
| HD787 | 12000 | 6384 | $5.50 \pm 21.50$ | Smith et al. (2004) |
| HD1522 | Oja | m41 | $6.22 \pm 0.05$ | Häggkvist & Oja (1970) |
| HD1522 | DDO | m42 | $6.25 \pm 0.05$ | McClure & Forrester (1981) |
| HD1522 | Oja | m42 | $5.83 \pm 0.05$ | Häggkvist & Oja (1970) |
| HD1522 | WBVR | B | $4.81 \pm 0.05$ | Kornilov et al. (1991) |
| HD1522 | Johnson | B | $4.64 \pm 0.05$ | Mermilliod (1986) |
| HD1522 | Johnson | B | $4.75 \pm 0.05$ | Gutierrez-Moreno & et al. (1966) |
| HD1522 | Johnson | B | $4.77 \pm 0.05$ | Cousins & Stoy (1962) |
| HD1522 | Johnson | B | $4.77 \pm 0.05$ | Johnson (1964) |
| HD1522 | Johnson | B | $4.77 \pm 0.05$ | Johnson et al. (1966) |
| HD1522 | Johnson | B | $4.77 \pm 0.05$ | Jennens & Helfer (1975) |
| HD1522 | Johnson | B | $4.77 \pm 0.05$ | Ducati (2002) |
| HD1522 | Johnson | B | $4.78 \pm 0.05$ | Cousins & Stoy (1962) |
| HD1522 | Johnson | B | $4.78 \pm 0.05$ | Engels et al. (1981) |
| HD1522 | 13c | m45 | $4.40 \pm 0.05$ | Johnson & Mitchell (1995) |
| HD1522 | DDO | m45 | $5.27 \pm 0.05$ | McClure & Forrester (1981) |
| HD1522 | Oja | m45 | $4.47 \pm 0.05$ | Häggkvist & Oja (1970) |
| HD1522 | Vilnius | Y | $4.46 \pm 0.05$ | Jasevicius et al. (1990) |
| HD1522 | DDO | m48 | $3.99 \pm 0.05$ | McClure & Forrester (1981) |
| HD1522 | Vilnius | Z | $3.94 \pm 0.05$ | Jasevicius et al. (1990) |
| HD1522 | 13c | m52 | $3.86 \pm 0.05$ | Johnson & Mitchell (1995) |
| HD1522 | WBVR | V | $3.57 \pm 0.05$ | Kornilov et al. (1991) |
| HD1522 | Vilnius | V | $3.56 \pm 0.05$ | Jasevicius et al. (1990) |
| HD1522 | Johnson | V | $3.46 \pm 0.05$ | Mermilliod (1986) |

**Table 21** *(continued)*

| Star ID | System/Wvlen | Band/Bandpass | Value | Reference |
|---------|--------------|---------------|-------|-----------|
| HD1522 | Johnson | V | $3.54 \pm 0.05$ | Johnson (1964) |
| HD1522 | Johnson | V | $3.54 \pm 0.05$ | Gutierrez-Moreno & et al. (1966) |
| HD1522 | Johnson | V | $3.55 \pm 0.05$ | Cousins & Stoy (1962) |
| HD1522 | Johnson | V | $3.55 \pm 0.05$ | Johnson et al. (1966) |
| HD1522 | Johnson | V | $3.55 \pm 0.05$ | Ducati (2002) |
| HD1522 | Johnson | V | $3.56 \pm 0.05$ | Cousins & Stoy (1962) |
| HD1522 | Johnson | V | $3.56 \pm 0.05$ | Jennens & Helfer (1975) |
| HD1522 | Johnson | V | $3.56 \pm 0.05$ | Engels et al. (1981) |
| HD1522 | 13c | m58 | $3.28 \pm 0.05$ | Johnson & Mitchell (1995) |
| HD1522 | 13c | m63 | $2.95 \pm 0.05$ | Johnson & Mitchell (1995) |
| HD1522 | Vilnius | S | $2.75 \pm 0.05$ | Jasevicius et al. (1990) |
| HD1522 | WBVR | R | $2.72 \pm 0.05$ | Kornilov et al. (1991) |
| HD1522 | 13c | m72 | $2.66 \pm 0.05$ | Johnson & Mitchell (1995) |
| HD1522 | 13c | m80 | $2.41 \pm 0.05$ | Johnson & Mitchell (1995) |
| HD1522 | 13c | m86 | $2.26 \pm 0.05$ | Johnson & Mitchell (1995) |
| HD1522 | 13c | m99 | $2.08 \pm 0.05$ | Johnson & Mitchell (1995) |
| HD1522 | 13c | m110 | $1.90 \pm 0.05$ | Johnson & Mitchell (1995) |
| HD1522 | 1250 | 310 | $368.50 \pm 13.70$ | Smith et al. (2004) |
| HD1522 | Johnson | J | $1.66 \pm 0.05$ | Johnson et al. (1966) |
| HD1522 | Johnson | J | $1.66 \pm 0.05$ | Glass (1974) |
| HD1522 | Johnson | J | $1.66 \pm 0.05$ | Engels et al. (1981) |
| HD1522 | Johnson | J | $1.66 \pm 0.05$ | Ducati (2002) |
| HD1522 | Johnson | J | $1.67 \pm 0.05$ | Engels et al. (1981) |
| HD1522 | Johnson | H | $1.07 \pm 0.05$ | Glass (1974) |
| HD1522 | Johnson | H | $1.09 \pm 0.05$ | Ducati (2002) |
| HD1522 | Johnson | H | $1.11 \pm 0.05$ | Engels et al. (1981) |
| HD1522 | Johnson | H | $1.12 \pm 0.05$ | Engels et al. (1981) |
| HD1522 | 2200 | 361 | $269.50 \pm 6.40$ | Smith et al. (2004) |
| HD1522 | Johnson | K | $0.92 \pm 0.05$ | Glass (1974) |
| HD1522 | Johnson | K | $0.94 \pm 0.04$ | Neugebauer & Leighton (1969) |
| HD1522 | Johnson | K | $0.95 \pm 0.05$ | Johnson et al. (1966) |
| HD1522 | Johnson | K | $0.95 \pm 0.05$ | Ducati (2002) |
| HD1522 | Johnson | K | $0.97 \pm 0.05$ | Engels et al. (1981) |
| HD1522 | Johnson | L | $0.80 \pm 0.05$ | Glass (1974) |
| HD1522 | Johnson | L | $0.83 \pm 0.05$ | Ducati (2002) |
| HD1522 | Johnson | L | $0.86 \pm 0.05$ | Johnson et al. (1966) |
| HD1522 | Johnson | L | $0.91 \pm 0.05$ | Engels et al. (1981) |
| HD1522 | 3500 | 898 | $130.20 \pm 5.10$ | Smith et al. (2004) |
| HD1522 | Johnson | N | $-0.44 \pm 0.05$ | Ducati (2002) |
| HD1522 | 4900 | 712 | $62.20 \pm 5.80$ | Smith et al. (2004) |
| HD1522 | Johnson | M | $1.08 \pm 0.05$ | Ducati (2002) |

**Table 21** *(continued)*

| Star ID | System/Wvlen | Band/Bandpass | Value | Reference |
|---------|--------------|---------------|-------|-----------|
| HD1522 | 12000 | 6384 | $16.40 \pm 29.80$ | Smith et al. (2004) |
| HD1632 | WBVR | W | $9.34 \pm 0.05$ | Kornilov et al. (1991) |
| HD1632 | Oja | m41 | $8.98 \pm 0.05$ | Häggkvist & Oja (1970) |
| HD1632 | Oja | m42 | $8.76 \pm 0.05$ | Häggkvist & Oja (1970) |
| HD1632 | KronComet | COp | $8.03 \pm 0.13$ | This work |
| HD1632 | WBVR | B | $7.46 \pm 0.05$ | Kornilov et al. (1991) |
| HD1632 | Johnson | B | $7.39 \pm 0.05$ | Haggkvist & Oja (1970) |
| HD1632 | KronComet | Bc | $7.22 \pm 0.03$ | This work |
| HD1632 | Oja | m45 | $7.05 \pm 0.05$ | Häggkvist & Oja (1970) |
| HD1632 | KronComet | C2 | $6.27 \pm 0.03$ | This work |
| HD1632 | KronComet | Gc | $5.98 \pm 0.02$ | This work |
| HD1632 | WBVR | V | $5.81 \pm 0.05$ | Kornilov et al. (1991) |
| HD1632 | Johnson | V | $5.79 \pm 0.05$ | Haggkvist & Oja (1970) |
| HD1632 | Johnson | V | $5.85 \pm 0.04$ | This work |
| HD1632 | WBVR | R | $4.52 \pm 0.05$ | Kornilov et al. (1991) |
| HD1632 | KronComet | Rc | $4.40 \pm 0.01$ | This work |
| HD1632 | 1250 | 310 | $133.10 \pm 9.30$ | Smith et al. (2004) |
| HD1632 | 2200 | 361 | $129.80 \pm 11.70$ | Smith et al. (2004) |
| HD1632 | Johnson | K | $1.80 \pm 0.06$ | Neugebauer & Leighton (1969) |
| HD1632 | 3500 | 898 | $63.80 \pm 6.70$ | Smith et al. (2004) |
| HD1632 | 4900 | 712 | $28.60 \pm 5.30$ | Smith et al. (2004) |
| HD1632 | 12000 | 6384 | $4.50 \pm 19.00$ | Smith et al. (2004) |
| HD2436 | KronComet | NH | $11.32 \pm 0.03$ | This work |
| HD2436 | KronComet | UVc | $10.87 \pm 0.03$ | This work |
| HD2436 | Vilnius | U | $11.56 \pm 0.05$ | Bartkevicius et al. (1973) |
| HD2436 | WBVR | W | $9.54 \pm 0.05$ | Kornilov et al. (1991) |
| HD2436 | Johnson | U | $9.64 \pm 0.05$ | Rybka (1969) |
| HD2436 | Vilnius | P | $10.72 \pm 0.05$ | Bartkevicius et al. (1973) |
| HD2436 | KronComet | CN | $10.30 \pm 0.02$ | This work |
| HD2436 | Vilnius | X | $9.30 \pm 0.05$ | Bartkevicius et al. (1973) |
| HD2436 | KronComet | COp | $8.45 \pm 0.02$ | This work |
| HD2436 | WBVR | B | $7.68 \pm 0.05$ | Kornilov et al. (1991) |
| HD2436 | Johnson | B | $7.64 \pm 0.05$ | Rybka (1969) |
| HD2436 | Johnson | B | $7.74 \pm 0.03$ | This work |
| HD2436 | KronComet | Bc | $7.56 \pm 0.02$ | This work |
| HD2436 | Vilnius | Y | $7.21 \pm 0.05$ | Bartkevicius et al. (1973) |
| HD2436 | KronComet | C2 | $6.60 \pm 0.05$ | This work |
| HD2436 | Vilnius | Z | $6.65 \pm 0.05$ | Bartkevicius et al. (1973) |
| HD2436 | KronComet | Gc | $6.29 \pm 0.03$ | This work |
| HD2436 | WBVR | V | $6.05 \pm 0.05$ | Kornilov et al. (1991) |
| HD2436 | Vilnius | V | $6.09 \pm 0.05$ | Bartkevicius et al. (1973) |

**Table 21** (continued)

| Star ID | System/Wvlen | Band/Bandpass | Value | Reference |
|---------|--------------|---------------|-------|-----------|
| HD2436 | Johnson | V | $6.06 \pm 0.05$ | Rybka (1969) |
| HD2436 | Johnson | V | $6.16 \pm 0.06$ | This work |
| HD2436 | Vilnius | S | $5.03 \pm 0.05$ | Bartkevicius et al. (1973) |
| HD2436 | WBVR | R | $4.83 \pm 0.05$ | Kornilov et al. (1991) |
| HD2436 | KronComet | Rc | $4.70 \pm 0.02$ | This work |
| HD2436 | 1250 | 310 | $83.60 \pm 7.50$ | Smith et al. (2004) |
| HD2436 | 2200 | 361 | $76.10 \pm 6.60$ | Smith et al. (2004) |
| HD2436 | Johnson | K | $2.21 \pm 0.05$ | Neugebauer & Leighton (1969) |
| HD2436 | 3500 | 898 | $37.90 \pm 4.50$ | Smith et al. (2004) |
| HD2436 | 4900 | 712 | $16.20 \pm 4.50$ | Smith et al. (2004) |
| HD2436 | 12000 | 6384 | $6.80 \pm 24.50$ | Smith et al. (2004) |
| HD3346 | DDO | m35 | $10.16 \pm 0.05$ | McClure & Forrester (1981) |
| HD3346 | WBVR | W | $8.67 \pm 0.05$ | Kornilov et al. (1991) |
| HD3346 | Johnson | U | $8.49 \pm 0.05$ | Mermilliod (1986) |
| HD3346 | Johnson | U | $8.70 \pm 0.05$ | Argue (1966) |
| HD3346 | DDO | m38 | $8.80 \pm 0.05$ | McClure & Forrester (1981) |
| HD3346 | DDO | m41 | $8.79 \pm 0.05$ | McClure & Forrester (1981) |
| HD3346 | Oja | m41 | $8.34 \pm 0.05$ | Häggkvist & Oja (1970) |
| HD3346 | DDO | m42 | $8.58 \pm 0.05$ | McClure & Forrester (1981) |
| HD3346 | Oja | m42 | $8.09 \pm 0.05$ | Häggkvist & Oja (1970) |
| HD3346 | KronComet | COp | $7.34 \pm 0.13$ | This work |
| HD3346 | WBVR | B | $6.79 \pm 0.05$ | Kornilov et al. (1991) |
| HD3346 | Johnson | B | $6.65 \pm 0.05$ | Mermilliod (1986) |
| HD3346 | Johnson | B | $6.73 \pm 0.05$ | Argue (1966) |
| HD3346 | KronComet | Bc | $6.55 \pm 0.04$ | This work |
| HD3346 | DDO | m45 | $7.22 \pm 0.05$ | McClure & Forrester (1981) |
| HD3346 | Oja | m45 | $6.40 \pm 0.05$ | Häggkvist & Oja (1970) |
| HD3346 | DDO | m48 | $5.81 \pm 0.05$ | McClure & Forrester (1981) |
| HD3346 | KronComet | C2 | $5.59 \pm 0.03$ | This work |
| HD3346 | KronComet | Gc | $5.29 \pm 0.02$ | This work |
| HD3346 | WBVR | V | $5.15 \pm 0.05$ | Kornilov et al. (1991) |
| HD3346 | Johnson | V | $5.08 \pm 0.05$ | Mermilliod (1986) |
| HD3346 | Johnson | V | $5.13 \pm 0.05$ | Argue (1966) |
| HD3346 | Johnson | V | $5.18 \pm 0.04$ | This work |
| HD3346 | WBVR | R | $3.88 \pm 0.05$ | Kornilov et al. (1991) |
| HD3346 | KronComet | Rc | $3.73 \pm 0.01$ | This work |
| HD3346 | 1250 | 310 | $207.00 \pm 15.00$ | Smith et al. (2004) |
| HD3346 | 2200 | 361 | $201.20 \pm 10.20$ | Smith et al. (2004) |
| HD3346 | Johnson | K | $1.20 \pm 0.05$ | Neugebauer & Leighton (1969) |
| HD3346 | 3500 | 898 | $99.30 \pm 8.20$ | Smith et al. (2004) |
| HD3346 | 4900 | 712 | $43.90 \pm 5.00$ | Smith et al. (2004) |

**Table 21** *(continued)*

| Star ID | System/Wvlen | Band/Bandpass | Value | Reference |
|---------|--------------|---------------|-------|-----------|
| HD3346 | 12000 | 6384 | $6.30 \pm 17.60$ | Smith et al. (2004) |
| HD3546 | 13c | m33 | $5.74 \pm 0.05$ | Johnson & Mitchell (1995) |
| HD3546 | Geneva | U | $6.16 \pm 0.08$ | Golay (1972) |
| HD3546 | Vilnius | U | $7.42 \pm 0.05$ | Bartkevicius & Metik (1969) |
| HD3546 | Vilnius | U | $7.48 \pm 0.05$ | Sperauskas et al. (1981) |
| HD3546 | 13c | m35 | $5.49 \pm 0.05$ | Johnson & Mitchell (1995) |
| HD3546 | DDO | m35 | $7.01 \pm 0.05$ | McClure & Forrester (1981) |
| HD3546 | Stromgren | u | $6.85 \pm 0.08$ | Olson (1974) |
| HD3546 | Stromgren | u | $6.97 \pm 0.08$ | Olsen (1993) |
| HD3546 | Stromgren | u | $6.97 \pm 0.08$ | Hauck & Mermilliod (1998) |
| HD3546 | Johnson | U | $5.67 \pm 0.05$ | Mermilliod (1986) |
| HD3546 | Johnson | U | $5.69 \pm 0.05$ | Jennens & Helfer (1975) |
| HD3546 | Johnson | U | $5.71 \pm 0.05$ | Roman (1955) |
| HD3546 | Johnson | U | $5.72 \pm 0.05$ | Johnson (1964) |
| HD3546 | Johnson | U | $5.72 \pm 0.05$ | Johnson et al. (1966) |
| HD3546 | Johnson | U | $5.72 \pm 0.05$ | Ducati (2002) |
| HD3546 | Geneva | B1 | $5.64 \pm 0.08$ | Golay (1972) |
| HD3546 | Oja | m41 | $6.17 \pm 0.05$ | Häggkvist & Oja (1970) |
| HD3546 | DDO | m42 | $6.57 \pm 0.05$ | McClure & Forrester (1981) |
| HD3546 | Oja | m42 | $6.18 \pm 0.05$ | Häggkvist & Oja (1970) |
| HD3546 | Geneva | B | $4.46 \pm 0.08$ | Golay (1972) |
| HD3546 | KronComet | COp | $5.41 \pm 0.13$ | This work |
| HD3546 | WBVR | B | $5.25 \pm 0.05$ | Kornilov et al. (1991) |
| HD3546 | Johnson | B | $5.19 \pm 0.05$ | Mermilliod (1986) |
| HD3546 | Johnson | B | $5.21 \pm 0.05$ | Häggkvist & Oja (1966) |
| HD3546 | Johnson | B | $5.23 \pm 0.05$ | Jennens & Helfer (1975) |
| HD3546 | Johnson | B | $5.25 \pm 0.05$ | Roman (1955) |
| HD3546 | Johnson | B | $5.25 \pm 0.05$ | Johnson (1964) |
| HD3546 | Johnson | B | $5.25 \pm 0.05$ | Johnson et al. (1966) |
| HD3546 | Johnson | B | $5.25 \pm 0.05$ | Ducati (2002) |
| HD3546 | KronComet | Bc | $5.04 \pm 0.03$ | This work |
| HD3546 | Geneva | B2 | $5.70 \pm 0.08$ | Golay (1972) |
| HD3546 | 13c | m45 | $4.99 \pm 0.05$ | Johnson & Mitchell (1995) |
| HD3546 | DDO | m45 | $5.81 \pm 0.05$ | McClure & Forrester (1981) |
| HD3546 | Oja | m45 | $5.00 \pm 0.05$ | Häggkvist & Oja (1970) |
| HD3546 | Vilnius | Y | $5.05 \pm 0.05$ | Bartkevicius & Metik (1969) |
| HD3546 | Vilnius | Y | $5.06 \pm 0.05$ | Sperauskas et al. (1981) |
| HD3546 | Stromgren | b | $4.91 \pm 0.08$ | Olson (1974) |
| HD3546 | Stromgren | b | $4.93 \pm 0.08$ | Olsen (1993) |
| HD3546 | Stromgren | b | $4.93 \pm 0.08$ | Hauck & Mermilliod (1998) |
| HD3546 | DDO | m48 | $4.67 \pm 0.05$ | McClure & Forrester (1981) |

**Table 21** *continued on next page*

**Table 21** *(continued)*

| Star ID | System/Wvlen | Band/Bandpass | Value | Reference |
|---------|--------------|---------------|-------|-----------|
| HD3546 | KronComet | C2 | $4.46 \pm 0.03$ | This work |
| HD3546 | Vilnius | Z | $4.63 \pm 0.05$ | Sperauskas et al. (1981) |
| HD3546 | Vilnius | Z | $4.64 \pm 0.05$ | Bartkevicius & Metik (1969) |
| HD3546 | 13c | m52 | $4.56 \pm 0.05$ | Johnson & Mitchell (1995) |
| HD3546 | KronComet | Gc | $4.36 \pm 0.02$ | This work |
| HD3546 | Geneva | V1 | $5.11 \pm 0.08$ | Golay (1972) |
| HD3546 | WBVR | V | $4.36 \pm 0.05$ | Kornilov et al. (1991) |
| HD3546 | Vilnius | V | $4.38 \pm 0.05$ | Bartkevicius & Metik (1969) |
| HD3546 | Vilnius | V | $4.38 \pm 0.05$ | Sperauskas et al. (1981) |
| HD3546 | Stromgren | y | $4.38 \pm 0.08$ | Olson (1974) |
| HD3546 | Stromgren | y | $4.38 \pm 0.08$ | Olsen (1993) |
| HD3546 | Stromgren | y | $4.38 \pm 0.08$ | Hauck & Mermilliod (1998) |
| HD3546 | Geneva | V | $4.35 \pm 0.08$ | Golay (1972) |
| HD3546 | Johnson | V | $4.31 \pm 0.05$ | Mermilliod (1986) |
| HD3546 | Johnson | V | $4.34 \pm 0.05$ | Häggkvist & Oja (1966) |
| HD3546 | Johnson | V | $4.36 \pm 0.05$ | Jennens & Helfer (1975) |
| HD3546 | Johnson | V | $4.37 \pm 0.05$ | Roman (1955) |
| HD3546 | Johnson | V | $4.38 \pm 0.05$ | Johnson et al. (1966) |
| HD3546 | Johnson | V | $4.38 \pm 0.05$ | Ducati (2002) |
| HD3546 | Johnson | V | $4.39 \pm 0.05$ | Johnson (1964) |
| HD3546 | Johnson | V | $4.43 \pm 0.06$ | This work |
| HD3546 | 13c | m58 | $4.15 \pm 0.05$ | Johnson & Mitchell (1995) |
| HD3546 | Geneva | G | $5.34 \pm 0.08$ | Golay (1972) |
| HD3546 | 13c | m63 | $3.88 \pm 0.05$ | Johnson & Mitchell (1995) |
| HD3546 | Vilnius | S | $3.66 \pm 0.05$ | Bartkevicius & Metik (1969) |
| HD3546 | Vilnius | S | $3.70 \pm 0.05$ | Sperauskas et al. (1981) |
| HD3546 | WBVR | R | $3.68 \pm 0.05$ | Kornilov et al. (1991) |
| HD3546 | KronComet | Rc | $3.39 \pm 0.01$ | This work |
| HD3546 | 13c | m72 | $3.62 \pm 0.05$ | Johnson & Mitchell (1995) |
| HD3546 | 13c | m80 | $3.39 \pm 0.05$ | Johnson & Mitchell (1995) |
| HD3546 | 13c | m86 | $3.30 \pm 0.05$ | Johnson & Mitchell (1995) |
| HD3546 | 13c | m99 | $3.15 \pm 0.05$ | Johnson & Mitchell (1995) |
| HD3546 | 13c | m110 | $2.98 \pm 0.05$ | Johnson & Mitchell (1995) |
| HD3546 | 1250 | 310 | $126.00 \pm 5.70$ | Smith et al. (2004) |
| HD3546 | Johnson | J | $2.71 \pm 0.05$ | Alonso et al. (1998) |
| HD3546 | Johnson | J | $2.83 \pm 0.05$ | Arribas & Martinez Roger (1987) |
| HD3546 | Johnson | J | $2.83 \pm 0.05$ | Ducati (2002) |
| HD3546 | Johnson | J | $2.84 \pm 0.05$ | Johnson et al. (1966) |
| HD3546 | Johnson | J | $2.84 \pm 0.05$ | Voelcker (1975) |
| HD3546 | Johnson | J | $2.84 \pm 0.05$ | Jameson & Akinci (1979) |
| HD3546 | Johnson | H | $2.27 \pm 0.05$ | Alonso et al. (1998) |

**Table 21** *(continued)*

| Star ID | System/Wvlen | Band/Bandpass | Value | Reference |
|---------|--------------|---------------|-------|-----------|
| HD3546 | Johnson | H | $2.33 \pm 0.05$ | Voelcker (1975) |
| HD3546 | Johnson | H | $2.33 \pm 0.05$ | Arribas & Martinez Roger (1987) |
| HD3546 | Johnson | H | $2.33 \pm 0.05$ | Ducati (2002) |
| HD3546 | 2200 | 361 | $82.00 \pm 5.70$ | Smith et al. (2004) |
| HD3546 | Johnson | K | $2.20 \pm 0.06$ | Neugebauer & Leighton (1969) |
| HD3546 | Johnson | K | $2.21 \pm 0.05$ | Johnson et al. (1966) |
| HD3546 | Johnson | K | $2.22 \pm 0.05$ | Ducati (2002) |
| HD3546 | Johnson | L | $2.13 \pm 0.05$ | Ducati (2002) |
| HD3546 | 3500 | 898 | $38.30 \pm 5.30$ | Smith et al. (2004) |
| HD3546 | 4900 | 712 | $18.40 \pm 5.30$ | Smith et al. (2004) |
| HD3546 | 12000 | 6384 | $-10.50 \pm 21.30$ | Smith et al. (2004) |
| HD3574 | Oja | m41 | $8.77 \pm 0.05$ | Häggkvist & Oja (1970) |
| HD3574 | Oja | m42 | $8.43 \pm 0.05$ | Häggkvist & Oja (1970) |
| HD3574 | KronComet | COp | $7.90 \pm 0.02$ | This work |
| HD3574 | WBVR | B | $7.16 \pm 0.05$ | Kornilov et al. (1991) |
| HD3574 | Johnson | B | $7.07 \pm 0.05$ | Haggkvist & Oja (1970) |
| HD3574 | Johnson | B | $7.18 \pm 0.02$ | This work |
| HD3574 | KronComet | Bc | $7.06 \pm 0.02$ | This work |
| HD3574 | Oja | m45 | $6.76 \pm 0.05$ | Häggkvist & Oja (1970) |
| HD3574 | Vilnius | Y | $6.59 \pm 0.05$ | Bartkevicius et al. (1973) |
| HD3574 | KronComet | C2 | $5.96 \pm 0.04$ | This work |
| HD3574 | Vilnius | Z | $5.96 \pm 0.05$ | Bartkevicius et al. (1973) |
| HD3574 | KronComet | Gc | $5.71 \pm 0.02$ | This work |
| HD3574 | WBVR | V | $5.45 \pm 0.05$ | Kornilov et al. (1991) |
| HD3574 | Vilnius | V | $5.43 \pm 0.05$ | Bartkevicius et al. (1973) |
| HD3574 | Johnson | V | $5.43 \pm 0.05$ | Haggkvist & Oja (1970) |
| HD3574 | Johnson | V | $5.54 \pm 0.03$ | This work |
| HD3574 | Vilnius | S | $4.36 \pm 0.05$ | Bartkevicius et al. (1973) |
| HD3574 | WBVR | R | $4.24 \pm 0.05$ | Kornilov et al. (1991) |
| HD3574 | KronComet | Rc | $4.07 \pm 0.02$ | This work |
| HD3574 | 1250 | 310 | $166.50 \pm 9.90$ | Smith et al. (2004) |
| HD3574 | 2200 | 361 | $158.50 \pm 7.60$ | Smith et al. (2004) |
| HD3574 | Johnson | K | $1.72 \pm 0.05$ | Neugebauer & Leighton (1969) |
| HD3574 | 3500 | 898 | $80.30 \pm 22.80$ | Smith et al. (2004) |
| HD3574 | 4900 | 712 | $34.30 \pm 7.40$ | Smith et al. (2004) |
| HD3574 | 12000 | 6384 | $4.70 \pm 18.60$ | Smith et al. (2004) |
| HD3627 | Geneva | B1 | $5.47 \pm 0.08$ | Golay (1972) |
| HD3627 | DDO | m41 | $6.48 \pm 0.05$ | McClure & Forrester (1981) |
| HD3627 | Oja | m41 | $6.04 \pm 0.05$ | Häggkvist & Oja (1970) |
| HD3627 | DDO | m42 | $6.14 \pm 0.05$ | McClure & Forrester (1981) |
| HD3627 | Oja | m42 | $5.68 \pm 0.05$ | Häggkvist & Oja (1970) |

**Table 21** *(continued)*

| Star ID | System/Wvlen | Band/Bandpass | Value | Reference |
|---------|--------------|---------------|-------|-----------|
| HD3627 | Geneva | B | $3.95 \pm 0.08$ | Golay (1972) |
| HD3627 | WBVR | B | $4.59 \pm 0.05$ | Kornilov et al. (1991) |
| HD3627 | Johnson | B | $4.48 \pm 0.05$ | Mermilliod (1986) |
| HD3627 | Johnson | B | $4.50 \pm 0.05$ | Bouigue (1959) |
| HD3627 | Johnson | B | $4.54 \pm 0.05$ | Argue (1966) |
| HD3627 | Johnson | B | $4.55 \pm 0.05$ | Häggkvist & Oja (1966) |
| HD3627 | Johnson | B | $4.56 \pm 0.05$ | Johnson et al. (1966) |
| HD3627 | Johnson | B | $4.56 \pm 0.05$ | Ducati (2002) |
| HD3627 | Johnson | B | $4.60 \pm 0.05$ | Johnson (1964) |
| HD3627 | Geneva | B2 | $4.99 \pm 0.08$ | Golay (1972) |
| HD3627 | 13c | m45 | $4.18 \pm 0.05$ | Johnson & Mitchell (1995) |
| HD3627 | DDO | m45 | $5.03 \pm 0.05$ | McClure & Forrester (1981) |
| HD3627 | Oja | m45 | $4.22 \pm 0.05$ | Häggkvist & Oja (1970) |
| HD3627 | Vilnius | Y | $4.18 \pm 0.05$ | Zdanavicius et al. (1969) |
| HD3627 | Vilnius | Y | $4.20 \pm 0.05$ | Dzervitis & Paupers (1981) |
| HD3627 | DDO | m48 | $3.74 \pm 0.05$ | McClure & Forrester (1981) |
| HD3627 | Vilnius | Z | $3.71 \pm 0.05$ | Zdanavicius et al. (1969) |
| HD3627 | Vilnius | Z | $3.71 \pm 0.05$ | Dzervitis & Paupers (1981) |
| HD3627 | 13c | m52 | $3.65 \pm 0.05$ | Johnson & Mitchell (1995) |
| HD3627 | Geneva | V1 | $4.10 \pm 0.08$ | Golay (1972) |
| HD3627 | WBVR | V | $3.28 \pm 0.05$ | Kornilov et al. (1991) |
| HD3627 | Vilnius | V | $3.28 \pm 0.05$ | Zdanavicius et al. (1969) |
| HD3627 | Vilnius | V | $3.28 \pm 0.05$ | Dzervitis & Paupers (1981) |
| HD3627 | Geneva | V | $3.30 \pm 0.08$ | Golay (1972) |
| HD3627 | Johnson | V | $3.19 \pm 0.05$ | Bouigue (1959) |
| HD3627 | Johnson | V | $3.22 \pm 0.05$ | Mermilliod (1986) |
| HD3627 | Johnson | V | $3.26 \pm 0.05$ | Argue (1966) |
| HD3627 | Johnson | V | $3.28 \pm 0.05$ | Johnson et al. (1966) |
| HD3627 | Johnson | V | $3.28 \pm 0.05$ | Ducati (2002) |
| HD3627 | Johnson | V | $3.29 \pm 0.05$ | Häggkvist & Oja (1966) |
| HD3627 | Johnson | V | $3.30 \pm 0.05$ | Johnson (1964) |
| HD3627 | 13c | m58 | $2.98 \pm 0.05$ | Johnson & Mitchell (1995) |
| HD3627 | Geneva | G | $4.20 \pm 0.08$ | Golay (1972) |
| HD3627 | 13c | m63 | $2.64 \pm 0.05$ | Johnson & Mitchell (1995) |
| HD3627 | Vilnius | S | $2.36 \pm 0.05$ | Zdanavicius et al. (1969) |
| HD3627 | Vilnius | S | $2.41 \pm 0.05$ | Dzervitis & Paupers (1981) |
| HD3627 | WBVR | R | $2.37 \pm 0.05$ | Kornilov et al. (1991) |
| HD3627 | KronComet | Rc | $2.11 \pm 0.01$ | This work |
| HD3627 | 13c | m72 | $2.32 \pm 0.05$ | Johnson & Mitchell (1995) |
| HD3627 | 13c | m80 | $2.04 \pm 0.05$ | Johnson & Mitchell (1995) |
| HD3627 | 13c | m86 | $1.91 \pm 0.05$ | Johnson & Mitchell (1995) |

**Table 21** *(continued)*

| Star ID | System/Wvlen | Band/Bandpass | Value | Reference |
|---------|--------------|---------------|-------|-----------|
| HD3627 | 13c | m99 | $1.70 \pm 0.05$ | Johnson & Mitchell (1995) |
| HD3627 | 13c | m110 | $1.47 \pm 0.05$ | Johnson & Mitchell (1995) |
| HD3627 | 1250 | 310 | $546.10 \pm 13.50$ | Smith et al. (2004) |
| HD3627 | Johnson | J | $1.12 \pm 0.05$ | Alonso et al. (1994) |
| HD3627 | Johnson | J | $1.12 \pm 0.05$ | Alonso et al. (1998) |
| HD3627 | Johnson | J | $1.15 \pm 0.05$ | Selby et al. (1988) |
| HD3627 | Johnson | J | $1.15 \pm 0.05$ | Blackwell et al. (1990) |
| HD3627 | Johnson | J | $1.18 \pm 0.05$ | Ducati (2002) |
| HD3627 | Johnson | J | $1.24 \pm 0.05$ | Johnson et al. (1966) |
| HD3627 | Johnson | J | $1.24 \pm 0.05$ | Shenavrin et al. (2011) |
| HD3627 | Johnson | H | $0.55 \pm 0.05$ | Alonso et al. (1994) |
| HD3627 | Johnson | H | $0.55 \pm 0.05$ | Alonso et al. (1998) |
| HD3627 | Johnson | H | $0.63 \pm 0.05$ | Shenavrin et al. (2011) |
| HD3627 | 2200 | 361 | $412.00 \pm 5.90$ | Smith et al. (2004) |
| HD3627 | Johnson | K | $0.43 \pm 0.05$ | Ducati (2002) |
| HD3627 | Johnson | K | $0.44 \pm 0.05$ | Neugebauer & Leighton (1969) |
| HD3627 | Johnson | K | $0.48 \pm 0.05$ | Johnson et al. (1966) |
| HD3627 | Johnson | K | $0.48 \pm 0.05$ | Shenavrin et al. (2011) |
| HD3627 | Johnson | L | $0.30 \pm 0.05$ | Johnson et al. (1966) |
| HD3627 | Johnson | L | $0.30 \pm 0.05$ | Ducati (2002) |
| HD3627 | 3500 | 898 | $199.70 \pm 6.20$ | Smith et al. (2004) |
| HD3627 | Johnson | N | $-0.01 \pm 0.05$ | Ducati (2002) |
| HD3627 | 4900 | 712 | $93.80 \pm 5.20$ | Smith et al. (2004) |
| HD3627 | 12000 | 6384 | $21.50 \pm 18.10$ | Smith et al. (2004) |
| HD5006 | KronComet | NH | $12.72 \pm 0.03$ | This work |
| HD5006 | KronComet | UVc | $12.28 \pm 0.03$ | This work |
| HD5006 | Johnson | U | $10.57 \pm 0.16$ | This work |
| HD5006 | KronComet | CN | $11.41 \pm 0.02$ | This work |
| HD5006 | KronComet | COp | $9.82 \pm 0.02$ | This work |
| HD5006 | Johnson | B | $9.21 \pm 0.03$ | This work |
| HD5006 | KronComet | Bc | $9.01 \pm 0.01$ | This work |
| HD5006 | KronComet | C2 | $7.87 \pm 0.04$ | This work |
| HD5006 | KronComet | Gc | $7.67 \pm 0.05$ | This work |
| HD5006 | Johnson | V | $7.57 \pm 0.06$ | This work |
| HD5006 | KronComet | Rc | $6.12 \pm 0.04$ | This work |
| HD5006 | 1250 | 310 | $71.70 \pm 7.90$ | Smith et al. (2004) |
| HD5006 | 2200 | 361 | $69.00 \pm 11.20$ | Smith et al. (2004) |
| HD5006 | Johnson | K | $2.37 \pm 0.08$ | Neugebauer & Leighton (1969) |
| HD5006 | 3500 | 898 | $32.10 \pm 6.00$ | Smith et al. (2004) |
| HD5006 | 4900 | 712 | $15.80 \pm 5.50$ | Smith et al. (2004) |
| HD5006 | 12000 | 6384 | $0.80 \pm 17.90$ | Smith et al. (2004) |

**Table 21** *(continued)*

| Star ID | System/Wvlen | Band/Bandpass | Value | Reference |
|---------|--------------|---------------|-------|-----------|
| HD5575 | Vilnius | U | $9.21 \pm 0.05$ | Zdanavicius et al. (1972) |
| HD5575 | DDO | m35 | $8.74 \pm 0.05$ | McClure & Forrester (1981) |
| HD5575 | WBVR | W | $7.29 \pm 0.05$ | Kornilov et al. (1991) |
| HD5575 | Johnson | U | $7.29 \pm 0.16$ | This work |
| HD5575 | Vilnius | P | $8.60 \pm 0.05$ | Zdanavicius et al. (1972) |
| HD5575 | DDO | m38 | $7.65 \pm 0.05$ | McClure & Forrester (1981) |
| HD5575 | Vilnius | X | $7.62 \pm 0.05$ | Zdanavicius et al. (1972) |
| HD5575 | DDO | m41 | $8.18 \pm 0.05$ | McClure & Forrester (1981) |
| HD5575 | Oja | m41 | $7.73 \pm 0.05$ | Häggkvist & Oja (1970) |
| HD5575 | DDO | m42 | $7.91 \pm 0.05$ | McClure & Forrester (1981) |
| HD5575 | Oja | m42 | $7.48 \pm 0.05$ | Häggkvist & Oja (1970) |
| HD5575 | WBVR | B | $6.55 \pm 0.05$ | Kornilov et al. (1991) |
| HD5575 | Johnson | B | $6.48 \pm 0.02$ | This work |
| HD5575 | Johnson | B | $6.50 \pm 0.05$ | Haggkvist & Oja (1970) |
| HD5575 | DDO | m45 | $7.05 \pm 0.05$ | McClure & Forrester (1981) |
| HD5575 | Oja | m45 | $6.25 \pm 0.05$ | Häggkvist & Oja (1970) |
| HD5575 | Vilnius | Y | $6.27 \pm 0.05$ | Zdanavicius et al. (1972) |
| HD5575 | DDO | m48 | $5.83 \pm 0.05$ | McClure & Forrester (1981) |
| HD5575 | Vilnius | Z | $5.76 \pm 0.05$ | Zdanavicius et al. (1972) |
| HD5575 | WBVR | V | $5.44 \pm 0.05$ | Kornilov et al. (1991) |
| HD5575 | Vilnius | V | $5.45 \pm 0.05$ | Zdanavicius et al. (1972) |
| HD5575 | Johnson | V | $5.42 \pm 0.05$ | Haggkvist & Oja (1970) |
| HD5575 | Johnson | V | $5.51 \pm 0.02$ | This work |
| HD5575 | Vilnius | S | $4.70 \pm 0.05$ | Zdanavicius et al. (1972) |
| HD5575 | WBVR | R | $4.68 \pm 0.05$ | Kornilov et al. (1991) |
| HD5575 | 1250 | 310 | $58.80 \pm 7.10$ | Smith et al. (2004) |
| HD5575 | 2200 | 361 | $39.70 \pm 5.90$ | Smith et al. (2004) |
| HD5575 | Johnson | K | $2.93 \pm 0.10$ | Neugebauer & Leighton (1969) |
| HD5575 | 3500 | 898 | $17.90 \pm 4.50$ | Smith et al. (2004) |
| HD5575 | 4900 | 712 | $8.30 \pm 4.90$ | Smith et al. (2004) |
| HD5575 | 12000 | 6384 | $-2.90 \pm 21.00$ | Smith et al. (2004) |
| HD6186 | 13c | m33 | $5.98 \pm 0.05$ | Johnson & Mitchell (1995) |
| HD6186 | Geneva | U | $6.45 \pm 0.08$ | Golay (1972) |
| HD6186 | Vilnius | U | $7.64 \pm 0.05$ | Kakaras et al. (1968) |
| HD6186 | Vilnius | U | $7.70 \pm 0.05$ | Zdanavicius et al. (1969) |
| HD6186 | Vilnius | U | $7.71 \pm 0.05$ | Jasevicius et al. (1990) |
| HD6186 | 13c | m35 | $5.75 \pm 0.05$ | Johnson & Mitchell (1995) |
| HD6186 | DDO | m35 | $7.24 \pm 0.05$ | McClure & Forrester (1981) |
| HD6186 | Johnson | U | $5.87 \pm 0.05$ | Mermilliod (1986) |
| HD6186 | Johnson | U | $5.90 \pm 0.05$ | Cousins (1962a) |
| HD6186 | Johnson | U | $5.92 \pm 0.05$ | Jennens & Helfer (1975) |

**Table 21** *(continued)*

| Star ID | System/Wvlen | Band/Bandpass | Value | Reference |
|---------|--------------|---------------|-------|-----------|
| HD6186 | Johnson | U | $5.94 \pm 0.05$ | Argue (1966) |
| HD6186 | Johnson | U | $5.94 \pm 0.05$ | Gutierrez-Moreno & et al. (1966) |
| HD6186 | Johnson | U | $5.95 \pm 0.05$ | Johnson et al. (1966) |
| HD6186 | Johnson | U | $5.95 \pm 0.05$ | Ducati (2002) |
| HD6186 | Geneva | B1 | $5.78 \pm 0.08$ | Golay (1972) |
| HD6186 | Oja | m41 | $6.29 \pm 0.05$ | Häggkvist & Oja (1970) |
| HD6186 | DDO | m42 | $6.61 \pm 0.05$ | McClure & Forrester (1981) |
| HD6186 | DDO | m42 | $6.61 \pm 0.05$ | Mermilliod & Nitschelm (1989) |
| HD6186 | Oja | m42 | $6.18 \pm 0.05$ | Häggkvist & Oja (1970) |
| HD6186 | Geneva | B | $4.52 \pm 0.08$ | Golay (1972) |
| HD6186 | KronComet | COp | $5.46 \pm 0.13$ | This work |
| HD6186 | WBVR | B | $5.25 \pm 0.05$ | Kornilov et al. (1991) |
| HD6186 | Johnson | B | $5.18 \pm 0.05$ | Mermilliod (1986) |
| HD6186 | Johnson | B | $5.21 \pm 0.05$ | Häggkvist & Oja (1966) |
| HD6186 | Johnson | B | $5.22 \pm 0.05$ | Cousins (1962a) |
| HD6186 | Johnson | B | $5.23 \pm 0.05$ | Argue (1966) |
| HD6186 | Johnson | B | $5.23 \pm 0.05$ | Jennens & Helfer (1975) |
| HD6186 | Johnson | B | $5.24 \pm 0.05$ | Johnson et al. (1966) |
| HD6186 | Johnson | B | $5.24 \pm 0.05$ | Ducati (2002) |
| HD6186 | Johnson | B | $5.26 \pm 0.05$ | Gutierrez-Moreno & et al. (1966) |
| HD6186 | KronComet | Bc | $5.06 \pm 0.03$ | This work |
| HD6186 | Geneva | B2 | $5.72 \pm 0.08$ | Golay (1972) |
| HD6186 | 13c | m45 | $4.98 \pm 0.05$ | Johnson & Mitchell (1995) |
| HD6186 | DDO | m45 | $5.79 \pm 0.05$ | McClure & Forrester (1981) |
| HD6186 | DDO | m45 | $5.79 \pm 0.05$ | Mermilliod & Nitschelm (1989) |
| HD6186 | Oja | m45 | $4.96 \pm 0.05$ | Häggkvist & Oja (1970) |
| HD6186 | Vilnius | Y | $5.02 \pm 0.05$ | Kakaras et al. (1968) |
| HD6186 | Vilnius | Y | $5.04 \pm 0.05$ | Jasevicius et al. (1990) |
| HD6186 | Vilnius | Y | $5.05 \pm 0.05$ | Zdanavicius et al. (1969) |
| HD6186 | DDO | m48 | $4.61 \pm 0.05$ | McClure & Forrester (1981) |
| HD6186 | DDO | m48 | $4.62 \pm 0.05$ | Mermilliod & Nitschelm (1989) |
| HD6186 | KronComet | C2 | $4.42 \pm 0.03$ | This work |
| HD6186 | Vilnius | Z | $4.57 \pm 0.05$ | Kakaras et al. (1968) |
| HD6186 | Vilnius | Z | $4.57 \pm 0.05$ | Zdanavicius et al. (1969) |
| HD6186 | Vilnius | Z | $4.59 \pm 0.05$ | Jasevicius et al. (1990) |
| HD6186 | 13c | m52 | $4.51 \pm 0.05$ | Johnson & Mitchell (1995) |
| HD6186 | KronComet | Gc | $4.31 \pm 0.02$ | This work |
| HD6186 | Geneva | V1 | $5.05 \pm 0.08$ | Golay (1972) |
| HD6186 | WBVR | V | $4.27 \pm 0.05$ | Kornilov et al. (1991) |
| HD6186 | Vilnius | V | $4.28 \pm 0.05$ | Kakaras et al. (1968) |
| HD6186 | Vilnius | V | $4.28 \pm 0.05$ | Zdanavicius et al. (1969) |

**Table 21** *continued on next page*

**Table 21** *(continued)*

| Star ID | System/Wvlen | Band/Bandpass | Value | Reference |
|---------|--------------|---------------|-------|-----------|
| HD6186 | Vilnius | V | $4.28 \pm 0.05$ | Jasevicius et al. (1990) |
| HD6186 | Geneva | V | $4.27 \pm 0.08$ | Golay (1972) |
| HD6186 | Johnson | V | $4.24 \pm 0.05$ | Mermilliod (1986) |
| HD6186 | Johnson | V | $4.25 \pm 0.05$ | Häggkvist & Oja (1966) |
| HD6186 | Johnson | V | $4.26 \pm 0.05$ | Cousins (1962a) |
| HD6186 | Johnson | V | $4.27 \pm 0.05$ | Jennens & Helfer (1975) |
| HD6186 | Johnson | V | $4.28 \pm 0.05$ | Johnson et al. (1966) |
| HD6186 | Johnson | V | $4.28 \pm 0.05$ | Gutierrez-Moreno & et al. (1966) |
| HD6186 | Johnson | V | $4.28 \pm 0.05$ | Ducati (2002) |
| HD6186 | Johnson | V | $4.29 \pm 0.05$ | Argue (1966) |
| HD6186 | Johnson | V | $4.41 \pm 0.07$ | This work |
| HD6186 | 13c | m58 | $4.07 \pm 0.05$ | Johnson & Mitchell (1995) |
| HD6186 | Geneva | G | $5.27 \pm 0.08$ | Golay (1972) |
| HD6186 | 13c | m63 | $3.74 \pm 0.05$ | Johnson & Mitchell (1995) |
| HD6186 | Vilnius | S | $3.54 \pm 0.05$ | Kakaras et al. (1968) |
| HD6186 | Vilnius | S | $3.57 \pm 0.05$ | Zdanavicius et al. (1969) |
| HD6186 | Vilnius | S | $3.59 \pm 0.05$ | Jasevicius et al. (1990) |
| HD6186 | WBVR | R | $3.55 \pm 0.05$ | Kornilov et al. (1991) |
| HD6186 | KronComet | Rc | $3.27 \pm 0.02$ | This work |
| HD6186 | 13c | m72 | $3.47 \pm 0.05$ | Johnson & Mitchell (1995) |
| HD6186 | 13c | m80 | $3.25 \pm 0.05$ | Johnson & Mitchell (1995) |
| HD6186 | 13c | m86 | $3.15 \pm 0.05$ | Johnson & Mitchell (1995) |
| HD6186 | 13c | m99 | $3.00 \pm 0.05$ | Johnson & Mitchell (1995) |
| HD6186 | 13c | m110 | $2.87 \pm 0.05$ | Johnson & Mitchell (1995) |
| HD6186 | 1250 | 310 | $151.90 \pm 6.50$ | Smith et al. (2004) |
| HD6186 | Johnson | J | $2.60 \pm 0.05$ | Johnson et al. (1966) |
| HD6186 | Johnson | J | $2.60 \pm 0.05$ | Ducati (2002) |
| HD6186 | Johnson | J | $2.60 \pm 0.05$ | Shenavrin et al. (2011) |
| HD6186 | Johnson | H | $2.12 \pm 0.05$ | Shenavrin et al. (2011) |
| HD6186 | 2200 | 361 | $101.90 \pm 5.30$ | Smith et al. (2004) |
| HD6186 | Johnson | K | $1.99 \pm 0.05$ | Neugebauer & Leighton (1969) |
| HD6186 | Johnson | K | $2.00 \pm 0.05$ | Johnson et al. (1966) |
| HD6186 | Johnson | K | $2.00 \pm 0.05$ | Ducati (2002) |
| HD6186 | Johnson | K | $2.00 \pm 0.05$ | Shenavrin et al. (2011) |
| HD6186 | 3500 | 898 | $46.90 \pm 4.50$ | Smith et al. (2004) |
| HD6186 | 4900 | 712 | $23.40 \pm 5.20$ | Smith et al. (2004) |
| HD6186 | 12000 | 6384 | $-7.90 \pm 35.10$ | Smith et al. (2004) |
| HD6262 | KronComet | C2 | $7.53 \pm 0.03$ | This work |
| HD6262 | KronComet | Gc | $7.33 \pm 0.02$ | This work |
| HD6262 | WBVR | V | $6.99 \pm 0.05$ | Kornilov et al. (1991) |
| HD6262 | Johnson | V | $7.08 \pm 0.01$ | Oja (1986) |

**Table 21** *(continued)*

| Star ID | System/Wvlen | Band/Bandpass | Value | Reference |
|---|---|---|---|---|
| HD6262 | Johnson | V | $7.08 \pm 0.05$ | Oja (1986) |
| HD6262 | Johnson | V | $7.25 \pm 0.04$ | This work |
| HD6262 | WBVR | R | $5.39 \pm 0.05$ | Kornilov et al. (1991) |
| HD6262 | KronComet | Rc | $5.76 \pm 0.03$ | This work |
| HD6262 | 2200 | 361 | $82.60 \pm 10.60$ | Smith et al. (2004) |
| HD6262 | Johnson | K | $2.18 \pm 0.08$ | Neugebauer & Leighton (1969) |
| HD6262 | 3500 | 898 | $39.50 \pm 10.50$ | Smith et al. (2004) |
| HD6262 | 4900 | 712 | $16.90 \pm 6.90$ | Smith et al. (2004) |
| HD6262 | 12000 | 6384 | $1.50 \pm 18.40$ | Smith et al. (2004) |
| HD6409 | KronComet | COp | $9.55 \pm 0.13$ | This work |
| HD6409 | KronComet | Bc | $8.87 \pm 0.03$ | This work |
| HD6409 | KronComet | C2 | $7.70 \pm 0.03$ | This work |
| HD6409 | KronComet | Gc | $7.57 \pm 0.02$ | This work |
| HD6409 | Johnson | V | $7.53 \pm 0.02$ | This work |
| HD6409 | KronComet | Rc | $6.01 \pm 0.01$ | This work |
| HD6409 | 1250 | 310 | $71.30 \pm 9.30$ | Smith et al. (2004) |
| HD6409 | Johnson | J | $3.28 \pm 0.05$ | Chen et al. (1998) |
| HD6409 | Johnson | H | $2.56 \pm 0.05$ | Chen et al. (1998) |
| HD6409 | 2200 | 361 | $75.00 \pm 7.40$ | Smith et al. (2004) |
| HD6409 | Johnson | K | $2.34 \pm 0.09$ | Neugebauer & Leighton (1969) |
| HD6409 | 3500 | 898 | $37.70 \pm 5.00$ | Smith et al. (2004) |
| HD6409 | 4900 | 712 | $15.80 \pm 5.50$ | Smith et al. (2004) |
| HD6409 | 12000 | 6384 | $3.10 \pm 22.60$ | Smith et al. (2004) |
| HD7000 | KronComet | COp | $10.43 \pm 0.13$ | This work |
| HD7000 | KronComet | Bc | $9.89 \pm 0.03$ | This work |
| HD7000 | KronComet | C2 | $8.56 \pm 0.02$ | This work |
| HD7000 | KronComet | Gc | $8.53 \pm 0.02$ | This work |
| HD7000 | Johnson | V | $8.49 \pm 0.04$ | This work |
| HD7000 | KronComet | Rc | $6.91 \pm 0.02$ | This work |
| HD7000 | 1250 | 310 | $49.90 \pm 8.10$ | Smith et al. (2004) |
| HD7000 | 2200 | 361 | $56.00 \pm 8.80$ | Smith et al. (2004) |
| HD7000 | Johnson | K | $2.60 \pm 0.09$ | Neugebauer & Leighton (1969) |
| HD7000 | 3500 | 898 | $27.60 \pm 6.90$ | Smith et al. (2004) |
| HD7000 | 4900 | 712 | $12.90 \pm 5.00$ | Smith et al. (2004) |
| HD7000 | 12000 | 6384 | $-0.30 \pm 21.30$ | Smith et al. (2004) |
| HD7087 | 13c | m33 | $6.55 \pm 0.05$ | Johnson & Mitchell (1995) |
| HD7087 | Geneva | U | $7.04 \pm 0.08$ | Golay (1972) |
| HD7087 | 13c | m35 | $6.36 \pm 0.05$ | Johnson & Mitchell (1995) |
| HD7087 | DDO | m35 | $7.83 \pm 0.05$ | Clariá et al. (2008) |
| HD7087 | WBVR | W | $6.37 \pm 0.05$ | Kornilov et al. (1991) |
| HD7087 | Johnson | U | $6.49 \pm 0.05$ | Argue (1966) |

**Table 21** (continued)

| Star ID | System/Wvlen | Band/Bandpass | Value | Reference |
|---------|--------------|---------------|-------|-----------|
| HD7087 | Johnson | U | $6.50 \pm 0.05$ | Johnson et al. (1966) |
| HD7087 | Johnson | U | $6.51 \pm 0.05$ | Johnson et al. (1966) |
| HD7087 | Johnson | U | $6.51 \pm 0.05$ | Mermilliod (1986) |
| HD7087 | 13c | m37 | $6.42 \pm 0.05$ | Johnson & Mitchell (1995) |
| HD7087 | DDO | m38 | $6.77 \pm 0.05$ | Clariá et al. (2008) |
| HD7087 | 13c | m40 | $6.27 \pm 0.05$ | Johnson & Mitchell (1995) |
| HD7087 | Geneva | B1 | $6.30 \pm 0.08$ | Golay (1972) |
| HD7087 | DDO | m41 | $7.33 \pm 0.05$ | Clariá et al. (2008) |
| HD7087 | Oja | m41 | $6.89 \pm 0.05$ | Häggkvist & Oja (1970) |
| HD7087 | DDO | m42 | $7.08 \pm 0.05$ | Clariá et al. (2008) |
| HD7087 | Oja | m42 | $6.64 \pm 0.05$ | Häggkvist & Oja (1970) |
| HD7087 | Geneva | B | $5.00 \pm 0.08$ | Golay (1972) |
| HD7087 | WBVR | B | $5.72 \pm 0.05$ | Kornilov et al. (1991) |
| HD7087 | Johnson | B | $5.66 \pm 0.05$ | Argue (1966) |
| HD7087 | Johnson | B | $5.68 \pm 0.05$ | Häggkvist & Oja (1966) |
| HD7087 | Johnson | B | $5.68 \pm 0.05$ | Moffett & Barnes (1979) |
| HD7087 | Johnson | B | $5.69 \pm 0.05$ | Johnson et al. (1966) |
| HD7087 | Johnson | B | $5.69 \pm 0.05$ | Mermilliod (1986) |
| HD7087 | Geneva | B2 | $6.14 \pm 0.08$ | Golay (1972) |
| HD7087 | 13c | m45 | $5.39 \pm 0.05$ | Johnson & Mitchell (1995) |
| HD7087 | DDO | m45 | $6.23 \pm 0.05$ | Clariá et al. (2008) |
| HD7087 | Oja | m45 | $5.41 \pm 0.05$ | Häggkvist & Oja (1970) |
| HD7087 | DDO | m48 | $5.03 \pm 0.05$ | Clariá et al. (2008) |
| HD7087 | 13c | m52 | $4.90 \pm 0.05$ | Johnson & Mitchell (1995) |
| HD7087 | Geneva | V1 | $5.44 \pm 0.08$ | Golay (1972) |
| HD7087 | WBVR | V | $4.67 \pm 0.05$ | Kornilov et al. (1991) |
| HD7087 | Geneva | V | $4.67 \pm 0.08$ | Golay (1972) |
| HD7087 | Johnson | V | $4.64 \pm 0.05$ | Argue (1966) |
| HD7087 | Johnson | V | $4.66 \pm 0.05$ | Häggkvist & Oja (1966) |
| HD7087 | Johnson | V | $4.66 \pm 0.05$ | Johnson et al. (1966) |
| HD7087 | Johnson | V | $4.66 \pm 0.05$ | Moffett & Barnes (1979) |
| HD7087 | Johnson | V | $4.67 \pm 0.05$ | Mermilliod (1986) |
| HD7087 | 13c | m58 | $4.42 \pm 0.05$ | Johnson & Mitchell (1995) |
| HD7087 | Geneva | G | $5.65 \pm 0.08$ | Golay (1972) |
| HD7087 | 13c | m63 | $4.14 \pm 0.05$ | Johnson & Mitchell (1995) |
| HD7087 | WBVR | R | $3.95 \pm 0.05$ | Kornilov et al. (1991) |
| HD7087 | 13c | m72 | $3.89 \pm 0.05$ | Johnson & Mitchell (1995) |
| HD7087 | 13c | m80 | $3.66 \pm 0.05$ | Johnson & Mitchell (1995) |
| HD7087 | 13c | m86 | $3.56 \pm 0.05$ | Johnson & Mitchell (1995) |
| HD7087 | 13c | m99 | $3.40 \pm 0.05$ | Johnson & Mitchell (1995) |
| HD7087 | 13c | m110 | $3.20 \pm 0.05$ | Johnson & Mitchell (1995) |

Table 21 *(continued)*

| Star ID | System/Wvlen | Band/Bandpass | Value | Reference |
|---------|--------------|---------------|-------|-----------|
| HD7087 | 1250 | 310 | $100.00 \pm 7.30$ | Smith et al. (2004) |
| HD7087 | 2200 | 361 | $66.70 \pm 9.20$ | Smith et al. (2004) |
| HD7087 | Johnson | K | $2.36 \pm 0.10$ | Neugebauer & Leighton (1969) |
| HD7087 | 3500 | 898 | $31.60 \pm 6.30$ | Smith et al. (2004) |
| HD7087 | 4900 | 712 | $16.20 \pm 4.80$ | Smith et al. (2004) |
| HD7087 | 12000 | 6384 | $0.70 \pm 23.10$ | Smith et al. (2004) |
| HD7318 | 13c | m33 | $6.60 \pm 0.05$ | Johnson & Mitchell (1995) |
| HD7318 | 13c | m35 | $6.42 \pm 0.05$ | Johnson & Mitchell (1995) |
| HD7318 | WBVR | W | $6.42 \pm 0.05$ | Kornilov et al. (1991) |
| HD7318 | Johnson | U | $6.53 \pm 0.05$ | Johnson (1964) |
| HD7318 | Johnson | U | $6.53 \pm 0.05$ | Johnson et al. (1966) |
| HD7318 | Johnson | U | $6.54 \pm 0.05$ | Lutz & Lutz (1977) |
| HD7318 | Johnson | U | $6.55 \pm 0.05$ | Argue (1966) |
| HD7318 | Johnson | U | $6.56 \pm 0.05$ | Mermilliod (1986) |
| HD7318 | Johnson | U | $6.57 \pm 0.05$ | Ducati (2002) |
| HD7318 | Johnson | U | $6.61 \pm 0.05$ | Fernie (1983) |
| HD7318 | 13c | m37 | $6.49 \pm 0.05$ | Johnson & Mitchell (1995) |
| HD7318 | 13c | m40 | $6.34 \pm 0.05$ | Johnson & Mitchell (1995) |
| HD7318 | Oja | m41 | $6.93 \pm 0.05$ | Häggkvist & Oja (1970) |
| HD7318 | Oja | m42 | $6.66 \pm 0.05$ | Häggkvist & Oja (1970) |
| HD7318 | KronComet | COp | $5.92 \pm 0.08$ | This work |
| HD7318 | WBVR | B | $5.74 \pm 0.05$ | Kornilov et al. (1991) |
| HD7318 | Johnson | B | $5.68 \pm 0.05$ | Argue (1966) |
| HD7318 | Johnson | B | $5.69 \pm 0.05$ | Johnson (1964) |
| HD7318 | Johnson | B | $5.69 \pm 0.05$ | Johnson et al. (1966) |
| HD7318 | Johnson | B | $5.69 \pm 0.05$ | Lutz & Lutz (1977) |
| HD7318 | Johnson | B | $5.69 \pm 0.05$ | Fernie (1983) |
| HD7318 | Johnson | B | $5.69 \pm 0.05$ | Ducati (2002) |
| HD7318 | Johnson | B | $5.70 \pm 0.05$ | Häggkvist & Oja (1966) |
| HD7318 | Johnson | B | $5.71 \pm 0.05$ | Mermilliod (1986) |
| HD7318 | KronComet | Bc | $5.48 \pm 0.02$ | This work |
| HD7318 | 13c | m45 | $5.41 \pm 0.05$ | Johnson & Mitchell (1995) |
| HD7318 | Oja | m45 | $5.43 \pm 0.05$ | Häggkvist & Oja (1970) |
| HD7318 | KronComet | C2 | $4.82 \pm 0.01$ | This work |
| HD7318 | 13c | m52 | $4.93 \pm 0.05$ | Johnson & Mitchell (1995) |
| HD7318 | KronComet | Gc | $4.72 \pm 0.02$ | This work |
| HD7318 | WBVR | V | $4.67 \pm 0.05$ | Kornilov et al. (1991) |
| HD7318 | Johnson | V | $4.64 \pm 0.05$ | Argue (1966) |
| HD7318 | Johnson | V | $4.64 \pm 0.05$ | Lutz & Lutz (1977) |
| HD7318 | Johnson | V | $4.65 \pm 0.05$ | Häggkvist & Oja (1966) |
| HD7318 | Johnson | V | $4.65 \pm 0.05$ | Fernie (1983) |

**Table 21** *(continued)*

| Star ID | System/Wvlen | Band/Bandpass | Value | Reference |
|---------|--------------|---------------|-------|-----------|
| HD7318 | Johnson | V | $4.66 \pm 0.05$ | Johnson (1964) |
| HD7318 | Johnson | V | $4.66 \pm 0.05$ | Johnson et al. (1966) |
| HD7318 | Johnson | V | $4.66 \pm 0.05$ | Ducati (2002) |
| HD7318 | Johnson | V | $4.68 \pm 0.08$ | This work |
| HD7318 | Johnson | V | $4.69 \pm 0.05$ | Mermilliod (1986) |
| HD7318 | 13c | m58 | $4.43 \pm 0.05$ | Johnson & Mitchell (1995) |
| HD7318 | 13c | m63 | $4.16 \pm 0.05$ | Johnson & Mitchell (1995) |
| HD7318 | WBVR | R | $3.94 \pm 0.05$ | Kornilov et al. (1991) |
| HD7318 | KronComet | Rc | $3.68 \pm 0.01$ | This work |
| HD7318 | 13c | m72 | $3.88 \pm 0.05$ | Johnson & Mitchell (1995) |
| HD7318 | 13c | m80 | $3.66 \pm 0.05$ | Johnson & Mitchell (1995) |
| HD7318 | 13c | m86 | $3.56 \pm 0.05$ | Johnson & Mitchell (1995) |
| HD7318 | 13c | m99 | $3.40 \pm 0.05$ | Johnson & Mitchell (1995) |
| HD7318 | 13c | m110 | $3.26 \pm 0.05$ | Johnson & Mitchell (1995) |
| HD7318 | 1250 | 310 | $107.90 \pm 8.20$ | Smith et al. (2004) |
| HD7318 | Johnson | J | $3.01 \pm 0.05$ | Johnson et al. (1966) |
| HD7318 | Johnson | J | $3.01 \pm 0.05$ | Voelcker (1975) |
| HD7318 | Johnson | J | $3.01 \pm 0.05$ | Ducati (2002) |
| HD7318 | Johnson | H | $2.53 \pm 0.05$ | Voelcker (1975) |
| HD7318 | Johnson | H | $2.53 \pm 0.05$ | Ducati (2002) |
| HD7318 | 2200 | 361 | $75.80 \pm 7.30$ | Smith et al. (2004) |
| HD7318 | Johnson | K | $2.37 \pm 0.08$ | Neugebauer & Leighton (1969) |
| HD7318 | Johnson | K | $2.38 \pm 0.05$ | Johnson et al. (1966) |
| HD7318 | Johnson | K | $2.38 \pm 0.05$ | Ducati (2002) |
| HD7318 | Johnson | L | $2.26 \pm 0.05$ | Ducati (2002) |
| HD7318 | 3500 | 898 | $31.70 \pm 11.40$ | Smith et al. (2004) |
| HD7318 | 4900 | 712 | $15.40 \pm 6.00$ | Smith et al. (2004) |
| HD7318 | 12000 | 6384 | $-3.90 \pm 21.10$ | Smith et al. (2004) |
| HD8126 | DDO | m35 | $9.66 \pm 0.05$ | McClure & Forrester (1981) |
| HD8126 | WBVR | W | $8.18 \pm 0.05$ | Kornilov et al. (1991) |
| HD8126 | DDO | m38 | $8.43 \pm 0.05$ | McClure & Forrester (1981) |
| HD8126 | DDO | m41 | $8.58 \pm 0.05$ | McClure & Forrester (1981) |
| HD8126 | Oja | m41 | $8.13 \pm 0.05$ | Häggkvist & Oja (1970) |
| HD8126 | DDO | m42 | $8.32 \pm 0.05$ | McClure & Forrester (1981) |
| HD8126 | Oja | m42 | $7.86 \pm 0.05$ | Häggkvist & Oja (1970) |
| HD8126 | KronComet | COp | $7.11 \pm 0.09$ | This work |
| HD8126 | WBVR | B | $6.68 \pm 0.05$ | Kornilov et al. (1991) |
| HD8126 | Johnson | B | $6.62 \pm 0.05$ | Haggkvist & Oja (1970) |
| HD8126 | KronComet | Bc | $6.41 \pm 0.02$ | This work |
| HD8126 | DDO | m45 | $7.10 \pm 0.05$ | McClure & Forrester (1981) |
| HD8126 | Oja | m45 | $6.30 \pm 0.05$ | Häggkvist & Oja (1970) |

**Table 21** *continued on next page*

Table 21 (continued)

| Star ID | System/Wvlen | Band/Bandpass | Value | Reference |
|---------|--------------|---------------|-------|-----------|
| HD8126 | DDO | m48 | $5.78 \pm 0.05$ | McClure & Forrester (1981) |
| HD8126 | KronComet | C2 | $5.62 \pm 0.01$ | This work |
| HD8126 | KronComet | Gc | $5.36 \pm 0.01$ | This work |
| HD8126 | WBVR | V | $5.24 \pm 0.05$ | Kornilov et al. (1991) |
| HD8126 | Johnson | V | $5.23 \pm 0.05$ | Häggkvist & Oja (1970) |
| HD8126 | Johnson | V | $5.26 \pm 0.07$ | This work |
| HD8126 | WBVR | R | $4.20 \pm 0.05$ | Kornilov et al. (1991) |
| HD8126 | KronComet | Rc | $3.95 \pm 0.01$ | This work |
| HD8126 | 1250 | 310 | $127.00 \pm 5.50$ | Smith et al. (2004) |
| HD8126 | 2200 | 361 | $110.60 \pm 5.50$ | Smith et al. (2004) |
| HD8126 | Johnson | K | $1.93 \pm 0.06$ | Neugebauer & Leighton (1969) |
| HD8126 | 3500 | 898 | $53.30 \pm 5.00$ | Smith et al. (2004) |
| HD8126 | 4900 | 712 | $25.10 \pm 4.50$ | Smith et al. (2004) |
| HD8126 | 12000 | 6384 | $-1.60 \pm 19.60$ | Smith et al. (2004) |
| HD9500 | WBVR | W | $10.39 \pm 0.05$ | Kornilov et al. (1991) |
| HD9500 | Johnson | U | $10.09 \pm 0.16$ | This work |
| HD9500 | Johnson | U | $10.48 \pm 0.05$ | Guetter & Hewitt (1984) |
| HD9500 | KronComet | COp | $9.12 \pm 0.08$ | This work |
| HD9500 | WBVR | B | $8.61 \pm 0.05$ | Kornilov et al. (1991) |
| HD9500 | Johnson | B | $8.48 \pm 0.02$ | This work |
| HD9500 | Johnson | B | $8.61 \pm 0.05$ | Guetter & Hewitt (1984) |
| HD9500 | KronComet | Bc | $8.42 \pm 0.03$ | This work |
| HD9500 | KronComet | C2 | $7.28 \pm 0.02$ | This work |
| HD9500 | KronComet | Gc | $7.13 \pm 0.03$ | This work |
| HD9500 | WBVR | V | $6.94 \pm 0.05$ | Kornilov et al. (1991) |
| HD9500 | Johnson | V | $7.00 \pm 0.05$ | Guetter & Hewitt (1984) |
| HD9500 | Johnson | V | $7.03 \pm 0.05$ | This work |
| HD9500 | WBVR | R | $5.29 \pm 0.05$ | Kornilov et al. (1991) |
| HD9500 | KronComet | Rc | $5.61 \pm 0.05$ | This work |
| HD9500 | 1250 | 310 | $100.40 \pm 9.70$ | Smith et al. (2004) |
| HD9500 | 2200 | 361 | $102.50 \pm 8.50$ | Smith et al. (2004) |
| HD9500 | Johnson | K | $1.87 \pm 0.07$ | Neugebauer & Leighton (1969) |
| HD9500 | 3500 | 898 | $52.40 \pm 5.60$ | Smith et al. (2004) |
| HD9500 | 4900 | 712 | $21.70 \pm 5.00$ | Smith et al. (2004) |
| HD9500 | 12000 | 6384 | $4.70 \pm 18.80$ | Smith et al. (2004) |
| HD9927 | DDO | m41 | $6.77 \pm 0.05$ | McClure & Forrester (1981) |
| HD9927 | Oja | m41 | $6.32 \pm 0.05$ | Häggkvist & Oja (1970) |
| HD9927 | DDO | m42 | $6.44 \pm 0.05$ | McClure & Forrester (1981) |
| HD9927 | Oja | m42 | $5.96 \pm 0.05$ | Häggkvist & Oja (1970) |
| HD9927 | WBVR | B | $4.90 \pm 0.05$ | Kornilov et al. (1991) |
| HD9927 | Johnson | B | $4.84 \pm 0.05$ | Johnson & Morgan (1953b) |

**Table 21** *(continued)*

| Star ID | System/Wvlen | Band/Bandpass | Value | Reference |
|---------|--------------|---------------|-------|-----------|
| HD9927 | Johnson | B | $4.84 \pm 0.05$ | Häggkvist & Oja (1966) |
| HD9927 | Johnson | B | $4.84 \pm 0.05$ | Argue (1966) |
| HD9927 | Johnson | B | $4.85 \pm 0.05$ | Johnson et al. (1966) |
| HD9927 | Johnson | B | $4.85 \pm 0.05$ | Moffett & Barnes (1979) |
| HD9927 | Johnson | B | $4.85 \pm 0.05$ | Mermilliod (1986) |
| HD9927 | Johnson | B | $4.85 \pm 0.05$ | Ducati (2002) |
| HD9927 | Johnson | B | $4.86 \pm 0.05$ | Johnson (1964) |
| HD9927 | Johnson | B | $4.86 \pm 0.05$ | Jennens & Helfer (1975) |
| HD9927 | 13c | m45 | $4.46 \pm 0.05$ | Johnson & Mitchell (1995) |
| HD9927 | DDO | m45 | $5.35 \pm 0.05$ | McClure & Forrester (1981) |
| HD9927 | Oja | m45 | $4.52 \pm 0.05$ | Häggkvist & Oja (1970) |
| HD9927 | Vilnius | Y | $4.53 \pm 0.05$ | Kazlauskas et al. (2005) |
| HD9927 | DDO | m48 | $4.05 \pm 0.05$ | McClure & Forrester (1981) |
| HD9927 | Vilnius | Z | $4.01 \pm 0.05$ | Kazlauskas et al. (2005) |
| HD9927 | 13c | m52 | $3.93 \pm 0.05$ | Johnson & Mitchell (1995) |
| HD9927 | WBVR | V | $3.59 \pm 0.05$ | Kornilov et al. (1991) |
| HD9927 | Vilnius | V | $3.60 \pm 0.05$ | Kazlauskas et al. (2005) |
| HD9927 | Johnson | V | $3.56 \pm 0.05$ | Johnson & Morgan (1953b) |
| HD9927 | Johnson | V | $3.56 \pm 0.05$ | Argue (1966) |
| HD9927 | Johnson | V | $3.57 \pm 0.05$ | Johnson et al. (1966) |
| HD9927 | Johnson | V | $3.57 \pm 0.05$ | Ducati (2002) |
| HD9927 | Johnson | V | $3.58 \pm 0.05$ | Johnson (1964) |
| HD9927 | Johnson | V | $3.58 \pm 0.05$ | Häggkvist & Oja (1966) |
| HD9927 | Johnson | V | $3.58 \pm 0.05$ | Jennens & Helfer (1975) |
| HD9927 | Johnson | V | $3.58 \pm 0.05$ | Moffett & Barnes (1979) |
| HD9927 | Johnson | V | $3.58 \pm 0.05$ | Mermilliod (1986) |
| HD9927 | 13c | m58 | $3.28 \pm 0.05$ | Johnson & Mitchell (1995) |
| HD9927 | 13c | m63 | $2.93 \pm 0.05$ | Johnson & Mitchell (1995) |
| HD9927 | Vilnius | S | $2.73 \pm 0.05$ | Kazlauskas et al. (2005) |
| HD9927 | WBVR | R | $2.69 \pm 0.05$ | Kornilov et al. (1991) |
| HD9927 | 13c | m72 | $2.60 \pm 0.05$ | Johnson & Mitchell (1995) |
| HD9927 | 13c | m80 | $2.33 \pm 0.05$ | Johnson & Mitchell (1995) |
| HD9927 | 13c | m86 | $2.22 \pm 0.05$ | Johnson & Mitchell (1995) |
| HD9927 | 13c | m99 | $2.03 \pm 0.05$ | Johnson & Mitchell (1995) |
| HD9927 | 13c | m110 | $1.82 \pm 0.05$ | Johnson & Mitchell (1995) |
| HD9927 | 1250 | 310 | $422.20 \pm 24.80$ | Smith et al. (2004) |
| HD9927 | Johnson | J | $1.44 \pm 0.05$ | Alonso et al. (1998) |
| HD9927 | Johnson | J | $1.46 \pm 0.05$ | Selby et al. (1988) |
| HD9927 | Johnson | J | $1.46 \pm 0.05$ | Blackwell et al. (1990) |
| HD9927 | Johnson | J | $1.48 \pm 0.05$ | Ducati (2002) |
| HD9927 | Johnson | J | $1.53 \pm 0.05$ | Johnson et al. (1966) |

**Table 21** *continued on next page*

**Table 21** *(continued)*

| Star ID | System/Wvlen | Band/Bandpass | Value | Reference |
|---------|-------------|---------------|-------|-----------|
| HD9927 | Johnson | J | $1.53 \pm 0.05$ | Shenavrin et al. (2011) |
| HD9927 | Johnson | H | $0.88 \pm 0.05$ | Alonso et al. (1998) |
| HD9927 | Johnson | H | $0.95 \pm 0.05$ | Shenavrin et al. (2011) |
| HD9927 | 2200 | 361 | $318.60 \pm 17.30$ | Smith et al. (2004) |
| HD9927 | Johnson | K | $0.74 \pm 0.05$ | Ducati (2002) |
| HD9927 | Johnson | K | $0.79 \pm 0.04$ | Neugebauer & Leighton (1969) |
| HD9927 | Johnson | K | $0.79 \pm 0.05$ | Johnson et al. (1966) |
| HD9927 | Johnson | K | $0.79 \pm 0.05$ | Shenavrin et al. (2011) |
| HD9927 | 3500 | 898 | $162.40 \pm 10.50$ | Smith et al. (2004) |
| HD9927 | 4900 | 712 | $72.60 \pm 6.20$ | Smith et al. (2004) |
| HD9927 | 12000 | 6384 | $19.10 \pm 19.40$ | Smith et al. (2004) |
| HD10380 | 13c | m33 | $7.64 \pm 0.05$ | Johnson & Mitchell (1995) |
| HD10380 | Vilnius | U | $9.20 \pm 0.05$ | Straizys et al. (1989a) |
| HD10380 | Vilnius | U | $9.31 \pm 0.05$ | Zdanavicius et al. (1969) |
| HD10380 | 13c | m35 | $7.26 \pm 0.05$ | Johnson & Mitchell (1995) |
| HD10380 | DDO | m35 | $8.75 \pm 0.05$ | Mermilliod & Nitschelm (1989) |
| HD10380 | DDO | m35 | $8.79 \pm 0.05$ | McClure & Forrester (1981) |
| HD10380 | Stromgren | u | $8.68 \pm 0.08$ | Olsen (1993) |
| HD10380 | Stromgren | u | $8.68 \pm 0.08$ | Hauck & Mermilliod (1998) |
| HD10380 | Johnson | U | $7.31 \pm 0.05$ | Mermilliod (1986) |
| HD10380 | Johnson | U | $7.35 \pm 0.05$ | Argue (1966) |
| HD10380 | Johnson | U | $7.36 \pm 0.05$ | Johnson et al. (1966) |
| HD10380 | Johnson | U | $7.36 \pm 0.05$ | Jennens & Helfer (1975) |
| HD10380 | Johnson | U | $7.36 \pm 0.05$ | Ducati (2002) |
| HD10380 | Johnson | U | $7.38 \pm 0.05$ | Cousins & Stoy (1962) |
| HD10380 | Johnson | U | $7.38 \pm 0.05$ | Johnson (1964) |
| HD10380 | Oja | m41 | $7.26 \pm 0.05$ | Häggkvist & Oja (1970) |
| HD10380 | DDO | m42 | $7.44 \pm 0.05$ | Mermilliod & Nitschelm (1989) |
| HD10380 | DDO | m42 | $7.47 \pm 0.05$ | McClure & Forrester (1981) |
| HD10380 | Oja | m42 | $6.99 \pm 0.05$ | Häggkvist & Oja (1970) |
| HD10380 | KronComet | COp | $6.28 \pm 0.09$ | This work |
| HD10380 | WBVR | B | $5.85 \pm 0.05$ | Kornilov et al. (1991) |
| HD10380 | Johnson | B | $5.78 \pm 0.05$ | Mermilliod (1986) |
| HD10380 | Johnson | B | $5.80 \pm 0.05$ | Cousins & Stoy (1962) |
| HD10380 | Johnson | B | $5.80 \pm 0.05$ | Häggkvist & Oja (1966) |
| HD10380 | Johnson | B | $5.80 \pm 0.05$ | Johnson et al. (1966) |
| HD10380 | Johnson | B | $5.80 \pm 0.05$ | Argue (1966) |
| HD10380 | Johnson | B | $5.81 \pm 0.05$ | Jennens & Helfer (1975) |
| HD10380 | Johnson | B | $5.81 \pm 0.05$ | Moffett & Barnes (1979) |
| HD10380 | Johnson | B | $5.81 \pm 0.05$ | Ducati (2002) |
| HD10380 | Johnson | B | $5.83 \pm 0.05$ | Glass (1974) |

**Table 21** *(continued)*

| Star ID | System/Wvlen | Band/Bandpass | Value | Reference |
|---------|--------------|---------------|-------|-----------|
| HD10380 | Johnson | B | $5.84 \pm 0.05$ | Johnson (1964) |
| HD10380 | KronComet | Bc | $5.63 \pm 0.02$ | This work |
| HD10380 | 13c | m45 | $5.39 \pm 0.05$ | Johnson & Mitchell (1995) |
| HD10380 | DDO | m45 | $6.29 \pm 0.05$ | Mermilliod & Nitschelm (1989) |
| HD10380 | DDO | m45 | $6.30 \pm 0.05$ | McClure & Forrester (1981) |
| HD10380 | Oja | m45 | $5.48 \pm 0.05$ | Häggkvist & Oja (1970) |
| HD10380 | Vilnius | Y | $5.40 \pm 0.05$ | Straizys et al. (1989a) |
| HD10380 | Vilnius | Y | $5.43 \pm 0.05$ | Zdanavicius et al. (1969) |
| HD10380 | Stromgren | b | $5.27 \pm 0.08$ | Olsen (1993) |
| HD10380 | Stromgren | b | $5.27 \pm 0.08$ | Hauck & Mermilliod (1998) |
| HD10380 | DDO | m48 | $4.97 \pm 0.05$ | Mermilliod & Nitschelm (1989) |
| HD10380 | DDO | m48 | $4.98 \pm 0.05$ | McClure & Forrester (1981) |
| HD10380 | KronComet | C2 | $4.81 \pm 0.01$ | This work |
| HD10380 | Vilnius | Z | $4.89 \pm 0.05$ | Zdanavicius et al. (1969) |
| HD10380 | Vilnius | Z | $4.89 \pm 0.05$ | Straizys et al. (1989a) |
| HD10380 | 13c | m52 | $4.82 \pm 0.05$ | Johnson & Mitchell (1995) |
| HD10380 | KronComet | Gc | $4.58 \pm 0.01$ | This work |
| HD10380 | WBVR | V | $4.44 \pm 0.05$ | Kornilov et al. (1991) |
| HD10380 | Vilnius | V | $4.44 \pm 0.05$ | Zdanavicius et al. (1969) |
| HD10380 | Vilnius | V | $4.44 \pm 0.05$ | Straizys et al. (1989a) |
| HD10380 | Stromgren | y | $4.44 \pm 0.08$ | Olsen (1993) |
| HD10380 | Stromgren | y | $4.44 \pm 0.08$ | Hauck & Mermilliod (1998) |
| HD10380 | Johnson | V | $4.42 \pm 0.05$ | Mermilliod (1986) |
| HD10380 | Johnson | V | $4.43 \pm 0.05$ | Cousins & Stoy (1962) |
| HD10380 | Johnson | V | $4.44 \pm 0.05$ | Häggkvist & Oja (1966) |
| HD10380 | Johnson | V | $4.44 \pm 0.05$ | Johnson et al. (1966) |
| HD10380 | Johnson | V | $4.44 \pm 0.05$ | Argue (1966) |
| HD10380 | Johnson | V | $4.44 \pm 0.05$ | Jennens & Helfer (1975) |
| HD10380 | Johnson | V | $4.44 \pm 0.05$ | Moffett & Barnes (1979) |
| HD10380 | Johnson | V | $4.44 \pm 0.05$ | Ducati (2002) |
| HD10380 | Johnson | V | $4.45 \pm 0.05$ | Johnson (1964) |
| HD10380 | Johnson | V | $4.45 \pm 0.05$ | Glass (1974) |
| HD10380 | Johnson | V | $4.49 \pm 0.07$ | This work |
| HD10380 | 13c | m58 | $4.12 \pm 0.05$ | Johnson & Mitchell (1995) |
| HD10380 | 13c | m63 | $3.72 \pm 0.05$ | Johnson & Mitchell (1995) |
| HD10380 | Vilnius | S | $3.46 \pm 0.05$ | Straizys et al. (1989a) |
| HD10380 | Vilnius | S | $3.50 \pm 0.05$ | Zdanavicius et al. (1969) |
| HD10380 | WBVR | R | $3.44 \pm 0.05$ | Kornilov et al. (1991) |
| HD10380 | KronComet | Rc | $3.18 \pm 0.01$ | This work |
| HD10380 | 13c | m72 | $3.38 \pm 0.05$ | Johnson & Mitchell (1995) |
| HD10380 | 13c | m80 | $3.06 \pm 0.05$ | Johnson & Mitchell (1995) |

**Table 21** *(continued)*

| Star ID | System/Wvlen | Band/Bandpass | Value | Reference |
|---------|--------------|---------------|-------|-----------|
| HD10380 | 13c | m86 | 2.91 ± 0.05 | Johnson & Mitchell (1995) |
| HD10380 | 13c | m99 | 2.68 ± 0.05 | Johnson & Mitchell (1995) |
| HD10380 | 13c | m110 | 2.44 ± 0.05 | Johnson & Mitchell (1995) |
| HD10380 | 1250 | 310 | 235.00 ± 7.10 | Smith et al. (2004) |
| HD10380 | Johnson | J | 2.04 ± 0.05 | Alonso et al. (1998) |
| HD10380 | Johnson | J | 2.13 ± 0.05 | Johnson et al. (1966) |
| HD10380 | Johnson | J | 2.13 ± 0.05 | Glass (1974) |
| HD10380 | Johnson | J | 2.14 ± 0.05 | Ducati (2002) |
| HD10380 | Johnson | J | 2.17 ± 0.05 | Carter (1990) |
| HD10380 | Johnson | H | 1.38 ± 0.05 | Carter (1990) |
| HD10380 | Johnson | H | 1.39 ± 0.05 | Alonso et al. (1998) |
| HD10380 | Johnson | H | 1.41 ± 0.05 | Ducati (2002) |
| HD10380 | Johnson | H | 1.44 ± 0.05 | Glass (1974) |
| HD10380 | 2200 | 361 | 195.60 ± 6.30 | Smith et al. (2004) |
| HD10380 | Johnson | K | 1.24 ± 0.05 | Johnson et al. (1966) |
| HD10380 | Johnson | K | 1.25 ± 0.05 | Ducati (2002) |
| HD10380 | Johnson | K | 1.26 ± 0.03 | Neugebauer & Leighton (1969) |
| HD10380 | Johnson | K | 1.26 ± 0.05 | Glass (1974) |
| HD10380 | Johnson | K | 1.27 ± 0.05 | Glass (1974) |
| HD10380 | Johnson | L | 1.12 ± 0.05 | Glass (1974) |
| HD10380 | Johnson | L | 1.14 ± 0.05 | Ducati (2002) |
| HD10380 | Johnson | L | 1.18 ± 0.05 | Glass (1974) |
| HD10380 | 3500 | 898 | 95.80 ± 4.70 | Smith et al. (2004) |
| HD10380 | 4900 | 712 | 43.70 ± 5.30 | Smith et al. (2004) |
| HD10380 | 12000 | 6384 | 5.70 ± 24.70 | Smith et al. (2004) |
| HD11928 | Oja | m41 | 8.91 ± 0.05 | Häggkvist & Oja (1970) |
| HD11928 | Oja | m42 | 8.73 ± 0.05 | Häggkvist & Oja (1970) |
| HD11928 | KronComet | COp | 8.03 ± 0.07 | This work |
| HD11928 | WBVR | B | 7.48 ± 0.05 | Kornilov et al. (1991) |
| HD11928 | Johnson | B | 7.38 ± 0.05 | This work |
| HD11928 | Johnson | B | 7.42 ± 0.05 | Haggkvist & Oja (1970) |
| HD11928 | Johnson | B | 7.42 ± 0.05 | Mermilliod (1986) |
| HD11928 | KronComet | Bc | 7.26 ± 0.01 | This work |
| HD11928 | Oja | m45 | 7.08 ± 0.05 | Häggkvist & Oja (1970) |
| HD11928 | Vilnius | Y | 6.89 ± 0.05 | Zdanavicius et al. (1972) |
| HD11928 | KronComet | C2 | 6.27 ± 0.01 | This work |
| HD11928 | Vilnius | Z | 6.22 ± 0.05 | Zdanavicius et al. (1972) |
| HD11928 | KronComet | Gc | 6.01 ± 0.02 | This work |
| HD11928 | WBVR | V | 5.86 ± 0.05 | Kornilov et al. (1991) |
| HD11928 | Vilnius | V | 5.68 ± 0.05 | Zdanavicius et al. (1972) |
| HD11928 | Johnson | V | 5.82 ± 0.05 | Haggkvist & Oja (1970) |

**Table 21** *continued on next page*

**Table 21** *(continued)*

| Star ID | System/Wvlen | Band/Bandpass | Value | Reference |
|---------|--------------|---------------|-------|-----------|
| HD11928 | Johnson | V | $5.85 \pm 0.05$ | Mermilliod (1986) |
| HD11928 | Johnson | V | $5.93 \pm 0.04$ | This work |
| HD11928 | Vilnius | S | $4.52 \pm 0.05$ | Zdanavicius et al. (1972) |
| HD11928 | WBVR | R | $4.40 \pm 0.05$ | Kornilov et al. (1991) |
| HD11928 | KronComet | Rc | $4.52 \pm 0.02$ | This work |
| HD11928 | 1250 | 310 | $169.20 \pm 6.60$ | Smith et al. (2004) |
| HD11928 | 2200 | 361 | $165.50 \pm 6.20$ | Smith et al. (2004) |
| HD11928 | Johnson | K | $1.43 \pm 0.05$ | Neugebauer & Leighton (1969) |
| HD11928 | 3500 | 898 | $84.50 \pm 4.20$ | Smith et al. (2004) |
| HD11928 | 4900 | 712 | $34.90 \pm 5.00$ | Smith et al. (2004) |
| HD11928 | 12000 | 6384 | $7.40 \pm 21.80$ | Smith et al. (2004) |
| HD12274 | Oja | m42 | $6.93 \pm 0.05$ | Häggkvist & Oja (1970) |
| HD12274 | WBVR | B | $5.62 \pm 0.05$ | Kornilov et al. (1991) |
| HD12274 | Johnson | B | $5.52 \pm 0.05$ | Mermilliod (1986) |
| HD12274 | Johnson | B | $5.55 \pm 0.05$ | Westerlund (1962) |
| HD12274 | Johnson | B | $5.57 \pm 0.05$ | Cousins & Stoy (1962) |
| HD12274 | Johnson | B | $5.57 \pm 0.05$ | Johnson et al. (1966) |
| HD12274 | Johnson | B | $5.58 \pm 0.05$ | Ducati (2002) |
| HD12274 | Johnson | B | $5.59 \pm 0.05$ | Cousins & Stoy (1962) |
| HD12274 | Johnson | B | $5.62 \pm 0.05$ | Johnson (1964) |
| HD12274 | Johnson | B | $5.62 \pm 0.05$ | Glass (1974) |
| HD12274 | 13c | m45 | $5.16 \pm 0.05$ | Johnson & Mitchell (1995) |
| HD12274 | Oja | m45 | $5.22 \pm 0.05$ | Häggkvist & Oja (1970) |
| HD12274 | 13c | m52 | $4.46 \pm 0.05$ | Johnson & Mitchell (1995) |
| HD12274 | WBVR | V | $4.00 \pm 0.05$ | Kornilov et al. (1991) |
| HD12274 | Johnson | V | $3.95 \pm 0.05$ | Mermilliod (1986) |
| HD12274 | Johnson | V | $3.98 \pm 0.05$ | Westerlund (1962) |
| HD12274 | Johnson | V | $3.98 \pm 0.05$ | Cousins & Stoy (1962) |
| HD12274 | Johnson | V | $4.00 \pm 0.05$ | Cousins & Stoy (1962) |
| HD12274 | Johnson | V | $4.01 \pm 0.05$ | Johnson et al. (1966) |
| HD12274 | Johnson | V | $4.02 \pm 0.05$ | Glass (1974) |
| HD12274 | Johnson | V | $4.02 \pm 0.05$ | Ducati (2002) |
| HD12274 | Johnson | V | $4.06 \pm 0.05$ | Johnson (1964) |
| HD12274 | 13c | m58 | $3.65 \pm 0.05$ | Johnson & Mitchell (1995) |
| HD12274 | 13c | m63 | $3.22 \pm 0.05$ | Johnson & Mitchell (1995) |
| HD12274 | WBVR | R | $2.71 \pm 0.05$ | Kornilov et al. (1991) |
| HD12274 | 13c | m72 | $2.67 \pm 0.05$ | Johnson & Mitchell (1995) |
| HD12274 | 13c | m80 | $2.27 \pm 0.05$ | Johnson & Mitchell (1995) |
| HD12274 | 13c | m86 | $2.07 \pm 0.05$ | Johnson & Mitchell (1995) |
| HD12274 | 13c | m99 | $1.75 \pm 0.05$ | Johnson & Mitchell (1995) |
| HD12274 | 13c | m110 | $1.46 \pm 0.05$ | Johnson & Mitchell (1995) |

Table 21 (continued)

| Star ID | System/Wvlen | Band/Bandpass | Value | Reference |
|---------|--------------|---------------|-------|-----------|
| HD12274 | Johnson | J | $1.11 \pm 0.05$ | Johnson et al. (1966) |
| HD12274 | Johnson | J | $1.11 \pm 0.05$ | Glass (1974) |
| HD12274 | Johnson | J | $1.12 \pm 0.05$ | Glass (1974) |
| HD12274 | Johnson | J | $1.12 \pm 0.05$ | Ducati (2002) |
| HD12274 | Johnson | J | $1.13 \pm 0.05$ | Carter (1990) |
| HD12274 | Johnson | J | $1.13 \pm 0.05$ | Feast et al. (1990) |
| HD12274 | Johnson | H | $0.26 \pm 0.05$ | Carter (1990) |
| HD12274 | Johnson | H | $0.26 \pm 0.05$ | Feast et al. (1990) |
| HD12274 | Johnson | H | $0.27 \pm 0.05$ | Ducati (2002) |
| HD12274 | Johnson | H | $0.30 \pm 0.05$ | Glass (1974) |
| HD12274 | 2200 | 361 | $647.90 \pm 83.90$ | Smith et al. (2004) |
| HD12274 | Johnson | K | $0.09 \pm 0.05$ | Glass (1974) |
| HD12274 | Johnson | K | $0.10 \pm 0.05$ | Johnson et al. (1966) |
| HD12274 | Johnson | K | $0.10 \pm 0.05$ | Glass (1974) |
| HD12274 | Johnson | K | $0.10 \pm 0.05$ | Ducati (2002) |
| HD12274 | Johnson | K | $0.17 \pm 0.05$ | Neugebauer & Leighton (1969) |
| HD12274 | Johnson | L | $-0.01 \pm 0.05$ | Glass (1974) |
| HD12274 | Johnson | L | $-0.03 \pm 0.05$ | Glass (1974) |
| HD12274 | Johnson | L | $-0.05 \pm 0.05$ | Ducati (2002) |
| HD12274 | 3500 | 898 | $364.10 \pm 39.60$ | Smith et al. (2004) |
| HD12274 | 4900 | 712 | $144.00 \pm 21.80$ | Smith et al. (2004) |
| HD12274 | 12000 | 6384 | $37.20 \pm 20.00$ | Smith et al. (2004) |
| HD12929 | Geneva | B1 | $3.91 \pm 0.08$ | Golay (1972) |
| HD12929 | DDO | m41 | $4.89 \pm 0.05$ | McClure & Forrester (1981) |
| HD12929 | DDO | m41 | $4.89 \pm 0.05$ | Mermilliod & Nitschelm (1989) |
| HD12929 | Oja | m41 | $4.45 \pm 0.05$ | Häggkvist & Oja (1970) |
| HD12929 | DDO | m42 | $4.65 \pm 0.05$ | McClure & Forrester (1981) |
| HD12929 | DDO | m42 | $4.65 \pm 0.05$ | Mermilliod & Nitschelm (1989) |
| HD12929 | Oja | m42 | $4.22 \pm 0.05$ | Häggkvist & Oja (1970) |
| HD12929 | Geneva | B | $2.51 \pm 0.08$ | Golay (1972) |
| HD12929 | WBVR | B | $3.20 \pm 0.05$ | Kornilov et al. (1991) |
| HD12929 | Johnson | B | $3.14 \pm 0.05$ | Johnson & Morgan (1953b) |
| HD12929 | Johnson | B | $3.14 \pm 0.05$ | Mermilliod (1986) |
| HD12929 | Johnson | B | $3.15 \pm 0.05$ | Sharpless (1952) |
| HD12929 | Johnson | B | $3.15 \pm 0.05$ | Johnson & Harris (1954) |
| HD12929 | Johnson | B | $3.15 \pm 0.05$ | de Vaucouleurs (1958) |
| HD12929 | Johnson | B | $3.15 \pm 0.05$ | Johnson (1964) |
| HD12929 | Johnson | B | $3.15 \pm 0.05$ | Johnson et al. (1966) |
| HD12929 | Johnson | B | $3.15 \pm 0.05$ | Lee (1970) |
| HD12929 | Johnson | B | $3.16 \pm 0.05$ | Serkowski (1961) |
| HD12929 | Johnson | B | $3.17 \pm 0.05$ | Gutierrez-Moreno & et al. (1966) |

**Table 21** *(continued)*

| Star ID | System/Wvlen | Band/Bandpass | Value | Reference |
|---------|-------------|---------------|-------|-----------|
| HD12929 | Johnson | B | $3.17 \pm 0.05$ | Moreno (1971) |
| HD12929 | Johnson | B | $3.17 \pm 0.05$ | Mendoza et al. (1978) |
| HD12929 | Johnson | B | $3.17 \pm 0.05$ | Ducati (2002) |
| HD12929 | Johnson | B | $3.18 \pm 0.05$ | Häggkvist & Oja (1966) |
| HD12929 | Johnson | B | $3.18 \pm 0.05$ | Appenzeller (1966) |
| HD12929 | Johnson | B | $3.20 \pm 0.05$ | Hogg (1958) |
| HD12929 | Geneva | B2 | $3.61 \pm 0.08$ | Golay (1972) |
| HD12929 | 13c | m45 | $2.82 \pm 0.05$ | Johnson & Mitchell (1995) |
| HD12929 | DDO | m45 | $3.67 \pm 0.05$ | McClure & Forrester (1981) |
| HD12929 | DDO | m45 | $3.67 \pm 0.05$ | Mermilliod & Nitschelm (1989) |
| HD12929 | Oja | m45 | $2.87 \pm 0.05$ | Häggkvist & Oja (1970) |
| HD12929 | Vilnius | Y | $2.83 \pm 0.05$ | Straizys et al. (1989a) |
| HD12929 | Vilnius | Y | $2.86 \pm 0.05$ | Zdanavicius et al. (1969) |
| HD12929 | DDO | m48 | $2.43 \pm 0.05$ | McClure & Forrester (1981) |
| HD12929 | DDO | m48 | $2.43 \pm 0.05$ | Mermilliod & Nitschelm (1989) |
| HD12929 | Vilnius | Z | $2.37 \pm 0.05$ | Zdanavicius et al. (1969) |
| HD12929 | Vilnius | Z | $2.37 \pm 0.05$ | Straizys et al. (1989a) |
| HD12929 | 13c | m52 | $2.32 \pm 0.05$ | Johnson & Mitchell (1995) |
| HD12929 | Geneva | V1 | $2.82 \pm 0.08$ | Golay (1972) |
| HD12929 | WBVR | V | $2.02 \pm 0.05$ | Kornilov et al. (1991) |
| HD12929 | Vilnius | V | $2.00 \pm 0.05$ | Zdanavicius et al. (1969) |
| HD12929 | Vilnius | V | $2.00 \pm 0.05$ | Straizys et al. (1989a) |
| HD12929 | Geneva | V | $2.02 \pm 0.08$ | Golay (1972) |
| HD12929 | Johnson | V | $1.98 \pm 0.05$ | Mermilliod (1986) |
| HD12929 | Johnson | V | $1.99 \pm 0.05$ | Johnson & Morgan (1953b) |
| HD12929 | Johnson | V | $2.00 \pm 0.05$ | Sharpless (1952) |
| HD12929 | Johnson | V | $2.00 \pm 0.05$ | Johnson & Harris (1954) |
| HD12929 | Johnson | V | $2.00 \pm 0.05$ | de Vaucouleurs (1958) |
| HD12929 | Johnson | V | $2.00 \pm 0.05$ | Johnson (1964) |
| HD12929 | Johnson | V | $2.00 \pm 0.05$ | Johnson et al. (1966) |
| HD12929 | Johnson | V | $2.00 \pm 0.05$ | Lee (1970) |
| HD12929 | Johnson | V | $2.00 \pm 0.05$ | Mendoza et al. (1978) |
| HD12929 | Johnson | V | $2.01 \pm 0.05$ | Ducati (2002) |
| HD12929 | Johnson | V | $2.02 \pm 0.05$ | Hogg (1958) |
| HD12929 | Johnson | V | $2.02 \pm 0.05$ | Serkowski (1961) |
| HD12929 | Johnson | V | $2.02 \pm 0.05$ | Gutierrez-Moreno & et al. (1966) |
| HD12929 | Johnson | V | $2.02 \pm 0.05$ | Moreno (1971) |
| HD12929 | Johnson | V | $2.03 \pm 0.05$ | Appenzeller (1966) |
| HD12929 | Johnson | V | $2.04 \pm 0.05$ | Häggkvist & Oja (1966) |
| HD12929 | 13c | m58 | $1.75 \pm 0.05$ | Johnson & Mitchell (1995) |
| HD12929 | Geneva | G | $2.98 \pm 0.08$ | Golay (1972) |

**Table 21** *(continued)*

| Star ID | System/Wvlen | Band/Bandpass | Value | Reference |
|---------|--------------|---------------|-------|-----------|
| HD12929 | 13c | m63 | $1.43 \pm 0.05$ | Johnson & Mitchell (1995) |
| HD12929 | Vilnius | S | $1.18 \pm 0.05$ | Straizys et al. (1989a) |
| HD12929 | Vilnius | S | $1.21 \pm 0.05$ | Zdanavicius et al. (1969) |
| HD12929 | WBVR | R | $1.18 \pm 0.05$ | Kornilov et al. (1991) |
| HD12929 | 13c | m72 | $1.13 \pm 0.05$ | Johnson & Mitchell (1995) |
| HD12929 | 13c | m80 | $0.87 \pm 0.05$ | Johnson & Mitchell (1995) |
| HD12929 | 13c | m86 | $0.74 \pm 0.05$ | Johnson & Mitchell (1995) |
| HD12929 | 13c | m99 | $0.56 \pm 0.05$ | Johnson & Mitchell (1995) |
| HD12929 | 13c | m110 | $0.35 \pm 0.05$ | Johnson & Mitchell (1995) |
| HD12929 | Johnson | J | $0.00 \pm 0.05$ | Alonso et al. (1994) |
| HD12929 | Johnson | J | $-0.02 \pm 0.05$ | Grasdalen (1974) |
| HD12929 | Johnson | J | $-0.04 \pm 0.05$ | Noguchi et al. (1981) |
| HD12929 | Johnson | J | $0.06 \pm 0.05$ | Ducati (2002) |
| HD12929 | Johnson | J | $0.09 \pm 0.05$ | Rydgren & Vrba (1983) |
| HD12929 | Johnson | J | $0.10 \pm 0.05$ | Johnson et al. (1966) |
| HD12929 | Johnson | J | $0.10 \pm 0.05$ | Lee (1970) |
| HD12929 | Johnson | J | $0.10 \pm 0.05$ | Voelcker (1975) |
| HD12929 | Johnson | J | $0.10 \pm 0.05$ | Ghosh et al. (1984) |
| HD12929 | Johnson | J | $0.10 \pm 0.05$ | Campins et al. (1985) |
| HD12929 | Johnson | H | $-0.49 \pm 0.05$ | Lee (1970) |
| HD12929 | Johnson | H | $-0.49 \pm 0.05$ | Ghosh et al. (1984) |
| HD12929 | Johnson | H | $-0.52 \pm 0.05$ | Rydgren & Vrba (1983) |
| HD12929 | Johnson | H | $-0.52 \pm 0.05$ | Ducati (2002) |
| HD12929 | Johnson | H | $-0.54 \pm 0.05$ | Noguchi et al. (1981) |
| HD12929 | Johnson | H | $-0.56 \pm 0.05$ | Alonso et al. (1994) |
| HD12929 | 2200 | 361 | $1,138.10 \pm 9.10$ | Smith et al. (2004) |
| HD12929 | Johnson | K | $-0.63 \pm 0.05$ | Neugebauer & Leighton (1969) |
| HD12929 | Johnson | K | $-0.63 \pm 0.05$ | Lee (1970) |
| HD12929 | Johnson | K | $-0.63 \pm 0.05$ | Ducati (2002) |
| HD12929 | Johnson | K | $-0.64 \pm 0.05$ | Johnson et al. (1966) |
| HD12929 | Johnson | L | $-0.73 \pm 0.05$ | Ducati (2002) |
| HD12929 | Johnson | L | $-0.74 \pm 0.05$ | Johnson et al. (1966) |
| HD12929 | Johnson | L | $-0.74 \pm 0.05$ | Lee (1970) |
| HD12929 | 3500 | 898 | $544.70 \pm 11.90$ | Smith et al. (2004) |
| HD12929 | Johnson | N | $-0.76 \pm 0.05$ | Ducati (2002) |
| HD12929 | 4900 | 712 | $262.70 \pm 6.20$ | Smith et al. (2004) |
| HD12929 | Johnson | M | $-0.50 \pm 0.05$ | Johnson (1964) |
| HD12929 | Johnson | M | $-0.63 \pm 0.05$ | Ducati (2002) |
| HD12929 | 12000 | 6384 | $55.70 \pm 22.70$ | Smith et al. (2004) |
| HD14146 | KronComet | COp | $8.90 \pm 0.04$ | This work |
| HD14146 | WBVR | B | $8.39 \pm 0.05$ | Kornilov et al. (1991) |

**Table 21** *continued on next page*

**Table 21** *(continued)*

| Star ID | System/Wvlen | Band/Bandpass | Value | Reference |
|---------|--------------|---------------|-------|-----------|
| HD14146 | Johnson | B | $8.36 \pm 0.17$ | This work |
| HD14146 | KronComet | Bc | $8.19 \pm 0.14$ | This work |
| HD14146 | KronComet | C2 | $7.15 \pm 0.13$ | This work |
| HD14146 | KronComet | Gc | $6.87 \pm 0.16$ | This work |
| HD14146 | WBVR | V | $6.64 \pm 0.05$ | Kornilov et al. (1991) |
| HD14146 | Johnson | V | $6.70 \pm 0.03$ | This work |
| HD14146 | WBVR | R | $5.24 \pm 0.05$ | Kornilov et al. (1991) |
| HD14146 | KronComet | Rc | $5.22 \pm 0.19$ | This work |
| HD14146 | 1250 | 310 | $74.90 \pm 11.40$ | Smith et al. (2004) |
| HD14146 | 2200 | 361 | $73.20 \pm 8.20$ | Smith et al. (2004) |
| HD14146 | Johnson | K | $2.31 \pm 0.06$ | Neugebauer & Leighton (1969) |
| HD14146 | 3500 | 898 | $45.20 \pm 13.90$ | Smith et al. (2004) |
| HD14146 | 4900 | 712 | $38.90 \pm 27.20$ | Smith et al. (2004) |
| HD14146 | 12000 | 6384 | $52.20 \pm 61.70$ | Smith et al. (2004) |
| HD14512 | KronComet | COp | $9.62 \pm 0.01$ | This work |
| HD14512 | Johnson | B | $9.16 \pm 0.02$ | This work |
| HD14512 | KronComet | Bc | $9.10 \pm 0.01$ | This work |
| HD14512 | KronComet | C2 | $7.74 \pm 0.01$ | This work |
| HD14512 | KronComet | Gc | $7.76 \pm 0.01$ | This work |
| HD14512 | Johnson | V | $7.67 \pm 0.01$ | This work |
| HD14512 | KronComet | Rc | $6.09 \pm 0.01$ | This work |
| HD14512 | 1250 | 310 | $119.50 \pm 6.80$ | Smith et al. (2004) |
| HD14512 | 2200 | 361 | $133.70 \pm 5.80$ | Smith et al. (2004) |
| HD14512 | Johnson | K | $1.61 \pm 0.05$ | Neugebauer & Leighton (1969) |
| HD14512 | 3500 | 898 | $70.10 \pm 4.50$ | Smith et al. (2004) |
| HD14512 | 4900 | 712 | $29.40 \pm 6.20$ | Smith et al. (2004) |
| HD14512 | 12000 | 6384 | $9.60 \pm 24.60$ | Smith et al. (2004) |
| HD14770 | KronComet | NH | $8.08 \pm 0.02$ | This work |
| HD14770 | KronComet | UVc | $7.54 \pm 0.01$ | This work |
| HD14770 | DDO | m35 | $8.24 \pm 0.05$ | McClure & Forrester (1981) |
| HD14770 | WBVR | W | $6.78 \pm 0.05$ | Kornilov et al. (1991) |
| HD14770 | Johnson | U | $6.62 \pm 0.19$ | This work |
| HD14770 | Johnson | U | $6.92 \pm 0.05$ | Argue (1966) |
| HD14770 | Johnson | U | $6.92 \pm 0.05$ | Mermilliod (1986) |
| HD14770 | Johnson | U | $6.93 \pm 0.05$ | Ducati (2002) |
| HD14770 | DDO | m38 | $7.18 \pm 0.05$ | McClure & Forrester (1981) |
| HD14770 | KronComet | CN | $7.48 \pm 0.11$ | This work |
| HD14770 | DDO | m41 | $7.75 \pm 0.05$ | McClure & Forrester (1981) |
| HD14770 | Oja | m41 | $7.30 \pm 0.05$ | Häggkvist & Oja (1970) |
| HD14770 | DDO | m42 | $7.53 \pm 0.05$ | McClure & Forrester (1981) |
| HD14770 | Oja | m42 | $7.11 \pm 0.05$ | Häggkvist & Oja (1970) |

**Table 21** *(continued)*

| Star ID | System/Wvlen | Band/Bandpass | Value | Reference |
|---------|--------------|---------------|-------|-----------|
| HD14770 | KronComet | COp | $6.24 \pm 0.10$ | This work |
| HD14770 | WBVR | B | $6.20 \pm 0.05$ | Kornilov et al. (1991) |
| HD14770 | Johnson | B | $6.07 \pm 0.11$ | This work |
| HD14770 | Johnson | B | $6.16 \pm 0.05$ | Argue (1966) |
| HD14770 | Johnson | B | $6.17 \pm 0.05$ | Mermilliod (1986) |
| HD14770 | Johnson | B | $6.17 \pm 0.05$ | Ducati (2002) |
| HD14770 | KronComet | Bc | $5.98 \pm 0.04$ | This work |
| HD14770 | DDO | m45 | $6.72 \pm 0.05$ | McClure & Forrester (1981) |
| HD14770 | Oja | m45 | $5.92 \pm 0.05$ | Häggkvist & Oja (1970) |
| HD14770 | DDO | m48 | $5.54 \pm 0.05$ | McClure & Forrester (1981) |
| HD14770 | KronComet | C2 | $5.34 \pm 0.02$ | This work |
| HD14770 | KronComet | Gc | $5.26 \pm 0.02$ | This work |
| HD14770 | WBVR | V | $5.20 \pm 0.05$ | Kornilov et al. (1991) |
| HD14770 | Johnson | V | $5.18 \pm 0.05$ | Argue (1966) |
| HD14770 | Johnson | V | $5.19 \pm 0.05$ | Ducati (2002) |
| HD14770 | Johnson | V | $5.20 \pm 0.05$ | Mermilliod (1986) |
| HD14770 | Johnson | V | $5.22 \pm 0.04$ | This work |
| HD14770 | WBVR | R | $4.49 \pm 0.05$ | Kornilov et al. (1991) |
| HD14770 | KronComet | Rc | $4.21 \pm 0.01$ | This work |
| HD14770 | Johnson | J | $3.54 \pm 0.05$ | Selby et al. (1988) |
| HD14770 | Johnson | J | $3.54 \pm 0.05$ | Ducati (2002) |
| HD14770 | Johnson | K | $2.88 \pm 0.08$ | Neugebauer & Leighton (1969) |
| HD14770 | Johnson | K | $2.97 \pm 0.05$ | Ducati (2002) |
| HD14872 | DDO | m41 | $8.31 \pm 0.05$ | McClure & Forrester (1981) |
| HD14872 | DDO | m41 | $8.31 \pm 0.05$ | Mermilliod & Nitschelm (1989) |
| HD14872 | Oja | m41 | $7.82 \pm 0.05$ | Häggkvist & Oja (1970) |
| HD14872 | DDO | m42 | $8.06 \pm 0.05$ | McClure & Forrester (1981) |
| HD14872 | DDO | m42 | $8.06 \pm 0.05$ | Mermilliod & Nitschelm (1989) |
| HD14872 | Oja | m42 | $7.56 \pm 0.05$ | Häggkvist & Oja (1970) |
| HD14872 | KronComet | COp | $6.84 \pm 0.10$ | This work |
| HD14872 | WBVR | B | $6.32 \pm 0.05$ | Kornilov et al. (1991) |
| HD14872 | Johnson | B | $6.18 \pm 0.11$ | This work |
| HD14872 | Johnson | B | $6.22 \pm 0.05$ | Mermilliod (1986) |
| HD14872 | Johnson | B | $6.23 \pm 0.05$ | Johnson et al. (1966) |
| HD14872 | Johnson | B | $6.24 \pm 0.05$ | Argue (1966) |
| HD14872 | Johnson | B | $6.25 \pm 0.05$ | Moffett & Barnes (1979) |
| HD14872 | Johnson | B | $6.28 \pm 0.05$ | Neckel (1974) |
| HD14872 | Johnson | B | $6.29 \pm 0.05$ | Häggkvist & Oja (1966) |
| HD14872 | KronComet | Bc | $6.14 \pm 0.04$ | This work |
| HD14872 | 13c | m45 | $5.82 \pm 0.05$ | Johnson & Mitchell (1995) |
| HD14872 | DDO | m45 | $6.74 \pm 0.05$ | McClure & Forrester (1981) |

**Table 21** *(continued)*

| Star ID | System/Wvlen | Band/Bandpass | Value | Reference |
|---------|--------------|---------------|-------|-----------|
| HD14872 | DDO | m45 | $6.74 \pm 0.05$ | Mermilliod & Nitschelm (1989) |
| HD14872 | Oja | m45 | $5.91 \pm 0.05$ | Häggkvist & Oja (1970) |
| HD14872 | DDO | m48 | $5.35 \pm 0.05$ | McClure & Forrester (1981) |
| HD14872 | DDO | m48 | $5.35 \pm 0.05$ | Mermilliod & Nitschelm (1989) |
| HD14872 | KronComet | C2 | $5.19 \pm 0.02$ | This work |
| HD14872 | 13c | m52 | $5.16 \pm 0.05$ | Johnson & Mitchell (1995) |
| HD14872 | KronComet | Gc | $4.93 \pm 0.02$ | This work |
| HD14872 | WBVR | V | $4.74 \pm 0.05$ | Kornilov et al. (1991) |
| HD14872 | Johnson | V | $4.70 \pm 0.05$ | Johnson et al. (1966) |
| HD14872 | Johnson | V | $4.70 \pm 0.05$ | Mermilliod (1986) |
| HD14872 | Johnson | V | $4.71 \pm 0.05$ | Argue (1966) |
| HD14872 | Johnson | V | $4.71 \pm 0.05$ | Moffett & Barnes (1979) |
| HD14872 | Johnson | V | $4.73 \pm 0.05$ | Neckel (1974) |
| HD14872 | Johnson | V | $4.75 \pm 0.05$ | Häggkvist & Oja (1966) |
| HD14872 | Johnson | V | $4.79 \pm 0.04$ | This work |
| HD14872 | 13c | m58 | $4.37 \pm 0.05$ | Johnson & Mitchell (1995) |
| HD14872 | 13c | m63 | $3.93 \pm 0.05$ | Johnson & Mitchell (1995) |
| HD14872 | WBVR | R | $3.56 \pm 0.05$ | Kornilov et al. (1991) |
| HD14872 | KronComet | Rc | $3.34 \pm 0.02$ | This work |
| HD14872 | 13c | m72 | $3.45 \pm 0.05$ | Johnson & Mitchell (1995) |
| HD14872 | 13c | m80 | $3.09 \pm 0.05$ | Johnson & Mitchell (1995) |
| HD14872 | 13c | m86 | $2.93 \pm 0.05$ | Johnson & Mitchell (1995) |
| HD14872 | 13c | m99 | $2.64 \pm 0.05$ | Johnson & Mitchell (1995) |
| HD14872 | 13c | m110 | $2.35 \pm 0.05$ | Johnson & Mitchell (1995) |
| HD14872 | Johnson | K | $1.13 \pm 0.05$ | Neugebauer & Leighton (1969) |
| HD14872 | 3500 | 898 | $119.10 \pm 21.90$ | Smith et al. (2004) |
| HD14872 | 4900 | 712 | $52.10 \pm 7.20$ | Smith et al. (2004) |
| HD14872 | 12000 | 6384 | $11.70 \pm 19.90$ | Smith et al. (2004) |
| HD14901 | KronComet | COp | $9.29 \pm 0.02$ | This work |
| HD14901 | Johnson | B | $8.90 \pm 0.05$ | This work |
| HD14901 | KronComet | Bc | $8.90 \pm 0.02$ | This work |
| HD14901 | KronComet | C2 | $7.39 \pm 0.03$ | This work |
| HD14901 | KronComet | Gc | $7.51 \pm 0.03$ | This work |
| HD14901 | Johnson | V | $7.41 \pm 0.03$ | This work |
| HD14901 | KronComet | Rc | $5.72 \pm 0.03$ | This work |
| HD14901 | 1250 | 310 | $224.00 \pm 9.20$ | Smith et al. (2004) |
| HD14901 | 2200 | 361 | $250.30 \pm 8.50$ | Smith et al. (2004) |
| HD14901 | Johnson | K | $0.97 \pm 0.05$ | Neugebauer & Leighton (1969) |
| HD14901 | 3500 | 898 | $131.20 \pm 6.70$ | Smith et al. (2004) |
| HD14901 | 4900 | 712 | $57.80 \pm 5.30$ | Smith et al. (2004) |
| HD14901 | 12000 | 6384 | $15.10 \pm 20.00$ | Smith et al. (2004) |

**Table 21** *(continued)*

| Star ID | System/Wvlen | Band/Bandpass | Value | Reference |
|---------|--------------|---------------|-------|-----------|
| HD15656 | Oja | m41 | $8.15 \pm 0.05$ | Häggkvist & Oja (1970) |
| HD15656 | DDO | m42 | $8.38 \pm 0.05$ | McClure & Forrester (1981) |
| HD15656 | Oja | m42 | $7.89 \pm 0.05$ | Häggkvist & Oja (1970) |
| HD15656 | KronComet | COp | $7.07 \pm 0.10$ | This work |
| HD15656 | WBVR | B | $6.67 \pm 0.05$ | Kornilov et al. (1991) |
| HD15656 | Johnson | B | $6.52 \pm 0.11$ | This work |
| HD15656 | Johnson | B | $6.62 \pm 0.05$ | Argue (1966) |
| HD15656 | Johnson | B | $6.63 \pm 0.05$ | Mermilliod (1986) |
| HD15656 | KronComet | Bc | $6.46 \pm 0.04$ | This work |
| HD15656 | DDO | m45 | $7.09 \pm 0.05$ | McClure & Forrester (1981) |
| HD15656 | Oja | m45 | $6.28 \pm 0.05$ | Häggkvist & Oja (1970) |
| HD15656 | DDO | m48 | $5.74 \pm 0.05$ | McClure & Forrester (1981) |
| HD15656 | KronComet | C2 | $5.60 \pm 0.03$ | This work |
| HD15656 | KronComet | Gc | $5.33 \pm 0.03$ | This work |
| HD15656 | WBVR | V | $5.15 \pm 0.05$ | Kornilov et al. (1991) |
| HD15656 | Johnson | V | $5.15 \pm 0.05$ | Argue (1966) |
| HD15656 | Johnson | V | $5.16 \pm 0.05$ | Mermilliod (1986) |
| HD15656 | Johnson | V | $5.21 \pm 0.04$ | This work |
| HD15656 | WBVR | R | $4.02 \pm 0.05$ | Kornilov et al. (1991) |
| HD15656 | KronComet | Rc | $3.80 \pm 0.02$ | This work |
| HD15656 | 1250 | 310 | $151.50 \pm 5.90$ | Smith et al. (2004) |
| HD15656 | 2200 | 361 | $133.00 \pm 5.10$ | Smith et al. (2004) |
| HD15656 | Johnson | K | $1.66 \pm 0.05$ | Neugebauer & Leighton (1969) |
| HD15656 | 3500 | 898 | $63.80 \pm 11.70$ | Smith et al. (2004) |
| HD15656 | 4900 | 712 | $28.90 \pm 4.90$ | Smith et al. (2004) |
| HD15656 | 12000 | 6384 | $4.90 \pm 20.80$ | Smith et al. (2004) |
| HD16396 | KronComet | NH | $10.27 \pm 0.01$ | This work |
| HD16396 | KronComet | UVc | $10.07 \pm 0.11$ | This work |
| HD16396 | WBVR | W | $8.91 \pm 0.05$ | Kornilov et al. (1991) |
| HD16396 | Johnson | U | $8.95 \pm 0.02$ | This work |
| HD16396 | KronComet | CN | $9.91 \pm 0.03$ | This work |
| HD16396 | KronComet | COp | $8.36 \pm 0.01$ | This work |
| HD16396 | WBVR | B | $8.04 \pm 0.05$ | Kornilov et al. (1991) |
| HD16396 | Johnson | B | $7.96 \pm 0.02$ | This work |
| HD16396 | KronComet | Bc | $7.81 \pm 0.01$ | This work |
| HD16396 | KronComet | C2 | $7.14 \pm 0.01$ | This work |
| HD16396 | KronComet | Gc | $6.99 \pm 0.01$ | This work |
| HD16396 | WBVR | V | $6.91 \pm 0.05$ | Kornilov et al. (1991) |
| HD16396 | Johnson | V | $6.93 \pm 0.01$ | This work |
| HD16396 | WBVR | R | $6.10 \pm 0.05$ | Kornilov et al. (1991) |
| HD16396 | KronComet | Rc | $5.85 \pm 0.01$ | This work |

**Table 21** *(continued)*

| Star ID | System/Wvlen | Band/Bandpass | Value | Reference |
|---------|--------------|---------------|-------|-----------|
| HD17228 | KronComet | NH | $9.06 \pm 0.05$ | This work |
| HD17228 | KronComet | UVc | $8.54 \pm 0.03$ | This work |
| HD17228 | DDO | m35 | $9.18 \pm 0.05$ | McClure & Forrester (1981) |
| HD17228 | WBVR | W | $7.72 \pm 0.05$ | Kornilov et al. (1991) |
| HD17228 | Johnson | U | $7.64 \pm 0.19$ | This work |
| HD17228 | Johnson | U | $7.85 \pm 0.01$ | Oja (1991) |
| HD17228 | Johnson | U | $7.88 \pm 0.05$ | Mermilliod (1986) |
| HD17228 | DDO | m38 | $8.16 \pm 0.05$ | McClure & Forrester (1981) |
| HD17228 | KronComet | CN | $8.51 \pm 0.10$ | This work |
| HD17228 | DDO | m41 | $8.74 \pm 0.05$ | McClure & Forrester (1981) |
| HD17228 | DDO | m42 | $8.57 \pm 0.05$ | McClure & Forrester (1981) |
| HD17228 | KronComet | COp | $7.32 \pm 0.10$ | This work |
| HD17228 | WBVR | B | $7.23 \pm 0.05$ | Kornilov et al. (1991) |
| HD17228 | Johnson | B | $7.16 \pm 0.12$ | This work |
| HD17228 | Johnson | B | $7.18 \pm 0.05$ | Haggkvist & Oja (1970) |
| HD17228 | Johnson | B | $7.19 \pm 0.01$ | Oja (1991) |
| HD17228 | Johnson | B | $7.26 \pm 0.05$ | Mermilliod (1986) |
| HD17228 | KronComet | Bc | $7.06 \pm 0.04$ | This work |
| HD17228 | DDO | m45 | $7.76 \pm 0.05$ | McClure & Forrester (1981) |
| HD17228 | DDO | m48 | $6.61 \pm 0.05$ | McClure & Forrester (1981) |
| HD17228 | KronComet | C2 | $6.43 \pm 0.03$ | This work |
| HD17228 | KronComet | Gc | $6.35 \pm 0.03$ | This work |
| HD17228 | WBVR | V | $6.28 \pm 0.05$ | Kornilov et al. (1991) |
| HD17228 | Johnson | V | $6.25 \pm 0.01$ | Oja (1991) |
| HD17228 | Johnson | V | $6.25 \pm 0.05$ | Haggkvist & Oja (1970) |
| HD17228 | Johnson | V | $6.32 \pm 0.04$ | This work |
| HD17228 | Johnson | V | $6.32 \pm 0.05$ | Mermilliod (1986) |
| HD17228 | WBVR | R | $5.59 \pm 0.05$ | Kornilov et al. (1991) |
| HD17228 | KronComet | Rc | $5.32 \pm 0.01$ | This work |
| HD17361 | 13c | m33 | $6.80 \pm 0.05$ | Johnson & Mitchell (1995) |
| HD17361 | Geneva | U | $7.26 \pm 0.08$ | Golay (1972) |
| HD17361 | 13c | m35 | $6.60 \pm 0.05$ | Johnson & Mitchell (1995) |
| HD17361 | DDO | m35 | $8.04 \pm 0.05$ | McClure & Forrester (1981) |
| HD17361 | WBVR | W | $6.58 \pm 0.05$ | Kornilov et al. (1991) |
| HD17361 | Johnson | U | $6.66 \pm 0.05$ | Johnson (1964) |
| HD17361 | Johnson | U | $6.66 \pm 0.05$ | Johnson et al. (1966) |
| HD17361 | Johnson | U | $6.66 \pm 0.05$ | Ducati (2002) |
| HD17361 | Johnson | U | $6.68 \pm 0.05$ | Mermilliod (1986) |
| HD17361 | Johnson | U | $6.69 \pm 0.05$ | Argue (1966) |
| HD17361 | Johnson | U | $6.69 \pm 0.05$ | Oja (1983) |
| HD17361 | Johnson | U | $6.71 \pm 0.01$ | Oja (1984) |

**Table 21** *(continued)*

| Star ID | System/Wvlen | Band/Bandpass | Value | Reference |
|---------|--------------|---------------|-------|-----------|
| HD17361 | Johnson | U | $6.71 \pm 0.05$ | Jennens & Helfer (1975) |
| HD17361 | Johnson | U | $6.71 \pm 0.05$ | Oja (1984) |
| HD17361 | Johnson | U | $6.71 \pm 0.05$ | Oja (1985b) |
| HD17361 | Johnson | U | $6.72 \pm 0.05$ | Oja (1986) |
| HD17361 | 13c | m37 | $6.66 \pm 0.05$ | Johnson & Mitchell (1995) |
| HD17361 | DDO | m38 | $6.92 \pm 0.05$ | McClure & Forrester (1981) |
| HD17361 | 13c | m40 | $6.34 \pm 0.05$ | Johnson & Mitchell (1995) |
| HD17361 | Geneva | B1 | $6.35 \pm 0.08$ | Golay (1972) |
| HD17361 | DDO | m41 | $7.35 \pm 0.05$ | McClure & Forrester (1981) |
| HD17361 | Oja | m41 | $6.91 \pm 0.05$ | Häggkvist & Oja (1970) |
| HD17361 | DDO | m42 | $7.09 \pm 0.05$ | McClure & Forrester (1981) |
| HD17361 | Oja | m42 | $6.65 \pm 0.05$ | Häggkvist & Oja (1970) |
| HD17361 | Geneva | B | $4.97 \pm 0.08$ | Golay (1972) |
| HD17361 | WBVR | B | $5.66 \pm 0.05$ | Kornilov et al. (1991) |
| HD17361 | Johnson | B | $5.61 \pm 0.05$ | Argue (1966) |
| HD17361 | Johnson | B | $5.61 \pm 0.05$ | Häggkvist & Oja (1969a) |
| HD17361 | Johnson | B | $5.61 \pm 0.05$ | Oja (1983) |
| HD17361 | Johnson | B | $5.62 \pm 0.05$ | Haggkvist & Oja (1970) |
| HD17361 | Johnson | B | $5.63 \pm 0.01$ | Oja (1993) |
| HD17361 | Johnson | B | $5.63 \pm 0.05$ | Johnson (1964) |
| HD17361 | Johnson | B | $5.63 \pm 0.05$ | Häggkvist & Oja (1966) |
| HD17361 | Johnson | B | $5.63 \pm 0.05$ | Johnson et al. (1966) |
| HD17361 | Johnson | B | $5.63 \pm 0.05$ | Oja (1984) |
| HD17361 | Johnson | B | $5.63 \pm 0.05$ | Oja (1985b) |
| HD17361 | Johnson | B | $5.63 \pm 0.05$ | Oja (1986) |
| HD17361 | Johnson | B | $5.63 \pm 0.05$ | Mermilliod (1986) |
| HD17361 | Johnson | B | $5.63 \pm 0.05$ | Ducati (2002) |
| HD17361 | Johnson | B | $5.64 \pm 0.05$ | Jennens & Helfer (1975) |
| HD17361 | Geneva | B2 | $6.07 \pm 0.08$ | Golay (1972) |
| HD17361 | 13c | m45 | $5.31 \pm 0.05$ | Johnson & Mitchell (1995) |
| HD17361 | DDO | m45 | $6.14 \pm 0.05$ | McClure & Forrester (1981) |
| HD17361 | Oja | m45 | $5.32 \pm 0.05$ | Häggkvist & Oja (1970) |
| HD17361 | DDO | m48 | $4.92 \pm 0.05$ | McClure & Forrester (1981) |
| HD17361 | 13c | m52 | $4.80 \pm 0.05$ | Johnson & Mitchell (1995) |
| HD17361 | Geneva | V1 | $5.32 \pm 0.08$ | Golay (1972) |
| HD17361 | WBVR | V | $4.53 \pm 0.05$ | Kornilov et al. (1991) |
| HD17361 | Geneva | V | $4.54 \pm 0.08$ | Golay (1972) |
| HD17361 | Johnson | V | $4.49 \pm 0.05$ | Oja (1983) |
| HD17361 | Johnson | V | $4.50 \pm 0.05$ | Argue (1966) |
| HD17361 | Johnson | V | $4.50 \pm 0.05$ | Mermilliod (1986) |
| HD17361 | Johnson | V | $4.51 \pm 0.01$ | Oja (1993) |

**Table 21** *continued on next page*

**Table 21** *(continued)*

| Star ID | System/Wvlen | Band/Bandpass | Value | Reference |
|---------|--------------|---------------|-------|-----------|
| HD17361 | Johnson | V | $4.51 \pm 0.05$ | Häggkvist & Oja (1969a) |
| HD17361 | Johnson | V | $4.51 \pm 0.05$ | Haggkvist & Oja (1970) |
| HD17361 | Johnson | V | $4.51 \pm 0.05$ | Oja (1984) |
| HD17361 | Johnson | V | $4.51 \pm 0.05$ | Oja (1985b) |
| HD17361 | Johnson | V | $4.51 \pm 0.05$ | Oja (1986) |
| HD17361 | Johnson | V | $4.52 \pm 0.05$ | Johnson (1964) |
| HD17361 | Johnson | V | $4.52 \pm 0.05$ | Häggkvist & Oja (1966) |
| HD17361 | Johnson | V | $4.52 \pm 0.05$ | Johnson et al. (1966) |
| HD17361 | Johnson | V | $4.52 \pm 0.05$ | Ducati (2002) |
| HD17361 | Johnson | V | $4.53 \pm 0.05$ | Jennens & Helfer (1975) |
| HD17361 | 13c | m58 | $4.25 \pm 0.05$ | Johnson & Mitchell (1995) |
| HD17361 | Geneva | G | $5.48 \pm 0.08$ | Golay (1972) |
| HD17361 | 13c | m63 | $3.96 \pm 0.05$ | Johnson & Mitchell (1995) |
| HD17361 | WBVR | R | $3.74 \pm 0.05$ | Kornilov et al. (1991) |
| HD17361 | 13c | m72 | $3.67 \pm 0.05$ | Johnson & Mitchell (1995) |
| HD17361 | 13c | m80 | $3.44 \pm 0.05$ | Johnson & Mitchell (1995) |
| HD17361 | 13c | m86 | $3.32 \pm 0.05$ | Johnson & Mitchell (1995) |
| HD17361 | 13c | m99 | $3.14 \pm 0.05$ | Johnson & Mitchell (1995) |
| HD17361 | 13c | m110 | $2.95 \pm 0.05$ | Johnson & Mitchell (1995) |
| HD17361 | 1250 | 310 | $136.40 \pm 8.20$ | Smith et al. (2004) |
| HD17361 | Johnson | J | $2.70 \pm 0.05$ | Johnson et al. (1966) |
| HD17361 | Johnson | J | $2.70 \pm 0.05$ | Ducati (2002) |
| HD17361 | 2200 | 361 | $96.00 \pm 7.60$ | Smith et al. (2004) |
| HD17361 | Johnson | K | $2.02 \pm 0.05$ | Johnson et al. (1966) |
| HD17361 | Johnson | K | $2.02 \pm 0.05$ | Ducati (2002) |
| HD17361 | Johnson | K | $2.04 \pm 0.08$ | Neugebauer & Leighton (1969) |
| HD17361 | 3500 | 898 | $45.70 \pm 5.10$ | Smith et al. (2004) |
| HD17361 | 4900 | 712 | $22.30 \pm 4.70$ | Smith et al. (2004) |
| HD17361 | 12000 | 6384 | $-3.40 \pm 22.90$ | Smith et al. (2004) |
| HD17709 | Oja | m41 | $7.67 \pm 0.05$ | Häggkvist & Oja (1970) |
| HD17709 | Oja | m42 | $7.46 \pm 0.05$ | Häggkvist & Oja (1970) |
| HD17709 | WBVR | B | $6.17 \pm 0.05$ | Kornilov et al. (1991) |
| HD17709 | Johnson | B | $6.08 \pm 0.05$ | Argue (1966) |
| HD17709 | Johnson | B | $6.09 \pm 0.05$ | Johnson et al. (1966) |
| HD17709 | Johnson | B | $6.09 \pm 0.05$ | Mermilliod (1986) |
| HD17709 | Johnson | B | $6.10 \pm 0.05$ | Ducati (2002) |
| HD17709 | Johnson | B | $6.11 \pm 0.05$ | Häggkvist & Oja (1966) |
| HD17709 | Johnson | B | $6.11 \pm 0.05$ | Neckel (1974) |
| HD17709 | Johnson | B | $6.13 \pm 0.05$ | Johnson (1964) |
| HD17709 | Oja | m45 | $5.76 \pm 0.05$ | Häggkvist & Oja (1970) |
| HD17709 | Vilnius | Y | $5.67 \pm 0.05$ | Kazlauskas et al. (2005) |

**Table 21** *(continued)*

| Star ID | System/Wvlen | Band/Bandpass | Value | Reference |
|---|---|---|---|---|
| HD17709 | Vilnius | Z | $5.09 \pm 0.05$ | Kazlauskas et al. (2005) |
| HD17709 | WBVR | V | $4.57 \pm 0.05$ | Kornilov et al. (1991) |
| HD17709 | Vilnius | V | $4.56 \pm 0.05$ | Kazlauskas et al. (2005) |
| HD17709 | Johnson | V | $4.53 \pm 0.05$ | Johnson (1964) |
| HD17709 | Johnson | V | $4.53 \pm 0.05$ | Johnson et al. (1966) |
| HD17709 | Johnson | V | $4.53 \pm 0.05$ | Argue (1966) |
| HD17709 | Johnson | V | $4.53 \pm 0.05$ | Ducati (2002) |
| HD17709 | Johnson | V | $4.55 \pm 0.05$ | Neckel (1974) |
| HD17709 | Johnson | V | $4.56 \pm 0.05$ | Häggkvist & Oja (1966) |
| HD17709 | Johnson | V | $4.56 \pm 0.05$ | Mermilliod (1986) |
| HD17709 | Vilnius | S | $3.47 \pm 0.05$ | Kazlauskas et al. (2005) |
| HD17709 | WBVR | R | $3.34 \pm 0.05$ | Kornilov et al. (1991) |
| HD17709 | 1250 | 310 | $331.80 \pm 18.30$ | Smith et al. (2004) |
| HD17709 | Johnson | J | $1.76 \pm 0.05$ | Johnson et al. (1966) |
| HD17709 | Johnson | J | $1.76 \pm 0.05$ | Voelcker (1975) |
| HD17709 | Johnson | J | $1.76 \pm 0.05$ | Ducati (2002) |
| HD17709 | Johnson | J | $1.76 \pm 0.05$ | Shenavrin et al. (2011) |
| HD17709 | Johnson | J | $1.89 \pm 0.05$ | Chen et al. (1982) |
| HD17709 | Johnson | H | $0.98 \pm 0.05$ | Shenavrin et al. (2011) |
| HD17709 | 2200 | 361 | $316.40 \pm 17.00$ | Smith et al. (2004) |
| HD17709 | Johnson | K | $0.72 \pm 0.05$ | Ducati (2002) |
| HD17709 | Johnson | K | $0.74 \pm 0.05$ | Neugebauer & Leighton (1969) |
| HD17709 | Johnson | K | $0.78 \pm 0.05$ | Johnson et al. (1966) |
| HD17709 | Johnson | K | $0.78 \pm 0.05$ | Shenavrin et al. (2011) |
| HD17709 | 3500 | 898 | $158.00 \pm 9.70$ | Smith et al. (2004) |
| HD17709 | 4900 | 712 | $68.50 \pm 5.80$ | Smith et al. (2004) |
| HD17709 | 12000 | 6384 | $15.30 \pm 21.80$ | Smith et al. (2004) |
| HD18322 | 13c | m33 | $6.05 \pm 0.05$ | Johnson & Mitchell (1995) |
| HD18322 | Geneva | U | $6.52 \pm 0.08$ | Golay (1972) |
| HD18322 | Vilnius | U | $7.80 \pm 0.05$ | Kazlauskas et al. (2005) |
| HD18322 | 13c | m35 | $5.83 \pm 0.05$ | Johnson & Mitchell (1995) |
| HD18322 | DDO | m35 | $7.32 \pm 0.05$ | McClure & Forrester (1981) |
| HD18322 | Johnson | U | $5.97 \pm 0.05$ | Johnson (1964) |
| HD18322 | Johnson | U | $5.97 \pm 0.05$ | Johnson et al. (1966) |
| HD18322 | Johnson | U | $5.97 \pm 0.05$ | Ducati (2002) |
| HD18322 | Johnson | U | $5.98 \pm 0.05$ | Johnson et al. (1966) |
| HD18322 | Johnson | U | $5.98 \pm 0.05$ | Gutierrez-Moreno & et al. (1966) |
| HD18322 | Johnson | U | $5.98 \pm 0.05$ | Mermilliod (1986) |
| HD18322 | Johnson | U | $5.99 \pm 0.05$ | Oja (1970) |
| HD18322 | Johnson | U | $6.00 \pm 0.05$ | Cousins & Stoy (1962) |
| HD18322 | Johnson | U | $6.01 \pm 0.05$ | Jennens & Helfer (1975) |

**Table 21** (continued)

| Star ID | System/Wvlen | Band/Bandpass | Value | Reference |
|---------|--------------|---------------|-------|-----------|
| HD18322 | Geneva | B1 | $5.67 \pm 0.08$ | Golay (1972) |
| HD18322 | Oja | m41 | $6.20 \pm 0.05$ | Häggkvist & Oja (1970) |
| HD18322 | DDO | m42 | $6.44 \pm 0.05$ | McClure & Forrester (1981) |
| HD18322 | Oja | m42 | $5.99 \pm 0.05$ | Häggkvist & Oja (1970) |
| HD18322 | Geneva | B | $4.31 \pm 0.08$ | Golay (1972) |
| HD18322 | WBVR | B | $5.00 \pm 0.05$ | Kornilov et al. (1991) |
| HD18322 | Johnson | B | $4.98 \pm 0.05$ | Mermilliod (1986) |
| HD18322 | Johnson | B | $4.99 \pm 0.05$ | Cousins & Stoy (1962) |
| HD18322 | Johnson | B | $4.99 \pm 0.05$ | Johnson (1964) |
| HD18322 | Johnson | B | $4.99 \pm 0.05$ | Johnson et al. (1966) |
| HD18322 | Johnson | B | $4.99 \pm 0.05$ | Gutierrez-Moreno & et al. (1966) |
| HD18322 | Johnson | B | $4.99 \pm 0.05$ | Oja (1970) |
| HD18322 | Johnson | B | $4.99 \pm 0.05$ | Ducati (2002) |
| HD18322 | Johnson | B | $5.00 \pm 0.05$ | Cousins & Stoy (1962) |
| HD18322 | Johnson | B | $5.00 \pm 0.05$ | Glass (1974) |
| HD18322 | Johnson | B | $5.00 \pm 0.05$ | Jennens & Helfer (1975) |
| HD18322 | Geneva | B2 | $5.44 \pm 0.08$ | Golay (1972) |
| HD18322 | 13c | m45 | $4.65 \pm 0.05$ | Johnson & Mitchell (1995) |
| HD18322 | DDO | m45 | $5.51 \pm 0.05$ | McClure & Forrester (1981) |
| HD18322 | Oja | m45 | $4.69 \pm 0.05$ | Häggkvist & Oja (1970) |
| HD18322 | Vilnius | Y | $4.73 \pm 0.05$ | Kazlauskas et al. (2005) |
| HD18322 | DDO | m48 | $4.29 \pm 0.05$ | McClure & Forrester (1981) |
| HD18322 | Vilnius | Z | $4.26 \pm 0.05$ | Kazlauskas et al. (2005) |
| HD18322 | 13c | m52 | $4.16 \pm 0.05$ | Johnson & Mitchell (1995) |
| HD18322 | Geneva | V1 | $4.70 \pm 0.08$ | Golay (1972) |
| HD18322 | WBVR | V | $3.90 \pm 0.05$ | Kornilov et al. (1991) |
| HD18322 | Vilnius | V | $3.92 \pm 0.05$ | Kazlauskas et al. (2005) |
| HD18322 | Geneva | V | $3.91 \pm 0.08$ | Golay (1972) |
| HD18322 | Johnson | V | $3.86 \pm 0.05$ | Johnson (1964) |
| HD18322 | Johnson | V | $3.87 \pm 0.05$ | Johnson et al. (1966) |
| HD18322 | Johnson | V | $3.87 \pm 0.05$ | Ducati (2002) |
| HD18322 | Johnson | V | $3.88 \pm 0.05$ | Mermilliod (1986) |
| HD18322 | Johnson | V | $3.89 \pm 0.05$ | Cousins & Stoy (1962) |
| HD18322 | Johnson | V | $3.89 \pm 0.05$ | Gutierrez-Moreno & et al. (1966) |
| HD18322 | Johnson | V | $3.89 \pm 0.05$ | Glass (1974) |
| HD18322 | Johnson | V | $3.90 \pm 0.05$ | Cousins & Stoy (1962) |
| HD18322 | Johnson | V | $3.90 \pm 0.05$ | Oja (1970) |
| HD18322 | Johnson | V | $3.90 \pm 0.05$ | Jennens & Helfer (1975) |
| HD18322 | 13c | m58 | $3.64 \pm 0.05$ | Johnson & Mitchell (1995) |
| HD18322 | Geneva | G | $4.87 \pm 0.08$ | Golay (1972) |
| HD18322 | 13c | m63 | $3.32 \pm 0.05$ | Johnson & Mitchell (1995) |

**Table 21** *(continued)*

| Star ID | System/Wvlen | Band/Bandpass | Value | Reference |
|---------|--------------|---------------|-------|-----------|
| HD18322 | Vilnius | S | $3.12 \pm 0.05$ | Kazlauskas et al. (2005) |
| HD18322 | WBVR | R | $3.09 \pm 0.05$ | Kornilov et al. (1991) |
| HD18322 | 13c | m72 | $3.06 \pm 0.05$ | Johnson & Mitchell (1995) |
| HD18322 | 13c | m80 | $2.82 \pm 0.05$ | Johnson & Mitchell (1995) |
| HD18322 | 13c | m86 | $2.71 \pm 0.05$ | Johnson & Mitchell (1995) |
| HD18322 | 13c | m99 | $2.53 \pm 0.05$ | Johnson & Mitchell (1995) |
| HD18322 | 13c | m110 | $2.34 \pm 0.05$ | Johnson & Mitchell (1995) |
| HD18322 | 1250 | 310 | $240.30 \pm 8.00$ | Smith et al. (2004) |
| HD18322 | Johnson | J | $1.99 \pm 0.05$ | Alonso et al. (1998) |
| HD18322 | Johnson | J | $2.04 \pm 0.01$ | Laney et al. (2012) |
| HD18322 | Johnson | J | $2.07 \pm 0.05$ | Johnson et al. (1966) |
| HD18322 | Johnson | J | $2.07 \pm 0.05$ | Glass (1974) |
| HD18322 | Johnson | J | $2.08 \pm 0.05$ | Glass (1974) |
| HD18322 | Johnson | J | $2.08 \pm 0.05$ | Wu & Wang (1985) |
| HD18322 | Johnson | J | $2.08 \pm 0.05$ | Ducati (2002) |
| HD18322 | Johnson | J | $2.10 \pm 0.05$ | Carter (1990) |
| HD18322 | Johnson | H | $1.47 \pm 0.05$ | Carter (1990) |
| HD18322 | Johnson | H | $1.47 \pm 0.05$ | Alonso et al. (1998) |
| HD18322 | Johnson | H | $1.49 \pm 0.05$ | Ducati (2002) |
| HD18322 | Johnson | H | $1.50 \pm 0.01$ | Laney et al. (2012) |
| HD18322 | Johnson | H | $1.50 \pm 0.05$ | Glass (1974) |
| HD18322 | Johnson | H | $1.50 \pm 0.05$ | Wu & Wang (1985) |
| HD18322 | 2200 | 361 | $172.30 \pm 5.00$ | Smith et al. (2004) |
| HD18322 | Johnson | K | $1.37 \pm 0.01$ | Laney et al. (2012) |
| HD18322 | Johnson | K | $1.38 \pm 0.05$ | Neugebauer & Leighton (1969) |
| HD18322 | Johnson | K | $1.38 \pm 0.05$ | Glass (1974) |
| HD18322 | Johnson | K | $1.38 \pm 0.05$ | Ducati (2002) |
| HD18322 | Johnson | K | $1.40 \pm 0.05$ | Johnson et al. (1966) |
| HD18322 | Johnson | L | $1.31 \pm 0.05$ | Glass (1974) |
| HD18322 | Johnson | L | $1.31 \pm 0.05$ | Ducati (2002) |
| HD18322 | Johnson | L | $1.33 \pm 0.05$ | Glass (1974) |
| HD18322 | 3500 | 898 | $81.80 \pm 4.80$ | Smith et al. (2004) |
| HD18322 | 4900 | 712 | $40.00 \pm 5.20$ | Smith et al. (2004) |
| HD18322 | 12000 | 6384 | $5.80 \pm 20.70$ | Smith et al. (2004) |
| HD18449 | 13c | m33 | $7.60 \pm 0.05$ | Johnson & Mitchell (1995) |
| HD18449 | 13c | m35 | $7.31 \pm 0.05$ | Johnson & Mitchell (1995) |
| HD18449 | DDO | m35 | $8.81 \pm 0.05$ | McClure & Forrester (1981) |
| HD18449 | WBVR | W | $7.35 \pm 0.05$ | Kornilov et al. (1991) |
| HD18449 | Johnson | U | $7.43 \pm 0.05$ | Mermilliod (1986) |
| HD18449 | Johnson | U | $7.45 \pm 0.05$ | Oja (1983) |
| HD18449 | Johnson | U | $7.46 \pm 0.01$ | Oja (1984) |

**Table 21** *(continued)*

| Star ID | System/Wvlen | Band/Bandpass | Value | Reference |
|---------|--------------|---------------|-------|-----------|
| HD18449 | Johnson | U | $7.46 \pm 0.05$ | Argue (1966) |
| HD18449 | Johnson | U | $7.46 \pm 0.05$ | Jennens & Helfer (1975) |
| HD18449 | Johnson | U | $7.46 \pm 0.05$ | Oja (1984) |
| HD18449 | Johnson | U | $7.46 \pm 0.05$ | Oja (1985b) |
| HD18449 | Johnson | U | $7.46 \pm 0.05$ | Oja (1985a) |
| HD18449 | Johnson | U | $7.47 \pm 0.05$ | Johnson et al. (1966) |
| HD18449 | Johnson | U | $7.47 \pm 0.05$ | Oja (1986) |
| HD18449 | 13c | m37 | $7.39 \pm 0.05$ | Johnson & Mitchell (1995) |
| HD18449 | DDO | m38 | $7.65 \pm 0.05$ | McClure & Forrester (1981) |
| HD18449 | 13c | m40 | $6.97 \pm 0.05$ | Johnson & Mitchell (1995) |
| HD18449 | DDO | m41 | $7.97 \pm 0.05$ | McClure & Forrester (1981) |
| HD18449 | Oja | m41 | $7.51 \pm 0.05$ | Häggkvist & Oja (1970) |
| HD18449 | DDO | m42 | $7.72 \pm 0.05$ | McClure & Forrester (1981) |
| HD18449 | Oja | m42 | $7.24 \pm 0.05$ | Häggkvist & Oja (1970) |
| HD18449 | WBVR | B | $6.21 \pm 0.05$ | Kornilov et al. (1991) |
| HD18449 | Johnson | B | $6.16 \pm 0.05$ | Argue (1966) |
| HD18449 | Johnson | B | $6.16 \pm 0.05$ | Häggkvist & Oja (1969a) |
| HD18449 | Johnson | B | $6.16 \pm 0.05$ | Haggkvist & Oja (1970) |
| HD18449 | Johnson | B | $6.16 \pm 0.05$ | Oja (1985a) |
| HD18449 | Johnson | B | $6.17 \pm 0.01$ | Oja (1993) |
| HD18449 | Johnson | B | $6.17 \pm 0.05$ | Häggkvist & Oja (1966) |
| HD18449 | Johnson | B | $6.17 \pm 0.05$ | Oja (1983) |
| HD18449 | Johnson | B | $6.17 \pm 0.05$ | Oja (1984) |
| HD18449 | Johnson | B | $6.17 \pm 0.05$ | Oja (1985b) |
| HD18449 | Johnson | B | $6.17 \pm 0.05$ | Mermilliod (1986) |
| HD18449 | Johnson | B | $6.18 \pm 0.05$ | Oja (1986) |
| HD18449 | Johnson | B | $6.19 \pm 0.05$ | Johnson et al. (1966) |
| HD18449 | Johnson | B | $6.19 \pm 0.05$ | Jennens & Helfer (1975) |
| HD18449 | 13c | m45 | $5.79 \pm 0.05$ | Johnson & Mitchell (1995) |
| HD18449 | DDO | m45 | $6.67 \pm 0.05$ | McClure & Forrester (1981) |
| HD18449 | Oja | m45 | $5.84 \pm 0.05$ | Häggkvist & Oja (1970) |
| HD18449 | DDO | m48 | $5.40 \pm 0.05$ | McClure & Forrester (1981) |
| HD18449 | 13c | m52 | $5.25 \pm 0.05$ | Johnson & Mitchell (1995) |
| HD18449 | WBVR | V | $4.95 \pm 0.05$ | Kornilov et al. (1991) |
| HD18449 | Johnson | V | $4.92 \pm 0.05$ | Oja (1984) |
| HD18449 | Johnson | V | $4.93 \pm 0.01$ | Oja (1993) |
| HD18449 | Johnson | V | $4.93 \pm 0.05$ | Argue (1966) |
| HD18449 | Johnson | V | $4.93 \pm 0.05$ | Häggkvist & Oja (1969a) |
| HD18449 | Johnson | V | $4.93 \pm 0.05$ | Haggkvist & Oja (1970) |
| HD18449 | Johnson | V | $4.93 \pm 0.05$ | Oja (1983) |
| HD18449 | Johnson | V | $4.93 \pm 0.05$ | Oja (1985b) |

**Table 21** *(continued)*

| Star ID | System/Wvlen | Band/Bandpass | Value | Reference |
|---------|--------------|---------------|-------|-----------|
| HD18449 | Johnson | V | $4.93 \pm 0.05$ | Oja (1985a) |
| HD18449 | Johnson | V | $4.94 \pm 0.05$ | Häggkvist & Oja (1966) |
| HD18449 | Johnson | V | $4.94 \pm 0.05$ | Johnson et al. (1966) |
| HD18449 | Johnson | V | $4.94 \pm 0.05$ | Oja (1986) |
| HD18449 | Johnson | V | $4.94 \pm 0.05$ | Mermilliod (1986) |
| HD18449 | Johnson | V | $4.95 \pm 0.05$ | Jennens & Helfer (1975) |
| HD18449 | 13c | m58 | $4.65 \pm 0.05$ | Johnson & Mitchell (1995) |
| HD18449 | 13c | m63 | $4.30 \pm 0.05$ | Johnson & Mitchell (1995) |
| HD18449 | WBVR | R | $4.06 \pm 0.05$ | Kornilov et al. (1991) |
| HD18449 | 13c | m72 | $4.03 \pm 0.05$ | Johnson & Mitchell (1995) |
| HD18449 | 13c | m80 | $3.73 \pm 0.05$ | Johnson & Mitchell (1995) |
| HD18449 | 13c | m86 | $3.61 \pm 0.05$ | Johnson & Mitchell (1995) |
| HD18449 | 13c | m99 | $3.40 \pm 0.05$ | Johnson & Mitchell (1995) |
| HD18449 | 13c | m110 | $3.19 \pm 0.05$ | Johnson & Mitchell (1995) |
| HD18449 | 1250 | 310 | $119.00 \pm 11.30$ | Smith et al. (2004) |
| HD18449 | 2200 | 361 | $90.90 \pm 6.70$ | Smith et al. (2004) |
| HD18449 | Johnson | K | $2.09 \pm 0.07$ | Neugebauer & Leighton (1969) |
| HD18449 | 3500 | 898 | $41.30 \pm 10.00$ | Smith et al. (2004) |
| HD18449 | 4900 | 712 | $18.60 \pm 5.20$ | Smith et al. (2004) |
| HD18449 | 12000 | 6384 | $-3.90 \pm 20.80$ | Smith et al. (2004) |
| HD19613 | KronComet | COp | $10.23 \pm 0.03$ | This work |
| HD19613 | Johnson | B | $9.98 \pm 0.02$ | This work |
| HD19613 | KronComet | C2 | $8.49 \pm 0.01$ | This work |
| HD19613 | KronComet | Gc | $8.70 \pm 0.01$ | This work |
| HD19613 | Johnson | V | $8.59 \pm 0.02$ | This work |
| HD19613 | KronComet | Rc | $6.93 \pm 0.01$ | This work |
| HD19613 | 1250 | 310 | $80.20 \pm 5.10$ | Smith et al. (2004) |
| HD19613 | 2200 | 361 | $90.20 \pm 5.00$ | Smith et al. (2004) |
| HD19613 | Johnson | K | $2.09 \pm 0.08$ | Neugebauer & Leighton (1969) |
| HD19613 | 3500 | 898 | $46.90 \pm 4.30$ | Smith et al. (2004) |
| HD19613 | 4900 | 712 | $19.90 \pm 4.90$ | Smith et al. (2004) |
| HD19613 | 12000 | 6384 | $9.50 \pm 17.30$ | Smith et al. (2004) |
| HD19787 | Oja | m41 | $6.57 \pm 0.05$ | Häggkvist & Oja (1970) |
| HD19787 | DDO | m42 | $6.78 \pm 0.05$ | McClure & Forrester (1981) |
| HD19787 | Oja | m42 | $6.34 \pm 0.05$ | Häggkvist & Oja (1970) |
| HD19787 | KronComet | COp | $5.60 \pm 0.06$ | This work |
| HD19787 | WBVR | B | $5.40 \pm 0.05$ | Kornilov et al. (1991) |
| HD19787 | Johnson | B | $5.31 \pm 0.05$ | This work |
| HD19787 | Johnson | B | $5.36 \pm 0.05$ | Gehrels et al. (1964) |
| HD19787 | Johnson | B | $5.36 \pm 0.05$ | Argue (1966) |
| HD19787 | Johnson | B | $5.37 \pm 0.05$ | Gutierrez-Moreno & et al. (1966) |

**Table 21** (continued)

| Star ID | System/Wvlen | Band/Bandpass | Value | Reference |
|---------|-------------|---------------|-------|-----------|
| HD19787 | Johnson | B | $5.38 \pm 0.05$ | Häggkvist & Oja (1966) |
| HD19787 | Johnson | B | $5.38 \pm 0.05$ | Jennens & Helfer (1975) |
| HD19787 | Johnson | B | $5.39 \pm 0.05$ | Mermilliod (1986) |
| HD19787 | Johnson | B | $5.40 \pm 0.05$ | Johnson et al. (1966) |
| HD19787 | Johnson | B | $5.40 \pm 0.05$ | Ducati (2002) |
| HD19787 | KronComet | Bc | $5.20 \pm 0.01$ | This work |
| HD19787 | 13c | m45 | $5.08 \pm 0.05$ | Johnson & Mitchell (1995) |
| HD19787 | DDO | m45 | $5.90 \pm 0.05$ | McClure & Forrester (1981) |
| HD19787 | Oja | m45 | $5.07 \pm 0.05$ | Häggkvist & Oja (1970) |
| HD19787 | DDO | m48 | $4.71 \pm 0.05$ | McClure & Forrester (1981) |
| HD19787 | KronComet | C2 | $4.56 \pm 0.01$ | This work |
| HD19787 | 13c | m52 | $4.60 \pm 0.05$ | Johnson & Mitchell (1995) |
| HD19787 | KronComet | Gc | $4.43 \pm 0.02$ | This work |
| HD19787 | WBVR | V | $4.35 \pm 0.05$ | Kornilov et al. (1991) |
| HD19787 | Johnson | V | $4.33 \pm 0.05$ | Gehrels et al. (1964) |
| HD19787 | Johnson | V | $4.33 \pm 0.05$ | Argue (1966) |
| HD19787 | Johnson | V | $4.34 \pm 0.05$ | Gutierrez-Moreno & et al. (1966) |
| HD19787 | Johnson | V | $4.35 \pm 0.05$ | Häggkvist & Oja (1966) |
| HD19787 | Johnson | V | $4.35 \pm 0.05$ | Jennens & Helfer (1975) |
| HD19787 | Johnson | V | $4.35 \pm 0.05$ | Mermilliod (1986) |
| HD19787 | Johnson | V | $4.37 \pm 0.05$ | Johnson et al. (1966) |
| HD19787 | Johnson | V | $4.37 \pm 0.05$ | Ducati (2002) |
| HD19787 | Johnson | V | $4.42 \pm 0.08$ | This work |
| HD19787 | 13c | m58 | $4.11 \pm 0.05$ | Johnson & Mitchell (1995) |
| HD19787 | 13c | m63 | $3.82 \pm 0.05$ | Johnson & Mitchell (1995) |
| HD19787 | WBVR | R | $3.61 \pm 0.05$ | Kornilov et al. (1991) |
| HD19787 | KronComet | Rc | $3.37 \pm 0.02$ | This work |
| HD19787 | 13c | m72 | $3.60 \pm 0.05$ | Johnson & Mitchell (1995) |
| HD19787 | 13c | m80 | $3.34 \pm 0.05$ | Johnson & Mitchell (1995) |
| HD19787 | 13c | m86 | $3.23 \pm 0.05$ | Johnson & Mitchell (1995) |
| HD19787 | 13c | m99 | $3.07 \pm 0.05$ | Johnson & Mitchell (1995) |
| HD19787 | 13c | m110 | $2.89 \pm 0.05$ | Johnson & Mitchell (1995) |
| HD19787 | Johnson | J | $2.68 \pm 0.01$ | Laney et al. (2012) |
| HD19787 | Johnson | J | $2.69 \pm 0.05$ | Johnson et al. (1966) |
| HD19787 | Johnson | J | $2.69 \pm 0.05$ | Voelcker (1975) |
| HD19787 | Johnson | J | $2.69 \pm 0.05$ | Ducati (2002) |
| HD19787 | Johnson | H | $2.16 \pm 0.01$ | Laney et al. (2012) |
| HD19787 | 2200 | 361 | $95.70 \pm 12.20$ | Smith et al. (2004) |
| HD19787 | Johnson | K | $2.05 \pm 0.01$ | Laney et al. (2012) |
| HD19787 | Johnson | K | $2.07 \pm 0.06$ | Neugebauer & Leighton (1969) |
| HD19787 | Johnson | K | $2.08 \pm 0.05$ | Johnson et al. (1966) |

**Table 21** *(continued)*

| Star ID | System/Wvlen | Band/Bandpass | Value | Reference |
|---------|--------------|---------------|-------|-----------|
| HD19787 | Johnson | K | $2.08 \pm 0.05$ | Ducati (2002) |
| HD19787 | 3500 | 898 | $41.80 \pm 8.50$ | Smith et al. (2004) |
| HD19787 | 4900 | 712 | $20.00 \pm 6.10$ | Smith et al. (2004) |
| HD19787 | 12000 | 6384 | $10.60 \pm 31.10$ | Smith et al. (2004) |
| HD20644 | DDO | m35 | $9.30 \pm 0.05$ | McClure & Forrester (1981) |
| HD20644 | DDO | m38 | $8.00 \pm 0.05$ | McClure & Forrester (1981) |
| HD20644 | DDO | m41 | $8.11 \pm 0.05$ | McClure & Forrester (1981) |
| HD20644 | DDO | m42 | $7.79 \pm 0.05$ | McClure & Forrester (1981) |
| HD20644 | Oja | m42 | $7.31 \pm 0.05$ | Häggkvist & Oja (1970) |
| HD20644 | Johnson | B | $6.00 \pm 0.05$ | Mermilliod (1986) |
| HD20644 | Johnson | B | $6.01 \pm 0.05$ | Johnson et al. (1966) |
| HD20644 | Johnson | B | $6.02 \pm 0.05$ | Argue (1966) |
| HD20644 | Johnson | B | $6.02 \pm 0.05$ | Ducati (2002) |
| HD20644 | Johnson | B | $6.05 \pm 0.05$ | Häggkvist & Oja (1966) |
| HD20644 | KronComet | Bc | $5.88 \pm 0.03$ | This work |
| HD20644 | 13c | m45 | $5.62 \pm 0.05$ | Johnson & Mitchell (1995) |
| HD20644 | DDO | m45 | $6.53 \pm 0.05$ | McClure & Forrester (1981) |
| HD20644 | Oja | m45 | $5.71 \pm 0.05$ | Häggkvist & Oja (1970) |
| HD20644 | DDO | m48 | $5.09 \pm 0.05$ | McClure & Forrester (1981) |
| HD20644 | KronComet | C2 | $4.89 \pm 0.02$ | This work |
| HD20644 | 13c | m52 | $4.91 \pm 0.05$ | Johnson & Mitchell (1995) |
| HD20644 | KronComet | Gc | $4.65 \pm 0.02$ | This work |
| HD20644 | WBVR | V | $4.47 \pm 0.05$ | Kornilov et al. (1991) |
| HD20644 | Johnson | V | $4.45 \pm 0.05$ | Argue (1966) |
| HD20644 | Johnson | V | $4.46 \pm 0.05$ | Mermilliod (1986) |
| HD20644 | Johnson | V | $4.47 \pm 0.05$ | Johnson et al. (1966) |
| HD20644 | Johnson | V | $4.47 \pm 0.05$ | Ducati (2002) |
| HD20644 | Johnson | V | $4.49 \pm 0.05$ | Häggkvist & Oja (1966) |
| HD20644 | Johnson | V | $4.58 \pm 0.09$ | This work |
| HD20644 | 13c | m58 | $4.10 \pm 0.05$ | Johnson & Mitchell (1995) |
| HD20644 | 13c | m63 | $3.66 \pm 0.05$ | Johnson & Mitchell (1995) |
| HD20644 | WBVR | R | $3.32 \pm 0.05$ | Kornilov et al. (1991) |
| HD20644 | KronComet | Rc | $3.07 \pm 0.01$ | This work |
| HD20644 | 13c | m72 | $3.23 \pm 0.05$ | Johnson & Mitchell (1995) |
| HD20644 | 13c | m80 | $2.88 \pm 0.05$ | Johnson & Mitchell (1995) |
| HD20644 | 13c | m86 | $2.72 \pm 0.05$ | Johnson & Mitchell (1995) |
| HD20644 | 13c | m99 | $2.38 \pm 0.05$ | Johnson & Mitchell (1995) |
| HD20644 | 13c | m110 | $2.09 \pm 0.05$ | Johnson & Mitchell (1995) |
| HD20644 | 1250 | 310 | $350.50 \pm 24.40$ | Smith et al. (2004) |
| HD20644 | Johnson | J | $1.74 \pm 0.05$ | Selby et al. (1988) |
| HD20644 | Johnson | J | $1.74 \pm 0.05$ | Ducati (2002) |

**Table 21** *continued on next page*

**Table 21** *(continued)*

| Star ID | System/Wvlen | Band/Bandpass | Value | Reference |
|---------|--------------|---------------|-------|-----------|
| HD20644 | 2200 | 361 | 325.20 ± 18.30 | Smith et al. (2004) |
| HD20644 | Johnson | K | 0.80 ± 0.05 | Ducati (2002) |
| HD20644 | Johnson | K | 0.88 ± 0.05 | Neugebauer & Leighton (1969) |
| HD20644 | 3500 | 898 | 166.00 ± 9.30 | Smith et al. (2004) |
| HD20644 | 4900 | 712 | 73.80 ± 5.50 | Smith et al. (2004) |
| HD20644 | 12000 | 6384 | 27.60 ± 23.80 | Smith et al. (2004) |
| HD20844 | KronComet | C2 | 7.86 ± 0.01 | This work |
| HD20844 | KronComet | Gc | 7.62 ± 0.01 | This work |
| HD20844 | WBVR | V | 7.38 ± 0.05 | Kornilov et al. (1991) |
| HD20844 | Johnson | V | 7.46 ± 0.01 | This work |
| HD20844 | WBVR | R | 5.50 ± 0.05 | Kornilov et al. (1991) |
| HD20844 | KronComet | Rc | 5.74 ± 0.01 | This work |
| HD20844 | Johnson | K | 1.78 ± 0.07 | Neugebauer & Leighton (1969) |
| HD21465 | KronComet | C2 | 7.70 ± 0.04 | This work |
| HD21465 | KronComet | Gc | 7.41 ± 0.03 | This work |
| HD21465 | WBVR | V | 7.06 ± 0.05 | Kornilov et al. (1991) |
| HD21465 | Johnson | V | 7.10 ± 0.05 | Häggkvist & Oja (1969b) |
| HD21465 | Johnson | V | 7.17 ± 0.03 | This work |
| HD21465 | WBVR | R | 5.53 ± 0.05 | Kornilov et al. (1991) |
| HD21465 | KronComet | Rc | 5.31 ± 0.02 | This work |
| HD21465 | Johnson | K | 2.38 ± 0.07 | Neugebauer & Leighton (1969) |
| HD21465 | 3500 | 898 | 34.90 ± 10.20 | Smith et al. (2004) |
| HD21465 | 4900 | 712 | 14.60 ± 6.90 | Smith et al. (2004) |
| HD25604 | Geneva | B1 | 6.10 ± 0.08 | Golay (1972) |
| HD25604 | Oja | m41 | 6.68 ± 0.05 | Häggkvist & Oja (1970) |
| HD25604 | DDO | m42 | 6.85 ± 0.05 | McClure & Forrester (1981) |
| HD25604 | Oja | m42 | 6.40 ± 0.05 | Häggkvist & Oja (1970) |
| HD25604 | Geneva | B | 4.75 ± 0.08 | Golay (1972) |
| HD25604 | KronComet | COp | 5.66 ± 0.11 | This work |
| HD25604 | WBVR | B | 5.45 ± 0.05 | Kornilov et al. (1991) |
| HD25604 | Johnson | B | 5.39 ± 0.05 | Gehrels et al. (1964) |
| HD25604 | Johnson | B | 5.42 ± 0.05 | Argue (1966) |
| HD25604 | Johnson | B | 5.43 ± 0.05 | Häggkvist & Oja (1966) |
| HD25604 | Johnson | B | 5.43 ± 0.05 | Mermilliod (1986) |
| HD25604 | Johnson | B | 5.44 ± 0.05 | Johnson et al. (1966) |
| HD25604 | Johnson | B | 5.44 ± 0.05 | Ducati (2002) |
| HD25604 | Johnson | B | 5.46 ± 0.05 | Johnson (1964) |
| HD25604 | KronComet | Bc | 5.20 ± 0.03 | This work |
| HD25604 | Geneva | B2 | 5.87 ± 0.08 | Golay (1972) |
| HD25604 | 13c | m45 | 5.12 ± 0.05 | Johnson & Mitchell (1995) |
| HD25604 | DDO | m45 | 5.94 ± 0.05 | McClure & Forrester (1981) |

<div align="center">**Table 21** *(continued)*</div>

| Star ID | System/Wvlen | Band/Bandpass | Value | Reference |
|---------|--------------|---------------|-------|-----------|
| HD25604 | Oja | m45 | $5.12 \pm 0.05$ | Häggkvist & Oja (1970) |
| HD25604 | DDO | m48 | $4.73 \pm 0.05$ | McClure & Forrester (1981) |
| HD25604 | KronComet | C2 | $4.55 \pm 0.02$ | This work |
| HD25604 | 13c | m52 | $4.64 \pm 0.05$ | Johnson & Mitchell (1995) |
| HD25604 | KronComet | Gc | $4.42 \pm 0.02$ | This work |
| HD25604 | Geneva | V1 | $5.15 \pm 0.08$ | Golay (1972) |
| HD25604 | WBVR | V | $4.36 \pm 0.05$ | Kornilov et al. (1991) |
| HD25604 | Geneva | V | $4.37 \pm 0.08$ | Golay (1972) |
| HD25604 | Johnson | V | $4.33 \pm 0.05$ | Gehrels et al. (1964) |
| HD25604 | Johnson | V | $4.35 \pm 0.05$ | Argue (1966) |
| HD25604 | Johnson | V | $4.36 \pm 0.05$ | Häggkvist & Oja (1966) |
| HD25604 | Johnson | V | $4.37 \pm 0.05$ | Johnson et al. (1966) |
| HD25604 | Johnson | V | $4.37 \pm 0.05$ | Ducati (2002) |
| HD25604 | Johnson | V | $4.38 \pm 0.05$ | Johnson (1964) |
| HD25604 | Johnson | V | $4.38 \pm 0.05$ | Mermilliod (1986) |
| HD25604 | Johnson | V | $4.38 \pm 0.06$ | This work |
| HD25604 | 13c | m58 | $4.10 \pm 0.05$ | Johnson & Mitchell (1995) |
| HD25604 | Geneva | G | $5.34 \pm 0.08$ | Golay (1972) |
| HD25604 | 13c | m63 | $3.82 \pm 0.05$ | Johnson & Mitchell (1995) |
| HD25604 | WBVR | R | $3.61 \pm 0.05$ | Kornilov et al. (1991) |
| HD25604 | KronComet | Rc | $3.34 \pm 0.01$ | This work |
| HD25604 | 13c | m72 | $3.54 \pm 0.05$ | Johnson & Mitchell (1995) |
| HD25604 | 13c | m80 | $3.31 \pm 0.05$ | Johnson & Mitchell (1995) |
| HD25604 | 13c | m86 | $3.21 \pm 0.05$ | Johnson & Mitchell (1995) |
| HD25604 | 13c | m99 | $3.03 \pm 0.05$ | Johnson & Mitchell (1995) |
| HD25604 | 13c | m110 | $2.82 \pm 0.05$ | Johnson & Mitchell (1995) |
| HD25604 | 1250 | 310 | $171.10 \pm 15.30$ | Smith et al. (2004) |
| HD25604 | Johnson | J | $2.55 \pm 0.05$ | Selby et al. (1988) |
| HD25604 | Johnson | J | $2.55 \pm 0.05$ | Blackwell et al. (1990) |
| HD25604 | Johnson | J | $2.59 \pm 0.05$ | Ducati (2002) |
| HD25604 | Johnson | J | $2.63 \pm 0.05$ | Johnson et al. (1966) |
| HD25604 | Johnson | J | $2.63 \pm 0.05$ | Voelcker (1975) |
| HD25604 | Johnson | J | $2.63 \pm 0.05$ | Shenavrin et al. (2011) |
| HD25604 | Johnson | H | $2.10 \pm 0.05$ | Shenavrin et al. (2011) |
| HD25604 | Johnson | H | $2.13 \pm 0.05$ | Voelcker (1975) |
| HD25604 | Johnson | H | $2.13 \pm 0.05$ | Ducati (2002) |
| HD25604 | 2200 | 361 | $119.60 \pm 10.90$ | Smith et al. (2004) |
| HD25604 | Johnson | K | $1.96 \pm 0.05$ | Ducati (2002) |
| HD25604 | Johnson | K | $1.97 \pm 0.05$ | Johnson et al. (1966) |
| HD25604 | Johnson | K | $1.97 \pm 0.05$ | Shenavrin et al. (2011) |
| HD25604 | Johnson | K | $2.01 \pm 0.07$ | Neugebauer & Leighton (1969) |

<div align="center">**Table 21** *continued on next page*</div>

**Table 21** *(continued)*

| Star ID | System/Wvlen | Band/Bandpass | Value | Reference |
|---------|--------------|---------------|-------|-----------|
| HD25604 | 3500 | 898 | $57.00 \pm 7.20$ | Smith et al. (2004) |
| HD25604 | 4900 | 712 | $27.20 \pm 5.60$ | Smith et al. (2004) |
| HD25604 | 12000 | 6384 | $25.60 \pm 42.40$ | Smith et al. (2004) |
| HD26605 | KronComet | NH | $9.47 \pm 0.02$ | This work |
| HD26605 | KronComet | UVc | $9.09 \pm 0.10$ | This work |
| HD26605 | DDO | m35 | $9.68 \pm 0.05$ | McClure & Forrester (1981) |
| HD26605 | WBVR | W | $8.22 \pm 0.05$ | Kornilov et al. (1991) |
| HD26605 | Johnson | U | $8.03 \pm 0.04$ | This work |
| HD26605 | Johnson | U | $8.35 \pm 0.01$ | Oja (1991) |
| HD26605 | DDO | m38 | $8.61 \pm 0.05$ | McClure & Forrester (1981) |
| HD26605 | KronComet | CN | $8.98 \pm 0.16$ | This work |
| HD26605 | DDO | m41 | $9.11 \pm 0.05$ | McClure & Forrester (1981) |
| HD26605 | DDO | m42 | $8.93 \pm 0.05$ | McClure & Forrester (1981) |
| HD26605 | KronComet | COp | $7.76 \pm 0.10$ | This work |
| HD26605 | WBVR | B | $7.54 \pm 0.05$ | Kornilov et al. (1991) |
| HD26605 | Johnson | B | $7.49 \pm 0.05$ | This work |
| HD26605 | Johnson | B | $7.50 \pm 0.01$ | Oja (1991) |
| HD26605 | KronComet | Bc | $7.33 \pm 0.03$ | This work |
| HD26605 | DDO | m45 | $8.05 \pm 0.05$ | McClure & Forrester (1981) |
| HD26605 | DDO | m48 | $6.85 \pm 0.05$ | McClure & Forrester (1981) |
| HD26605 | KronComet | C2 | $6.66 \pm 0.02$ | This work |
| HD26605 | KronComet | Gc | $6.54 \pm 0.03$ | This work |
| HD26605 | WBVR | V | $6.49 \pm 0.05$ | Kornilov et al. (1991) |
| HD26605 | Johnson | V | $6.45 \pm 0.01$ | Oja (1991) |
| HD26605 | Johnson | V | $6.50 \pm 0.05$ | This work |
| HD26605 | WBVR | R | $5.71 \pm 0.05$ | Kornilov et al. (1991) |
| HD26605 | KronComet | Rc | $5.45 \pm 0.02$ | This work |
| HD27308 | KronComet | C2 | $8.38 \pm 0.04$ | This work |
| HD27308 | KronComet | Gc | $8.14 \pm 0.03$ | This work |
| HD27308 | WBVR | V | $7.89 \pm 0.05$ | Kornilov et al. (1991) |
| HD27308 | Johnson | V | $7.97 \pm 0.04$ | This work |
| HD27308 | WBVR | R | $5.84 \pm 0.05$ | Kornilov et al. (1991) |
| HD27308 | KronComet | Rc | $6.13 \pm 0.03$ | This work |
| HD27308 | 1250 | 310 | $79.70 \pm 7.40$ | Smith et al. (2004) |
| HD27308 | 2200 | 361 | $105.20 \pm 14.00$ | Smith et al. (2004) |
| HD27308 | Johnson | K | $1.84 \pm 0.06$ | Neugebauer & Leighton (1969) |
| HD27308 | 3500 | 898 | $54.30 \pm 9.60$ | Smith et al. (2004) |
| HD27308 | 4900 | 712 | $25.40 \pm 5.80$ | Smith et al. (2004) |
| HD27308 | 12000 | 6384 | $20.50 \pm 27.80$ | Smith et al. (2004) |
| HD27348 | Vilnius | U | $8.39 \pm 0.05$ | Kazlauskas et al. (2005) |
| HD27348 | DDO | m35 | $7.87 \pm 0.05$ | McClure & Forrester (1981) |

**Table 21** *continued on next page*

**Table 21** (continued)

| Star ID | System/Wvlen | Band/Bandpass | Value | Reference |
|---------|--------------|---------------|-------|-----------|
| HD27348 | WBVR | W | $6.43 \pm 0.05$ | Kornilov et al. (1991) |
| HD27348 | Johnson | U | $6.55 \pm 0.05$ | Argue (1966) |
| HD27348 | Johnson | U | $6.57 \pm 0.05$ | McClure (1970) |
| HD27348 | Johnson | U | $6.59 \pm 0.05$ | Mermilliod (1986) |
| HD27348 | Vilnius | P | $7.82 \pm 0.05$ | Kazlauskas et al. (2005) |
| HD27348 | DDO | m38 | $6.83 \pm 0.05$ | McClure & Forrester (1981) |
| HD27348 | Vilnius | X | $6.89 \pm 0.05$ | Kazlauskas et al. (2005) |
| HD27348 | DDO | m41 | $7.44 \pm 0.05$ | McClure & Forrester (1981) |
| HD27348 | Oja | m41 | $6.97 \pm 0.05$ | Häggkvist & Oja (1970) |
| HD27348 | DDO | m42 | $7.23 \pm 0.05$ | McClure & Forrester (1981) |
| HD27348 | Oja | m42 | $6.78 \pm 0.05$ | Häggkvist & Oja (1970) |
| HD27348 | WBVR | B | $5.89 \pm 0.05$ | Kornilov et al. (1991) |
| HD27348 | Johnson | B | $5.86 \pm 0.05$ | Argue (1966) |
| HD27348 | Johnson | B | $5.87 \pm 0.05$ | Häggkvist & Oja (1966) |
| HD27348 | Johnson | B | $5.87 \pm 0.05$ | McClure (1970) |
| HD27348 | Johnson | B | $5.91 \pm 0.05$ | Mermilliod (1986) |
| HD27348 | DDO | m45 | $6.43 \pm 0.05$ | McClure & Forrester (1981) |
| HD27348 | Oja | m45 | $5.60 \pm 0.05$ | Häggkvist & Oja (1970) |
| HD27348 | Vilnius | Y | $5.68 \pm 0.05$ | Kazlauskas et al. (2005) |
| HD27348 | DDO | m48 | $5.27 \pm 0.05$ | McClure & Forrester (1981) |
| HD27348 | Vilnius | Z | $5.24 \pm 0.05$ | Kazlauskas et al. (2005) |
| HD27348 | WBVR | V | $4.94 \pm 0.05$ | Kornilov et al. (1991) |
| HD27348 | Vilnius | V | $4.95 \pm 0.05$ | Kazlauskas et al. (2005) |
| HD27348 | Johnson | V | $4.92 \pm 0.05$ | Argue (1966) |
| HD27348 | Johnson | V | $4.93 \pm 0.05$ | Häggkvist & Oja (1966) |
| HD27348 | Johnson | V | $4.93 \pm 0.05$ | McClure (1970) |
| HD27348 | Johnson | V | $4.95 \pm 0.05$ | Mermilliod (1986) |
| HD27348 | Vilnius | S | $4.25 \pm 0.05$ | Kazlauskas et al. (2005) |
| HD27348 | WBVR | R | $4.25 \pm 0.05$ | Kornilov et al. (1991) |
| HD27348 | 1250 | 310 | $83.00 \pm 12.20$ | Smith et al. (2004) |
| HD27348 | 2200 | 361 | $62.90 \pm 11.30$ | Smith et al. (2004) |
| HD27348 | Johnson | K | $2.76 \pm 0.08$ | Neugebauer & Leighton (1969) |
| HD27348 | 3500 | 898 | $25.80 \pm 8.00$ | Smith et al. (2004) |
| HD27348 | 4900 | 712 | $11.40 \pm 6.30$ | Smith et al. (2004) |
| HD27348 | 12000 | 6384 | $-34.30 \pm 24.20$ | Smith et al. (2004) |
| HD27697 | KronComet | NH | $6.59 \pm 0.02$ | This work |
| HD27697 | KronComet | UVc | $6.37 \pm 0.01$ | This work |
| HD27697 | Geneva | U | $6.07 \pm 0.08$ | Golay (1972) |
| HD27697 | Stromgren | u | $6.73 \pm 0.08$ | Olson (1974) |
| HD27697 | Stromgren | u | $6.79 \pm 0.08$ | Grønbech et al. (1976) |
| HD27697 | Stromgren | u | $6.80 \pm 0.08$ | Olsen (1983) |

**Table 21** (continued)

| Star ID | System/Wvlen | Band/Bandpass | Value | Reference |
|---------|--------------|---------------|-------|-----------|
| HD27697 | Stromgren | u | $6.80 \pm 0.08$ | Gray & Olsen (1991) |
| HD27697 | Stromgren | u | $6.80 \pm 0.08$ | Olsen (1993) |
| HD27697 | Stromgren | u | $6.81 \pm 0.08$ | Hauck & Mermilliod (1998) |
| HD27697 | Stromgren | u | $6.82 \pm 0.08$ | Crawford & Barnes (1970) |
| HD27697 | Stromgren | u | $6.82 \pm 0.08$ | Reglero et al. (1987) |
| HD27697 | Stromgren | u | $6.83 \pm 0.08$ | Crawford & Perry (1966) |
| HD27697 | Johnson | U | $5.36 \pm 0.03$ | This work |
| HD27697 | Johnson | U | $5.49 \pm 0.05$ | Jennens & Helfer (1975) |
| HD27697 | Johnson | U | $5.55 \pm 0.05$ | Johnson & Morgan (1953b) |
| HD27697 | Johnson | U | $5.55 \pm 0.05$ | Shao (1964) |
| HD27697 | Johnson | U | $5.56 \pm 0.05$ | Johnson & Harris (1954) |
| HD27697 | Johnson | U | $5.56 \pm 0.05$ | Johnson & Knuckles (1955) |
| HD27697 | Johnson | U | $5.56 \pm 0.05$ | Johnson (1964) |
| HD27697 | Johnson | U | $5.56 \pm 0.05$ | Ducati (2002) |
| HD27697 | Johnson | U | $5.57 \pm 0.05$ | Johnson et al. (1966) |
| HD27697 | Johnson | U | $5.57 \pm 0.05$ | Argue (1966) |
| HD27697 | Johnson | U | $5.57 \pm 0.05$ | Mermilliod (1986) |
| HD27697 | KronComet | CN | $6.45 \pm 0.01$ | This work |
| HD27697 | Geneva | B1 | $5.34 \pm 0.08$ | Golay (1972) |
| HD27697 | Stromgren | v | $5.35 \pm 0.08$ | Olson (1974) |
| HD27697 | Stromgren | v | $5.38 \pm 0.08$ | Crawford & Barnes (1970) |
| HD27697 | Stromgren | v | $5.38 \pm 0.08$ | Grønbech et al. (1976) |
| HD27697 | Stromgren | v | $5.38 \pm 0.08$ | Olsen (1983) |
| HD27697 | Stromgren | v | $5.38 \pm 0.08$ | Gray & Olsen (1991) |
| HD27697 | Stromgren | v | $5.38 \pm 0.08$ | Olsen (1993) |
| HD27697 | Stromgren | v | $5.38 \pm 0.08$ | Hauck & Mermilliod (1998) |
| HD27697 | Stromgren | v | $5.39 \pm 0.08$ | Crawford & Perry (1966) |
| HD27697 | Stromgren | v | $5.39 \pm 0.08$ | Reglero et al. (1987) |
| HD27697 | Geneva | B | $4.04 \pm 0.08$ | Golay (1972) |
| HD27697 | KronComet | COp | $4.96 \pm 0.01$ | This work |
| HD27697 | Johnson | B | $4.58 \pm 0.01$ | This work |
| HD27697 | Johnson | B | $4.69 \pm 0.05$ | Jennens & Helfer (1975) |
| HD27697 | Johnson | B | $4.71 \pm 0.05$ | Johnson & Morgan (1953b) |
| HD27697 | Johnson | B | $4.71 \pm 0.05$ | Shao (1964) |
| HD27697 | Johnson | B | $4.74 \pm 0.05$ | Johnson & Harris (1954) |
| HD27697 | Johnson | B | $4.74 \pm 0.05$ | Johnson (1964) |
| HD27697 | Johnson | B | $4.74 \pm 0.05$ | Argue (1966) |
| HD27697 | Johnson | B | $4.74 \pm 0.05$ | Coleman (1982) |
| HD27697 | Johnson | B | $4.74 \pm 0.05$ | Mermilliod (1986) |
| HD27697 | Johnson | B | $4.74 \pm 0.05$ | Ducati (2002) |
| HD27697 | Johnson | B | $4.75 \pm 0.05$ | Johnson & Knuckles (1955) |

**Table 21** (continued)

| Star ID | System/Wvlen | Band/Bandpass | Value | Reference |
|---------|--------------|---------------|-------|-----------|
| HD27697 | Johnson | B | $4.75 \pm 0.05$ | Gehrels et al. (1964) |
| HD27697 | Johnson | B | $4.75 \pm 0.05$ | Häggkvist & Oja (1966) |
| HD27697 | Johnson | B | $4.75 \pm 0.05$ | Johnson et al. (1966) |
| HD27697 | KronComet | Bc | $4.53 \pm 0.01$ | This work |
| HD27697 | Geneva | B2 | $5.19 \pm 0.08$ | Golay (1972) |
| HD27697 | Stromgren | b | $4.34 \pm 0.08$ | Olson (1974) |
| HD27697 | Stromgren | b | $4.35 \pm 0.08$ | Grønbech et al. (1976) |
| HD27697 | Stromgren | b | $4.35 \pm 0.08$ | Reglero et al. (1987) |
| HD27697 | Stromgren | b | $4.36 \pm 0.08$ | Crawford & Perry (1966) |
| HD27697 | Stromgren | b | $4.36 \pm 0.08$ | Crawford & Barnes (1970) |
| HD27697 | Stromgren | b | $4.36 \pm 0.08$ | Olsen (1983) |
| HD27697 | Stromgren | b | $4.36 \pm 0.08$ | Gray & Olsen (1991) |
| HD27697 | Stromgren | b | $4.36 \pm 0.08$ | Olsen (1993) |
| HD27697 | Stromgren | b | $4.36 \pm 0.08$ | Hauck & Mermilliod (1998) |
| HD27697 | KronComet | C2 | $3.87 \pm 0.01$ | This work |
| HD27697 | KronComet | Gc | $3.80 \pm 0.01$ | This work |
| HD27697 | Geneva | V1 | $4.54 \pm 0.08$ | Golay (1972) |
| HD27697 | Stromgren | y | $3.76 \pm 0.08$ | Crawford & Perry (1966) |
| HD27697 | Stromgren | y | $3.76 \pm 0.08$ | Crawford & Barnes (1970) |
| HD27697 | Stromgren | y | $3.76 \pm 0.08$ | Olson (1974) |
| HD27697 | Stromgren | y | $3.76 \pm 0.08$ | Grønbech et al. (1976) |
| HD27697 | Stromgren | y | $3.76 \pm 0.08$ | Olsen (1983) |
| HD27697 | Stromgren | y | $3.76 \pm 0.08$ | Reglero et al. (1987) |
| HD27697 | Stromgren | y | $3.76 \pm 0.08$ | Gray & Olsen (1991) |
| HD27697 | Stromgren | y | $3.76 \pm 0.08$ | Olsen (1993) |
| HD27697 | Stromgren | y | $3.76 \pm 0.08$ | Hauck & Mermilliod (1998) |
| HD27697 | Geneva | V | $3.77 \pm 0.08$ | Golay (1972) |
| HD27697 | Johnson | V | $3.69 \pm 0.01$ | This work |
| HD27697 | Johnson | V | $3.73 \pm 0.05$ | Johnson & Morgan (1953b) |
| HD27697 | Johnson | V | $3.73 \pm 0.05$ | Shao (1964) |
| HD27697 | Johnson | V | $3.75 \pm 0.05$ | Mermilliod (1986) |
| HD27697 | Johnson | V | $3.76 \pm 0.05$ | Johnson & Harris (1954) |
| HD27697 | Johnson | V | $3.76 \pm 0.05$ | Gehrels et al. (1964) |
| HD27697 | Johnson | V | $3.76 \pm 0.05$ | Johnson (1964) |
| HD27697 | Johnson | V | $3.76 \pm 0.05$ | Häggkvist & Oja (1966) |
| HD27697 | Johnson | V | $3.76 \pm 0.05$ | Johnson et al. (1966) |
| HD27697 | Johnson | V | $3.76 \pm 0.05$ | Argue (1966) |
| HD27697 | Johnson | V | $3.76 \pm 0.05$ | Coleman (1982) |
| HD27697 | Johnson | V | $3.76 \pm 0.05$ | Ducati (2002) |
| HD27697 | Johnson | V | $3.77 \pm 0.05$ | Johnson & Knuckles (1955) |
| HD27697 | Johnson | V | $3.77 \pm 0.05$ | Jennens & Helfer (1975) |

**Table 21** *continued on next page*

**Table 21** *(continued)*

| Star ID | System/Wvlen | Band/Bandpass | Value | Reference |
|---------|--------------|---------------|-------|-----------|
| HD27697 | Geneva | G | $4.76 \pm 0.08$ | Golay (1972) |
| HD27697 | KronComet | Rc | $2.77 \pm 0.01$ | This work |
| HD27697 | Johnson | J | $2.12 \pm 0.05$ | Selby et al. (1988) |
| HD27697 | Johnson | J | $2.12 \pm 0.05$ | Blackwell et al. (1990) |
| HD27697 | Johnson | J | $2.15 \pm 0.05$ | Ducati (2002) |
| HD27697 | Johnson | J | $2.19 \pm 0.05$ | Iriarte (1970) |
| HD27697 | Johnson | J | $2.23 \pm 0.05$ | Johnson et al. (1966) |
| HD27697 | Johnson | H | $1.75 \pm 0.05$ | Iriarte (1970) |
| HD27697 | 2200 | 361 | $155.40 \pm 7.60$ | Smith et al. (2004) |
| HD27697 | Johnson | K | $1.59 \pm 0.05$ | Ducati (2002) |
| HD27697 | Johnson | K | $1.60 \pm 0.05$ | Neugebauer & Leighton (1969) |
| HD27697 | Johnson | K | $1.62 \pm 0.05$ | Iriarte (1970) |
| HD27697 | Johnson | K | $1.64 \pm 0.05$ | Johnson et al. (1966) |
| HD27697 | Johnson | L | $1.43 \pm 0.05$ | Iriarte (1970) |
| HD27697 | Johnson | L | $1.55 \pm 0.05$ | Johnson et al. (1966) |
| HD27697 | Johnson | L | $1.55 \pm 0.05$ | Ducati (2002) |
| HD27697 | 3500 | 898 | $69.00 \pm 7.80$ | Smith et al. (2004) |
| HD27697 | Johnson | N | $0.32 \pm 0.05$ | Johnson (1964) |
| HD27697 | 4900 | 712 | $33.80 \pm 4.50$ | Smith et al. (2004) |
| HD27697 | 12000 | 6384 | $-0.90 \pm 25.70$ | Smith et al. (2004) |
| HD28100 | 13c | m33 | $6.49 \pm 0.05$ | Johnson & Mitchell (1995) |
| HD28100 | Geneva | U | $6.91 \pm 0.08$ | Golay (1972) |
| HD28100 | 13c | m35 | $6.23 \pm 0.05$ | Johnson & Mitchell (1995) |
| HD28100 | DDO | m35 | $7.71 \pm 0.05$ | McClure & Forrester (1981) |
| HD28100 | WBVR | W | $6.24 \pm 0.05$ | Kornilov et al. (1991) |
| HD28100 | Johnson | U | $6.39 \pm 0.05$ | Johnson et al. (1966) |
| HD28100 | Johnson | U | $6.39 \pm 0.05$ | Argue (1966) |
| HD28100 | Johnson | U | $6.39 \pm 0.05$ | Ducati (2002) |
| HD28100 | Johnson | U | $6.42 \pm 0.05$ | Mermilliod (1986) |
| HD28100 | 13c | m37 | $6.28 \pm 0.05$ | Johnson & Mitchell (1995) |
| HD28100 | DDO | m38 | $6.65 \pm 0.05$ | McClure & Forrester (1981) |
| HD28100 | 13c | m40 | $6.20 \pm 0.05$ | Johnson & Mitchell (1995) |
| HD28100 | Geneva | B1 | $6.23 \pm 0.08$ | Golay (1972) |
| HD28100 | DDO | m41 | $7.23 \pm 0.05$ | McClure & Forrester (1981) |
| HD28100 | Oja | m41 | $6.78 \pm 0.05$ | Häggkvist & Oja (1970) |
| HD28100 | DDO | m42 | $7.02 \pm 0.05$ | McClure & Forrester (1981) |
| HD28100 | Oja | m42 | $6.58 \pm 0.05$ | Häggkvist & Oja (1970) |
| HD28100 | Geneva | B | $4.96 \pm 0.08$ | Golay (1972) |
| HD28100 | WBVR | B | $5.68 \pm 0.05$ | Kornilov et al. (1991) |
| HD28100 | Johnson | B | $5.67 \pm 0.05$ | Häggkvist & Oja (1966) |
| HD28100 | Johnson | B | $5.67 \pm 0.05$ | Johnson et al. (1966) |

**Table 21** *(continued)*

| Star ID | System/Wvlen | Band/Bandpass | Value | Reference |
|---------|-------------|---------------|-------|-----------|
| HD28100 | Johnson | B | $5.67 \pm 0.05$ | Argue (1966) |
| HD28100 | Johnson | B | $5.67 \pm 0.05$ | Ducati (2002) |
| HD28100 | Johnson | B | $5.69 \pm 0.05$ | Mermilliod (1986) |
| HD28100 | Geneva | B2 | $6.13 \pm 0.08$ | Golay (1972) |
| HD28100 | 13c | m45 | $5.39 \pm 0.05$ | Johnson & Mitchell (1995) |
| HD28100 | DDO | m45 | $6.22 \pm 0.05$ | McClure & Forrester (1981) |
| HD28100 | Oja | m45 | $5.41 \pm 0.05$ | Häggkvist & Oja (1970) |
| HD28100 | DDO | m48 | $5.04 \pm 0.05$ | McClure & Forrester (1981) |
| HD28100 | 13c | m52 | $4.92 \pm 0.05$ | Johnson & Mitchell (1995) |
| HD28100 | Geneva | V1 | $5.46 \pm 0.08$ | Golay (1972) |
| HD28100 | WBVR | V | $4.69 \pm 0.05$ | Kornilov et al. (1991) |
| HD28100 | Geneva | V | $4.70 \pm 0.08$ | Golay (1972) |
| HD28100 | Johnson | V | $4.69 \pm 0.05$ | Häggkvist & Oja (1966) |
| HD28100 | Johnson | V | $4.69 \pm 0.05$ | Johnson et al. (1966) |
| HD28100 | Johnson | V | $4.69 \pm 0.05$ | Argue (1966) |
| HD28100 | Johnson | V | $4.69 \pm 0.05$ | Ducati (2002) |
| HD28100 | Johnson | V | $4.70 \pm 0.05$ | Mermilliod (1986) |
| HD28100 | 13c | m58 | $4.45 \pm 0.05$ | Johnson & Mitchell (1995) |
| HD28100 | Geneva | G | $5.69 \pm 0.08$ | Golay (1972) |
| HD28100 | 13c | m63 | $4.18 \pm 0.05$ | Johnson & Mitchell (1995) |
| HD28100 | WBVR | R | $3.98 \pm 0.05$ | Kornilov et al. (1991) |
| HD28100 | 13c | m72 | $3.91 \pm 0.05$ | Johnson & Mitchell (1995) |
| HD28100 | 13c | m80 | $3.69 \pm 0.05$ | Johnson & Mitchell (1995) |
| HD28100 | 13c | m86 | $3.58 \pm 0.05$ | Johnson & Mitchell (1995) |
| HD28100 | 13c | m99 | $3.42 \pm 0.05$ | Johnson & Mitchell (1995) |
| HD28100 | 13c | m110 | $3.27 \pm 0.05$ | Johnson & Mitchell (1995) |
| HD28100 | Johnson | J | $3.01 \pm 0.05$ | Selby et al. (1988) |
| HD28100 | Johnson | J | $3.01 \pm 0.05$ | Blackwell et al. (1990) |
| HD28100 | Johnson | J | $3.01 \pm 0.05$ | Ducati (2002) |
| HD28100 | Johnson | K | $2.44 \pm 0.05$ | Ducati (2002) |
| HD28100 | Johnson | K | $2.44 \pm 0.09$ | Neugebauer & Leighton (1969) |
| HD28292 | WBVR | W | $7.08 \pm 0.05$ | Kornilov et al. (1991) |
| HD28292 | Johnson | U | $7.13 \pm 0.05$ | Mermilliod (1986) |
| HD28292 | Johnson | U | $7.19 \pm 0.05$ | Argue (1966) |
| HD28292 | Johnson | U | $7.20 \pm 0.05$ | Gutierrez-Moreno & et al. (1966) |
| HD28292 | Oja | m41 | $7.37 \pm 0.05$ | Häggkvist & Oja (1970) |
| HD28292 | Oja | m42 | $7.11 \pm 0.05$ | Häggkvist & Oja (1970) |
| HD28292 | KronComet | COp | $6.43 \pm 0.13$ | This work |
| HD28292 | WBVR | B | $6.13 \pm 0.05$ | Kornilov et al. (1991) |
| HD28292 | Johnson | B | $6.09 \pm 0.05$ | Argue (1966) |
| HD28292 | Johnson | B | $6.11 \pm 0.05$ | Gutierrez-Moreno & et al. (1966) |

**Table 21** *(continued)*

| Star ID | System/Wvlen | Band/Bandpass | Value | Reference |
|---------|--------------|---------------|-------|-----------|
| HD28292 | Johnson | B | $6.12 \pm 0.05$ | Mermilliod (1986) |
| HD28292 | KronComet | Bc | $5.91 \pm 0.03$ | This work |
| HD28292 | Oja | m45 | $5.77 \pm 0.05$ | Häggkvist & Oja (1970) |
| HD28292 | KronComet | C2 | $5.24 \pm 0.03$ | This work |
| HD28292 | KronComet | Gc | $5.07 \pm 0.02$ | This work |
| HD28292 | WBVR | V | $4.97 \pm 0.05$ | Kornilov et al. (1991) |
| HD28292 | Johnson | V | $4.96 \pm 0.05$ | Argue (1966) |
| HD28292 | Johnson | V | $4.97 \pm 0.05$ | Gutierrez-Moreno & et al. (1966) |
| HD28292 | Johnson | V | $4.98 \pm 0.05$ | Mermilliod (1986) |
| HD28292 | Johnson | V | $5.00 \pm 0.04$ | This work |
| HD28292 | WBVR | R | $4.14 \pm 0.05$ | Kornilov et al. (1991) |
| HD28292 | KronComet | Rc | $3.89 \pm 0.02$ | This work |
| HD28292 | Johnson | J | $3.04 \pm 0.01$ | Laney et al. (2012) |
| HD28292 | Johnson | H | $2.48 \pm 0.01$ | Laney et al. (2012) |
| HD28292 | Johnson | K | $2.30 \pm 0.07$ | Neugebauer & Leighton (1969) |
| HD28292 | Johnson | K | $2.34 \pm 0.01$ | Laney et al. (2012) |
| HD28292 | 3500 | 898 | $130.60 \pm 64.20$ | Smith et al. (2004) |
| HD28292 | 12000 | 6384 | $11.20 \pm 30.40$ | Smith et al. (2004) |
| HD28305 | KronComet | NH | $6.53 \pm 0.02$ | This work |
| HD28305 | KronComet | UVc | $6.32 \pm 0.02$ | This work |
| HD28305 | Stromgren | u | $6.68 \pm 0.08$ | Crawford & Perry (1966) |
| HD28305 | Stromgren | u | $6.69 \pm 0.08$ | Grønbech et al. (1976) |
| HD28305 | Stromgren | u | $6.69 \pm 0.08$ | Olsen (1983) |
| HD28305 | Stromgren | u | $6.69 \pm 0.08$ | Olsen (1993) |
| HD28305 | Stromgren | u | $6.69 \pm 0.08$ | Hauck & Mermilliod (1998) |
| HD28305 | Stromgren | u | $6.70 \pm 0.08$ | Crawford & Barnes (1970) |
| HD28305 | Stromgren | u | $6.70 \pm 0.08$ | Gray & Olsen (1991) |
| HD28305 | Johnson | U | $5.13 \pm 0.06$ | This work |
| HD28305 | Johnson | U | $5.41 \pm 0.05$ | Johnson & Knuckles (1955) |
| HD28305 | Johnson | U | $5.41 \pm 0.05$ | Argue (1966) |
| HD28305 | Johnson | U | $5.42 \pm 0.05$ | Johnson et al. (1966) |
| HD28305 | Johnson | U | $5.42 \pm 0.05$ | Ducati (2002) |
| HD28305 | Johnson | U | $5.44 \pm 0.05$ | Johnson & Harris (1954) |
| HD28305 | Johnson | U | $5.44 \pm 0.05$ | Hogg (1958) |
| HD28305 | Johnson | U | $5.44 \pm 0.05$ | Grant (1959) |
| HD28305 | Johnson | U | $5.44 \pm 0.05$ | Shao (1964) |
| HD28305 | Johnson | U | $5.44 \pm 0.05$ | Johnson (1964) |
| HD28305 | Johnson | U | $5.45 \pm 0.05$ | Johnson & Morgan (1953b) |
| HD28305 | Johnson | U | $5.47 \pm 0.05$ | Jennens & Helfer (1975) |
| HD28305 | KronComet | CN | $6.30 \pm 0.06$ | This work |
| HD28305 | Stromgren | v | $5.20 \pm 0.08$ | Crawford & Perry (1966) |

Table 21 (continued)

| Star ID | System/Wvlen | Band/Bandpass | Value | Reference |
|---------|-------------|---------------|-------|-----------|
| HD28305 | Stromgren | v | $5.21 \pm 0.08$ | Crawford & Barnes (1970) |
| HD28305 | Stromgren | v | $5.21 \pm 0.08$ | Grønbech et al. (1976) |
| HD28305 | Stromgren | v | $5.21 \pm 0.08$ | Olsen (1983) |
| HD28305 | Stromgren | v | $5.21 \pm 0.08$ | Hauck & Mermilliod (1998) |
| HD28305 | Stromgren | v | $5.22 \pm 0.08$ | Gray & Olsen (1991) |
| HD28305 | Stromgren | v | $5.22 \pm 0.08$ | Olsen (1993) |
| HD28305 | KronComet | COp | $4.74 \pm 0.05$ | This work |
| HD28305 | Johnson | B | $4.37 \pm 0.06$ | This work |
| HD28305 | Johnson | B | $4.53 \pm 0.05$ | Johnson & Knuckles (1955) |
| HD28305 | Johnson | B | $4.53 \pm 0.05$ | Gehrels et al. (1964) |
| HD28305 | Johnson | B | $4.53 \pm 0.05$ | Argue (1966) |
| HD28305 | Johnson | B | $4.54 \pm 0.05$ | Coleman (1982) |
| HD28305 | Johnson | B | $4.54 \pm 0.05$ | Ducati (2002) |
| HD28305 | Johnson | B | $4.55 \pm 0.05$ | Häggkvist & Oja (1966) |
| HD28305 | Johnson | B | $4.55 \pm 0.05$ | Johnson et al. (1966) |
| HD28305 | Johnson | B | $4.56 \pm 0.05$ | Johnson & Harris (1954) |
| HD28305 | Johnson | B | $4.56 \pm 0.05$ | Hogg (1958) |
| HD28305 | Johnson | B | $4.56 \pm 0.05$ | Grant (1959) |
| HD28305 | Johnson | B | $4.56 \pm 0.05$ | Shao (1964) |
| HD28305 | Johnson | B | $4.56 \pm 0.05$ | Jennens & Helfer (1975) |
| HD28305 | Johnson | B | $4.58 \pm 0.05$ | Johnson & Morgan (1953b) |
| HD28305 | Johnson | B | $4.58 \pm 0.05$ | Johnson (1964) |
| HD28305 | KronComet | Bc | $4.33 \pm 0.05$ | This work |
| HD28305 | Stromgren | b | $4.14 \pm 0.08$ | Crawford & Perry (1966) |
| HD28305 | Stromgren | b | $4.14 \pm 0.08$ | Grønbech et al. (1976) |
| HD28305 | Stromgren | b | $4.14 \pm 0.08$ | Olsen (1993) |
| HD28305 | Stromgren | b | $4.15 \pm 0.08$ | Crawford & Barnes (1970) |
| HD28305 | Stromgren | b | $4.15 \pm 0.08$ | Olsen (1983) |
| HD28305 | Stromgren | b | $4.15 \pm 0.08$ | Gray & Olsen (1991) |
| HD28305 | Stromgren | b | $4.15 \pm 0.08$ | Hauck & Mermilliod (1998) |
| HD28305 | KronComet | C2 | $3.66 \pm 0.05$ | This work |
| HD28305 | KronComet | Gc | $3.59 \pm 0.04$ | This work |
| HD28305 | Stromgren | y | $3.53 \pm 0.08$ | Crawford & Perry (1966) |
| HD28305 | Stromgren | y | $3.53 \pm 0.08$ | Crawford & Barnes (1970) |
| HD28305 | Stromgren | y | $3.53 \pm 0.08$ | Grønbech et al. (1976) |
| HD28305 | Stromgren | y | $3.53 \pm 0.08$ | Olsen (1983) |
| HD28305 | Stromgren | y | $3.53 \pm 0.08$ | Gray & Olsen (1991) |
| HD28305 | Stromgren | y | $3.53 \pm 0.08$ | Olsen (1993) |
| HD28305 | Stromgren | y | $3.53 \pm 0.08$ | Hauck & Mermilliod (1998) |
| HD28305 | Johnson | V | $3.52 \pm 0.05$ | Johnson & Knuckles (1955) |
| HD28305 | Johnson | V | $3.52 \pm 0.05$ | Gehrels et al. (1964) |

Table 21 continued on next page

**Table 21** *(continued)*

| Star ID | System/Wvlen | Band/Bandpass | Value | Reference |
|---------|--------------|---------------|-------|-----------|
| HD28305 | Johnson | V | $3.52 \pm 0.05$ | Argue (1966) |
| HD28305 | Johnson | V | $3.53 \pm 0.05$ | Coleman (1982) |
| HD28305 | Johnson | V | $3.53 \pm 0.05$ | Ducati (2002) |
| HD28305 | Johnson | V | $3.54 \pm 0.05$ | Johnson & Harris (1954) |
| HD28305 | Johnson | V | $3.54 \pm 0.05$ | Hogg (1958) |
| HD28305 | Johnson | V | $3.54 \pm 0.05$ | Grant (1959) |
| HD28305 | Johnson | V | $3.54 \pm 0.05$ | Shao (1964) |
| HD28305 | Johnson | V | $3.54 \pm 0.05$ | Häggkvist & Oja (1966) |
| HD28305 | Johnson | V | $3.54 \pm 0.05$ | Johnson et al. (1966) |
| HD28305 | Johnson | V | $3.54 \pm 0.10$ | This work |
| HD28305 | Johnson | V | $3.55 \pm 0.05$ | Johnson & Morgan (1953b) |
| HD28305 | Johnson | V | $3.55 \pm 0.05$ | Johnson (1964) |
| HD28305 | Johnson | V | $3.55 \pm 0.05$ | Jennens & Helfer (1975) |
| HD28305 | KronComet | Rc | $2.54 \pm 0.02$ | This work |
| HD28305 | 1250 | 310 | $275.60 \pm 9.00$ | Smith et al. (2004) |
| HD28305 | Johnson | J | $1.88 \pm 0.05$ | Selby et al. (1988) |
| HD28305 | Johnson | J | $1.88 \pm 0.05$ | Blackwell et al. (1990) |
| HD28305 | Johnson | J | $1.90 \pm 0.05$ | Ducati (2002) |
| HD28305 | Johnson | J | $1.93 \pm 0.05$ | Iriarte (1970) |
| HD28305 | Johnson | J | $1.94 \pm 0.05$ | Johnson et al. (1966) |
| HD28305 | Johnson | J | $1.94 \pm 0.05$ | Shenavrin et al. (2011) |
| HD28305 | Johnson | H | $1.45 \pm 0.05$ | Shenavrin et al. (2011) |
| HD28305 | Johnson | H | $1.46 \pm 0.05$ | Iriarte (1970) |
| HD28305 | 2200 | 361 | $184.30 \pm 7.20$ | Smith et al. (2004) |
| HD28305 | Johnson | K | $1.31 \pm 0.05$ | Neugebauer & Leighton (1969) |
| HD28305 | Johnson | K | $1.32 \pm 0.05$ | Ducati (2002) |
| HD28305 | Johnson | K | $1.33 \pm 0.05$ | Johnson et al. (1966) |
| HD28305 | Johnson | K | $1.33 \pm 0.05$ | Shenavrin et al. (2011) |
| HD28305 | Johnson | K | $1.38 \pm 0.05$ | Iriarte (1970) |
| HD28305 | Johnson | L | $1.21 \pm 0.05$ | Iriarte (1970) |
| HD28305 | 3500 | 898 | $88.90 \pm 6.30$ | Smith et al. (2004) |
| HD28305 | 4900 | 712 | $43.30 \pm 5.10$ | Smith et al. (2004) |
| HD28305 | 12000 | 6384 | $20.60 \pm 33.00$ | Smith et al. (2004) |
| HD28307 | Stromgren | u | $6.74 \pm 0.08$ | Crawford & Perry (1966) |
| HD28307 | Stromgren | u | $6.75 \pm 0.08$ | Crawford & Barnes (1970) |
| HD28307 | Stromgren | u | $6.77 \pm 0.08$ | Gray & Olsen (1991) |
| HD28307 | Stromgren | u | $6.77 \pm 0.08$ | Hauck & Mermilliod (1998) |
| HD28307 | Stromgren | u | $6.78 \pm 0.08$ | Grønbech et al. (1976) |
| HD28307 | Stromgren | u | $6.79 \pm 0.08$ | Olsen (1983) |
| HD28307 | Johnson | U | $5.48 \pm 0.05$ | Gutierrez-Moreno & et al. (1966) |
| HD28307 | Johnson | U | $5.49 \pm 0.05$ | Johnson et al. (1966) |

| Star ID | System/Wvlen | Band/Bandpass | Value | Reference |
|---------|--------------|---------------|-------|-----------|
| HD28307 | Johnson | U | $5.50 \pm 0.05$ | Johnson et al. (1966) |
| HD28307 | Johnson | U | $5.51 \pm 0.01$ | Oja (1993) |
| HD28307 | Johnson | U | $5.52 \pm 0.05$ | Tolbert (1964) |
| HD28307 | Johnson | U | $5.52 \pm 0.05$ | Ducati (2002) |
| HD28307 | Johnson | U | $5.53 \pm 0.05$ | Argue (1966) |
| HD28307 | Johnson | U | $5.55 \pm 0.05$ | Johnson & Knuckles (1955) |
| HD28307 | Johnson | U | $5.55 \pm 0.05$ | Johnson (1964) |
| HD28307 | Johnson | U | $5.56 \pm 0.05$ | de Vaucouleurs (1959) |
| HD28307 | Johnson | U | $5.59 \pm 0.05$ | Jennens & Helfer (1975) |
| HD28307 | Stromgren | v | $5.39 \pm 0.08$ | Crawford & Perry (1966) |
| HD28307 | Stromgren | v | $5.39 \pm 0.08$ | Crawford & Barnes (1970) |
| HD28307 | Stromgren | v | $5.40 \pm 0.08$ | Gray & Olsen (1991) |
| HD28307 | Stromgren | v | $5.40 \pm 0.08$ | Hauck & Mermilliod (1998) |
| HD28307 | Stromgren | v | $5.41 \pm 0.08$ | Grønbech et al. (1976) |
| HD28307 | Stromgren | v | $5.41 \pm 0.08$ | Olsen (1983) |
| HD28307 | Johnson | B | $4.78 \pm 0.01$ | Oja (1993) |
| HD28307 | Johnson | B | $4.78 \pm 0.05$ | Johnson et al. (1966) |
| HD28307 | Johnson | B | $4.78 \pm 0.05$ | Gutierrez-Moreno & et al. (1966) |
| HD28307 | Johnson | B | $4.79 \pm 0.05$ | Ducati (2002) |
| HD28307 | Johnson | B | $4.80 \pm 0.05$ | Tolbert (1964) |
| HD28307 | Johnson | B | $4.80 \pm 0.05$ | Häggkvist & Oja (1966) |
| HD28307 | Johnson | B | $4.80 \pm 0.05$ | Argue (1966) |
| HD28307 | Johnson | B | $4.80 \pm 0.05$ | Coleman (1982) |
| HD28307 | Johnson | B | $4.81 \pm 0.05$ | Johnson & Knuckles (1955) |
| HD28307 | Johnson | B | $4.81 \pm 0.05$ | de Vaucouleurs (1959) |
| HD28307 | Johnson | B | $4.81 \pm 0.05$ | Johnson (1964) |
| HD28307 | Johnson | B | $4.84 \pm 0.05$ | Jennens & Helfer (1975) |
| HD28307 | Stromgren | b | $4.42 \pm 0.08$ | Crawford & Perry (1966) |
| HD28307 | Stromgren | b | $4.42 \pm 0.08$ | Crawford & Barnes (1970) |
| HD28307 | Stromgren | b | $4.42 \pm 0.08$ | Grønbech et al. (1976) |
| HD28307 | Stromgren | b | $4.42 \pm 0.08$ | Gray & Olsen (1991) |
| HD28307 | Stromgren | b | $4.42 \pm 0.08$ | Hauck & Mermilliod (1998) |
| HD28307 | Stromgren | b | $4.43 \pm 0.08$ | Olsen (1983) |
| HD28307 | Stromgren | y | $3.84 \pm 0.08$ | Crawford & Perry (1966) |
| HD28307 | Stromgren | y | $3.84 \pm 0.08$ | Crawford & Barnes (1970) |
| HD28307 | Stromgren | y | $3.84 \pm 0.08$ | Grønbech et al. (1976) |
| HD28307 | Stromgren | y | $3.84 \pm 0.08$ | Olsen (1983) |
| HD28307 | Stromgren | y | $3.84 \pm 0.08$ | Gray & Olsen (1991) |
| HD28307 | Stromgren | y | $3.84 \pm 0.08$ | Hauck & Mermilliod (1998) |
| HD28307 | Johnson | V | $3.83 \pm 0.05$ | Johnson et al. (1966) |
| HD28307 | Johnson | V | $3.83 \pm 0.05$ | Gutierrez-Moreno & et al. (1966) |

**Table 21** *(continued)*

| Star ID | System/Wvlen | Band/Bandpass | Value | Reference |
|---------|--------------|---------------|-------|-----------|
| HD28307 | Johnson | V | $3.84 \pm 0.01$ | Oja (1993) |
| HD28307 | Johnson | V | $3.84 \pm 0.05$ | Argue (1966) |
| HD28307 | Johnson | V | $3.84 \pm 0.05$ | Ducati (2002) |
| HD28307 | Johnson | V | $3.85 \pm 0.05$ | Johnson & Knuckles (1955) |
| HD28307 | Johnson | V | $3.85 \pm 0.05$ | de Vaucouleurs (1959) |
| HD28307 | Johnson | V | $3.85 \pm 0.05$ | Tolbert (1964) |
| HD28307 | Johnson | V | $3.85 \pm 0.05$ | Johnson (1964) |
| HD28307 | Johnson | V | $3.85 \pm 0.05$ | Häggkvist & Oja (1966) |
| HD28307 | Johnson | V | $3.85 \pm 0.05$ | Coleman (1982) |
| HD28307 | Johnson | V | $3.88 \pm 0.05$ | Jennens & Helfer (1975) |
| HD28307 | Johnson | J | $2.29 \pm 0.05$ | Johnson et al. (1966) |
| HD28307 | Johnson | J | $2.29 \pm 0.05$ | Iriarte (1970) |
| HD28307 | Johnson | J | $2.29 \pm 0.05$ | Ducati (2002) |
| HD28307 | Johnson | J | $2.29 \pm 0.05$ | Shenavrin et al. (2011) |
| HD28307 | Johnson | H | $1.80 \pm 0.05$ | Iriarte (1970) |
| HD28307 | Johnson | H | $1.84 \pm 0.05$ | Shenavrin et al. (2011) |
| HD28307 | Johnson | K | $1.64 \pm 0.06$ | Neugebauer & Leighton (1969) |
| HD28307 | Johnson | K | $1.72 \pm 0.05$ | Iriarte (1970) |
| HD28307 | Johnson | K | $1.73 \pm 0.05$ | Johnson et al. (1966) |
| HD28307 | Johnson | K | $1.73 \pm 0.05$ | Ducati (2002) |
| HD28307 | Johnson | K | $1.73 \pm 0.05$ | Shenavrin et al. (2011) |
| HD28307 | Johnson | L | $1.65 \pm 0.05$ | Iriarte (1970) |
| HD28307 | 3500 | 898 | $201.50 \pm 35.20$ | Smith et al. (2004) |
| HD28307 | 12000 | 6384 | $46.50 \pm 27.60$ | Smith et al. (2004) |
| HD28581 | KronComet | COp | $9.39 \pm 0.10$ | This work |
| HD28581 | WBVR | B | $8.87 \pm 0.05$ | Kornilov et al. (1991) |
| HD28581 | Johnson | B | $8.68 \pm 0.05$ | This work |
| HD28581 | Johnson | B | $8.77 \pm 0.05$ | Landolt (1967) |
| HD28581 | Johnson | B | $8.79 \pm 0.05$ | Slutskij et al. (1980) |
| HD28581 | KronComet | Bc | $8.69 \pm 0.03$ | This work |
| HD28581 | KronComet | C2 | $7.58 \pm 0.03$ | This work |
| HD28581 | KronComet | Gc | $7.34 \pm 0.03$ | This work |
| HD28581 | WBVR | V | $7.08 \pm 0.05$ | Kornilov et al. (1991) |
| HD28581 | Johnson | V | $7.04 \pm 0.05$ | Landolt (1967) |
| HD28581 | Johnson | V | $7.13 \pm 0.05$ | Slutskij et al. (1980) |
| HD28581 | Johnson | V | $7.17 \pm 0.04$ | This work |
| HD28581 | WBVR | R | $5.72 \pm 0.05$ | Kornilov et al. (1991) |
| HD28581 | KronComet | Rc | $5.47 \pm 0.02$ | This work |
| HD28581 | 1250 | 310 | $44.80 \pm 7.90$ | Smith et al. (2004) |
| HD28581 | 2200 | 361 | $41.00 \pm 7.80$ | Smith et al. (2004) |
| HD28581 | Johnson | K | $2.81 \pm 0.10$ | Neugebauer & Leighton (1969) |

**Table 21** *(continued)*

| Star ID | System/Wvlen | Band/Bandpass | Value | Reference |
|---------|--------------|---------------|-------|-----------|
| HD28581 | 3500 | 898 | $20.50 \pm 6.60$ | Smith et al. (2004) |
| HD28581 | 4900 | 712 | $10.00 \pm 6.40$ | Smith et al. (2004) |
| HD28581 | 12000 | 6384 | $11.70 \pm 38.90$ | Smith et al. (2004) |
| HD28595 | KronComet | C2 | $6.79 \pm 0.03$ | This work |
| HD28595 | KronComet | Gc | $6.52 \pm 0.02$ | This work |
| HD28595 | WBVR | V | $6.33 \pm 0.05$ | Kornilov et al. (1991) |
| HD28595 | Johnson | V | $6.31 \pm 0.05$ | Mermilliod (1986) |
| HD28595 | Johnson | V | $6.34 \pm 0.05$ | Roman (1955) |
| HD28595 | Johnson | V | $6.40 \pm 0.04$ | This work |
| HD28595 | WBVR | R | $4.75 \pm 0.05$ | Kornilov et al. (1991) |
| HD28595 | KronComet | Rc | $4.84 \pm 0.02$ | This work |
| HD28595 | Johnson | K | $1.52 \pm 0.06$ | Neugebauer & Leighton (1969) |
| HD29094 | Oja | m41 | $6.69 \pm 0.05$ | Häggkvist & Oja (1970) |
| HD29094 | Oja | m42 | $6.42 \pm 0.05$ | Häggkvist & Oja (1970) |
| HD29094 | Geneva | B | $4.83 \pm 0.08$ | Golay (1972) |
| HD29094 | WBVR | B | $5.48 \pm 0.05$ | Kornilov et al. (1991) |
| HD29094 | Johnson | B | $5.44 \pm 0.05$ | Argue (1966) |
| HD29094 | Johnson | B | $5.46 \pm 0.05$ | Häggkvist & Oja (1966) |
| HD29094 | Johnson | B | $5.49 \pm 0.05$ | Johnson et al. (1966) |
| HD29094 | Johnson | B | $5.51 \pm 0.05$ | Johnson (1964) |
| HD29094 | Geneva | B2 | $6.00 \pm 0.08$ | Golay (1972) |
| HD29094 | 13c | m45 | $5.16 \pm 0.05$ | Johnson & Mitchell (1995) |
| HD29094 | Oja | m45 | $5.23 \pm 0.05$ | Häggkvist & Oja (1970) |
| HD29094 | 13c | m52 | $4.53 \pm 0.05$ | Johnson & Mitchell (1995) |
| HD29094 | Geneva | V1 | $5.07 \pm 0.08$ | Golay (1972) |
| HD29094 | WBVR | V | $4.25 \pm 0.05$ | Kornilov et al. (1991) |
| HD29094 | Geneva | V | $4.29 \pm 0.08$ | Golay (1972) |
| HD29094 | Johnson | V | $4.22 \pm 0.05$ | Häggkvist & Oja (1966) |
| HD29094 | Johnson | V | $4.22 \pm 0.05$ | Argue (1966) |
| HD29094 | Johnson | V | $4.26 \pm 0.05$ | Ducati (2002) |
| HD29094 | Johnson | V | $4.27 \pm 0.05$ | Johnson et al. (1966) |
| HD29094 | Johnson | V | $4.28 \pm 0.05$ | Johnson (1964) |
| HD29094 | 13c | m58 | $3.93 \pm 0.05$ | Johnson & Mitchell (1995) |
| HD29094 | Geneva | G | $5.18 \pm 0.08$ | Golay (1972) |
| HD29094 | 13c | m63 | $3.58 \pm 0.05$ | Johnson & Mitchell (1995) |
| HD29094 | WBVR | R | $3.33 \pm 0.05$ | Kornilov et al. (1991) |
| HD29094 | 13c | m72 | $3.27 \pm 0.05$ | Johnson & Mitchell (1995) |
| HD29094 | 13c | m80 | $3.00 \pm 0.05$ | Johnson & Mitchell (1995) |
| HD29094 | 13c | m86 | $2.85 \pm 0.05$ | Johnson & Mitchell (1995) |
| HD29094 | 13c | m99 | $2.61 \pm 0.05$ | Johnson & Mitchell (1995) |
| HD29094 | 13c | m110 | $2.39 \pm 0.05$ | Johnson & Mitchell (1995) |

**Table 21** *(continued)*

| Star ID | System/Wvlen | Band/Bandpass | Value | Reference |
|---------|--------------|---------------|-------|-----------|
| HD29094 | Johnson | H | $1.48 \pm 0.05$ | Shenavrin et al. (2011) |
| HD29094 | Johnson | H | $1.53 \pm 0.05$ | Voelcker (1975) |
| HD29094 | Johnson | H | $1.53 \pm 0.05$ | Ducati (2002) |
| HD29094 | 2200 | 361 | $160.50 \pm 9.00$ | Smith et al. (2004) |
| HD29094 | Johnson | K | $1.32 \pm 0.04$ | Neugebauer & Leighton (1969) |
| HD29094 | Johnson | K | $1.33 \pm 0.05$ | Johnson et al. (1966) |
| HD29094 | Johnson | K | $1.33 \pm 0.05$ | Ducati (2002) |
| HD29094 | Johnson | K | $1.33 \pm 0.05$ | Shenavrin et al. (2011) |
| HD29094 | Johnson | L | $1.17 \pm 0.05$ | Johnson et al. (1966) |
| HD29094 | Johnson | L | $1.17 \pm 0.05$ | Ducati (2002) |
| HD29094 | 3500 | 898 | $80.10 \pm 11.20$ | Smith et al. (2004) |
| HD29094 | 4900 | 712 | $38.30 \pm 6.80$ | Smith et al. (2004) |
| HD29094 | 12000 | 6384 | $-19.80 \pm 23.50$ | Smith et al. (2004) |
| HD30504 | Oja | m41 | $7.81 \pm 0.05$ | Häggkvist & Oja (1970) |
| HD30504 | DDO | m42 | $8.05 \pm 0.05$ | McClure & Forrester (1981) |
| HD30504 | Oja | m42 | $7.58 \pm 0.05$ | Häggkvist & Oja (1970) |
| HD30504 | KronComet | COp | $6.95 \pm 0.01$ | This work |
| HD30504 | WBVR | B | $6.39 \pm 0.05$ | Kornilov et al. (1991) |
| HD30504 | Johnson | B | $6.28 \pm 0.01$ | This work |
| HD30504 | Johnson | B | $6.31 \pm 0.05$ | Argue (1966) |
| HD30504 | Johnson | B | $6.31 \pm 0.05$ | Mermilliod (1986) |
| HD30504 | Johnson | B | $6.33 \pm 0.05$ | Häggkvist & Oja (1966) |
| HD30504 | Johnson | B | $6.35 \pm 0.05$ | Jennens & Helfer (1975) |
| HD30504 | Johnson | B | $6.35 \pm 0.05$ | Fernie (1983) |
| HD30504 | KronComet | Bc | $6.18 \pm 0.01$ | This work |
| HD30504 | DDO | m45 | $6.80 \pm 0.05$ | McClure & Forrester (1981) |
| HD30504 | Oja | m45 | $5.98 \pm 0.05$ | Häggkvist & Oja (1970) |
| HD30504 | DDO | m48 | $5.45 \pm 0.05$ | McClure & Forrester (1981) |
| HD30504 | KronComet | C2 | $5.32 \pm 0.01$ | This work |
| HD30504 | KronComet | Gc | $5.05 \pm 0.01$ | This work |
| HD30504 | WBVR | V | $4.89 \pm 0.05$ | Kornilov et al. (1991) |
| HD30504 | Johnson | V | $4.87 \pm 0.05$ | Argue (1966) |
| HD30504 | Johnson | V | $4.88 \pm 0.05$ | Mermilliod (1986) |
| HD30504 | Johnson | V | $4.89 \pm 0.05$ | Häggkvist & Oja (1966) |
| HD30504 | Johnson | V | $4.89 \pm 0.05$ | Fernie (1983) |
| HD30504 | Johnson | V | $4.90 \pm 0.05$ | Jennens & Helfer (1975) |
| HD30504 | WBVR | R | $3.79 \pm 0.05$ | Kornilov et al. (1991) |
| HD30504 | KronComet | Rc | $3.59 \pm 0.01$ | This work |
| HD30504 | 1250 | 310 | $194.10 \pm 8.40$ | Smith et al. (2004) |
| HD30504 | 2200 | 361 | $181.00 \pm 4.70$ | Smith et al. (2004) |
| HD30504 | Johnson | K | $1.46 \pm 0.04$ | Neugebauer & Leighton (1969) |

**Table 21** *(continued)*

| Star ID | System/Wvlen | Band/Bandpass | Value | Reference |
|---------|--------------|---------------|-------|-----------|
| HD30504 | 3500 | 898 | $81.70 \pm 12.90$ | Smith et al. (2004) |
| HD30504 | 4900 | 712 | $38.10 \pm 5.60$ | Smith et al. (2004) |
| HD30504 | 12000 | 6384 | $19.20 \pm 24.40$ | Smith et al. (2004) |
| HD30605 | KronComet | COp | $8.11 \pm 0.07$ | This work |
| HD30605 | WBVR | B | $7.75 \pm 0.05$ | Kornilov et al. (1991) |
| HD30605 | Johnson | B | $7.68 \pm 0.05$ | Argue (1966) |
| HD30605 | KronComet | Bc | $7.54 \pm 0.05$ | This work |
| HD30605 | KronComet | C2 | $6.48 \pm 0.02$ | This work |
| HD30605 | KronComet | Gc | $6.28 \pm 0.02$ | This work |
| HD30605 | WBVR | V | $6.08 \pm 0.05$ | Kornilov et al. (1991) |
| HD30605 | Johnson | V | $6.08 \pm 0.05$ | Argue (1966) |
| HD30605 | Johnson | V | $6.12 \pm 0.07$ | This work |
| HD30605 | WBVR | R | $4.88 \pm 0.05$ | Kornilov et al. (1991) |
| HD30605 | KronComet | Rc | $4.62 \pm 0.02$ | This work |
| HD30605 | Johnson | K | $2.41 \pm 0.09$ | Neugebauer & Leighton (1969) |
| HD30605 | 3500 | 898 | $69.20 \pm 14.00$ | Smith et al. (2004) |
| HD30605 | 4900 | 712 | $31.30 \pm 8.40$ | Smith et al. (2004) |
| HD30605 | 12000 | 6384 | $4.70 \pm 25.40$ | Smith et al. (2004) |
| HD30834 | Geneva | B1 | $7.14 \pm 0.08$ | Golay (1972) |
| HD30834 | Vilnius | X | $7.62 \pm 0.05$ | Kazlauskas et al. (2005) |
| HD30834 | DDO | m41 | $8.08 \pm 0.05$ | McClure & Forrester (1981) |
| HD30834 | Oja | m41 | $7.59 \pm 0.05$ | Häggkvist & Oja (1970) |
| HD30834 | DDO | m42 | $7.82 \pm 0.05$ | McClure & Forrester (1981) |
| HD30834 | Oja | m42 | $7.33 \pm 0.05$ | Häggkvist & Oja (1970) |
| HD30834 | Geneva | B | $5.61 \pm 0.08$ | Golay (1972) |
| HD30834 | WBVR | B | $6.23 \pm 0.05$ | Kornilov et al. (1991) |
| HD30834 | Johnson | B | $6.13 \pm 0.05$ | Miczaika (1954) |
| HD30834 | Johnson | B | $6.16 \pm 0.05$ | Sanwal et al. (1973) |
| HD30834 | Johnson | B | $6.18 \pm 0.05$ | Johnson & Knuckles (1957) |
| HD30834 | Johnson | B | $6.18 \pm 0.05$ | Häggkvist & Oja (1966) |
| HD30834 | Johnson | B | $6.18 \pm 0.05$ | Johnson et al. (1966) |
| HD30834 | Johnson | B | $6.19 \pm 0.05$ | Mermilliod (1986) |
| HD30834 | Johnson | B | $6.20 \pm 0.05$ | Moffett & Barnes (1980) |
| HD30834 | Johnson | B | $6.21 \pm 0.05$ | O'Connell (1964) |
| HD30834 | Johnson | B | $6.22 \pm 0.05$ | Shao (1964) |
| HD30834 | Geneva | B2 | $6.64 \pm 0.08$ | Golay (1972) |
| HD30834 | DDO | m45 | $6.69 \pm 0.05$ | McClure & Forrester (1981) |
| HD30834 | Oja | m45 | $5.86 \pm 0.05$ | Häggkvist & Oja (1970) |
| HD30834 | Vilnius | Y | $5.82 \pm 0.05$ | Kazlauskas et al. (2005) |
| HD30834 | DDO | m48 | $5.34 \pm 0.05$ | McClure & Forrester (1981) |
| HD30834 | Vilnius | Z | $5.24 \pm 0.05$ | Kazlauskas et al. (2005) |

**Table 21** *(continued)*

| Star ID | System/Wvlen | Band/Bandpass | Value | Reference |
|---------|--------------|---------------|-------|-----------|
| HD30834 | Geneva | V1 | $5.60 \pm 0.08$ | Golay (1972) |
| HD30834 | WBVR | V | $4.78 \pm 0.05$ | Kornilov et al. (1991) |
| HD30834 | Vilnius | V | $4.79 \pm 0.05$ | Kazlauskas et al. (2005) |
| HD30834 | Geneva | V | $4.79 \pm 0.08$ | Golay (1972) |
| HD30834 | Johnson | V | $4.71 \pm 0.05$ | Miczaika (1954) |
| HD30834 | Johnson | V | $4.76 \pm 0.05$ | Sanwal et al. (1973) |
| HD30834 | Johnson | V | $4.77 \pm 0.05$ | Johnson & Knuckles (1957) |
| HD30834 | Johnson | V | $4.77 \pm 0.05$ | Häggkvist & Oja (1966) |
| HD30834 | Johnson | V | $4.77 \pm 0.05$ | Johnson et al. (1966) |
| HD30834 | Johnson | V | $4.78 \pm 0.05$ | Mermilliod (1986) |
| HD30834 | Johnson | V | $4.79 \pm 0.05$ | O'Connell (1964) |
| HD30834 | Johnson | V | $4.79 \pm 0.05$ | Moffett & Barnes (1980) |
| HD30834 | Johnson | V | $4.80 \pm 0.05$ | Shao (1964) |
| HD30834 | Geneva | G | $5.67 \pm 0.08$ | Golay (1972) |
| HD30834 | Vilnius | S | $3.83 \pm 0.05$ | Kazlauskas et al. (2005) |
| HD30834 | WBVR | R | $3.74 \pm 0.05$ | Kornilov et al. (1991) |
| HD30834 | 1250 | 310 | $221.20 \pm 8.60$ | Smith et al. (2004) |
| HD30834 | 2200 | 361 | $187.00 \pm 8.70$ | Smith et al. (2004) |
| HD30834 | Johnson | K | $1.48 \pm 0.07$ | Neugebauer & Leighton (1969) |
| HD30834 | 3500 | 898 | $89.00 \pm 10.40$ | Smith et al. (2004) |
| HD30834 | 4900 | 712 | $42.80 \pm 5.80$ | Smith et al. (2004) |
| HD30834 | 12000 | 6384 | $15.60 \pm 24.20$ | Smith et al. (2004) |
| HD31139 | Oja | m41 | $8.45 \pm 0.05$ | Häggkvist & Oja (1970) |
| HD31139 | Oja | m42 | $8.27 \pm 0.05$ | Häggkvist & Oja (1970) |
| HD31139 | Oja | m45 | $6.58 \pm 0.05$ | Häggkvist & Oja (1970) |
| HD31139 | Vilnius | Y | $6.52 \pm 0.05$ | Zdanavicius et al. (1972) |
| HD31139 | Vilnius | Z | $5.86 \pm 0.05$ | Zdanavicius et al. (1972) |
| HD31139 | WBVR | V | $5.35 \pm 0.05$ | Kornilov et al. (1991) |
| HD31139 | Vilnius | V | $5.33 \pm 0.05$ | Zdanavicius et al. (1972) |
| HD31139 | Johnson | V | $5.32 \pm 0.05$ | Cousins (1962a) |
| HD31139 | Johnson | V | $5.32 \pm 0.05$ | Argue (1966) |
| HD31139 | Johnson | V | $5.33 \pm 0.05$ | Johnson et al. (1966) |
| HD31139 | Johnson | V | $5.35 \pm 0.05$ | Mermilliod (1986) |
| HD31139 | Vilnius | S | $4.17 \pm 0.05$ | Zdanavicius et al. (1972) |
| HD31139 | WBVR | R | $3.93 \pm 0.05$ | Kornilov et al. (1991) |
| HD31139 | 1250 | 310 | $269.70 \pm 16.10$ | Smith et al. (2004) |
| HD31139 | 2200 | 361 | $254.40 \pm 8.50$ | Smith et al. (2004) |
| HD31139 | Johnson | K | $1.03 \pm 0.06$ | Neugebauer & Leighton (1969) |
| HD31139 | 3500 | 898 | $126.70 \pm 6.10$ | Smith et al. (2004) |
| HD31139 | 4900 | 712 | $54.20 \pm 6.00$ | Smith et al. (2004) |
| HD31139 | 12000 | 6384 | $4.60 \pm 20.30$ | Smith et al. (2004) |

**Table 21** *(continued)*

| Star ID | System/Wvlen | Band/Bandpass | Value | Reference |
|---------|--------------|---------------|-------|-----------|
| HD31421 | 13c | m33 | $6.51 \pm 0.05$ | Johnson & Mitchell (1995) |
| HD31421 | 13c | m35 | $6.24 \pm 0.05$ | Johnson & Mitchell (1995) |
| HD31421 | DDO | m35 | $7.70 \pm 0.05$ | McClure & Forrester (1981) |
| HD31421 | WBVR | W | $6.24 \pm 0.05$ | Kornilov et al. (1991) |
| HD31421 | Johnson | U | $6.31 \pm 0.05$ | Johnson et al. (1966) |
| HD31421 | Johnson | U | $6.31 \pm 0.05$ | Ducati (2002) |
| HD31421 | Johnson | U | $6.36 \pm 0.05$ | Argue (1966) |
| HD31421 | Johnson | U | $6.37 \pm 0.05$ | Mermilliod (1986) |
| HD31421 | 13c | m37 | $6.33 \pm 0.05$ | Johnson & Mitchell (1995) |
| HD31421 | DDO | m38 | $6.59 \pm 0.05$ | McClure & Forrester (1981) |
| HD31421 | 13c | m40 | $6.00 \pm 0.05$ | Johnson & Mitchell (1995) |
| HD31421 | DDO | m41 | $6.96 \pm 0.05$ | McClure & Forrester (1981) |
| HD31421 | Oja | m41 | $6.50 \pm 0.05$ | Häggkvist & Oja (1970) |
| HD31421 | DDO | m42 | $6.75 \pm 0.05$ | McClure & Forrester (1981) |
| HD31421 | Oja | m42 | $6.29 \pm 0.05$ | Häggkvist & Oja (1970) |
| HD31421 | WBVR | B | $5.27 \pm 0.05$ | Kornilov et al. (1991) |
| HD31421 | Johnson | B | $5.21 \pm 0.05$ | Johnson et al. (1966) |
| HD31421 | Johnson | B | $5.21 \pm 0.05$ | Ducati (2002) |
| HD31421 | Johnson | B | $5.23 \pm 0.05$ | Argue (1966) |
| HD31421 | Johnson | B | $5.23 \pm 0.05$ | Mermilliod (1986) |
| HD31421 | Johnson | B | $5.25 \pm 0.05$ | Häggkvist & Oja (1966) |
| HD31421 | 13c | m45 | $4.93 \pm 0.05$ | Johnson & Mitchell (1995) |
| HD31421 | DDO | m45 | $5.75 \pm 0.05$ | McClure & Forrester (1981) |
| HD31421 | Oja | m45 | $4.93 \pm 0.05$ | Häggkvist & Oja (1970) |
| HD31421 | DDO | m48 | $4.51 \pm 0.05$ | McClure & Forrester (1981) |
| HD31421 | 13c | m52 | $4.41 \pm 0.05$ | Johnson & Mitchell (1995) |
| HD31421 | WBVR | V | $4.08 \pm 0.05$ | Kornilov et al. (1991) |
| HD31421 | Johnson | V | $4.06 \pm 0.05$ | Johnson et al. (1966) |
| HD31421 | Johnson | V | $4.06 \pm 0.05$ | Mermilliod (1986) |
| HD31421 | Johnson | V | $4.06 \pm 0.05$ | Ducati (2002) |
| HD31421 | Johnson | V | $4.07 \pm 0.05$ | Argue (1966) |
| HD31421 | Johnson | V | $4.09 \pm 0.05$ | Häggkvist & Oja (1966) |
| HD31421 | 13c | m58 | $3.85 \pm 0.05$ | Johnson & Mitchell (1995) |
| HD31421 | 13c | m63 | $3.52 \pm 0.05$ | Johnson & Mitchell (1995) |
| HD31421 | WBVR | R | $3.23 \pm 0.05$ | Kornilov et al. (1991) |
| HD31421 | 13c | m72 | $3.20 \pm 0.05$ | Johnson & Mitchell (1995) |
| HD31421 | 13c | m80 | $2.95 \pm 0.05$ | Johnson & Mitchell (1995) |
| HD31421 | 13c | m86 | $2.82 \pm 0.05$ | Johnson & Mitchell (1995) |
| HD31421 | 13c | m99 | $2.62 \pm 0.05$ | Johnson & Mitchell (1995) |
| HD31421 | 13c | m110 | $2.40 \pm 0.05$ | Johnson & Mitchell (1995) |
| HD31421 | 1250 | 310 | $231.30 \pm 7.30$ | Smith et al. (2004) |

**Table 21** *continued on next page*

**Table 21** *(continued)*

| Star ID | System/Wvlen | Band/Bandpass | Value | Reference |
|---------|--------------|---------------|-------|-----------|
| HD31421 | Johnson | J | $2.12 \pm 0.05$ | Johnson et al. (1966) |
| HD31421 | Johnson | J | $2.12 \pm 0.05$ | Wu & Wang (1985) |
| HD31421 | Johnson | J | $2.12 \pm 0.05$ | Ducati (2002) |
| HD31421 | Johnson | H | $1.43 \pm 0.05$ | Wu & Wang (1985) |
| HD31421 | Johnson | H | $1.43 \pm 0.05$ | Ducati (2002) |
| HD31421 | 2200 | 361 | $172.70 \pm 5.90$ | Smith et al. (2004) |
| HD31421 | Johnson | K | $1.36 \pm 0.05$ | Johnson et al. (1966) |
| HD31421 | Johnson | K | $1.36 \pm 0.05$ | Ducati (2002) |
| HD31421 | Johnson | K | $1.41 \pm 0.04$ | Neugebauer & Leighton (1969) |
| HD31421 | 3500 | 898 | $81.60 \pm 8.40$ | Smith et al. (2004) |
| HD31421 | 4900 | 712 | $39.00 \pm 4.80$ | Smith et al. (2004) |
| HD31421 | 12000 | 6384 | $-10.20 \pm 27.80$ | Smith et al. (2004) |
| HD33463 | KronComet | C2 | $6.86 \pm 0.03$ | This work |
| HD33463 | KronComet | Gc | $6.63 \pm 0.03$ | This work |
| HD33463 | WBVR | V | $6.41 \pm 0.05$ | Kornilov et al. (1991) |
| HD33463 | Johnson | V | $6.38 \pm 0.05$ | Neckel (1974) |
| HD33463 | Johnson | V | $6.49 \pm 0.04$ | This work |
| HD33463 | WBVR | R | $4.81 \pm 0.05$ | Kornilov et al. (1991) |
| HD33463 | KronComet | Rc | $4.86 \pm 0.12$ | This work |
| HD33463 | 2200 | 361 | $165.30 \pm 17.10$ | Smith et al. (2004) |
| HD33463 | Johnson | K | $1.59 \pm 0.08$ | Neugebauer & Leighton (1969) |
| HD33463 | 3500 | 898 | $82.90 \pm 14.80$ | Smith et al. (2004) |
| HD33463 | 4900 | 712 | $38.10 \pm 11.30$ | Smith et al. (2004) |
| HD33463 | 12000 | 6384 | $8.30 \pm 28.30$ | Smith et al. (2004) |
| HD34334 | Geneva | B1 | $6.66 \pm 0.08$ | Golay (1972) |
| HD34334 | DDO | m41 | $7.59 \pm 0.05$ | Clark & McClure (1979) |
| HD34334 | DDO | m41 | $7.59 \pm 0.05$ | McClure & Forrester (1981) |
| HD34334 | Stromgren | v | $6.66 \pm 0.08$ | Olsen (1993) |
| HD34334 | Stromgren | v | $6.66 \pm 0.08$ | Hauck & Mermilliod (1998) |
| HD34334 | DDO | m42 | $7.39 \pm 0.05$ | Clark & McClure (1979) |
| HD34334 | DDO | m42 | $7.39 \pm 0.05$ | McClure & Forrester (1981) |
| HD34334 | Geneva | B | $5.20 \pm 0.08$ | Golay (1972) |
| HD34334 | KronComet | COp | $6.21 \pm 0.06$ | This work |
| HD34334 | WBVR | B | $5.85 \pm 0.05$ | Kornilov et al. (1991) |
| HD34334 | Johnson | B | $5.76 \pm 0.06$ | This work |
| HD34334 | Johnson | B | $5.79 \pm 0.05$ | Ljunggren & Oja (1965) |
| HD34334 | Johnson | B | $5.79 \pm 0.05$ | Argue (1966) |
| HD34334 | Johnson | B | $5.80 \pm 0.05$ | Cuffey (1973) |
| HD34334 | Johnson | B | $5.81 \pm 0.05$ | Johnson et al. (1966) |
| HD34334 | Johnson | B | $5.83 \pm 0.05$ | Mermilliod (1986) |
| HD34334 | KronComet | Bc | $5.65 \pm 0.03$ | This work |

**Table 21** *(continued)*

| Star ID | System/Wvlen | Band/Bandpass | Value | Reference |
|---------|-------------|---------------|-------|-----------|
| HD34334 | Geneva | B2 | $6.27 \pm 0.08$ | Golay (1972) |
| HD34334 | DDO | m45 | $6.31 \pm 0.05$ | Clark & McClure (1979) |
| HD34334 | DDO | m45 | $6.31 \pm 0.05$ | McClure & Forrester (1981) |
| HD34334 | Oja | m45 | $5.51 \pm 0.05$ | Häggkvist & Oja (1970) |
| HD34334 | Stromgren | b | $5.34 \pm 0.08$ | Olsen (1993) |
| HD34334 | Stromgren | b | $5.34 \pm 0.08$ | Hauck & Mermilliod (1998) |
| HD34334 | DDO | m48 | $5.04 \pm 0.05$ | Clark & McClure (1979) |
| HD34334 | DDO | m48 | $5.04 \pm 0.05$ | McClure & Forrester (1981) |
| HD34334 | KronComet | C2 | $4.91 \pm 0.02$ | This work |
| HD34334 | KronComet | Gc | $4.67 \pm 0.04$ | This work |
| HD34334 | Geneva | V1 | $5.37 \pm 0.08$ | Golay (1972) |
| HD34334 | WBVR | V | $4.55 \pm 0.05$ | Kornilov et al. (1991) |
| HD34334 | Stromgren | y | $4.55 \pm 0.08$ | Olsen (1993) |
| HD34334 | Stromgren | y | $4.55 \pm 0.08$ | Hauck & Mermilliod (1998) |
| HD34334 | Geneva | V | $4.57 \pm 0.08$ | Golay (1972) |
| HD34334 | Johnson | V | $4.53 \pm 0.05$ | Cuffey (1973) |
| HD34334 | Johnson | V | $4.54 \pm 0.05$ | Ljunggren & Oja (1965) |
| HD34334 | Johnson | V | $4.54 \pm 0.05$ | Johnson et al. (1966) |
| HD34334 | Johnson | V | $4.54 \pm 0.05$ | Argue (1966) |
| HD34334 | Johnson | V | $4.57 \pm 0.05$ | Mermilliod (1986) |
| HD34334 | Johnson | V | $4.62 \pm 0.04$ | This work |
| HD34334 | Geneva | G | $5.47 \pm 0.08$ | Golay (1972) |
| HD34334 | WBVR | R | $3.60 \pm 0.05$ | Kornilov et al. (1991) |
| HD34334 | KronComet | Rc | $3.36 \pm 0.05$ | This work |
| HD34334 | Johnson | J | $2.26 \pm 0.05$ | Alonso et al. (1998) |
| HD34334 | Johnson | H | $1.62 \pm 0.05$ | Alonso et al. (1998) |
| HD34334 | Johnson | K | $1.43 \pm 0.06$ | Neugebauer & Leighton (1969) |
| HD34334 | 3500 | 898 | $60.70 \pm 10.40$ | Smith et al. (2004) |
| HD34334 | 4900 | 712 | $25.70 \pm 7.40$ | Smith et al. (2004) |
| HD34559 | DDO | m35 | $7.90 \pm 0.05$ | McClure & Forrester (1981) |
| HD34559 | WBVR | W | $6.45 \pm 0.05$ | Kornilov et al. (1991) |
| HD34559 | Johnson | U | $6.47 \pm 0.05$ | Argue (1963) |
| HD34559 | Johnson | U | $6.52 \pm 0.05$ | Mermilliod (1986) |
| HD34559 | Johnson | U | $6.58 \pm 0.05$ | Argue (1966) |
| HD34559 | DDO | m38 | $6.87 \pm 0.05$ | McClure & Forrester (1981) |
| HD34559 | DDO | m41 | $7.44 \pm 0.05$ | McClure & Forrester (1981) |
| HD34559 | Oja | m41 | $6.99 \pm 0.05$ | Häggkvist & Oja (1970) |
| HD34559 | DDO | m42 | $7.24 \pm 0.05$ | McClure & Forrester (1981) |
| HD34559 | Oja | m42 | $6.81 \pm 0.05$ | Häggkvist & Oja (1970) |
| HD34559 | WBVR | B | $5.92 \pm 0.05$ | Kornilov et al. (1991) |
| HD34559 | Johnson | B | $5.81 \pm 0.05$ | Argue (1963) |

**Table 21** *(continued)*

| Star ID | System/Wvlen | Band/Bandpass | Value | Reference |
|---------|--------------|---------------|-------|-----------|
| HD34559 | Johnson | B | $5.84 \pm 0.05$ | Mermilliod (1986) |
| HD34559 | Johnson | B | $5.88 \pm 0.05$ | Häggkvist & Oja (1966) |
| HD34559 | Johnson | B | $5.88 \pm 0.05$ | Argue (1966) |
| HD34559 | DDO | m45 | $6.44 \pm 0.05$ | McClure & Forrester (1981) |
| HD34559 | Oja | m45 | $5.62 \pm 0.05$ | Häggkvist & Oja (1970) |
| HD34559 | DDO | m48 | $5.29 \pm 0.05$ | McClure & Forrester (1981) |
| HD34559 | WBVR | V | $4.97 \pm 0.05$ | Kornilov et al. (1991) |
| HD34559 | Johnson | V | $4.88 \pm 0.05$ | Argue (1963) |
| HD34559 | Johnson | V | $4.90 \pm 0.05$ | Mermilliod (1986) |
| HD34559 | Johnson | V | $4.94 \pm 0.05$ | Häggkvist & Oja (1966) |
| HD34559 | Johnson | V | $4.95 \pm 0.05$ | Argue (1966) |
| HD34559 | WBVR | R | $4.28 \pm 0.05$ | Kornilov et al. (1991) |
| HD34559 | 1250 | 310 | $64.60 \pm 6.20$ | Smith et al. (2004) |
| HD34559 | 2200 | 361 | $35.70 \pm 9.00$ | Smith et al. (2004) |
| HD34559 | Johnson | K | $2.89 \pm 0.07$ | Neugebauer & Leighton (1969) |
| HD34559 | 3500 | 898 | $14.80 \pm 6.00$ | Smith et al. (2004) |
| HD34559 | 4900 | 712 | $7.60 \pm 6.50$ | Smith et al. (2004) |
| HD34559 | 12000 | 6384 | $36.30 \pm 36.50$ | Smith et al. (2004) |
| HD34577 | KronComet | COp | $9.85 \pm 0.07$ | This work |
| HD34577 | Johnson | B | $9.05 \pm 0.06$ | This work |
| HD34577 | Johnson | B | $9.18 \pm 0.05$ | Neckel (1974) |
| HD34577 | KronComet | Bc | $9.06 \pm 0.07$ | This work |
| HD34577 | KronComet | C2 | $7.76 \pm 0.04$ | This work |
| HD34577 | KronComet | Gc | $7.61 \pm 0.04$ | This work |
| HD34577 | Johnson | V | $7.35 \pm 0.05$ | Neckel (1974) |
| HD34577 | Johnson | V | $7.48 \pm 0.06$ | This work |
| HD34577 | KronComet | Rc | $5.79 \pm 0.04$ | This work |
| HD34577 | Johnson | K | $1.93 \pm 0.06$ | Neugebauer & Leighton (1969) |
| HD34577 | 3500 | 898 | $76.80 \pm 15.00$ | Smith et al. (2004) |
| HD34577 | 12000 | 6384 | $-10.70 \pm 33.30$ | Smith et al. (2004) |
| HD35497 | 13c | m33 | $0.85 \pm 0.05$ | Johnson & Mitchell (1995) |
| HD35497 | Geneva | U | $1.40 \pm 0.08$ | Golay (1972) |
| HD35497 | Vilnius | U | $2.72 \pm 0.05$ | Zdanavicius et al. (1969) |
| HD35497 | Vilnius | U | $2.72 \pm 0.05$ | Meistas & Zitkevicius (1976) |
| HD35497 | 13c | m35 | $0.91 \pm 0.05$ | Johnson & Mitchell (1995) |
| HD35497 | Stromgren | u | $2.24 \pm 0.08$ | Cameron (1966) |
| HD35497 | Stromgren | u | $2.24 \pm 0.08$ | Johansen & Gyldenkerne (1970) |
| HD35497 | Stromgren | u | $2.24 \pm 0.08$ | Crawford & Barnes (1970) |
| HD35497 | Stromgren | u | $2.24 \pm 0.08$ | Hauck & Mermilliod (1998) |
| HD35497 | Stromgren | u | $2.25 \pm 0.08$ | Crawford et al. (1966) |
| HD35497 | WBVR | W | $0.94 \pm 0.05$ | Kornilov et al. (1991) |

**Table 21** *(continued)*

| Star ID | System/Wvlen | Band/Bandpass | Value | Reference |
|---------|--------------|---------------|-------|-----------|
| HD35497 | Johnson | U | $1.03 \pm 0.05$ | Johnson & Morgan (1953b) |
| HD35497 | Johnson | U | $1.03 \pm 0.05$ | Johnson & Harris (1954) |
| HD35497 | Johnson | U | $1.03 \pm 0.05$ | Johnson et al. (1966) |
| HD35497 | Johnson | U | $1.03 \pm 0.05$ | Ducati (2002) |
| HD35497 | Johnson | U | $1.04 \pm 0.05$ | Johnson (1965b) |
| HD35497 | 13c | m37 | $1.09 \pm 0.05$ | Johnson & Mitchell (1995) |
| HD35497 | Vilnius | P | $2.34 \pm 0.05$ | Zdanavicius et al. (1969) |
| HD35497 | Vilnius | P | $2.34 \pm 0.05$ | Meistas & Zitkevicius (1976) |
| HD35497 | 13c | m40 | $1.47 \pm 0.05$ | Johnson & Mitchell (1995) |
| HD35497 | Geneva | B1 | $1.35 \pm 0.08$ | Golay (1972) |
| HD35497 | Vilnius | X | $1.90 \pm 0.05$ | Zdanavicius et al. (1969) |
| HD35497 | Vilnius | X | $1.90 \pm 0.05$ | Meistas & Zitkevicius (1976) |
| HD35497 | Oja | m41 | $2.16 \pm 0.05$ | Häggkvist & Oja (1970) |
| HD35497 | Stromgren | v | $1.64 \pm 0.08$ | Crawford et al. (1966) |
| HD35497 | Stromgren | v | $1.64 \pm 0.08$ | Cameron (1966) |
| HD35497 | Stromgren | v | $1.64 \pm 0.08$ | Johansen & Gyldenkerne (1970) |
| HD35497 | Stromgren | v | $1.64 \pm 0.08$ | Crawford & Barnes (1970) |
| HD35497 | Stromgren | v | $1.64 \pm 0.08$ | Hauck & Mermilliod (1998) |
| HD35497 | Oja | m42 | $2.21 \pm 0.05$ | Häggkvist & Oja (1970) |
| HD35497 | Geneva | B | $0.52 \pm 0.08$ | Golay (1972) |
| HD35497 | WBVR | B | $1.53 \pm 0.05$ | Kornilov et al. (1991) |
| HD35497 | Johnson | B | $1.51 \pm 0.05$ | Johnson & Morgan (1953b) |
| HD35497 | Johnson | B | $1.52 \pm 0.05$ | Johnson & Harris (1954) |
| HD35497 | Johnson | B | $1.52 \pm 0.05$ | Johnson et al. (1966) |
| HD35497 | Johnson | B | $1.52 \pm 0.05$ | Ducati (2002) |
| HD35497 | Johnson | B | $1.53 \pm 0.05$ | Johnson (1965b) |
| HD35497 | Johnson | B | $1.54 \pm 0.05$ | Häggkvist & Oja (1966) |
| HD35497 | Johnson | B | $1.62 \pm 0.05$ | Engels et al. (1981) |
| HD35497 | Geneva | B2 | $2.12 \pm 0.08$ | Golay (1972) |
| HD35497 | 13c | m45 | $1.61 \pm 0.05$ | Johnson & Mitchell (1995) |
| HD35497 | Oja | m45 | $1.56 \pm 0.05$ | Häggkvist & Oja (1970) |
| HD35497 | Vilnius | Y | $1.75 \pm 0.05$ | Zdanavicius et al. (1969) |
| HD35497 | Vilnius | Y | $1.75 \pm 0.05$ | Meistas & Zitkevicius (1976) |
| HD35497 | Stromgren | b | $1.60 \pm 0.08$ | Crawford et al. (1966) |
| HD35497 | Stromgren | b | $1.60 \pm 0.08$ | Cameron (1966) |
| HD35497 | Stromgren | b | $1.60 \pm 0.08$ | Johansen & Gyldenkerne (1970) |
| HD35497 | Stromgren | b | $1.60 \pm 0.08$ | Crawford & Barnes (1970) |
| HD35497 | Stromgren | b | $1.60 \pm 0.08$ | Hauck & Mermilliod (1998) |
| HD35497 | Vilnius | Z | $1.70 \pm 0.05$ | Zdanavicius et al. (1969) |
| HD35497 | Vilnius | Z | $1.70 \pm 0.05$ | Meistas & Zitkevicius (1976) |
| HD35497 | 13c | m52 | $1.65 \pm 0.05$ | Johnson & Mitchell (1995) |

**Table 21** *(continued)*

| Star ID | System/Wvlen | Band/Bandpass | Value | Reference |
|---------|--------------|---------------|-------|-----------|
| HD35497 | Geneva | V1 | $2.36 \pm 0.08$ | Golay (1972) |
| HD35497 | WBVR | V | $1.65 \pm 0.05$ | Kornilov et al. (1991) |
| HD35497 | Vilnius | V | $1.65 \pm 0.05$ | Zdanavicius et al. (1969) |
| HD35497 | Vilnius | V | $1.65 \pm 0.05$ | Meistas & Zitkevicius (1976) |
| HD35497 | Stromgren | y | $1.65 \pm 0.08$ | Crawford et al. (1966) |
| HD35497 | Stromgren | y | $1.65 \pm 0.08$ | Cameron (1966) |
| HD35497 | Stromgren | y | $1.65 \pm 0.08$ | Johansen & Gyldenkerne (1970) |
| HD35497 | Stromgren | y | $1.65 \pm 0.08$ | Crawford & Barnes (1970) |
| HD35497 | Stromgren | y | $1.65 \pm 0.08$ | Hauck & Mermilliod (1998) |
| HD35497 | Geneva | V | $1.65 \pm 0.08$ | Golay (1972) |
| HD35497 | Johnson | V | $1.64 \pm 0.05$ | Johnson & Morgan (1953b) |
| HD35497 | Johnson | V | $1.65 \pm 0.05$ | Johnson & Harris (1954) |
| HD35497 | Johnson | V | $1.65 \pm 0.05$ | Johnson et al. (1966) |
| HD35497 | Johnson | V | $1.65 \pm 0.05$ | Ducati (2002) |
| HD35497 | Johnson | V | $1.66 \pm 0.05$ | Johnson (1965b) |
| HD35497 | Johnson | V | $1.68 \pm 0.05$ | Häggkvist & Oja (1966) |
| HD35497 | Johnson | V | $1.68 \pm 0.05$ | Engels et al. (1981) |
| HD35497 | 13c | m58 | $1.70 \pm 0.05$ | Johnson & Mitchell (1995) |
| HD35497 | Geneva | G | $2.89 \pm 0.08$ | Golay (1972) |
| HD35497 | 13c | m63 | $1.70 \pm 0.05$ | Johnson & Mitchell (1995) |
| HD35497 | Vilnius | S | $1.55 \pm 0.05$ | Zdanavicius et al. (1969) |
| HD35497 | Vilnius | S | $1.55 \pm 0.05$ | Meistas & Zitkevicius (1976) |
| HD35497 | WBVR | R | $1.73 \pm 0.05$ | Kornilov et al. (1991) |
| HD35497 | 13c | m72 | $1.77 \pm 0.05$ | Johnson & Mitchell (1995) |
| HD35497 | 13c | m80 | $1.78 \pm 0.05$ | Johnson & Mitchell (1995) |
| HD35497 | 13c | m86 | $1.80 \pm 0.05$ | Johnson & Mitchell (1995) |
| HD35497 | 13c | m99 | $1.87 \pm 0.05$ | Johnson & Mitchell (1995) |
| HD35497 | 13c | m110 | $1.87 \pm 0.05$ | Johnson & Mitchell (1995) |
| HD35497 | 1250 | 310 | $255.60 \pm 8.70$ | Smith et al. (2004) |
| HD35497 | Johnson | J | $1.87 \pm 0.05$ | Alonso et al. (1994) |
| HD35497 | Johnson | J | $1.87 \pm 0.05$ | Alonso et al. (1998) |
| HD35497 | Johnson | J | $1.88 \pm 0.05$ | Selby et al. (1988) |
| HD35497 | Johnson | J | $1.95 \pm 0.05$ | Johnson (1965c) |
| HD35497 | Johnson | J | $1.95 \pm 0.05$ | Ducati (2002) |
| HD35497 | Johnson | J | $1.96 \pm 0.05$ | Engels et al. (1981) |
| HD35497 | Johnson | J | $1.96 \pm 0.05$ | Leitherer & Wolf (1984) |
| HD35497 | Johnson | J | $1.97 \pm 0.05$ | Johnson et al. (1966) |
| HD35497 | Johnson | J | $1.97 \pm 0.05$ | Voelcker (1975) |
| HD35497 | Johnson | J | $1.97 \pm 0.05$ | Persi et al. (1979) |
| HD35497 | Johnson | J | $1.97 \pm 0.05$ | Bergeat & Lunel (1980) |
| HD35497 | Johnson | J | $1.97 \pm 0.05$ | Shenavrin et al. (2011) |

**Table 21** *(continued)*

| Star ID | System/Wvlen | Band/Bandpass | Value | Reference |
|---------|--------------|---------------|-------|-----------|
| HD35497 | Johnson | H | $1.93 \pm 0.05$ | Bergeat & Lunel (1980) |
| HD35497 | Johnson | H | $1.93 \pm 0.05$ | Alonso et al. (1994) |
| HD35497 | Johnson | H | $1.94 \pm 0.05$ | Alonso et al. (1998) |
| HD35497 | Johnson | H | $1.98 \pm 0.05$ | Engels et al. (1981) |
| HD35497 | Johnson | H | $1.98 \pm 0.05$ | Leitherer & Wolf (1984) |
| HD35497 | Johnson | H | $1.98 \pm 0.05$ | Ducati (2002) |
| HD35497 | Johnson | H | $2.00 \pm 0.05$ | Persi et al. (1979) |
| HD35497 | Johnson | H | $2.00 \pm 0.05$ | Gnedin et al. (1981) |
| HD35497 | Johnson | H | $2.00 \pm 0.05$ | Gnedin et al. (1982) |
| HD35497 | Johnson | H | $2.02 \pm 0.05$ | Engels et al. (1981) |
| HD35497 | Johnson | H | $2.03 \pm 0.05$ | Shenavrin et al. (2011) |
| HD35497 | 2200 | 361 | $95.60 \pm 12.80$ | Smith et al. (2004) |
| HD35497 | Johnson | K | $2.02 \pm 0.05$ | Engels et al. (1981) |
| HD35497 | Johnson | K | $2.03 \pm 0.05$ | Ducati (2002) |
| HD35497 | Johnson | K | $2.03 \pm 0.06$ | Neugebauer & Leighton (1969) |
| HD35497 | Johnson | K | $2.05 \pm 0.05$ | Johnson et al. (1966) |
| HD35497 | Johnson | K | $2.05 \pm 0.05$ | Shenavrin et al. (2011) |
| HD35497 | Johnson | L | $1.99 \pm 0.05$ | Ducati (2002) |
| HD35497 | Johnson | L | $2.03 \pm 0.05$ | Engels et al. (1981) |
| HD35497 | 3500 | 898 | $39.00 \pm 12.80$ | Smith et al. (2004) |
| HD35497 | Johnson | N | $2.27 \pm 0.05$ | Ducati (2002) |
| HD35497 | 4900 | 712 | $20.70 \pm 5.90$ | Smith et al. (2004) |
| HD35497 | Johnson | M | $2.11 \pm 0.05$ | Ducati (2002) |
| HD35497 | 12000 | 6384 | $-5.50 \pm 35.30$ | Smith et al. (2004) |
| HD35620 | Geneva | U | $8.82 \pm 0.08$ | Golay (1972) |
| HD35620 | Vilnius | U | $10.01 \pm 0.05$ | Kakaras et al. (1968) |
| HD35620 | DDO | m35 | $9.53 \pm 0.05$ | McClure & Forrester (1981) |
| HD35620 | Stromgren | u | $9.46 \pm 0.08$ | Olsen (1993) |
| HD35620 | Stromgren | u | $9.46 \pm 0.08$ | Hauck & Mermilliod (1998) |
| HD35620 | WBVR | W | $8.07 \pm 0.05$ | Kornilov et al. (1991) |
| HD35620 | Johnson | U | $8.12 \pm 0.05$ | Argue (1966) |
| HD35620 | Johnson | U | $8.14 \pm 0.05$ | Mermilliod (1986) |
| HD35620 | Johnson | U | $8.15 \pm 0.05$ | Jennens & Helfer (1975) |
| HD35620 | Johnson | U | $8.16 \pm 0.05$ | McClure (1970) |
| HD35620 | Vilnius | P | $9.34 \pm 0.05$ | Kakaras et al. (1968) |
| HD35620 | DDO | m38 | $8.29 \pm 0.05$ | McClure & Forrester (1981) |
| HD35620 | Geneva | B1 | $7.54 \pm 0.08$ | Golay (1972) |
| HD35620 | Vilnius | X | $7.99 \pm 0.05$ | Kakaras et al. (1968) |
| HD35620 | DDO | m41 | $8.55 \pm 0.05$ | McClure & Forrester (1981) |
| HD35620 | Oja | m41 | $8.15 \pm 0.05$ | Häggkvist & Oja (1970) |
| HD35620 | Stromgren | v | $7.58 \pm 0.08$ | Olsen (1993) |

**Table 21** *(continued)*

| Star ID | System/Wvlen | Band/Bandpass | Value | Reference |
|---------|--------------|---------------|-------|-----------|
| HD35620 | Stromgren | v | 7.58 ± 0.08 | Hauck & Mermilliod (1998) |
| HD35620 | DDO | m42 | 8.15 ± 0.05 | McClure & Forrester (1981) |
| HD35620 | Oja | m42 | 7.71 ± 0.08 | Häggkvist & Oja (1970) |
| HD35620 | Geneva | B | 5.95 ± 0.08 | Golay (1972) |
| HD35620 | WBVR | B | 6.53 ± 0.05 | Kornilov et al. (1991) |
| HD35620 | Johnson | B | 6.46 ± 0.05 | Argue (1966) |
| HD35620 | Johnson | B | 6.48 ± 0.05 | McClure (1970) |
| HD35620 | Johnson | B | 6.48 ± 0.05 | Mermilliod (1986) |
| HD35620 | Johnson | B | 6.49 ± 0.05 | Ljunggren & Oja (1965) |
| HD35620 | Johnson | B | 6.50 ± 0.05 | Jennens & Helfer (1975) |
| HD35620 | Geneva | B2 | 6.93 ± 0.08 | Golay (1972) |
| HD35620 | DDO | m45 | 6.96 ± 0.05 | McClure & Forrester (1981) |
| HD35620 | Oja | m45 | 6.17 ± 0.05 | Häggkvist & Oja (1970) |
| HD35620 | Vilnius | Y | 6.10 ± 0.05 | Kakaras et al. (1968) |
| HD35620 | Stromgren | b | 5.94 ± 0.08 | Olsen (1993) |
| HD35620 | Stromgren | b | 5.94 ± 0.08 | Hauck & Mermilliod (1998) |
| HD35620 | DDO | m48 | 5.59 ± 0.05 | McClure & Forrester (1981) |
| HD35620 | Vilnius | Z | 5.58 ± 0.15 | Kakaras et al. (1968) |
| HD35620 | Geneva | V1 | 5.92 ± 0.08 | Golay (1972) |
| HD35620 | WBVR | V | 5.08 ± 0.05 | Kornilov et al. (1991) |
| HD35620 | Vilnius | V | 5.10 ± 0.05 | Kakaras et al. (1968) |
| HD35620 | Stromgren | y | 5.07 ± 0.08 | Olsen (1993) |
| HD35620 | Stromgren | y | 5.07 ± 0.08 | Hauck & Mermilliod (1998) |
| HD35620 | Geneva | V | 5.12 ± 0.08 | Golay (1972) |
| HD35620 | Johnson | V | 5.06 ± 0.05 | Argue (1966) |
| HD35620 | Johnson | V | 5.08 ± 0.05 | Ljunggren & Oja (1965) |
| HD35620 | Johnson | V | 5.08 ± 0.05 | McClure (1970) |
| HD35620 | Johnson | V | 5.08 ± 0.05 | Mermilliod (1986) |
| HD35620 | Johnson | V | 5.09 ± 0.05 | Jennens & Helfer (1975) |
| HD35620 | Geneva | G | 5.98 ± 0.08 | Golay (1972) |
| HD35620 | Vilnius | S | 4.14 ± 0.05 | Kakaras et al. (1968) |
| HD35620 | WBVR | R | 4.10 ± 0.05 | Kornilov et al. (1991) |
| HD35620 | Johnson | J | 2.72 ± 0.05 | Alonso et al. (1998) |
| HD35620 | Johnson | H | 2.10 ± 0.05 | Alonso et al. (1998) |
| HD35620 | Johnson | K | 1.93 ± 0.06 | Neugebauer & Leighton (1969) |
| HD35620 | 3500 | 898 | 202.40 ± 80.70 | Smith et al. (2004) |
| HD35620 | 4900 | 712 | 33.70 ± 8.00 | Smith et al. (2004) |
| HD35620 | 12000 | 6384 | 132.70 ± 55.90 | Smith et al. (2004) |
| HD37329 | KronComet | NH | 9.27 ± 0.04 | This work |
| HD37329 | KronComet | UVc | 9.04 ± 0.06 | This work |
| HD37329 | DDO | m35 | 9.49 ± 0.05 | McClure & Forrester (1981) |

**Table 21** (continued)

| Star ID | System/Wvlen | Band/Bandpass | Value | Reference |
|---|---|---|---|---|
| HD37329 | WBVR | W | $8.03 \pm 0.05$ | Kornilov et al. (1991) |
| HD37329 | Johnson | U | $7.94 \pm 0.06$ | This work |
| HD37329 | Johnson | U | $8.17 \pm 0.01$ | Oja (1991) |
| HD37329 | Johnson | U | $8.18 \pm 0.05$ | Guetter & Hewitt (1984) |
| HD37329 | DDO | m38 | $8.45 \pm 0.05$ | McClure & Forrester (1981) |
| HD37329 | KronComet | CN | $9.00 \pm 0.09$ | This work |
| HD37329 | DDO | m41 | $8.97 \pm 0.05$ | McClure & Forrester (1981) |
| HD37329 | DDO | m42 | $8.80 \pm 0.05$ | McClure & Forrester (1981) |
| HD37329 | KronComet | COp | $7.66 \pm 0.06$ | This work |
| HD37329 | WBVR | B | $7.44 \pm 0.05$ | Kornilov et al. (1991) |
| HD37329 | Johnson | B | $7.30 \pm 0.05$ | This work |
| HD37329 | Johnson | B | $7.41 \pm 0.01$ | Oja (1991) |
| HD37329 | Johnson | B | $7.41 \pm 0.05$ | Guetter & Hewitt (1984) |
| HD37329 | KronComet | Bc | $7.23 \pm 0.05$ | This work |
| HD37329 | DDO | m45 | $7.96 \pm 0.05$ | McClure & Forrester (1981) |
| HD37329 | DDO | m48 | $6.79 \pm 0.05$ | McClure & Forrester (1981) |
| HD37329 | KronComet | C2 | $6.58 \pm 0.03$ | This work |
| HD37329 | KronComet | Gc | $6.51 \pm 0.03$ | This work |
| HD37329 | WBVR | V | $6.45 \pm 0.05$ | Kornilov et al. (1991) |
| HD37329 | Johnson | V | $6.43 \pm 0.01$ | Oja (1991) |
| HD37329 | Johnson | V | $6.43 \pm 0.05$ | Guetter & Hewitt (1984) |
| HD37329 | Johnson | V | $6.46 \pm 0.05$ | This work |
| HD37329 | WBVR | R | $5.72 \pm 0.05$ | Kornilov et al. (1991) |
| HD37329 | KronComet | Rc | $5.45 \pm 0.02$ | This work |
| HD37329 | Johnson | H | $4.33 \pm 0.05$ | Gnedin et al. (1982) |
| HD38152 | KronComet | COp | $10.69 \pm 0.11$ | This work |
| HD38152 | Johnson | B | $10.12 \pm 0.04$ | This work |
| HD38152 | KronComet | Bc | $10.12 \pm 0.07$ | This work |
| HD38152 | KronComet | C2 | $8.69 \pm 0.05$ | This work |
| HD38152 | KronComet | Gc | $8.64 \pm 0.05$ | This work |
| HD38152 | Johnson | V | $8.57 \pm 0.05$ | This work |
| HD38152 | KronComet | Rc | $6.89 \pm 0.06$ | This work |
| HD38152 | 1250 | 310 | $47.60 \pm 12.10$ | Smith et al. (2004) |
| HD38152 | 2200 | 361 | $55.80 \pm 11.80$ | Smith et al. (2004) |
| HD38152 | Johnson | K | $2.48 \pm 0.09$ | Neugebauer & Leighton (1969) |
| HD38152 | 3500 | 898 | $26.10 \pm 11.80$ | Smith et al. (2004) |
| HD38152 | 4900 | 712 | $10.90 \pm 10.70$ | Smith et al. (2004) |
| HD38152 | 12000 | 6384 | $17.80 \pm 81.00$ | Smith et al. (2004) |
| HD38656 | 13c | m33 | $6.21 \pm 0.05$ | Johnson & Mitchell (1995) |
| HD38656 | Geneva | U | $6.66 \pm 0.08$ | Golay (1972) |
| HD38656 | Vilnius | U | $7.94 \pm 0.05$ | Kazlauskas et al. (2005) |

**Table 21** *(continued)*

| Star ID | System/Wvlen | Band/Bandpass | Value | Reference |
|---------|--------------|---------------|-------|-----------|
| HD38656 | 13c | m35 | $6.00 \pm 0.05$ | Johnson & Mitchell (1995) |
| HD38656 | DDO | m35 | $7.48 \pm 0.05$ | McClure & Forrester (1981) |
| HD38656 | WBVR | W | $6.01 \pm 0.05$ | Kornilov et al. (1991) |
| HD38656 | Johnson | U | $6.10 \pm 0.05$ | Argue (1963) |
| HD38656 | Johnson | U | $6.12 \pm 0.05$ | Jennens & Helfer (1975) |
| HD38656 | Johnson | U | $6.14 \pm 0.05$ | Argue (1966) |
| HD38656 | Johnson | U | $6.15 \pm 0.05$ | Oja (1985a) |
| HD38656 | Johnson | U | $6.15 \pm 0.05$ | Oja (1986) |
| HD38656 | Johnson | U | $6.16 \pm 0.01$ | Oja (1984) |
| HD38656 | Johnson | U | $6.16 \pm 0.05$ | Johnson & Knuckles (1957) |
| HD38656 | Johnson | U | $6.16 \pm 0.05$ | Johnson et al. (1966) |
| HD38656 | Johnson | U | $6.16 \pm 0.05$ | Oja (1984) |
| HD38656 | Johnson | U | $6.16 \pm 0.05$ | Oja (1985b) |
| HD38656 | Johnson | U | $6.16 \pm 0.05$ | Oja (1985a) |
| HD38656 | Johnson | U | $6.16 \pm 0.05$ | Ducati (2002) |
| HD38656 | Johnson | U | $6.17 \pm 0.05$ | Johnson et al. (1966) |
| HD38656 | Johnson | U | $6.17 \pm 0.05$ | Oja (1983) |
| HD38656 | 13c | m37 | $6.11 \pm 0.05$ | Johnson & Mitchell (1995) |
| HD38656 | Vilnius | P | $7.37 \pm 0.05$ | Kazlauskas et al. (2005) |
| HD38656 | DDO | m38 | $6.44 \pm 0.05$ | McClure & Forrester (1981) |
| HD38656 | 13c | m40 | $6.00 \pm 0.05$ | Johnson & Mitchell (1995) |
| HD38656 | Geneva | B1 | $6.01 \pm 0.08$ | Golay (1972) |
| HD38656 | Vilnius | X | $6.46 \pm 0.05$ | Kazlauskas et al. (2005) |
| HD38656 | DDO | m41 | $7.00 \pm 0.05$ | McClure & Forrester (1981) |
| HD38656 | Oja | m41 | $6.57 \pm 0.05$ | Häggkvist & Oja (1970) |
| HD38656 | DDO | m42 | $6.84 \pm 0.05$ | McClure & Forrester (1981) |
| HD38656 | Oja | m42 | $6.45 \pm 0.05$ | Häggkvist & Oja (1970) |
| HD38656 | Geneva | B | $4.75 \pm 0.08$ | Golay (1972) |
| HD38656 | WBVR | B | $5.50 \pm 0.05$ | Kornilov et al. (1991) |
| HD38656 | Johnson | B | $5.42 \pm 0.05$ | Argue (1963) |
| HD38656 | Johnson | B | $5.45 \pm 0.01$ | Oja (1993) |
| HD38656 | Johnson | B | $5.45 \pm 0.05$ | Ljunggren & Oja (1965) |
| HD38656 | Johnson | B | $5.45 \pm 0.05$ | Argue (1966) |
| HD38656 | Johnson | B | $5.46 \pm 0.05$ | Miczaika (1954) |
| HD38656 | Johnson | B | $5.46 \pm 0.05$ | Häggkvist & Oja (1969a) |
| HD38656 | Johnson | B | $5.46 \pm 0.05$ | Haggkvist & Oja (1970) |
| HD38656 | Johnson | B | $5.46 \pm 0.05$ | Oja (1984) |
| HD38656 | Johnson | B | $5.46 \pm 0.05$ | Oja (1985a) |
| HD38656 | Johnson | B | $5.46 \pm 0.05$ | Oja (1986) |
| HD38656 | Johnson | B | $5.47 \pm 0.05$ | Johnson & Knuckles (1957) |
| HD38656 | Johnson | B | $5.47 \pm 0.05$ | Häggkvist & Oja (1966) |

**Table 21** *continued on next page*

**Table 21** *(continued)*

| Star ID | System/Wvlen | Band/Bandpass | Value | Reference |
|---------|--------------|---------------|-------|-----------|
| HD38656 | Johnson | B | $5.47 \pm 0.05$ | Johnson et al. (1966) |
| HD38656 | Johnson | B | $5.47 \pm 0.05$ | Oja (1983) |
| HD38656 | Johnson | B | $5.47 \pm 0.05$ | Oja (1985b) |
| HD38656 | Johnson | B | $5.47 \pm 0.05$ | Ducati (2002) |
| HD38656 | Johnson | B | $5.48 \pm 0.05$ | Jennens & Helfer (1975) |
| HD38656 | Geneva | B2 | $5.92 \pm 0.08$ | Golay (1972) |
| HD38656 | 13c | m45 | $5.22 \pm 0.05$ | Johnson & Mitchell (1995) |
| HD38656 | DDO | m45 | $6.03 \pm 0.05$ | McClure & Forrester (1981) |
| HD38656 | Oja | m45 | $5.24 \pm 0.05$ | Häggkvist & Oja (1970) |
| HD38656 | Vilnius | Y | $5.27 \pm 0.05$ | Kazlauskas et al. (2005) |
| HD38656 | DDO | m48 | $4.87 \pm 0.05$ | McClure & Forrester (1981) |
| HD38656 | Vilnius | Z | $4.82 \pm 0.05$ | Kazlauskas et al. (2005) |
| HD38656 | 13c | m52 | $4.76 \pm 0.05$ | Johnson & Mitchell (1995) |
| HD38656 | Geneva | V1 | $5.29 \pm 0.08$ | Golay (1972) |
| HD38656 | WBVR | V | $4.53 \pm 0.05$ | Kornilov et al. (1991) |
| HD38656 | Vilnius | V | $4.53 \pm 0.05$ | Kazlauskas et al. (2005) |
| HD38656 | Geneva | V | $4.52 \pm 0.08$ | Golay (1972) |
| HD38656 | Johnson | V | $4.48 \pm 0.05$ | Argue (1963) |
| HD38656 | Johnson | V | $4.50 \pm 0.01$ | Oja (1993) |
| HD38656 | Johnson | V | $4.50 \pm 0.05$ | Argue (1966) |
| HD38656 | Johnson | V | $4.50 \pm 0.05$ | Oja (1984) |
| HD38656 | Johnson | V | $4.50 \pm 0.05$ | Oja (1985a) |
| HD38656 | Johnson | V | $4.51 \pm 0.05$ | Häggkvist & Oja (1969a) |
| HD38656 | Johnson | V | $4.51 \pm 0.05$ | Oja (1983) |
| HD38656 | Johnson | V | $4.51 \pm 0.05$ | Oja (1985b) |
| HD38656 | Johnson | V | $4.51 \pm 0.05$ | Oja (1985a) |
| HD38656 | Johnson | V | $4.51 \pm 0.05$ | Oja (1986) |
| HD38656 | Johnson | V | $4.52 \pm 0.05$ | Miczaika (1954) |
| HD38656 | Johnson | V | $4.52 \pm 0.05$ | Ljunggren & Oja (1965) |
| HD38656 | Johnson | V | $4.52 \pm 0.05$ | Häggkvist & Oja (1966) |
| HD38656 | Johnson | V | $4.52 \pm 0.05$ | Haggkvist & Oja (1970) |
| HD38656 | Johnson | V | $4.53 \pm 0.05$ | Johnson & Knuckles (1957) |
| HD38656 | Johnson | V | $4.53 \pm 0.05$ | Johnson et al. (1966) |
| HD38656 | Johnson | V | $4.53 \pm 0.05$ | Jennens & Helfer (1975) |
| HD38656 | Johnson | V | $4.53 \pm 0.05$ | Ducati (2002) |
| HD38656 | 13c | m58 | $4.31 \pm 0.05$ | Johnson & Mitchell (1995) |
| HD38656 | Geneva | G | $5.50 \pm 0.08$ | Golay (1972) |
| HD38656 | 13c | m63 | $4.04 \pm 0.05$ | Johnson & Mitchell (1995) |
| HD38656 | Vilnius | S | $3.83 \pm 0.05$ | Kazlauskas et al. (2005) |
| HD38656 | WBVR | R | $3.82 \pm 0.05$ | Kornilov et al. (1991) |
| HD38656 | 13c | m72 | $3.79 \pm 0.05$ | Johnson & Mitchell (1995) |

**Table 21** *(continued)*

| Star ID | System/Wvlen | Band/Bandpass | Value | Reference |
|---------|--------------|---------------|-------|-----------|
| HD38656 | 13c | m80 | $3.60 \pm 0.05$ | Johnson & Mitchell (1995) |
| HD38656 | 13c | m86 | $3.50 \pm 0.05$ | Johnson & Mitchell (1995) |
| HD38656 | 13c | m99 | $3.34 \pm 0.05$ | Johnson & Mitchell (1995) |
| HD38656 | 13c | m110 | $3.18 \pm 0.05$ | Johnson & Mitchell (1995) |
| HD38656 | Johnson | J | $2.86 \pm 0.05$ | Selby et al. (1988) |
| HD38656 | Johnson | J | $2.86 \pm 0.05$ | Blackwell et al. (1990) |
| HD38656 | Johnson | J | $2.89 \pm 0.05$ | Ducati (2002) |
| HD38656 | Johnson | J | $2.92 \pm 0.05$ | Johnson et al. (1966) |
| HD38656 | Johnson | J | $2.92 \pm 0.05$ | Voelcker (1975) |
| HD38656 | Johnson | J | $2.92 \pm 0.05$ | Ghosh et al. (1984) |
| HD38656 | Johnson | H | $2.43 \pm 0.05$ | Voelcker (1975) |
| HD38656 | Johnson | H | $2.43 \pm 0.05$ | Ghosh et al. (1984) |
| HD38656 | Johnson | H | $2.43 \pm 0.05$ | Ducati (2002) |
| HD38656 | Johnson | K | $2.32 \pm 0.05$ | Ducati (2002) |
| HD38656 | Johnson | K | $2.34 \pm 0.05$ | Johnson et al. (1966) |
| HD38656 | Johnson | K | $2.36 \pm 0.06$ | Neugebauer & Leighton (1969) |
| HD38656 | Johnson | L | $2.22 \pm 0.05$ | Ducati (2002) |
| HD39003 | 13c | m33 | $6.26 \pm 0.05$ | Johnson & Mitchell (1995) |
| HD39003 | Geneva | U | $6.72 \pm 0.08$ | Golay (1972) |
| HD39003 | Vilnius | U | $7.99 \pm 0.05$ | Kazlauskas et al. (2005) |
| HD39003 | 13c | m35 | $6.05 \pm 0.05$ | Johnson & Mitchell (1995) |
| HD39003 | DDO | m35 | $7.50 \pm 0.05$ | McClure & Forrester (1981) |
| HD39003 | Stromgren | u | $7.40 \pm 0.08$ | Olsen (1993) |
| HD39003 | Stromgren | u | $7.40 \pm 0.08$ | Hauck & Mermilliod (1998) |
| HD39003 | WBVR | W | $6.04 \pm 0.05$ | Kornilov et al. (1991) |
| HD39003 | Johnson | U | $6.15 \pm 0.05$ | Argue (1966) |
| HD39003 | Johnson | U | $6.16 \pm 0.01$ | Oja (1984) |
| HD39003 | Johnson | U | $6.16 \pm 0.05$ | Oja (1985a) |
| HD39003 | Johnson | U | $6.17 \pm 0.05$ | Oja (1983) |
| HD39003 | Johnson | U | $6.17 \pm 0.05$ | Oja (1984) |
| HD39003 | Johnson | U | $6.17 \pm 0.05$ | Oja (1985b) |
| HD39003 | Johnson | U | $6.17 \pm 0.05$ | Mermilliod (1986) |
| HD39003 | Johnson | U | $6.18 \pm 0.05$ | Oja (1986) |
| HD39003 | Johnson | U | $6.20 \pm 0.05$ | Johnson et al. (1966) |
| HD39003 | Johnson | U | $6.20 \pm 0.05$ | Ducati (2002) |
| HD39003 | 13c | m37 | $6.12 \pm 0.05$ | Johnson & Mitchell (1995) |
| HD39003 | Vilnius | P | $7.37 \pm 0.05$ | Kazlauskas et al. (2005) |
| HD39003 | DDO | m38 | $6.39 \pm 0.05$ | McClure & Forrester (1981) |
| HD39003 | 13c | m40 | $5.81 \pm 0.05$ | Johnson & Mitchell (1995) |
| HD39003 | Geneva | B1 | $5.84 \pm 0.08$ | Golay (1972) |
| HD39003 | Vilnius | X | $6.28 \pm 0.05$ | Kazlauskas et al. (2005) |

Table 21 (continued)

| Star ID | System/Wvlen | Band/Bandpass | Value | Reference |
|---------|--------------|---------------|-------|-----------|
| HD39003 | DDO | m41 | $6.86 \pm 0.05$ | McClure & Forrester (1981) |
| HD39003 | Oja | m41 | $6.44 \pm 0.05$ | Häggkvist & Oja (1970) |
| HD39003 | Stromgren | v | $5.85 \pm 0.08$ | Olsen (1993) |
| HD39003 | Stromgren | v | $5.85 \pm 0.08$ | Hauck & Mermilliod (1998) |
| HD39003 | DDO | m42 | $6.56 \pm 0.05$ | McClure & Forrester (1981) |
| HD39003 | Oja | m42 | $6.13 \pm 0.05$ | Häggkvist & Oja (1970) |
| HD39003 | Geneva | B | $4.46 \pm 0.08$ | Golay (1972) |
| HD39003 | WBVR | B | $5.13 \pm 0.05$ | Kornilov et al. (1991) |
| HD39003 | Johnson | B | $5.07 \pm 0.05$ | Argue (1966) |
| HD39003 | Johnson | B | $5.08 \pm 0.05$ | Mermilliod (1986) |
| HD39003 | Johnson | B | $5.09 \pm 0.01$ | Oja (1993) |
| HD39003 | Johnson | B | $5.09 \pm 0.05$ | Ljunggren & Oja (1965) |
| HD39003 | Johnson | B | $5.09 \pm 0.05$ | Häggkvist & Oja (1969a) |
| HD39003 | Johnson | B | $5.09 \pm 0.05$ | Haggkvist & Oja (1970) |
| HD39003 | Johnson | B | $5.09 \pm 0.05$ | Oja (1984) |
| HD39003 | Johnson | B | $5.09 \pm 0.05$ | Oja (1985b) |
| HD39003 | Johnson | B | $5.09 \pm 0.05$ | Oja (1985a) |
| HD39003 | Johnson | B | $5.10 \pm 0.05$ | Häggkvist & Oja (1966) |
| HD39003 | Johnson | B | $5.10 \pm 0.05$ | Oja (1983) |
| HD39003 | Johnson | B | $5.10 \pm 0.05$ | Oja (1986) |
| HD39003 | Johnson | B | $5.11 \pm 0.05$ | Johnson et al. (1966) |
| HD39003 | Johnson | B | $5.11 \pm 0.05$ | Ducati (2002) |
| HD39003 | Geneva | B2 | $5.55 \pm 0.08$ | Golay (1972) |
| HD39003 | 13c | m45 | $4.78 \pm 0.05$ | Johnson & Mitchell (1995) |
| HD39003 | DDO | m45 | $5.62 \pm 0.05$ | McClure & Forrester (1981) |
| HD39003 | Oja | m45 | $4.82 \pm 0.05$ | Häggkvist & Oja (1970) |
| HD39003 | Vilnius | Y | $4.82 \pm 0.05$ | Kazlauskas et al. (2005) |
| HD39003 | Stromgren | b | $4.64 \pm 0.08$ | Olsen (1993) |
| HD39003 | Stromgren | b | $4.64 \pm 0.08$ | Hauck & Mermilliod (1998) |
| HD39003 | DDO | m48 | $4.38 \pm 0.05$ | McClure & Forrester (1981) |
| HD39003 | Vilnius | Z | $4.32 \pm 0.05$ | Kazlauskas et al. (2005) |
| HD39003 | 13c | m52 | $4.26 \pm 0.05$ | Johnson & Mitchell (1995) |
| HD39003 | Geneva | V1 | $4.77 \pm 0.08$ | Golay (1972) |
| HD39003 | WBVR | V | $3.99 \pm 0.05$ | Kornilov et al. (1991) |
| HD39003 | Vilnius | V | $3.99 \pm 0.05$ | Kazlauskas et al. (2005) |
| HD39003 | Stromgren | y | $3.95 \pm 0.08$ | Olsen (1993) |
| HD39003 | Stromgren | y | $3.95 \pm 0.08$ | Hauck & Mermilliod (1998) |
| HD39003 | Geneva | V | $3.98 \pm 0.08$ | Golay (1972) |
| HD39003 | Johnson | V | $3.94 \pm 0.05$ | Oja (1985a) |
| HD39003 | Johnson | V | $3.95 \pm 0.01$ | Oja (1993) |
| HD39003 | Johnson | V | $3.95 \pm 0.05$ | Argue (1966) |

**Table 21** *(continued)*

| Star ID | System/Wvlen | Band/Bandpass | Value | Reference |
|---------|--------------|---------------|-------|-----------|
| HD39003 | Johnson | V | $3.95 \pm 0.05$ | Oja (1983) |
| HD39003 | Johnson | V | $3.95 \pm 0.05$ | Oja (1984) |
| HD39003 | Johnson | V | $3.95 \pm 0.05$ | Mermilliod (1986) |
| HD39003 | Johnson | V | $3.96 \pm 0.05$ | Oja (1985b) |
| HD39003 | Johnson | V | $3.96 \pm 0.05$ | Oja (1986) |
| HD39003 | Johnson | V | $3.97 \pm 0.05$ | Ljunggren & Oja (1965) |
| HD39003 | Johnson | V | $3.97 \pm 0.05$ | Häggkvist & Oja (1966) |
| HD39003 | Johnson | V | $3.97 \pm 0.05$ | Johnson et al. (1966) |
| HD39003 | Johnson | V | $3.97 \pm 0.05$ | Häggkvist & Oja (1969a) |
| HD39003 | Johnson | V | $3.97 \pm 0.05$ | Haggkvist & Oja (1970) |
| HD39003 | Johnson | V | $3.97 \pm 0.05$ | Ducati (2002) |
| HD39003 | 13c | m58 | $3.74 \pm 0.05$ | Johnson & Mitchell (1995) |
| HD39003 | Geneva | G | $4.94 \pm 0.08$ | Golay (1972) |
| HD39003 | 13c | m63 | $3.42 \pm 0.05$ | Johnson & Mitchell (1995) |
| HD39003 | Vilnius | S | $3.20 \pm 0.05$ | Kazlauskas et al. (2005) |
| HD39003 | WBVR | R | $3.19 \pm 0.05$ | Kornilov et al. (1991) |
| HD39003 | 13c | m72 | $3.16 \pm 0.05$ | Johnson & Mitchell (1995) |
| HD39003 | 13c | m80 | $2.92 \pm 0.05$ | Johnson & Mitchell (1995) |
| HD39003 | 13c | m86 | $2.80 \pm 0.05$ | Johnson & Mitchell (1995) |
| HD39003 | 13c | m99 | $2.63 \pm 0.05$ | Johnson & Mitchell (1995) |
| HD39003 | 13c | m110 | $2.39 \pm 0.05$ | Johnson & Mitchell (1995) |
| HD39003 | Johnson | J | $2.06 \pm 0.05$ | Alonso et al. (1998) |
| HD39003 | Johnson | J | $2.18 \pm 0.05$ | Johnson et al. (1966) |
| HD39003 | Johnson | J | $2.18 \pm 0.05$ | Ducati (2002) |
| HD39003 | Johnson | J | $2.18 \pm 0.05$ | Shenavrin et al. (2011) |
| HD39003 | Johnson | H | $1.55 \pm 0.05$ | Alonso et al. (1998) |
| HD39003 | Johnson | H | $1.64 \pm 0.05$ | Shenavrin et al. (2011) |
| HD39003 | Johnson | K | $1.47 \pm 0.05$ | Neugebauer & Leighton (1969) |
| HD39003 | Johnson | K | $1.51 \pm 0.05$ | Johnson et al. (1966) |
| HD39003 | Johnson | K | $1.51 \pm 0.05$ | Ducati (2002) |
| HD39003 | Johnson | K | $1.51 \pm 0.05$ | Shenavrin et al. (2011) |
| HD39003 | 3500 | 898 | $89.50 \pm 19.70$ | Smith et al. (2004) |
| HD39003 | 4900 | 712 | $37.70 \pm 5.60$ | Smith et al. (2004) |
| HD39003 | 12000 | 6384 | $1.80 \pm 25.20$ | Smith et al. (2004) |
| HD39045 | Geneva | B1 | $9.29 \pm 0.08$ | Golay (1972) |
| HD39045 | Geneva | B | $7.68 \pm 0.08$ | Golay (1972) |
| HD39045 | WBVR | B | $8.12 \pm 0.05$ | Kornilov et al. (1991) |
| HD39045 | Johnson | B | $8.00 \pm 0.05$ | Johnson et al. (1966) |
| HD39045 | Johnson | B | $8.00 \pm 0.05$ | Mermilliod (1986) |
| HD39045 | Johnson | B | $8.00 \pm 0.05$ | Ducati (2002) |
| HD39045 | Johnson | B | $8.01 \pm 0.05$ | Lee (1970) |

**Table 21** *(continued)*

| Star ID | System/Wvlen | Band/Bandpass | Value | Reference |
|---------|--------------|---------------|-------|-----------|
| HD39045 | Johnson | B | $8.01 \pm 0.05$ | Neckel (1974) |
| HD39045 | Johnson | B | $8.02 \pm 0.05$ | Roman (1955) |
| HD39045 | Geneva | B2 | $8.62 \pm 0.08$ | Golay (1972) |
| HD39045 | Geneva | V1 | $7.26 \pm 0.08$ | Golay (1972) |
| HD39045 | WBVR | V | $6.34 \pm 0.05$ | Kornilov et al. (1991) |
| HD39045 | Geneva | V | $6.45 \pm 0.08$ | Golay (1972) |
| HD39045 | Johnson | V | $6.24 \pm 0.05$ | Mermilliod (1986) |
| HD39045 | Johnson | V | $6.25 \pm 0.05$ | Johnson et al. (1966) |
| HD39045 | Johnson | V | $6.25 \pm 0.05$ | Neckel (1974) |
| HD39045 | Johnson | V | $6.26 \pm 0.05$ | Ducati (2002) |
| HD39045 | Johnson | V | $6.28 \pm 0.05$ | Roman (1955) |
| HD39045 | Johnson | V | $6.28 \pm 0.05$ | Lee (1970) |
| HD39045 | Geneva | G | $7.34 \pm 0.08$ | Golay (1972) |
| HD39045 | WBVR | R | $4.66 \pm 0.05$ | Kornilov et al. (1991) |
| HD39045 | Johnson | J | $2.40 \pm 0.05$ | Lee (1970) |
| HD39045 | Johnson | J | $2.43 \pm 0.05$ | Ducati (2002) |
| HD39045 | Johnson | J | $2.47 \pm 0.05$ | Johnson et al. (1966) |
| HD39045 | Johnson | H | $1.49 \pm 0.05$ | Lee (1970) |
| HD39045 | Johnson | H | $1.49 \pm 0.05$ | Ducati (2002) |
| HD39045 | 2200 | 361 | $215.30 \pm 13.50$ | Smith et al. (2004) |
| HD39045 | Johnson | K | $1.19 \pm 0.06$ | Neugebauer & Leighton (1969) |
| HD39045 | Johnson | K | $1.26 \pm 0.05$ | Lee (1970) |
| HD39045 | Johnson | K | $1.29 \pm 0.05$ | Ducati (2002) |
| HD39045 | Johnson | K | $1.33 \pm 0.05$ | Johnson et al. (1966) |
| HD39045 | Johnson | L | $1.08 \pm 0.05$ | Lee (1970) |
| HD39045 | Johnson | L | $1.08 \pm 0.05$ | Ducati (2002) |
| HD39045 | 3500 | 898 | $109.70 \pm 10.80$ | Smith et al. (2004) |
| HD39045 | 4900 | 712 | $46.10 \pm 6.70$ | Smith et al. (2004) |
| HD39045 | 12000 | 6384 | $0.10 \pm 25.50$ | Smith et al. (2004) |
| HD39225 | WBVR | B | $7.61 \pm 0.05$ | Kornilov et al. (1991) |
| HD39225 | Johnson | B | $7.57 \pm 0.05$ | Roman (1955) |
| HD39225 | Johnson | B | $7.58 \pm 0.05$ | Neckel (1974) |
| HD39225 | Johnson | B | $7.64 \pm 0.05$ | Fernie (1983) |
| HD39225 | Vilnius | Y | $7.21 \pm 0.05$ | Jasevicius et al. (1990) |
| HD39225 | Vilnius | Z | $6.51 \pm 0.05$ | Jasevicius et al. (1990) |
| HD39225 | WBVR | V | $5.97 \pm 0.05$ | Kornilov et al. (1991) |
| HD39225 | Vilnius | V | $5.98 \pm 0.05$ | Jasevicius et al. (1990) |
| HD39225 | Johnson | V | $5.98 \pm 0.05$ | Roman (1955) |
| HD39225 | Johnson | V | $5.98 \pm 0.05$ | Neckel (1974) |
| HD39225 | Johnson | V | $6.02 \pm 0.05$ | Fernie (1983) |
| HD39225 | Vilnius | S | $4.84 \pm 0.05$ | Jasevicius et al. (1990) |

**Table 21** *(continued)*

| Star ID | System/Wvlen | Band/Bandpass | Value | Reference |
|---------|--------------|---------------|-------|-----------|
| HD39225 | WBVR | R | $4.63 \pm 0.05$ | Kornilov et al. (1991) |
| HD39225 | 1250 | 310 | $111.60 \pm 10.60$ | Smith et al. (2004) |
| HD39225 | 2200 | 361 | $111.90 \pm 9.90$ | Smith et al. (2004) |
| HD39225 | Johnson | K | $1.82 \pm 0.07$ | Neugebauer & Leighton (1969) |
| HD39225 | 3500 | 898 | $57.30 \pm 9.60$ | Smith et al. (2004) |
| HD39225 | 4900 | 712 | $27.20 \pm 6.00$ | Smith et al. (2004) |
| HD39225 | 12000 | 6384 | $4.60 \pm 27.40$ | Smith et al. (2004) |
| HD39732 | KronComet | C2 | $7.71 \pm 0.02$ | This work |
| HD39732 | KronComet | Gc | $7.48 \pm 0.03$ | This work |
| HD39732 | WBVR | V | $7.39 \pm 0.05$ | Kornilov et al. (1991) |
| HD39732 | Johnson | V | $7.40 \pm 0.06$ | This work |
| HD39732 | WBVR | R | $5.56 \pm 0.05$ | Kornilov et al. (1991) |
| HD39732 | KronComet | Rc | $5.67 \pm 0.05$ | This work |
| HD39732 | 1250 | 310 | $96.80 \pm 8.50$ | Smith et al. (2004) |
| HD39732 | 2200 | 361 | $119.70 \pm 9.00$ | Smith et al. (2004) |
| HD39732 | Johnson | K | $1.89 \pm 0.06$ | Neugebauer & Leighton (1969) |
| HD39732 | 3500 | 898 | $56.20 \pm 8.70$ | Smith et al. (2004) |
| HD39732 | 4900 | 712 | $25.50 \pm 5.20$ | Smith et al. (2004) |
| HD40441 | KronComet | COp | $9.02 \pm 0.03$ | This work |
| HD40441 | WBVR | B | $8.37 \pm 0.05$ | Kornilov et al. (1991) |
| HD40441 | Johnson | B | $8.22 \pm 0.03$ | This work |
| HD40441 | Johnson | B | $8.34 \pm 0.05$ | Janes (1979) |
| HD40441 | KronComet | Bc | $8.16 \pm 0.01$ | This work |
| HD40441 | KronComet | C2 | $7.16 \pm 0.01$ | This work |
| HD40441 | KronComet | Gc | $6.88 \pm 0.02$ | This work |
| HD40441 | WBVR | V | $6.70 \pm 0.05$ | Kornilov et al. (1991) |
| HD40441 | Johnson | V | $6.72 \pm 0.05$ | Janes (1979) |
| HD40441 | Johnson | V | $6.78 \pm 0.02$ | This work |
| HD40441 | WBVR | R | $5.42 \pm 0.05$ | Kornilov et al. (1991) |
| HD40441 | KronComet | Rc | $5.28 \pm 0.03$ | This work |
| HD40441 | Johnson | K | $2.73 \pm 0.09$ | Neugebauer & Leighton (1969) |
| HD41116 | Geneva | U | $6.01 \pm 0.08$ | Golay (1972) |
| HD41116 | WBVR | W | $5.38 \pm 0.05$ | Kornilov et al. (1991) |
| HD41116 | Johnson | U | $5.54 \pm 0.05$ | Johnson (1964) |
| HD41116 | Johnson | U | $5.54 \pm 0.05$ | Johnson et al. (1966) |
| HD41116 | Johnson | U | $5.54 \pm 0.05$ | Mermilliod (1986) |
| HD41116 | Johnson | U | $5.54 \pm 0.05$ | Ducati (2002) |
| HD41116 | Geneva | B1 | $5.44 \pm 0.08$ | Golay (1972) |
| HD41116 | Oja | m41 | $6.03 \pm 0.05$ | Häggkvist & Oja (1970) |
| HD41116 | Oja | m42 | $5.87 \pm 0.05$ | Häggkvist & Oja (1970) |
| HD41116 | Geneva | B | $4.26 \pm 0.08$ | Golay (1972) |

**Table 21** *continued on next page*

**Table 21** *(continued)*

| Star ID | System/Wvlen | Band/Bandpass | Value | Reference |
|---------|--------------|---------------|-------|-----------|
| HD41116 | WBVR | B | $5.01 \pm 0.05$ | Kornilov et al. (1991) |
| HD41116 | Johnson | B | $4.94 \pm 0.05$ | Giclas (1954) |
| HD41116 | Johnson | B | $4.98 \pm 0.05$ | Häggkvist & Oja (1966) |
| HD41116 | Johnson | B | $5.01 \pm 0.05$ | Mermilliod (1986) |
| HD41116 | Johnson | B | $5.02 \pm 0.05$ | Johnson (1964) |
| HD41116 | Johnson | B | $5.02 \pm 0.05$ | Johnson et al. (1966) |
| HD41116 | Johnson | B | $5.02 \pm 0.05$ | Ducati (2002) |
| HD41116 | Geneva | B2 | $5.48 \pm 0.08$ | Golay (1972) |
| HD41116 | Oja | m45 | $4.78 \pm 0.05$ | Häggkvist & Oja (1970) |
| HD41116 | Geneva | V1 | $4.95 \pm 0.08$ | Golay (1972) |
| HD41116 | WBVR | V | $4.17 \pm 0.05$ | Kornilov et al. (1991) |
| HD41116 | Geneva | V | $4.18 \pm 0.08$ | Golay (1972) |
| HD41116 | Johnson | V | $4.14 \pm 0.05$ | Giclas (1954) |
| HD41116 | Johnson | V | $4.15 \pm 0.05$ | Johnson (1964) |
| HD41116 | Johnson | V | $4.15 \pm 0.05$ | Johnson et al. (1966) |
| HD41116 | Johnson | V | $4.15 \pm 0.05$ | Ducati (2002) |
| HD41116 | Johnson | V | $4.16 \pm 0.05$ | Häggkvist & Oja (1966) |
| HD41116 | Johnson | V | $4.18 \pm 0.05$ | Mermilliod (1986) |
| HD41116 | Geneva | G | $5.18 \pm 0.08$ | Golay (1972) |
| HD41116 | WBVR | R | $3.53 \pm 0.05$ | Kornilov et al. (1991) |
| HD41116 | 1250 | 310 | $118.70 \pm 2.90$ | Smith et al. (2004) |
| HD41116 | Johnson | J | $2.71 \pm 0.05$ | Johnson et al. (1966) |
| HD41116 | Johnson | J | $2.71 \pm 0.05$ | Ducati (2002) |
| HD41116 | Johnson | J | $2.71 \pm 0.05$ | Shenavrin et al. (2011) |
| HD41116 | Johnson | H | $2.25 \pm 0.05$ | Shenavrin et al. (2011) |
| HD41116 | 2200 | 361 | $89.20 \pm 50.00$ | Smith et al. (2004) |
| HD41116 | Johnson | K | $2.13 \pm 0.05$ | Johnson et al. (1966) |
| HD41116 | Johnson | K | $2.13 \pm 0.05$ | Ducati (2002) |
| HD41116 | Johnson | K | $2.13 \pm 0.05$ | Shenavrin et al. (2011) |
| HD41116 | Johnson | K | $2.21 \pm 0.06$ | Neugebauer & Leighton (1969) |
| HD41116 | 3500 | 898 | $50.80 \pm 58.30$ | Smith et al. (2004) |
| HD41116 | 4900 | 712 | $25.70 \pm 995.00$ | Smith et al. (2004) |
| HD41116 | 12000 | 6384 | $39.20 \pm 9,378.10$ | Smith et al. (2004) |
| HD42049 | WBVR | W | $9.47 \pm 0.05$ | Kornilov et al. (1991) |
| HD42049 | Oja | m41 | $9.22 \pm 0.05$ | Häggkvist & Oja (1970) |
| HD42049 | Oja | m42 | $8.92 \pm 0.05$ | Häggkvist & Oja (1970) |
| HD42049 | WBVR | B | $7.62 \pm 0.05$ | Kornilov et al. (1991) |
| HD42049 | Johnson | B | $7.56 \pm 0.05$ | Haggkvist & Oja (1970) |
| HD42049 | Oja | m45 | $7.22 \pm 0.05$ | Häggkvist & Oja (1970) |
| HD42049 | WBVR | V | $5.95 \pm 0.05$ | Kornilov et al. (1991) |
| HD42049 | Johnson | V | $5.93 \pm 0.05$ | Haggkvist & Oja (1970) |

**Table 21** *continued on next page*

**Table 21** *(continued)*

| Star ID | System/Wvlen | Band/Bandpass | Value | Reference |
|---|---|---|---|---|
| HD42049 | WBVR | R | $4.70 \pm 0.05$ | Kornilov et al. (1991) |
| HD42049 | Johnson | K | $2.08 \pm 0.05$ | Neugebauer & Leighton (1969) |
| HD43039 | 13c | m33 | $6.19 \pm 0.05$ | Johnson & Mitchell (1995) |
| HD43039 | Geneva | U | $6.68 \pm 0.08$ | Golay (1972) |
| HD43039 | Vilnius | U | $7.95 \pm 0.05$ | Bartkevicius & Sperauskas (1974) |
| HD43039 | 13c | m35 | $5.96 \pm 0.05$ | Johnson & Mitchell (1995) |
| HD43039 | DDO | m35 | $7.47 \pm 0.05$ | McClure & Forrester (1981) |
| HD43039 | DDO | m35 | $7.47 \pm 0.05$ | Mermilliod & Nitschelm (1989) |
| HD43039 | Stromgren | u | $7.43 \pm 0.08$ | Olsen (1993) |
| HD43039 | Stromgren | u | $7.43 \pm 0.08$ | Hauck & Mermilliod (1998) |
| HD43039 | Johnson | U | $6.14 \pm 0.05$ | Roman (1955) |
| HD43039 | Johnson | U | $6.15 \pm 0.05$ | Szabados (1977) |
| HD43039 | Johnson | U | $6.16 \pm 0.05$ | Johnson (1964) |
| HD43039 | Johnson | U | $6.16 \pm 0.05$ | Argue (1966) |
| HD43039 | Johnson | U | $6.17 \pm 0.05$ | Johnson et al. (1966) |
| HD43039 | Johnson | U | $6.17 \pm 0.05$ | Jennens & Helfer (1975) |
| HD43039 | Johnson | U | $6.17 \pm 0.05$ | Mermilliod (1986) |
| HD43039 | Johnson | U | $6.17 \pm 0.05$ | Ducati (2002) |
| HD43039 | Johnson | U | $6.18 \pm 0.05$ | Fernie (1969) |
| HD43039 | Johnson | U | $6.21 \pm 0.05$ | Cuffey (1973) |
| HD43039 | Geneva | B1 | $5.96 \pm 0.08$ | Golay (1972) |
| HD43039 | Oja | m41 | $6.46 \pm 0.05$ | Häggkvist & Oja (1970) |
| HD43039 | DDO | m42 | $6.76 \pm 0.05$ | McClure & Forrester (1981) |
| HD43039 | DDO | m42 | $6.76 \pm 0.05$ | Mermilliod & Nitschelm (1989) |
| HD43039 | Oja | m42 | $6.33 \pm 0.05$ | Häggkvist & Oja (1970) |
| HD43039 | Geneva | B | $4.66 \pm 0.08$ | Golay (1972) |
| HD43039 | WBVR | B | $5.38 \pm 0.05$ | Kornilov et al. (1991) |
| HD43039 | Johnson | B | $5.33 \pm 0.05$ | Häggkvist & Oja (1966) |
| HD43039 | Johnson | B | $5.34 \pm 0.05$ | Argue (1966) |
| HD43039 | Johnson | B | $5.35 \pm 0.05$ | Fernie (1969) |
| HD43039 | Johnson | B | $5.35 \pm 0.05$ | Szabados (1977) |
| HD43039 | Johnson | B | $5.36 \pm 0.05$ | Johnson et al. (1966) |
| HD43039 | Johnson | B | $5.36 \pm 0.05$ | Jennens & Helfer (1975) |
| HD43039 | Johnson | B | $5.36 \pm 0.05$ | Moffett & Barnes (1980) |
| HD43039 | Johnson | B | $5.36 \pm 0.05$ | Mermilliod (1986) |
| HD43039 | Johnson | B | $5.36 \pm 0.05$ | Ducati (2002) |
| HD43039 | Johnson | B | $5.37 \pm 0.05$ | Roman (1955) |
| HD43039 | Johnson | B | $5.37 \pm 0.05$ | Johnson (1964) |
| HD43039 | Johnson | B | $5.40 \pm 0.05$ | Cuffey (1973) |
| HD43039 | Geneva | B2 | $5.83 \pm 0.08$ | Golay (1972) |
| HD43039 | 13c | m45 | $5.04 \pm 0.05$ | Johnson & Mitchell (1995) |

**Table 21** *(continued)*

| Star ID | System/Wvlen | Band/Bandpass | Value | Reference |
|---------|--------------|---------------|-------|-----------|
| HD43039 | DDO | m45 | $5.89 \pm 0.05$ | McClure & Forrester (1981) |
| HD43039 | DDO | m45 | $5.89 \pm 0.05$ | Mermilliod & Nitschelm (1989) |
| HD43039 | Oja | m45 | $5.08 \pm 0.05$ | Häggkvist & Oja (1970) |
| HD43039 | Vilnius | Y | $5.10 \pm 0.05$ | Bartkevicius & Sperauskas (1974) |
| HD43039 | Stromgren | b | $4.98 \pm 0.08$ | Olsen (1993) |
| HD43039 | Stromgren | b | $4.98 \pm 0.08$ | Hauck & Mermilliod (1998) |
| HD43039 | DDO | m48 | $4.70 \pm 0.05$ | McClure & Forrester (1981) |
| HD43039 | DDO | m48 | $4.70 \pm 0.05$ | Mermilliod & Nitschelm (1989) |
| HD43039 | Vilnius | Z | $4.64 \pm 0.05$ | Bartkevicius & Sperauskas (1974) |
| HD43039 | 13c | m52 | $4.56 \pm 0.05$ | Johnson & Mitchell (1995) |
| HD43039 | Geneva | V1 | $5.13 \pm 0.08$ | Golay (1972) |
| HD43039 | WBVR | V | $4.34 \pm 0.05$ | Kornilov et al. (1991) |
| HD43039 | Vilnius | V | $4.33 \pm 0.05$ | Bartkevicius & Sperauskas (1974) |
| HD43039 | Stromgren | y | $4.35 \pm 0.08$ | Olsen (1993) |
| HD43039 | Stromgren | y | $4.35 \pm 0.08$ | Hauck & Mermilliod (1998) |
| HD43039 | Geneva | V | $4.34 \pm 0.08$ | Golay (1972) |
| HD43039 | Johnson | V | $4.32 \pm 0.05$ | Häggkvist & Oja (1966) |
| HD43039 | Johnson | V | $4.32 \pm 0.05$ | Argue (1966) |
| HD43039 | Johnson | V | $4.33 \pm 0.05$ | Johnson (1964) |
| HD43039 | Johnson | V | $4.33 \pm 0.05$ | Szabados (1977) |
| HD43039 | Johnson | V | $4.33 \pm 0.05$ | Mermilliod (1986) |
| HD43039 | Johnson | V | $4.34 \pm 0.05$ | Roman (1955) |
| HD43039 | Johnson | V | $4.34 \pm 0.05$ | Fernie (1969) |
| HD43039 | Johnson | V | $4.34 \pm 0.05$ | Jennens & Helfer (1975) |
| HD43039 | Johnson | V | $4.34 \pm 0.05$ | Moffett & Barnes (1980) |
| HD43039 | Johnson | V | $4.35 \pm 0.05$ | Johnson et al. (1966) |
| HD43039 | Johnson | V | $4.35 \pm 0.05$ | Ducati (2002) |
| HD43039 | Johnson | V | $4.37 \pm 0.05$ | Cuffey (1973) |
| HD43039 | 13c | m58 | $4.08 \pm 0.05$ | Johnson & Mitchell (1995) |
| HD43039 | Geneva | G | $5.33 \pm 0.08$ | Golay (1972) |
| HD43039 | 13c | m63 | $3.77 \pm 0.05$ | Johnson & Mitchell (1995) |
| HD43039 | Vilnius | S | $3.59 \pm 0.05$ | Bartkevicius & Sperauskas (1974) |
| HD43039 | WBVR | R | $3.58 \pm 0.05$ | Kornilov et al. (1991) |
| HD43039 | 13c | m72 | $3.51 \pm 0.05$ | Johnson & Mitchell (1995) |
| HD43039 | 13c | m80 | $3.27 \pm 0.05$ | Johnson & Mitchell (1995) |
| HD43039 | 13c | m86 | $3.16 \pm 0.05$ | Johnson & Mitchell (1995) |
| HD43039 | 13c | m99 | $3.00 \pm 0.05$ | Johnson & Mitchell (1995) |
| HD43039 | 13c | m110 | $2.79 \pm 0.05$ | Johnson & Mitchell (1995) |
| HD43039 | 1250 | 310 | $155.00 \pm 9.70$ | Smith et al. (2004) |
| HD43039 | Johnson | J | $2.53 \pm 0.05$ | Selby et al. (1988) |
| HD43039 | Johnson | J | $2.53 \pm 0.05$ | Blackwell et al. (1990) |

**Table 21** (continued)

| Star ID | System/Wvlen | Band/Bandpass | Value | Reference |
|---------|--------------|---------------|-------|-----------|
| HD43039 | Johnson | J | $2.57 \pm 0.05$ | Ducati (2002) |
| HD43039 | Johnson | J | $2.59 \pm 0.05$ | Johnson et al. (1966) |
| HD43039 | Johnson | J | $2.59 \pm 0.05$ | Voelcker (1975) |
| HD43039 | Johnson | J | $2.59 \pm 0.05$ | Glass (1975) |
| HD43039 | Johnson | J | $2.59 \pm 0.05$ | Arribas & Martinez Roger (1987) |
| HD43039 | Johnson | H | $2.02 \pm 0.05$ | Arribas & Martinez Roger (1987) |
| HD43039 | Johnson | H | $2.05 \pm 0.05$ | Ducati (2002) |
| HD43039 | Johnson | H | $2.08 \pm 0.05$ | Voelcker (1975) |
| HD43039 | 2200 | 361 | $115.20 \pm 12.50$ | Smith et al. (2004) |
| HD43039 | Johnson | K | $1.92 \pm 0.05$ | Ducati (2002) |
| HD43039 | Johnson | K | $1.94 \pm 0.05$ | Johnson et al. (1966) |
| HD43039 | Johnson | K | $1.94 \pm 0.05$ | Glass (1975) |
| HD43039 | Johnson | K | $2.06 \pm 0.07$ | Neugebauer & Leighton (1969) |
| HD43039 | Johnson | L | $1.82 \pm 0.05$ | Ducati (2002) |
| HD43039 | 3500 | 898 | $54.90 \pm 8.40$ | Smith et al. (2004) |
| HD43039 | 4900 | 712 | $26.00 \pm 6.00$ | Smith et al. (2004) |
| HD43039 | 12000 | 6384 | $13.70 \pm 27.10$ | Smith et al. (2004) |
| HD46709 | KronComet | NH | $10.63 \pm 0.00$ | This work |
| HD46709 | KronComet | UVc | $10.27 \pm 0.00$ | This work |
| HD46709 | DDO | m35 | $10.65 \pm 0.05$ | McClure & Forrester (1981) |
| HD46709 | WBVR | W | $9.15 \pm 0.05$ | Kornilov et al. (1991) |
| HD46709 | Johnson | U | $9.08 \pm 0.02$ | This work |
| HD46709 | Johnson | U | $9.17 \pm 0.05$ | Mermilliod (1986) |
| HD46709 | Johnson | U | $9.25 \pm 0.05$ | McClure (1970) |
| HD46709 | DDO | m38 | $9.38 \pm 0.05$ | McClure & Forrester (1981) |
| HD46709 | KronComet | CN | $9.93 \pm 0.00$ | This work |
| HD46709 | DDO | m41 | $9.52 \pm 0.05$ | McClure & Forrester (1981) |
| HD46709 | Oja | m41 | $9.00 \pm 0.05$ | Häggkvist & Oja (1970) |
| HD46709 | DDO | m42 | $9.16 \pm 0.05$ | McClure & Forrester (1981) |
| HD46709 | Oja | m42 | $8.62 \pm 0.05$ | Häggkvist & Oja (1970) |
| HD46709 | KronComet | COp | $8.03 \pm 0.01$ | This work |
| HD46709 | WBVR | B | $7.44 \pm 0.05$ | Kornilov et al. (1991) |
| HD46709 | Johnson | B | $7.33 \pm 0.00$ | This work |
| HD46709 | Johnson | B | $7.37 \pm 0.05$ | Haggkvist & Oja (1970) |
| HD46709 | Johnson | B | $7.38 \pm 0.05$ | Mermilliod (1986) |
| HD46709 | Johnson | B | $7.40 \pm 0.05$ | McClure (1970) |
| HD46709 | KronComet | Bc | $7.25 \pm 0.01$ | This work |
| HD46709 | DDO | m45 | $7.93 \pm 0.05$ | McClure & Forrester (1981) |
| HD46709 | Oja | m45 | $7.04 \pm 0.05$ | Häggkvist & Oja (1970) |
| HD46709 | DDO | m48 | $6.51 \pm 0.05$ | McClure & Forrester (1981) |
| HD46709 | KronComet | C2 | $6.30 \pm 0.00$ | This work |

**Table 21** *(continued)*

| Star ID | System/Wvlen | Band/Bandpass | Value | Reference |
|---------|-------------|---------------|-------|-----------|
| HD46709 | KronComet | Gc | $6.11 \pm 0.00$ | This work |
| HD46709 | WBVR | V | $5.91 \pm 0.05$ | Kornilov et al. (1991) |
| HD46709 | Johnson | V | $5.87 \pm 0.05$ | Haggkvist & Oja (1970) |
| HD46709 | Johnson | V | $5.88 \pm 0.05$ | Mermilliod (1986) |
| HD46709 | Johnson | V | $5.89 \pm 0.05$ | McClure (1970) |
| HD46709 | Johnson | V | $6.00 \pm 0.01$ | This work |
| HD46709 | WBVR | R | $4.84 \pm 0.05$ | Kornilov et al. (1991) |
| HD46709 | KronComet | Rc | $4.61 \pm 0.01$ | This work |
| HD46709 | 1250 | 310 | $70.60 \pm 12.00$ | Smith et al. (2004) |
| HD46709 | Johnson | J | $3.27 \pm 0.05$ | Iijima & Ishida (1978) |
| HD46709 | Johnson | H | $2.61 \pm 0.05$ | Iijima & Ishida (1978) |
| HD46709 | 2200 | 361 | $60.00 \pm 11.10$ | Smith et al. (2004) |
| HD46709 | Johnson | K | $2.58 \pm 0.08$ | Neugebauer & Leighton (1969) |
| HD46709 | 3500 | 898 | $22.40 \pm 9.00$ | Smith et al. (2004) |
| HD46709 | 4900 | 712 | $10.60 \pm 6.90$ | Smith et al. (2004) |
| HD46709 | 12000 | 6384 | $-30.60 \pm 37.40$ | Smith et al. (2004) |
| HD48450 | Vilnius | U | $10.55 \pm 0.05$ | Zdanavicius et al. (1972) |
| HD48450 | WBVR | W | $8.52 \pm 0.05$ | Kornilov et al. (1991) |
| HD48450 | Johnson | U | $8.53 \pm 0.05$ | Rybka (1969) |
| HD48450 | Johnson | U | $8.53 \pm 0.05$ | Ducati (2002) |
| HD48450 | Johnson | U | $8.61 \pm 0.05$ | Mermilliod (1986) |
| HD48450 | Vilnius | P | $9.80 \pm 0.05$ | Zdanavicius et al. (1972) |
| HD48450 | Vilnius | X | $8.47 \pm 0.05$ | Zdanavicius et al. (1972) |
| HD48450 | Oja | m41 | $8.41 \pm 0.05$ | Häggkvist & Oja (1970) |
| HD48450 | Oja | m42 | $8.11 \pm 0.05$ | Häggkvist & Oja (1970) |
| HD48450 | WBVR | B | $6.91 \pm 0.05$ | Kornilov et al. (1991) |
| HD48450 | Johnson | B | $6.88 \pm 0.05$ | Mermilliod (1986) |
| HD48450 | Johnson | B | $6.89 \pm 0.05$ | Rybka (1969) |
| HD48450 | Johnson | B | $6.89 \pm 0.05$ | Ducati (2002) |
| HD48450 | Oja | m45 | $6.52 \pm 0.05$ | Häggkvist & Oja (1970) |
| HD48450 | Vilnius | Y | $6.51 \pm 0.05$ | Zdanavicius et al. (1972) |
| HD48450 | Vilnius | Z | $5.95 \pm 0.05$ | Zdanavicius et al. (1972) |
| HD48450 | WBVR | V | $5.43 \pm 0.05$ | Kornilov et al. (1991) |
| HD48450 | Vilnius | V | $5.48 \pm 0.05$ | Zdanavicius et al. (1972) |
| HD48450 | Johnson | V | $5.44 \pm 0.05$ | Rybka (1969) |
| HD48450 | Johnson | V | $5.44 \pm 0.05$ | Ducati (2002) |
| HD48450 | Johnson | V | $5.45 \pm 0.05$ | Mermilliod (1986) |
| HD48450 | Vilnius | S | $4.49 \pm 0.05$ | Zdanavicius et al. (1972) |
| HD48450 | WBVR | R | $4.35 \pm 0.05$ | Kornilov et al. (1991) |
| HD48450 | Johnson | J | $2.90 \pm 0.05$ | McWilliam & Lambert (1984) |
| HD48450 | Johnson | J | $2.90 \pm 0.05$ | Ducati (2002) |

**Table 21** *(continued)*

| Star ID | System/Wvlen | Band/Bandpass | Value | Reference |
|---------|--------------|---------------|-------|-----------|
| HD48450 | 2200 | 361 | $82.20 \pm 5.70$ | Smith et al. (2004) |
| HD48450 | Johnson | K | $2.00 \pm 0.05$ | Ducati (2002) |
| HD48450 | Johnson | K | $2.16 \pm 0.07$ | Neugebauer & Leighton (1969) |
| HD48450 | 3500 | 898 | $39.90 \pm 13.40$ | Smith et al. (2004) |
| HD48450 | 4900 | 712 | $16.80 \pm 5.60$ | Smith et al. (2004) |
| HD48450 | 12000 | 6384 | $1.60 \pm 25.30$ | Smith et al. (2004) |
| HD49738 | Vilnius | U | $10.45 \pm 0.05$ | Zdanavicius et al. (1972) |
| HD49738 | WBVR | W | $8.41 \pm 0.05$ | Kornilov et al. (1991) |
| HD49738 | Vilnius | P | $9.69 \pm 0.05$ | Zdanavicius et al. (1972) |
| HD49738 | Vilnius | X | $8.45 \pm 0.05$ | Zdanavicius et al. (1972) |
| HD49738 | DDO | m41 | $8.90 \pm 0.05$ | Eggen (1989) |
| HD49738 | Oja | m41 | $8.49 \pm 0.05$ | Häggkvist & Oja (1970) |
| HD49738 | DDO | m42 | $8.56 \pm 0.05$ | Eggen (1989) |
| HD49738 | Oja | m42 | $8.12 \pm 0.05$ | Häggkvist & Oja (1970) |
| HD49738 | WBVR | B | $7.06 \pm 0.05$ | Kornilov et al. (1991) |
| HD49738 | Johnson | B | $6.99 \pm 0.05$ | Haggkvist & Oja (1970) |
| HD49738 | DDO | m45 | $7.48 \pm 0.05$ | Eggen (1989) |
| HD49738 | Oja | m45 | $6.67 \pm 0.05$ | Häggkvist & Oja (1970) |
| HD49738 | Vilnius | Y | $6.67 \pm 0.05$ | Zdanavicius et al. (1972) |
| HD49738 | DDO | m48 | $6.16 \pm 0.05$ | Eggen (1989) |
| HD49738 | Vilnius | Z | $6.11 \pm 0.05$ | Zdanavicius et al. (1972) |
| HD49738 | WBVR | V | $5.70 \pm 0.05$ | Kornilov et al. (1991) |
| HD49738 | Vilnius | V | $5.74 \pm 0.05$ | Zdanavicius et al. (1972) |
| HD49738 | Johnson | V | $5.65 \pm 0.05$ | Haggkvist & Oja (1970) |
| HD49738 | Vilnius | S | $4.85 \pm 0.05$ | Zdanavicius et al. (1972) |
| HD49738 | WBVR | R | $4.78 \pm 0.05$ | Kornilov et al. (1991) |
| HD49738 | 1250 | 310 | $63.80 \pm 10.80$ | Smith et al. (2004) |
| HD49738 | 2200 | 361 | $53.60 \pm 9.30$ | Smith et al. (2004) |
| HD49738 | Johnson | K | $2.87 \pm 0.08$ | Neugebauer & Leighton (1969) |
| HD49738 | 3500 | 898 | $33.40 \pm 17.90$ | Smith et al. (2004) |
| HD49738 | 4900 | 712 | $21.30 \pm 6.90$ | Smith et al. (2004) |
| HD49738 | 12000 | 6384 | $6.30 \pm 25.80$ | Smith et al. (2004) |
| HD49968 | Oja | m41 | $8.72 \pm 0.05$ | Häggkvist & Oja (1970) |
| HD49968 | Oja | m42 | $8.39 \pm 0.05$ | Häggkvist & Oja (1970) |
| HD49968 | KronComet | COp | $7.79 \pm 0.04$ | This work |
| HD49968 | WBVR | B | $7.18 \pm 0.05$ | Kornilov et al. (1991) |
| HD49968 | Johnson | B | $7.08 \pm 0.07$ | This work |
| HD49968 | Johnson | B | $7.11 \pm 0.05$ | Haggkvist & Oja (1970) |
| HD49968 | Johnson | B | $7.15 \pm 0.05$ | Mermilliod (1986) |
| HD49968 | KronComet | Bc | $6.99 \pm 0.03$ | This work |
| HD49968 | Oja | m45 | $6.76 \pm 0.05$ | Häggkvist & Oja (1970) |

**Table 21** *(continued)*

| Star ID | System/Wvlen | Band/Bandpass | Value | Reference |
|---------|--------------|---------------|-------|-----------|
| HD49968 | KronComet | C2 | $6.16 \pm 0.04$ | This work |
| HD49968 | KronComet | Gc | $5.87 \pm 0.04$ | This work |
| HD49968 | WBVR | V | $5.68 \pm 0.05$ | Kornilov et al. (1991) |
| HD49968 | Johnson | V | $5.66 \pm 0.05$ | Haggkvist & Oja (1970) |
| HD49968 | Johnson | V | $5.68 \pm 0.05$ | Mermilliod (1986) |
| HD49968 | Johnson | V | $5.77 \pm 0.07$ | This work |
| HD49968 | WBVR | R | $4.59 \pm 0.05$ | Kornilov et al. (1991) |
| HD49968 | KronComet | Rc | $4.41 \pm 0.05$ | This work |
| HD49968 | 1250 | 310 | $79.80 \pm 6.70$ | Smith et al. (2004) |
| HD49968 | 2200 | 361 | $68.40 \pm 8.20$ | Smith et al. (2004) |
| HD49968 | Johnson | K | $2.22 \pm 0.09$ | Neugebauer & Leighton (1969) |
| HD49968 | 3500 | 898 | $33.00 \pm 5.70$ | Smith et al. (2004) |
| HD49968 | 4900 | 712 | $14.20 \pm 5.80$ | Smith et al. (2004) |
| HD49968 | 12000 | 6384 | $14.10 \pm 40.60$ | Smith et al. (2004) |
| HD52960 | Vilnius | U | $10.04 \pm 0.05$ | Kazlauskas et al. (2005) |
| HD52960 | WBVR | W | $8.05 \pm 0.05$ | Kornilov et al. (1991) |
| HD52960 | Johnson | U | $8.11 \pm 0.05$ | Argue (1963) |
| HD52960 | Johnson | U | $8.11 \pm 0.05$ | Mermilliod (1986) |
| HD52960 | Johnson | U | $8.14 \pm 0.05$ | Argue (1966) |
| HD52960 | Johnson | U | $8.15 \pm 0.05$ | McClure (1970) |
| HD52960 | Vilnius | P | $9.32 \pm 0.05$ | Kazlauskas et al. (2005) |
| HD52960 | Vilnius | X | $8.00 \pm 0.05$ | Kazlauskas et al. (2005) |
| HD52960 | DDO | m41 | $8.50 \pm 0.05$ | McClure & Forrester (1981) |
| HD52960 | Oja | m41 | $8.04 \pm 0.05$ | Häggkvist & Oja (1970) |
| HD52960 | DDO | m42 | $8.19 \pm 0.05$ | McClure & Forrester (1981) |
| HD52960 | Oja | m42 | $7.70 \pm 0.05$ | Häggkvist & Oja (1970) |
| HD52960 | WBVR | B | $6.56 \pm 0.05$ | Kornilov et al. (1991) |
| HD52960 | Johnson | B | $6.37 \pm 0.06$ | This work |
| HD52960 | Johnson | B | $6.48 \pm 0.05$ | Argue (1963) |
| HD52960 | Johnson | B | $6.52 \pm 0.05$ | Argue (1966) |
| HD52960 | Johnson | B | $6.53 \pm 0.05$ | McClure (1970) |
| HD52960 | Johnson | B | $6.53 \pm 0.05$ | Mermilliod (1986) |
| HD52960 | KronComet | Bc | $6.25 \pm 0.04$ | This work |
| HD52960 | DDO | m45 | $7.01 \pm 0.05$ | McClure & Forrester (1981) |
| HD52960 | Oja | m45 | $6.17 \pm 0.05$ | Häggkvist & Oja (1970) |
| HD52960 | Vilnius | Y | $6.15 \pm 0.05$ | Kazlauskas et al. (2005) |
| HD52960 | DDO | m48 | $5.67 \pm 0.05$ | McClure & Forrester (1981) |
| HD52960 | KronComet | C2 | $5.39 \pm 0.05$ | This work |
| HD52960 | Vilnius | Z | $5.60 \pm 0.05$ | Kazlauskas et al. (2005) |
| HD52960 | KronComet | Gc | $5.21 \pm 0.03$ | This work |
| HD52960 | WBVR | V | $5.13 \pm 0.05$ | Kornilov et al. (1991) |

**Table 21** *(continued)*

| Star ID | System/Wvlen | Band/Bandpass | Value | Reference |
|---------|--------------|---------------|-------|-----------|
| HD52960 | Vilnius | V | $5.15 \pm 0.05$ | Kazlauskas et al. (2005) |
| HD52960 | Johnson | V | $5.10 \pm 0.05$ | Argue (1963) |
| HD52960 | Johnson | V | $5.13 \pm 0.05$ | Argue (1966) |
| HD52960 | Johnson | V | $5.14 \pm 0.05$ | McClure (1970) |
| HD52960 | Johnson | V | $5.14 \pm 0.05$ | Mermilliod (1986) |
| HD52960 | Johnson | V | $5.17 \pm 0.03$ | This work |
| HD52960 | Vilnius | S | $4.21 \pm 0.05$ | Kazlauskas et al. (2005) |
| HD52960 | WBVR | R | $4.13 \pm 0.05$ | Kornilov et al. (1991) |
| HD52960 | Johnson | J | $2.70 \pm 0.05$ | Alonso et al. (1998) |
| HD52960 | Johnson | H | $2.08 \pm 0.05$ | Alonso et al. (1998) |
| HD52960 | 2200 | 361 | $110.70 \pm 18.00$ | Smith et al. (2004) |
| HD52960 | Johnson | K | $1.93 \pm 0.05$ | Neugebauer & Leighton (1969) |
| HD52960 | 3500 | 898 | $67.60 \pm 13.70$ | Smith et al. (2004) |
| HD52960 | 4900 | 712 | $29.40 \pm 8.30$ | Smith et al. (2004) |
| HD52960 | 12000 | 6384 | $10.80 \pm 23.60$ | Smith et al. (2004) |
| HD54719 | Oja | m41 | $7.12 \pm 0.05$ | Häggkvist & Oja (1970) |
| HD54719 | Oja | m42 | $6.72 \pm 0.05$ | Häggkvist & Oja (1970) |
| HD54719 | KronComet | COp | $5.74 \pm 0.44$ | This work |
| HD54719 | WBVR | B | $5.69 \pm 0.05$ | Kornilov et al. (1991) |
| HD54719 | Johnson | B | $5.54 \pm 0.08$ | This work |
| HD54719 | Johnson | B | $5.63 \pm 0.05$ | Argue (1963) |
| HD54719 | Johnson | B | $5.64 \pm 0.05$ | Argue (1966) |
| HD54719 | Johnson | B | $5.66 \pm 0.05$ | Häggkvist & Oja (1966) |
| HD54719 | Johnson | B | $5.67 \pm 0.05$ | Johnson (1964) |
| HD54719 | Johnson | B | $5.68 \pm 0.05$ | Johnson et al. (1966) |
| HD54719 | Johnson | B | $5.68 \pm 0.05$ | Moffett & Barnes (1979) |
| HD54719 | Johnson | B | $5.68 \pm 0.05$ | Ducati (2002) |
| HD54719 | KronComet | Bc | $5.45 \pm 0.10$ | This work |
| HD54719 | 13c | m45 | $5.27 \pm 0.05$ | Johnson & Mitchell (1995) |
| HD54719 | Oja | m45 | $5.33 \pm 0.05$ | Häggkvist & Oja (1970) |
| HD54719 | Vilnius | Y | $5.33 \pm 0.05$ | Kazlauskas et al. (2005) |
| HD54719 | KronComet | C2 | $4.67 \pm 0.04$ | This work |
| HD54719 | Vilnius | Z | $4.80 \pm 0.05$ | Kazlauskas et al. (2005) |
| HD54719 | 13c | m52 | $4.71 \pm 0.05$ | Johnson & Mitchell (1995) |
| HD54719 | KronComet | Gc | $4.52 \pm 0.08$ | This work |
| HD54719 | WBVR | V | $4.40 \pm 0.05$ | Kornilov et al. (1991) |
| HD54719 | Vilnius | V | $4.40 \pm 0.05$ | Kazlauskas et al. (2005) |
| HD54719 | Johnson | V | $4.37 \pm 0.05$ | Argue (1963) |
| HD54719 | Johnson | V | $4.38 \pm 0.05$ | Argue (1966) |
| HD54719 | Johnson | V | $4.40 \pm 0.05$ | Häggkvist & Oja (1966) |
| HD54719 | Johnson | V | $4.41 \pm 0.05$ | Moffett & Barnes (1979) |

**Table 21** *(continued)*

| Star ID | System/Wvlen | Band/Bandpass | Value | Reference |
|---------|--------------|---------------|-------|-----------|
| HD54719 | Johnson | V | $4.42 \pm 0.05$ | Johnson (1964) |
| HD54719 | Johnson | V | $4.42 \pm 0.05$ | Johnson et al. (1966) |
| HD54719 | Johnson | V | $4.42 \pm 0.05$ | Ducati (2002) |
| HD54719 | Johnson | V | $4.45 \pm 0.03$ | This work |
| HD54719 | 13c | m58 | $4.10 \pm 0.05$ | Johnson & Mitchell (1995) |
| HD54719 | 13c | m63 | $3.76 \pm 0.05$ | Johnson & Mitchell (1995) |
| HD54719 | Vilnius | S | $3.55 \pm 0.05$ | Kazlauskas et al. (2005) |
| HD54719 | WBVR | R | $3.52 \pm 0.05$ | Kornilov et al. (1991) |
| HD54719 | KronComet | Rc | $3.26 \pm 0.08$ | This work |
| HD54719 | 13c | m72 | $3.47 \pm 0.05$ | Johnson & Mitchell (1995) |
| HD54719 | 13c | m80 | $3.20 \pm 0.05$ | Johnson & Mitchell (1995) |
| HD54719 | 13c | m86 | $3.05 \pm 0.05$ | Johnson & Mitchell (1995) |
| HD54719 | 13c | m99 | $2.86 \pm 0.05$ | Johnson & Mitchell (1995) |
| HD54719 | 13c | m110 | $2.62 \pm 0.05$ | Johnson & Mitchell (1995) |
| HD54719 | 1250 | 310 | $196.10 \pm 8.80$ | Smith et al. (2004) |
| HD54719 | Johnson | J | $2.38 \pm 0.05$ | Johnson et al. (1966) |
| HD54719 | Johnson | J | $2.38 \pm 0.05$ | Voelcker (1975) |
| HD54719 | Johnson | J | $2.38 \pm 0.05$ | Ducati (2002) |
| HD54719 | Johnson | H | $1.79 \pm 0.05$ | Voelcker (1975) |
| HD54719 | Johnson | H | $1.79 \pm 0.05$ | Ducati (2002) |
| HD54719 | 2200 | 361 | $141.20 \pm 6.60$ | Smith et al. (2004) |
| HD54719 | Johnson | K | $1.59 \pm 0.05$ | Johnson et al. (1966) |
| HD54719 | Johnson | K | $1.59 \pm 0.05$ | Ducati (2002) |
| HD54719 | Johnson | K | $1.71 \pm 0.06$ | Neugebauer & Leighton (1969) |
| HD54719 | Johnson | L | $1.44 \pm 0.05$ | Ducati (2002) |
| HD54719 | 3500 | 898 | $68.80 \pm 7.30$ | Smith et al. (2004) |
| HD54719 | 4900 | 712 | $31.80 \pm 6.10$ | Smith et al. (2004) |
| HD54719 | 12000 | 6384 | $2.70 \pm 25.70$ | Smith et al. (2004) |
| HD57669 | DDO | m35 | $9.04 \pm 0.05$ | McClure & Forrester (1981) |
| HD57669 | WBVR | W | $7.56 \pm 0.05$ | Kornilov et al. (1991) |
| HD57669 | Johnson | U | $7.65 \pm 0.05$ | Argue (1963) |
| HD57669 | Johnson | U | $7.67 \pm 0.05$ | McClure (1970) |
| HD57669 | Johnson | U | $7.68 \pm 0.05$ | Mermilliod (1986) |
| HD57669 | Johnson | U | $7.80 \pm 0.05$ | Fernie (1983) |
| HD57669 | DDO | m38 | $7.87 \pm 0.05$ | McClure & Forrester (1981) |
| HD57669 | DDO | m41 | $8.33 \pm 0.05$ | McClure & Forrester (1981) |
| HD57669 | Oja | m41 | $7.87 \pm 0.05$ | Häggkvist & Oja (1970) |
| HD57669 | DDO | m42 | $7.95 \pm 0.05$ | McClure & Forrester (1981) |
| HD57669 | Oja | m42 | $7.46 \pm 0.05$ | Häggkvist & Oja (1970) |
| HD57669 | WBVR | B | $6.49 \pm 0.05$ | Kornilov et al. (1991) |
| HD57669 | Johnson | B | $6.41 \pm 0.05$ | Argue (1963) |

**Table 21** *(continued)*

| Star ID | System/Wvlen | Band/Bandpass | Value | Reference |
|---------|--------------|---------------|-------|-----------|
| HD57669 | Johnson | B | $6.43 \pm 0.05$ | McClure (1970) |
| HD57669 | Johnson | B | $6.47 \pm 0.05$ | Mermilliod (1986) |
| HD57669 | Johnson | B | $6.53 \pm 0.05$ | Fernie (1983) |
| HD57669 | DDO | m45 | $6.98 \pm 0.05$ | McClure & Forrester (1981) |
| HD57669 | Oja | m45 | $6.12 \pm 0.05$ | Häggkvist & Oja (1970) |
| HD57669 | DDO | m48 | $5.67 \pm 0.05$ | McClure & Forrester (1981) |
| HD57669 | WBVR | V | $5.22 \pm 0.05$ | Kornilov et al. (1991) |
| HD57669 | Johnson | V | $5.17 \pm 0.05$ | Argue (1963) |
| HD57669 | Johnson | V | $5.20 \pm 0.05$ | McClure (1970) |
| HD57669 | Johnson | V | $5.22 \pm 0.05$ | Mermilliod (1986) |
| HD57669 | Johnson | V | $5.27 \pm 0.05$ | Fernie (1983) |
| HD57669 | WBVR | R | $4.37 \pm 0.05$ | Kornilov et al. (1991) |
| HD57669 | 1250 | 310 | $82.80 \pm 5.30$ | Smith et al. (2004) |
| HD57669 | 2200 | 361 | $59.50 \pm 5.70$ | Smith et al. (2004) |
| HD57669 | Johnson | K | $2.49 \pm 0.05$ | Neugebauer & Leighton (1969) |
| HD57669 | 3500 | 898 | $27.30 \pm 11.00$ | Smith et al. (2004) |
| HD57669 | 4900 | 712 | $13.50 \pm 4.80$ | Smith et al. (2004) |
| HD57669 | 12000 | 6384 | $-1.40 \pm 21.10$ | Smith et al. (2004) |
| HD60136 | KronComet | COp | $9.08 \pm 0.12$ | This work |
| HD60136 | Johnson | B | $8.46 \pm 0.01$ | Oja (1991) |
| HD60136 | Johnson | B | $8.47 \pm 0.11$ | This work |
| HD60136 | KronComet | Bc | $8.33 \pm 0.11$ | This work |
| HD60136 | KronComet | C2 | $7.13 \pm 0.08$ | This work |
| HD60136 | KronComet | Gc | $6.97 \pm 0.08$ | This work |
| HD60136 | Johnson | V | $6.83 \pm 0.01$ | Oja (1991) |
| HD60136 | Johnson | V | $6.93 \pm 0.13$ | This work |
| HD60136 | KronComet | Rc | $5.45 \pm 0.06$ | This work |
| HD60136 | 1250 | 310 | $157.00 \pm 20.70$ | Smith et al. (2004) |
| HD60136 | 2200 | 361 | $152.40 \pm 20.30$ | Smith et al. (2004) |
| HD60136 | Johnson | K | $1.81 \pm 0.06$ | Neugebauer & Leighton (1969) |
| HD60136 | 3500 | 898 | $70.60 \pm 14.20$ | Smith et al. (2004) |
| HD60136 | 4900 | 712 | $29.00 \pm 6.00$ | Smith et al. (2004) |
| HD60136 | 12000 | 6384 | $9.30 \pm 20.30$ | Smith et al. (2004) |
| HD61294 | Geneva | B1 | $8.59 \pm 0.08$ | Golay (1972) |
| HD61294 | Oja | m41 | $8.99 \pm 0.05$ | Häggkvist & Oja (1970) |
| HD61294 | Oja | m42 | $8.73 \pm 0.05$ | Häggkvist & Oja (1970) |
| HD61294 | Geneva | B | $6.88 \pm 0.08$ | Golay (1972) |
| HD61294 | KronComet | COp | $8.07 \pm 0.00$ | This work |
| HD61294 | WBVR | B | $7.43 \pm 0.05$ | Kornilov et al. (1991) |
| HD61294 | Johnson | B | $7.35 \pm 0.04$ | This work |
| HD61294 | Johnson | B | $7.36 \pm 0.05$ | Haggkvist & Oja (1970) |

**Table 21** *continued on next page*

**Table 21** *(continued)*

| Star ID | System/Wvlen | Band/Bandpass | Value | Reference |
|---------|--------------|---------------|-------|-----------|
| HD61294 | Johnson | B | $7.36 \pm 0.05$ | Ducati (2002) |
| HD61294 | Johnson | B | $7.42 \pm 0.05$ | Mermilliod (1986) |
| HD61294 | KronComet | Bc | $7.23 \pm 0.01$ | This work |
| HD61294 | Geneva | B2 | $7.84 \pm 0.08$ | Golay (1972) |
| HD61294 | Oja | m45 | $7.03 \pm 0.05$ | Häggkvist & Oja (1970) |
| HD61294 | KronComet | C2 | $6.20 \pm 0.01$ | This work |
| HD61294 | KronComet | Gc | $5.92 \pm 0.01$ | This work |
| HD61294 | Geneva | V1 | $6.61 \pm 0.08$ | Golay (1972) |
| HD61294 | WBVR | V | $5.77 \pm 0.05$ | Kornilov et al. (1991) |
| HD61294 | Geneva | V | $5.77 \pm 0.08$ | Golay (1972) |
| HD61294 | Johnson | V | $5.73 \pm 0.05$ | Haggkvist & Oja (1970) |
| HD61294 | Johnson | V | $5.73 \pm 0.05$ | Ducati (2002) |
| HD61294 | Johnson | V | $5.78 \pm 0.05$ | Mermilliod (1986) |
| HD61294 | Johnson | V | $5.82 \pm 0.06$ | This work |
| HD61294 | Geneva | G | $6.62 \pm 0.08$ | Golay (1972) |
| HD61294 | WBVR | R | $4.48 \pm 0.05$ | Kornilov et al. (1991) |
| HD61294 | KronComet | Rc | $4.34 \pm 0.01$ | This work |
| HD61294 | Johnson | J | $2.75 \pm 0.05$ | McWilliam & Lambert (1984) |
| HD61294 | Johnson | J | $2.75 \pm 0.05$ | Ducati (2002) |
| HD61294 | Johnson | K | $1.70 \pm 0.05$ | Ducati (2002) |
| HD61294 | Johnson | K | $1.81 \pm 0.06$ | Neugebauer & Leighton (1969) |
| HD61913 | Geneva | U | $9.67 \pm 0.08$ | Golay (1972) |
| HD61913 | Vilnius | U | $10.89 \pm 0.05$ | Kazlauskas et al. (2005) |
| HD61913 | WBVR | W | $9.03 \pm 0.05$ | Kornilov et al. (1991) |
| HD61913 | Johnson | U | $9.05 \pm 0.05$ | Mermilliod (1986) |
| HD61913 | Johnson | U | $9.11 \pm 0.05$ | Johnson & Knuckles (1957) |
| HD61913 | Johnson | U | $9.11 \pm 0.05$ | Johnson et al. (1966) |
| HD61913 | Johnson | U | $9.11 \pm 0.05$ | Nicolet (1978) |
| HD61913 | Johnson | U | $9.11 \pm 0.05$ | Ducati (2002) |
| HD61913 | Vilnius | P | $9.94 \pm 0.05$ | Kazlauskas et al. (2005) |
| HD61913 | KronComet | CN | $9.37 \pm 0.05$ | This work |
| HD61913 | Geneva | B1 | $8.18 \pm 0.08$ | Golay (1972) |
| HD61913 | Vilnius | X | $8.72 \pm 0.05$ | Kazlauskas et al. (2005) |
| HD61913 | Oja | m41 | $8.59 \pm 0.05$ | Häggkvist & Oja (1970) |
| HD61913 | Oja | m42 | $8.43 \pm 0.05$ | Häggkvist & Oja (1970) |
| HD61913 | Geneva | B | $6.62 \pm 0.08$ | Golay (1972) |
| HD61913 | KronComet | COp | $7.82 \pm 0.05$ | This work |
| HD61913 | WBVR | B | $7.22 \pm 0.05$ | Kornilov et al. (1991) |
| HD61913 | Johnson | B | $7.17 \pm 0.05$ | Mermilliod (1986) |
| HD61913 | Johnson | B | $7.19 \pm 0.04$ | This work |
| HD61913 | Johnson | B | $7.20 \pm 0.05$ | Johnson & Knuckles (1957) |

**Table 21** *(continued)*

| Star ID | System/Wvlen | Band/Bandpass | Value | Reference |
|---------|--------------|---------------|-------|-----------|
| HD61913 | Johnson | B | $7.20 \pm 0.05$ | Johnson et al. (1966) |
| HD61913 | Johnson | B | $7.20 \pm 0.05$ | Nicolet (1978) |
| HD61913 | Johnson | B | $7.20 \pm 0.05$ | Ducati (2002) |
| HD61913 | Johnson | B | $7.41 \pm 0.05$ | Miczaika (1954) |
| HD61913 | KronComet | Bc | $7.07 \pm 0.01$ | This work |
| HD61913 | Geneva | B2 | $7.62 \pm 0.08$ | Golay (1972) |
| HD61913 | Oja | m45 | $6.90 \pm 0.05$ | Häggkvist & Oja (1970) |
| HD61913 | Vilnius | Y | $6.77 \pm 0.05$ | Kazlauskas et al. (2005) |
| HD61913 | KronComet | C2 | $5.86 \pm 0.02$ | This work |
| HD61913 | Vilnius | Z | $6.00 \pm 0.05$ | Kazlauskas et al. (2005) |
| HD61913 | KronComet | Gc | $5.72 \pm 0.02$ | This work |
| HD61913 | Geneva | V1 | $6.33 \pm 0.08$ | Golay (1972) |
| HD61913 | WBVR | V | $5.53 \pm 0.05$ | Kornilov et al. (1991) |
| HD61913 | Vilnius | V | $5.53 \pm 0.05$ | Kazlauskas et al. (2005) |
| HD61913 | Geneva | V | $5.53 \pm 0.08$ | Golay (1972) |
| HD61913 | Johnson | V | $5.52 \pm 0.05$ | Mermilliod (1986) |
| HD61913 | Johnson | V | $5.56 \pm 0.05$ | Johnson & Knuckles (1957) |
| HD61913 | Johnson | V | $5.56 \pm 0.05$ | Johnson et al. (1966) |
| HD61913 | Johnson | V | $5.56 \pm 0.05$ | Nicolet (1978) |
| HD61913 | Johnson | V | $5.56 \pm 0.05$ | Ducati (2002) |
| HD61913 | Johnson | V | $5.61 \pm 0.05$ | This work |
| HD61913 | Johnson | V | $5.72 \pm 0.05$ | Miczaika (1954) |
| HD61913 | Geneva | G | $6.45 \pm 0.08$ | Golay (1972) |
| HD61913 | Vilnius | S | $4.31 \pm 0.05$ | Kazlauskas et al. (2005) |
| HD61913 | WBVR | R | $3.92 \pm 0.05$ | Kornilov et al. (1991) |
| HD61913 | KronComet | Rc | $4.15 \pm 0.02$ | This work |
| HD61913 | Johnson | J | $1.65 \pm 0.05$ | Chen et al. (1998) |
| HD61913 | Johnson | J | $1.74 \pm 0.05$ | McWilliam & Lambert (1984) |
| HD61913 | Johnson | J | $1.74 \pm 0.05$ | Ducati (2002) |
| HD61913 | Johnson | H | $1.05 \pm 0.05$ | Chen et al. (1998) |
| HD61913 | 2200 | 361 | $326.40 \pm 10.30$ | Smith et al. (2004) |
| HD61913 | Johnson | K | $0.60 \pm 0.05$ | Ducati (2002) |
| HD61913 | Johnson | K | $0.68 \pm 0.04$ | Neugebauer & Leighton (1969) |
| HD61913 | 3500 | 898 | $175.10 \pm 18.10$ | Smith et al. (2004) |
| HD61913 | 4900 | 712 | $75.20 \pm 9.80$ | Smith et al. (2004) |
| HD61913 | 12000 | 6384 | $21.10 \pm 22.60$ | Smith et al. (2004) |
| HD62044 | KronComet | NH | $7.44 \pm 0.05$ | This work |
| HD62044 | KronComet | UVc | $7.22 \pm 0.05$ | This work |
| HD62044 | Geneva | U | $6.89 \pm 0.08$ | Golay (1972) |
| HD62044 | DDO | m35 | $7.72 \pm 0.05$ | McClure & Forrester (1981) |
| HD62044 | Stromgren | u | $7.70 \pm 0.08$ | Reglero et al. (1987) |

**Table 21** *(continued)*

| Star ID | System/Wvlen | Band/Bandpass | Value | Reference |
|---------|--------------|---------------|-------|-----------|
| HD62044 | Stromgren | u | $7.70 \pm 0.08$ | Hauck & Mermilliod (1998) |
| HD62044 | WBVR | W | $6.19 \pm 0.05$ | Kornilov et al. (1991) |
| HD62044 | Johnson | U | $6.28 \pm 0.05$ | Mermilliod (1986) |
| HD62044 | Johnson | U | $6.32 \pm 0.05$ | Argue (1963) |
| HD62044 | Johnson | U | $6.32 \pm 0.05$ | Jennens & Helfer (1975) |
| HD62044 | Johnson | U | $6.38 \pm 0.05$ | Johnson et al. (1966) |
| HD62044 | Johnson | U | $6.38 \pm 0.05$ | Ducati (2002) |
| HD62044 | DDO | m38 | $6.61 \pm 0.05$ | McClure & Forrester (1981) |
| HD62044 | KronComet | CN | $7.17 \pm 0.04$ | This work |
| HD62044 | Geneva | B1 | $6.09 \pm 0.08$ | Golay (1972) |
| HD62044 | DDO | m41 | $7.13 \pm 0.05$ | McClure & Forrester (1981) |
| HD62044 | Oja | m41 | $6.61 \pm 0.05$ | Häggkvist & Oja (1970) |
| HD62044 | Stromgren | v | $6.20 \pm 0.08$ | Reglero et al. (1987) |
| HD62044 | Stromgren | v | $6.20 \pm 0.08$ | Hauck & Mermilliod (1998) |
| HD62044 | DDO | m42 | $6.89 \pm 0.05$ | McClure & Forrester (1981) |
| HD62044 | Oja | m42 | $6.37 \pm 0.05$ | Häggkvist & Oja (1970) |
| HD62044 | Geneva | B | $4.73 \pm 0.08$ | Golay (1972) |
| HD62044 | KronComet | COp | $5.76 \pm 0.03$ | This work |
| HD62044 | WBVR | B | $5.37 \pm 0.05$ | Kornilov et al. (1991) |
| HD62044 | Johnson | B | $5.31 \pm 0.05$ | Mermilliod (1986) |
| HD62044 | Johnson | B | $5.34 \pm 0.05$ | Argue (1963) |
| HD62044 | Johnson | B | $5.36 \pm 0.05$ | Häggkvist & Oja (1966) |
| HD62044 | Johnson | B | $5.37 \pm 0.05$ | Jennens & Helfer (1975) |
| HD62044 | Johnson | B | $5.38 \pm 0.06$ | This work |
| HD62044 | Johnson | B | $5.41 \pm 0.05$ | Johnson et al. (1966) |
| HD62044 | Johnson | B | $5.41 \pm 0.05$ | Ducati (2002) |
| HD62044 | KronComet | Bc | $5.22 \pm 0.04$ | This work |
| HD62044 | Geneva | B2 | $5.84 \pm 0.08$ | Golay (1972) |
| HD62044 | DDO | m45 | $5.93 \pm 0.05$ | McClure & Forrester (1981) |
| HD62044 | Oja | m45 | $5.06 \pm 0.05$ | Häggkvist & Oja (1970) |
| HD62044 | Stromgren | b | $4.98 \pm 0.08$ | Reglero et al. (1987) |
| HD62044 | Stromgren | b | $4.98 \pm 0.08$ | Hauck & Mermilliod (1998) |
| HD62044 | DDO | m48 | $4.71 \pm 0.05$ | McClure & Forrester (1981) |
| HD62044 | KronComet | C2 | $4.54 \pm 0.04$ | This work |
| HD62044 | KronComet | Gc | $4.36 \pm 0.03$ | This work |
| HD62044 | Geneva | V1 | $5.07 \pm 0.08$ | Golay (1972) |
| HD62044 | WBVR | V | $4.22 \pm 0.05$ | Kornilov et al. (1991) |
| HD62044 | Stromgren | y | $4.29 \pm 0.08$ | Reglero et al. (1987) |
| HD62044 | Stromgren | y | $4.29 \pm 0.08$ | Hauck & Mermilliod (1998) |
| HD62044 | Geneva | V | $4.27 \pm 0.08$ | Golay (1972) |
| HD62044 | Johnson | V | $4.20 \pm 0.05$ | Mermilliod (1986) |

**Table 21** *(continued)*

| Star ID | System/Wvlen | Band/Bandpass | Value | Reference |
|---------|--------------|---------------|-------|-----------|
| HD62044 | Johnson | V | $4.22 \pm 0.05$ | Argue (1963) |
| HD62044 | Johnson | V | $4.24 \pm 0.05$ | Jennens & Helfer (1975) |
| HD62044 | Johnson | V | $4.25 \pm 0.05$ | Häggkvist & Oja (1966) |
| HD62044 | Johnson | V | $4.25 \pm 0.06$ | This work |
| HD62044 | Johnson | V | $4.29 \pm 0.05$ | Johnson et al. (1966) |
| HD62044 | Johnson | V | $4.29 \pm 0.05$ | Ducati (2002) |
| HD62044 | Geneva | G | $5.21 \pm 0.08$ | Golay (1972) |
| HD62044 | WBVR | R | $3.37 \pm 0.05$ | Kornilov et al. (1991) |
| HD62044 | KronComet | Rc | $3.16 \pm 0.01$ | This work |
| HD62044 | 1250 | 310 | $208.30 \pm 5.80$ | Smith et al. (2004) |
| HD62044 | Johnson | J | $2.35 \pm 0.05$ | Johnson et al. (1966) |
| HD62044 | Johnson | J | $2.35 \pm 0.05$ | Ducati (2002) |
| HD62044 | 2200 | 361 | $156.30 \pm 5.50$ | Smith et al. (2004) |
| HD62044 | Johnson | K | $1.65 \pm 0.05$ | Johnson et al. (1966) |
| HD62044 | Johnson | K | $1.65 \pm 0.05$ | Neugebauer & Leighton (1969) |
| HD62044 | Johnson | K | $1.65 \pm 0.05$ | Ducati (2002) |
| HD62044 | 3500 | 898 | $75.50 \pm 4.50$ | Smith et al. (2004) |
| HD62044 | 4900 | 712 | $38.60 \pm 8.60$ | Smith et al. (2004) |
| HD62044 | 12000 | 6384 | $-1.30 \pm 26.40$ | Smith et al. (2004) |
| HD62285 | DDO | m35 | $10.18 \pm 0.05$ | McClure & Forrester (1981) |
| HD62285 | DDO | m42 | $8.66 \pm 0.05$ | McClure & Forrester (1981) |
| HD62285 | Oja | m42 | $8.18 \pm 0.05$ | Häggkvist & Oja (1970) |
| HD62285 | KronComet | COp | $7.49 \pm 0.01$ | This work |
| HD62285 | WBVR | B | $6.89 \pm 0.05$ | Kornilov et al. (1991) |
| HD62285 | Johnson | B | $6.80 \pm 0.04$ | This work |
| HD62285 | Johnson | B | $6.81 \pm 0.05$ | Argue (1963) |
| HD62285 | Johnson | B | $6.83 \pm 0.05$ | Mermilliod (1986) |
| HD62285 | Johnson | B | $6.85 \pm 0.05$ | Argue (1966) |
| HD62285 | Johnson | B | $6.87 \pm 0.05$ | Bakos (1968) |
| HD62285 | KronComet | Bc | $6.66 \pm 0.01$ | This work |
| HD62285 | DDO | m45 | $7.33 \pm 0.05$ | McClure & Forrester (1981) |
| HD62285 | Oja | m45 | $6.50 \pm 0.05$ | Häggkvist & Oja (1970) |
| HD62285 | DDO | m48 | $5.94 \pm 0.05$ | McClure & Forrester (1981) |
| HD62285 | KronComet | C2 | $5.74 \pm 0.01$ | This work |
| HD62285 | KronComet | Gc | $5.47 \pm 0.01$ | This work |
| HD62285 | WBVR | V | $5.31 \pm 0.05$ | Kornilov et al. (1991) |
| HD62285 | Johnson | V | $5.28 \pm 0.05$ | Argue (1963) |
| HD62285 | Johnson | V | $5.31 \pm 0.05$ | Argue (1966) |
| HD62285 | Johnson | V | $5.31 \pm 0.05$ | Bakos (1968) |
| HD62285 | Johnson | V | $5.32 \pm 0.05$ | Mermilliod (1986) |
| HD62285 | Johnson | V | $5.34 \pm 0.06$ | This work |

| Star ID | System/Wvlen | Band/Bandpass | Value | Reference |
|---------|--------------|---------------|-------|-----------|
| HD62285 | WBVR | R | $4.13 \pm 0.05$ | Kornilov et al. (1991) |
| HD62285 | KronComet | Rc | $3.93 \pm 0.01$ | This work |
| HD62285 | 1250 | 310 | $155.80 \pm 10.70$ | Smith et al. (2004) |
| HD62285 | 2200 | 361 | $140.10 \pm 10.50$ | Smith et al. (2004) |
| HD62285 | Johnson | K | $1.66 \pm 0.06$ | Neugebauer & Leighton (1969) |
| HD62285 | 3500 | 898 | $68.60 \pm 6.80$ | Smith et al. (2004) |
| HD62285 | 4900 | 712 | $30.20 \pm 6.50$ | Smith et al. (2004) |
| HD62285 | 12000 | 6384 | $2.70 \pm 25.20$ | Smith et al. (2004) |
| HD62345 | 13c | m33 | $5.23 \pm 0.05$ | Johnson & Mitchell (1995) |
| HD62345 | KronComet | NH | $6.30 \pm 0.01$ | This work |
| HD62345 | KronComet | UVc | $6.06 \pm 0.01$ | This work |
| HD62345 | Geneva | U | $5.71 \pm 0.08$ | Golay (1972) |
| HD62345 | Vilnius | U | $6.96 \pm 0.05$ | Zdanavicius et al. (1969) |
| HD62345 | 13c | m35 | $5.05 \pm 0.05$ | Johnson & Mitchell (1995) |
| HD62345 | DDO | m35 | $6.53 \pm 0.05$ | McClure & Forrester (1981) |
| HD62345 | Stromgren | u | $6.43 \pm 0.08$ | Philip & Philip (1973) |
| HD62345 | Stromgren | u | $6.44 \pm 0.08$ | Gray & Olsen (1991) |
| HD62345 | Stromgren | u | $6.44 \pm 0.08$ | Olsen (1993) |
| HD62345 | Stromgren | u | $6.44 \pm 0.08$ | Hauck & Mermilliod (1998) |
| HD62345 | Stromgren | u | $6.45 \pm 0.08$ | Crawford & Barnes (1970) |
| HD62345 | Stromgren | u | $6.59 \pm 0.08$ | Williams (1966) |
| HD62345 | WBVR | W | $5.04 \pm 0.05$ | Kornilov et al. (1991) |
| HD62345 | Johnson | U | $5.17 \pm 0.05$ | Argue (1963) |
| HD62345 | Johnson | U | $5.18 \pm 0.05$ | Johnson & Morgan (1953b) |
| HD62345 | Johnson | U | $5.18 \pm 0.05$ | Johnson & Harris (1954) |
| HD62345 | Johnson | U | $5.18 \pm 0.05$ | Johnson et al. (1966) |
| HD62345 | Johnson | U | $5.19 \pm 0.05$ | Johnson et al. (1966) |
| HD62345 | Johnson | U | $5.19 \pm 0.05$ | Argue (1966) |
| HD62345 | Johnson | U | $5.19 \pm 0.05$ | Ducati (2002) |
| HD62345 | Johnson | U | $5.21 \pm 0.05$ | Jennens & Helfer (1975) |
| HD62345 | Johnson | U | $5.22 \pm 0.05$ | Appenzeller (1966) |
| HD62345 | Johnson | U | $5.24 \pm 0.05$ | Mermilliod (1986) |
| HD62345 | 13c | m37 | $5.14 \pm 0.05$ | Johnson & Mitchell (1995) |
| HD62345 | Vilnius | P | $6.37 \pm 0.05$ | Zdanavicius et al. (1969) |
| HD62345 | DDO | m38 | $5.49 \pm 0.05$ | McClure & Forrester (1981) |
| HD62345 | KronComet | CN | $6.10 \pm 0.01$ | This work |
| HD62345 | 13c | m40 | $5.03 \pm 0.05$ | Johnson & Mitchell (1995) |
| HD62345 | Geneva | B1 | $5.06 \pm 0.08$ | Golay (1972) |
| HD62345 | Vilnius | X | $5.45 \pm 0.05$ | Zdanavicius et al. (1969) |
| HD62345 | DDO | m41 | $6.08 \pm 0.05$ | McClure & Forrester (1981) |
| HD62345 | Oja | m41 | $5.61 \pm 0.05$ | Häggkvist & Oja (1970) |

**Table 21** *(continued)*

| Star ID | System/Wvlen | Band/Bandpass | Value | Reference |
|---------|--------------|---------------|-------|-----------|
| HD62345 | Stromgren | v | $5.09 \pm 0.08$ | Philip & Philip (1973) |
| HD62345 | Stromgren | v | $5.09 \pm 0.08$ | Gray & Olsen (1991) |
| HD62345 | Stromgren | v | $5.09 \pm 0.08$ | Olsen (1993) |
| HD62345 | Stromgren | v | $5.09 \pm 0.08$ | Hauck & Mermilliod (1998) |
| HD62345 | Stromgren | v | $5.10 \pm 0.08$ | Crawford & Barnes (1970) |
| HD62345 | Stromgren | v | $5.22 \pm 0.08$ | Williams (1966) |
| HD62345 | DDO | m42 | $5.88 \pm 0.05$ | McClure & Forrester (1981) |
| HD62345 | Oja | m42 | $5.41 \pm 0.05$ | Häggkvist & Oja (1970) |
| HD62345 | Geneva | B | $3.80 \pm 0.08$ | Golay (1972) |
| HD62345 | KronComet | COp | $4.75 \pm 0.01$ | This work |
| HD62345 | WBVR | B | $4.53 \pm 0.05$ | Kornilov et al. (1991) |
| HD62345 | Johnson | B | $4.47 \pm 0.05$ | Oja (1963) |
| HD62345 | Johnson | B | $4.48 \pm 0.04$ | This work |
| HD62345 | Johnson | B | $4.49 \pm 0.05$ | Argue (1963) |
| HD62345 | Johnson | B | $4.49 \pm 0.05$ | Johnson et al. (1966) |
| HD62345 | Johnson | B | $4.49 \pm 0.05$ | Neckel (1974) |
| HD62345 | Johnson | B | $4.49 \pm 0.05$ | Ducati (2002) |
| HD62345 | Johnson | B | $4.50 \pm 0.05$ | Johnson & Morgan (1953b) |
| HD62345 | Johnson | B | $4.50 \pm 0.05$ | Johnson & Harris (1954) |
| HD62345 | Johnson | B | $4.50 \pm 0.05$ | Serkowski (1961) |
| HD62345 | Johnson | B | $4.50 \pm 0.05$ | Argue (1966) |
| HD62345 | Johnson | B | $4.51 \pm 0.05$ | Ljunggren & Oja (1965) |
| HD62345 | Johnson | B | $4.51 \pm 0.05$ | Appenzeller (1966) |
| HD62345 | Johnson | B | $4.51 \pm 0.05$ | Jennens & Helfer (1975) |
| HD62345 | Johnson | B | $4.54 \pm 0.05$ | Mermilliod (1986) |
| HD62345 | KronComet | Bc | $4.32 \pm 0.01$ | This work |
| HD62345 | Geneva | B2 | $4.98 \pm 0.08$ | Golay (1972) |
| HD62345 | 13c | m45 | $4.25 \pm 0.05$ | Johnson & Mitchell (1995) |
| HD62345 | DDO | m45 | $5.08 \pm 0.05$ | McClure & Forrester (1981) |
| HD62345 | Oja | m45 | $4.24 \pm 0.05$ | Häggkvist & Oja (1970) |
| HD62345 | Vilnius | Y | $4.27 \pm 0.05$ | Zdanavicius et al. (1969) |
| HD62345 | Stromgren | b | $4.14 \pm 0.08$ | Crawford & Barnes (1970) |
| HD62345 | Stromgren | b | $4.14 \pm 0.08$ | Philip & Philip (1973) |
| HD62345 | Stromgren | b | $4.14 \pm 0.08$ | Gray & Olsen (1991) |
| HD62345 | Stromgren | b | $4.14 \pm 0.08$ | Olsen (1993) |
| HD62345 | Stromgren | b | $4.14 \pm 0.08$ | Hauck & Mermilliod (1998) |
| HD62345 | Stromgren | b | $4.21 \pm 0.08$ | Williams (1966) |
| HD62345 | DDO | m48 | $3.92 \pm 0.05$ | McClure & Forrester (1981) |
| HD62345 | KronComet | C2 | $3.70 \pm 0.01$ | This work |
| HD62345 | Vilnius | Z | $3.84 \pm 0.05$ | Zdanavicius et al. (1969) |
| HD62345 | 13c | m52 | $3.81 \pm 0.05$ | Johnson & Mitchell (1995) |

**Table 21** *continued on next page*

**Table 21** (continued)

| Star ID | System/Wvlen | Band/Bandpass | Value | Reference |
|---------|--------------|---------------|-------|-----------|
| HD62345 | KronComet | Gc | $3.61 \pm 0.01$ | This work |
| HD62345 | Geneva | V1 | $4.35 \pm 0.08$ | Golay (1972) |
| HD62345 | WBVR | V | $3.57 \pm 0.05$ | Kornilov et al. (1991) |
| HD62345 | Vilnius | V | $3.57 \pm 0.05$ | Zdanavicius et al. (1969) |
| HD62345 | Stromgren | y | $3.57 \pm 0.08$ | Williams (1966) |
| HD62345 | Stromgren | y | $3.57 \pm 0.08$ | Crawford & Barnes (1970) |
| HD62345 | Stromgren | y | $3.57 \pm 0.08$ | Philip & Philip (1973) |
| HD62345 | Stromgren | y | $3.57 \pm 0.08$ | Gray & Olsen (1991) |
| HD62345 | Stromgren | y | $3.57 \pm 0.08$ | Olsen (1993) |
| HD62345 | Stromgren | y | $3.57 \pm 0.08$ | Hauck & Mermilliod (1998) |
| HD62345 | Geneva | V | $3.59 \pm 0.08$ | Golay (1972) |
| HD62345 | Johnson | V | $3.54 \pm 0.05$ | Oja (1963) |
| HD62345 | Johnson | V | $3.55 \pm 0.05$ | Neckel (1974) |
| HD62345 | Johnson | V | $3.56 \pm 0.05$ | Argue (1963) |
| HD62345 | Johnson | V | $3.56 \pm 0.05$ | Argue (1966) |
| HD62345 | Johnson | V | $3.57 \pm 0.05$ | Johnson & Morgan (1953b) |
| HD62345 | Johnson | V | $3.57 \pm 0.05$ | Johnson & Harris (1954) |
| HD62345 | Johnson | V | $3.57 \pm 0.05$ | Serkowski (1961) |
| HD62345 | Johnson | V | $3.57 \pm 0.05$ | Ljunggren & Oja (1965) |
| HD62345 | Johnson | V | $3.57 \pm 0.05$ | Johnson et al. (1966) |
| HD62345 | Johnson | V | $3.57 \pm 0.05$ | Appenzeller (1966) |
| HD62345 | Johnson | V | $3.57 \pm 0.05$ | Jennens & Helfer (1975) |
| HD62345 | Johnson | V | $3.57 \pm 0.05$ | Ducati (2002) |
| HD62345 | Johnson | V | $3.60 \pm 0.05$ | Mermilliod (1986) |
| HD62345 | Johnson | V | $3.60 \pm 0.21$ | This work |
| HD62345 | 13c | m58 | $3.39 \pm 0.05$ | Johnson & Mitchell (1995) |
| HD62345 | Geneva | G | $4.58 \pm 0.08$ | Golay (1972) |
| HD62345 | 13c | m63 | $3.10 \pm 0.05$ | Johnson & Mitchell (1995) |
| HD62345 | WBVR | R | $2.90 \pm 0.05$ | Kornilov et al. (1991) |
| HD62345 | KronComet | Rc | $2.63 \pm 0.01$ | This work |
| HD62345 | 13c | m72 | $2.89 \pm 0.05$ | Johnson & Mitchell (1995) |
| HD62345 | 13c | m80 | $2.67 \pm 0.05$ | Johnson & Mitchell (1995) |
| HD62345 | 13c | m86 | $2.59 \pm 0.05$ | Johnson & Mitchell (1995) |
| HD62345 | 13c | m99 | $2.46 \pm 0.05$ | Johnson & Mitchell (1995) |
| HD62345 | 13c | m110 | $2.29 \pm 0.05$ | Johnson & Mitchell (1995) |
| HD62345 | 1250 | 310 | $253.20 \pm 3.60$ | Smith et al. (2004) |
| HD62345 | Johnson | J | $1.97 \pm 0.05$ | Alonso et al. (1998) |
| HD62345 | Johnson | J | $2.02 \pm 0.05$ | Johnson et al. (1966) |
| HD62345 | Johnson | J | $2.02 \pm 0.05$ | Ducati (2002) |
| HD62345 | Johnson | J | $2.02 \pm 0.05$ | Shenavrin et al. (2011) |
| HD62345 | Johnson | H | $1.56 \pm 0.05$ | Alonso et al. (1998) |

**Table 21** *(continued)*

| Star ID | System/Wvlen | Band/Bandpass | Value | Reference |
|---------|--------------|---------------|-------|-----------|
| HD62345 | Johnson | H | $1.57 \pm 0.05$ | Shenavrin et al. (2011) |
| HD62345 | 2200 | 361 | $160.00 \pm 6.00$ | Smith et al. (2004) |
| HD62345 | Johnson | K | $1.46 \pm 0.05$ | Johnson et al. (1966) |
| HD62345 | Johnson | K | $1.46 \pm 0.05$ | Ducati (2002) |
| HD62345 | Johnson | K | $1.46 \pm 0.05$ | Shenavrin et al. (2011) |
| HD62345 | Johnson | K | $1.48 \pm 0.06$ | Neugebauer & Leighton (1969) |
| HD62345 | 3500 | 898 | $73.00 \pm 14.50$ | Smith et al. (2004) |
| HD62345 | 4900 | 712 | $37.60 \pm 4.70$ | Smith et al. (2004) |
| HD62345 | 12000 | 6384 | $15.60 \pm 28.20$ | Smith et al. (2004) |
| HD62509 | Geneva | U | $3.51 \pm 0.08$ | Golay (1972) |
| HD62509 | Vilnius | U | $4.55 \pm 0.05$ | Zdanavicius et al. (1969) |
| HD62509 | Vilnius | U | $4.70 \pm 0.05$ | Kakaras et al. (1968) |
| HD62509 | DDO | m35 | $4.30 \pm 0.05$ | McClure & Forrester (1981) |
| HD62509 | DDO | m35 | $4.30 \pm 0.05$ | Mermilliod & Nitschelm (1989) |
| HD62509 | WBVR | W | $2.83 \pm 0.05$ | Kornilov et al. (1991) |
| HD62509 | Johnson | U | $2.99 \pm 0.05$ | Johnson & Morgan (1953b) |
| HD62509 | Johnson | U | $3.00 \pm 0.05$ | Johnson (1964) |
| HD62509 | Johnson | U | $3.00 \pm 0.05$ | Johnson et al. (1966) |
| HD62509 | Johnson | U | $3.00 \pm 0.05$ | Lee (1970) |
| HD62509 | Johnson | U | $3.00 \pm 0.05$ | Ducati (2002) |
| HD62509 | Johnson | U | $3.01 \pm 0.05$ | Cowley et al. (1967) |
| HD62509 | Vilnius | P | $3.97 \pm 0.05$ | Zdanavicius et al. (1969) |
| HD62509 | Vilnius | P | $4.12 \pm 0.05$ | Kakaras et al. (1968) |
| HD62509 | DDO | m38 | $3.24 \pm 0.05$ | McClure & Forrester (1981) |
| HD62509 | DDO | m38 | $3.24 \pm 0.05$ | Mermilliod & Nitschelm (1989) |
| HD62509 | 13c | m40 | $2.72 \pm 0.05$ | Johnson & Mitchell (1995) |
| HD62509 | Geneva | B1 | $2.77 \pm 0.08$ | Golay (1972) |
| HD62509 | Vilnius | X | $3.14 \pm 0.05$ | Kakaras et al. (1968) |
| HD62509 | Vilnius | X | $3.18 \pm 0.05$ | Zdanavicius et al. (1969) |
| HD62509 | DDO | m41 | $3.75 \pm 0.05$ | McClure & Forrester (1981) |
| HD62509 | DDO | m41 | $3.75 \pm 0.05$ | Mermilliod & Nitschelm (1989) |
| HD62509 | Oja | m41 | $3.31 \pm 0.05$ | Häggkvist & Oja (1970) |
| HD62509 | DDO | m42 | $3.54 \pm 0.05$ | McClure & Forrester (1981) |
| HD62509 | DDO | m42 | $3.54 \pm 0.05$ | Mermilliod & Nitschelm (1989) |
| HD62509 | Oja | m42 | $3.09 \pm 0.05$ | Häggkvist & Oja (1970) |
| HD62509 | Geneva | B | $1.45 \pm 0.08$ | Golay (1972) |
| HD62509 | KronComet | COp | $2.40 \pm 0.01$ | This work |
| HD62509 | WBVR | B | $2.16 \pm 0.05$ | Kornilov et al. (1991) |
| HD62509 | Johnson | B | $2.10 \pm 0.05$ | Oja (1963) |
| HD62509 | Johnson | B | $2.11 \pm 0.05$ | Bouigue (1959) |
| HD62509 | Johnson | B | $2.14 \pm 0.05$ | Häggkvist & Oja (1966) |

**Table 21** *continued on next page*

**Table 21** (continued)

| Star ID | System/Wvlen | Band/Bandpass | Value | Reference |
|---------|--------------|---------------|-------|-----------|
| HD62509 | Johnson | B | $2.14 \pm 0.05$ | Johnson et al. (1966) |
| HD62509 | Johnson | B | $2.14 \pm 0.05$ | Lee (1970) |
| HD62509 | Johnson | B | $2.14 \pm 0.05$ | Ducati (2002) |
| HD62509 | Johnson | B | $2.15 \pm 0.05$ | Johnson & Morgan (1953b) |
| HD62509 | Johnson | B | $2.15 \pm 0.05$ | Johnson (1964) |
| HD62509 | Johnson | B | $2.16 \pm 0.05$ | Cowley et al. (1967) |
| HD62509 | Johnson | B | $2.24 \pm 0.16$ | This work |
| HD62509 | KronComet | Bc | $1.91 \pm 0.01$ | This work |
| HD62509 | Geneva | B2 | $2.61 \pm 0.08$ | Golay (1972) |
| HD62509 | 13c | m45 | $1.84 \pm 0.05$ | Johnson & Mitchell (1995) |
| HD62509 | DDO | m45 | $2.67 \pm 0.05$ | McClure & Forrester (1981) |
| HD62509 | DDO | m45 | $2.67 \pm 0.05$ | Mermilliod & Nitschelm (1989) |
| HD62509 | Oja | m45 | $1.86 \pm 0.05$ | Häggkvist & Oja (1970) |
| HD62509 | Vilnius | Y | $1.88 \pm 0.05$ | Kakaras et al. (1968) |
| HD62509 | Vilnius | Y | $1.90 \pm 0.05$ | Zdanavicius et al. (1969) |
| HD62509 | DDO | m48 | $1.50 \pm 0.05$ | McClure & Forrester (1981) |
| HD62509 | DDO | m48 | $1.50 \pm 0.05$ | Mermilliod & Nitschelm (1989) |
| HD62509 | KronComet | C2 | $1.25 \pm 0.01$ | This work |
| HD62509 | Vilnius | Z | $1.46 \pm 0.05$ | Kakaras et al. (1968) |
| HD62509 | Vilnius | Z | $1.46 \pm 0.05$ | Zdanavicius et al. (1969) |
| HD62509 | 13c | m52 | $1.39 \pm 0.05$ | Johnson & Mitchell (1995) |
| HD62509 | KronComet | Gc | $1.17 \pm 0.01$ | This work |
| HD62509 | Geneva | V1 | $1.94 \pm 0.08$ | Golay (1972) |
| HD62509 | WBVR | V | $1.14 \pm 0.05$ | Kornilov et al. (1991) |
| HD62509 | Vilnius | V | $1.14 \pm 0.05$ | Kakaras et al. (1968) |
| HD62509 | Vilnius | V | $1.14 \pm 0.05$ | Zdanavicius et al. (1969) |
| HD62509 | Geneva | V | $1.16 \pm 0.08$ | Golay (1972) |
| HD62509 | Johnson | V | $1.14 \pm 0.05$ | Johnson et al. (1966) |
| HD62509 | Johnson | V | $1.14 \pm 0.05$ | Lee (1970) |
| HD62509 | Johnson | V | $1.14 \pm 0.05$ | Ducati (2002) |
| HD62509 | Johnson | V | $1.15 \pm 0.05$ | Johnson & Morgan (1953b) |
| HD62509 | Johnson | V | $1.15 \pm 0.05$ | Johnson (1964) |
| HD62509 | Johnson | V | $1.15 \pm 0.05$ | Häggkvist & Oja (1966) |
| HD62509 | Johnson | V | $1.15 \pm 0.05$ | Cowley et al. (1967) |
| HD62509 | Johnson | V | $1.16 \pm 0.05$ | Bouigue (1959) |
| HD62509 | 13c | m58 | $0.91 \pm 0.05$ | Johnson & Mitchell (1995) |
| HD62509 | Geneva | G | $2.15 \pm 0.08$ | Golay (1972) |
| HD62509 | Alexander | m608 | $-0.82 \pm 0.05$ | Alexander et al. (1983) |
| HD62509 | 13c | m63 | $0.61 \pm 0.05$ | Johnson & Mitchell (1995) |
| HD62509 | Vilnius | S | $0.38 \pm 0.05$ | Kakaras et al. (1968) |
| HD62509 | Vilnius | S | $0.41 \pm 0.05$ | Zdanavicius et al. (1969) |

**Table 21** *(continued)*

| Star ID | System/Wvlen | Band/Bandpass | Value | Reference |
|---------|--------------|---------------|-------|-----------|
| HD62509 | Alexander | m683 | $-0.16 \pm 0.05$ | Alexander et al. (1983) |
| HD62509 | WBVR | R | $0.42 \pm 0.05$ | Kornilov et al. (1991) |
| HD62509 | Alexander | m710 | $0.03 \pm 0.05$ | Alexander et al. (1983) |
| HD62509 | KronComet | Rc | $0.12 \pm 0.01$ | This work |
| HD62509 | 13c | m72 | $0.36 \pm 0.05$ | Johnson & Mitchell (1995) |
| HD62509 | Alexander | m746 | $0.42 \pm 0.05$ | Alexander et al. (1983) |
| HD62509 | 13c | m80 | $0.14 \pm 0.05$ | Johnson & Mitchell (1995) |
| HD62509 | 13c | m86 | $0.05 \pm 0.05$ | Johnson & Mitchell (1995) |
| HD62509 | 13c | m99 | $-0.10 \pm 0.05$ | Johnson & Mitchell (1995) |
| HD62509 | 13c | m110 | $-0.31 \pm 0.05$ | Johnson & Mitchell (1995) |
| HD62509 | Johnson | J | $-0.44 \pm 0.05$ | Bergeat & Lunel (1980) |
| HD62509 | Johnson | J | $-0.49 \pm 0.05$ | Low & Johnson (1964) |
| HD62509 | Johnson | J | $-0.49 \pm 0.05$ | Johnson et al. (1966) |
| HD62509 | Johnson | J | $-0.49 \pm 0.05$ | Lee (1970) |
| HD62509 | Johnson | J | $-0.49 \pm 0.05$ | Campins et al. (1985) |
| HD62509 | Johnson | J | $-0.51 \pm 0.05$ | Rydgren & Vrba (1983) |
| HD62509 | Johnson | J | $-0.52 \pm 0.05$ | Ducati (2002) |
| HD62509 | Johnson | J | $-0.56 \pm 0.05$ | Kenyon (1988) |
| HD62509 | Johnson | J | $-0.59 \pm 0.05$ | Selby et al. (1988) |
| HD62509 | Johnson | J | $-0.59 \pm 0.05$ | Blackwell et al. (1990) |
| HD62509 | Johnson | H | $-0.96 \pm 0.05$ | Bergeat & Lunel (1980) |
| HD62509 | Johnson | H | $-0.97 \pm 0.05$ | Lee (1970) |
| HD62509 | Johnson | H | $-0.99 \pm 0.05$ | Rydgren & Vrba (1983) |
| HD62509 | Johnson | H | $-1.00 \pm 0.05$ | Ney & Merrill (1980) |
| HD62509 | Johnson | H | $-1.00 \pm 0.05$ | Ducati (2002) |
| HD62509 | Johnson | H | $-1.01 \pm 0.05$ | Strecker et al. (1979) |
| HD62509 | Johnson | H | $-1.01 \pm 0.05$ | Noguchi et al. (1981) |
| HD62509 | Johnson | H | $-1.02 \pm 0.05$ | Phillips et al. (1980) |
| HD62509 | Johnson | H | $-1.02 \pm 0.05$ | Alonso et al. (1994) |
| HD62509 | Johnson | H | $-1.05 \pm 0.05$ | Kenyon (1988) |
| HD62509 | Johnson | K | $-1.09 \pm 0.05$ | Johnson et al. (1966) |
| HD62509 | Johnson | K | $-1.11 \pm 0.05$ | Lee (1970) |
| HD62509 | Johnson | K | $-1.11 \pm 0.05$ | Ducati (2002) |
| HD62509 | Johnson | K | $-1.12 \pm 0.05$ | Neugebauer & Leighton (1969) |
| HD62509 | Johnson | L | $-1.17 \pm 0.05$ | Johnson et al. (1966) |
| HD62509 | Johnson | L | $-1.19 \pm 0.05$ | Ducati (2002) |
| HD62509 | Johnson | L | $-1.22 \pm 0.05$ | Lee (1970) |
| HD62509 | 3500 | 898 | $819.90 \pm 13.30$ | Smith et al. (2004) |
| HD62509 | Johnson | N | $-1.23 \pm 0.05$ | Ducati (2002) |
| HD62509 | Johnson | N | $-1.24 \pm 0.05$ | Low & Johnson (1964) |
| HD62509 | Johnson | N | $-1.24 \pm 0.05$ | Johnson (1964) |

**Table 21** *(continued)*

| Star ID | System/Wvlen | Band/Bandpass | Value | Reference |
|---------|--------------|---------------|-------|-----------|
| HD62509 | 4900 | 712 | $409.30 \pm 5.70$ | Smith et al. (2004) |
| HD62509 | Johnson | M | $-0.90 \pm 0.05$ | Johnson (1964) |
| HD62509 | Johnson | M | $-1.12 \pm 0.05$ | Ducati (2002) |
| HD62509 | 12000 | 6384 | $85.40 \pm 21.90$ | Smith et al. (2004) |
| HD62721 | Oja | m41 | $7.80 \pm 0.05$ | Häggkvist & Oja (1970) |
| HD62721 | DDO | m42 | $8.07 \pm 0.05$ | McClure & Forrester (1981) |
| HD62721 | DDO | m42 | $8.07 \pm 0.05$ | Mermilliod & Nitschelm (1989) |
| HD62721 | Oja | m42 | $7.58 \pm 0.05$ | Häggkvist & Oja (1970) |
| HD62721 | KronComet | COp | $6.83 \pm 0.09$ | This work |
| HD62721 | WBVR | B | $6.37 \pm 0.05$ | Kornilov et al. (1991) |
| HD62721 | Johnson | B | $6.30 \pm 0.05$ | Argue (1966) |
| HD62721 | Johnson | B | $6.31 \pm 0.05$ | Häggkvist & Oja (1966) |
| HD62721 | Johnson | B | $6.31 \pm 0.05$ | Neckel (1974) |
| HD62721 | Johnson | B | $6.33 \pm 0.05$ | Johnson et al. (1966) |
| HD62721 | Johnson | B | $6.33 \pm 0.05$ | Ducati (2002) |
| HD62721 | Johnson | B | $6.34 \pm 0.05$ | Johnson (1964) |
| HD62721 | Johnson | B | $6.35 \pm 0.05$ | Roman (1955) |
| HD62721 | Johnson | B | $6.35 \pm 0.05$ | Mermilliod (1986) |
| HD62721 | KronComet | Bc | $6.10 \pm 0.04$ | This work |
| HD62721 | 13c | m45 | $5.90 \pm 0.05$ | Johnson & Mitchell (1995) |
| HD62721 | DDO | m45 | $6.78 \pm 0.05$ | McClure & Forrester (1981) |
| HD62721 | DDO | m45 | $6.78 \pm 0.05$ | Mermilliod & Nitschelm (1989) |
| HD62721 | Oja | m45 | $5.94 \pm 0.05$ | Häggkvist & Oja (1970) |
| HD62721 | Vilnius | Y | $5.92 \pm 0.05$ | Kazlauskas et al. (2005) |
| HD62721 | DDO | m48 | $5.46 \pm 0.05$ | McClure & Forrester (1981) |
| HD62721 | DDO | m48 | $5.46 \pm 0.05$ | Mermilliod & Nitschelm (1989) |
| HD62721 | KronComet | C2 | $5.31 \pm 0.03$ | This work |
| HD62721 | Vilnius | Z | $5.39 \pm 0.05$ | Kazlauskas et al. (2005) |
| HD62721 | 13c | m52 | $5.30 \pm 0.05$ | Johnson & Mitchell (1995) |
| HD62721 | KronComet | Gc | $5.00 \pm 0.03$ | This work |
| HD62721 | WBVR | V | $4.88 \pm 0.05$ | Kornilov et al. (1991) |
| HD62721 | Vilnius | V | $4.87 \pm 0.05$ | Kazlauskas et al. (2005) |
| HD62721 | Johnson | V | $4.86 \pm 0.05$ | Johnson (1964) |
| HD62721 | Johnson | V | $4.86 \pm 0.05$ | Neckel (1974) |
| HD62721 | Johnson | V | $4.87 \pm 0.05$ | Johnson et al. (1966) |
| HD62721 | Johnson | V | $4.87 \pm 0.05$ | Argue (1966) |
| HD62721 | Johnson | V | $4.87 \pm 0.05$ | Ducati (2002) |
| HD62721 | Johnson | V | $4.88 \pm 0.05$ | Häggkvist & Oja (1966) |
| HD62721 | Johnson | V | $4.90 \pm 0.05$ | Roman (1955) |
| HD62721 | Johnson | V | $4.90 \pm 0.05$ | Mermilliod (1986) |
| HD62721 | Johnson | V | $4.97 \pm 0.07$ | This work |

**Table 21** *(continued)*

| Star ID | System/Wvlen | Band/Bandpass | Value | Reference |
|---------|--------------|---------------|-------|-----------|
| HD62721 | 13c | m58 | $4.55 \pm 0.05$ | Johnson & Mitchell (1995) |
| HD62721 | 13c | m63 | $4.12 \pm 0.05$ | Johnson & Mitchell (1995) |
| HD62721 | Vilnius | S | $3.85 \pm 0.05$ | Kazlauskas et al. (2005) |
| HD62721 | WBVR | R | $3.74 \pm 0.05$ | Kornilov et al. (1991) |
| HD62721 | KronComet | Rc | $3.56 \pm 0.02$ | This work |
| HD62721 | 13c | m72 | $3.67 \pm 0.05$ | Johnson & Mitchell (1995) |
| HD62721 | 13c | m80 | $3.30 \pm 0.05$ | Johnson & Mitchell (1995) |
| HD62721 | 13c | m86 | $3.15 \pm 0.05$ | Johnson & Mitchell (1995) |
| HD62721 | 13c | m99 | $2.87 \pm 0.05$ | Johnson & Mitchell (1995) |
| HD62721 | 13c | m110 | $2.59 \pm 0.05$ | Johnson & Mitchell (1995) |
| HD62721 | 1250 | 310 | $203.10 \pm 6.90$ | Smith et al. (2004) |
| HD62721 | Johnson | J | $2.27 \pm 0.05$ | Alonso et al. (1998) |
| HD62721 | Johnson | J | $2.28 \pm 0.05$ | Johnson et al. (1966) |
| HD62721 | Johnson | J | $2.28 \pm 0.05$ | Ducati (2002) |
| HD62721 | Johnson | H | $1.55 \pm 0.05$ | Alonso et al. (1998) |
| HD62721 | 2200 | 361 | $179.60 \pm 6.50$ | Smith et al. (2004) |
| HD62721 | Johnson | K | $1.32 \pm 0.06$ | Neugebauer & Leighton (1969) |
| HD62721 | Johnson | K | $1.33 \pm 0.05$ | Johnson et al. (1966) |
| HD62721 | Johnson | K | $1.33 \pm 0.05$ | Ducati (2002) |
| HD62721 | 3500 | 898 | $88.10 \pm 4.80$ | Smith et al. (2004) |
| HD62721 | 4900 | 712 | $39.20 \pm 5.20$ | Smith et al. (2004) |
| HD62721 | 12000 | 6384 | $4.10 \pm 25.20$ | Smith et al. (2004) |
| HD65345 | DDO | m35 | $8.24 \pm 0.05$ | McClure & Forrester (1981) |
| HD65345 | Stromgren | u | $8.16 \pm 0.08$ | Olsen (1993) |
| HD65345 | Stromgren | u | $8.16 \pm 0.08$ | Hauck & Mermilliod (1998) |
| HD65345 | WBVR | W | $6.79 \pm 0.05$ | Kornilov et al. (1991) |
| HD65345 | Johnson | U | $6.91 \pm 0.05$ | Roman (1955) |
| HD65345 | Johnson | U | $6.91 \pm 0.05$ | Johnson et al. (1966) |
| HD65345 | Johnson | U | $6.93 \pm 0.05$ | Cousins (1963b) |
| HD65345 | Johnson | U | $6.99 \pm 0.05$ | Mermilliod (1986) |
| HD65345 | DDO | m38 | $7.19 \pm 0.05$ | McClure & Forrester (1981) |
| HD65345 | DDO | m41 | $7.76 \pm 0.05$ | McClure & Forrester (1981) |
| HD65345 | Oja | m41 | $7.32 \pm 0.05$ | Häggkvist & Oja (1970) |
| HD65345 | Stromgren | v | $6.79 \pm 0.08$ | Olsen (1993) |
| HD65345 | Stromgren | v | $6.79 \pm 0.08$ | Hauck & Mermilliod (1998) |
| HD65345 | DDO | m42 | $7.57 \pm 0.05$ | McClure & Forrester (1981) |
| HD65345 | Oja | m42 | $7.15 \pm 0.05$ | Häggkvist & Oja (1970) |
| HD65345 | WBVR | B | $6.25 \pm 0.05$ | Kornilov et al. (1991) |
| HD65345 | Johnson | B | $6.19 \pm 0.05$ | Roman (1955) |
| HD65345 | Johnson | B | $6.19 \pm 0.05$ | Johnson et al. (1966) |
| HD65345 | Johnson | B | $6.24 \pm 0.05$ | Cousins (1963b) |

**Table 21** *continued on next page*

**Table 21** *(continued)*

| Star ID | System/Wvlen | Band/Bandpass | Value | Reference |
|---------|--------------|---------------|-------|-----------|
| HD65345 | Johnson | B | $6.28 \pm 0.05$ | Mermilliod (1986) |
| HD65345 | DDO | m45 | $6.78 \pm 0.05$ | McClure & Forrester (1981) |
| HD65345 | Oja | m45 | $5.98 \pm 0.05$ | Häggkvist & Oja (1970) |
| HD65345 | Stromgren | b | $5.87 \pm 0.08$ | Olsen (1993) |
| HD65345 | Stromgren | b | $5.87 \pm 0.08$ | Hauck & Mermilliod (1998) |
| HD65345 | DDO | m48 | $5.63 \pm 0.05$ | McClure & Forrester (1981) |
| HD65345 | WBVR | V | $5.31 \pm 0.05$ | Kornilov et al. (1991) |
| HD65345 | Stromgren | y | $5.30 \pm 0.08$ | Olsen (1993) |
| HD65345 | Stromgren | y | $5.30 \pm 0.08$ | Hauck & Mermilliod (1998) |
| HD65345 | Johnson | V | $5.28 \pm 0.05$ | Roman (1955) |
| HD65345 | Johnson | V | $5.28 \pm 0.05$ | Johnson et al. (1966) |
| HD65345 | Johnson | V | $5.30 \pm 0.05$ | Cousins (1963b) |
| HD65345 | Johnson | V | $5.34 \pm 0.05$ | Mermilliod (1986) |
| HD65345 | WBVR | R | $4.61 \pm 0.05$ | Kornilov et al. (1991) |
| HD65345 | 1250 | 310 | $60.70 \pm 13.60$ | Smith et al. (2004) |
| HD65345 | 2200 | 361 | $39.90 \pm 15.60$ | Smith et al. (2004) |
| HD65345 | 3500 | 898 | $20.90 \pm 11.40$ | Smith et al. (2004) |
| HD65345 | 4900 | 712 | $9.20 \pm 6.10$ | Smith et al. (2004) |
| HD65345 | 12000 | 6384 | $-1.70 \pm 18.70$ | Smith et al. (2004) |
| HD65759 | DDO | m35 | $9.70 \pm 0.05$ | McClure & Forrester (1981) |
| HD65759 | WBVR | W | $8.23 \pm 0.05$ | Kornilov et al. (1991) |
| HD65759 | DDO | m38 | $8.51 \pm 0.05$ | McClure & Forrester (1981) |
| HD65759 | DDO | m41 | $8.82 \pm 0.05$ | McClure & Forrester (1981) |
| HD65759 | Oja | m41 | $8.36 \pm 0.05$ | Häggkvist & Oja (1970) |
| HD65759 | DDO | m42 | $8.48 \pm 0.05$ | McClure & Forrester (1981) |
| HD65759 | Oja | m42 | $8.01 \pm 0.05$ | Häggkvist & Oja (1970) |
| HD65759 | KronComet | COp | $7.24 \pm 0.09$ | This work |
| HD65759 | WBVR | B | $6.93 \pm 0.05$ | Kornilov et al. (1991) |
| HD65759 | Johnson | B | $6.86 \pm 0.05$ | Haggkvist & Oja (1970) |
| HD65759 | KronComet | Bc | $6.68 \pm 0.04$ | This work |
| HD65759 | DDO | m45 | $7.41 \pm 0.05$ | McClure & Forrester (1981) |
| HD65759 | Oja | m45 | $6.59 \pm 0.05$ | Häggkvist & Oja (1970) |
| HD65759 | DDO | m48 | $6.09 \pm 0.05$ | McClure & Forrester (1981) |
| HD65759 | KronComet | C2 | $5.86 \pm 0.03$ | This work |
| HD65759 | KronComet | Gc | $5.71 \pm 0.03$ | This work |
| HD65759 | WBVR | V | $5.60 \pm 0.05$ | Kornilov et al. (1991) |
| HD65759 | Johnson | V | $5.55 \pm 0.05$ | Haggkvist & Oja (1970) |
| HD65759 | Johnson | V | $5.62 \pm 0.07$ | This work |
| HD65759 | WBVR | R | $4.68 \pm 0.05$ | Kornilov et al. (1991) |
| HD65759 | KronComet | Rc | $4.45 \pm 0.02$ | This work |
| HD65759 | 1250 | 310 | $58.40 \pm 5.30$ | Smith et al. (2004) |

**Table 21** *(continued)*

| Star ID | System/Wvlen | Band/Bandpass | Value | Reference |
|---------|--------------|---------------|-------|-----------|
| HD65759 | 2200 | 361 | $42.70 \pm 5.40$ | Smith et al. (2004) |
| HD65759 | Johnson | K | $2.57 \pm 0.09$ | Neugebauer & Leighton (1969) |
| HD65759 | 3500 | 898 | $21.20 \pm 5.40$ | Smith et al. (2004) |
| HD65759 | 4900 | 712 | $9.90 \pm 5.00$ | Smith et al. (2004) |
| HD65759 | 12000 | 6384 | $-7.00 \pm 21.20$ | Smith et al. (2004) |
| HD66216 | WBVR | W | $7.05 \pm 0.05$ | Kornilov et al. (1991) |
| HD66216 | Johnson | U | $7.13 \pm 0.05$ | Argue (1963) |
| HD66216 | Johnson | U | $7.14 \pm 0.05$ | Argue (1966) |
| HD66216 | Johnson | U | $7.21 \pm 0.05$ | Mermilliod (1986) |
| HD66216 | Oja | m41 | $7.38 \pm 0.05$ | Häggkvist & Oja (1970) |
| HD66216 | Oja | m42 | $7.11 \pm 0.05$ | Häggkvist & Oja (1970) |
| HD66216 | KronComet | COp | $6.33 \pm 0.09$ | This work |
| HD66216 | WBVR | B | $6.11 \pm 0.05$ | Kornilov et al. (1991) |
| HD66216 | Johnson | B | $6.04 \pm 0.05$ | Argue (1963) |
| HD66216 | Johnson | B | $6.05 \pm 0.05$ | Argue (1966) |
| HD66216 | Johnson | B | $6.08 \pm 0.05$ | Häggkvist & Oja (1966) |
| HD66216 | Johnson | B | $6.12 \pm 0.05$ | Mermilliod (1986) |
| HD66216 | KronComet | Bc | $5.83 \pm 0.04$ | This work |
| HD66216 | Oja | m45 | $5.76 \pm 0.05$ | Häggkvist & Oja (1970) |
| HD66216 | KronComet | C2 | $5.16 \pm 0.03$ | This work |
| HD66216 | KronComet | Gc | $5.01 \pm 0.03$ | This work |
| HD66216 | WBVR | V | $4.95 \pm 0.05$ | Kornilov et al. (1991) |
| HD66216 | Johnson | V | $4.92 \pm 0.05$ | Argue (1963) |
| HD66216 | Johnson | V | $4.94 \pm 0.05$ | Argue (1966) |
| HD66216 | Johnson | V | $4.95 \pm 0.05$ | Häggkvist & Oja (1966) |
| HD66216 | Johnson | V | $4.97 \pm 0.07$ | This work |
| HD66216 | Johnson | V | $4.98 \pm 0.05$ | Mermilliod (1986) |
| HD66216 | WBVR | R | $4.13 \pm 0.05$ | Kornilov et al. (1991) |
| HD66216 | KronComet | Rc | $3.89 \pm 0.02$ | This work |
| HD66216 | 1250 | 310 | $103.80 \pm 6.50$ | Smith et al. (2004) |
| HD66216 | 2200 | 361 | $75.70 \pm 5.90$ | Smith et al. (2004) |
| HD66216 | Johnson | K | $2.39 \pm 0.08$ | Neugebauer & Leighton (1969) |
| HD66216 | 3500 | 898 | $35.50 \pm 4.10$ | Smith et al. (2004) |
| HD66216 | 4900 | 712 | $16.80 \pm 5.00$ | Smith et al. (2004) |
| HD66216 | 12000 | 6384 | $-0.30 \pm 24.90$ | Smith et al. (2004) |
| HD67743 | KronComet | COp | $9.48 \pm 0.08$ | This work |
| HD67743 | Johnson | B | $8.90 \pm 0.05$ | This work |
| HD67743 | KronComet | Bc | $8.78 \pm 0.07$ | This work |
| HD67743 | KronComet | C2 | $7.66 \pm 0.04$ | This work |
| HD67743 | KronComet | Gc | $7.55 \pm 0.03$ | This work |
| HD67743 | Johnson | V | $7.49 \pm 0.05$ | This work |

**Table 21** *(continued)*

| Star ID | System/Wvlen | Band/Bandpass | Value | Reference |
|---------|--------------|---------------|-------|-----------|
| HD67743 | KronComet | Rc | $6.03 \pm 0.01$ | This work |
| HD67743 | 1250 | 310 | $81.00 \pm 6.10$ | Smith et al. (2004) |
| HD67743 | 2200 | 361 | $73.50 \pm 11.10$ | Smith et al. (2004) |
| HD67743 | Johnson | K | $2.56 \pm 0.08$ | Neugebauer & Leighton (1969) |
| HD67743 | 3500 | 898 | $39.30 \pm 6.50$ | Smith et al. (2004) |
| HD67743 | 4900 | 712 | $15.10 \pm 6.00$ | Smith et al. (2004) |
| HD67743 | 12000 | 6384 | $-10.90 \pm 26.70$ | Smith et al. (2004) |
| HD74442 | 13c | m33 | $6.08 \pm 0.05$ | Johnson & Mitchell (1995) |
| HD74442 | Geneva | U | $6.55 \pm 0.08$ | Golay (1972) |
| HD74442 | Vilnius | U | $7.72 \pm 0.05$ | Kazlauskas et al. (2005) |
| HD74442 | 13c | m35 | $5.89 \pm 0.05$ | Johnson & Mitchell (1995) |
| HD74442 | DDO | m35 | $7.35 \pm 0.05$ | McClure & Forrester (1981) |
| HD74442 | WBVR | W | $5.88 \pm 0.05$ | Kornilov et al. (1991) |
| HD74442 | Johnson | U | $5.99 \pm 0.05$ | Gutierrez-Moreno & et al. (1966) |
| HD74442 | Johnson | U | $6.01 \pm 0.05$ | Argue (1963) |
| HD74442 | Johnson | U | $6.01 \pm 0.05$ | Johnson et al. (1966) |
| HD74442 | Johnson | U | $6.01 \pm 0.05$ | Ducati (2002) |
| HD74442 | Johnson | U | $6.06 \pm 0.05$ | Mermilliod (1986) |
| HD74442 | 13c | m37 | $5.97 \pm 0.05$ | Johnson & Mitchell (1995) |
| HD74442 | Vilnius | P | $7.13 \pm 0.05$ | Kazlauskas et al. (2005) |
| HD74442 | DDO | m38 | $6.25 \pm 0.05$ | McClure & Forrester (1981) |
| HD74442 | 13c | m40 | $5.70 \pm 0.05$ | Johnson & Mitchell (1995) |
| HD74442 | Geneva | B1 | $5.72 \pm 0.08$ | Golay (1972) |
| HD74442 | Vilnius | X | $6.06 \pm 0.05$ | Kazlauskas et al. (2005) |
| HD74442 | DDO | m41 | $6.72 \pm 0.05$ | McClure & Forrester (1981) |
| HD74442 | Oja | m41 | $6.30 \pm 0.05$ | Häggkvist & Oja (1970) |
| HD74442 | DDO | m42 | $6.47 \pm 0.05$ | McClure & Forrester (1981) |
| HD74442 | Oja | m42 | $6.02 \pm 0.05$ | Häggkvist & Oja (1970) |
| HD74442 | Geneva | B | $4.36 \pm 0.08$ | Golay (1972) |
| HD74442 | WBVR | B | $5.05 \pm 0.05$ | Kornilov et al. (1991) |
| HD74442 | Johnson | B | $5.01 \pm 0.05$ | Ljunggren & Oja (1965) |
| HD74442 | Johnson | B | $5.01 \pm 0.05$ | Gutierrez-Moreno & et al. (1966) |
| HD74442 | Johnson | B | $5.02 \pm 0.05$ | Johnson et al. (1966) |
| HD74442 | Johnson | B | $5.02 \pm 0.05$ | Ducati (2002) |
| HD74442 | Johnson | B | $5.03 \pm 0.05$ | Argue (1963) |
| HD74442 | Johnson | B | $5.04 \pm 0.05$ | Mermilliod (1986) |
| HD74442 | Geneva | B2 | $5.49 \pm 0.08$ | Golay (1972) |
| HD74442 | 13c | m45 | $4.72 \pm 0.05$ | Johnson & Mitchell (1995) |
| HD74442 | DDO | m45 | $5.54 \pm 0.05$ | McClure & Forrester (1981) |
| HD74442 | Oja | m45 | $4.72 \pm 0.05$ | Häggkvist & Oja (1970) |
| HD74442 | Vilnius | Y | $4.70 \pm 0.05$ | Kazlauskas et al. (2005) |

**Table 21** *(continued)*

| Star ID | System/Wvlen | Band/Bandpass | Value | Reference |
|---------|--------------|---------------|-------|-----------|
| HD74442 | DDO | m48 | $4.33 \pm 0.05$ | McClure & Forrester (1981) |
| HD74442 | Vilnius | Z | $4.21 \pm 0.05$ | Kazlauskas et al. (2005) |
| HD74442 | 13c | m52 | $4.24 \pm 0.05$ | Johnson & Mitchell (1995) |
| HD74442 | Geneva | V1 | $4.74 \pm 0.08$ | Golay (1972) |
| HD74442 | WBVR | V | $3.95 \pm 0.05$ | Kornilov et al. (1991) |
| HD74442 | Vilnius | V | $3.87 \pm 0.05$ | Kazlauskas et al. (2005) |
| HD74442 | Geneva | V | $3.96 \pm 0.08$ | Golay (1972) |
| HD74442 | Johnson | V | $3.92 \pm 0.05$ | Gutierrez-Moreno & et al. (1966) |
| HD74442 | Johnson | V | $3.94 \pm 0.05$ | Argue (1963) |
| HD74442 | Johnson | V | $3.94 \pm 0.05$ | Ljunggren & Oja (1965) |
| HD74442 | Johnson | V | $3.94 \pm 0.05$ | Johnson et al. (1966) |
| HD74442 | Johnson | V | $3.94 \pm 0.05$ | Ducati (2002) |
| HD74442 | Johnson | V | $3.95 \pm 0.05$ | Mermilliod (1986) |
| HD74442 | 13c | m58 | $3.71 \pm 0.05$ | Johnson & Mitchell (1995) |
| HD74442 | Geneva | G | $4.92 \pm 0.08$ | Golay (1972) |
| HD74442 | 13c | m63 | $3.40 \pm 0.05$ | Johnson & Mitchell (1995) |
| HD74442 | Vilnius | S | $3.10 \pm 0.05$ | Kazlauskas et al. (2005) |
| HD74442 | WBVR | R | $3.17 \pm 0.05$ | Kornilov et al. (1991) |
| HD74442 | KronComet | Rc | $2.91 \pm 0.04$ | This work |
| HD74442 | 13c | m72 | $3.16 \pm 0.05$ | Johnson & Mitchell (1995) |
| HD74442 | 13c | m80 | $2.90 \pm 0.05$ | Johnson & Mitchell (1995) |
| HD74442 | 13c | m86 | $2.79 \pm 0.05$ | Johnson & Mitchell (1995) |
| HD74442 | 13c | m99 | $2.63 \pm 0.05$ | Johnson & Mitchell (1995) |
| HD74442 | 13c | m110 | $2.43 \pm 0.05$ | Johnson & Mitchell (1995) |
| HD74442 | 1250 | 310 | $232.80 \pm 11.10$ | Smith et al. (2004) |
| HD74442 | Johnson | J | $2.14 \pm 0.01$ | Laney et al. (2012) |
| HD74442 | Johnson | J | $2.19 \pm 0.05$ | Johnson et al. (1966) |
| HD74442 | Johnson | J | $2.19 \pm 0.05$ | Wu & Wang (1985) |
| HD74442 | Johnson | J | $2.19 \pm 0.05$ | Ducati (2002) |
| HD74442 | Johnson | H | $1.60 \pm 0.01$ | Laney et al. (2012) |
| HD74442 | Johnson | H | $1.68 \pm 0.05$ | Wu & Wang (1985) |
| HD74442 | Johnson | H | $1.68 \pm 0.05$ | Ducati (2002) |
| HD74442 | 2200 | 361 | $163.80 \pm 7.80$ | Smith et al. (2004) |
| HD74442 | Johnson | K | $1.48 \pm 0.05$ | Ducati (2002) |
| HD74442 | Johnson | K | $1.49 \pm 0.01$ | Laney et al. (2012) |
| HD74442 | Johnson | K | $1.51 \pm 0.05$ | Johnson et al. (1966) |
| HD74442 | Johnson | K | $1.52 \pm 0.05$ | Neugebauer & Leighton (1969) |
| HD74442 | 3500 | 898 | $77.00 \pm 4.90$ | Smith et al. (2004) |
| HD74442 | 4900 | 712 | $39.10 \pm 4.80$ | Smith et al. (2004) |
| HD74442 | 12000 | 6384 | $7.50 \pm 46.90$ | Smith et al. (2004) |
| HD75506 | Geneva | U | $7.31 \pm 0.08$ | Golay (1972) |

**Table 21** *(continued)*

| Star ID | System/Wvlen | Band/Bandpass | Value | Reference |
|---------|--------------|---------------|-------|-----------|
| HD75506 | DDO | m35 | $8.13 \pm 0.05$ | McClure & Forrester (1981) |
| HD75506 | WBVR | W | $6.66 \pm 0.05$ | Kornilov et al. (1991) |
| HD75506 | Johnson | U | $6.80 \pm 0.05$ | Argue (1966) |
| HD75506 | Johnson | U | $6.83 \pm 0.05$ | Argue (1963) |
| HD75506 | Johnson | U | $6.88 \pm 0.05$ | Mermilliod (1986) |
| HD75506 | DDO | m38 | $7.11 \pm 0.05$ | McClure & Forrester (1981) |
| HD75506 | Geneva | B1 | $6.68 \pm 0.08$ | Golay (1972) |
| HD75506 | DDO | m41 | $7.65 \pm 0.05$ | McClure & Forrester (1981) |
| HD75506 | Oja | m41 | $7.21 \pm 0.05$ | Häggkvist & Oja (1970) |
| HD75506 | DDO | m42 | $7.52 \pm 0.05$ | McClure & Forrester (1981) |
| HD75506 | Oja | m42 | $7.08 \pm 0.05$ | Häggkvist & Oja (1970) |
| HD75506 | Geneva | B | $5.42 \pm 0.08$ | Golay (1972) |
| HD75506 | WBVR | B | $6.14 \pm 0.05$ | Kornilov et al. (1991) |
| HD75506 | Johnson | B | $6.09 \pm 0.05$ | Häggkvist & Oja (1966) |
| HD75506 | Johnson | B | $6.13 \pm 0.05$ | Argue (1966) |
| HD75506 | Johnson | B | $6.14 \pm 0.05$ | Argue (1963) |
| HD75506 | Johnson | B | $6.18 \pm 0.05$ | Mermilliod (1986) |
| HD75506 | KronComet | Bc | $5.86 \pm 0.05$ | This work |
| HD75506 | Geneva | B2 | $6.59 \pm 0.08$ | Golay (1972) |
| HD75506 | DDO | m45 | $6.69 \pm 0.05$ | McClure & Forrester (1981) |
| HD75506 | Oja | m45 | $5.87 \pm 0.05$ | Häggkvist & Oja (1970) |
| HD75506 | DDO | m48 | $5.52 \pm 0.05$ | McClure & Forrester (1981) |
| HD75506 | KronComet | C2 | $5.24 \pm 0.03$ | This work |
| HD75506 | KronComet | Gc | $5.13 \pm 0.04$ | This work |
| HD75506 | Geneva | V1 | $5.93 \pm 0.08$ | Golay (1972) |
| HD75506 | WBVR | V | $5.16 \pm 0.05$ | Kornilov et al. (1991) |
| HD75506 | Geneva | V | $5.16 \pm 0.08$ | Golay (1972) |
| HD75506 | Johnson | V | $5.13 \pm 0.05$ | Häggkvist & Oja (1966) |
| HD75506 | Johnson | V | $5.15 \pm 0.05$ | Argue (1966) |
| HD75506 | Johnson | V | $5.16 \pm 0.05$ | Argue (1963) |
| HD75506 | Johnson | V | $5.21 \pm 0.05$ | Mermilliod (1986) |
| HD75506 | Johnson | V | $5.26 \pm 0.06$ | This work |
| HD75506 | Geneva | G | $6.15 \pm 0.08$ | Golay (1972) |
| HD75506 | Alexander | m608 | $3.17 \pm 0.05$ | Alexander et al. (1983) |
| HD75506 | Alexander | m683 | $3.81 \pm 0.05$ | Alexander et al. (1983) |
| HD75506 | WBVR | R | $4.44 \pm 0.05$ | Kornilov et al. (1991) |
| HD75506 | Alexander | m710 | $4.00 \pm 0.05$ | Alexander et al. (1983) |
| HD75506 | KronComet | Rc | $4.14 \pm 0.02$ | This work |
| HD75506 | Alexander | m746 | $4.40 \pm 0.05$ | Alexander et al. (1983) |
| HD75506 | 1250 | 310 | $66.20 \pm 5.90$ | Smith et al. (2004) |
| HD75506 | 2200 | 361 | $43.70 \pm 5.70$ | Smith et al. (2004) |

**Table 21** *(continued)*

| Star ID | System/Wvlen | Band/Bandpass | Value | Reference |
|---------|--------------|---------------|-------|-----------|
| HD75506 | Johnson | K | $2.91 \pm 0.08$ | Neugebauer & Leighton (1969) |
| HD75506 | 3500 | 898 | $20.20 \pm 5.10$ | Smith et al. (2004) |
| HD75506 | 4900 | 712 | $8.40 \pm 5.20$ | Smith et al. (2004) |
| HD75506 | 12000 | 6384 | $-19.70 \pm 17.40$ | Smith et al. (2004) |
| HD76294 | 13c | m33 | $4.97 \pm 0.05$ | Johnson & Mitchell (1995) |
| HD76294 | 13c | m35 | $4.79 \pm 0.05$ | Johnson & Mitchell (1995) |
| HD76294 | DDO | m35 | $6.19 \pm 0.05$ | McClure & Forrester (1981) |
| HD76294 | WBVR | W | $4.76 \pm 0.05$ | Kornilov et al. (1991) |
| HD76294 | Johnson | U | $4.87 \pm 0.05$ | McClure (1970) |
| HD76294 | Johnson | U | $4.89 \pm 0.05$ | Mermilliod (1986) |
| HD76294 | Johnson | U | $4.90 \pm 0.05$ | Cousins (1962b) |
| HD76294 | Johnson | U | $4.90 \pm 0.05$ | Argue (1963) |
| HD76294 | Johnson | U | $4.91 \pm 0.05$ | Johnson et al. (1966) |
| HD76294 | Johnson | U | $4.92 \pm 0.05$ | Johnson et al. (1966) |
| HD76294 | Johnson | U | $4.92 \pm 0.05$ | Jennens & Helfer (1975) |
| HD76294 | Johnson | U | $4.92 \pm 0.05$ | Ducati (2002) |
| HD76294 | 13c | m37 | $4.87 \pm 0.05$ | Johnson & Mitchell (1995) |
| HD76294 | DDO | m38 | $5.14 \pm 0.05$ | McClure & Forrester (1981) |
| HD76294 | DDO | m38 | $5.18 \pm 0.05$ | Mermilliod & Nitschelm (1989) |
| HD76294 | 13c | m40 | $4.70 \pm 0.05$ | Johnson & Mitchell (1995) |
| HD76294 | DDO | m41 | $5.72 \pm 0.05$ | McClure & Forrester (1981) |
| HD76294 | DDO | m41 | $5.74 \pm 0.05$ | Mermilliod & Nitschelm (1989) |
| HD76294 | Oja | m41 | $5.31 \pm 0.05$ | Häggkvist & Oja (1970) |
| HD76294 | DDO | m42 | $5.48 \pm 0.05$ | McClure & Forrester (1981) |
| HD76294 | DDO | m42 | $5.51 \pm 0.05$ | Mermilliod & Nitschelm (1989) |
| HD76294 | Oja | m42 | $5.06 \pm 0.05$ | Häggkvist & Oja (1970) |
| HD76294 | KronComet | COp | $4.33 \pm 0.06$ | This work |
| HD76294 | WBVR | B | $4.14 \pm 0.05$ | Kornilov et al. (1991) |
| HD76294 | Johnson | B | $4.08 \pm 0.05$ | McClure (1970) |
| HD76294 | Johnson | B | $4.09 \pm 0.05$ | Häggkvist & Oja (1966) |
| HD76294 | Johnson | B | $4.10 \pm 0.05$ | Johnson et al. (1966) |
| HD76294 | Johnson | B | $4.10 \pm 0.05$ | Jennens & Helfer (1975) |
| HD76294 | Johnson | B | $4.10 \pm 0.05$ | Ducati (2002) |
| HD76294 | Johnson | B | $4.11 \pm 0.05$ | Argue (1963) |
| HD76294 | Johnson | B | $4.11 \pm 0.05$ | Mermilliod (1986) |
| HD76294 | Johnson | B | $4.12 \pm 0.05$ | Cousins (1962b) |
| HD76294 | KronComet | Bc | $3.88 \pm 0.05$ | This work |
| HD76294 | 13c | m45 | $3.85 \pm 0.05$ | Johnson & Mitchell (1995) |
| HD76294 | DDO | m45 | $4.65 \pm 0.05$ | McClure & Forrester (1981) |
| HD76294 | DDO | m45 | $4.66 \pm 0.05$ | Mermilliod & Nitschelm (1989) |
| HD76294 | Oja | m45 | $3.84 \pm 0.05$ | Häggkvist & Oja (1970) |

**Table 21** *(continued)*

| Star ID | System/Wvlen | Band/Bandpass | Value | Reference |
|---------|-------------|---------------|-------|-----------|
| HD76294 | DDO | m48 | $3.47 \pm 0.05$ | McClure & Forrester (1981) |
| HD76294 | DDO | m48 | $3.48 \pm 0.05$ | Mermilliod & Nitschelm (1989) |
| HD76294 | KronComet | C2 | $3.24 \pm 0.03$ | This work |
| HD76294 | 13c | m52 | $3.38 \pm 0.05$ | Johnson & Mitchell (1995) |
| HD76294 | KronComet | Gc | $3.14 \pm 0.04$ | This work |
| HD76294 | WBVR | V | $3.12 \pm 0.05$ | Kornilov et al. (1991) |
| HD76294 | Johnson | V | $3.10 \pm 0.05$ | Häggkvist & Oja (1966) |
| HD76294 | Johnson | V | $3.10 \pm 0.05$ | Johnson et al. (1966) |
| HD76294 | Johnson | V | $3.10 \pm 0.05$ | McClure (1970) |
| HD76294 | Johnson | V | $3.10 \pm 0.05$ | Jennens & Helfer (1975) |
| HD76294 | Johnson | V | $3.10 \pm 0.05$ | Ducati (2002) |
| HD76294 | Johnson | V | $3.11 \pm 0.05$ | Argue (1963) |
| HD76294 | Johnson | V | $3.11 \pm 0.05$ | Mermilliod (1986) |
| HD76294 | Johnson | V | $3.12 \pm 0.05$ | Cousins (1962b) |
| HD76294 | Johnson | V | $3.17 \pm 0.06$ | This work |
| HD76294 | 13c | m58 | $2.92 \pm 0.05$ | Johnson & Mitchell (1995) |
| HD76294 | 13c | m63 | $2.63 \pm 0.05$ | Johnson & Mitchell (1995) |
| HD76294 | WBVR | R | $2.40 \pm 0.05$ | Kornilov et al. (1991) |
| HD76294 | KronComet | Rc | $2.13 \pm 0.02$ | This work |
| HD76294 | 13c | m72 | $2.40 \pm 0.05$ | Johnson & Mitchell (1995) |
| HD76294 | 13c | m80 | $2.16 \pm 0.05$ | Johnson & Mitchell (1995) |
| HD76294 | 13c | m86 | $2.06 \pm 0.05$ | Johnson & Mitchell (1995) |
| HD76294 | 13c | m99 | $1.91 \pm 0.05$ | Johnson & Mitchell (1995) |
| HD76294 | 13c | m110 | $1.74 \pm 0.05$ | Johnson & Mitchell (1995) |
| HD76294 | 1250 | 310 | $444.20 \pm 13.40$ | Smith et al. (2004) |
| HD76294 | Johnson | J | $1.42 \pm 0.05$ | Alonso et al. (1998) |
| HD76294 | Johnson | J | $1.48 \pm 0.05$ | Johnson et al. (1966) |
| HD76294 | Johnson | J | $1.48 \pm 0.05$ | Ducati (2002) |
| HD76294 | 2200 | 361 | $305.90 \pm 12.80$ | Smith et al. (2004) |
| HD76294 | Johnson | K | $0.87 \pm 0.05$ | Johnson et al. (1966) |
| HD76294 | Johnson | K | $0.87 \pm 0.05$ | Ducati (2002) |
| HD76294 | Johnson | K | $0.89 \pm 0.04$ | Neugebauer & Leighton (1969) |
| HD76294 | Johnson | L | $0.79 \pm 0.05$ | Johnson et al. (1966) |
| HD76294 | Johnson | L | $0.79 \pm 0.05$ | Ducati (2002) |
| HD76294 | 3500 | 898 | $145.20 \pm 8.40$ | Smith et al. (2004) |
| HD76294 | 4900 | 712 | $71.80 \pm 5.40$ | Smith et al. (2004) |
| HD76294 | 12000 | 6384 | $16.70 \pm 23.70$ | Smith et al. (2004) |
| HD76830 | WBVR | W | $9.67 \pm 0.05$ | Kornilov et al. (1991) |
| HD76830 | Johnson | U | $9.56 \pm 0.05$ | Mermilliod (1986) |
| HD76830 | Johnson | U | $9.62 \pm 0.05$ | Roman (1955) |
| HD76830 | Johnson | U | $9.62 \pm 0.05$ | Johnson et al. (1966) |

**Table 21** *(continued)*

| Star ID | System/Wvlen | Band/Bandpass | Value | Reference |
|---------|--------------|---------------|-------|-----------|
| HD76830 | Johnson | U | $9.62 \pm 0.05$ | Ducati (2002) |
| HD76830 | KronComet | COp | $8.43 \pm 0.06$ | This work |
| HD76830 | WBVR | B | $7.96 \pm 0.05$ | Kornilov et al. (1991) |
| HD76830 | Johnson | B | $7.92 \pm 0.05$ | Mermilliod (1986) |
| HD76830 | Johnson | B | $7.93 \pm 0.05$ | Roman (1955) |
| HD76830 | Johnson | B | $7.93 \pm 0.05$ | Johnson et al. (1966) |
| HD76830 | Johnson | B | $7.93 \pm 0.05$ | Ducati (2002) |
| HD76830 | KronComet | Bc | $7.72 \pm 0.05$ | This work |
| HD76830 | KronComet | C2 | $6.59 \pm 0.03$ | This work |
| HD76830 | KronComet | Gc | $6.48 \pm 0.04$ | This work |
| HD76830 | WBVR | V | $6.36 \pm 0.05$ | Kornilov et al. (1991) |
| HD76830 | Johnson | V | $6.38 \pm 0.05$ | Roman (1955) |
| HD76830 | Johnson | V | $6.38 \pm 0.05$ | Johnson et al. (1966) |
| HD76830 | Johnson | V | $6.38 \pm 0.05$ | Mermilliod (1986) |
| HD76830 | Johnson | V | $6.38 \pm 0.05$ | Ducati (2002) |
| HD76830 | Johnson | V | $6.44 \pm 0.06$ | This work |
| HD76830 | WBVR | R | $4.65 \pm 0.05$ | Kornilov et al. (1991) |
| HD76830 | KronComet | Rc | $5.04 \pm 0.02$ | This work |
| HD76830 | 1250 | 310 | $196.10 \pm 5.20$ | Smith et al. (2004) |
| HD76830 | Johnson | J | $2.28 \pm 0.05$ | McWilliam & Lambert (1984) |
| HD76830 | Johnson | J | $2.28 \pm 0.05$ | Ducati (2002) |
| HD76830 | 2200 | 361 | $202.20 \pm 5.90$ | Smith et al. (2004) |
| HD76830 | Johnson | K | $1.18 \pm 0.05$ | Ducati (2002) |
| HD76830 | Johnson | K | $1.21 \pm 0.05$ | Neugebauer & Leighton (1969) |
| HD76830 | 3500 | 898 | $103.20 \pm 4.00$ | Smith et al. (2004) |
| HD76830 | 4900 | 712 | $43.70 \pm 5.60$ | Smith et al. (2004) |
| HD76830 | 12000 | 6384 | $11.20 \pm 36.30$ | Smith et al. (2004) |
| HD79554 | Vilnius | U | $9.95 \pm 0.05$ | Zdanavicius et al. (1972) |
| HD79554 | DDO | m35 | $9.44 \pm 0.05$ | McClure & Forrester (1981) |
| HD79554 | WBVR | W | $7.96 \pm 0.05$ | Kornilov et al. (1991) |
| HD79554 | Johnson | U | $7.96 \pm 0.05$ | Rybka (1969) |
| HD79554 | Johnson | U | $8.06 \pm 0.05$ | Gutierrez-Moreno & et al. (1966) |
| HD79554 | Vilnius | P | $9.23 \pm 0.05$ | Zdanavicius et al. (1972) |
| HD79554 | DDO | m38 | $8.26 \pm 0.05$ | McClure & Forrester (1981) |
| HD79554 | Vilnius | X | $8.04 \pm 0.05$ | Zdanavicius et al. (1972) |
| HD79554 | DDO | m41 | $8.51 \pm 0.05$ | McClure & Forrester (1981) |
| HD79554 | Oja | m41 | $8.03 \pm 0.05$ | Häggkvist & Oja (1970) |
| HD79554 | DDO | m42 | $8.27 \pm 0.05$ | McClure & Forrester (1981) |
| HD79554 | Oja | m42 | $7.81 \pm 0.05$ | Häggkvist & Oja (1970) |
| HD79554 | WBVR | B | $6.71 \pm 0.05$ | Kornilov et al. (1991) |
| HD79554 | Johnson | B | $6.65 \pm 0.05$ | Rybka (1969) |

Table 21 (continued)

| Star ID | System/Wvlen | Band/Bandpass | Value | Reference |
|---------|--------------|---------------|-------|-----------|
| HD79554 | Johnson | B | $6.67 \pm 0.05$ | Gutierrez-Moreno & et al. (1966) |
| HD79554 | DDO | m45 | $7.18 \pm 0.05$ | McClure & Forrester (1981) |
| HD79554 | Oja | m45 | $6.34 \pm 0.05$ | Häggkvist & Oja (1970) |
| HD79554 | Vilnius | Y | $6.32 \pm 0.05$ | Zdanavicius et al. (1972) |
| HD79554 | DDO | m48 | $5.87 \pm 0.05$ | McClure & Forrester (1981) |
| HD79554 | Vilnius | Z | $5.76 \pm 0.05$ | Zdanavicius et al. (1972) |
| HD79554 | WBVR | V | $5.36 \pm 0.05$ | Kornilov et al. (1991) |
| HD79554 | Vilnius | V | $5.36 \pm 0.05$ | Zdanavicius et al. (1972) |
| HD79554 | Johnson | V | $5.33 \pm 0.05$ | Rybka (1969) |
| HD79554 | Johnson | V | $5.35 \pm 0.05$ | Gutierrez-Moreno & et al. (1966) |
| HD79554 | Johnson | V | $5.39 \pm 0.06$ | This work |
| HD79554 | Vilnius | S | $4.45 \pm 0.05$ | Zdanavicius et al. (1972) |
| HD79554 | WBVR | R | $4.40 \pm 0.05$ | Kornilov et al. (1991) |
| HD79554 | 1250 | 310 | $93.60 \pm 4.20$ | Smith et al. (2004) |
| HD79554 | 2200 | 361 | $72.90 \pm 5.20$ | Smith et al. (2004) |
| HD79554 | Johnson | K | $2.41 \pm 0.07$ | Neugebauer & Leighton (1969) |
| HD79554 | 3500 | 898 | $34.30 \pm 4.60$ | Smith et al. (2004) |
| HD79554 | 4900 | 712 | $15.30 \pm 5.60$ | Smith et al. (2004) |
| HD79554 | 12000 | 6384 | $2.70 \pm 29.80$ | Smith et al. (2004) |
| HD81817 | Oja | m41 | $7.38 \pm 0.05$ | Häggkvist & Oja (1970) |
| HD81817 | Oja | m42 | $7.00 \pm 0.05$ | Häggkvist & Oja (1970) |
| HD81817 | WBVR | B | $5.82 \pm 0.05$ | Kornilov et al. (1991) |
| HD81817 | Johnson | B | $5.76 \pm 0.05$ | Häggkvist & Oja (1966) |
| HD81817 | Johnson | B | $5.76 \pm 0.05$ | Mermilliod (1986) |
| HD81817 | Johnson | B | $5.78 \pm 0.05$ | Johnson et al. (1966) |
| HD81817 | 13c | m45 | $5.32 \pm 0.05$ | Johnson & Mitchell (1995) |
| HD81817 | Oja | m45 | $5.45 \pm 0.05$ | Häggkvist & Oja (1970) |
| HD81817 | Vilnius | Y | $5.34 \pm 0.05$ | Cernis et al. (1989) |
| HD81817 | Vilnius | Z | $4.72 \pm 0.05$ | Cernis et al. (1989) |
| HD81817 | 13c | m52 | $4.66 \pm 0.05$ | Johnson & Mitchell (1995) |
| HD81817 | WBVR | V | $4.29 \pm 0.05$ | Kornilov et al. (1991) |
| HD81817 | Vilnius | V | $4.25 \pm 0.05$ | Cernis et al. (1989) |
| HD81817 | Johnson | V | $4.26 \pm 0.05$ | Mermilliod (1986) |
| HD81817 | Johnson | V | $4.27 \pm 0.05$ | Häggkvist & Oja (1966) |
| HD81817 | Johnson | V | $4.30 \pm 0.05$ | Johnson et al. (1966) |
| HD81817 | 13c | m58 | $3.93 \pm 0.05$ | Johnson & Mitchell (1995) |
| HD81817 | 13c | m63 | $3.55 \pm 0.05$ | Johnson & Mitchell (1995) |
| HD81817 | Vilnius | S | $3.27 \pm 0.05$ | Cernis et al. (1989) |
| HD81817 | WBVR | R | $3.24 \pm 0.05$ | Kornilov et al. (1991) |
| HD81817 | 13c | m72 | $3.18 \pm 0.05$ | Johnson & Mitchell (1995) |
| HD81817 | 13c | m80 | $2.85 \pm 0.05$ | Johnson & Mitchell (1995) |

**Table 21** *(continued)*

| Star ID | System/Wvlen | Band/Bandpass | Value | Reference |
|---------|--------------|---------------|-------|-----------|
| HD81817 | 13c | m86 | $2.70 \pm 0.05$ | Johnson & Mitchell (1995) |
| HD81817 | 13c | m99 | $2.47 \pm 0.05$ | Johnson & Mitchell (1995) |
| HD81817 | 13c | m110 | $2.23 \pm 0.05$ | Johnson & Mitchell (1995) |
| HD81817 | 1250 | 310 | $287.70 \pm 7.50$ | Smith et al. (2004) |
| HD81817 | 2200 | 361 | $237.70 \pm 5.70$ | Smith et al. (2004) |
| HD81817 | 3500 | 898 | $119.50 \pm 4.90$ | Smith et al. (2004) |
| HD81817 | 4900 | 712 | $52.00 \pm 4.90$ | Smith et al. (2004) |
| HD81817 | 12000 | 6384 | $14.40 \pm 16.10$ | Smith et al. (2004) |
| HD82198 | Geneva | U | $9.54 \pm 0.08$ | Golay (1972) |
| HD82198 | WBVR | W | $8.75 \pm 0.05$ | Kornilov et al. (1991) |
| HD82198 | Johnson | U | $8.71 \pm 0.05$ | Ducati (2002) |
| HD82198 | Johnson | U | $8.72 \pm 0.05$ | Mermilliod (1986) |
| HD82198 | Geneva | B1 | $8.04 \pm 0.08$ | Golay (1972) |
| HD82198 | Oja | m41 | $8.45 \pm 0.05$ | Häggkvist & Oja (1970) |
| HD82198 | Oja | m42 | $8.24 \pm 0.05$ | Häggkvist & Oja (1970) |
| HD82198 | Geneva | B | $6.39 \pm 0.08$ | Golay (1972) |
| HD82198 | WBVR | B | $6.96 \pm 0.05$ | Kornilov et al. (1991) |
| HD82198 | Johnson | B | $6.90 \pm 0.05$ | Haggkvist & Oja (1970) |
| HD82198 | Johnson | B | $6.90 \pm 0.05$ | Ducati (2002) |
| HD82198 | Johnson | B | $6.91 \pm 0.05$ | Mermilliod (1986) |
| HD82198 | Geneva | B2 | $7.36 \pm 0.08$ | Golay (1972) |
| HD82198 | Oja | m45 | $6.55 \pm 0.05$ | Häggkvist & Oja (1970) |
| HD82198 | Geneva | V1 | $6.22 \pm 0.08$ | Golay (1972) |
| HD82198 | WBVR | V | $5.39 \pm 0.05$ | Kornilov et al. (1991) |
| HD82198 | Geneva | V | $5.40 \pm 0.08$ | Golay (1972) |
| HD82198 | Johnson | V | $5.37 \pm 0.05$ | Haggkvist & Oja (1970) |
| HD82198 | Johnson | V | $5.37 \pm 0.05$ | Ducati (2002) |
| HD82198 | Johnson | V | $5.40 \pm 0.05$ | Mermilliod (1986) |
| HD82198 | Johnson | V | $5.40 \pm 0.06$ | This work |
| HD82198 | Geneva | G | $6.25 \pm 0.08$ | Golay (1972) |
| HD82198 | WBVR | R | $4.16 \pm 0.05$ | Kornilov et al. (1991) |
| HD82198 | Johnson | J | $2.49 \pm 0.05$ | McWilliam & Lambert (1984) |
| HD82198 | Johnson | J | $2.49 \pm 0.05$ | Ducati (2002) |
| HD82198 | 2200 | 361 | $144.00 \pm 18.90$ | Smith et al. (2004) |
| HD82198 | Johnson | K | $1.47 \pm 0.05$ | Ducati (2002) |
| HD82198 | Johnson | K | $1.53 \pm 0.03$ | Neugebauer & Leighton (1969) |
| HD82198 | 3500 | 898 | $68.20 \pm 13.30$ | Smith et al. (2004) |
| HD82198 | 4900 | 712 | $29.50 \pm 6.00$ | Smith et al. (2004) |
| HD82198 | 12000 | 6384 | $3.70 \pm 21.80$ | Smith et al. (2004) |
| HD82381 | Stromgren | u | $9.28 \pm 0.08$ | Olsen (1982) |
| HD82381 | Stromgren | u | $9.28 \pm 0.08$ | Hauck & Mermilliod (1998) |

**Table 21** *(continued)*

| Star ID | System/Wvlen | Band/Bandpass | Value | Reference |
|---------|--------------|---------------|-------|-----------|
| HD82381 | WBVR | W | $7.84 \pm 0.05$ | Kornilov et al. (1991) |
| HD82381 | Johnson | U | $7.62 \pm 0.05$ | Nicolet (1978) |
| HD82381 | Johnson | U | $7.64 \pm 0.05$ | Landolt (1968) |
| HD82381 | Johnson | U | $7.93 \pm 0.05$ | Lutz & Lutz (1977) |
| HD82381 | Johnson | U | $7.96 \pm 0.05$ | Mermilliod (1986) |
| HD82381 | Johnson | U | $7.97 \pm 0.05$ | Argue (1963) |
| HD82381 | Johnson | U | $7.97 \pm 0.05$ | Nicolet (1978) |
| HD82381 | Johnson | U | $7.97 \pm 0.05$ | Ducati (2002) |
| HD82381 | Johnson | U | $7.98 \pm 0.05$ | Eggen (1963) |
| HD82381 | DDO | m41 | $8.33 \pm 0.05$ | Eggen (1989) |
| HD82381 | Oja | m41 | $7.85 \pm 0.05$ | Häggkvist & Oja (1970) |
| HD82381 | Stromgren | v | $7.39 \pm 0.08$ | Olsen (1982) |
| HD82381 | Stromgren | v | $7.39 \pm 0.08$ | Hauck & Mermilliod (1998) |
| HD82381 | DDO | m42 | $8.06 \pm 0.05$ | Eggen (1989) |
| HD82381 | Oja | m42 | $7.58 \pm 0.05$ | Häggkvist & Oja (1970) |
| HD82381 | WBVR | B | $6.46 \pm 0.05$ | Kornilov et al. (1991) |
| HD82381 | Johnson | B | $6.11 \pm 0.05$ | Nicolet (1978) |
| HD82381 | Johnson | B | $6.13 \pm 0.05$ | Landolt (1968) |
| HD82381 | Johnson | B | $6.42 \pm 0.05$ | Lutz & Lutz (1977) |
| HD82381 | Johnson | B | $6.43 \pm 0.05$ | Mermilliod (1986) |
| HD82381 | Johnson | B | $6.44 \pm 0.05$ | Eggen (1963) |
| HD82381 | Johnson | B | $6.44 \pm 0.05$ | Argue (1963) |
| HD82381 | Johnson | B | $6.44 \pm 0.05$ | Nicolet (1978) |
| HD82381 | Johnson | B | $6.44 \pm 0.05$ | Ducati (2002) |
| HD82381 | DDO | m45 | $6.93 \pm 0.05$ | Eggen (1989) |
| HD82381 | Oja | m45 | $6.09 \pm 0.05$ | Häggkvist & Oja (1970) |
| HD82381 | Stromgren | b | $5.91 \pm 0.08$ | Olsen (1982) |
| HD82381 | Stromgren | b | $5.91 \pm 0.08$ | Hauck & Mermilliod (1998) |
| HD82381 | DDO | m48 | $5.60 \pm 0.05$ | Eggen (1989) |
| HD82381 | WBVR | V | $5.06 \pm 0.05$ | Kornilov et al. (1991) |
| HD82381 | Stromgren | y | $5.07 \pm 0.08$ | Olsen (1982) |
| HD82381 | Stromgren | y | $5.07 \pm 0.08$ | Hauck & Mermilliod (1998) |
| HD82381 | Johnson | V | $5.06 \pm 0.05$ | Lutz & Lutz (1977) |
| HD82381 | Johnson | V | $5.07 \pm 0.05$ | Argue (1963) |
| HD82381 | Johnson | V | $5.07 \pm 0.05$ | Nicolet (1978) |
| HD82381 | Johnson | V | $5.07 \pm 0.05$ | Ducati (2002) |
| HD82381 | Johnson | V | $5.08 \pm 0.05$ | Eggen (1963) |
| HD82381 | Johnson | V | $5.08 \pm 0.05$ | Mermilliod (1986) |
| HD82381 | Johnson | V | $5.10 \pm 0.06$ | This work |
| HD82381 | Johnson | V | $5.12 \pm 0.05$ | Nicolet (1978) |
| HD82381 | Johnson | V | $5.14 \pm 0.05$ | Landolt (1968) |

**Table 21** (continued)

| Star ID | System/Wvlen | Band/Bandpass | Value | Reference |
|---------|--------------|---------------|-------|-----------|
| HD82381 | WBVR | R | $4.08 \pm 0.05$ | Kornilov et al. (1991) |
| HD82381 | 1250 | 310 | $126.90 \pm 4.50$ | Smith et al. (2004) |
| HD82381 | Johnson | J | $2.72 \pm 0.05$ | Ridgway et al. (1980b) |
| HD82381 | Johnson | J | $2.72 \pm 0.05$ | Ducati (2002) |
| HD82381 | Johnson | H | $2.04 \pm 0.05$ | Ridgway et al. (1980b) |
| HD82381 | Johnson | H | $2.04 \pm 0.05$ | Ducati (2002) |
| HD82381 | 2200 | 361 | $103.60 \pm 5.70$ | Smith et al. (2004) |
| HD82381 | Johnson | K | $1.84 \pm 0.06$ | Neugebauer & Leighton (1969) |
| HD82381 | Johnson | K | $1.90 \pm 0.05$ | Ducati (2002) |
| HD82381 | Johnson | L | $1.82 \pm 0.05$ | Ducati (2002) |
| HD82381 | 3500 | 898 | $49.90 \pm 5.80$ | Smith et al. (2004) |
| HD82381 | 4900 | 712 | $22.50 \pm 5.00$ | Smith et al. (2004) |
| HD82381 | Johnson | M | $2.01 \pm 0.05$ | Ducati (2002) |
| HD82381 | 12000 | 6384 | $7.90 \pm 24.10$ | Smith et al. (2004) |
| HD82635 | 13c | m33 | $6.06 \pm 0.05$ | Johnson & Mitchell (1995) |
| HD82635 | Geneva | U | $6.58 \pm 0.08$ | Golay (1972) |
| HD82635 | Vilnius | U | $7.75 \pm 0.05$ | Sudzius et al. (1970) |
| HD82635 | 13c | m35 | $5.89 \pm 0.05$ | Johnson & Mitchell (1995) |
| HD82635 | DDO | m35 | $7.38 \pm 0.05$ | McClure & Forrester (1981) |
| HD82635 | DDO | m35 | $7.38 \pm 0.05$ | Mermilliod & Nitschelm (1989) |
| HD82635 | Stromgren | u | $7.35 \pm 0.08$ | Olsen (1983) |
| HD82635 | Stromgren | u | $7.36 \pm 0.08$ | Hauck & Mermilliod (1998) |
| HD82635 | Stromgren | u | $7.37 \pm 0.08$ | Crawford & Barnes (1970) |
| HD82635 | WBVR | W | $5.91 \pm 0.05$ | Kornilov et al. (1991) |
| HD82635 | Johnson | U | $6.07 \pm 0.05$ | Mermilliod (1986) |
| HD82635 | Johnson | U | $6.09 \pm 0.05$ | Argue (1963) |
| HD82635 | Johnson | U | $6.10 \pm 0.05$ | Johnson et al. (1966) |
| HD82635 | 13c | m37 | $5.98 \pm 0.05$ | Johnson & Mitchell (1995) |
| HD82635 | Vilnius | P | $7.21 \pm 0.05$ | Sudzius et al. (1970) |
| HD82635 | DDO | m38 | $6.35 \pm 0.05$ | McClure & Forrester (1981) |
| HD82635 | DDO | m38 | $6.35 \pm 0.05$ | Mermilliod & Nitschelm (1989) |
| HD82635 | 13c | m40 | $5.93 \pm 0.05$ | Johnson & Mitchell (1995) |
| HD82635 | Geneva | B1 | $5.98 \pm 0.08$ | Golay (1972) |
| HD82635 | Vilnius | X | $6.35 \pm 0.05$ | Sudzius et al. (1970) |
| HD82635 | DDO | m41 | $6.97 \pm 0.05$ | McClure & Forrester (1981) |
| HD82635 | DDO | m41 | $6.97 \pm 0.05$ | Mermilliod & Nitschelm (1989) |
| HD82635 | Oja | m41 | $6.51 \pm 0.05$ | Häggkvist & Oja (1970) |
| HD82635 | Stromgren | v | $6.07 \pm 0.08$ | Crawford & Barnes (1970) |
| HD82635 | Stromgren | v | $6.07 \pm 0.08$ | Olsen (1983) |
| HD82635 | Stromgren | v | $6.07 \pm 0.08$ | Hauck & Mermilliod (1998) |
| HD82635 | DDO | m42 | $6.80 \pm 0.05$ | McClure & Forrester (1981) |

**Table 21** continued on next page

**Table 21** *(continued)*

| Star ID | System/Wvlen | Band/Bandpass | Value | Reference |
|---------|--------------|---------------|-------|-----------|
| HD82635 | DDO | m42 | $6.80 \pm 0.05$ | Mermilliod & Nitschelm (1989) |
| HD82635 | Oja | m42 | $6.35 \pm 0.05$ | Häggkvist & Oja (1970) |
| HD82635 | Geneva | B | $4.75 \pm 0.08$ | Golay (1972) |
| HD82635 | WBVR | B | $5.47 \pm 0.05$ | Kornilov et al. (1991) |
| HD82635 | Johnson | B | $5.45 \pm 0.05$ | Häggkvist & Oja (1966) |
| HD82635 | Johnson | B | $5.47 \pm 0.05$ | Johnson et al. (1966) |
| HD82635 | Johnson | B | $5.47 \pm 0.05$ | Mermilliod (1986) |
| HD82635 | Johnson | B | $5.49 \pm 0.05$ | Argue (1963) |
| HD82635 | Geneva | B2 | $5.95 \pm 0.08$ | Golay (1972) |
| HD82635 | 13c | m45 | $5.20 \pm 0.05$ | Johnson & Mitchell (1995) |
| HD82635 | DDO | m45 | $6.03 \pm 0.05$ | McClure & Forrester (1981) |
| HD82635 | DDO | m45 | $6.03 \pm 0.05$ | Mermilliod & Nitschelm (1989) |
| HD82635 | Oja | m45 | $5.19 \pm 0.05$ | Häggkvist & Oja (1970) |
| HD82635 | Vilnius | Y | $5.22 \pm 0.05$ | Sudzius et al. (1970) |
| HD82635 | Stromgren | b | $5.16 \pm 0.08$ | Crawford & Barnes (1970) |
| HD82635 | Stromgren | b | $5.16 \pm 0.08$ | Olsen (1983) |
| HD82635 | Stromgren | b | $5.16 \pm 0.08$ | Hauck & Mermilliod (1998) |
| HD82635 | DDO | m48 | $4.88 \pm 0.05$ | McClure & Forrester (1981) |
| HD82635 | DDO | m48 | $4.88 \pm 0.05$ | Mermilliod & Nitschelm (1989) |
| HD82635 | Vilnius | Z | $4.82 \pm 0.05$ | Sudzius et al. (1970) |
| HD82635 | 13c | m52 | $4.76 \pm 0.05$ | Johnson & Mitchell (1995) |
| HD82635 | Geneva | V1 | $5.35 \pm 0.08$ | Golay (1972) |
| HD82635 | WBVR | V | $4.54 \pm 0.05$ | Kornilov et al. (1991) |
| HD82635 | Vilnius | V | $4.55 \pm 0.05$ | Sudzius et al. (1970) |
| HD82635 | Stromgren | y | $4.60 \pm 0.08$ | Crawford & Barnes (1970) |
| HD82635 | Stromgren | y | $4.60 \pm 0.08$ | Olsen (1983) |
| HD82635 | Stromgren | y | $4.60 \pm 0.08$ | Hauck & Mermilliod (1998) |
| HD82635 | Geneva | V | $4.58 \pm 0.08$ | Golay (1972) |
| HD82635 | Johnson | V | $4.54 \pm 0.05$ | Häggkvist & Oja (1966) |
| HD82635 | Johnson | V | $4.54 \pm 0.06$ | This work |
| HD82635 | Johnson | V | $4.55 \pm 0.05$ | Johnson et al. (1966) |
| HD82635 | Johnson | V | $4.56 \pm 0.05$ | Mermilliod (1986) |
| HD82635 | Johnson | V | $4.57 \pm 0.05$ | Argue (1963) |
| HD82635 | 13c | m58 | $4.34 \pm 0.05$ | Johnson & Mitchell (1995) |
| HD82635 | Geneva | G | $5.58 \pm 0.08$ | Golay (1972) |
| HD82635 | Alexander | m608 | $2.61 \pm 0.05$ | Alexander et al. (1983) |
| HD82635 | 13c | m63 | $4.06 \pm 0.05$ | Johnson & Mitchell (1995) |
| HD82635 | Vilnius | S | $3.85 \pm 0.05$ | Sudzius et al. (1970) |
| HD82635 | Alexander | m683 | $3.27 \pm 0.05$ | Alexander et al. (1983) |
| HD82635 | WBVR | R | $3.88 \pm 0.05$ | Kornilov et al. (1991) |
| HD82635 | Alexander | m710 | $3.47 \pm 0.05$ | Alexander et al. (1983) |

**Table 21** *(continued)*

| Star ID | System/Wvlen | Band/Bandpass | Value | Reference |
|---------|--------------|---------------|-------|-----------|
| HD82635 | 13c | m72 | $3.87 \pm 0.05$ | Johnson & Mitchell (1995) |
| HD82635 | Alexander | m746 | $3.88 \pm 0.05$ | Alexander et al. (1983) |
| HD82635 | 13c | m80 | $3.67 \pm 0.05$ | Johnson & Mitchell (1995) |
| HD82635 | 13c | m86 | $3.57 \pm 0.05$ | Johnson & Mitchell (1995) |
| HD82635 | 13c | m99 | $3.44 \pm 0.05$ | Johnson & Mitchell (1995) |
| HD82635 | 13c | m110 | $3.25 \pm 0.05$ | Johnson & Mitchell (1995) |
| HD82635 | 1250 | 310 | $144.60 \pm 20.40$ | Smith et al. (2004) |
| HD82635 | 2200 | 361 | $100.90 \pm 28.60$ | Smith et al. (2004) |
| HD82635 | Johnson | K | $2.49 \pm 0.05$ | Neugebauer & Leighton (1969) |
| HD82635 | 3500 | 898 | $47.20 \pm 16.60$ | Smith et al. (2004) |
| HD82741 | 13c | m33 | $6.57 \pm 0.05$ | Johnson & Mitchell (1995) |
| HD82741 | KronComet | NH | $7.45 \pm 0.17$ | This work |
| HD82741 | KronComet | UVc | $7.23 \pm 0.18$ | This work |
| HD82741 | Geneva | U | $7.06 \pm 0.08$ | Golay (1972) |
| HD82741 | 13c | m35 | $6.35 \pm 0.05$ | Johnson & Mitchell (1995) |
| HD82741 | DDO | m35 | $7.89 \pm 0.05$ | McClure & Forrester (1981) |
| HD82741 | WBVR | W | $6.41 \pm 0.05$ | Kornilov et al. (1991) |
| HD82741 | Johnson | U | $6.08 \pm 0.17$ | This work |
| HD82741 | Johnson | U | $6.56 \pm 0.05$ | Argue (1963) |
| HD82741 | Johnson | U | $6.56 \pm 0.05$ | Johnson et al. (1966) |
| HD82741 | Johnson | U | $6.58 \pm 0.05$ | Mermilliod (1986) |
| HD82741 | 13c | m37 | $6.46 \pm 0.05$ | Johnson & Mitchell (1995) |
| HD82741 | DDO | m38 | $6.84 \pm 0.05$ | McClure & Forrester (1981) |
| HD82741 | KronComet | CN | $7.20 \pm 0.13$ | This work |
| HD82741 | 13c | m40 | $6.34 \pm 0.05$ | Johnson & Mitchell (1995) |
| HD82741 | Geneva | B1 | $6.37 \pm 0.08$ | Golay (1972) |
| HD82741 | DDO | m41 | $7.36 \pm 0.05$ | McClure & Forrester (1981) |
| HD82741 | Oja | m41 | $6.87 \pm 0.05$ | Häggkvist & Oja (1970) |
| HD82741 | DDO | m42 | $7.21 \pm 0.05$ | McClure & Forrester (1981) |
| HD82741 | Oja | m42 | $6.74 \pm 0.05$ | Häggkvist & Oja (1970) |
| HD82741 | Geneva | B | $5.10 \pm 0.08$ | Golay (1972) |
| HD82741 | KronComet | COp | $5.84 \pm 0.10$ | This work |
| HD82741 | WBVR | B | $5.82 \pm 0.05$ | Kornilov et al. (1991) |
| HD82741 | Johnson | B | $5.74 \pm 0.10$ | This work |
| HD82741 | Johnson | B | $5.79 \pm 0.05$ | Häggkvist & Oja (1966) |
| HD82741 | Johnson | B | $5.80 \pm 0.05$ | Argue (1963) |
| HD82741 | Johnson | B | $5.80 \pm 0.05$ | Johnson et al. (1966) |
| HD82741 | Johnson | B | $5.81 \pm 0.05$ | Mermilliod (1986) |
| HD82741 | KronComet | Bc | $5.56 \pm 0.09$ | This work |
| HD82741 | Geneva | B2 | $6.26 \pm 0.08$ | Golay (1972) |
| HD82741 | 13c | m45 | $5.51 \pm 0.05$ | Johnson & Mitchell (1995) |

**Table 21** *(continued)*

| Star ID | System/Wvlen | Band/Bandpass | Value | Reference |
|---------|--------------|---------------|-------|-----------|
| HD82741 | DDO | m45 | $6.35 \pm 0.05$ | McClure & Forrester (1981) |
| HD82741 | Oja | m45 | $5.51 \pm 0.05$ | Häggkvist & Oja (1970) |
| HD82741 | DDO | m48 | $5.18 \pm 0.05$ | McClure & Forrester (1981) |
| HD82741 | KronComet | C2 | $4.93 \pm 0.05$ | This work |
| HD82741 | 13c | m52 | $5.04 \pm 0.05$ | Johnson & Mitchell (1995) |
| HD82741 | KronComet | Gc | $4.82 \pm 0.07$ | This work |
| HD82741 | Geneva | V1 | $5.58 \pm 0.08$ | Golay (1972) |
| HD82741 | WBVR | V | $4.82 \pm 0.05$ | Kornilov et al. (1991) |
| HD82741 | Geneva | V | $4.82 \pm 0.08$ | Golay (1972) |
| HD82741 | Johnson | V | $4.80 \pm 0.05$ | Häggkvist & Oja (1966) |
| HD82741 | Johnson | V | $4.81 \pm 0.05$ | Argue (1963) |
| HD82741 | Johnson | V | $4.81 \pm 0.05$ | Johnson et al. (1966) |
| HD82741 | Johnson | V | $4.82 \pm 0.05$ | Mermilliod (1986) |
| HD82741 | Johnson | V | $5.07 \pm 0.18$ | This work |
| HD82741 | 13c | m58 | $4.56 \pm 0.05$ | Johnson & Mitchell (1995) |
| HD82741 | Geneva | G | $5.79 \pm 0.08$ | Golay (1972) |
| HD82741 | 13c | m63 | $4.27 \pm 0.05$ | Johnson & Mitchell (1995) |
| HD82741 | WBVR | R | $4.08 \pm 0.05$ | Kornilov et al. (1991) |
| HD82741 | KronComet | Rc | $3.80 \pm 0.04$ | This work |
| HD82741 | 13c | m72 | $4.05 \pm 0.05$ | Johnson & Mitchell (1995) |
| HD82741 | 13c | m80 | $3.82 \pm 0.05$ | Johnson & Mitchell (1995) |
| HD82741 | 13c | m86 | $3.72 \pm 0.05$ | Johnson & Mitchell (1995) |
| HD82741 | 13c | m99 | $3.56 \pm 0.05$ | Johnson & Mitchell (1995) |
| HD82741 | 13c | m110 | $3.34 \pm 0.05$ | Johnson & Mitchell (1995) |
| HD82741 | 1250 | 310 | $92.40 \pm 7.60$ | Smith et al. (2004) |
| HD82741 | 2200 | 361 | $60.20 \pm 5.90$ | Smith et al. (2004) |
| HD82741 | Johnson | K | $2.54 \pm 0.11$ | Neugebauer & Leighton (1969) |
| HD82741 | 3500 | 898 | $27.90 \pm 4.60$ | Smith et al. (2004) |
| HD82741 | 4900 | 712 | $13.50 \pm 5.20$ | Smith et al. (2004) |
| HD82741 | 12000 | 6384 | $2.80 \pm 19.70$ | Smith et al. (2004) |
| HD84181 | KronComet | COp | $10.13 \pm 0.02$ | This work |
| HD84181 | Johnson | B | $9.35 \pm 0.01$ | This work |
| HD84181 | KronComet | Bc | $9.27 \pm 0.02$ | This work |
| HD84181 | KronComet | C2 | $8.21 \pm 0.01$ | This work |
| HD84181 | KronComet | Gc | $7.95 \pm 0.01$ | This work |
| HD84181 | Johnson | V | $7.90 \pm 0.01$ | This work |
| HD84181 | KronComet | Rc | $6.43 \pm 0.01$ | This work |
| HD84181 | 1250 | 310 | $32.30 \pm 5.30$ | Smith et al. (2004) |
| HD84181 | 3500 | 898 | $17.80 \pm 8.40$ | Smith et al. (2004) |
| HD84181 | 4900 | 712 | $5.50 \pm 5.30$ | Smith et al. (2004) |
| HD84181 | 12000 | 6384 | $151.10 \pm 110.40$ | Smith et al. (2004) |

**Table 21** *continued on next page*

**Table 21** *(continued)*

| Star ID | System/Wvlen | Band/Bandpass | Value | Reference |
|---------|--------------|---------------|-------|-----------|
| HD84914 | KronComet | COp | $8.87 \pm 0.02$ | This work |
| HD84914 | WBVR | B | $8.25 \pm 0.05$ | Kornilov et al. (1991) |
| HD84914 | Johnson | B | $8.11 \pm 0.07$ | This work |
| HD84914 | Johnson | B | $8.18 \pm 0.05$ | Guetter & Hewitt (1984) |
| HD84914 | KronComet | Bc | $8.02 \pm 0.03$ | This work |
| HD84914 | KronComet | C2 | $7.07 \pm 0.03$ | This work |
| HD84914 | KronComet | Gc | $6.77 \pm 0.03$ | This work |
| HD84914 | WBVR | V | $6.63 \pm 0.05$ | Kornilov et al. (1991) |
| HD84914 | Johnson | V | $6.61 \pm 0.05$ | Guetter & Hewitt (1984) |
| HD84914 | Johnson | V | $6.66 \pm 0.05$ | This work |
| HD84914 | WBVR | R | $5.35 \pm 0.05$ | Kornilov et al. (1991) |
| HD84914 | KronComet | Rc | $5.24 \pm 0.03$ | This work |
| HD84914 | 1250 | 310 | $55.30 \pm 6.40$ | Smith et al. (2004) |
| HD84914 | 2200 | 361 | $52.60 \pm 5.30$ | Smith et al. (2004) |
| HD84914 | Johnson | K | $2.69 \pm 0.07$ | Neugebauer & Leighton (1969) |
| HD84914 | 3500 | 898 | $24.00 \pm 6.20$ | Smith et al. (2004) |
| HD84914 | 4900 | 712 | $10.50 \pm 5.70$ | Smith et al. (2004) |
| HD84914 | 12000 | 6384 | $1.50 \pm 19.00$ | Smith et al. (2004) |
| HD85503 | 13c | m33 | $6.65 \pm 0.05$ | Johnson & Mitchell (1995) |
| HD85503 | KronComet | NH | $7.53 \pm 0.11$ | This work |
| HD85503 | KronComet | UVc | $7.32 \pm 0.14$ | This work |
| HD85503 | Geneva | U | $7.11 \pm 0.08$ | Golay (1972) |
| HD85503 | Vilnius | U | $8.41 \pm 0.05$ | Kazlauskas et al. (2005) |
| HD85503 | 13c | m35 | $6.44 \pm 0.05$ | Johnson & Mitchell (1995) |
| HD85503 | DDO | m35 | $7.87 \pm 0.05$ | McClure & Forrester (1981) |
| HD85503 | DDO | m35 | $7.87 \pm 0.05$ | Mermilliod & Nitschelm (1989) |
| HD85503 | Stromgren | u | $7.83 \pm 0.08$ | Olsen (1993) |
| HD85503 | Stromgren | u | $7.83 \pm 0.08$ | Hauck & Mermilliod (1998) |
| HD85503 | WBVR | W | $6.41 \pm 0.05$ | Kornilov et al. (1991) |
| HD85503 | Johnson | U | $5.88 \pm 0.06$ | This work |
| HD85503 | Johnson | U | $6.48 \pm 0.05$ | Argue (1963) |
| HD85503 | Johnson | U | $6.49 \pm 0.05$ | Johnson et al. (1966) |
| HD85503 | Johnson | U | $6.49 \pm 0.05$ | McClure (1970) |
| HD85503 | Johnson | U | $6.49 \pm 0.05$ | Mermilliod (1986) |
| HD85503 | Johnson | U | $6.50 \pm 0.05$ | Johnson et al. (1966) |
| HD85503 | Johnson | U | $6.50 \pm 0.05$ | Ducati (2002) |
| HD85503 | 13c | m37 | $6.46 \pm 0.05$ | Johnson & Mitchell (1995) |
| HD85503 | Vilnius | P | $7.73 \pm 0.05$ | Kazlauskas et al. (2005) |
| HD85503 | DDO | m38 | $6.64 \pm 0.05$ | McClure & Forrester (1981) |
| HD85503 | DDO | m38 | $6.64 \pm 0.05$ | Mermilliod & Nitschelm (1989) |
| HD85503 | KronComet | CN | $7.26 \pm 0.16$ | This work |

**Table 21** *(continued)*

| Star ID | System/Wvlen | Band/Bandpass | Value | Reference |
|---------|--------------|---------------|-------|-----------|
| HD85503 | 13c | m40 | $5.98 \pm 0.05$ | Johnson & Mitchell (1995) |
| HD85503 | Geneva | B1 | $5.98 \pm 0.08$ | Golay (1972) |
| HD85503 | Vilnius | X | $6.45 \pm 0.05$ | Kazlauskas et al. (2005) |
| HD85503 | DDO | m41 | $7.04 \pm 0.05$ | McClure & Forrester (1981) |
| HD85503 | DDO | m41 | $7.04 \pm 0.05$ | Mermilliod & Nitschelm (1989) |
| HD85503 | Oja | m41 | $6.61 \pm 0.05$ | Häggkvist & Oja (1970) |
| HD85503 | Stromgren | v | $6.04 \pm 0.08$ | Olsen (1993) |
| HD85503 | Stromgren | v | $6.04 \pm 0.08$ | Hauck & Mermilliod (1998) |
| HD85503 | DDO | m42 | $6.64 \pm 0.05$ | McClure & Forrester (1981) |
| HD85503 | DDO | m42 | $6.64 \pm 0.05$ | Mermilliod & Nitschelm (1989) |
| HD85503 | Oja | m42 | $6.18 \pm 0.05$ | Häggkvist & Oja (1970) |
| HD85503 | Geneva | B | $4.49 \pm 0.08$ | Golay (1972) |
| HD85503 | KronComet | COp | $5.29 \pm 0.11$ | This work |
| HD85503 | WBVR | B | $5.14 \pm 0.05$ | Kornilov et al. (1991) |
| HD85503 | Johnson | B | $4.83 \pm 0.10$ | This work |
| HD85503 | Johnson | B | $5.10 \pm 0.05$ | Argue (1963) |
| HD85503 | Johnson | B | $5.10 \pm 0.05$ | Johnson et al. (1966) |
| HD85503 | Johnson | B | $5.10 \pm 0.05$ | McClure (1970) |
| HD85503 | Johnson | B | $5.10 \pm 0.05$ | Ducati (2002) |
| HD85503 | Johnson | B | $5.11 \pm 0.05$ | Häggkvist & Oja (1966) |
| HD85503 | Johnson | B | $5.11 \pm 0.05$ | Mermilliod (1986) |
| HD85503 | KronComet | Bc | $4.83 \pm 0.04$ | This work |
| HD85503 | Geneva | B2 | $5.53 \pm 0.08$ | Golay (1972) |
| HD85503 | 13c | m45 | $4.72 \pm 0.05$ | Johnson & Mitchell (1995) |
| HD85503 | DDO | m45 | $5.60 \pm 0.05$ | McClure & Forrester (1981) |
| HD85503 | DDO | m45 | $5.60 \pm 0.05$ | Mermilliod & Nitschelm (1989) |
| HD85503 | Oja | m45 | $4.76 \pm 0.05$ | Häggkvist & Oja (1970) |
| HD85503 | Vilnius | Y | $4.83 \pm 0.05$ | Kazlauskas et al. (2005) |
| HD85503 | Stromgren | b | $4.64 \pm 0.08$ | Olsen (1993) |
| HD85503 | Stromgren | b | $4.64 \pm 0.08$ | Hauck & Mermilliod (1998) |
| HD85503 | DDO | m48 | $4.33 \pm 0.05$ | McClure & Forrester (1981) |
| HD85503 | DDO | m48 | $4.33 \pm 0.05$ | Mermilliod & Nitschelm (1989) |
| HD85503 | KronComet | C2 | $4.16 \pm 0.05$ | This work |
| HD85503 | Vilnius | Z | $4.32 \pm 0.05$ | Kazlauskas et al. (2005) |
| HD85503 | 13c | m52 | $4.23 \pm 0.05$ | Johnson & Mitchell (1995) |
| HD85503 | KronComet | Gc | $3.97 \pm 0.01$ | This work |
| HD85503 | Geneva | V1 | $4.69 \pm 0.08$ | Golay (1972) |
| HD85503 | WBVR | V | $3.88 \pm 0.05$ | Kornilov et al. (1991) |
| HD85503 | Vilnius | V | $3.91 \pm 0.05$ | Kazlauskas et al. (2005) |
| HD85503 | Stromgren | y | $3.88 \pm 0.08$ | Olsen (1993) |
| HD85503 | Stromgren | y | $3.88 \pm 0.08$ | Hauck & Mermilliod (1998) |

**Table 21** *(continued)*

| Star ID | System/Wvlen | Band/Bandpass | Value | Reference |
|---------|--------------|---------------|-------|-----------|
| HD85503 | Geneva | V | $3.90 \pm 0.08$ | Golay (1972) |
| HD85503 | Johnson | V | $3.87 \pm 0.05$ | Argue (1963) |
| HD85503 | Johnson | V | $3.88 \pm 0.05$ | Johnson et al. (1966) |
| HD85503 | Johnson | V | $3.88 \pm 0.05$ | McClure (1970) |
| HD85503 | Johnson | V | $3.88 \pm 0.05$ | Mermilliod (1986) |
| HD85503 | Johnson | V | $3.88 \pm 0.05$ | Ducati (2002) |
| HD85503 | Johnson | V | $3.90 \pm 0.05$ | Häggkvist & Oja (1966) |
| HD85503 | Johnson | V | $3.91 \pm 0.10$ | This work |
| HD85503 | 13c | m58 | $3.60 \pm 0.05$ | Johnson & Mitchell (1995) |
| HD85503 | Geneva | G | $4.81 \pm 0.08$ | Golay (1972) |
| HD85503 | 13c | m63 | $3.27 \pm 0.05$ | Johnson & Mitchell (1995) |
| HD85503 | Vilnius | S | $3.09 \pm 0.05$ | Kazlauskas et al. (2005) |
| HD85503 | WBVR | R | $3.04 \pm 0.05$ | Kornilov et al. (1991) |
| HD85503 | KronComet | Rc | $2.78 \pm 0.01$ | This work |
| HD85503 | 13c | m72 | $3.01 \pm 0.05$ | Johnson & Mitchell (1995) |
| HD85503 | 13c | m80 | $2.73 \pm 0.05$ | Johnson & Mitchell (1995) |
| HD85503 | 13c | m86 | $2.61 \pm 0.05$ | Johnson & Mitchell (1995) |
| HD85503 | 13c | m99 | $2.42 \pm 0.05$ | Johnson & Mitchell (1995) |
| HD85503 | 13c | m110 | $2.21 \pm 0.05$ | Johnson & Mitchell (1995) |
| HD85503 | 1250 | 310 | $281.70 \pm 11.20$ | Smith et al. (2004) |
| HD85503 | Johnson | J | $1.87 \pm 0.05$ | Alonso et al. (1998) |
| HD85503 | Johnson | J | $1.93 \pm 0.05$ | Johnson et al. (1966) |
| HD85503 | Johnson | J | $1.93 \pm 0.05$ | Ducati (2002) |
| HD85503 | Johnson | J | $1.93 \pm 0.05$ | Shenavrin et al. (2011) |
| HD85503 | Johnson | H | $1.36 \pm 0.05$ | Alonso et al. (1998) |
| HD85503 | Johnson | H | $1.36 \pm 0.05$ | Shenavrin et al. (2011) |
| HD85503 | 2200 | 361 | $202.10 \pm 7.20$ | Smith et al. (2004) |
| HD85503 | Johnson | K | $1.20 \pm 0.04$ | Neugebauer & Leighton (1969) |
| HD85503 | Johnson | K | $1.22 \pm 0.05$ | Johnson et al. (1966) |
| HD85503 | Johnson | K | $1.22 \pm 0.05$ | Ducati (2002) |
| HD85503 | Johnson | K | $1.22 \pm 0.05$ | Shenavrin et al. (2011) |
| HD85503 | 3500 | 898 | $98.70 \pm 5.20$ | Smith et al. (2004) |
| HD85503 | 4900 | 712 | $45.80 \pm 5.40$ | Smith et al. (2004) |
| HD85503 | 12000 | 6384 | $13.20 \pm 20.20$ | Smith et al. (2004) |
| HD87046 | KronComet | COp | $9.59 \pm 0.01$ | This work |
| HD87046 | Johnson | B | $8.92 \pm 0.01$ | This work |
| HD87046 | KronComet | Bc | $8.84 \pm 0.01$ | This work |
| HD87046 | KronComet | C2 | $7.68 \pm 0.01$ | This work |
| HD87046 | KronComet | Gc | $7.52 \pm 0.01$ | This work |
| HD87046 | Johnson | V | $7.48 \pm 0.01$ | This work |
| HD87046 | KronComet | Rc | $6.00 \pm 0.01$ | This work |

Table 21 (continued)

| Star ID | System/Wvlen | Band/Bandpass | Value | Reference |
|---------|-------------|---------------|-------|-----------|
| HD87046 | 1250 | 310 | $61.00 \pm 5.10$ | Smith et al. (2004) |
| HD87046 | 2200 | 361 | $62.90 \pm 7.00$ | Smith et al. (2004) |
| HD87046 | Johnson | K | $2.41 \pm 0.07$ | Neugebauer & Leighton (1969) |
| HD87046 | 3500 | 898 | $31.60 \pm 5.30$ | Smith et al. (2004) |
| HD87046 | 4900 | 712 | $12.80 \pm 4.80$ | Smith et al. (2004) |
| HD87046 | 12000 | 6384 | $-0.10 \pm 19.10$ | Smith et al. (2004) |
| HD87837 | Geneva | B1 | $6.89 \pm 0.08$ | Golay (1972) |
| HD87837 | Oja | m41 | $7.34 \pm 0.05$ | Häggkvist & Oja (1970) |
| HD87837 | DDO | m42 | $7.55 \pm 0.05$ | McClure & Forrester (1981) |
| HD87837 | Oja | m42 | $7.07 \pm 0.05$ | Häggkvist & Oja (1970) |
| HD87837 | Geneva | B | $5.27 \pm 0.08$ | Golay (1972) |
| HD87837 | WBVR | B | $5.85 \pm 0.05$ | Kornilov et al. (1991) |
| HD87837 | Johnson | B | $5.81 \pm 0.05$ | Argue (1963) |
| HD87837 | Johnson | B | $5.81 \pm 0.05$ | Gutierrez-Moreno & et al. (1966) |
| HD87837 | Johnson | B | $5.81 \pm 0.05$ | Barnes et al. (1978) |
| HD87837 | Johnson | B | $5.82 \pm 0.05$ | Häggkvist & Oja (1966) |
| HD87837 | Johnson | B | $5.82 \pm 0.05$ | Johnson et al. (1966) |
| HD87837 | Johnson | B | $5.82 \pm 0.05$ | Sturch & Helfer (1972) |
| HD87837 | Johnson | B | $5.82 \pm 0.05$ | Ducati (2002) |
| HD87837 | Johnson | B | $5.84 \pm 0.05$ | Mermilliod (1986) |
| HD87837 | Geneva | B2 | $6.25 \pm 0.08$ | Golay (1972) |
| HD87837 | 13c | m45 | $5.37 \pm 0.05$ | Johnson & Mitchell (1995) |
| HD87837 | DDO | m45 | $6.28 \pm 0.05$ | McClure & Forrester (1981) |
| HD87837 | Oja | m45 | $5.44 \pm 0.05$ | Häggkvist & Oja (1970) |
| HD87837 | DDO | m48 | $4.94 \pm 0.05$ | McClure & Forrester (1981) |
| HD87837 | 13c | m52 | $4.76 \pm 0.05$ | Johnson & Mitchell (1995) |
| HD87837 | Geneva | V1 | $5.21 \pm 0.08$ | Golay (1972) |
| HD87837 | WBVR | V | $4.36 \pm 0.05$ | Kornilov et al. (1991) |
| HD87837 | Geneva | V | $4.40 \pm 0.08$ | Golay (1972) |
| HD87837 | Johnson | V | $4.35 \pm 0.05$ | Argue (1963) |
| HD87837 | Johnson | V | $4.36 \pm 0.05$ | Gutierrez-Moreno & et al. (1966) |
| HD87837 | Johnson | V | $4.37 \pm 0.05$ | Johnson et al. (1966) |
| HD87837 | Johnson | V | $4.38 \pm 0.05$ | Barnes et al. (1978) |
| HD87837 | Johnson | V | $4.38 \pm 0.05$ | Ducati (2002) |
| HD87837 | Johnson | V | $4.39 \pm 0.05$ | Häggkvist & Oja (1966) |
| HD87837 | Johnson | V | $4.39 \pm 0.05$ | Sturch & Helfer (1972) |
| HD87837 | Johnson | V | $4.40 \pm 0.05$ | Mermilliod (1986) |
| HD87837 | Johnson | V | $4.43 \pm 0.06$ | This work |
| HD87837 | 13c | m58 | $4.00 \pm 0.05$ | Johnson & Mitchell (1995) |
| HD87837 | Geneva | G | $5.25 \pm 0.08$ | Golay (1972) |
| HD87837 | 13c | m63 | $3.60 \pm 0.05$ | Johnson & Mitchell (1995) |

**Table 21** *(continued)*

| Star ID | System/Wvlen | Band/Bandpass | Value | Reference |
|---------|--------------|---------------|-------|-----------|
| HD87837 | WBVR | R | $3.29 \pm 0.05$ | Kornilov et al. (1991) |
| HD87837 | 13c | m72 | $3.22 \pm 0.05$ | Johnson & Mitchell (1995) |
| HD87837 | 13c | m80 | $2.87 \pm 0.05$ | Johnson & Mitchell (1995) |
| HD87837 | 13c | m86 | $2.72 \pm 0.05$ | Johnson & Mitchell (1995) |
| HD87837 | 13c | m99 | $2.47 \pm 0.05$ | Johnson & Mitchell (1995) |
| HD87837 | 13c | m110 | $2.21 \pm 0.05$ | Johnson & Mitchell (1995) |
| HD87837 | 1250 | 310 | $285.70 \pm 11.50$ | Smith et al. (2004) |
| HD87837 | Johnson | J | $1.88 \pm 0.05$ | Ridgway et al. (1980b) |
| HD87837 | Johnson | J | $1.89 \pm 0.05$ | Kenyon (1988) |
| HD87837 | Johnson | J | $1.89 \pm 0.05$ | Ducati (2002) |
| HD87837 | Johnson | J | $1.91 \pm 0.05$ | Johnson et al. (1966) |
| HD87837 | Johnson | J | $1.91 \pm 0.05$ | Shenavrin et al. (2011) |
| HD87837 | Johnson | H | $1.17 \pm 0.05$ | Kenyon (1988) |
| HD87837 | Johnson | H | $1.18 \pm 0.05$ | Ducati (2002) |
| HD87837 | Johnson | H | $1.19 \pm 0.05$ | Ridgway et al. (1980b) |
| HD87837 | Johnson | H | $1.21 \pm 0.05$ | Shenavrin et al. (2011) |
| HD87837 | 2200 | 361 | $238.40 \pm 8.40$ | Smith et al. (2004) |
| HD87837 | Johnson | K | $1.03 \pm 0.07$ | Neugebauer & Leighton (1969) |
| HD87837 | Johnson | K | $1.04 \pm 0.05$ | Johnson et al. (1966) |
| HD87837 | Johnson | K | $1.04 \pm 0.05$ | Ducati (2002) |
| HD87837 | Johnson | K | $1.04 \pm 0.05$ | Shenavrin et al. (2011) |
| HD87837 | Johnson | L | $0.92 \pm 0.05$ | Ducati (2002) |
| HD87837 | 3500 | 898 | $118.90 \pm 9.30$ | Smith et al. (2004) |
| HD87837 | 4900 | 712 | $53.80 \pm 7.40$ | Smith et al. (2004) |
| HD87837 | Johnson | M | $1.15 \pm 0.05$ | Ducati (2002) |
| HD87837 | 12000 | 6384 | $18.30 \pm 22.70$ | Smith et al. (2004) |
| HD90254 | WBVR | W | $9.22 \pm 0.05$ | Kornilov et al. (1991) |
| HD90254 | Johnson | U | $9.15 \pm 0.05$ | Mermilliod (1986) |
| HD90254 | Johnson | U | $9.18 \pm 0.05$ | Cousins (1963b) |
| HD90254 | Johnson | U | $9.19 \pm 0.05$ | Johnson et al. (1966) |
| HD90254 | Johnson | U | $9.20 \pm 0.05$ | Ducati (2002) |
| HD90254 | Oja | m41 | $8.78 \pm 0.05$ | Häggkvist & Oja (1970) |
| HD90254 | Oja | m42 | $8.59 \pm 0.05$ | Häggkvist & Oja (1970) |
| HD90254 | WBVR | B | $7.29 \pm 0.05$ | Kornilov et al. (1991) |
| HD90254 | Johnson | B | $7.20 \pm 0.05$ | Mermilliod (1986) |
| HD90254 | Johnson | B | $7.22 \pm 0.05$ | Cousins (1963b) |
| HD90254 | Johnson | B | $7.23 \pm 0.05$ | Johnson et al. (1966) |
| HD90254 | Johnson | B | $7.23 \pm 0.05$ | Ducati (2002) |
| HD90254 | Oja | m45 | $6.89 \pm 0.05$ | Häggkvist & Oja (1970) |
| HD90254 | WBVR | V | $5.61 \pm 0.05$ | Kornilov et al. (1991) |
| HD90254 | Johnson | V | $5.60 \pm 0.05$ | Cousins (1963b) |

**Table 21** *(continued)*

| Star ID | System/Wvlen | Band/Bandpass | Value | Reference |
|---------|--------------|---------------|-------|-----------|
| HD90254 | Johnson | V | $5.60 \pm 0.05$ | Mermilliod (1986) |
| HD90254 | Johnson | V | $5.61 \pm 0.05$ | Johnson et al. (1966) |
| HD90254 | Johnson | V | $5.61 \pm 0.05$ | Ducati (2002) |
| HD90254 | Johnson | V | $5.69 \pm 0.06$ | This work |
| HD90254 | WBVR | R | $4.22 \pm 0.05$ | Kornilov et al. (1991) |
| HD90254 | 1250 | 310 | $183.90 \pm 5.90$ | Smith et al. (2004) |
| HD90254 | Johnson | J | $2.34 \pm 0.05$ | McWilliam & Lambert (1984) |
| HD90254 | Johnson | J | $2.34 \pm 0.05$ | Ducati (2002) |
| HD90254 | 2200 | 361 | $178.50 \pm 6.30$ | Smith et al. (2004) |
| HD90254 | Johnson | K | $1.25 \pm 0.05$ | Ducati (2002) |
| HD90254 | Johnson | K | $1.35 \pm 0.05$ | Neugebauer & Leighton (1969) |
| HD90254 | 3500 | 898 | $89.60 \pm 6.00$ | Smith et al. (2004) |
| HD90254 | 4900 | 712 | $39.00 \pm 4.70$ | Smith et al. (2004) |
| HD90254 | 12000 | 6384 | $10.50 \pm 31.10$ | Smith et al. (2004) |
| HD92620 | WBVR | W | $9.48 \pm 0.05$ | Kornilov et al. (1991) |
| HD92620 | Johnson | U | $9.38 \pm 0.05$ | Mermilliod (1986) |
| HD92620 | Johnson | U | $9.47 \pm 0.05$ | Ducati (2002) |
| HD92620 | Johnson | U | $9.49 \pm 0.05$ | Smak (1964) |
| HD92620 | Oja | m41 | $9.01 \pm 0.05$ | Häggkvist & Oja (1970) |
| HD92620 | Oja | m42 | $8.91 \pm 0.05$ | Häggkvist & Oja (1970) |
| HD92620 | WBVR | B | $7.69 \pm 0.05$ | Kornilov et al. (1991) |
| HD92620 | Johnson | B | $7.59 \pm 0.05$ | Mermilliod (1986) |
| HD92620 | Johnson | B | $7.63 \pm 0.05$ | Häggkvist & Oja (1969b) |
| HD92620 | Johnson | B | $7.64 \pm 0.05$ | Ducati (2002) |
| HD92620 | Johnson | B | $7.66 \pm 0.05$ | Smak (1964) |
| HD92620 | Oja | m45 | $7.39 \pm 0.05$ | Häggkvist & Oja (1970) |
| HD92620 | WBVR | V | $6.05 \pm 0.05$ | Kornilov et al. (1991) |
| HD92620 | Johnson | V | $5.98 \pm 0.05$ | Mermilliod (1986) |
| HD92620 | Johnson | V | $6.00 \pm 0.05$ | Häggkvist & Oja (1969b) |
| HD92620 | Johnson | V | $6.02 \pm 0.05$ | Ducati (2002) |
| HD92620 | Johnson | V | $6.04 \pm 0.05$ | Smak (1964) |
| HD92620 | Johnson | V | $6.08 \pm 0.06$ | This work |
| HD92620 | Alexander | m608 | $3.91 \pm 0.05$ | Alexander et al. (1983) |
| HD92620 | Alexander | m683 | $4.37 \pm 0.05$ | Alexander et al. (1983) |
| HD92620 | WBVR | R | $4.39 \pm 0.05$ | Kornilov et al. (1991) |
| HD92620 | Alexander | m710 | $4.55 \pm 0.05$ | Alexander et al. (1983) |
| HD92620 | Alexander | m746 | $3.99 \pm 0.05$ | Alexander et al. (1983) |
| HD92620 | 1250 | 310 | $226.80 \pm 9.60$ | Smith et al. (2004) |
| HD92620 | Johnson | J | $2.10 \pm 0.05$ | McWilliam & Lambert (1984) |
| HD92620 | Johnson | J | $2.10 \pm 0.05$ | Ducati (2002) |
| HD92620 | Johnson | J | $2.14 \pm 0.05$ | Kerschbaum & Hron (1994) |

**Table 21** *(continued)*

| Star ID | System/Wvlen | Band/Bandpass | Value | Reference |
|---------|--------------|---------------|-------|-----------|
| HD92620 | Johnson | H | $1.34 \pm 0.05$ | Kerschbaum & Hron (1994) |
| HD92620 | 2200 | 361 | $234.60 \pm 6.60$ | Smith et al. (2004) |
| HD92620 | Johnson | K | $1.01 \pm 0.05$ | Ducati (2002) |
| HD92620 | Johnson | K | $1.02 \pm 0.04$ | Neugebauer & Leighton (1969) |
| HD92620 | 3500 | 898 | $121.10 \pm 5.40$ | Smith et al. (2004) |
| HD92620 | 4900 | 712 | $51.40 \pm 5.40$ | Smith et al. (2004) |
| HD92620 | 12000 | 6384 | $15.70 \pm 20.40$ | Smith et al. (2004) |
| HD93287 | KronComet | COp | $10.42 \pm 0.02$ | This work |
| HD93287 | Johnson | B | $9.73 \pm 0.01$ | This work |
| HD93287 | KronComet | Bc | $9.67 \pm 0.02$ | This work |
| HD93287 | KronComet | C2 | $8.43 \pm 0.01$ | This work |
| HD93287 | KronComet | Gc | $8.28 \pm 0.01$ | This work |
| HD93287 | Johnson | V | $8.25 \pm 0.01$ | This work |
| HD93287 | KronComet | Rc | $6.73 \pm 0.01$ | This work |
| HD93287 | 1250 | 310 | $33.90 \pm 4.10$ | Smith et al. (2004) |
| HD93287 | 2200 | 361 | $37.30 \pm 5.00$ | Smith et al. (2004) |
| HD93287 | Johnson | K | $2.99 \pm 0.07$ | Neugebauer & Leighton (1969) |
| HD93287 | 3500 | 898 | $18.60 \pm 8.00$ | Smith et al. (2004) |
| HD93287 | 4900 | 712 | $8.00 \pm 4.60$ | Smith et al. (2004) |
| HD93287 | 12000 | 6384 | $0.00 \pm 17.80$ | Smith et al. (2004) |
| HD94252 | KronComet | COp | $9.64 \pm 0.01$ | This work |
| HD94252 | Johnson | B | $8.90 \pm 0.01$ | This work |
| HD94252 | KronComet | Bc | $8.82 \pm 0.01$ | This work |
| HD94252 | KronComet | C2 | $7.77 \pm 0.01$ | This work |
| HD94252 | KronComet | Gc | $7.55 \pm 0.01$ | This work |
| HD94252 | Johnson | V | $7.49 \pm 0.01$ | This work |
| HD94252 | KronComet | Rc | $6.05 \pm 0.01$ | This work |
| HD94252 | 1250 | 310 | $42.40 \pm 5.80$ | Smith et al. (2004) |
| HD94252 | 2200 | 361 | $41.90 \pm 5.70$ | Smith et al. (2004) |
| HD94252 | Johnson | K | $2.85 \pm 0.09$ | Neugebauer & Leighton (1969) |
| HD94252 | 3500 | 898 | $21.60 \pm 4.30$ | Smith et al. (2004) |
| HD94252 | 4900 | 712 | $9.20 \pm 5.10$ | Smith et al. (2004) |
| HD94252 | 12000 | 6384 | $-0.20 \pm 22.80$ | Smith et al. (2004) |
| HD94264 | 13c | m33 | $5.80 \pm 0.05$ | Johnson & Mitchell (1995) |
| HD94264 | Geneva | U | $6.31 \pm 0.08$ | Golay (1972) |
| HD94264 | Vilnius | U | $7.59 \pm 0.05$ | Zdanavicius et al. (1969) |
| HD94264 | Vilnius | U | $7.64 \pm 0.05$ | Kazlauskas et al. (2005) |
| HD94264 | 13c | m35 | $5.60 \pm 0.05$ | Johnson & Mitchell (1995) |
| HD94264 | DDO | m35 | $7.08 \pm 0.05$ | McClure & Forrester (1981) |
| HD94264 | WBVR | W | $5.64 \pm 0.05$ | Kornilov et al. (1991) |
| HD94264 | Johnson | U | $5.73 \pm 0.05$ | Argue (1966) |

**Table 21** (continued)

| Star ID | System/Wvlen | Band/Bandpass | Value | Reference |
|---------|--------------|---------------|-------|-----------|
| HD94264 | Johnson | U | $5.76 \pm 0.05$ | Argue (1963) |
| HD94264 | Johnson | U | $5.78 \pm 0.05$ | Johnson & Morgan (1953b) |
| HD94264 | Johnson | U | $5.78 \pm 0.05$ | Mermilliod (1986) |
| HD94264 | Johnson | U | $5.79 \pm 0.05$ | Johnson et al. (1966) |
| HD94264 | Johnson | U | $5.79 \pm 0.05$ | Jennens & Helfer (1975) |
| HD94264 | Johnson | U | $5.79 \pm 0.05$ | Ducati (2002) |
| HD94264 | 13c | m37 | $5.72 \pm 0.05$ | Johnson & Mitchell (1995) |
| HD94264 | Vilnius | P | $7.01 \pm 0.05$ | Zdanavicius et al. (1969) |
| HD94264 | Vilnius | P | $7.05 \pm 0.05$ | Kazlauskas et al. (2005) |
| HD94264 | DDO | m38 | $6.02 \pm 0.05$ | McClure & Forrester (1981) |
| HD94264 | 13c | m40 | $5.46 \pm 0.05$ | Johnson & Mitchell (1995) |
| HD94264 | Geneva | B1 | $5.53 \pm 0.08$ | Golay (1972) |
| HD94264 | Vilnius | X | $5.98 \pm 0.05$ | Zdanavicius et al. (1969) |
| HD94264 | Vilnius | X | $6.01 \pm 0.05$ | Kazlauskas et al. (2005) |
| HD94264 | DDO | m41 | $6.48 \pm 0.05$ | McClure & Forrester (1981) |
| HD94264 | Oja | m41 | $6.06 \pm 0.05$ | Häggkvist & Oja (1970) |
| HD94264 | DDO | m42 | $6.29 \pm 0.05$ | McClure & Forrester (1981) |
| HD94264 | Oja | m42 | $5.87 \pm 0.05$ | Häggkvist & Oja (1970) |
| HD94264 | Geneva | B | $4.20 \pm 0.08$ | Golay (1972) |
| HD94264 | WBVR | B | $4.88 \pm 0.05$ | Kornilov et al. (1991) |
| HD94264 | Johnson | B | $4.83 \pm 0.05$ | Miczaika (1954) |
| HD94264 | Johnson | B | $4.83 \pm 0.05$ | Argue (1966) |
| HD94264 | Johnson | B | $4.83 \pm 0.05$ | Mermilliod (1986) |
| HD94264 | Johnson | B | $4.84 \pm 0.05$ | Häggkvist & Oja (1966) |
| HD94264 | Johnson | B | $4.87 \pm 0.05$ | Johnson & Morgan (1953b) |
| HD94264 | Johnson | B | $4.87 \pm 0.05$ | Argue (1963) |
| HD94264 | Johnson | B | $4.87 \pm 0.05$ | Johnson et al. (1966) |
| HD94264 | Johnson | B | $4.87 \pm 0.05$ | Jennens & Helfer (1975) |
| HD94264 | Johnson | B | $4.87 \pm 0.05$ | Ducati (2002) |
| HD94264 | Geneva | B2 | $5.34 \pm 0.08$ | Golay (1972) |
| HD94264 | 13c | m45 | $4.54 \pm 0.05$ | Johnson & Mitchell (1995) |
| HD94264 | DDO | m45 | $5.37 \pm 0.05$ | McClure & Forrester (1981) |
| HD94264 | Oja | m45 | $4.59 \pm 0.05$ | Häggkvist & Oja (1970) |
| HD94264 | Vilnius | Y | $4.63 \pm 0.05$ | Zdanavicius et al. (1969) |
| HD94264 | Vilnius | Y | $4.67 \pm 0.05$ | Kazlauskas et al. (2005) |
| HD94264 | DDO | m48 | $4.19 \pm 0.05$ | McClure & Forrester (1981) |
| HD94264 | Vilnius | Z | $4.18 \pm 0.05$ | Zdanavicius et al. (1969) |
| HD94264 | Vilnius | Z | $4.20 \pm 0.05$ | Kazlauskas et al. (2005) |
| HD94264 | 13c | m52 | $4.06 \pm 0.05$ | Johnson & Mitchell (1995) |
| HD94264 | Geneva | V1 | $4.62 \pm 0.08$ | Golay (1972) |
| HD94264 | WBVR | V | $3.80 \pm 0.05$ | Kornilov et al. (1991) |

**Table 21** *(continued)*

| Star ID | System/Wvlen | Band/Bandpass | Value | Reference |
|---------|--------------|---------------|-------|-----------|
| HD94264 | Vilnius | V | $3.83 \pm 0.05$ | Zdanavicius et al. (1969) |
| HD94264 | Vilnius | V | $3.86 \pm 0.05$ | Kazlauskas et al. (2005) |
| HD94264 | Geneva | V | $3.83 \pm 0.08$ | Golay (1972) |
| HD94264 | Johnson | V | $3.79 \pm 0.05$ | Miczaika (1954) |
| HD94264 | Johnson | V | $3.79 \pm 0.05$ | Argue (1963) |
| HD94264 | Johnson | V | $3.79 \pm 0.05$ | Häggkvist & Oja (1966) |
| HD94264 | Johnson | V | $3.79 \pm 0.05$ | Argue (1966) |
| HD94264 | Johnson | V | $3.80 \pm 0.05$ | Mermilliod (1986) |
| HD94264 | Johnson | V | $3.83 \pm 0.05$ | Johnson et al. (1966) |
| HD94264 | Johnson | V | $3.83 \pm 0.05$ | Jennens & Helfer (1975) |
| HD94264 | Johnson | V | $3.83 \pm 0.05$ | Ducati (2002) |
| HD94264 | Johnson | V | $3.84 \pm 0.05$ | Johnson & Morgan (1953b) |
| HD94264 | 13c | m58 | $3.57 \pm 0.05$ | Johnson & Mitchell (1995) |
| HD94264 | Geneva | G | $4.79 \pm 0.08$ | Golay (1972) |
| HD94264 | 13c | m63 | $3.26 \pm 0.05$ | Johnson & Mitchell (1995) |
| HD94264 | Vilnius | S | $3.09 \pm 0.05$ | Zdanavicius et al. (1969) |
| HD94264 | Vilnius | S | $3.09 \pm 0.05$ | Kazlauskas et al. (2005) |
| HD94264 | WBVR | R | $3.03 \pm 0.05$ | Kornilov et al. (1991) |
| HD94264 | 13c | m72 | $3.01 \pm 0.05$ | Johnson & Mitchell (1995) |
| HD94264 | 13c | m80 | $2.76 \pm 0.05$ | Johnson & Mitchell (1995) |
| HD94264 | 13c | m86 | $2.66 \pm 0.05$ | Johnson & Mitchell (1995) |
| HD94264 | 13c | m99 | $2.49 \pm 0.05$ | Johnson & Mitchell (1995) |
| HD94264 | 13c | m110 | $2.34 \pm 0.05$ | Johnson & Mitchell (1995) |
| HD94264 | 1250 | 310 | $255.50 \pm 9.10$ | Smith et al. (2004) |
| HD94264 | Johnson | J | $1.98 \pm 0.05$ | Alonso et al. (1998) |
| HD94264 | Johnson | J | $2.07 \pm 0.05$ | Johnson et al. (1966) |
| HD94264 | Johnson | J | $2.07 \pm 0.05$ | Ducati (2002) |
| HD94264 | Johnson | H | $1.46 \pm 0.05$ | Alonso et al. (1998) |
| HD94264 | 2200 | 361 | $183.80 \pm 12.30$ | Smith et al. (2004) |
| HD94264 | Johnson | K | $1.37 \pm 0.04$ | Neugebauer & Leighton (1969) |
| HD94264 | Johnson | K | $1.39 \pm 0.05$ | Johnson et al. (1966) |
| HD94264 | Johnson | K | $1.39 \pm 0.05$ | Ducati (2002) |
| HD94264 | 3500 | 898 | $89.20 \pm 8.80$ | Smith et al. (2004) |
| HD94264 | 4900 | 712 | $42.70 \pm 5.80$ | Smith et al. (2004) |
| HD94264 | 12000 | 6384 | $7.80 \pm 17.20$ | Smith et al. (2004) |
| HD94336 | KronComet | COp | $9.28 \pm 0.01$ | This work |
| HD94336 | WBVR | B | $8.68 \pm 0.05$ | Kornilov et al. (1991) |
| HD94336 | Johnson | B | $8.38 \pm 0.05$ | Fernie (1983) |
| HD94336 | Johnson | B | $8.60 \pm 0.01$ | This work |
| HD94336 | Johnson | B | $8.63 \pm 0.01$ | Oja (1991) |
| HD94336 | KronComet | Bc | $8.49 \pm 0.01$ | This work |

**Table 21** *(continued)*

| Star ID | System/Wvlen | Band/Bandpass | Value | Reference |
|---------|--------------|---------------|-------|-----------|
| HD94336 | KronComet | C2 | $7.41 \pm 0.01$ | This work |
| HD94336 | KronComet | Gc | $7.23 \pm 0.01$ | This work |
| HD94336 | WBVR | V | $7.06 \pm 0.05$ | Kornilov et al. (1991) |
| HD94336 | Johnson | V | $6.91 \pm 0.05$ | Fernie (1983) |
| HD94336 | Johnson | V | $7.06 \pm 0.01$ | Oja (1991) |
| HD94336 | Johnson | V | $7.19 \pm 0.01$ | This work |
| HD94336 | WBVR | R | $5.52 \pm 0.05$ | Kornilov et al. (1991) |
| HD94336 | KronComet | Rc | $5.73 \pm 0.01$ | This work |
| HD94336 | 1250 | 310 | $70.10 \pm 4.20$ | Smith et al. (2004) |
| HD94336 | 2200 | 361 | $68.30 \pm 5.70$ | Smith et al. (2004) |
| HD94336 | Johnson | K | $2.35 \pm 0.06$ | Neugebauer & Leighton (1969) |
| HD94336 | 3500 | 898 | $33.80 \pm 4.90$ | Smith et al. (2004) |
| HD94336 | 4900 | 712 | $14.80 \pm 4.80$ | Smith et al. (2004) |
| HD94336 | 12000 | 6384 | $4.90 \pm 18.50$ | Smith et al. (2004) |
| HD94600 | DDO | m35 | $8.46 \pm 0.05$ | McClure & Forrester (1981) |
| HD94600 | WBVR | W | $7.00 \pm 0.05$ | Kornilov et al. (1991) |
| HD94600 | Johnson | U | $7.14 \pm 0.05$ | Mermilliod (1986) |
| HD94600 | Johnson | U | $7.15 \pm 0.05$ | Argue (1963) |
| HD94600 | DDO | m38 | $7.37 \pm 0.05$ | McClure & Forrester (1981) |
| HD94600 | DDO | m41 | $7.81 \pm 0.05$ | McClure & Forrester (1981) |
| HD94600 | Oja | m41 | $7.37 \pm 0.05$ | Häggkvist & Oja (1970) |
| HD94600 | DDO | m42 | $7.59 \pm 0.05$ | McClure & Forrester (1981) |
| HD94600 | Oja | m42 | $7.15 \pm 0.05$ | Häggkvist & Oja (1970) |
| HD94600 | WBVR | B | $6.16 \pm 0.05$ | Kornilov et al. (1991) |
| HD94600 | Johnson | B | $6.10 \pm 0.05$ | Häggkvist & Oja (1966) |
| HD94600 | Johnson | B | $6.12 \pm 0.05$ | Mermilliod (1986) |
| HD94600 | Johnson | B | $6.15 \pm 0.05$ | Argue (1963) |
| HD94600 | DDO | m45 | $6.66 \pm 0.05$ | McClure & Forrester (1981) |
| HD94600 | Oja | m45 | $5.84 \pm 0.05$ | Häggkvist & Oja (1970) |
| HD94600 | DDO | m48 | $5.44 \pm 0.05$ | McClure & Forrester (1981) |
| HD94600 | WBVR | V | $5.03 \pm 0.05$ | Kornilov et al. (1991) |
| HD94600 | Johnson | V | $5.01 \pm 0.05$ | Häggkvist & Oja (1966) |
| HD94600 | Johnson | V | $5.02 \pm 0.05$ | Mermilliod (1986) |
| HD94600 | Johnson | V | $5.04 \pm 0.05$ | Argue (1963) |
| HD94600 | WBVR | R | $4.23 \pm 0.05$ | Kornilov et al. (1991) |
| HD94600 | 1250 | 310 | $104.20 \pm 13.70$ | Smith et al. (2004) |
| HD94600 | 2200 | 361 | $80.50 \pm 12.40$ | Smith et al. (2004) |
| HD94600 | Johnson | K | $2.45 \pm 0.06$ | Neugebauer & Leighton (1969) |
| HD94600 | 3500 | 898 | $37.40 \pm 7.00$ | Smith et al. (2004) |
| HD94600 | 4900 | 712 | $18.90 \pm 5.30$ | Smith et al. (2004) |
| HD94600 | 12000 | 6384 | $3.10 \pm 18.10$ | Smith et al. (2004) |

**Table 21** *(continued)*

| Star ID | System/Wvlen | Band/Bandpass | Value | Reference |
|---------|--------------|---------------|-------|-----------|
| HD95212 | WBVR | B | $6.99 \pm 0.05$ | Kornilov et al. (1991) |
| HD95212 | Johnson | B | $6.94 \pm 0.05$ | Häggkvist & Oja (1969b) |
| HD95212 | DDO | m45 | $7.42 \pm 0.05$ | McClure & Forrester (1981) |
| HD95212 | Oja | m45 | $6.60 \pm 0.05$ | Häggkvist & Oja (1970) |
| HD95212 | Vilnius | Y | $6.53 \pm 0.05$ | Zdanavicius et al. (1972) |
| HD95212 | DDO | m48 | $6.05 \pm 0.05$ | McClure & Forrester (1981) |
| HD95212 | Vilnius | Z | $5.97 \pm 0.05$ | Zdanavicius et al. (1972) |
| HD95212 | WBVR | V | $5.46 \pm 0.05$ | Kornilov et al. (1991) |
| HD95212 | Vilnius | V | $5.49 \pm 0.05$ | Zdanavicius et al. (1972) |
| HD95212 | Johnson | V | $5.47 \pm 0.05$ | Häggkvist & Oja (1969b) |
| HD95212 | Vilnius | S | $4.48 \pm 0.05$ | Zdanavicius et al. (1972) |
| HD95212 | WBVR | R | $4.36 \pm 0.05$ | Kornilov et al. (1991) |
| HD95212 | 1250 | 310 | $120.30 \pm 9.40$ | Smith et al. (2004) |
| HD95212 | 2200 | 361 | $103.10 \pm 6.90$ | Smith et al. (2004) |
| HD95212 | Johnson | K | $2.00 \pm 0.06$ | Neugebauer & Leighton (1969) |
| HD95212 | 3500 | 898 | $50.30 \pm 4.90$ | Smith et al. (2004) |
| HD95212 | 4900 | 712 | $22.50 \pm 4.70$ | Smith et al. (2004) |
| HD95212 | 12000 | 6384 | $6.20 \pm 16.70$ | Smith et al. (2004) |
| HD95345 | Geneva | U | $7.70 \pm 0.08$ | Golay (1972) |
| HD95345 | Vilnius | U | $8.90 \pm 0.05$ | Sleivyte (1987) |
| HD95345 | Vilnius | U | $8.90 \pm 0.05$ | Forbes et al. (1993) |
| HD95345 | DDO | m35 | $8.45 \pm 0.05$ | McClure & Forrester (1981) |
| HD95345 | DDO | m35 | $8.45 \pm 0.05$ | Dean (1981) |
| HD95345 | DDO | m35 | $8.46 \pm 0.05$ | Cousins & Caldwell (1996) |
| HD95345 | Stromgren | u | $8.42 \pm 0.08$ | Pilachowski (1978) |
| HD95345 | Stromgren | u | $8.42 \pm 0.08$ | Hauck & Mermilliod (1998) |
| HD95345 | WBVR | W | $6.99 \pm 0.05$ | Kornilov et al. (1991) |
| HD95345 | Johnson | U | $7.10 \pm 0.05$ | Gutierrez-Moreno & et al. (1966) |
| HD95345 | Johnson | U | $7.11 \pm 0.05$ | Cousins (1963a) |
| HD95345 | Johnson | U | $7.11 \pm 0.05$ | Cousins (1984) |
| HD95345 | Johnson | U | $7.12 \pm 0.05$ | Johnson et al. (1966) |
| HD95345 | Johnson | U | $7.13 \pm 0.05$ | Argue (1963) |
| HD95345 | Johnson | U | $7.16 \pm 0.05$ | Mermilliod (1986) |
| HD95345 | Vilnius | P | $8.26 \pm 0.05$ | Forbes et al. (1993) |
| HD95345 | Vilnius | P | $8.28 \pm 0.05$ | Sleivyte (1987) |
| HD95345 | DDO | m38 | $7.34 \pm 0.05$ | McClure & Forrester (1981) |
| HD95345 | DDO | m38 | $7.34 \pm 0.05$ | Dean (1981) |
| HD95345 | DDO | m38 | $7.34 \pm 0.05$ | Mermilliod & Nitschelm (1989) |
| HD95345 | DDO | m38 | $7.35 \pm 0.05$ | Cousins & Caldwell (1996) |
| HD95345 | Geneva | B1 | $6.76 \pm 0.08$ | Golay (1972) |
| HD95345 | Vilnius | X | $7.18 \pm 0.05$ | Sleivyte (1987) |

**Table 21** (continued)

| Star ID | System/Wvlen | Band/Bandpass | Value | Reference |
|---------|--------------|---------------|-------|-----------|
| HD95345 | Vilnius | X | $7.18 \pm 0.05$ | Forbes et al. (1993) |
| HD95345 | DDO | m41 | $7.73 \pm 0.05$ | McClure & Forrester (1981) |
| HD95345 | DDO | m41 | $7.73 \pm 0.05$ | Dean (1981) |
| HD95345 | DDO | m41 | $7.73 \pm 0.05$ | Mermilliod & Nitschelm (1989) |
| HD95345 | DDO | m41 | $7.73 \pm 0.05$ | Cousins & Caldwell (1996) |
| HD95345 | Oja | m41 | $7.25 \pm 0.05$ | Häggkvist & Oja (1970) |
| HD95345 | Stromgren | v | $6.78 \pm 0.08$ | Pilachowski (1978) |
| HD95345 | Stromgren | v | $6.78 \pm 0.08$ | Hauck & Mermilliod (1998) |
| HD95345 | DDO | m42 | $7.49 \pm 0.05$ | Cousins & Caldwell (1996) |
| HD95345 | DDO | m42 | $7.50 \pm 0.05$ | McClure & Forrester (1981) |
| HD95345 | DDO | m42 | $7.50 \pm 0.05$ | Dean (1981) |
| HD95345 | DDO | m42 | $7.50 \pm 0.05$ | Mermilliod & Nitschelm (1989) |
| HD95345 | Oja | m42 | $7.03 \pm 0.05$ | Häggkvist & Oja (1970) |
| HD95345 | Geneva | B | $5.37 \pm 0.08$ | Golay (1972) |
| HD95345 | WBVR | B | $6.03 \pm 0.05$ | Kornilov et al. (1991) |
| HD95345 | Johnson | B | $5.99 \pm 0.05$ | Cousins (1963a) |
| HD95345 | Johnson | B | $5.99 \pm 0.05$ | Gehrels et al. (1964) |
| HD95345 | Johnson | B | $5.99 \pm 0.05$ | Gutierrez-Moreno & et al. (1966) |
| HD95345 | Johnson | B | $6.00 \pm 0.05$ | Johnson et al. (1966) |
| HD95345 | Johnson | B | $6.01 \pm 0.05$ | Argue (1963) |
| HD95345 | Johnson | B | $6.01 \pm 0.05$ | Cousins (1984) |
| HD95345 | Johnson | B | $6.02 \pm 0.05$ | Häggkvist & Oja (1966) |
| HD95345 | Johnson | B | $6.04 \pm 0.05$ | Mermilliod (1986) |
| HD95345 | Geneva | B2 | $6.47 \pm 0.08$ | Golay (1972) |
| HD95345 | DDO | m45 | $6.53 \pm 0.05$ | McClure & Forrester (1981) |
| HD95345 | DDO | m45 | $6.53 \pm 0.05$ | Dean (1981) |
| HD95345 | DDO | m45 | $6.53 \pm 0.05$ | Mermilliod & Nitschelm (1989) |
| HD95345 | DDO | m45 | $6.53 \pm 0.05$ | Cousins & Caldwell (1996) |
| HD95345 | Oja | m45 | $5.66 \pm 0.05$ | Häggkvist & Oja (1970) |
| HD95345 | Vilnius | Y | $5.70 \pm 0.05$ | Sleivyte (1987) |
| HD95345 | Vilnius | Y | $5.70 \pm 0.05$ | Forbes et al. (1993) |
| HD95345 | Stromgren | b | $5.56 \pm 0.08$ | Pilachowski (1978) |
| HD95345 | Stromgren | b | $5.56 \pm 0.08$ | Hauck & Mermilliod (1998) |
| HD95345 | DDO | m48 | $5.28 \pm 0.05$ | Cousins & Caldwell (1996) |
| HD95345 | DDO | m48 | $5.29 \pm 0.05$ | McClure & Forrester (1981) |
| HD95345 | DDO | m48 | $5.29 \pm 0.05$ | Dean (1981) |
| HD95345 | DDO | m48 | $5.29 \pm 0.05$ | Mermilliod & Nitschelm (1989) |
| HD95345 | Vilnius | Z | $5.19 \pm 0.05$ | Forbes et al. (1993) |
| HD95345 | Vilnius | Z | $5.20 \pm 0.05$ | Sleivyte (1987) |
| HD95345 | Geneva | V1 | $5.66 \pm 0.08$ | Golay (1972) |
| HD95345 | WBVR | V | $4.85 \pm 0.05$ | Kornilov et al. (1991) |

**Table 21** *(continued)*

| Star ID | System/Wvlen | Band/Bandpass | Value | Reference |
|---------|--------------|---------------|-------|-----------|
| HD95345 | Vilnius | V | $4.84 \pm 0.05$ | Forbes et al. (1993) |
| HD95345 | Vilnius | V | $4.85 \pm 0.05$ | Sleivyte (1987) |
| HD95345 | Stromgren | y | $4.85 \pm 0.08$ | Pilachowski (1978) |
| HD95345 | Stromgren | y | $4.85 \pm 0.08$ | Hauck & Mermilliod (1998) |
| HD95345 | Geneva | V | $4.88 \pm 0.08$ | Golay (1972) |
| HD95345 | Johnson | V | $4.83 \pm 0.05$ | Cousins (1963a) |
| HD95345 | Johnson | V | $4.83 \pm 0.05$ | Gutierrez-Moreno & et al. (1966) |
| HD95345 | Johnson | V | $4.84 \pm 0.05$ | Argue (1963) |
| HD95345 | Johnson | V | $4.84 \pm 0.05$ | Häggkvist & Oja (1966) |
| HD95345 | Johnson | V | $4.84 \pm 0.05$ | Johnson et al. (1966) |
| HD95345 | Johnson | V | $4.85 \pm 0.05$ | Gehrels et al. (1964) |
| HD95345 | Johnson | V | $4.85 \pm 0.05$ | Cousins (1984) |
| HD95345 | Johnson | V | $4.87 \pm 0.05$ | Mermilliod (1986) |
| HD95345 | Geneva | G | $5.82 \pm 0.08$ | Golay (1972) |
| HD95345 | Vilnius | S | $4.04 \pm 0.05$ | Sleivyte (1987) |
| HD95345 | Vilnius | S | $4.04 \pm 0.05$ | Forbes et al. (1993) |
| HD95345 | WBVR | R | $4.01 \pm 0.05$ | Kornilov et al. (1991) |
| HD95345 | 1250 | 310 | $111.90 \pm 4.80$ | Smith et al. (2004) |
| HD95345 | Johnson | J | $2.85 \pm 0.05$ | Alonso et al. (1998) |
| HD95345 | Johnson | H | $2.29 \pm 0.05$ | Alonso et al. (1998) |
| HD95345 | 2200 | 361 | $82.50 \pm 5.30$ | Smith et al. (2004) |
| HD95345 | Johnson | K | $2.24 \pm 0.06$ | Neugebauer & Leighton (1969) |
| HD95345 | 3500 | 898 | $39.20 \pm 4.30$ | Smith et al. (2004) |
| HD95345 | 4900 | 712 | $19.30 \pm 4.80$ | Smith et al. (2004) |
| HD95345 | 12000 | 6384 | $-10.00 \pm 24.50$ | Smith et al. (2004) |
| HD96274 | KronComet | NH | $12.11 \pm 0.19$ | This work |
| HD96274 | KronComet | UVc | $11.73 \pm 0.14$ | This work |
| HD96274 | Johnson | U | $10.23 \pm 0.02$ | This work |
| HD96274 | KronComet | CN | $10.86 \pm 0.05$ | This work |
| HD96274 | KronComet | COp | $9.29 \pm 0.02$ | This work |
| HD96274 | Johnson | B | $8.52 \pm 0.01$ | This work |
| HD96274 | KronComet | Bc | $8.43 \pm 0.02$ | This work |
| HD96274 | KronComet | C2 | $7.40 \pm 0.01$ | This work |
| HD96274 | KronComet | Gc | $7.17 \pm 0.01$ | This work |
| HD96274 | Johnson | V | $7.10 \pm 0.01$ | This work |
| HD96274 | KronComet | Rc | $5.65 \pm 0.01$ | This work |
| HD96274 | 1250 | 310 | $61.70 \pm 5.60$ | Smith et al. (2004) |
| HD96274 | 2200 | 361 | $59.60 \pm 5.40$ | Smith et al. (2004) |
| HD96274 | Johnson | K | $2.33 \pm 0.09$ | Neugebauer & Leighton (1969) |
| HD96274 | 3500 | 898 | $29.80 \pm 5.00$ | Smith et al. (2004) |
| HD96274 | 4900 | 712 | $12.40 \pm 5.30$ | Smith et al. (2004) |

Table 21 *(continued)*

| Star ID | System/Wvlen | Band/Bandpass | Value | Reference |
|---------|--------------|---------------|-------|-----------|
| HD96274 | 12000 | 6384 | $2.60 \pm 24.20$ | Smith et al. (2004) |
| HD96833 | 13c | m33 | $5.38 \pm 0.05$ | Johnson & Mitchell (1995) |
| HD96833 | Geneva | U | $5.83 \pm 0.08$ | Golay (1972) |
| HD96833 | Vilnius | U | $7.05 \pm 0.05$ | Sudzius et al. (1970) |
| HD96833 | 13c | m35 | $5.15 \pm 0.05$ | Johnson & Mitchell (1995) |
| HD96833 | DDO | m35 | $6.61 \pm 0.05$ | McClure & Forrester (1981) |
| HD96833 | Stromgren | u | $6.56 \pm 0.08$ | Crawford & Barnes (1970) |
| HD96833 | Stromgren | u | $6.56 \pm 0.08$ | Hauck & Mermilliod (1998) |
| HD96833 | WBVR | W | $5.16 \pm 0.05$ | Kornilov et al. (1991) |
| HD96833 | Johnson | U | $5.24 \pm 0.05$ | Johnson & Morgan (1953b) |
| HD96833 | Johnson | U | $5.25 \pm 0.05$ | Argue (1966) |
| HD96833 | Johnson | U | $5.26 \pm 0.05$ | Argue (1963) |
| HD96833 | Johnson | U | $5.27 \pm 0.05$ | Johnson (1964) |
| HD96833 | Johnson | U | $5.27 \pm 0.05$ | Johnson et al. (1966) |
| HD96833 | Johnson | U | $5.27 \pm 0.05$ | Jennens & Helfer (1975) |
| HD96833 | Johnson | U | $5.27 \pm 0.05$ | Ducati (2002) |
| HD96833 | 13c | m37 | $5.23 \pm 0.05$ | Johnson & Mitchell (1995) |
| HD96833 | Vilnius | P | $6.43 \pm 0.05$ | Sudzius et al. (1970) |
| HD96833 | DDO | m38 | $5.50 \pm 0.05$ | McClure & Forrester (1981) |
| HD96833 | 13c | m40 | $4.89 \pm 0.05$ | Johnson & Mitchell (1995) |
| HD96833 | Geneva | B1 | $4.92 \pm 0.08$ | Golay (1972) |
| HD96833 | Vilnius | X | $5.32 \pm 0.05$ | Sudzius et al. (1970) |
| HD96833 | DDO | m41 | $5.92 \pm 0.05$ | McClure & Forrester (1981) |
| HD96833 | Oja | m41 | $5.47 \pm 0.05$ | Häggkvist & Oja (1970) |
| HD96833 | Stromgren | v | $4.94 \pm 0.08$ | Crawford & Barnes (1970) |
| HD96833 | Stromgren | v | $4.94 \pm 0.08$ | Hauck & Mermilliod (1998) |
| HD96833 | DDO | m42 | $5.63 \pm 0.05$ | McClure & Forrester (1981) |
| HD96833 | Oja | m42 | $5.18 \pm 0.05$ | Häggkvist & Oja (1970) |
| HD96833 | Geneva | B | $3.52 \pm 0.08$ | Golay (1972) |
| HD96833 | WBVR | B | $4.19 \pm 0.05$ | Kornilov et al. (1991) |
| HD96833 | Johnson | B | $4.14 \pm 0.05$ | Johnson & Morgan (1953b) |
| HD96833 | Johnson | B | $4.14 \pm 0.05$ | Argue (1966) |
| HD96833 | Johnson | B | $4.15 \pm 0.05$ | Oja (1963) |
| HD96833 | Johnson | B | $4.15 \pm 0.05$ | Argue (1963) |
| HD96833 | Johnson | B | $4.15 \pm 0.05$ | Johnson (1964) |
| HD96833 | Johnson | B | $4.15 \pm 0.05$ | Johnson et al. (1966) |
| HD96833 | Johnson | B | $4.15 \pm 0.05$ | Jennens & Helfer (1975) |
| HD96833 | Johnson | B | $4.15 \pm 0.05$ | Ducati (2002) |
| HD96833 | Johnson | B | $4.16 \pm 0.05$ | Häggkvist & Oja (1966) |
| HD96833 | Geneva | B2 | $4.62 \pm 0.08$ | Golay (1972) |
| HD96833 | 13c | m45 | $3.82 \pm 0.05$ | Johnson & Mitchell (1995) |

**Table 21** (continued)

| Star ID | System/Wvlen | Band/Bandpass | Value | Reference |
|---|---|---|---|---|
| HD96833 | DDO | m45 | $4.67 \pm 0.05$ | McClure & Forrester (1981) |
| HD96833 | Oja | m45 | $3.85 \pm 0.05$ | Häggkvist & Oja (1970) |
| HD96833 | Vilnius | Y | $3.85 \pm 0.05$ | Sudzius et al. (1970) |
| HD96833 | Stromgren | b | $3.71 \pm 0.08$ | Crawford & Barnes (1970) |
| HD96833 | Stromgren | b | $3.71 \pm 0.08$ | Hauck & Mermilliod (1998) |
| HD96833 | DDO | m48 | $3.43 \pm 0.05$ | McClure & Forrester (1981) |
| HD96833 | Vilnius | Z | $3.37 \pm 0.05$ | Sudzius et al. (1970) |
| HD96833 | 13c | m52 | $3.31 \pm 0.05$ | Johnson & Mitchell (1995) |
| HD96833 | Geneva | V1 | $3.82 \pm 0.08$ | Golay (1972) |
| HD96833 | WBVR | V | $3.02 \pm 0.05$ | Kornilov et al. (1991) |
| HD96833 | Vilnius | V | $3.01 \pm 0.05$ | Sudzius et al. (1970) |
| HD96833 | Stromgren | y | $3.01 \pm 0.08$ | Crawford & Barnes (1970) |
| HD96833 | Stromgren | y | $3.01 \pm 0.08$ | Hauck & Mermilliod (1998) |
| HD96833 | Geneva | V | $3.02 \pm 0.08$ | Golay (1972) |
| HD96833 | Johnson | V | $2.98 \pm 0.05$ | Argue (1966) |
| HD96833 | Johnson | V | $3.00 \pm 0.05$ | Argue (1963) |
| HD96833 | Johnson | V | $3.01 \pm 0.05$ | Johnson & Morgan (1953b) |
| HD96833 | Johnson | V | $3.01 \pm 0.05$ | Oja (1963) |
| HD96833 | Johnson | V | $3.01 \pm 0.05$ | Johnson (1964) |
| HD96833 | Johnson | V | $3.01 \pm 0.05$ | Häggkvist & Oja (1966) |
| HD96833 | Johnson | V | $3.01 \pm 0.05$ | Johnson et al. (1966) |
| HD96833 | Johnson | V | $3.01 \pm 0.05$ | Jennens & Helfer (1975) |
| HD96833 | Johnson | V | $3.01 \pm 0.05$ | Ducati (2002) |
| HD96833 | 13c | m58 | $2.76 \pm 0.05$ | Johnson & Mitchell (1995) |
| HD96833 | Geneva | G | $3.98 \pm 0.08$ | Golay (1972) |
| HD96833 | 13c | m63 | $2.44 \pm 0.05$ | Johnson & Mitchell (1995) |
| HD96833 | Vilnius | S | $2.21 \pm 0.05$ | Sudzius et al. (1970) |
| HD96833 | WBVR | R | $2.20 \pm 0.05$ | Kornilov et al. (1991) |
| HD96833 | 13c | m72 | $2.23 \pm 0.05$ | Johnson & Mitchell (1995) |
| HD96833 | 13c | m80 | $1.92 \pm 0.05$ | Johnson & Mitchell (1995) |
| HD96833 | 13c | m86 | $1.81 \pm 0.05$ | Johnson & Mitchell (1995) |
| HD96833 | 13c | m99 | $1.62 \pm 0.05$ | Johnson & Mitchell (1995) |
| HD96833 | 13c | m110 | $1.44 \pm 0.05$ | Johnson & Mitchell (1995) |
| HD96833 | 1250 | 310 | $578.50 \pm 12.10$ | Smith et al. (2004) |
| HD96833 | Johnson | J | $1.16 \pm 0.05$ | Johnson et al. (1966) |
| HD96833 | Johnson | J | $1.16 \pm 0.05$ | Ducati (2002) |
| HD96833 | Johnson | J | $1.16 \pm 0.05$ | Shenavrin et al. (2011) |
| HD96833 | Johnson | H | $0.58 \pm 0.05$ | Shenavrin et al. (2011) |
| HD96833 | 2200 | 361 | $416.70 \pm 6.50$ | Smith et al. (2004) |
| HD96833 | Johnson | K | $0.39 \pm 0.04$ | Neugebauer & Leighton (1969) |
| HD96833 | Johnson | K | $0.43 \pm 0.05$ | Shenavrin et al. (2011) |

**Table 21** *(continued)*

| Star ID | System/Wvlen | Band/Bandpass | Value | Reference |
|---------|--------------|---------------|-------|-----------|
| HD96833 | Johnson | K | $0.44 \pm 0.05$ | Johnson et al. (1966) |
| HD96833 | Johnson | K | $0.44 \pm 0.05$ | Ducati (2002) |
| HD96833 | Johnson | L | $0.32 \pm 0.05$ | Johnson et al. (1966) |
| HD96833 | Johnson | L | $0.32 \pm 0.05$ | Ducati (2002) |
| HD96833 | 3500 | 898 | $200.50 \pm 9.70$ | Smith et al. (2004) |
| HD96833 | 4900 | 712 | $96.10 \pm 4.80$ | Smith et al. (2004) |
| HD96833 | 12000 | 6384 | $23.30 \pm 18.30$ | Smith et al. (2004) |
| HD100236 | KronComet | COp | $9.91 \pm 0.06$ | This work |
| HD100236 | Johnson | B | $9.39 \pm 0.05$ | This work |
| HD100236 | KronComet | Bc | $9.29 \pm 0.05$ | This work |
| HD100236 | KronComet | C2 | $8.14 \pm 0.04$ | This work |
| HD100236 | KronComet | Gc | $8.03 \pm 0.03$ | This work |
| HD100236 | Johnson | V | $8.04 \pm 0.04$ | This work |
| HD100236 | KronComet | Rc | $6.57 \pm 0.03$ | This work |
| HD100236 | 1250 | 310 | $47.70 \pm 4.10$ | Smith et al. (2004) |
| HD100236 | 2200 | 361 | $54.80 \pm 11.40$ | Smith et al. (2004) |
| HD100236 | Johnson | K | $2.70 \pm 0.09$ | Neugebauer & Leighton (1969) |
| HD100236 | 3500 | 898 | $28.90 \pm 7.60$ | Smith et al. (2004) |
| HD100236 | 4900 | 712 | $13.30 \pm 5.80$ | Smith et al. (2004) |
| HD100236 | 12000 | 6384 | $0.90 \pm 22.50$ | Smith et al. (2004) |
| HD102224 | 13c | m33 | $6.15 \pm 0.05$ | Johnson & Mitchell (1995) |
| HD102224 | Geneva | U | $6.59 \pm 0.08$ | Golay (1972) |
| HD102224 | Vilnius | U | $8.00 \pm 0.05$ | Kazlauskas et al. (2005) |
| HD102224 | 13c | m35 | $5.86 \pm 0.05$ | Johnson & Mitchell (1995) |
| HD102224 | DDO | m35 | $7.37 \pm 0.05$ | McClure & Forrester (1981) |
| HD102224 | WBVR | W | $5.89 \pm 0.05$ | Kornilov et al. (1991) |
| HD102224 | Johnson | U | $5.99 \pm 0.05$ | Schild (1973) |
| HD102224 | Johnson | U | $6.04 \pm 0.05$ | Argue (1963) |
| HD102224 | Johnson | U | $6.06 \pm 0.05$ | Johnson et al. (1966) |
| HD102224 | Johnson | U | $6.06 \pm 0.05$ | Jennens & Helfer (1975) |
| HD102224 | Johnson | U | $6.06 \pm 0.05$ | Ducati (2002) |
| HD102224 | Johnson | U | $6.07 \pm 0.05$ | Johnson et al. (1966) |
| HD102224 | 13c | m37 | $5.95 \pm 0.05$ | Johnson & Mitchell (1995) |
| HD102224 | Vilnius | P | $7.35 \pm 0.05$ | Kazlauskas et al. (2005) |
| HD102224 | DDO | m38 | $6.24 \pm 0.05$ | McClure & Forrester (1981) |
| HD102224 | 13c | m40 | $5.61 \pm 0.05$ | Johnson & Mitchell (1995) |
| HD102224 | Geneva | B1 | $5.66 \pm 0.08$ | Golay (1972) |
| HD102224 | Vilnius | X | $6.23 \pm 0.05$ | Kazlauskas et al. (2005) |
| HD102224 | DDO | m41 | $6.61 \pm 0.05$ | McClure & Forrester (1981) |
| HD102224 | Oja | m41 | $6.14 \pm 0.05$ | Häggkvist & Oja (1970) |
| HD102224 | DDO | m42 | $6.40 \pm 0.05$ | McClure & Forrester (1981) |

**Table 21** *(continued)*

| Star ID | System/Wvlen | Band/Bandpass | Value | Reference |
|---------|--------------|---------------|-------|-----------|
| HD102224 | Oja | m42 | $5.94 \pm 0.05$ | Häggkvist & Oja (1970) |
| HD102224 | Geneva | B | $4.26 \pm 0.08$ | Golay (1972) |
| HD102224 | WBVR | B | $4.91 \pm 0.05$ | Kornilov et al. (1991) |
| HD102224 | Johnson | B | $4.87 \pm 0.05$ | Schild (1973) |
| HD102224 | Johnson | B | $4.88 \pm 0.05$ | Häggkvist & Oja (1966) |
| HD102224 | Johnson | B | $4.90 \pm 0.05$ | Argue (1963) |
| HD102224 | Johnson | B | $4.90 \pm 0.05$ | Johnson et al. (1966) |
| HD102224 | Johnson | B | $4.90 \pm 0.05$ | Jennens & Helfer (1975) |
| HD102224 | Johnson | B | $4.90 \pm 0.05$ | Ducati (2002) |
| HD102224 | Geneva | B2 | $5.36 \pm 0.08$ | Golay (1972) |
| HD102224 | 13c | m45 | $4.54 \pm 0.05$ | Johnson & Mitchell (1995) |
| HD102224 | DDO | m45 | $5.40 \pm 0.05$ | McClure & Forrester (1981) |
| HD102224 | Oja | m45 | $4.56 \pm 0.05$ | Häggkvist & Oja (1970) |
| HD102224 | Vilnius | Y | $4.74 \pm 0.05$ | Kazlauskas et al. (2005) |
| HD102224 | DDO | m48 | $4.14 \pm 0.05$ | McClure & Forrester (1981) |
| HD102224 | Vilnius | Z | $4.22 \pm 0.05$ | Kazlauskas et al. (2005) |
| HD102224 | 13c | m52 | $4.00 \pm 0.05$ | Johnson & Mitchell (1995) |
| HD102224 | Geneva | V1 | $4.51 \pm 0.08$ | Golay (1972) |
| HD102224 | WBVR | V | $3.70 \pm 0.05$ | Kornilov et al. (1991) |
| HD102224 | Vilnius | V | $3.85 \pm 0.05$ | Kazlauskas et al. (2005) |
| HD102224 | Geneva | V | $3.72 \pm 0.08$ | Golay (1972) |
| HD102224 | Johnson | V | $3.69 \pm 0.05$ | Häggkvist & Oja (1966) |
| HD102224 | Johnson | V | $3.70 \pm 0.05$ | Argue (1963) |
| HD102224 | Johnson | V | $3.70 \pm 0.05$ | Schild (1973) |
| HD102224 | Johnson | V | $3.72 \pm 0.05$ | Johnson et al. (1966) |
| HD102224 | Johnson | V | $3.72 \pm 0.05$ | Jennens & Helfer (1975) |
| HD102224 | Johnson | V | $3.72 \pm 0.05$ | Ducati (2002) |
| HD102224 | 13c | m58 | $3.44 \pm 0.05$ | Johnson & Mitchell (1995) |
| HD102224 | Geneva | G | $4.66 \pm 0.08$ | Golay (1972) |
| HD102224 | 13c | m63 | $3.09 \pm 0.05$ | Johnson & Mitchell (1995) |
| HD102224 | Vilnius | S | $3.02 \pm 0.05$ | Kazlauskas et al. (2005) |
| HD102224 | WBVR | R | $2.83 \pm 0.05$ | Kornilov et al. (1991) |
| HD102224 | 13c | m72 | $2.80 \pm 0.05$ | Johnson & Mitchell (1995) |
| HD102224 | 13c | m80 | $2.53 \pm 0.05$ | Johnson & Mitchell (1995) |
| HD102224 | 13c | m86 | $2.39 \pm 0.05$ | Johnson & Mitchell (1995) |
| HD102224 | 13c | m99 | $2.21 \pm 0.05$ | Johnson & Mitchell (1995) |
| HD102224 | 13c | m110 | $2.01 \pm 0.05$ | Johnson & Mitchell (1995) |
| HD102224 | 1250 | 310 | $352.70 \pm 8.40$ | Smith et al. (2004) |
| HD102224 | Johnson | J | $1.60 \pm 0.05$ | Noguchi et al. (1981) |
| HD102224 | Johnson | J | $1.63 \pm 0.05$ | Alonso et al. (1998) |
| HD102224 | Johnson | J | $1.66 \pm 0.05$ | Ducati (2002) |

**Table 21** *(continued)*

| Star ID | System/Wvlen | Band/Bandpass | Value | Reference |
|---------|--------------|---------------|-------|-----------|
| HD102224 | Johnson | J | $1.72 \pm 0.05$ | Johnson et al. (1966) |
| HD102224 | Johnson | H | $1.02 \pm 0.05$ | Noguchi et al. (1981) |
| HD102224 | Johnson | H | $1.02 \pm 0.05$ | Ducati (2002) |
| HD102224 | 2200 | 361 | $267.30 \pm 5.50$ | Smith et al. (2004) |
| HD102224 | Johnson | K | $0.92 \pm 0.03$ | Neugebauer & Leighton (1969) |
| HD102224 | Johnson | K | $0.93 \pm 0.05$ | Ducati (2002) |
| HD102224 | Johnson | K | $0.95 \pm 0.05$ | Johnson et al. (1966) |
| HD102224 | Johnson | L | $0.78 \pm 0.05$ | Johnson et al. (1966) |
| HD102224 | Johnson | L | $0.78 \pm 0.05$ | Ducati (2002) |
| HD102224 | 3500 | 898 | $129.10 \pm 4.80$ | Smith et al. (2004) |
| HD102224 | 4900 | 712 | $60.50 \pm 5.20$ | Smith et al. (2004) |
| HD102224 | 12000 | 6384 | $16.10 \pm 16.80$ | Smith et al. (2004) |
| HD104207 | Johnson | U | $10.14 \pm 0.05$ | Mermilliod (1986) |
| HD104207 | WBVR | B | $8.50 \pm 0.05$ | Kornilov et al. (1991) |
| HD104207 | Johnson | B | $8.52 \pm 0.02$ | Yoss & Griffin (1997) |
| HD104207 | Johnson | B | $8.52 \pm 0.05$ | Haggkvist & Oja (1973) |
| HD104207 | Johnson | B | $8.58 \pm 0.05$ | Mermilliod (1986) |
| HD104207 | WBVR | V | $6.87 \pm 0.05$ | Kornilov et al. (1991) |
| HD104207 | Johnson | V | $6.94 \pm 0.05$ | Haggkvist & Oja (1973) |
| HD104207 | Johnson | V | $6.95 \pm 0.02$ | Yoss & Griffin (1997) |
| HD104207 | Johnson | V | $7.00 \pm 0.05$ | Mermilliod (1986) |
| HD104207 | Johnson | V | $7.00 \pm 0.07$ | This work |
| HD104207 | WBVR | R | $4.98 \pm 0.05$ | Kornilov et al. (1991) |
| HD104207 | 1250 | 310 | $167.50 \pm 5.20$ | Smith et al. (2004) |
| HD104207 | Johnson | J | $2.46 \pm 0.05$ | Kerschbaum & Hron (1994) |
| HD104207 | Johnson | H | $1.64 \pm 0.05$ | Kerschbaum & Hron (1994) |
| HD104207 | 2200 | 361 | $178.70 \pm 5.20$ | Smith et al. (2004) |
| HD104207 | Johnson | K | $1.30 \pm 0.04$ | Neugebauer & Leighton (1969) |
| HD104207 | 3500 | 898 | $92.80 \pm 7.80$ | Smith et al. (2004) |
| HD104207 | 4900 | 712 | $40.00 \pm 4.40$ | Smith et al. (2004) |
| HD104207 | 12000 | 6384 | $7.60 \pm 20.40$ | Smith et al. (2004) |
| HD104575 | KronComet | NH | $12.71 \pm 0.02$ | This work |
| HD104575 | KronComet | UVc | $12.33 \pm 0.03$ | This work |
| HD104575 | Johnson | U | $10.83 \pm 0.02$ | This work |
| HD104575 | KronComet | CN | $11.82 \pm 0.05$ | This work |
| HD104575 | KronComet | COp | $9.90 \pm 0.02$ | This work |
| HD104575 | Johnson | B | $9.20 \pm 0.01$ | This work |
| HD104575 | KronComet | Bc | $9.10 \pm 0.01$ | This work |
| HD104575 | KronComet | C2 | $7.97 \pm 0.01$ | This work |
| HD104575 | KronComet | Gc | $7.80 \pm 0.01$ | This work |
| HD104575 | Johnson | V | $7.78 \pm 0.01$ | This work |

**Table 21** *(continued)*

| Star ID | System/Wvlen | Band/Bandpass | Value | Reference |
|---------|--------------|---------------|-------|-----------|
| HD104575 | KronComet | Rc | 6.29 ± 0.01 | This work |
| HD104575 | 1250 | 310 | 50.20 ± 6.30 | Smith et al. (2004) |
| HD104575 | 2200 | 361 | 50.10 ± 5.20 | Smith et al. (2004) |
| HD104575 | Johnson | K | 2.81 ± 0.10 | Neugebauer & Leighton (1969) |
| HD104575 | 3500 | 898 | 25.20 ± 4.40 | Smith et al. (2004) |
| HD104575 | 4900 | 712 | 11.10 ± 4.90 | Smith et al. (2004) |
| HD104575 | 12000 | 6384 | 3.90 ± 21.60 | Smith et al. (2004) |
| HD104831 | KronComet | COp | 9.41 ± 0.01 | This work |
| HD104831 | Johnson | B | 8.66 ± 0.01 | This work |
| HD104831 | KronComet | Bc | 8.59 ± 0.01 | This work |
| HD104831 | KronComet | C2 | 7.57 ± 0.01 | This work |
| HD104831 | KronComet | Gc | 7.29 ± 0.01 | This work |
| HD104831 | Johnson | V | 7.23 ± 0.01 | This work |
| HD104831 | KronComet | Rc | 5.77 ± 0.01 | This work |
| HD104831 | 1250 | 310 | 43.10 ± 4.60 | Smith et al. (2004) |
| HD104831 | 2200 | 361 | 42.10 ± 5.30 | Smith et al. (2004) |
| HD104831 | Johnson | K | 2.93 ± 0.09 | Neugebauer & Leighton (1969) |
| HD104831 | 3500 | 898 | 19.30 ± 4.40 | Smith et al. (2004) |
| HD104831 | 4900 | 712 | 9.00 ± 4.90 | Smith et al. (2004) |
| HD104831 | 12000 | 6384 | 0.90 ± 24.00 | Smith et al. (2004) |
| HD104979 | 13c | m33 | 5.70 ± 0.05 | Johnson & Mitchell (1995) |
| HD104979 | Geneva | U | 6.21 ± 0.08 | Golay (1972) |
| HD104979 | Vilnius | U | 7.41 ± 0.05 | Straizys et al. (1989a) |
| HD104979 | Vilnius | U | 7.51 ± 0.05 | Forbes et al. (1993) |
| HD104979 | 13c | m35 | 5.50 ± 0.05 | Johnson & Mitchell (1995) |
| HD104979 | DDO | m35 | 7.02 ± 0.05 | Lu et al. (1983) |
| HD104979 | Stromgren | u | 6.95 ± 0.08 | Pilachowski (1978) |
| HD104979 | Stromgren | u | 6.96 ± 0.08 | Olsen (1993) |
| HD104979 | Stromgren | u | 6.96 ± 0.08 | Hauck & Mermilliod (1998) |
| HD104979 | WBVR | W | 5.57 ± 0.05 | Kornilov et al. (1991) |
| HD104979 | Johnson | U | 5.71 ± 0.05 | Gutierrez-Moreno & et al. (1966) |
| HD104979 | Johnson | U | 5.72 ± 0.05 | Cousins (1963c) |
| HD104979 | Johnson | U | 5.73 ± 0.05 | Piirola (1976) |
| HD104979 | Johnson | U | 5.75 ± 0.05 | Argue (1963) |
| HD104979 | Johnson | U | 5.75 ± 0.05 | Johnson et al. (1966) |
| HD104979 | Johnson | U | 5.75 ± 0.05 | Jennens & Helfer (1975) |
| HD104979 | Johnson | U | 5.75 ± 0.05 | Ducati (2002) |
| HD104979 | 13c | m37 | 5.65 ± 0.05 | Johnson & Mitchell (1995) |
| HD104979 | Vilnius | P | 6.90 ± 0.05 | Straizys et al. (1989a) |
| HD104979 | Vilnius | P | 6.94 ± 0.05 | Forbes et al. (1993) |
| HD104979 | DDO | m38 | 6.04 ± 0.05 | Lu et al. (1983) |

**Table 21** *(continued)*

| Star ID | System/Wvlen | Band/Bandpass | Value | Reference |
|---------|--------------|---------------|-------|-----------|
| HD104979 | 13c | m40 | $5.62 \pm 0.05$ | Johnson & Mitchell (1995) |
| HD104979 | Geneva | B1 | $5.66 \pm 0.08$ | Golay (1972) |
| HD104979 | Vilnius | X | $6.07 \pm 0.05$ | Straizys et al. (1989a) |
| HD104979 | Vilnius | X | $6.10 \pm 0.05$ | Forbes et al. (1993) |
| HD104979 | DDO | m41 | $6.63 \pm 0.05$ | Lu et al. (1983) |
| HD104979 | Oja | m41 | $6.17 \pm 0.05$ | Häggkvist & Oja (1970) |
| HD104979 | Stromgren | v | $5.68 \pm 0.08$ | Olsen (1993) |
| HD104979 | Stromgren | v | $5.68 \pm 0.08$ | Hauck & Mermilliod (1998) |
| HD104979 | Stromgren | v | $5.69 \pm 0.08$ | Pilachowski (1978) |
| HD104979 | DDO | m42 | $6.51 \pm 0.05$ | Lu et al. (1983) |
| HD104979 | Oja | m42 | $6.07 \pm 0.05$ | Häggkvist & Oja (1970) |
| HD104979 | Geneva | B | $4.41 \pm 0.08$ | Golay (1972) |
| HD104979 | WBVR | B | $5.12 \pm 0.05$ | Kornilov et al. (1991) |
| HD104979 | Johnson | B | $5.08 \pm 0.05$ | Häggkvist & Oja (1966) |
| HD104979 | Johnson | B | $5.08 \pm 0.05$ | Piirola (1976) |
| HD104979 | Johnson | B | $5.09 \pm 0.05$ | Cousins (1963c) |
| HD104979 | Johnson | B | $5.09 \pm 0.05$ | Gutierrez-Moreno & et al. (1966) |
| HD104979 | Johnson | B | $5.11 \pm 0.05$ | Johnson et al. (1966) |
| HD104979 | Johnson | B | $5.11 \pm 0.05$ | Jennens & Helfer (1975) |
| HD104979 | Johnson | B | $5.11 \pm 0.05$ | Ducati (2002) |
| HD104979 | Johnson | B | $5.13 \pm 0.05$ | Argue (1963) |
| HD104979 | Geneva | B2 | $5.59 \pm 0.08$ | Golay (1972) |
| HD104979 | 13c | m45 | $4.81 \pm 0.05$ | Johnson & Mitchell (1995) |
| HD104979 | DDO | m45 | $5.66 \pm 0.05$ | Lu et al. (1983) |
| HD104979 | Oja | m45 | $4.83 \pm 0.05$ | Häggkvist & Oja (1970) |
| HD104979 | Vilnius | Y | $4.86 \pm 0.05$ | Straizys et al. (1989a) |
| HD104979 | Vilnius | Y | $4.88 \pm 0.05$ | Forbes et al. (1993) |
| HD104979 | Stromgren | b | $4.71 \pm 0.08$ | Pilachowski (1978) |
| HD104979 | Stromgren | b | $4.72 \pm 0.08$ | Olsen (1993) |
| HD104979 | Stromgren | b | $4.72 \pm 0.08$ | Hauck & Mermilliod (1998) |
| HD104979 | DDO | m48 | $4.48 \pm 0.05$ | Lu et al. (1983) |
| HD104979 | Vilnius | Z | $4.40 \pm 0.05$ | Straizys et al. (1989a) |
| HD104979 | Vilnius | Z | $4.44 \pm 0.05$ | Forbes et al. (1993) |
| HD104979 | 13c | m52 | $4.35 \pm 0.05$ | Johnson & Mitchell (1995) |
| HD104979 | Geneva | V1 | $4.92 \pm 0.08$ | Golay (1972) |
| HD104979 | WBVR | V | $4.12 \pm 0.05$ | Kornilov et al. (1991) |
| HD104979 | Vilnius | V | $4.11 \pm 0.05$ | Straizys et al. (1989a) |
| HD104979 | Vilnius | V | $4.15 \pm 0.05$ | Forbes et al. (1993) |
| HD104979 | Stromgren | y | $4.12 \pm 0.08$ | Pilachowski (1978) |
| HD104979 | Stromgren | y | $4.12 \pm 0.08$ | Olsen (1993) |
| HD104979 | Stromgren | y | $4.12 \pm 0.08$ | Hauck & Mermilliod (1998) |

**Table 21** *(continued)*

| Star ID | System/Wvlen | Band/Bandpass | Value | Reference |
|---------|--------------|---------------|-------|-----------|
| HD104979 | Geneva | V | $4.14 \pm 0.08$ | Golay (1972) |
| HD104979 | Johnson | V | $4.10 \pm 0.05$ | Häggkvist & Oja (1966) |
| HD104979 | Johnson | V | $4.11 \pm 0.05$ | Gutierrez-Moreno & et al. (1966) |
| HD104979 | Johnson | V | $4.12 \pm 0.05$ | Cousins (1963c) |
| HD104979 | Johnson | V | $4.12 \pm 0.05$ | Johnson et al. (1966) |
| HD104979 | Johnson | V | $4.12 \pm 0.05$ | Jennens & Helfer (1975) |
| HD104979 | Johnson | V | $4.12 \pm 0.05$ | Piirola (1976) |
| HD104979 | Johnson | V | $4.12 \pm 0.05$ | Ducati (2002) |
| HD104979 | Johnson | V | $4.14 \pm 0.05$ | Argue (1963) |
| HD104979 | Johnson | V | $4.17 \pm 0.07$ | This work |
| HD104979 | 13c | m58 | $3.90 \pm 0.05$ | Johnson & Mitchell (1995) |
| HD104979 | Geneva | G | $5.14 \pm 0.08$ | Golay (1972) |
| HD104979 | 13c | m63 | $3.62 \pm 0.05$ | Johnson & Mitchell (1995) |
| HD104979 | Vilnius | S | $3.41 \pm 0.05$ | Straizys et al. (1989a) |
| HD104979 | Vilnius | S | $3.44 \pm 0.05$ | Forbes et al. (1993) |
| HD104979 | WBVR | R | $3.42 \pm 0.05$ | Kornilov et al. (1991) |
| HD104979 | 13c | m72 | $3.39 \pm 0.05$ | Johnson & Mitchell (1995) |
| HD104979 | 13c | m80 | $3.16 \pm 0.05$ | Johnson & Mitchell (1995) |
| HD104979 | 13c | m86 | $3.07 \pm 0.05$ | Johnson & Mitchell (1995) |
| HD104979 | 13c | m99 | $2.93 \pm 0.05$ | Johnson & Mitchell (1995) |
| HD104979 | 13c | m110 | $2.79 \pm 0.05$ | Johnson & Mitchell (1995) |
| HD104979 | 1250 | 310 | $167.30 \pm 6.90$ | Smith et al. (2004) |
| HD104979 | Johnson | J | $2.41 \pm 0.05$ | Frogel et al. (1978) |
| HD104979 | Johnson | J | $2.41 \pm 0.05$ | Kenyon & Gallagher (1983) |
| HD104979 | Johnson | J | $2.42 \pm 0.05$ | Alonso et al. (1998) |
| HD104979 | Johnson | J | $2.43 \pm 0.05$ | Ducati (2002) |
| HD104979 | Johnson | J | $2.45 \pm 0.05$ | Ghosh et al. (1984) |
| HD104979 | Johnson | J | $2.46 \pm 0.01$ | Laney et al. (2012) |
| HD104979 | Johnson | J | $2.48 \pm 0.05$ | Johnson et al. (1966) |
| HD104979 | Johnson | H | $1.94 \pm 0.05$ | Frogel et al. (1978) |
| HD104979 | Johnson | H | $1.94 \pm 0.05$ | Kenyon & Gallagher (1983) |
| HD104979 | Johnson | H | $1.94 \pm 0.05$ | Ducati (2002) |
| HD104979 | Johnson | H | $1.95 \pm 0.05$ | Ghosh et al. (1984) |
| HD104979 | Johnson | H | $1.96 \pm 0.05$ | Alonso et al. (1998) |
| HD104979 | Johnson | H | $1.99 \pm 0.01$ | Laney et al. (2012) |
| HD104979 | 2200 | 361 | $111.60 \pm 5.20$ | Smith et al. (2004) |
| HD104979 | Johnson | K | $1.84 \pm 0.06$ | Neugebauer & Leighton (1969) |
| HD104979 | Johnson | K | $1.87 \pm 0.01$ | Laney et al. (2012) |
| HD104979 | Johnson | K | $1.89 \pm 0.05$ | Ducati (2002) |
| HD104979 | Johnson | K | $1.90 \pm 0.05$ | Johnson et al. (1966) |
| HD104979 | 3500 | 898 | $52.20 \pm 4.80$ | Smith et al. (2004) |

Table 21 *(continued)*

| Star ID | System/Wvlen | Band/Bandpass | Value | Reference |
|---------|--------------|---------------|-------|-----------|
| HD104979 | 4900 | 712 | $25.60 \pm 5.00$ | Smith et al. (2004) |
| HD104979 | 12000 | 6384 | $5.90 \pm 24.80$ | Smith et al. (2004) |
| HD105943 | WBVR | W | $9.48 \pm 0.05$ | Kornilov et al. (1991) |
| HD105943 | Oja | m41 | $9.27 \pm 0.05$ | Häggkvist & Oja (1970) |
| HD105943 | Oja | m42 | $8.91 \pm 0.05$ | Häggkvist & Oja (1970) |
| HD105943 | WBVR | B | $7.68 \pm 0.05$ | Kornilov et al. (1991) |
| HD105943 | Johnson | B | $7.62 \pm 0.05$ | Haggkvist & Oja (1970) |
| HD105943 | Oja | m45 | $7.28 \pm 0.05$ | Häggkvist & Oja (1970) |
| HD105943 | WBVR | V | $6.00 \pm 0.05$ | Kornilov et al. (1991) |
| HD105943 | Johnson | V | $6.00 \pm 0.05$ | Haggkvist & Oja (1970) |
| HD105943 | WBVR | R | $4.84 \pm 0.05$ | Kornilov et al. (1991) |
| HD105943 | 1250 | 310 | $80.90 \pm 5.00$ | Smith et al. (2004) |
| HD105943 | 2200 | 361 | $71.00 \pm 5.00$ | Smith et al. (2004) |
| HD105943 | 3500 | 898 | $35.70 \pm 5.60$ | Smith et al. (2004) |
| HD105943 | 4900 | 712 | $15.80 \pm 4.60$ | Smith et al. (2004) |
| HD105943 | 12000 | 6384 | $2.40 \pm 16.50$ | Smith et al. (2004) |
| HD106714 | KronComet | NH | $7.68 \pm 0.01$ | This work |
| HD106714 | KronComet | UVc | $7.43 \pm 0.01$ | This work |
| HD106714 | Stromgren | u | $7.66 \pm 0.08$ | Peña et al. (1993) |
| HD106714 | Johnson | U | $6.60 \pm 0.05$ | Argue (1963) |
| HD106714 | Johnson | U | $6.64 \pm 0.05$ | Mermilliod (1986) |
| HD106714 | Johnson | U | $6.68 \pm 0.01$ | This work |
| HD106714 | KronComet | CN | $7.42 \pm 0.02$ | This work |
| HD106714 | Stromgren | v | $6.37 \pm 0.08$ | Peña et al. (1993) |
| HD106714 | KronComet | COp | $6.17 \pm 0.01$ | This work |
| HD106714 | Johnson | B | $5.81 \pm 0.02$ | Yoss & Griffin (1997) |
| HD106714 | Johnson | B | $5.84 \pm 0.02$ | This work |
| HD106714 | Johnson | B | $5.89 \pm 0.05$ | Häggkvist & Oja (1966) |
| HD106714 | Johnson | B | $5.92 \pm 0.05$ | Argue (1963) |
| HD106714 | Johnson | B | $5.95 \pm 0.05$ | Mermilliod (1986) |
| HD106714 | KronComet | Bc | $5.73 \pm 0.01$ | This work |
| HD106714 | Stromgren | b | $5.46 \pm 0.08$ | Peña et al. (1993) |
| HD106714 | KronComet | C2 | $5.09 \pm 0.01$ | This work |
| HD106714 | KronComet | Gc | $4.99 \pm 0.01$ | This work |
| HD106714 | Stromgren | y | $4.86 \pm 0.08$ | Peña et al. (1993) |
| HD106714 | Johnson | V | $4.86 \pm 0.02$ | Yoss & Griffin (1997) |
| HD106714 | Johnson | V | $4.93 \pm 0.05$ | Häggkvist & Oja (1966) |
| HD106714 | Johnson | V | $4.95 \pm 0.05$ | Argue (1963) |
| HD106714 | Johnson | V | $5.00 \pm 0.05$ | Mermilliod (1986) |
| HD106714 | KronComet | Rc | $3.97 \pm 0.01$ | This work |
| HD106714 | 1250 | 310 | $73.60 \pm 6.70$ | Smith et al. (2004) |

**Table 21** *(continued)*

| Star ID | System/Wvlen | Band/Bandpass | Value | Reference |
|---------|-------------|---------------|-------|-----------|
| HD106714 | 2200 | 361 | $47.70 \pm 5.00$ | Smith et al. (2004) |
| HD106714 | Johnson | K | $2.62 \pm 0.09$ | Neugebauer & Leighton (1969) |
| HD106714 | 3500 | 898 | $22.60 \pm 4.00$ | Smith et al. (2004) |
| HD106714 | 4900 | 712 | $11.10 \pm 4.40$ | Smith et al. (2004) |
| HD106714 | 12000 | 6384 | $3.00 \pm 19.10$ | Smith et al. (2004) |
| HD107256 | KronComet | COp | $10.32 \pm 0.00$ | This work |
| HD107256 | Johnson | B | $9.79 \pm 0.05$ | Haggkvist & Oja (1973) |
| HD107256 | Johnson | B | $9.86 \pm 0.02$ | This work |
| HD107256 | KronComet | Bc | $9.76 \pm 0.01$ | This work |
| HD107256 | KronComet | C2 | $8.47 \pm 0.01$ | This work |
| HD107256 | KronComet | Gc | $8.51 \pm 0.00$ | This work |
| HD107256 | Johnson | V | $8.30 \pm 0.05$ | Haggkvist & Oja (1973) |
| HD107256 | Johnson | V | $8.48 \pm 0.01$ | This work |
| HD107256 | KronComet | Rc | $6.94 \pm 0.01$ | This work |
| HD107256 | 1250 | 310 | $53.50 \pm 4.70$ | Smith et al. (2004) |
| HD107256 | 2200 | 361 | $55.80 \pm 5.10$ | Smith et al. (2004) |
| HD107256 | Johnson | K | $2.58 \pm 0.08$ | Neugebauer & Leighton (1969) |
| HD107256 | 3500 | 898 | $29.10 \pm 4.20$ | Smith et al. (2004) |
| HD107256 | 4900 | 712 | $11.20 \pm 4.90$ | Smith et al. (2004) |
| HD107256 | 12000 | 6384 | $4.30 \pm 20.90$ | Smith et al. (2004) |
| HD107274 | DDO | m35 | $10.34 \pm 0.05$ | McClure & Forrester (1981) |
| HD107274 | WBVR | W | $8.88 \pm 0.05$ | Kornilov et al. (1991) |
| HD107274 | Johnson | U | $8.86 \pm 0.05$ | Mermilliod (1986) |
| HD107274 | Johnson | U | $8.92 \pm 0.05$ | Ducati (2002) |
| HD107274 | Johnson | U | $8.95 \pm 0.05$ | Schild (1973) |
| HD107274 | DDO | m38 | $8.98 \pm 0.05$ | McClure & Forrester (1981) |
| HD107274 | DDO | m41 | $8.94 \pm 0.05$ | McClure & Forrester (1981) |
| HD107274 | Oja | m41 | $8.48 \pm 0.05$ | Häggkvist & Oja (1970) |
| HD107274 | DDO | m42 | $8.77 \pm 0.05$ | McClure & Forrester (1981) |
| HD107274 | Oja | m42 | $8.27 \pm 0.05$ | Häggkvist & Oja (1970) |
| HD107274 | WBVR | B | $6.96 \pm 0.05$ | Kornilov et al. (1991) |
| HD107274 | Johnson | B | $6.87 \pm 0.05$ | Häggkvist & Oja (1969b) |
| HD107274 | Johnson | B | $6.89 \pm 0.05$ | Mermilliod (1986) |
| HD107274 | Johnson | B | $6.95 \pm 0.05$ | Ducati (2002) |
| HD107274 | Johnson | B | $6.99 \pm 0.05$ | Schild (1973) |
| HD107274 | DDO | m45 | $7.39 \pm 0.05$ | McClure & Forrester (1981) |
| HD107274 | Oja | m45 | $6.56 \pm 0.05$ | Häggkvist & Oja (1970) |
| HD107274 | DDO | m48 | $6.00 \pm 0.05$ | McClure & Forrester (1981) |
| HD107274 | WBVR | V | $5.30 \pm 0.05$ | Kornilov et al. (1991) |
| HD107274 | Johnson | V | $5.26 \pm 0.05$ | Mermilliod (1986) |
| HD107274 | Johnson | V | $5.28 \pm 0.05$ | Häggkvist & Oja (1969b) |

**Table 21** *(continued)*

| Star ID | System/Wvlen | Band/Bandpass | Value | Reference |
|---------|--------------|---------------|-------|-----------|
| HD107274 | Johnson | V | $5.29 \pm 0.05$ | Ducati (2002) |
| HD107274 | Johnson | V | $5.30 \pm 0.05$ | Schild (1973) |
| HD107274 | WBVR | R | $3.95 \pm 0.05$ | Kornilov et al. (1991) |
| HD107274 | 1250 | 310 | $219.00 \pm 5.80$ | Smith et al. (2004) |
| HD107274 | Johnson | J | $2.13 \pm 0.05$ | McWilliam & Lambert (1984) |
| HD107274 | Johnson | J | $2.13 \pm 0.05$ | Ducati (2002) |
| HD107274 | Johnson | J | $2.14 \pm 0.05$ | Kenyon & Gallagher (1983) |
| HD107274 | Johnson | H | $1.33 \pm 0.05$ | Kenyon & Gallagher (1983) |
| HD107274 | Johnson | H | $1.33 \pm 0.05$ | Ducati (2002) |
| HD107274 | 2200 | 361 | $211.30 \pm 5.10$ | Smith et al. (2004) |
| HD107274 | Johnson | K | $1.11 \pm 0.05$ | Ducati (2002) |
| HD107274 | Johnson | K | $1.17 \pm 0.04$ | Neugebauer & Leighton (1969) |
| HD107274 | 3500 | 898 | $106.30 \pm 4.50$ | Smith et al. (2004) |
| HD107274 | 4900 | 712 | $46.60 \pm 4.70$ | Smith et al. (2004) |
| HD107274 | 12000 | 6384 | $11.90 \pm 16.10$ | Smith et al. (2004) |
| HD109282 | Stromgren | u | $12.20 \pm 0.08$ | Peña et al. (1993) |
| HD109282 | Stromgren | u | $12.20 \pm 0.08$ | Hauck & Mermilliod (1998) |
| HD109282 | Johnson | U | $10.63 \pm 0.02$ | This work |
| HD109282 | Johnson | U | $10.76 \pm 0.05$ | Haggkvist & Oja (1973) |
| HD109282 | Stromgren | v | $10.03 \pm 0.08$ | Peña et al. (1993) |
| HD109282 | Stromgren | v | $10.03 \pm 0.08$ | Hauck & Mermilliod (1998) |
| HD109282 | KronComet | COp | $9.62 \pm 0.01$ | This work |
| HD109282 | Johnson | B | $8.88 \pm 0.01$ | This work |
| HD109282 | Johnson | B | $8.91 \pm 0.05$ | Haggkvist & Oja (1973) |
| HD109282 | Johnson | B | $8.94 \pm 0.02$ | Yoss & Griffin (1997) |
| HD109282 | KronComet | Bc | $8.83 \pm 0.01$ | This work |
| HD109282 | Stromgren | b | $8.44 \pm 0.08$ | Peña et al. (1993) |
| HD109282 | Stromgren | b | $8.44 \pm 0.08$ | Hauck & Mermilliod (1998) |
| HD109282 | KronComet | C2 | $7.64 \pm 0.01$ | This work |
| HD109282 | KronComet | Gc | $7.46 \pm 0.01$ | This work |
| HD109282 | Stromgren | y | $7.29 \pm 0.08$ | Peña et al. (1993) |
| HD109282 | Stromgren | y | $7.29 \pm 0.08$ | Hauck & Mermilliod (1998) |
| HD109282 | Johnson | V | $7.26 \pm 0.05$ | Haggkvist & Oja (1973) |
| HD109282 | Johnson | V | $7.29 \pm 0.02$ | Yoss & Griffin (1997) |
| HD109282 | Johnson | V | $7.41 \pm 0.01$ | This work |
| HD109282 | KronComet | Rc | $5.91 \pm 0.01$ | This work |
| HD109282 | 1250 | 310 | $74.90 \pm 7.70$ | Smith et al. (2004) |
| HD109282 | 2200 | 361 | $69.90 \pm 7.30$ | Smith et al. (2004) |
| HD109282 | Johnson | K | $2.45 \pm 0.06$ | Neugebauer & Leighton (1969) |
| HD109282 | 3500 | 898 | $35.20 \pm 4.50$ | Smith et al. (2004) |
| HD109282 | 4900 | 712 | $15.60 \pm 4.70$ | Smith et al. (2004) |

**Table 21** *(continued)*

| Star ID | System/Wvlen | Band/Bandpass | Value | Reference |
|---------|--------------|---------------|-------|-----------|
| HD109282 | 12000 | 6384 | $5.70 \pm 17.70$ | Smith et al. (2004) |
| HD109317 | KronComet | NH | $8.11 \pm 0.07$ | This work |
| HD109317 | KronComet | UVc | $7.90 \pm 0.10$ | This work |
| HD109317 | DDO | m35 | $8.60 \pm 0.05$ | Yoss (1977) |
| HD109317 | WBVR | W | $7.12 \pm 0.05$ | Kornilov et al. (1991) |
| HD109317 | Johnson | U | $6.64 \pm 0.05$ | This work |
| HD109317 | Johnson | U | $7.26 \pm 0.05$ | Helfer & Sturch (1970) |
| HD109317 | Johnson | U | $7.28 \pm 0.05$ | Argue (1963) |
| HD109317 | Johnson | U | $7.29 \pm 0.05$ | Mermilliod (1986) |
| HD109317 | DDO | m38 | $7.53 \pm 0.05$ | Yoss (1977) |
| HD109317 | KronComet | CN | $7.96 \pm 0.12$ | This work |
| HD109317 | DDO | m41 | $8.04 \pm 0.05$ | Yoss (1977) |
| HD109317 | Oja | m41 | $7.59 \pm 0.05$ | Häggkvist & Oja (1970) |
| HD109317 | DDO | m42 | $7.84 \pm 0.05$ | Yoss (1977) |
| HD109317 | Oja | m42 | $7.40 \pm 0.05$ | Häggkvist & Oja (1970) |
| HD109317 | KronComet | COp | $6.56 \pm 0.09$ | This work |
| HD109317 | WBVR | B | $6.45 \pm 0.05$ | Kornilov et al. (1991) |
| HD109317 | Johnson | B | $6.22 \pm 0.11$ | This work |
| HD109317 | Johnson | B | $6.39 \pm 0.02$ | Yoss & Griffin (1997) |
| HD109317 | Johnson | B | $6.41 \pm 0.05$ | Ljunggren (1966) |
| HD109317 | Johnson | B | $6.43 \pm 0.05$ | Helfer & Sturch (1970) |
| HD109317 | Johnson | B | $6.45 \pm 0.05$ | Argue (1963) |
| HD109317 | Johnson | B | $6.45 \pm 0.05$ | Mermilliod (1986) |
| HD109317 | KronComet | Bc | $6.17 \pm 0.03$ | This work |
| HD109317 | DDO | m45 | $6.98 \pm 0.05$ | Yoss (1977) |
| HD109317 | Oja | m45 | $6.15 \pm 0.05$ | Häggkvist & Oja (1970) |
| HD109317 | DDO | m48 | $5.80 \pm 0.05$ | Yoss (1977) |
| HD109317 | KronComet | C2 | $5.54 \pm 0.03$ | This work |
| HD109317 | KronComet | Gc | $5.43 \pm 0.02$ | This work |
| HD109317 | WBVR | V | $5.42 \pm 0.05$ | Kornilov et al. (1991) |
| HD109317 | Johnson | V | $5.38 \pm 0.02$ | Yoss & Griffin (1997) |
| HD109317 | Johnson | V | $5.40 \pm 0.05$ | Helfer & Sturch (1970) |
| HD109317 | Johnson | V | $5.41 \pm 0.05$ | Ljunggren (1966) |
| HD109317 | Johnson | V | $5.42 \pm 0.04$ | This work |
| HD109317 | Johnson | V | $5.43 \pm 0.05$ | Argue (1963) |
| HD109317 | Johnson | V | $5.43 \pm 0.05$ | Mermilliod (1986) |
| HD109317 | WBVR | R | $4.68 \pm 0.05$ | Kornilov et al. (1991) |
| HD109317 | KronComet | Rc | $4.41 \pm 0.01$ | This work |
| HD109317 | 1250 | 310 | $78.40 \pm 6.90$ | Smith et al. (2004) |
| HD109317 | 2200 | 361 | $53.50 \pm 7.50$ | Smith et al. (2004) |
| HD109317 | Johnson | K | $2.93 \pm 0.08$ | Neugebauer & Leighton (1969) |

**Table 21** *(continued)*

| Star ID | System/Wvlen | Band/Bandpass | Value | Reference |
|---------|--------------|---------------|-------|-----------|
| HD109317 | 3500 | 898 | $24.20 \pm 6.60$ | Smith et al. (2004) |
| HD109317 | 4900 | 712 | $11.10 \pm 5.20$ | Smith et al. (2004) |
| HD109317 | 12000 | 6384 | $-9.40 \pm 17.40$ | Smith et al. (2004) |
| HD110296 | Geneva | B1 | $10.43 \pm 0.08$ | Golay (1972) |
| HD110296 | Oja | m41 | $10.87 \pm 0.05$ | Häggkvist & Oja (1970) |
| HD110296 | Oja | m42 | $10.67 \pm 0.05$ | Häggkvist & Oja (1970) |
| HD110296 | Geneva | B | $8.78 \pm 0.08$ | Golay (1972) |
| HD110296 | KronComet | COp | $9.82 \pm 0.09$ | This work |
| HD110296 | Johnson | B | $9.05 \pm 0.10$ | This work |
| HD110296 | Johnson | B | $9.33 \pm 0.05$ | Ljunggren (1966) |
| HD110296 | Johnson | B | $9.35 \pm 0.02$ | Yoss & Griffin (1997) |
| HD110296 | KronComet | Bc | $9.07 \pm 0.01$ | This work |
| HD110296 | Geneva | B2 | $9.73 \pm 0.08$ | Golay (1972) |
| HD110296 | Oja | m45 | $9.00 \pm 0.05$ | Häggkvist & Oja (1970) |
| HD110296 | KronComet | C2 | $8.22 \pm 0.01$ | This work |
| HD110296 | KronComet | Gc | $7.92 \pm 0.01$ | This work |
| HD110296 | Geneva | V1 | $8.60 \pm 0.08$ | Golay (1972) |
| HD110296 | Geneva | V | $7.78 \pm 0.08$ | Golay (1972) |
| HD110296 | Johnson | V | $7.78 \pm 0.05$ | Ljunggren (1966) |
| HD110296 | Johnson | V | $7.80 \pm 0.02$ | Yoss & Griffin (1997) |
| HD110296 | Johnson | V | $7.83 \pm 0.04$ | This work |
| HD110296 | Geneva | G | $8.63 \pm 0.08$ | Golay (1972) |
| HD110296 | KronComet | Rc | $6.45 \pm 0.02$ | This work |
| HD111067 | KronComet | NH | $9.06 \pm 0.06$ | This work |
| HD111067 | KronComet | UVc | $8.76 \pm 0.11$ | This work |
| HD111067 | DDO | m35 | $9.37 \pm 0.05$ | Yoss (1977) |
| HD111067 | WBVR | W | $7.91 \pm 0.05$ | Kornilov et al. (1991) |
| HD111067 | Johnson | U | $7.36 \pm 0.05$ | This work |
| HD111067 | Johnson | U | $8.01 \pm 0.05$ | Argue (1963) |
| HD111067 | Johnson | U | $8.07 \pm 0.05$ | Mermilliod (1986) |
| HD111067 | DDO | m38 | $8.17 \pm 0.05$ | Yoss (1977) |
| HD111067 | KronComet | CN | $8.62 \pm 0.12$ | This work |
| HD111067 | DDO | m41 | $8.38 \pm 0.05$ | Yoss (1977) |
| HD111067 | Oja | m41 | $7.91 \pm 0.05$ | Häggkvist & Oja (1970) |
| HD111067 | DDO | m42 | $8.10 \pm 0.05$ | Yoss (1977) |
| HD111067 | Oja | m42 | $7.61 \pm 0.05$ | Häggkvist & Oja (1970) |
| HD111067 | KronComet | COp | $6.77 \pm 0.10$ | This work |
| HD111067 | WBVR | B | $6.50 \pm 0.05$ | Kornilov et al. (1991) |
| HD111067 | Johnson | B | $6.20 \pm 0.10$ | This work |
| HD111067 | Johnson | B | $6.46 \pm 0.02$ | Yoss & Griffin (1997) |
| HD111067 | Johnson | B | $6.47 \pm 0.05$ | Argue (1963) |

**Table 21** *continued on next page*

**Table 21** *(continued)*

| Star ID | System/Wvlen | Band/Bandpass | Value | Reference |
|---------|--------------|---------------|-------|-----------|
| HD111067 | Johnson | B | $6.54 \pm 0.05$ | Mermilliod (1986) |
| HD111067 | KronComet | Bc | $6.19 \pm 0.03$ | This work |
| HD111067 | DDO | m45 | $6.94 \pm 0.05$ | Yoss (1977) |
| HD111067 | Oja | m45 | $6.09 \pm 0.05$ | Häggkvist & Oja (1970) |
| HD111067 | DDO | m48 | $5.64 \pm 0.05$ | Yoss (1977) |
| HD111067 | KronComet | C2 | $5.49 \pm 0.03$ | This work |
| HD111067 | KronComet | Gc | $5.22 \pm 0.02$ | This work |
| HD111067 | WBVR | V | $5.12 \pm 0.05$ | Kornilov et al. (1991) |
| HD111067 | Johnson | V | $5.12 \pm 0.02$ | Yoss & Griffin (1997) |
| HD111067 | Johnson | V | $5.12 \pm 0.05$ | Argue (1963) |
| HD111067 | Johnson | V | $5.13 \pm 0.05$ | This work |
| HD111067 | Johnson | V | $5.18 \pm 0.05$ | Mermilliod (1986) |
| HD111067 | WBVR | R | $4.14 \pm 0.05$ | Kornilov et al. (1991) |
| HD111067 | KronComet | Rc | $3.89 \pm 0.02$ | This work |
| HD111067 | 1250 | 310 | $118.20 \pm 5.30$ | Smith et al. (2004) |
| HD111067 | 2200 | 361 | $94.80 \pm 4.60$ | Smith et al. (2004) |
| HD111067 | Johnson | K | $2.01 \pm 0.04$ | Neugebauer & Leighton (1969) |
| HD111067 | 3500 | 898 | $46.20 \pm 6.10$ | Smith et al. (2004) |
| HD111067 | 4900 | 712 | $20.90 \pm 5.00$ | Smith et al. (2004) |
| HD111067 | 12000 | 6384 | $3.50 \pm 19.30$ | Smith et al. (2004) |
| HD113226 | Geneva | B1 | $4.33 \pm 0.08$ | Golay (1972) |
| HD113226 | DDO | m41 | $5.34 \pm 0.05$ | Clark & McClure (1979) |
| HD113226 | Oja | m41 | $4.88 \pm 0.05$ | Häggkvist & Oja (1970) |
| HD113226 | DDO | m42 | $5.12 \pm 0.05$ | Clark & McClure (1979) |
| HD113226 | Oja | m42 | $4.66 \pm 0.05$ | Häggkvist & Oja (1970) |
| HD113226 | Geneva | B | $3.06 \pm 0.08$ | Golay (1972) |
| HD113226 | WBVR | B | $3.78 \pm 0.05$ | Kornilov et al. (1991) |
| HD113226 | Johnson | B | $3.56 \pm 0.05$ | Fernie (1983) |
| HD113226 | Johnson | B | $3.71 \pm 0.05$ | Ducati (2002) |
| HD113226 | Johnson | B | $3.74 \pm 0.05$ | Popper (1959) |
| HD113226 | Johnson | B | $3.74 \pm 0.05$ | Ljunggren & Oja (1965) |
| HD113226 | Johnson | B | $3.76 \pm 0.05$ | Argue (1963) |
| HD113226 | Johnson | B | $3.77 \pm 0.05$ | Johnson (1964) |
| HD113226 | Johnson | B | $3.77 \pm 0.05$ | Gutierrez-Moreno & et al. (1966) |
| HD113226 | Johnson | B | $3.77 \pm 0.05$ | Mermilliod (1986) |
| HD113226 | Johnson | B | $3.78 \pm 0.05$ | Johnson et al. (1966) |
| HD113226 | Geneva | B2 | $4.24 \pm 0.08$ | Golay (1972) |
| HD113226 | 13c | m45 | $3.51 \pm 0.05$ | Johnson & Mitchell (1995) |
| HD113226 | DDO | m45 | $4.33 \pm 0.05$ | Clark & McClure (1979) |
| HD113226 | Oja | m45 | $3.49 \pm 0.05$ | Häggkvist & Oja (1970) |
| HD113226 | Vilnius | Y | $3.57 \pm 0.05$ | Zdanavicius et al. (1969) |

**Table 21** *continued on next page*

| Star ID | System/Wvlen | Band/Bandpass | Value | Reference |
|---|---|---|---|---|
| HD113226 | DDO | m48 | $3.19 \pm 0.05$ | Clark & McClure (1979) |
| HD113226 | Vilnius | Z | $3.10 \pm 0.05$ | Zdanavicius et al. (1969) |
| HD113226 | 13c | m52 | $3.08 \pm 0.05$ | Johnson & Mitchell (1995) |
| HD113226 | Geneva | V1 | $3.63 \pm 0.08$ | Golay (1972) |
| HD113226 | WBVR | V | $2.83 \pm 0.05$ | Kornilov et al. (1991) |
| HD113226 | Vilnius | V | $2.84 \pm 0.05$ | Zdanavicius et al. (1969) |
| HD113226 | Geneva | V | $2.85 \pm 0.08$ | Golay (1972) |
| HD113226 | Johnson | V | $2.70 \pm 0.05$ | Fernie (1983) |
| HD113226 | Johnson | V | $2.79 \pm 0.05$ | Ducati (2002) |
| HD113226 | Johnson | V | $2.82 \pm 0.05$ | Popper (1959) |
| HD113226 | Johnson | V | $2.82 \pm 0.05$ | Argue (1963) |
| HD113226 | Johnson | V | $2.83 \pm 0.05$ | Ljunggren & Oja (1965) |
| HD113226 | Johnson | V | $2.83 \pm 0.05$ | Gutierrez-Moreno & et al. (1966) |
| HD113226 | Johnson | V | $2.83 \pm 0.05$ | Mermilliod (1986) |
| HD113226 | Johnson | V | $2.84 \pm 0.05$ | Johnson (1964) |
| HD113226 | Johnson | V | $2.84 \pm 0.05$ | Johnson et al. (1966) |
| HD113226 | 13c | m58 | $2.64 \pm 0.05$ | Johnson & Mitchell (1995) |
| HD113226 | Geneva | G | $3.86 \pm 0.08$ | Golay (1972) |
| HD113226 | 13c | m63 | $2.37 \pm 0.05$ | Johnson & Mitchell (1995) |
| HD113226 | Vilnius | S | $2.14 \pm 0.05$ | Zdanavicius et al. (1969) |
| HD113226 | WBVR | R | $2.17 \pm 0.05$ | Kornilov et al. (1991) |
| HD113226 | KronComet | Rc | $1.89 \pm 0.02$ | This work |
| HD113226 | 13c | m72 | $2.14 \pm 0.05$ | Johnson & Mitchell (1995) |
| HD113226 | 13c | m80 | $1.93 \pm 0.05$ | Johnson & Mitchell (1995) |
| HD113226 | 13c | m86 | $1.85 \pm 0.05$ | Johnson & Mitchell (1995) |
| HD113226 | 13c | m99 | $1.73 \pm 0.05$ | Johnson & Mitchell (1995) |
| HD113226 | 13c | m110 | $1.57 \pm 0.05$ | Johnson & Mitchell (1995) |
| HD113226 | 1250 | 310 | $488.20 \pm 24.60$ | Smith et al. (2004) |
| HD113226 | Johnson | J | $1.23 \pm 0.05$ | Noguchi et al. (1981) |
| HD113226 | Johnson | J | $1.29 \pm 0.05$ | Alonso et al. (1998) |
| HD113226 | Johnson | J | $1.29 \pm 0.05$ | Ducati (2002) |
| HD113226 | Johnson | J | $1.31 \pm 0.01$ | Laney et al. (2012) |
| HD113226 | Johnson | J | $1.36 \pm 0.05$ | Johnson et al. (1966) |
| HD113226 | Johnson | J | $1.36 \pm 0.05$ | Shenavrin et al. (2011) |
| HD113226 | Johnson | H | $0.77 \pm 0.05$ | Noguchi et al. (1981) |
| HD113226 | Johnson | H | $0.77 \pm 0.05$ | Ducati (2002) |
| HD113226 | Johnson | H | $0.87 \pm 0.05$ | Alonso et al. (1998) |
| HD113226 | Johnson | H | $0.89 \pm 0.01$ | Laney et al. (2012) |
| HD113226 | Johnson | H | $0.91 \pm 0.05$ | Shenavrin et al. (2011) |
| HD113226 | 2200 | 361 | $306.50 \pm 19.10$ | Smith et al. (2004) |
| HD113226 | Johnson | K | $0.75 \pm 0.05$ | Neugebauer & Leighton (1969) |

**Table 21** *(continued)*

| Star ID | System/Wvlen | Band/Bandpass | Value | Reference |
|---------|--------------|---------------|-------|-----------|
| HD113226 | Johnson | K | $0.79 \pm 0.01$ | Laney et al. (2012) |
| HD113226 | Johnson | K | $0.80 \pm 0.05$ | Johnson et al. (1966) |
| HD113226 | Johnson | K | $0.80 \pm 0.05$ | Ducati (2002) |
| HD113226 | Johnson | K | $0.80 \pm 0.05$ | Shenavrin et al. (2011) |
| HD113226 | Johnson | L | $0.75 \pm 0.05$ | Johnson et al. (1966) |
| HD113226 | Johnson | L | $0.75 \pm 0.05$ | Ducati (2002) |
| HD113226 | 3500 | 898 | $152.90 \pm 19.50$ | Smith et al. (2004) |
| HD113226 | 4900 | 712 | $75.10 \pm 10.30$ | Smith et al. (2004) |
| HD113226 | 12000 | 6384 | $15.80 \pm 21.40$ | Smith et al. (2004) |
| HD114780 | WBVR | W | $9.15 \pm 0.05$ | Kornilov et al. (1991) |
| HD114780 | Oja | m41 | $8.78 \pm 0.05$ | Häggkvist & Oja (1970) |
| HD114780 | Oja | m42 | $8.58 \pm 0.05$ | Häggkvist & Oja (1970) |
| HD114780 | KronComet | COp | $7.79 \pm 0.11$ | This work |
| HD114780 | WBVR | B | $7.35 \pm 0.05$ | Kornilov et al. (1991) |
| HD114780 | Johnson | B | $7.12 \pm 0.12$ | This work |
| HD114780 | Johnson | B | $7.28 \pm 0.05$ | Häggkvist & Oja (1969b) |
| HD114780 | Johnson | B | $7.28 \pm 0.05$ | Ducati (2002) |
| HD114780 | KronComet | Bc | $7.08 \pm 0.03$ | This work |
| HD114780 | Oja | m45 | $6.93 \pm 0.05$ | Häggkvist & Oja (1970) |
| HD114780 | KronComet | C2 | $6.22 \pm 0.04$ | This work |
| HD114780 | KronComet | Gc | $5.92 \pm 0.01$ | This work |
| HD114780 | WBVR | V | $5.78 \pm 0.05$ | Kornilov et al. (1991) |
| HD114780 | Johnson | V | $5.77 \pm 0.05$ | Häggkvist & Oja (1969b) |
| HD114780 | Johnson | V | $5.77 \pm 0.05$ | Ducati (2002) |
| HD114780 | Johnson | V | $5.82 \pm 0.05$ | This work |
| HD114780 | WBVR | R | $4.54 \pm 0.05$ | Kornilov et al. (1991) |
| HD114780 | KronComet | Rc | $4.47 \pm 0.01$ | This work |
| HD114780 | 1250 | 310 | $112.70 \pm 4.20$ | Smith et al. (2004) |
| HD114780 | Johnson | J | $2.86 \pm 0.05$ | McWilliam & Lambert (1984) |
| HD114780 | Johnson | J | $2.86 \pm 0.05$ | Ducati (2002) |
| HD114780 | 2200 | 361 | $105.00 \pm 4.40$ | Smith et al. (2004) |
| HD114780 | Johnson | K | $1.85 \pm 0.05$ | Ducati (2002) |
| HD114780 | Johnson | K | $1.92 \pm 0.05$ | Neugebauer & Leighton (1969) |
| HD114780 | 3500 | 898 | $58.10 \pm 12.90$ | Smith et al. (2004) |
| HD114780 | 4900 | 712 | $23.60 \pm 6.90$ | Smith et al. (2004) |
| HD114780 | 12000 | 6384 | $8.00 \pm 20.80$ | Smith et al. (2004) |
| HD115723 | DDO | m41 | $9.10 \pm 0.05$ | Yoss (1977) |
| HD115723 | Oja | m41 | $8.64 \pm 0.05$ | Häggkvist & Oja (1970) |
| HD115723 | DDO | m42 | $8.81 \pm 0.05$ | Yoss (1977) |
| HD115723 | Oja | m42 | $8.32 \pm 0.05$ | Häggkvist & Oja (1970) |
| HD115723 | KronComet | COp | $7.71 \pm 0.00$ | This work |

**Table 21** *(continued)*

| Star ID | System/Wvlen | Band/Bandpass | Value | Reference |
|---------|--------------|---------------|-------|-----------|
| HD115723 | WBVR | B | $7.22 \pm 0.05$ | Kornilov et al. (1991) |
| HD115723 | Johnson | B | $7.11 \pm 0.02$ | This work |
| HD115723 | Johnson | B | $7.14 \pm 0.02$ | Yoss & Griffin (1997) |
| HD115723 | Johnson | B | $7.17 \pm 0.05$ | Häggkvist & Oja (1969b) |
| HD115723 | KronComet | Bc | $7.00 \pm 0.01$ | This work |
| HD115723 | DDO | m45 | $7.64 \pm 0.05$ | Yoss (1977) |
| HD115723 | Oja | m45 | $6.80 \pm 0.05$ | Häggkvist & Oja (1970) |
| HD115723 | DDO | m48 | $6.33 \pm 0.05$ | Yoss (1977) |
| HD115723 | KronComet | C2 | $6.21 \pm 0.01$ | This work |
| HD115723 | KronComet | Gc | $5.98 \pm 0.00$ | This work |
| HD115723 | WBVR | V | $5.81 \pm 0.05$ | Kornilov et al. (1991) |
| HD115723 | Johnson | V | $5.78 \pm 0.02$ | Yoss & Griffin (1997) |
| HD115723 | Johnson | V | $5.82 \pm 0.05$ | Häggkvist & Oja (1969b) |
| HD115723 | Johnson | V | $5.89 \pm 0.01$ | This work |
| HD115723 | WBVR | R | $4.83 \pm 0.05$ | Kornilov et al. (1991) |
| HD115723 | KronComet | Rc | $4.60 \pm 0.01$ | This work |
| HD115723 | 1250 | 310 | $60.70 \pm 5.00$ | Smith et al. (2004) |
| HD115723 | 2200 | 361 | $48.10 \pm 4.80$ | Smith et al. (2004) |
| HD115723 | Johnson | K | $2.62 \pm 0.07$ | Neugebauer & Leighton (1969) |
| HD115723 | 3500 | 898 | $23.20 \pm 4.40$ | Smith et al. (2004) |
| HD115723 | 4900 | 712 | $10.90 \pm 5.10$ | Smith et al. (2004) |
| HD115723 | 12000 | 6384 | $2.80 \pm 17.70$ | Smith et al. (2004) |
| HD116207 | KronComet | COp | $9.78 \pm 0.05$ | This work |
| HD116207 | Johnson | B | $9.19 \pm 0.05$ | This work |
| HD116207 | KronComet | Bc | $9.09 \pm 0.03$ | This work |
| HD116207 | KronComet | C2 | $7.85 \pm 0.02$ | This work |
| HD116207 | KronComet | Gc | $7.78 \pm 0.02$ | This work |
| HD116207 | Johnson | V | $7.76 \pm 0.02$ | This work |
| HD116207 | KronComet | Rc | $6.24 \pm 0.02$ | This work |
| HD116207 | 1250 | 310 | $98.20 \pm 7.60$ | Smith et al. (2004) |
| HD116207 | 2200 | 361 | $99.00 \pm 6.90$ | Smith et al. (2004) |
| HD116207 | Johnson | K | $2.23 \pm 0.06$ | Neugebauer & Leighton (1969) |
| HD116207 | 3500 | 898 | $49.90 \pm 5.00$ | Smith et al. (2004) |
| HD116207 | 4900 | 712 | $21.30 \pm 5.00$ | Smith et al. (2004) |
| HD116207 | 12000 | 6384 | $6.80 \pm 21.80$ | Smith et al. (2004) |
| HD118669 | KronComet | COp | $9.67 \pm 0.03$ | This work |
| HD118669 | Johnson | B | $9.08 \pm 0.08$ | This work |
| HD118669 | KronComet | Bc | $9.00 \pm 0.03$ | This work |
| HD118669 | KronComet | C2 | $7.88 \pm 0.04$ | This work |
| HD118669 | KronComet | Gc | $7.75 \pm 0.04$ | This work |
| HD118669 | Johnson | V | $7.72 \pm 0.03$ | This work |

**Table 21** *(continued)*

| Star ID | System/Wvlen | Band/Bandpass | Value | Reference |
|---------|--------------|---------------|-------|-----------|
| HD118669 | KronComet | Rc | $6.29 \pm 0.04$ | This work |
| HD118669 | 1250 | 310 | $49.50 \pm 7.00$ | Smith et al. (2004) |
| HD118669 | 2200 | 361 | $50.10 \pm 6.20$ | Smith et al. (2004) |
| HD118669 | Johnson | K | $2.65 \pm 0.07$ | Neugebauer & Leighton (1969) |
| HD118669 | 3500 | 898 | $24.60 \pm 4.40$ | Smith et al. (2004) |
| HD118669 | 4900 | 712 | $11.10 \pm 5.00$ | Smith et al. (2004) |
| HD118669 | 12000 | 6384 | $-9.70 \pm 15.80$ | Smith et al. (2004) |
| HD119584 | KronComet | COp | $8.14 \pm 0.03$ | This work |
| HD119584 | WBVR | B | $7.61 \pm 0.05$ | Kornilov et al. (1991) |
| HD119584 | Johnson | B | $7.49 \pm 0.03$ | This work |
| HD119584 | Johnson | B | $7.54 \pm 0.02$ | Yoss & Griffin (1997) |
| HD119584 | Johnson | B | $7.55 \pm 0.05$ | Häggkvist & Oja (1969b) |
| HD119584 | Johnson | B | $7.59 \pm 0.05$ | Mermilliod (1986) |
| HD119584 | KronComet | Bc | $7.38 \pm 0.01$ | This work |
| HD119584 | KronComet | C2 | $6.55 \pm 0.01$ | This work |
| HD119584 | KronComet | Gc | $6.29 \pm 0.01$ | This work |
| HD119584 | WBVR | V | $6.12 \pm 0.05$ | Kornilov et al. (1991) |
| HD119584 | Johnson | V | $6.09 \pm 0.02$ | Yoss & Griffin (1997) |
| HD119584 | Johnson | V | $6.13 \pm 0.05$ | Häggkvist & Oja (1969b) |
| HD119584 | Johnson | V | $6.14 \pm 0.05$ | Mermilliod (1986) |
| HD119584 | Johnson | V | $6.17 \pm 0.01$ | This work |
| HD119584 | WBVR | R | $5.04 \pm 0.05$ | Kornilov et al. (1991) |
| HD119584 | KronComet | Rc | $4.81 \pm 0.01$ | This work |
| HD119584 | 1250 | 310 | $55.40 \pm 5.90$ | Smith et al. (2004) |
| HD119584 | 2200 | 361 | $50.60 \pm 7.90$ | Smith et al. (2004) |
| HD119584 | Johnson | K | $2.64 \pm 0.07$ | Neugebauer & Leighton (1969) |
| HD119584 | 3500 | 898 | $24.30 \pm 6.00$ | Smith et al. (2004) |
| HD119584 | 4900 | 712 | $11.20 \pm 5.00$ | Smith et al. (2004) |
| HD119584 | 12000 | 6384 | $1.80 \pm 18.50$ | Smith et al. (2004) |
| HD120819 | Oja | m41 | $9.07 \pm 0.05$ | Häggkvist & Oja (1970) |
| HD120819 | Oja | m42 | $8.88 \pm 0.05$ | Häggkvist & Oja (1970) |
| HD120819 | KronComet | COp | $8.22 \pm 0.03$ | This work |
| HD120819 | WBVR | B | $7.57 \pm 0.05$ | Kornilov et al. (1991) |
| HD120819 | Johnson | B | $7.40 \pm 0.04$ | This work |
| HD120819 | Johnson | B | $7.48 \pm 0.05$ | Mermilliod (1986) |
| HD120819 | Johnson | B | $7.49 \pm 0.05$ | Ljunggren & Oja (1965) |
| HD120819 | Johnson | B | $7.49 \pm 0.05$ | Ducati (2002) |
| HD120819 | KronComet | Bc | $7.34 \pm 0.01$ | This work |
| HD120819 | Oja | m45 | $7.16 \pm 0.05$ | Häggkvist & Oja (1970) |
| HD120819 | Vilnius | Y | $7.07 \pm 0.05$ | Straizys & Meistas (1989) |
| HD120819 | KronComet | C2 | $6.35 \pm 0.01$ | This work |

**Table 21** *(continued)*

| Star ID | System/Wvlen | Band/Bandpass | Value | Reference |
|---------|--------------|---------------|-------|-----------|
| HD120819 | Vilnius | Z | $6.44 \pm 0.05$ | Straizys & Meistas (1989) |
| HD120819 | KronComet | Gc | $6.05 \pm 0.01$ | This work |
| HD120819 | WBVR | V | $5.90 \pm 0.05$ | Kornilov et al. (1991) |
| HD120819 | Vilnius | V | $5.87 \pm 0.05$ | Straizys & Meistas (1989) |
| HD120819 | Johnson | V | $5.87 \pm 0.05$ | Ljunggren & Oja (1965) |
| HD120819 | Johnson | V | $5.87 \pm 0.05$ | Ducati (2002) |
| HD120819 | Johnson | V | $5.88 \pm 0.05$ | Mermilliod (1986) |
| HD120819 | Johnson | V | $5.97 \pm 0.01$ | This work |
| HD120819 | Vilnius | S | $4.75 \pm 0.05$ | Straizys & Meistas (1989) |
| HD120819 | WBVR | R | $4.56 \pm 0.05$ | Kornilov et al. (1991) |
| HD120819 | KronComet | Rc | $4.50 \pm 0.01$ | This work |
| HD120819 | Johnson | J | $2.72 \pm 0.05$ | McWilliam & Lambert (1984) |
| HD120819 | Johnson | J | $2.72 \pm 0.05$ | Ducati (2002) |
| HD120819 | Johnson | K | $1.67 \pm 0.05$ | Ducati (2002) |
| HD120819 | Johnson | K | $1.69 \pm 0.04$ | Neugebauer & Leighton (1969) |
| HD121860 | KronComet | COp | $9.52 \pm 0.03$ | This work |
| HD121860 | Johnson | B | $8.87 \pm 0.05$ | This work |
| HD121860 | KronComet | Bc | $8.75 \pm 0.01$ | This work |
| HD121860 | KronComet | C2 | $7.61 \pm 0.01$ | This work |
| HD121860 | KronComet | Gc | $7.47 \pm 0.01$ | This work |
| HD121860 | Johnson | V | $7.43 \pm 0.01$ | This work |
| HD121860 | KronComet | Rc | $5.96 \pm 0.01$ | This work |
| HD121860 | 1250 | 310 | $70.80 \pm 7.70$ | Smith et al. (2004) |
| HD121860 | 2200 | 361 | $75.50 \pm 8.10$ | Smith et al. (2004) |
| HD121860 | Johnson | K | $2.26 \pm 0.05$ | Neugebauer & Leighton (1969) |
| HD121860 | 3500 | 898 | $38.20 \pm 5.50$ | Smith et al. (2004) |
| HD121860 | 4900 | 712 | $15.90 \pm 5.20$ | Smith et al. (2004) |
| HD121860 | 12000 | 6384 | $-18.90 \pm 19.60$ | Smith et al. (2004) |
| HD122316 | KronComet | COp | $10.47 \pm 0.05$ | This work |
| HD122316 | Johnson | B | $10.29 \pm 0.05$ | Haggkvist & Oja (1973) |
| HD122316 | Johnson | B | $10.29 \pm 0.06$ | This work |
| HD122316 | KronComet | Bc | $10.60 \pm 0.08$ | This work |
| HD122316 | KronComet | C2 | $8.82 \pm 0.02$ | This work |
| HD122316 | KronComet | Gc | $9.13 \pm 0.08$ | This work |
| HD122316 | Johnson | V | $8.80 \pm 0.05$ | Haggkvist & Oja (1973) |
| HD122316 | Johnson | V | $8.82 \pm 0.11$ | This work |
| HD122316 | KronComet | Rc | $6.96 \pm 0.06$ | This work |
| HD122316 | 1250 | 310 | $200.20 \pm 16.10$ | Smith et al. (2004) |
| HD122316 | 2200 | 361 | $235.60 \pm 5.80$ | Smith et al. (2004) |
| HD122316 | Johnson | K | $1.10 \pm 0.03$ | Neugebauer & Leighton (1969) |
| HD122316 | 3500 | 898 | $129.10 \pm 6.80$ | Smith et al. (2004) |

**Table 21** *(continued)*

| Star ID | System/Wvlen | Band/Bandpass | Value | Reference |
|---------|--------------|---------------|-------|-----------|
| HD122316 | 4900 | 712 | $60.50 \pm 4.60$ | Smith et al. (2004) |
| HD122316 | 12000 | 6384 | $33.40 \pm 16.60$ | Smith et al. (2004) |
| HD126307 | KronComet | COp | $8.64 \pm 0.00$ | This work |
| HD126307 | WBVR | B | $8.04 \pm 0.05$ | Kornilov et al. (1991) |
| HD126307 | Johnson | B | $7.90 \pm 0.01$ | This work |
| HD126307 | Johnson | B | $7.99 \pm 0.05$ | Bakos (1968) |
| HD126307 | KronComet | Bc | $7.85 \pm 0.00$ | This work |
| HD126307 | KronComet | C2 | $6.85 \pm 0.01$ | This work |
| HD126307 | KronComet | Gc | $6.55 \pm 0.01$ | This work |
| HD126307 | WBVR | V | $6.39 \pm 0.05$ | Kornilov et al. (1991) |
| HD126307 | Johnson | V | $6.38 \pm 0.05$ | Bakos (1968) |
| HD126307 | Johnson | V | $6.48 \pm 0.01$ | This work |
| HD126307 | WBVR | R | $5.14 \pm 0.05$ | Kornilov et al. (1991) |
| HD126307 | KronComet | Rc | $4.99 \pm 0.01$ | This work |
| HD126307 | 2200 | 361 | $93.20 \pm 9.80$ | Smith et al. (2004) |
| HD126307 | Johnson | K | $2.48 \pm 0.11$ | Neugebauer & Leighton (1969) |
| HD126307 | 3500 | 898 | $46.60 \pm 6.50$ | Smith et al. (2004) |
| HD126307 | 4900 | 712 | $20.70 \pm 4.80$ | Smith et al. (2004) |
| HD126307 | 12000 | 6384 | $-18.50 \pm 17.00$ | Smith et al. (2004) |
| HD127093 | KronComet | COp | $8.81 \pm 0.07$ | This work |
| HD127093 | WBVR | B | $8.33 \pm 0.05$ | Kornilov et al. (1991) |
| HD127093 | Johnson | B | $8.16 \pm 0.06$ | This work |
| HD127093 | Johnson | B | $8.28 \pm 0.05$ | Bakos (1968) |
| HD127093 | KronComet | Bc | $8.08 \pm 0.05$ | This work |
| HD127093 | KronComet | C2 | $6.91 \pm 0.13$ | This work |
| HD127093 | KronComet | Gc | $6.79 \pm 0.03$ | This work |
| HD127093 | WBVR | V | $6.71 \pm 0.05$ | Kornilov et al. (1991) |
| HD127093 | Johnson | V | $6.69 \pm 0.05$ | Bakos (1968) |
| HD127093 | Johnson | V | $6.72 \pm 0.04$ | This work |
| HD127093 | WBVR | R | $5.03 \pm 0.05$ | Kornilov et al. (1991) |
| HD127093 | KronComet | Rc | $5.27 \pm 0.04$ | This work |
| HD127093 | 1250 | 310 | $126.20 \pm 98.50$ | Smith et al. (2004) |
| HD127093 | 2200 | 361 | $130.00 \pm 238.10$ | Smith et al. (2004) |
| HD127093 | Johnson | K | $1.72 \pm 0.05$ | Neugebauer & Leighton (1969) |
| HD127093 | 3500 | 898 | $67.00 \pm 4.80$ | Smith et al. (2004) |
| HD127093 | 4900 | 712 | $28.30 \pm 4.60$ | Smith et al. (2004) |
| HD127093 | 12000 | 6384 | $7.50 \pm 42.60$ | Smith et al. (2004) |
| HD128902 | Vilnius | U | $10.89 \pm 0.05$ | Straizys & Meistas (1989) |
| HD128902 | DDO | m35 | $10.36 \pm 0.05$ | McClure & Forrester (1981) |
| HD128902 | WBVR | W | $8.87 \pm 0.05$ | Kornilov et al. (1991) |
| HD128902 | Johnson | U | $8.81 \pm 0.05$ | Rybka (1969) |

**Table 21** *continued on next page*

**Table 21** *(continued)*

| Star ID | System/Wvlen | Band/Bandpass | Value | Reference |
|---------|--------------|---------------|-------|-----------|
| HD128902 | Johnson | U | $8.85 \pm 0.05$ | Argue (1963) |
| HD128902 | Johnson | U | $8.87 \pm 0.05$ | Roman (1955) |
| HD128902 | Johnson | U | $8.87 \pm 0.05$ | Johnson et al. (1966) |
| HD128902 | Vilnius | P | $10.09 \pm 0.05$ | Straizys & Meistas (1989) |
| HD128902 | DDO | m38 | $9.06 \pm 0.05$ | McClure & Forrester (1981) |
| HD128902 | Vilnius | X | $8.73 \pm 0.05$ | Straizys & Meistas (1989) |
| HD128902 | DDO | m41 | $9.14 \pm 0.05$ | McClure & Forrester (1981) |
| HD128902 | Oja | m41 | $8.63 \pm 0.05$ | Häggkvist & Oja (1970) |
| HD128902 | DDO | m42 | $8.93 \pm 0.05$ | McClure & Forrester (1981) |
| HD128902 | Oja | m42 | $8.40 \pm 0.05$ | Häggkvist & Oja (1970) |
| HD128902 | WBVR | B | $7.28 \pm 0.05$ | Kornilov et al. (1991) |
| HD128902 | Johnson | B | $7.17 \pm 0.05$ | Argue (1963) |
| HD128902 | Johnson | B | $7.18 \pm 0.05$ | Roman (1955) |
| HD128902 | Johnson | B | $7.18 \pm 0.05$ | Johnson et al. (1966) |
| HD128902 | Johnson | B | $7.20 \pm 0.05$ | Rybka (1969) |
| HD128902 | DDO | m45 | $7.73 \pm 0.05$ | McClure & Forrester (1981) |
| HD128902 | Oja | m45 | $6.87 \pm 0.05$ | Häggkvist & Oja (1970) |
| HD128902 | Vilnius | Y | $6.79 \pm 0.05$ | Straizys & Meistas (1989) |
| HD128902 | DDO | m48 | $6.35 \pm 0.05$ | McClure & Forrester (1981) |
| HD128902 | Vilnius | Z | $6.20 \pm 0.05$ | Straizys & Meistas (1989) |
| HD128902 | WBVR | V | $5.75 \pm 0.05$ | Kornilov et al. (1991) |
| HD128902 | Vilnius | V | $5.70 \pm 0.05$ | Straizys & Meistas (1989) |
| HD128902 | Johnson | V | $5.69 \pm 0.05$ | Argue (1963) |
| HD128902 | Johnson | V | $5.70 \pm 0.05$ | Roman (1955) |
| HD128902 | Johnson | V | $5.70 \pm 0.05$ | Johnson et al. (1966) |
| HD128902 | Johnson | V | $5.72 \pm 0.05$ | Rybka (1969) |
| HD128902 | Vilnius | S | $4.68 \pm 0.05$ | Straizys & Meistas (1989) |
| HD128902 | WBVR | R | $4.62 \pm 0.05$ | Kornilov et al. (1991) |
| HD128902 | 1250 | 310 | $90.50 \pm 5.10$ | Smith et al. (2004) |
| HD128902 | 2200 | 361 | $81.20 \pm 4.90$ | Smith et al. (2004) |
| HD128902 | Johnson | K | $2.20 \pm 0.05$ | Neugebauer & Leighton (1969) |
| HD128902 | 3500 | 898 | $39.70 \pm 4.00$ | Smith et al. (2004) |
| HD128902 | 4900 | 712 | $18.50 \pm 4.70$ | Smith et al. (2004) |
| HD128902 | 12000 | 6384 | $3.80 \pm 16.20$ | Smith et al. (2004) |
| HD129312 | Geneva | U | $7.13 \pm 0.08$ | Golay (1972) |
| HD129312 | DDO | m35 | $7.96 \pm 0.05$ | McClure & Forrester (1981) |
| HD129312 | DDO | m35 | $7.96 \pm 0.05$ | Mermilliod & Nitschelm (1989) |
| HD129312 | DDO | m35 | $7.98 \pm 0.05$ | Cousins & Caldwell (1996) |
| HD129312 | Stromgren | u | $7.92 \pm 0.08$ | Pilachowski (1978) |
| HD129312 | Stromgren | u | $7.92 \pm 0.08$ | Hauck & Mermilliod (1998) |
| HD129312 | Stromgren | u | $7.93 \pm 0.08$ | Olsen (1993) |

**Table 21** *(continued)*

| Star ID | System/Wvlen | Band/Bandpass | Value | Reference |
|---------|--------------|---------------|-------|-----------|
| HD129312 | WBVR | W | $6.44 \pm 0.05$ | Kornilov et al. (1991) |
| HD129312 | Johnson | U | $6.62 \pm 0.05$ | Argue (1963) |
| HD129312 | Johnson | U | $6.62 \pm 0.05$ | Gutierrez-Moreno & et al. (1966) |
| HD129312 | Johnson | U | $6.63 \pm 0.05$ | Cousins (1964a) |
| HD129312 | Johnson | U | $6.63 \pm 0.05$ | Cousins (1984) |
| HD129312 | Johnson | U | $6.64 \pm 0.05$ | Jennens & Helfer (1975) |
| HD129312 | DDO | m38 | $6.88 \pm 0.05$ | McClure & Forrester (1981) |
| HD129312 | DDO | m38 | $6.88 \pm 0.05$ | Mermilliod & Nitschelm (1989) |
| HD129312 | DDO | m38 | $6.90 \pm 0.05$ | Cousins & Caldwell (1996) |
| HD129312 | Geneva | B1 | $6.46 \pm 0.08$ | Golay (1972) |
| HD129312 | DDO | m41 | $7.49 \pm 0.05$ | McClure & Forrester (1981) |
| HD129312 | DDO | m41 | $7.49 \pm 0.05$ | Mermilliod & Nitschelm (1989) |
| HD129312 | DDO | m41 | $7.50 \pm 0.05$ | Cousins & Caldwell (1996) |
| HD129312 | Oja | m41 | $7.00 \pm 0.05$ | Häggkvist & Oja (1970) |
| HD129312 | Stromgren | v | $6.52 \pm 0.08$ | Pilachowski (1978) |
| HD129312 | Stromgren | v | $6.52 \pm 0.08$ | Olsen (1993) |
| HD129312 | Stromgren | v | $6.52 \pm 0.08$ | Hauck & Mermilliod (1998) |
| HD129312 | DDO | m42 | $7.24 \pm 0.05$ | Cousins & Caldwell (1996) |
| HD129312 | DDO | m42 | $7.25 \pm 0.05$ | McClure & Forrester (1981) |
| HD129312 | DDO | m42 | $7.25 \pm 0.05$ | Mermilliod & Nitschelm (1989) |
| HD129312 | Oja | m42 | $6.76 \pm 0.05$ | Häggkvist & Oja (1970) |
| HD129312 | Geneva | B | $5.18 \pm 0.08$ | Golay (1972) |
| HD129312 | WBVR | B | $5.85 \pm 0.08$ | Kornilov et al. (1991) |
| HD129312 | Johnson | B | $5.84 \pm 0.05$ | Häggkvist & Oja (1966) |
| HD129312 | Johnson | B | $5.86 \pm 0.05$ | Cousins (1964a) |
| HD129312 | Johnson | B | $5.86 \pm 0.05$ | Gutierrez-Moreno & et al. (1966) |
| HD129312 | Johnson | B | $5.87 \pm 0.05$ | Argue (1963) |
| HD129312 | Johnson | B | $5.87 \pm 0.05$ | Cousins (1984) |
| HD129312 | Johnson | B | $5.88 \pm 0.05$ | Jennens & Helfer (1975) |
| HD129312 | Geneva | B2 | $6.33 \pm 0.08$ | Golay (1972) |
| HD129312 | DDO | m45 | $6.43 \pm 0.05$ | McClure & Forrester (1981) |
| HD129312 | DDO | m45 | $6.43 \pm 0.05$ | Mermilliod & Nitschelm (1989) |
| HD129312 | DDO | m45 | $6.43 \pm 0.05$ | Cousins & Caldwell (1996) |
| HD129312 | Oja | m45 | $5.57 \pm 0.05$ | Häggkvist & Oja (1970) |
| HD129312 | Stromgren | b | $5.49 \pm 0.08$ | Pilachowski (1978) |
| HD129312 | Stromgren | b | $5.49 \pm 0.08$ | Olsen (1993) |
| HD129312 | Stromgren | b | $5.49 \pm 0.08$ | Hauck & Mermilliod (1998) |
| HD129312 | DDO | m48 | $5.24 \pm 0.05$ | McClure & Forrester (1981) |
| HD129312 | DDO | m48 | $5.24 \pm 0.05$ | Mermilliod & Nitschelm (1989) |
| HD129312 | DDO | m48 | $5.24 \pm 0.05$ | Cousins & Caldwell (1996) |
| HD129312 | Geneva | V1 | $5.65 \pm 0.08$ | Golay (1972) |

**Table 21** *(continued)*

| Star ID | System/Wvlen | Band/Bandpass | Value | Reference |
|---------|--------------|---------------|-------|-----------|
| HD129312 | WBVR | V | $4.85 \pm 0.05$ | Kornilov et al. (1991) |
| HD129312 | Stromgren | y | $4.87 \pm 0.08$ | Pilachowski (1978) |
| HD129312 | Stromgren | y | $4.87 \pm 0.08$ | Olsen (1993) |
| HD129312 | Stromgren | y | $4.87 \pm 0.08$ | Hauck & Mermilliod (1998) |
| HD129312 | Geneva | V | $4.88 \pm 0.08$ | Golay (1972) |
| HD129312 | Johnson | V | $4.85 \pm 0.05$ | Häggkvist & Oja (1966) |
| HD129312 | Johnson | V | $4.85 \pm 0.05$ | Gutierrez-Moreno & et al. (1966) |
| HD129312 | Johnson | V | $4.86 \pm 0.05$ | Argue (1963) |
| HD129312 | Johnson | V | $4.86 \pm 0.05$ | Cousins (1964a) |
| HD129312 | Johnson | V | $4.87 \pm 0.05$ | Jennens & Helfer (1975) |
| HD129312 | Johnson | V | $4.87 \pm 0.05$ | Cousins (1984) |
| HD129312 | Geneva | G | $5.87 \pm 0.08$ | Golay (1972) |
| HD129312 | WBVR | R | $4.13 \pm 0.05$ | Kornilov et al. (1991) |
| HD129312 | 1250 | 310 | $82.50 \pm 7.70$ | Smith et al. (2004) |
| HD129312 | Johnson | J | $3.17 \pm 0.05$ | Alonso et al. (1998) |
| HD129312 | Johnson | H | $2.72 \pm 0.05$ | Alonso et al. (1998) |
| HD129312 | 2200 | 361 | $52.90 \pm 7.70$ | Smith et al. (2004) |
| HD129312 | Johnson | K | $2.56 \pm 0.06$ | Neugebauer & Leighton (1969) |
| HD129312 | 3500 | 898 | $25.20 \pm 6.00$ | Smith et al. (2004) |
| HD129312 | 4900 | 712 | $11.50 \pm 4.80$ | Smith et al. (2004) |
| HD129312 | 12000 | 6384 | $0.80 \pm 19.80$ | Smith et al. (2004) |
| HD130084 | KronComet | COp | $8.56 \pm 0.04$ | This work |
| HD130084 | WBVR | B | $7.93 \pm 0.05$ | Kornilov et al. (1991) |
| HD130084 | Johnson | B | $7.62 \pm 0.10$ | This work |
| HD130084 | Johnson | B | $7.82 \pm 0.05$ | Mermilliod (1986) |
| HD130084 | Johnson | B | $7.86 \pm 0.05$ | Häggkvist & Oja (1969b) |
| HD130084 | Johnson | B | $7.86 \pm 0.05$ | Ducati (2002) |
| HD130084 | KronComet | Bc | $7.67 \pm 0.01$ | This work |
| HD130084 | KronComet | C2 | $6.71 \pm 0.02$ | This work |
| HD130084 | KronComet | Gc | $6.46 \pm 0.01$ | This work |
| HD130084 | WBVR | V | $6.29 \pm 0.05$ | Kornilov et al. (1991) |
| HD130084 | Johnson | V | $6.24 \pm 0.05$ | Mermilliod (1986) |
| HD130084 | Johnson | V | $6.28 \pm 0.05$ | Häggkvist & Oja (1969b) |
| HD130084 | Johnson | V | $6.28 \pm 0.05$ | Ducati (2002) |
| HD130084 | Johnson | V | $6.33 \pm 0.03$ | This work |
| HD130084 | WBVR | R | $4.92 \pm 0.05$ | Kornilov et al. (1991) |
| HD130084 | KronComet | Rc | $4.92 \pm 0.01$ | This work |
| HD130084 | 1250 | 310 | $90.00 \pm 14.80$ | Smith et al. (2004) |
| HD130084 | Johnson | J | $3.07 \pm 0.05$ | McWilliam & Lambert (1984) |
| HD130084 | Johnson | J | $3.07 \pm 0.05$ | Ducati (2002) |
| HD130084 | 2200 | 361 | $86.60 \pm 5.80$ | Smith et al. (2004) |

**Table 21** (continued)

| Star ID | System/Wvlen | Band/Bandpass | Value | Reference |
|---------|--------------|---------------|-------|-----------|
| HD130084 | Johnson | K | $2.03 \pm 0.05$ | Neugebauer & Leighton (1969) |
| HD130084 | Johnson | K | $2.05 \pm 0.05$ | Ducati (2002) |
| HD130084 | 3500 | 898 | $43.10 \pm 5.70$ | Smith et al. (2004) |
| HD130084 | 4900 | 712 | $18.80 \pm 5.30$ | Smith et al. (2004) |
| HD130084 | 12000 | 6384 | $1.40 \pm 24.80$ | Smith et al. (2004) |
| HD131507 | DDO | m35 | $9.85 \pm 0.05$ | Clark & McClure (1979) |
| HD131507 | WBVR | W | $8.39 \pm 0.05$ | Kornilov et al. (1991) |
| HD131507 | Johnson | U | $8.42 \pm 0.05$ | Roman (1955) |
| HD131507 | Johnson | U | $8.42 \pm 0.05$ | Argue (1963) |
| HD131507 | Johnson | U | $8.42 \pm 0.05$ | Johnson et al. (1966) |
| HD131507 | DDO | m38 | $8.64 \pm 0.05$ | Clark & McClure (1979) |
| HD131507 | DDO | m41 | $8.82 \pm 0.05$ | Clark & McClure (1979) |
| HD131507 | Oja | m41 | $8.30 \pm 0.05$ | Häggkvist & Oja (1970) |
| HD131507 | DDO | m42 | $8.54 \pm 0.05$ | Clark & McClure (1979) |
| HD131507 | Oja | m42 | $8.04 \pm 0.05$ | Häggkvist & Oja (1970) |
| HD131507 | WBVR | B | $6.91 \pm 0.05$ | Kornilov et al. (1991) |
| HD131507 | Johnson | B | $6.82 \pm 0.05$ | Roman (1955) |
| HD131507 | Johnson | B | $6.82 \pm 0.05$ | Argue (1963) |
| HD131507 | Johnson | B | $6.82 \pm 0.05$ | Johnson et al. (1966) |
| HD131507 | DDO | m45 | $7.34 \pm 0.05$ | Clark & McClure (1979) |
| HD131507 | Oja | m45 | $6.49 \pm 0.05$ | Häggkvist & Oja (1970) |
| HD131507 | DDO | m48 | $6.03 \pm 0.05$ | Clark & McClure (1979) |
| HD131507 | WBVR | V | $5.50 \pm 0.05$ | Kornilov et al. (1991) |
| HD131507 | Johnson | V | $5.44 \pm 0.05$ | Argue (1963) |
| HD131507 | Johnson | V | $5.46 \pm 0.05$ | Roman (1955) |
| HD131507 | Johnson | V | $5.46 \pm 0.05$ | Johnson et al. (1966) |
| HD131507 | WBVR | R | $4.48 \pm 0.05$ | Kornilov et al. (1991) |
| HD131507 | 1250 | 310 | $93.60 \pm 6.70$ | Smith et al. (2004) |
| HD131507 | 2200 | 361 | $75.50 \pm 5.90$ | Smith et al. (2004) |
| HD131507 | Johnson | K | $2.24 \pm 0.05$ | Neugebauer & Leighton (1969) |
| HD131507 | 3500 | 898 | $36.70 \pm 4.30$ | Smith et al. (2004) |
| HD131507 | 4900 | 712 | $16.80 \pm 4.80$ | Smith et al. (2004) |
| HD131507 | 12000 | 6384 | $4.30 \pm 15.50$ | Smith et al. (2004) |
| HD133124 | WBVR | B | $6.36 \pm 0.05$ | Kornilov et al. (1991) |
| HD133124 | Johnson | B | $6.30 \pm 0.05$ | Argue (1963) |
| HD133124 | Johnson | B | $6.30 \pm 0.05$ | Ljunggren & Oja (1965) |
| HD133124 | Johnson | B | $6.31 \pm 0.05$ | Ducati (2002) |
| HD133124 | Johnson | B | $6.32 \pm 0.05$ | Johnson et al. (1966) |
| HD133124 | Johnson | B | $6.32 \pm 0.05$ | Jennens & Helfer (1975) |
| HD133124 | Johnson | B | $6.32 \pm 0.05$ | Mermilliod (1986) |
| HD133124 | 13c | m45 | $5.85 \pm 0.05$ | Johnson & Mitchell (1995) |

**Table 21** *(continued)*

| Star ID | System/Wvlen | Band/Bandpass | Value | Reference |
|---------|--------------|---------------|-------|-----------|
| HD133124 | 13c | m52 | $5.22 \pm 0.05$ | Johnson & Mitchell (1995) |
| HD133124 | WBVR | V | $4.81 \pm 0.05$ | Kornilov et al. (1991) |
| HD133124 | Johnson | V | $4.79 \pm 0.05$ | Argue (1963) |
| HD133124 | Johnson | V | $4.80 \pm 0.05$ | Ljunggren & Oja (1965) |
| HD133124 | Johnson | V | $4.81 \pm 0.05$ | Ducati (2002) |
| HD133124 | Johnson | V | $4.82 \pm 0.05$ | Johnson et al. (1966) |
| HD133124 | Johnson | V | $4.82 \pm 0.05$ | Jennens & Helfer (1975) |
| HD133124 | Johnson | V | $4.82 \pm 0.05$ | Mermilliod (1986) |
| HD133124 | 13c | m58 | $4.44 \pm 0.05$ | Johnson & Mitchell (1995) |
| HD133124 | 13c | m63 | $4.01 \pm 0.05$ | Johnson & Mitchell (1995) |
| HD133124 | WBVR | R | $3.67 \pm 0.05$ | Kornilov et al. (1991) |
| HD133124 | 13c | m72 | $3.56 \pm 0.05$ | Johnson & Mitchell (1995) |
| HD133124 | 13c | m80 | $3.19 \pm 0.05$ | Johnson & Mitchell (1995) |
| HD133124 | 13c | m86 | $3.05 \pm 0.05$ | Johnson & Mitchell (1995) |
| HD133124 | 13c | m99 | $2.78 \pm 0.05$ | Johnson & Mitchell (1995) |
| HD133124 | 13c | m110 | $2.56 \pm 0.05$ | Johnson & Mitchell (1995) |
| HD133124 | 1250 | 310 | $224.60 \pm 7.50$ | Smith et al. (2004) |
| HD133124 | Johnson | J | $2.17 \pm 0.05$ | Selby et al. (1988) |
| HD133124 | Johnson | J | $2.17 \pm 0.05$ | Blackwell et al. (1990) |
| HD133124 | Johnson | J | $2.17 \pm 0.05$ | Ducati (2002) |
| HD133124 | 2200 | 361 | $194.20 \pm 10.10$ | Smith et al. (2004) |
| HD133124 | Johnson | K | $1.23 \pm 0.05$ | Ducati (2002) |
| HD133124 | Johnson | K | $1.30 \pm 0.04$ | Neugebauer & Leighton (1969) |
| HD133124 | 3500 | 898 | $95.60 \pm 6.50$ | Smith et al. (2004) |
| HD133124 | 4900 | 712 | $43.00 \pm 5.40$ | Smith et al. (2004) |
| HD133124 | 12000 | 6384 | $9.10 \pm 17.70$ | Smith et al. (2004) |
| HD133208 | 13c | m33 | $5.22 \pm 0.05$ | Johnson & Mitchell (1995) |
| HD133208 | Geneva | U | $5.65 \pm 0.08$ | Golay (1972) |
| HD133208 | Vilnius | U | $6.84 \pm 0.05$ | Straizys et al. (1989a) |
| HD133208 | Vilnius | U | $6.92 \pm 0.05$ | Zdanavicius et al. (1969) |
| HD133208 | 13c | m35 | $5.03 \pm 0.05$ | Johnson & Mitchell (1995) |
| HD133208 | DDO | m35 | $6.47 \pm 0.05$ | McClure & Forrester (1981) |
| HD133208 | WBVR | W | $4.99 \pm 0.05$ | Kornilov et al. (1991) |
| HD133208 | Johnson | U | $5.16 \pm 0.05$ | Argue (1963) |
| HD133208 | Johnson | U | $5.18 \pm 0.05$ | Johnson et al. (1966) |
| HD133208 | Johnson | U | $5.19 \pm 0.05$ | Johnson et al. (1966) |
| HD133208 | Johnson | U | $5.21 \pm 0.05$ | Ducati (2002) |
| HD133208 | Johnson | U | $5.24 \pm 0.05$ | Fernie (1983) |
| HD133208 | 13c | m37 | $5.05 \pm 0.05$ | Johnson & Mitchell (1995) |
| HD133208 | Vilnius | P | $6.27 \pm 0.05$ | Straizys et al. (1989a) |
| HD133208 | Vilnius | P | $6.34 \pm 0.05$ | Zdanavicius et al. (1969) |

**Table 21** *continued on next page*

**Table 21** (continued)

| Star ID | System/Wvlen | Band/Bandpass | Value | Reference |
|---------|--------------|---------------|-------|-----------|
| HD133208 | DDO | m38 | $5.42 \pm 0.05$ | McClure & Forrester (1981) |
| HD133208 | 13c | m40 | $4.97 \pm 0.05$ | Johnson & Mitchell (1995) |
| HD133208 | Geneva | B1 | $4.99 \pm 0.08$ | Golay (1972) |
| HD133208 | Vilnius | X | $5.37 \pm 0.05$ | Straizys et al. (1989a) |
| HD133208 | Vilnius | X | $5.42 \pm 0.05$ | Zdanavicius et al. (1969) |
| HD133208 | DDO | m41 | $6.00 \pm 0.05$ | McClure & Forrester (1981) |
| HD133208 | Oja | m41 | $5.55 \pm 0.05$ | Häggkvist & Oja (1970) |
| HD133208 | DDO | m42 | $5.80 \pm 0.05$ | McClure & Forrester (1981) |
| HD133208 | Oja | m42 | $5.36 \pm 0.05$ | Häggkvist & Oja (1970) |
| HD133208 | Geneva | B | $3.73 \pm 0.08$ | Golay (1972) |
| HD133208 | WBVR | B | $4.47 \pm 0.05$ | Kornilov et al. (1991) |
| HD133208 | Johnson | B | $4.43 \pm 0.05$ | Häggkvist & Oja (1966) |
| HD133208 | Johnson | B | $4.45 \pm 0.05$ | Argue (1963) |
| HD133208 | Johnson | B | $4.47 \pm 0.05$ | Johnson et al. (1966) |
| HD133208 | Johnson | B | $4.47 \pm 0.05$ | Ducati (2002) |
| HD133208 | Johnson | B | $4.49 \pm 0.05$ | Fernie (1983) |
| HD133208 | Geneva | B2 | $4.91 \pm 0.08$ | Golay (1972) |
| HD133208 | 13c | m45 | $4.18 \pm 0.05$ | Johnson & Mitchell (1995) |
| HD133208 | DDO | m45 | $5.01 \pm 0.05$ | McClure & Forrester (1981) |
| HD133208 | Oja | m45 | $4.18 \pm 0.05$ | Häggkvist & Oja (1970) |
| HD133208 | Vilnius | Y | $4.21 \pm 0.05$ | Straizys et al. (1989a) |
| HD133208 | Vilnius | Y | $4.24 \pm 0.05$ | Zdanavicius et al. (1969) |
| HD133208 | DDO | m48 | $3.84 \pm 0.05$ | McClure & Forrester (1981) |
| HD133208 | Vilnius | Z | $3.77 \pm 0.05$ | Zdanavicius et al. (1969) |
| HD133208 | Vilnius | Z | $3.77 \pm 0.05$ | Straizys et al. (1989a) |
| HD133208 | 13c | m52 | $3.73 \pm 0.05$ | Johnson & Mitchell (1995) |
| HD133208 | Geneva | V1 | $4.27 \pm 0.05$ | Golay (1972) |
| HD133208 | WBVR | V | $3.49 \pm 0.05$ | Kornilov et al. (1991) |
| HD133208 | Vilnius | V | $3.50 \pm 0.05$ | Zdanavicius et al. (1969) |
| HD133208 | Vilnius | V | $3.50 \pm 0.05$ | Straizys et al. (1989a) |
| HD133208 | Geneva | V | $3.50 \pm 0.08$ | Golay (1972) |
| HD133208 | Johnson | V | $3.47 \pm 0.05$ | Häggkvist & Oja (1966) |
| HD133208 | Johnson | V | $3.49 \pm 0.05$ | Argue (1963) |
| HD133208 | Johnson | V | $3.50 \pm 0.05$ | Johnson et al. (1966) |
| HD133208 | Johnson | V | $3.52 \pm 0.05$ | Ducati (2002) |
| HD133208 | Johnson | V | $3.55 \pm 0.05$ | Fernie (1983) |
| HD133208 | 13c | m58 | $3.29 \pm 0.05$ | Johnson & Mitchell (1995) |
| HD133208 | Geneva | G | $4.49 \pm 0.08$ | Golay (1972) |
| HD133208 | 13c | m63 | $3.01 \pm 0.05$ | Johnson & Mitchell (1995) |
| HD133208 | Vilnius | S | $2.79 \pm 0.05$ | Straizys et al. (1989a) |
| HD133208 | Vilnius | S | $2.82 \pm 0.05$ | Zdanavicius et al. (1969) |

**Table 21** (continued)

| Star ID | System/Wvlen | Band/Bandpass | Value | Reference |
|---------|-------------|---------------|-------|-----------|
| HD133208 | WBVR | R | $2.81 \pm 0.05$ | Kornilov et al. (1991) |
| HD133208 | 13c | m72 | $2.80 \pm 0.05$ | Johnson & Mitchell (1995) |
| HD133208 | 13c | m80 | $2.56 \pm 0.05$ | Johnson & Mitchell (1995) |
| HD133208 | 13c | m86 | $2.47 \pm 0.05$ | Johnson & Mitchell (1995) |
| HD133208 | 13c | m99 | $2.34 \pm 0.05$ | Johnson & Mitchell (1995) |
| HD133208 | 13c | m110 | $2.22 \pm 0.05$ | Johnson & Mitchell (1995) |
| HD133208 | 1250 | 310 | $278.20 \pm 8.70$ | Smith et al. (2004) |
| HD133208 | Johnson | J | $1.88 \pm 0.05$ | Castor & Simon (1983) |
| HD133208 | Johnson | J | $1.89 \pm 0.05$ | Blackwell et al. (1979) |
| HD133208 | Johnson | J | $1.90 \pm 0.05$ | Ducati (2002) |
| HD133208 | Johnson | J | $1.91 \pm 0.05$ | Selby et al. (1988) |
| HD133208 | Johnson | J | $1.91 \pm 0.05$ | Blackwell et al. (1990) |
| HD133208 | Johnson | J | $1.93 \pm 0.05$ | Johnson et al. (1966) |
| HD133208 | Johnson | J | $1.93 \pm 0.05$ | Bergeat et al. (1981) |
| HD133208 | Johnson | J | $1.93 \pm 0.05$ | Shenavrin et al. (2011) |
| HD133208 | Johnson | H | $1.42 \pm 0.05$ | Castor & Simon (1983) |
| HD133208 | Johnson | H | $1.43 \pm 0.05$ | Blackwell et al. (1979) |
| HD133208 | Johnson | H | $1.43 \pm 0.05$ | Ducati (2002) |
| HD133208 | Johnson | H | $1.45 \pm 0.05$ | Bergeat et al. (1981) |
| HD133208 | Johnson | H | $1.46 \pm 0.05$ | Shenavrin et al. (2011) |
| HD133208 | 2200 | 361 | $179.70 \pm 5.70$ | Smith et al. (2004) |
| HD133208 | Johnson | K | $1.31 \pm 0.03$ | Neugebauer & Leighton (1969) |
| HD133208 | Johnson | K | $1.34 \pm 0.05$ | Johnson et al. (1966) |
| HD133208 | Johnson | K | $1.34 \pm 0.05$ | Ducati (2002) |
| HD133208 | Johnson | K | $1.34 \pm 0.05$ | Shenavrin et al. (2011) |
| HD133208 | Johnson | L | $1.28 \pm 0.05$ | Ducati (2002) |
| HD133208 | 3500 | 898 | $84.60 \pm 4.60$ | Smith et al. (2004) |
| HD133208 | 4900 | 712 | $41.90 \pm 4.80$ | Smith et al. (2004) |
| HD133208 | Johnson | M | $1.38 \pm 0.05$ | Ducati (2002) |
| HD133208 | 12000 | 6384 | $10.10 \pm 17.00$ | Smith et al. (2004) |
| HD133485 | KronComet | NH | $9.47 \pm 0.10$ | This work |
| HD133485 | KronComet | UVc | $9.23 \pm 0.07$ | This work |
| HD133485 | DDO | m35 | $9.78 \pm 0.05$ | McClure & Forrester (1981) |
| HD133485 | WBVR | W | $8.31 \pm 0.05$ | Kornilov et al. (1991) |
| HD133485 | Johnson | U | $8.03 \pm 0.17$ | This work |
| HD133485 | DDO | m38 | $8.73 \pm 0.05$ | McClure & Forrester (1981) |
| HD133485 | KronComet | CN | $9.31 \pm 0.02$ | This work |
| HD133485 | DDO | m41 | $9.22 \pm 0.05$ | McClure & Forrester (1981) |
| HD133485 | DDO | m42 | $9.05 \pm 0.05$ | McClure & Forrester (1981) |
| HD133485 | KronComet | COp | $7.89 \pm 0.04$ | This work |
| HD133485 | WBVR | B | $7.65 \pm 0.05$ | Kornilov et al. (1991) |

**Table 21** *(continued)*

| Star ID | System/Wvlen | Band/Bandpass | Value | Reference |
|---------|-------------|---------------|-------|-----------|
| HD133485 | Johnson | B | $7.37 \pm 0.08$ | This work |
| HD133485 | Johnson | B | $7.61 \pm 0.05$ | Häggkvist & Oja (1969b) |
| HD133485 | KronComet | Bc | $7.41 \pm 0.02$ | This work |
| HD133485 | DDO | m45 | $8.17 \pm 0.05$ | McClure & Forrester (1981) |
| HD133485 | DDO | m48 | $6.98 \pm 0.05$ | McClure & Forrester (1981) |
| HD133485 | KronComet | C2 | $6.73 \pm 0.02$ | This work |
| HD133485 | KronComet | Gc | $6.63 \pm 0.01$ | This work |
| HD133485 | WBVR | V | $6.60 \pm 0.05$ | Kornilov et al. (1991) |
| HD133485 | Johnson | V | $6.59 \pm 0.03$ | This work |
| HD133485 | Johnson | V | $6.59 \pm 0.05$ | Häggkvist & Oja (1969b) |
| HD133485 | WBVR | R | $5.84 \pm 0.05$ | Kornilov et al. (1991) |
| HD133485 | KronComet | Rc | $5.56 \pm 0.01$ | This work |
| HD133582 | 13c | m33 | $7.24 \pm 0.05$ | Johnson & Mitchell (1995) |
| HD133582 | KronComet | NH | $8.11 \pm 0.04$ | This work |
| HD133582 | KronComet | UVc | $7.82 \pm 0.05$ | This work |
| HD133582 | 13c | m35 | $6.93 \pm 0.05$ | Johnson & Mitchell (1995) |
| HD133582 | DDO | m35 | $8.46 \pm 0.05$ | McClure & Forrester (1981) |
| HD133582 | WBVR | W | $7.01 \pm 0.05$ | Kornilov et al. (1991) |
| HD133582 | Johnson | U | $6.46 \pm 0.05$ | This work |
| HD133582 | Johnson | U | $7.09 \pm 0.05$ | Argue (1963) |
| HD133582 | Johnson | U | $7.12 \pm 0.05$ | Johnson et al. (1966) |
| HD133582 | Johnson | U | $7.12 \pm 0.05$ | Ducati (2002) |
| HD133582 | 13c | m37 | $7.00 \pm 0.05$ | Johnson & Mitchell (1995) |
| HD133582 | DDO | m38 | $7.30 \pm 0.05$ | McClure & Forrester (1981) |
| HD133582 | KronComet | CN | $7.75 \pm 0.10$ | This work |
| HD133582 | 13c | m40 | $6.59 \pm 0.05$ | Johnson & Mitchell (1995) |
| HD133582 | DDO | m41 | $7.60 \pm 0.05$ | McClure & Forrester (1981) |
| HD133582 | Oja | m41 | $7.17 \pm 0.05$ | Häggkvist & Oja (1970) |
| HD133582 | DDO | m42 | $7.34 \pm 0.05$ | McClure & Forrester (1981) |
| HD133582 | Oja | m42 | $6.87 \pm 0.05$ | Häggkvist & Oja (1970) |
| HD133582 | KronComet | COp | $6.00 \pm 0.07$ | This work |
| HD133582 | WBVR | B | $5.82 \pm 0.05$ | Kornilov et al. (1991) |
| HD133582 | Johnson | B | $5.57 \pm 0.11$ | This work |
| HD133582 | Johnson | B | $5.77 \pm 0.05$ | Argue (1963) |
| HD133582 | Johnson | B | $5.77 \pm 0.05$ | Ljunggren & Oja (1965) |
| HD133582 | Johnson | B | $5.78 \pm 0.05$ | Johnson et al. (1966) |
| HD133582 | Johnson | B | $5.78 \pm 0.05$ | Ducati (2002) |
| HD133582 | KronComet | Bc | $5.52 \pm 0.03$ | This work |
| HD133582 | 13c | m45 | $5.38 \pm 0.05$ | Johnson & Mitchell (1995) |
| HD133582 | DDO | m45 | $6.28 \pm 0.05$ | McClure & Forrester (1981) |
| HD133582 | Oja | m45 | $5.45 \pm 0.05$ | Häggkvist & Oja (1970) |

**Table 21** *(continued)*

| Star ID | System/Wvlen | Band/Bandpass | Value | Reference |
|---------|--------------|---------------|-------|-----------|
| HD133582 | DDO | m48 | $5.00 \pm 0.05$ | McClure & Forrester (1981) |
| HD133582 | KronComet | C2 | $4.77 \pm 0.03$ | This work |
| HD133582 | 13c | m52 | $4.82 \pm 0.05$ | Johnson & Mitchell (1995) |
| HD133582 | KronComet | Gc | $4.58 \pm 0.02$ | This work |
| HD133582 | WBVR | V | $4.53 \pm 0.05$ | Kornilov et al. (1991) |
| HD133582 | Johnson | V | $4.51 \pm 0.04$ | This work |
| HD133582 | Johnson | V | $4.51 \pm 0.05$ | Argue (1963) |
| HD133582 | Johnson | V | $4.52 \pm 0.05$ | Ljunggren & Oja (1965) |
| HD133582 | Johnson | V | $4.55 \pm 0.05$ | Johnson et al. (1966) |
| HD133582 | Johnson | V | $4.55 \pm 0.05$ | Ducati (2002) |
| HD133582 | 13c | m58 | $4.21 \pm 0.05$ | Johnson & Mitchell (1995) |
| HD133582 | 13c | m63 | $3.84 \pm 0.05$ | Johnson & Mitchell (1995) |
| HD133582 | WBVR | R | $3.61 \pm 0.05$ | Kornilov et al. (1991) |
| HD133582 | KronComet | Rc | $3.32 \pm 0.02$ | This work |
| HD133582 | 13c | m72 | $3.60 \pm 0.05$ | Johnson & Mitchell (1995) |
| HD133582 | 13c | m80 | $3.31 \pm 0.05$ | Johnson & Mitchell (1995) |
| HD133582 | 13c | m86 | $3.18 \pm 0.05$ | Johnson & Mitchell (1995) |
| HD133582 | 13c | m99 | $2.97 \pm 0.05$ | Johnson & Mitchell (1995) |
| HD133582 | 13c | m110 | $2.74 \pm 0.05$ | Johnson & Mitchell (1995) |
| HD133582 | 1250 | 310 | $205.90 \pm 9.40$ | Smith et al. (2004) |
| HD133582 | Johnson | J | $2.41 \pm 0.05$ | Johnson et al. (1966) |
| HD133582 | Johnson | J | $2.41 \pm 0.05$ | Ducati (2002) |
| HD133582 | 2200 | 361 | $161.20 \pm 9.10$ | Smith et al. (2004) |
| HD133582 | Johnson | K | $1.61 \pm 0.05$ | Neugebauer & Leighton (1969) |
| HD133582 | Johnson | K | $1.64 \pm 0.05$ | Johnson et al. (1966) |
| HD133582 | Johnson | K | $1.64 \pm 0.05$ | Ducati (2002) |
| HD133582 | 3500 | 898 | $77.90 \pm 6.90$ | Smith et al. (2004) |
| HD133582 | 4900 | 712 | $36.00 \pm 5.10$ | Smith et al. (2004) |
| HD133582 | 12000 | 6384 | $9.10 \pm 16.90$ | Smith et al. (2004) |
| HD135722 | 13c | m33 | $5.15 \pm 0.05$ | Johnson & Mitchell (1995) |
| HD135722 | KronComet | NH | $5.97 \pm 0.02$ | This work |
| HD135722 | KronComet | UVc | $5.72 \pm 0.02$ | This work |
| HD135722 | Vilnius | U | $6.91 \pm 0.05$ | Zdanavicius et al. (1969) |
| HD135722 | 13c | m35 | $4.94 \pm 0.05$ | Johnson & Mitchell (1995) |
| HD135722 | DDO | m35 | $6.42 \pm 0.05$ | McClure & Forrester (1981) |
| HD135722 | Stromgren | u | $6.35 \pm 0.08$ | Crawford & Barnes (1970) |
| HD135722 | Stromgren | u | $6.36 \pm 0.08$ | Piirola (1976) |
| HD135722 | Stromgren | u | $6.36 \pm 0.08$ | Hauck & Mermilliod (1998) |
| HD135722 | WBVR | W | $4.95 \pm 0.05$ | Kornilov et al. (1991) |
| HD135722 | Johnson | U | $4.49 \pm 0.05$ | This work |
| HD135722 | Johnson | U | $5.08 \pm 0.05$ | Tolbert (1964) |

**Table 21** *(continued)*

| Star ID | System/Wvlen | Band/Bandpass | Value | Reference |
|---------|--------------|---------------|-------|-----------|
| HD135722 | Johnson | U | $5.09 \pm 0.05$ | Jennens & Helfer (1975) |
| HD135722 | Johnson | U | $5.11 \pm 0.05$ | Argue (1963) |
| HD135722 | Johnson | U | $5.11 \pm 0.05$ | Mermilliod (1986) |
| HD135722 | Johnson | U | $5.12 \pm 0.05$ | Johnson et al. (1966) |
| HD135722 | Johnson | U | $5.12 \pm 0.05$ | Ducati (2002) |
| HD135722 | Johnson | U | $5.14 \pm 0.05$ | Johnson & Morgan (1953b) |
| HD135722 | Johnson | U | $5.14 \pm 0.05$ | Johnson (1953) |
| HD135722 | 13c | m37 | $5.06 \pm 0.05$ | Johnson & Mitchell (1995) |
| HD135722 | Vilnius | P | $6.33 \pm 0.05$ | Zdanavicius et al. (1969) |
| HD135722 | DDO | m38 | $5.39 \pm 0.05$ | McClure & Forrester (1981) |
| HD135722 | KronComet | CN | $5.78 \pm 0.02$ | This work |
| HD135722 | 13c | m40 | $4.95 \pm 0.05$ | Johnson & Mitchell (1995) |
| HD135722 | Vilnius | X | $5.44 \pm 0.05$ | Zdanavicius et al. (1969) |
| HD135722 | DDO | m41 | $5.94 \pm 0.05$ | McClure & Forrester (1981) |
| HD135722 | Oja | m41 | $5.45 \pm 0.05$ | Häggkvist & Oja (1970) |
| HD135722 | Stromgren | v | $5.01 \pm 0.08$ | Crawford & Barnes (1970) |
| HD135722 | Stromgren | v | $5.01 \pm 0.08$ | Hauck & Mermilliod (1998) |
| HD135722 | Stromgren | v | $5.02 \pm 0.08$ | Piirola (1976) |
| HD135722 | DDO | m42 | $5.82 \pm 0.05$ | McClure & Forrester (1981) |
| HD135722 | Oja | m42 | $5.37 \pm 0.05$ | Häggkvist & Oja (1970) |
| HD135722 | KronComet | COp | $4.50 \pm 0.09$ | This work |
| HD135722 | WBVR | B | $4.46 \pm 0.05$ | Kornilov et al. (1991) |
| HD135722 | Johnson | B | $4.16 \pm 0.09$ | This work |
| HD135722 | Johnson | B | $4.40 \pm 0.05$ | Oja (1963) |
| HD135722 | Johnson | B | $4.41 \pm 0.05$ | Ljunggren & Oja (1965) |
| HD135722 | Johnson | B | $4.42 \pm 0.05$ | Tolbert (1964) |
| HD135722 | Johnson | B | $4.43 \pm 0.05$ | Jennens & Helfer (1975) |
| HD135722 | Johnson | B | $4.43 \pm 0.05$ | Mermilliod (1986) |
| HD135722 | Johnson | B | $4.44 \pm 0.05$ | Johnson et al. (1966) |
| HD135722 | Johnson | B | $4.44 \pm 0.05$ | Ducati (2002) |
| HD135722 | Johnson | B | $4.45 \pm 0.05$ | Johnson & Morgan (1953b) |
| HD135722 | Johnson | B | $4.45 \pm 0.05$ | Johnson (1953) |
| HD135722 | Johnson | B | $4.45 \pm 0.05$ | Argue (1963) |
| HD135722 | KronComet | Bc | $4.17 \pm 0.01$ | This work |
| HD135722 | 13c | m45 | $4.19 \pm 0.05$ | Johnson & Mitchell (1995) |
| HD135722 | DDO | m45 | $5.00 \pm 0.05$ | McClure & Forrester (1981) |
| HD135722 | Oja | m45 | $4.15 \pm 0.05$ | Häggkvist & Oja (1970) |
| HD135722 | Vilnius | Y | $4.25 \pm 0.05$ | Zdanavicius et al. (1969) |
| HD135722 | Stromgren | b | $4.08 \pm 0.08$ | Crawford & Barnes (1970) |
| HD135722 | Stromgren | b | $4.08 \pm 0.08$ | Piirola (1976) |
| HD135722 | Stromgren | b | $4.08 \pm 0.08$ | Hauck & Mermilliod (1998) |

**Table 21** *continued on next page*

**Table 21** *(continued)*

| Star ID | System/Wvlen | Band/Bandpass | Value | Reference |
|---------|--------------|---------------|-------|-----------|
| HD135722 | DDO | m48 | $3.84 \pm 0.05$ | McClure & Forrester (1981) |
| HD135722 | KronComet | C2 | $3.55 \pm 0.04$ | This work |
| HD135722 | Vilnius | Z | $3.79 \pm 0.05$ | Zdanavicius et al. (1969) |
| HD135722 | 13c | m52 | $3.73 \pm 0.05$ | Johnson & Mitchell (1995) |
| HD135722 | KronComet | Gc | $3.46 \pm 0.01$ | This work |
| HD135722 | WBVR | V | $3.48 \pm 0.05$ | Kornilov et al. (1991) |
| HD135722 | Vilnius | V | $3.49 \pm 0.05$ | Zdanavicius et al. (1969) |
| HD135722 | Stromgren | y | $3.49 \pm 0.08$ | Crawford & Barnes (1970) |
| HD135722 | Stromgren | y | $3.49 \pm 0.08$ | Piirola (1976) |
| HD135722 | Stromgren | y | $3.49 \pm 0.08$ | Hauck & Mermilliod (1998) |
| HD135722 | Johnson | V | $3.45 \pm 0.05$ | Ljunggren & Oja (1965) |
| HD135722 | Johnson | V | $3.46 \pm 0.05$ | Tolbert (1964) |
| HD135722 | Johnson | V | $3.47 \pm 0.05$ | Oja (1963) |
| HD135722 | Johnson | V | $3.48 \pm 0.05$ | Jennens & Helfer (1975) |
| HD135722 | Johnson | V | $3.48 \pm 0.05$ | Mermilliod (1986) |
| HD135722 | Johnson | V | $3.49 \pm 0.04$ | This work |
| HD135722 | Johnson | V | $3.49 \pm 0.05$ | Argue (1963) |
| HD135722 | Johnson | V | $3.49 \pm 0.05$ | Johnson et al. (1966) |
| HD135722 | Johnson | V | $3.49 \pm 0.05$ | Ducati (2002) |
| HD135722 | Johnson | V | $3.50 \pm 0.05$ | Johnson & Morgan (1953b) |
| HD135722 | Johnson | V | $3.50 \pm 0.05$ | Johnson (1953) |
| HD135722 | 13c | m58 | $3.29 \pm 0.05$ | Johnson & Mitchell (1995) |
| HD135722 | 13c | m63 | $3.00 \pm 0.05$ | Johnson & Mitchell (1995) |
| HD135722 | Vilnius | S | $2.78 \pm 0.05$ | Zdanavicius et al. (1969) |
| HD135722 | WBVR | R | $2.76 \pm 0.05$ | Kornilov et al. (1991) |
| HD135722 | KronComet | Rc | $2.45 \pm 0.01$ | This work |
| HD135722 | 13c | m72 | $2.75 \pm 0.05$ | Johnson & Mitchell (1995) |
| HD135722 | 13c | m80 | $2.50 \pm 0.05$ | Johnson & Mitchell (1995) |
| HD135722 | 13c | m86 | $2.40 \pm 0.05$ | Johnson & Mitchell (1995) |
| HD135722 | 13c | m99 | $2.25 \pm 0.05$ | Johnson & Mitchell (1995) |
| HD135722 | 13c | m110 | $2.11 \pm 0.05$ | Johnson & Mitchell (1995) |
| HD135722 | 1250 | 310 | $309.40 \pm 10.20$ | Smith et al. (2004) |
| HD135722 | Johnson | J | $1.77 \pm 0.05$ | Blackwell et al. (1979) |
| HD135722 | Johnson | J | $1.77 \pm 0.05$ | Alonso et al. (1998) |
| HD135722 | Johnson | J | $1.78 \pm 0.05$ | Selby et al. (1988) |
| HD135722 | Johnson | J | $1.78 \pm 0.05$ | Blackwell et al. (1990) |
| HD135722 | Johnson | J | $1.80 \pm 0.05$ | Arribas & Martinez Roger (1987) |
| HD135722 | Johnson | J | $1.80 \pm 0.05$ | Ducati (2002) |
| HD135722 | Johnson | J | $1.87 \pm 0.05$ | Johnson et al. (1966) |
| HD135722 | Johnson | H | $1.26 \pm 0.05$ | Blackwell et al. (1979) |
| HD135722 | Johnson | H | $1.27 \pm 0.05$ | Ducati (2002) |

**Table 21** *(continued)*

| Star ID | System/Wvlen | Band/Bandpass | Value | Reference |
|---------|--------------|---------------|-------|-----------|
| HD135722 | Johnson | H | $1.28 \pm 0.05$ | Arribas & Martinez Roger (1987) |
| HD135722 | 2200 | 361 | $206.30 \pm 6.70$ | Smith et al. (2004) |
| HD135722 | Johnson | K | $1.14 \pm 0.03$ | Neugebauer & Leighton (1969) |
| HD135722 | Johnson | K | $1.19 \pm 0.05$ | Ducati (2002) |
| HD135722 | Johnson | K | $1.22 \pm 0.05$ | Johnson et al. (1966) |
| HD135722 | Johnson | L | $0.96 \pm 0.05$ | Johnson et al. (1966) |
| HD135722 | Johnson | L | $1.07 \pm 0.05$ | Ducati (2002) |
| HD135722 | 3500 | 898 | $97.50 \pm 5.50$ | Smith et al. (2004) |
| HD135722 | 4900 | 712 | $48.30 \pm 5.20$ | Smith et al. (2004) |
| HD135722 | Johnson | M | $1.23 \pm 0.05$ | Ducati (2002) |
| HD135722 | 12000 | 6384 | $11.10 \pm 17.00$ | Smith et al. (2004) |
| HD136404 | KronComet | NH | $12.73 \pm 0.10$ | This work |
| HD136404 | KronComet | UVc | $12.23 \pm 0.11$ | This work |
| HD136404 | Johnson | U | $10.48 \pm 0.18$ | This work |
| HD136404 | Johnson | U | $11.14 \pm 0.05$ | Mermilliod (1986) |
| HD136404 | KronComet | CN | $11.42 \pm 0.08$ | This work |
| HD136404 | KronComet | COp | $9.87 \pm 0.06$ | This work |
| HD136404 | Johnson | B | $9.04 \pm 0.09$ | This work |
| HD136404 | Johnson | B | $9.19 \pm 0.05$ | Mermilliod (1986) |
| HD136404 | KronComet | Bc | $9.01 \pm 0.04$ | This work |
| HD136404 | KronComet | C2 | $7.90 \pm 0.08$ | This work |
| HD136404 | KronComet | Gc | $7.63 \pm 0.01$ | This work |
| HD136404 | Johnson | V | $7.53 \pm 0.05$ | Mermilliod (1986) |
| HD136404 | Johnson | V | $7.57 \pm 0.04$ | This work |
| HD136404 | KronComet | Rc | $6.11 \pm 0.04$ | This work |
| HD136404 | 1250 | 310 | $38.00 \pm 6.60$ | Smith et al. (2004) |
| HD136404 | 2200 | 361 | $39.60 \pm 5.00$ | Smith et al. (2004) |
| HD136404 | Johnson | K | $2.84 \pm 0.08$ | Neugebauer & Leighton (1969) |
| HD136404 | 3500 | 898 | $46.70 \pm 27.90$ | Smith et al. (2004) |
| HD136404 | 4900 | 712 | $26.00 \pm 20.10$ | Smith et al. (2004) |
| HD136404 | 12000 | 6384 | $10.20 \pm 23.00$ | Smith et al. (2004) |
| HD137071 | Oja | m41 | $8.67 \pm 0.05$ | Häggkvist & Oja (1970) |
| HD137071 | DDO | m42 | $9.03 \pm 0.05$ | McClure & Forrester (1981) |
| HD137071 | Oja | m42 | $8.40 \pm 0.05$ | Häggkvist & Oja (1970) |
| HD137071 | KronComet | COp | $7.83 \pm 0.05$ | This work |
| HD137071 | WBVR | B | $7.23 \pm 0.05$ | Kornilov et al. (1991) |
| HD137071 | Johnson | B | $7.08 \pm 0.44$ | This work |
| HD137071 | Johnson | B | $7.10 \pm 0.05$ | Haggkvist & Oja (1970) |
| HD137071 | KronComet | Bc | $7.04 \pm 0.02$ | This work |
| HD137071 | DDO | m45 | $7.71 \pm 0.05$ | McClure & Forrester (1981) |
| HD137071 | Oja | m45 | $6.80 \pm 0.05$ | Häggkvist & Oja (1970) |

**Table 21** *(continued)*

| Star ID | System/Wvlen | Band/Bandpass | Value | Reference |
|---------|--------------|---------------|-------|-----------|
| HD137071 | Vilnius | Y | $6.69 \pm 0.05$ | Bartkevicius et al. (1973) |
| HD137071 | DDO | m48 | $6.23 \pm 0.05$ | McClure & Forrester (1981) |
| HD137071 | KronComet | C2 | $5.97 \pm 0.03$ | This work |
| HD137071 | Vilnius | Z | $6.03 \pm 0.05$ | Bartkevicius et al. (1973) |
| HD137071 | KronComet | Gc | $5.74 \pm 0.01$ | This work |
| HD137071 | WBVR | V | $5.55 \pm 0.05$ | Kornilov et al. (1991) |
| HD137071 | Vilnius | V | $5.51 \pm 0.05$ | Bartkevicius et al. (1973) |
| HD137071 | Johnson | V | $5.50 \pm 0.05$ | Haggkvist & Oja (1970) |
| HD137071 | Johnson | V | $5.61 \pm 0.05$ | This work |
| HD137071 | Vilnius | S | $4.43 \pm 0.05$ | Bartkevicius et al. (1973) |
| HD137071 | WBVR | R | $4.33 \pm 0.05$ | Kornilov et al. (1991) |
| HD137071 | KronComet | Rc | $4.15 \pm 0.01$ | This work |
| HD137071 | 1250 | 310 | $135.00 \pm 4.60$ | Smith et al. (2004) |
| HD137071 | 2200 | 361 | $123.40 \pm 4.40$ | Smith et al. (2004) |
| HD137071 | Johnson | K | $1.73 \pm 0.04$ | Neugebauer & Leighton (1969) |
| HD137071 | 3500 | 898 | $62.90 \pm 4.30$ | Smith et al. (2004) |
| HD137071 | 4900 | 712 | $26.90 \pm 4.60$ | Smith et al. (2004) |
| HD137071 | 12000 | 6384 | $-4.80 \pm 16.30$ | Smith et al. (2004) |
| HD137853 | Oja | m41 | $9.23 \pm 0.05$ | Häggkvist & Oja (1970) |
| HD137853 | Oja | m42 | $9.03 \pm 0.05$ | Häggkvist & Oja (1970) |
| HD137853 | KronComet | COp | $8.39 \pm 0.03$ | This work |
| HD137853 | WBVR | B | $7.72 \pm 0.05$ | Kornilov et al. (1991) |
| HD137853 | Johnson | B | $7.54 \pm 0.04$ | This work |
| HD137853 | Johnson | B | $7.64 \pm 0.05$ | Haggkvist & Oja (1970) |
| HD137853 | Johnson | B | $7.64 \pm 0.05$ | Ducati (2002) |
| HD137853 | KronComet | Bc | $7.51 \pm 0.02$ | This work |
| HD137853 | Oja | m45 | $7.32 \pm 0.05$ | Häggkvist & Oja (1970) |
| HD137853 | KronComet | C2 | $6.47 \pm 0.09$ | This work |
| HD137853 | KronComet | Gc | $6.19 \pm 0.01$ | This work |
| HD137853 | WBVR | V | $6.02 \pm 0.05$ | Kornilov et al. (1991) |
| HD137853 | Johnson | V | $6.02 \pm 0.05$ | Haggkvist & Oja (1970) |
| HD137853 | Johnson | V | $6.02 \pm 0.05$ | Ducati (2002) |
| HD137853 | Johnson | V | $6.06 \pm 0.04$ | This work |
| HD137853 | WBVR | R | $4.66 \pm 0.05$ | Kornilov et al. (1991) |
| HD137853 | KronComet | Rc | $4.58 \pm 0.03$ | This work |
| HD137853 | 1250 | 310 | $118.60 \pm 6.60$ | Smith et al. (2004) |
| HD137853 | Johnson | J | $2.84 \pm 0.05$ | McWilliam & Lambert (1984) |
| HD137853 | Johnson | J | $2.84 \pm 0.05$ | Ducati (2002) |
| HD137853 | 2200 | 361 | $115.10 \pm 5.70$ | Smith et al. (2004) |
| HD137853 | Johnson | K | $1.76 \pm 0.05$ | Ducati (2002) |
| HD137853 | Johnson | K | $1.82 \pm 0.04$ | Neugebauer & Leighton (1969) |

**Table 21** *continued on next page*

**Table 21** *(continued)*

| Star ID | System/Wvlen | Band/Bandpass | Value | Reference |
|---------|--------------|---------------|-------|-----------|
| HD137853 | 3500 | 898 | $56.80 \pm 4.70$ | Smith et al. (2004) |
| HD137853 | 4900 | 712 | $24.50 \pm 4.80$ | Smith et al. (2004) |
| HD137853 | 12000 | 6384 | $6.30 \pm 17.30$ | Smith et al. (2004) |
| HD138481 | 13c | m33 | $8.88 \pm 0.05$ | Johnson & Mitchell (1995) |
| HD138481 | 13c | m35 | $8.36 \pm 0.05$ | Johnson & Mitchell (1995) |
| HD138481 | DDO | m35 | $9.98 \pm 0.05$ | McClure & Forrester (1981) |
| HD138481 | Johnson | U | $8.42 \pm 0.05$ | Argue (1963) |
| HD138481 | Johnson | U | $8.49 \pm 0.05$ | Mermilliod (1986) |
| HD138481 | Johnson | U | $8.52 \pm 0.05$ | Johnson & Knuckles (1957) |
| HD138481 | Johnson | U | $8.52 \pm 0.05$ | Johnson et al. (1966) |
| HD138481 | Oja | m41 | $8.13 \pm 0.05$ | Häggkvist & Oja (1970) |
| HD138481 | DDO | m42 | $8.49 \pm 0.05$ | McClure & Forrester (1981) |
| HD138481 | Oja | m42 | $7.90 \pm 0.05$ | Häggkvist & Oja (1970) |
| HD138481 | WBVR | B | $6.68 \pm 0.05$ | Kornilov et al. (1991) |
| HD138481 | Johnson | B | $6.60 \pm 0.05$ | Argue (1963) |
| HD138481 | Johnson | B | $6.61 \pm 0.05$ | Johnson & Knuckles (1957) |
| HD138481 | Johnson | B | $6.61 \pm 0.05$ | Johnson et al. (1966) |
| HD138481 | Johnson | B | $6.62 \pm 0.05$ | Mermilliod (1986) |
| HD138481 | Johnson | B | $6.64 \pm 0.05$ | Häggkvist & Oja (1966) |
| HD138481 | Johnson | B | $6.66 \pm 0.05$ | Miczaika (1954) |
| HD138481 | 13c | m45 | $6.20 \pm 0.05$ | Johnson & Mitchell (1995) |
| HD138481 | DDO | m45 | $7.14 \pm 0.05$ | McClure & Forrester (1981) |
| HD138481 | Oja | m45 | $6.26 \pm 0.05$ | Häggkvist & Oja (1970) |
| HD138481 | DDO | m48 | $5.72 \pm 0.05$ | McClure & Forrester (1981) |
| HD138481 | 13c | m52 | $5.48 \pm 0.05$ | Johnson & Mitchell (1995) |
| HD138481 | WBVR | V | $5.03 \pm 0.05$ | Kornilov et al. (1991) |
| HD138481 | Johnson | V | $5.00 \pm 0.05$ | Argue (1963) |
| HD138481 | Johnson | V | $5.02 \pm 0.05$ | Johnson & Knuckles (1957) |
| HD138481 | Johnson | V | $5.02 \pm 0.05$ | Johnson et al. (1966) |
| HD138481 | Johnson | V | $5.03 \pm 0.05$ | Häggkvist & Oja (1966) |
| HD138481 | Johnson | V | $5.03 \pm 0.05$ | Mermilliod (1986) |
| HD138481 | Johnson | V | $5.07 \pm 0.05$ | Miczaika (1954) |
| HD138481 | 13c | m58 | $4.68 \pm 0.05$ | Johnson & Mitchell (1995) |
| HD138481 | 13c | m63 | $4.23 \pm 0.05$ | Johnson & Mitchell (1995) |
| HD138481 | WBVR | R | $3.79 \pm 0.05$ | Kornilov et al. (1991) |
| HD138481 | 13c | m72 | $3.73 \pm 0.05$ | Johnson & Mitchell (1995) |
| HD138481 | 13c | m80 | $3.32 \pm 0.05$ | Johnson & Mitchell (1995) |
| HD138481 | 13c | m86 | $3.15 \pm 0.05$ | Johnson & Mitchell (1995) |
| HD138481 | 13c | m99 | $2.85 \pm 0.05$ | Johnson & Mitchell (1995) |
| HD138481 | 13c | m110 | $2.62 \pm 0.05$ | Johnson & Mitchell (1995) |
| HD138481 | 1250 | 310 | $239.60 \pm 8.00$ | Smith et al. (2004) |

**Table 21** *(continued)*

| Star ID | System/Wvlen | Band/Bandpass | Value | Reference |
|---------|--------------|---------------|-------|-----------|
| HD138481 | 2200 | 361 | $213.80 \pm 5.60$ | Smith et al. (2004) |
| HD138481 | Johnson | K | $1.14 \pm 0.04$ | Neugebauer & Leighton (1969) |
| HD138481 | 3500 | 898 | $104.50 \pm 5.20$ | Smith et al. (2004) |
| HD138481 | 4900 | 712 | $46.60 \pm 5.10$ | Smith et al. (2004) |
| HD138481 | 12000 | 6384 | $9.20 \pm 16.00$ | Smith et al. (2004) |
| HD139153 | Oja | m41 | $8.33 \pm 0.05$ | Häggkvist & Oja (1970) |
| HD139153 | Oja | m42 | $8.11 \pm 0.05$ | Häggkvist & Oja (1970) |
| HD139153 | KronComet | COp | $7.49 \pm 0.03$ | This work |
| HD139153 | WBVR | B | $6.86 \pm 0.05$ | Kornilov et al. (1991) |
| HD139153 | Johnson | B | $6.64 \pm 0.05$ | This work |
| HD139153 | Johnson | B | $6.75 \pm 0.05$ | Mermilliod (1986) |
| HD139153 | Johnson | B | $6.75 \pm 0.05$ | Ducati (2002) |
| HD139153 | Johnson | B | $6.76 \pm 0.05$ | Walker (1971) |
| HD139153 | KronComet | Bc | $6.62 \pm 0.02$ | This work |
| HD139153 | Oja | m45 | $6.39 \pm 0.05$ | Häggkvist & Oja (1970) |
| HD139153 | KronComet | C2 | $5.59 \pm 0.10$ | This work |
| HD139153 | KronComet | Gc | $5.30 \pm 0.01$ | This work |
| HD139153 | WBVR | V | $5.17 \pm 0.05$ | Kornilov et al. (1991) |
| HD139153 | Johnson | V | $5.11 \pm 0.05$ | Ducati (2002) |
| HD139153 | Johnson | V | $5.12 \pm 0.05$ | Walker (1971) |
| HD139153 | Johnson | V | $5.13 \pm 0.05$ | Mermilliod (1986) |
| HD139153 | Johnson | V | $5.19 \pm 0.04$ | This work |
| HD139153 | WBVR | R | $3.82 \pm 0.05$ | Kornilov et al. (1991) |
| HD139153 | KronComet | Rc | $3.72 \pm 0.03$ | This work |
| HD139153 | 1250 | 310 | $246.10 \pm 4.10$ | Smith et al. (2004) |
| HD139153 | Johnson | J | $2.02 \pm 0.05$ | Ducati (2002) |
| HD139153 | 2200 | 361 | $242.10 \pm 7.80$ | Smith et al. (2004) |
| HD139153 | Johnson | K | $0.93 \pm 0.03$ | Neugebauer & Leighton (1969) |
| HD139153 | Johnson | K | $0.95 \pm 0.05$ | Ducati (2002) |
| HD139153 | 3500 | 898 | $118.20 \pm 24.50$ | Smith et al. (2004) |
| HD139153 | 4900 | 712 | $52.30 \pm 5.00$ | Smith et al. (2004) |
| HD139153 | 12000 | 6384 | $11.20 \pm 16.60$ | Smith et al. (2004) |
| HD139374 | KronComet | COp | $10.22 \pm 0.05$ | This work |
| HD139374 | Johnson | B | $9.66 \pm 0.06$ | This work |
| HD139374 | KronComet | Bc | $9.59 \pm 0.02$ | This work |
| HD139374 | KronComet | C2 | $8.45 \pm 0.03$ | This work |
| HD139374 | KronComet | Gc | $8.37 \pm 0.01$ | This work |
| HD139374 | Johnson | V | $8.32 \pm 0.03$ | This work |
| HD139374 | KronComet | Rc | $6.89 \pm 0.01$ | This work |
| HD139374 | 1250 | 310 | $26.60 \pm 5.40$ | Smith et al. (2004) |
| HD139374 | 2200 | 361 | $35.80 \pm 5.30$ | Smith et al. (2004) |

**Table 21** *(continued)*

| Star ID | System/Wvlen | Band/Bandpass | Value | Reference |
|---------|--------------|---------------|-------|-----------|
| HD139374 | Johnson | K | $2.92 \pm 0.06$ | Neugebauer & Leighton (1969) |
| HD139374 | 3500 | 898 | $17.90 \pm 4.80$ | Smith et al. (2004) |
| HD139374 | 4900 | 712 | $7.20 \pm 6.10$ | Smith et al. (2004) |
| HD139374 | 12000 | 6384 | $-0.50 \pm 15.90$ | Smith et al. (2004) |
| HD139971 | KronComet | COp | $9.83 \pm 0.08$ | This work |
| HD139971 | Johnson | B | $9.26 \pm 0.08$ | This work |
| HD139971 | KronComet | Bc | $9.15 \pm 0.06$ | This work |
| HD139971 | KronComet | C2 | $7.97 \pm 0.13$ | This work |
| HD139971 | KronComet | Gc | $7.84 \pm 0.04$ | This work |
| HD139971 | Johnson | V | $7.82 \pm 0.05$ | This work |
| HD139971 | KronComet | Rc | $6.33 \pm 0.05$ | This work |
| HD139971 | 1250 | 310 | $49.50 \pm 4.70$ | Smith et al. (2004) |
| HD139971 | 2200 | 361 | $51.50 \pm 5.00$ | Smith et al. (2004) |
| HD139971 | Johnson | K | $2.62 \pm 0.06$ | Neugebauer & Leighton (1969) |
| HD139971 | 3500 | 898 | $27.20 \pm 3.90$ | Smith et al. (2004) |
| HD139971 | 4900 | 712 | $11.60 \pm 4.80$ | Smith et al. (2004) |
| HD139971 | 12000 | 6384 | $2.20 \pm 16.10$ | Smith et al. (2004) |
| HD142176 | KronComet | COp | $9.50 \pm 0.04$ | This work |
| HD142176 | WBVR | B | $8.92 \pm 0.05$ | Kornilov et al. (1991) |
| HD142176 | Johnson | B | $8.78 \pm 0.05$ | This work |
| HD142176 | KronComet | Bc | $8.69 \pm 0.02$ | This work |
| HD142176 | KronComet | C2 | $7.82 \pm 0.10$ | This work |
| HD142176 | KronComet | Gc | $7.53 \pm 0.01$ | This work |
| HD142176 | WBVR | V | $7.37 \pm 0.05$ | Kornilov et al. (1991) |
| HD142176 | Johnson | V | $7.42 \pm 0.04$ | This work |
| HD142176 | WBVR | R | $6.22 \pm 0.05$ | Kornilov et al. (1991) |
| HD142176 | KronComet | Rc | $5.98 \pm 0.03$ | This work |
| HD143107 | 13c | m33 | $6.78 \pm 0.05$ | Johnson & Mitchell (1995) |
| HD143107 | Vilnius | U | $8.44 \pm 0.05$ | Sudzius et al. (1970) |
| HD143107 | 13c | m35 | $6.49 \pm 0.05$ | Johnson & Mitchell (1995) |
| HD143107 | DDO | m35 | $8.00 \pm 0.05$ | McClure & Forrester (1981) |
| HD143107 | DDO | m35 | $8.00 \pm 0.05$ | Mermilliod & Nitschelm (1989) |
| HD143107 | Stromgren | u | $7.92 \pm 0.08$ | Olsen (1993) |
| HD143107 | Stromgren | u | $7.93 \pm 0.08$ | Gray & Olsen (1991) |
| HD143107 | Stromgren | u | $7.94 \pm 0.08$ | Crawford & Barnes (1970) |
| HD143107 | Stromgren | u | $7.94 \pm 0.08$ | Philip & Philip (1973) |
| HD143107 | Stromgren | u | $7.94 \pm 0.08$ | Piirola (1976) |
| HD143107 | Stromgren | u | $7.94 \pm 0.08$ | Reglero et al. (1987) |
| HD143107 | Stromgren | u | $7.94 \pm 0.08$ | Fabregat & Reglero (1990) |
| HD143107 | Stromgren | u | $7.94 \pm 0.08$ | Hauck & Mermilliod (1998) |
| HD143107 | WBVR | W | $6.52 \pm 0.05$ | Kornilov et al. (1991) |

**Table 21** *(continued)*

| Star ID | System/Wvlen | Band/Bandpass | Value | Reference |
|---------|--------------|---------------|-------|-----------|
| HD143107 | Johnson | U | $6.61 \pm 0.05$ | Piirola (1976) |
| HD143107 | Johnson | U | $6.62 \pm 0.05$ | Oosterhoff (1960) |
| HD143107 | Johnson | U | $6.63 \pm 0.05$ | Oja (1985a) |
| HD143107 | Johnson | U | $6.64 \pm 0.05$ | Oja (1983) |
| HD143107 | Johnson | U | $6.64 \pm 0.05$ | Oja (1985b) |
| HD143107 | Johnson | U | $6.65 \pm 0.01$ | Oja (1984) |
| HD143107 | Johnson | U | $6.65 \pm 0.05$ | Oja (1984) |
| HD143107 | Johnson | U | $6.66 \pm 0.05$ | Johnson & Morgan (1953b) |
| HD143107 | Johnson | U | $6.66 \pm 0.05$ | Johnson & Harris (1954) |
| HD143107 | Johnson | U | $6.66 \pm 0.05$ | Argue (1963) |
| HD143107 | Johnson | U | $6.66 \pm 0.05$ | Johnson (1964) |
| HD143107 | Johnson | U | $6.66 \pm 0.05$ | Johnson et al. (1966) |
| HD143107 | Johnson | U | $6.66 \pm 0.05$ | Fernie (1969) |
| HD143107 | Johnson | U | $6.66 \pm 0.05$ | Lee (1970) |
| HD143107 | Johnson | U | $6.66 \pm 0.05$ | Jennens & Helfer (1975) |
| HD143107 | Johnson | U | $6.66 \pm 0.05$ | Nicolet (1978) |
| HD143107 | Johnson | U | $6.66 \pm 0.05$ | Mendoza et al. (1978) |
| HD143107 | Johnson | U | $6.66 \pm 0.05$ | Oja (1986) |
| HD143107 | Johnson | U | $6.67 \pm 0.05$ | Mermilliod (1986) |
| HD143107 | Johnson | U | $6.67 \pm 0.05$ | Ducati (2002) |
| HD143107 | 13c | m37 | $6.57 \pm 0.05$ | Johnson & Mitchell (1995) |
| HD143107 | Vilnius | P | $7.82 \pm 0.05$ | Sudzius et al. (1970) |
| HD143107 | DDO | m38 | $6.85 \pm 0.05$ | McClure & Forrester (1981) |
| HD143107 | DDO | m38 | $6.85 \pm 0.05$ | Mermilliod & Nitschelm (1989) |
| HD143107 | 13c | m40 | $6.17 \pm 0.05$ | Johnson & Mitchell (1995) |
| HD143107 | Vilnius | X | $6.62 \pm 0.05$ | Sudzius et al. (1970) |
| HD143107 | DDO | m41 | $7.17 \pm 0.05$ | McClure & Forrester (1981) |
| HD143107 | DDO | m41 | $7.17 \pm 0.05$ | Mermilliod & Nitschelm (1989) |
| HD143107 | Oja | m41 | $6.71 \pm 0.05$ | Häggkvist & Oja (1970) |
| HD143107 | Stromgren | v | $6.20 \pm 0.08$ | Crawford & Barnes (1970) |
| HD143107 | Stromgren | v | $6.20 \pm 0.08$ | Philip & Philip (1973) |
| HD143107 | Stromgren | v | $6.20 \pm 0.08$ | Piirola (1976) |
| HD143107 | Stromgren | v | $6.20 \pm 0.08$ | Fabregat & Reglero (1990) |
| HD143107 | Stromgren | v | $6.20 \pm 0.08$ | Olsen (1993) |
| HD143107 | Stromgren | v | $6.20 \pm 0.08$ | Hauck & Mermilliod (1998) |
| HD143107 | Stromgren | v | $6.21 \pm 0.08$ | Reglero et al. (1987) |
| HD143107 | Stromgren | v | $6.21 \pm 0.08$ | Gray & Olsen (1991) |
| HD143107 | DDO | m42 | $6.91 \pm 0.05$ | McClure & Forrester (1981) |
| HD143107 | DDO | m42 | $6.91 \pm 0.05$ | Mermilliod & Nitschelm (1989) |
| HD143107 | Oja | m42 | $6.45 \pm 0.05$ | Häggkvist & Oja (1970) |
| HD143107 | WBVR | B | $5.41 \pm 0.05$ | Kornilov et al. (1991) |

**Table 21** *(continued)*

| Star ID | System/Wvlen | Band/Bandpass | Value | Reference |
|---------|--------------|---------------|-------|-----------|
| HD143107 | Johnson | B | $5.35 \pm 0.05$ | Oja (1963) |
| HD143107 | Johnson | B | $5.35 \pm 0.05$ | Piirola (1976) |
| HD143107 | Johnson | B | $5.35 \pm 0.05$ | Oja (1985a) |
| HD143107 | Johnson | B | $5.36 \pm 0.01$ | Oja (1993) |
| HD143107 | Johnson | B | $5.36 \pm 0.05$ | Ljunggren & Oja (1965) |
| HD143107 | Johnson | B | $5.36 \pm 0.05$ | Fernie (1969) |
| HD143107 | Johnson | B | $5.36 \pm 0.05$ | Oja (1983) |
| HD143107 | Johnson | B | $5.36 \pm 0.05$ | Oja (1985b) |
| HD143107 | Johnson | B | $5.37 \pm 0.05$ | Oosterhoff (1960) |
| HD143107 | Johnson | B | $5.37 \pm 0.05$ | Oja (1984) |
| HD143107 | Johnson | B | $5.37 \pm 0.05$ | Oja (1986) |
| HD143107 | Johnson | B | $5.38 \pm 0.05$ | Johnson & Morgan (1953b) |
| HD143107 | Johnson | B | $5.38 \pm 0.05$ | Johnson & Harris (1954) |
| HD143107 | Johnson | B | $5.38 \pm 0.05$ | Johnson (1964) |
| HD143107 | Johnson | B | $5.38 \pm 0.05$ | Johnson et al. (1966) |
| HD143107 | Johnson | B | $5.38 \pm 0.05$ | Lee (1970) |
| HD143107 | Johnson | B | $5.38 \pm 0.05$ | Jennens & Helfer (1975) |
| HD143107 | Johnson | B | $5.38 \pm 0.05$ | Nicolet (1978) |
| HD143107 | Johnson | B | $5.38 \pm 0.05$ | Mendoza et al. (1978) |
| HD143107 | Johnson | B | $5.39 \pm 0.05$ | Argue (1963) |
| HD143107 | Johnson | B | $5.39 \pm 0.05$ | Moffett & Barnes (1979) |
| HD143107 | Johnson | B | $5.39 \pm 0.05$ | Mermilliod (1986) |
| HD143107 | Johnson | B | $5.39 \pm 0.05$ | Ducati (2002) |
| HD143107 | Johnson | B | $5.41 \pm 0.05$ | Miczaika (1954) |
| HD143107 | 13c | m45 | $5.00 \pm 0.05$ | Johnson & Mitchell (1995) |
| HD143107 | DDO | m45 | $5.88 \pm 0.05$ | McClure & Forrester (1981) |
| HD143107 | DDO | m45 | $5.88 \pm 0.05$ | Mermilliod & Nitschelm (1989) |
| HD143107 | Oja | m45 | $5.04 \pm 0.05$ | Häggkvist & Oja (1970) |
| HD143107 | Vilnius | Y | $5.03 \pm 0.05$ | Sudzius et al. (1970) |
| HD143107 | Stromgren | b | $4.88 \pm 0.08$ | Crawford & Barnes (1970) |
| HD143107 | Stromgren | b | $4.88 \pm 0.08$ | Philip & Philip (1973) |
| HD143107 | Stromgren | b | $4.88 \pm 0.08$ | Piirola (1976) |
| HD143107 | Stromgren | b | $4.88 \pm 0.08$ | Reglero et al. (1987) |
| HD143107 | Stromgren | b | $4.88 \pm 0.08$ | Fabregat & Reglero (1990) |
| HD143107 | Stromgren | b | $4.88 \pm 0.08$ | Olsen (1993) |
| HD143107 | Stromgren | b | $4.88 \pm 0.08$ | Hauck & Mermilliod (1998) |
| HD143107 | Stromgren | b | $4.89 \pm 0.08$ | Gray & Olsen (1991) |
| HD143107 | DDO | m48 | $4.61 \pm 0.05$ | McClure & Forrester (1981) |
| HD143107 | DDO | m48 | $4.61 \pm 0.05$ | Mermilliod & Nitschelm (1989) |
| HD143107 | Vilnius | Z | $4.52 \pm 0.05$ | Sudzius et al. (1970) |
| HD143107 | 13c | m52 | $4.46 \pm 0.05$ | Johnson & Mitchell (1995) |

**Table 21** *(continued)*

| Star ID | System/Wvlen | Band/Bandpass | Value | Reference |
|---------|--------------|---------------|-------|-----------|
| HD143107 | WBVR | V | $4.14 \pm 0.05$ | Kornilov et al. (1991) |
| HD143107 | Vilnius | V | $4.15 \pm 0.05$ | Sudzius et al. (1970) |
| HD143107 | Stromgren | y | $4.13 \pm 0.08$ | Crawford & Barnes (1970) |
| HD143107 | Stromgren | y | $4.13 \pm 0.08$ | Philip & Philip (1973) |
| HD143107 | Stromgren | y | $4.13 \pm 0.08$ | Piirola (1976) |
| HD143107 | Stromgren | y | $4.13 \pm 0.08$ | Reglero et al. (1987) |
| HD143107 | Stromgren | y | $4.13 \pm 0.08$ | Fabregat & Reglero (1990) |
| HD143107 | Stromgren | y | $4.13 \pm 0.08$ | Gray & Olsen (1991) |
| HD143107 | Stromgren | y | $4.13 \pm 0.08$ | Olsen (1993) |
| HD143107 | Stromgren | y | $4.13 \pm 0.08$ | Hauck & Mermilliod (1998) |
| HD143107 | Johnson | V | $4.12 \pm 0.05$ | Oja (1963) |
| HD143107 | Johnson | V | $4.12 \pm 0.05$ | Piirola (1976) |
| HD143107 | Johnson | V | $4.12 \pm 0.05$ | Oja (1983) |
| HD143107 | Johnson | V | $4.12 \pm 0.05$ | Oja (1985b) |
| HD143107 | Johnson | V | $4.13 \pm 0.01$ | Oja (1993) |
| HD143107 | Johnson | V | $4.13 \pm 0.05$ | Ljunggren & Oja (1965) |
| HD143107 | Johnson | V | $4.13 \pm 0.05$ | Oja (1984) |
| HD143107 | Johnson | V | $4.13 \pm 0.05$ | Oja (1985a) |
| HD143107 | Johnson | V | $4.13 \pm 0.05$ | Oja (1986) |
| HD143107 | Johnson | V | $4.14 \pm 0.05$ | Oosterhoff (1960) |
| HD143107 | Johnson | V | $4.14 \pm 0.05$ | Fernie (1969) |
| HD143107 | Johnson | V | $4.15 \pm 0.05$ | Johnson & Morgan (1953b) |
| HD143107 | Johnson | V | $4.15 \pm 0.05$ | Johnson & Harris (1954) |
| HD143107 | Johnson | V | $4.15 \pm 0.05$ | Argue (1963) |
| HD143107 | Johnson | V | $4.15 \pm 0.05$ | Johnson (1964) |
| HD143107 | Johnson | V | $4.15 \pm 0.05$ | Johnson et al. (1966) |
| HD143107 | Johnson | V | $4.15 \pm 0.05$ | Lee (1970) |
| HD143107 | Johnson | V | $4.15 \pm 0.05$ | Jennens & Helfer (1975) |
| HD143107 | Johnson | V | $4.15 \pm 0.05$ | Nicolet (1978) |
| HD143107 | Johnson | V | $4.15 \pm 0.05$ | Mendoza et al. (1978) |
| HD143107 | Johnson | V | $4.15 \pm 0.05$ | Moffett & Barnes (1979) |
| HD143107 | Johnson | V | $4.15 \pm 0.05$ | Mermilliod (1986) |
| HD143107 | Johnson | V | $4.16 \pm 0.05$ | Ducati (2002) |
| HD143107 | Johnson | V | $4.19 \pm 0.05$ | Miczaika (1954) |
| HD143107 | 13c | m58 | $3.86 \pm 0.05$ | Johnson & Mitchell (1995) |
| HD143107 | 13c | m63 | $3.51 \pm 0.05$ | Johnson & Mitchell (1995) |
| HD143107 | Vilnius | S | $3.31 \pm 0.05$ | Sudzius et al. (1970) |
| HD143107 | WBVR | R | $3.25 \pm 0.05$ | Kornilov et al. (1991) |
| HD143107 | 13c | m72 | $3.21 \pm 0.05$ | Johnson & Mitchell (1995) |
| HD143107 | 13c | m80 | $2.94 \pm 0.05$ | Johnson & Mitchell (1995) |
| HD143107 | 13c | m86 | $2.81 \pm 0.05$ | Johnson & Mitchell (1995) |

**Table 21** *(continued)*

| Star ID | System/Wvlen | Band/Bandpass | Value | Reference |
|---------|--------------|---------------|-------|-----------|
| HD143107 | 13c | m99 | $2.60 \pm 0.05$ | Johnson & Mitchell (1995) |
| HD143107 | 13c | m110 | $2.42 \pm 0.05$ | Johnson & Mitchell (1995) |
| HD143107 | 1250 | 310 | $245.00 \pm 7.20$ | Smith et al. (2004) |
| HD143107 | Johnson | J | $2.09 \pm 0.05$ | Johnson et al. (1966) |
| HD143107 | Johnson | J | $2.09 \pm 0.05$ | Lee (1970) |
| HD143107 | Johnson | J | $2.09 \pm 0.05$ | Ducati (2002) |
| HD143107 | Johnson | J | $2.09 \pm 0.05$ | Shenavrin et al. (2011) |
| HD143107 | Johnson | H | $1.42 \pm 0.05$ | Lee (1970) |
| HD143107 | Johnson | H | $1.42 \pm 0.05$ | Ducati (2002) |
| HD143107 | Johnson | H | $1.60 \pm 0.05$ | Shenavrin et al. (2011) |
| HD143107 | 2200 | 361 | $185.40 \pm 6.00$ | Smith et al. (2004) |
| HD143107 | Johnson | K | $1.28 \pm 0.05$ | Neugebauer & Leighton (1969) |
| HD143107 | Johnson | K | $1.30 \pm 0.05$ | Johnson et al. (1966) |
| HD143107 | Johnson | K | $1.30 \pm 0.05$ | Lee (1970) |
| HD143107 | Johnson | K | $1.30 \pm 0.05$ | Ducati (2002) |
| HD143107 | Johnson | K | $1.30 \pm 0.05$ | Shenavrin et al. (2011) |
| HD143107 | Johnson | L | $1.13 \pm 0.05$ | Johnson et al. (1966) |
| HD143107 | Johnson | L | $1.14 \pm 0.05$ | Ducati (2002) |
| HD143107 | Johnson | L | $1.15 \pm 0.05$ | Lee (1970) |
| HD143107 | 3500 | 898 | $89.50 \pm 4.80$ | Smith et al. (2004) |
| HD143107 | 4900 | 712 | $42.00 \pm 4.80$ | Smith et al. (2004) |
| HD143107 | 12000 | 6384 | $8.30 \pm 16.50$ | Smith et al. (2004) |
| HD144065 | KronComet | COp | $10.56 \pm 0.03$ | This work |
| HD144065 | Johnson | B | $9.88 \pm 0.05$ | This work |
| HD144065 | KronComet | Bc | $9.78 \pm 0.02$ | This work |
| HD144065 | KronComet | C2 | $8.50 \pm 0.02$ | This work |
| HD144065 | KronComet | Gc | $8.35 \pm 0.01$ | This work |
| HD144065 | Johnson | V | $8.27 \pm 0.02$ | This work |
| HD144065 | KronComet | Rc | $6.69 \pm 0.01$ | This work |
| HD144065 | 1250 | 310 | $41.30 \pm 11.00$ | Smith et al. (2004) |
| HD144065 | 2200 | 361 | $40.90 \pm 9.20$ | Smith et al. (2004) |
| HD144065 | Johnson | K | $2.96 \pm 0.08$ | Neugebauer & Leighton (1969) |
| HD144065 | 3500 | 898 | $21.00 \pm 8.60$ | Smith et al. (2004) |
| HD144065 | 4900 | 712 | $9.10 \pm 5.10$ | Smith et al. (2004) |
| HD144065 | 12000 | 6384 | $0.30 \pm 19.50$ | Smith et al. (2004) |
| HD147749 | WBVR | W | $8.80 \pm 0.05$ | Kornilov et al. (1991) |
| HD147749 | Johnson | U | $8.74 \pm 0.05$ | Mermilliod (1986) |
| HD147749 | Johnson | U | $8.74 \pm 0.05$ | Ducati (2002) |
| HD147749 | KronComet | CN | $8.86 \pm 0.03$ | This work |
| HD147749 | Oja | m41 | $8.35 \pm 0.05$ | Häggkvist & Oja (1970) |
| HD147749 | Oja | m42 | $8.18 \pm 0.05$ | Häggkvist & Oja (1970) |

**Table 21** *(continued)*

| Star ID | System/Wvlen | Band/Bandpass | Value | Reference |
|---------|-------------|---------------|-------|-----------|
| HD147749 | KronComet | COp | $7.26 \pm 0.04$ | This work |
| HD147749 | WBVR | B | $6.90 \pm 0.05$ | Kornilov et al. (1991) |
| HD147749 | Johnson | B | $6.55 \pm 0.09$ | This work |
| HD147749 | Johnson | B | $6.80 \pm 0.05$ | Mermilliod (1986) |
| HD147749 | Johnson | B | $6.80 \pm 0.05$ | Ducati (2002) |
| HD147749 | KronComet | Bc | $6.60 \pm 0.04$ | This work |
| HD147749 | Oja | m45 | $6.49 \pm 0.05$ | Häggkvist & Oja (1970) |
| HD147749 | KronComet | C2 | $5.56 \pm 0.04$ | This work |
| HD147749 | KronComet | Gc | $5.34 \pm 0.02$ | This work |
| HD147749 | WBVR | V | $5.22 \pm 0.05$ | Kornilov et al. (1991) |
| HD147749 | Johnson | V | $5.20 \pm 0.05$ | Mermilliod (1986) |
| HD147749 | Johnson | V | $5.20 \pm 0.05$ | Ducati (2002) |
| HD147749 | Johnson | V | $5.24 \pm 0.04$ | This work |
| HD147749 | WBVR | R | $3.82 \pm 0.05$ | Kornilov et al. (1991) |
| HD147749 | KronComet | Rc | $3.79 \pm 0.02$ | This work |
| HD147749 | Johnson | J | $1.92 \pm 0.05$ | McWilliam & Lambert (1984) |
| HD147749 | Johnson | J | $1.92 \pm 0.05$ | Ducati (2002) |
| HD147749 | 2200 | 361 | $386.70 \pm 7.60$ | Smith et al. (2004) |
| HD147749 | Johnson | K | $0.84 \pm 0.05$ | Ducati (2002) |
| HD147749 | Johnson | K | $0.85 \pm 0.21$ | Neugebauer & Leighton (1969) |
| HD147749 | 3500 | 898 | $197.30 \pm 9.30$ | Smith et al. (2004) |
| HD147749 | 4900 | 712 | $83.00 \pm 7.70$ | Smith et al. (2004) |
| HD147749 | 12000 | 6384 | $20.80 \pm 17.50$ | Smith et al. (2004) |
| HD148897 | Vilnius | U | $9.58 \pm 0.05$ | Jasevicius et al. (1990) |
| HD148897 | Vilnius | U | $9.60 \pm 0.05$ | Sudzius et al. (1970) |
| HD148897 | Vilnius | U | $9.63 \pm 0.05$ | Sleivyte (1985) |
| HD148897 | DDO | m35 | $9.13 \pm 0.05$ | McClure & Forrester (1981) |
| HD148897 | Stromgren | u | $9.01 \pm 0.08$ | Arellano Ferro (1990) |
| HD148897 | Stromgren | u | $9.01 \pm 0.08$ | Hauck & Mermilliod (1998) |
| HD148897 | WBVR | W | $7.61 \pm 0.05$ | Kornilov et al. (1991) |
| HD148897 | Johnson | U | $7.69 \pm 0.05$ | Fernie (1983) |
| HD148897 | Johnson | U | $7.69 \pm 0.05$ | Mermilliod (1986) |
| HD148897 | Johnson | U | $7.75 \pm 0.05$ | Argue (1963) |
| HD148897 | Vilnius | P | $8.79 \pm 0.05$ | Jasevicius et al. (1990) |
| HD148897 | Vilnius | P | $8.80 \pm 0.05$ | Sudzius et al. (1970) |
| HD148897 | Vilnius | P | $8.82 \pm 0.05$ | Sleivyte (1985) |
| HD148897 | DDO | m38 | $7.87 \pm 0.05$ | McClure & Forrester (1981) |
| HD148897 | DDO | m38 | $7.87 \pm 0.05$ | Mermilliod & Nitschelm (1989) |
| HD148897 | Vilnius | X | $7.72 \pm 0.05$ | Sudzius et al. (1970) |
| HD148897 | Vilnius | X | $7.74 \pm 0.05$ | Jasevicius et al. (1990) |
| HD148897 | Vilnius | X | $7.75 \pm 0.05$ | Sleivyte (1985) |

**Table 21** *(continued)*

| Star ID | System/Wvlen | Band/Bandpass | Value | Reference |
|---------|--------------|---------------|-------|-----------|
| HD148897 | DDO | m41 | 8.16 ± 0.05 | McClure & Forrester (1981) |
| HD148897 | DDO | m41 | 8.16 ± 0.05 | Mermilliod & Nitschelm (1989) |
| HD148897 | Oja | m41 | 7.68 ± 0.05 | Häggkvist & Oja (1970) |
| HD148897 | Stromgren | v | 7.27 ± 0.08 | Arellano Ferro et al. (1990) |
| HD148897 | Stromgren | v | 7.27 ± 0.08 | Hauck & Mermilliod (1998) |
| HD148897 | DDO | m42 | 8.01 ± 0.05 | McClure & Forrester (1981) |
| HD148897 | DDO | m42 | 8.01 ± 0.05 | Mermilliod & Nitschelm (1989) |
| HD148897 | Oja | m42 | 7.58 ± 0.05 | Häggkvist & Oja (1970) |
| HD148897 | WBVR | B | 6.56 ± 0.05 | Kornilov et al. (1991) |
| HD148897 | Johnson | B | 6.49 ± 0.05 | Mermilliod (1986) |
| HD148897 | Johnson | B | 6.53 ± 0.05 | Argue (1963) |
| HD148897 | Johnson | B | 6.53 ± 0.05 | Fernie (1983) |
| HD148897 | DDO | m45 | 7.08 ± 0.05 | McClure & Forrester (1981) |
| HD148897 | DDO | m45 | 7.08 ± 0.05 | Mermilliod & Nitschelm (1989) |
| HD148897 | Oja | m45 | 6.26 ± 0.05 | Häggkvist & Oja (1970) |
| HD148897 | Vilnius | Y | 6.20 ± 0.05 | Jasevicius et al. (1990) |
| HD148897 | Vilnius | Y | 6.21 ± 0.05 | Sudzius et al. (1970) |
| HD148897 | Vilnius | Y | 6.24 ± 0.05 | Sleivyte (1985) |
| HD148897 | Stromgren | b | 6.07 ± 0.08 | Arellano Ferro et al. (1990) |
| HD148897 | Stromgren | b | 6.07 ± 0.08 | Hauck & Mermilliod (1998) |
| HD148897 | DDO | m48 | 5.76 ± 0.05 | McClure & Forrester (1981) |
| HD148897 | DDO | m48 | 5.76 ± 0.05 | Mermilliod & Nitschelm (1989) |
| HD148897 | Vilnius | Z | 5.59 ± 0.05 | Jasevicius et al. (1990) |
| HD148897 | Vilnius | Z | 5.60 ± 0.05 | Sudzius et al. (1970) |
| HD148897 | Vilnius | Z | 5.63 ± 0.05 | Sleivyte (1985) |
| HD148897 | WBVR | V | 5.25 ± 0.05 | Kornilov et al. (1991) |
| HD148897 | Vilnius | V | 5.23 ± 0.05 | Jasevicius et al. (1990) |
| HD148897 | Vilnius | V | 5.25 ± 0.05 | Sudzius et al. (1970) |
| HD148897 | Vilnius | V | 5.27 ± 0.05 | Sleivyte (1985) |
| HD148897 | Stromgren | y | 5.25 ± 0.08 | Arellano Ferro et al. (1990) |
| HD148897 | Stromgren | y | 5.25 ± 0.08 | Hauck & Mermilliod (1998) |
| HD148897 | Johnson | V | 5.23 ± 0.05 | Fernie (1983) |
| HD148897 | Johnson | V | 5.24 ± 0.05 | Argue (1963) |
| HD148897 | Johnson | V | 5.24 ± 0.05 | Mermilliod (1986) |
| HD148897 | Vilnius | S | 4.30 ± 0.05 | Jasevicius et al. (1990) |
| HD148897 | Vilnius | S | 4.37 ± 0.05 | Sudzius et al. (1970) |
| HD148897 | Vilnius | S | 4.38 ± 0.05 | Sleivyte (1985) |
| HD148897 | WBVR | R | 4.28 ± 0.05 | Kornilov et al. (1991) |
| HD148897 | 1250 | 310 | 101.20 ± 5.20 | Smith et al. (2004) |
| HD148897 | 2200 | 361 | 83.40 ± 5.40 | Smith et al. (2004) |
| HD148897 | Johnson | K | 2.12 ± 0.05 | Neugebauer & Leighton (1969) |

**Table 21** *(continued)*

| Star ID | System/Wvlen | Band/Bandpass | Value | Reference |
|---------|--------------|---------------|-------|-----------|
| HD148897 | 3500 | 898 | $40.50 \pm 4.20$ | Smith et al. (2004) |
| HD148897 | 4900 | 712 | $19.50 \pm 4.70$ | Smith et al. (2004) |
| HD148897 | 12000 | 6384 | $4.20 \pm 16.50$ | Smith et al. (2004) |
| HD150047 | KronComet | COp | $9.14 \pm 0.00$ | This work |
| HD150047 | WBVR | B | $8.68 \pm 0.05$ | Kornilov et al. (1991) |
| HD150047 | Johnson | B | $8.67 \pm 0.03$ | This work |
| HD150047 | KronComet | Bc | $8.65 \pm 0.00$ | This work |
| HD150047 | KronComet | C2 | $7.26 \pm 0.01$ | This work |
| HD150047 | KronComet | Gc | $7.31 \pm 0.01$ | This work |
| HD150047 | WBVR | V | $7.07 \pm 0.05$ | Kornilov et al. (1991) |
| HD150047 | Johnson | V | $7.27 \pm 0.03$ | This work |
| HD150047 | WBVR | R | $5.07 \pm 0.05$ | Kornilov et al. (1991) |
| HD150047 | KronComet | Rc | $5.64 \pm 0.04$ | This work |
| HD150047 | 1250 | 310 | $182.10 \pm 9.60$ | Smith et al. (2004) |
| HD150047 | 2200 | 361 | $198.40 \pm 11.60$ | Smith et al. (2004) |
| HD150047 | Johnson | K | $1.19 \pm 0.03$ | Neugebauer & Leighton (1969) |
| HD150047 | 3500 | 898 | $102.60 \pm 8.30$ | Smith et al. (2004) |
| HD150047 | 4900 | 712 | $44.00 \pm 4.90$ | Smith et al. (2004) |
| HD150047 | 12000 | 6384 | $9.40 \pm 16.70$ | Smith et al. (2004) |
| HD150450 | KronComet | UVc | $9.28 \pm 0.07$ | Beichman et al. (1988) |
| HD150450 | Johnson | U | $8.03 \pm 0.30$ | This work |
| HD150450 | Johnson | U | $8.21 \pm 0.05$ | Ducati (2002) |
| HD150450 | Johnson | U | $8.23 \pm 0.05$ | Mermilliod (1986) |
| HD150450 | KronComet | CN | $8.65 \pm 0.04$ | This work |
| HD150450 | Oja | m41 | $7.97 \pm 0.05$ | Häggkvist & Oja (1970) |
| HD150450 | Oja | m42 | $7.79 \pm 0.05$ | Häggkvist & Oja (1970) |
| HD150450 | KronComet | COp | $7.11 \pm 0.03$ | This work |
| HD150450 | WBVR | B | $6.51 \pm 0.05$ | Kornilov et al. (1991) |
| HD150450 | Johnson | B | $6.35 \pm 0.17$ | This work |
| HD150450 | Johnson | B | $6.45 \pm 0.05$ | Häggkvist & Oja (1966) |
| HD150450 | Johnson | B | $6.45 \pm 0.05$ | Ducati (2002) |
| HD150450 | Johnson | B | $6.47 \pm 0.05$ | Mermilliod (1986) |
| HD150450 | KronComet | Bc | $6.26 \pm 0.04$ | This work |
| HD150450 | Oja | m45 | $6.12 \pm 0.05$ | Häggkvist & Oja (1970) |
| HD150450 | KronComet | C2 | $5.25 \pm 0.02$ | This work |
| HD150450 | KronComet | Gc | $4.99 \pm 0.01$ | This work |
| HD150450 | WBVR | V | $4.89 \pm 0.05$ | Kornilov et al. (1991) |
| HD150450 | Johnson | V | $4.90 \pm 0.05$ | Häggkvist & Oja (1966) |
| HD150450 | Johnson | V | $4.90 \pm 0.05$ | Mermilliod (1986) |
| HD150450 | Johnson | V | $4.90 \pm 0.05$ | Ducati (2002) |
| HD150450 | Johnson | V | $5.15 \pm 0.02$ | This work |

**Table 21** *(continued)*

| Star ID | System/Wvlen | Band/Bandpass | Value | Reference |
|---------|--------------|---------------|-------|-----------|
| HD150450 | WBVR | R | $3.47 \pm 0.05$ | Kornilov et al. (1991) |
| HD150450 | KronComet | Rc | $3.51 \pm 0.01$ | This work |
| HD150450 | 2200 | 361 | $546.00 \pm 6.50$ | Smith et al. (2004) |
| HD150450 | Johnson | K | $0.29 \pm 0.05$ | Neugebauer & Leighton (1969) |
| HD150450 | Johnson | K | $0.44 \pm 0.05$ | Ducati (2002) |
| HD150450 | 3500 | 898 | $276.50 \pm 11.30$ | Smith et al. (2004) |
| HD150450 | 4900 | 712 | $117.90 \pm 5.80$ | Smith et al. (2004) |
| HD150450 | 12000 | 6384 | $31.80 \pm 16.10$ | Smith et al. (2004) |
| HD150997 | 13c | m33 | $4.99 \pm 0.05$ | Johnson & Mitchell (1995) |
| HD150997 | Geneva | U | $5.48 \pm 0.08$ | Golay (1972) |
| HD150997 | Vilnius | U | $6.79 \pm 0.05$ | Zdanavicius et al. (1969) |
| HD150997 | 13c | m35 | $4.82 \pm 0.05$ | Johnson & Mitchell (1995) |
| HD150997 | DDO | m35 | $6.29 \pm 0.05$ | McClure & Forrester (1981) |
| HD150997 | WBVR | W | $4.85 \pm 0.05$ | Kornilov et al. (1991) |
| HD150997 | Johnson | U | $4.96 \pm 0.05$ | Mermilliod (1986) |
| HD150997 | Johnson | U | $4.98 \pm 0.05$ | Jennens & Helfer (1975) |
| HD150997 | Johnson | U | $5.02 \pm 0.05$ | Johnson et al. (1966) |
| HD150997 | Johnson | U | $5.03 \pm 0.05$ | Johnson et al. (1966) |
| HD150997 | Johnson | U | $5.03 \pm 0.05$ | Ducati (2002) |
| HD150997 | Johnson | U | $5.13 \pm 0.05$ | Argue (1963) |
| HD150997 | 13c | m37 | $4.95 \pm 0.05$ | Johnson & Mitchell (1995) |
| HD150997 | Vilnius | P | $6.25 \pm 0.05$ | Zdanavicius et al. (1969) |
| HD150997 | DDO | m38 | $5.29 \pm 0.05$ | McClure & Forrester (1981) |
| HD150997 | 13c | m40 | $4.88 \pm 0.05$ | Johnson & Mitchell (1995) |
| HD150997 | Geneva | B1 | $4.90 \pm 0.08$ | Golay (1972) |
| HD150997 | Vilnius | X | $5.37 \pm 0.05$ | Zdanavicius et al. (1969) |
| HD150997 | DDO | m41 | $5.90 \pm 0.05$ | McClure & Forrester (1981) |
| HD150997 | Oja | m41 | $5.42 \pm 0.05$ | Häggkvist & Oja (1970) |
| HD150997 | DDO | m42 | $5.75 \pm 0.05$ | McClure & Forrester (1981) |
| HD150997 | Oja | m42 | $5.31 \pm 0.05$ | Häggkvist & Oja (1970) |
| HD150997 | Geneva | B | $3.67 \pm 0.08$ | Golay (1972) |
| HD150997 | WBVR | B | $4.42 \pm 0.05$ | Kornilov et al. (1991) |
| HD150997 | Johnson | B | $4.38 \pm 0.05$ | Häggkvist & Oja (1966) |
| HD150997 | Johnson | B | $4.38 \pm 0.05$ | Mermilliod (1986) |
| HD150997 | Johnson | B | $4.39 \pm 0.05$ | Jennens & Helfer (1975) |
| HD150997 | Johnson | B | $4.42 \pm 0.05$ | Johnson et al. (1966) |
| HD150997 | Johnson | B | $4.42 \pm 0.05$ | Ducati (2002) |
| HD150997 | Johnson | B | $4.54 \pm 0.05$ | Argue (1963) |
| HD150997 | Geneva | B2 | $4.87 \pm 0.08$ | Golay (1972) |
| HD150997 | 13c | m45 | $4.15 \pm 0.05$ | Johnson & Mitchell (1995) |
| HD150997 | DDO | m45 | $4.97 \pm 0.05$ | McClure & Forrester (1981) |

**Table 21** *(continued)*

| Star ID | System/Wvlen | Band/Bandpass | Value | Reference |
|---|---|---|---|---|
| HD150997 | Oja | m45 | $4.13 \pm 0.05$ | Häggkvist & Oja (1970) |
| HD150997 | Vilnius | Y | $4.22 \pm 0.05$ | Zdanavicius et al. (1969) |
| HD150997 | DDO | m48 | $3.83 \pm 0.05$ | McClure & Forrester (1981) |
| HD150997 | Vilnius | Z | $3.77 \pm 0.05$ | Zdanavicius et al. (1969) |
| HD150997 | 13c | m52 | $3.73 \pm 0.05$ | Johnson & Mitchell (1995) |
| HD150997 | Geneva | V1 | $4.26 \pm 0.08$ | Golay (1972) |
| HD150997 | WBVR | V | $3.49 \pm 0.05$ | Kornilov et al. (1991) |
| HD150997 | Vilnius | V | $3.50 \pm 0.05$ | Zdanavicius et al. (1969) |
| HD150997 | Geneva | V | $3.48 \pm 0.08$ | Golay (1972) |
| HD150997 | Johnson | V | $3.47 \pm 0.05$ | Häggkvist & Oja (1966) |
| HD150997 | Johnson | V | $3.47 \pm 0.05$ | Mermilliod (1986) |
| HD150997 | Johnson | V | $3.48 \pm 0.05$ | Jennens & Helfer (1975) |
| HD150997 | Johnson | V | $3.50 \pm 0.05$ | Johnson et al. (1966) |
| HD150997 | Johnson | V | $3.50 \pm 0.05$ | Ducati (2002) |
| HD150997 | Johnson | V | $3.61 \pm 0.05$ | Argue (1963) |
| HD150997 | 13c | m58 | $3.31 \pm 0.05$ | Johnson & Mitchell (1995) |
| HD150997 | Geneva | G | $4.49 \pm 0.08$ | Golay (1972) |
| HD150997 | 13c | m63 | $3.02 \pm 0.05$ | Johnson & Mitchell (1995) |
| HD150997 | Vilnius | S | $2.83 \pm 0.05$ | Zdanavicius et al. (1969) |
| HD150997 | WBVR | R | $2.81 \pm 0.05$ | Kornilov et al. (1991) |
| HD150997 | 13c | m72 | $2.82 \pm 0.05$ | Johnson & Mitchell (1995) |
| HD150997 | 13c | m80 | $2.61 \pm 0.05$ | Johnson & Mitchell (1995) |
| HD150997 | 13c | m86 | $2.52 \pm 0.05$ | Johnson & Mitchell (1995) |
| HD150997 | 13c | m99 | $2.37 \pm 0.05$ | Johnson & Mitchell (1995) |
| HD150997 | 13c | m110 | $2.24 \pm 0.05$ | Johnson & Mitchell (1995) |
| HD150997 | 1250 | 310 | $273.60 \pm 8.70$ | Smith et al. (2004) |
| HD150997 | Johnson | J | $1.87 \pm 0.05$ | Alonso et al. (1998) |
| HD150997 | Johnson | J | $1.98 \pm 0.05$ | Johnson et al. (1966) |
| HD150997 | Johnson | J | $1.98 \pm 0.05$ | Ducati (2002) |
| HD150997 | Johnson | J | $1.98 \pm 0.05$ | Shenavrin et al. (2011) |
| HD150997 | Johnson | H | $1.44 \pm 0.05$ | Alonso et al. (1998) |
| HD150997 | Johnson | H | $1.48 \pm 0.05$ | Shenavrin et al. (2011) |
| HD150997 | 2200 | 361 | $177.90 \pm 9.20$ | Smith et al. (2004) |
| HD150997 | Johnson | K | $1.30 \pm 0.03$ | Neugebauer & Leighton (1969) |
| HD150997 | Johnson | K | $1.35 \pm 0.05$ | Johnson et al. (1966) |
| HD150997 | Johnson | K | $1.35 \pm 0.05$ | Ducati (2002) |
| HD150997 | Johnson | K | $1.35 \pm 0.05$ | Shenavrin et al. (2011) |
| HD150997 | Johnson | L | $1.25 \pm 0.05$ | Johnson et al. (1966) |
| HD150997 | Johnson | L | $1.25 \pm 0.05$ | Ducati (2002) |
| HD150997 | 3500 | 898 | $83.00 \pm 5.20$ | Smith et al. (2004) |
| HD150997 | 4900 | 712 | $41.90 \pm 4.90$ | Smith et al. (2004) |

**Table 21** *(continued)*

| Star ID | System/Wvlen | Band/Bandpass | Value | Reference |
|---------|-------------|---------------|-------|-----------|
| HD150997 | 12000 | 6384 | $8.80 \pm 16.20$ | Smith et al. (2004) |
| HD151732 | Oja | m41 | $8.65 \pm 0.05$ | Häggkvist & Oja (1970) |
| HD151732 | Oja | m42 | $8.59 \pm 0.05$ | Häggkvist & Oja (1970) |
| HD151732 | KronComet | COp | $7.92 \pm 0.07$ | This work |
| HD151732 | WBVR | B | $7.45 \pm 0.05$ | Kornilov et al. (1991) |
| HD151732 | Johnson | B | $7.39 \pm 0.17$ | This work |
| HD151732 | Johnson | B | $7.43 \pm 0.05$ | Mermilliod (1986) |
| HD151732 | Johnson | B | $7.48 \pm 0.05$ | Haggkvist & Oja (1970) |
| HD151732 | Johnson | B | $7.48 \pm 0.05$ | Ducati (2002) |
| HD151732 | KronComet | Bc | $7.31 \pm 0.07$ | This work |
| HD151732 | Oja | m45 | $7.20 \pm 0.05$ | Häggkvist & Oja (1970) |
| HD151732 | KronComet | C2 | $6.05 \pm 0.06$ | This work |
| HD151732 | KronComet | Gc | $6.00 \pm 0.05$ | This work |
| HD151732 | WBVR | V | $5.82 \pm 0.05$ | Kornilov et al. (1991) |
| HD151732 | Johnson | V | $5.83 \pm 0.05$ | Mermilliod (1986) |
| HD151732 | Johnson | V | $5.87 \pm 0.05$ | Haggkvist & Oja (1970) |
| HD151732 | Johnson | V | $5.87 \pm 0.05$ | Ducati (2002) |
| HD151732 | Johnson | V | $5.98 \pm 0.08$ | This work |
| HD151732 | WBVR | R | $4.04 \pm 0.05$ | Kornilov et al. (1991) |
| HD151732 | KronComet | Rc | $4.47 \pm 0.05$ | This work |
| HD151732 | 1250 | 310 | $351.30 \pm 8.30$ | Smith et al. (2004) |
| HD151732 | Johnson | J | $1.64 \pm 0.05$ | McWilliam & Lambert (1984) |
| HD151732 | Johnson | J | $1.64 \pm 0.05$ | Ducati (2002) |
| HD151732 | Johnson | J | $1.66 \pm 0.05$ | Kerschbaum et al. (1996) |
| HD151732 | Johnson | H | $0.84 \pm 0.05$ | Kerschbaum et al. (1996) |
| HD151732 | 2200 | 361 | $370.60 \pm 6.80$ | Smith et al. (2004) |
| HD151732 | Johnson | K | $0.58 \pm 0.03$ | Neugebauer & Leighton (1969) |
| HD151732 | 3500 | 898 | $191.50 \pm 6.00$ | Smith et al. (2004) |
| HD151732 | 4900 | 712 | $81.00 \pm 5.20$ | Smith et al. (2004) |
| HD151732 | 12000 | 6384 | $21.70 \pm 16.10$ | Smith et al. (2004) |
| HD152173 | Oja | m41 | $8.98 \pm 0.05$ | Häggkvist & Oja (1970) |
| HD152173 | Oja | m42 | $8.73 \pm 0.05$ | Häggkvist & Oja (1970) |
| HD152173 | KronComet | COp | $8.02 \pm 0.02$ | This work |
| HD152173 | WBVR | B | $7.40 \pm 0.05$ | Kornilov et al. (1991) |
| HD152173 | Johnson | B | $7.26 \pm 0.04$ | This work |
| HD152173 | Johnson | B | $7.31 \pm 0.05$ | Haggkvist & Oja (1970) |
| HD152173 | Johnson | B | $7.32 \pm 0.05$ | Ducati (2002) |
| HD152173 | Johnson | B | $7.36 \pm 0.05$ | Mermilliod (1986) |
| HD152173 | KronComet | Bc | $7.22 \pm 0.01$ | This work |
| HD152173 | Oja | m45 | $7.04 \pm 0.05$ | Häggkvist & Oja (1970) |
| HD152173 | KronComet | C2 | $6.20 \pm 0.01$ | This work |

**Table 21** *(continued)*

| Star ID | System/Wvlen | Band/Bandpass | Value | Reference |
|---------|--------------|---------------|-------|-----------|
| HD152173 | KronComet | Gc | $5.94 \pm 0.01$ | This work |
| HD152173 | WBVR | V | $5.74 \pm 0.05$ | Kornilov et al. (1991) |
| HD152173 | Johnson | V | $5.72 \pm 0.05$ | Haggkvist & Oja (1970) |
| HD152173 | Johnson | V | $5.72 \pm 0.05$ | Ducati (2002) |
| HD152173 | Johnson | V | $5.74 \pm 0.05$ | Mermilliod (1986) |
| HD152173 | Johnson | V | $5.81 \pm 0.01$ | This work |
| HD152173 | WBVR | R | $4.48 \pm 0.05$ | Kornilov et al. (1991) |
| HD152173 | KronComet | Rc | $4.35 \pm 0.01$ | Cutri et al. (2003b) |
| HD152173 | Johnson | J | $2.82 \pm 0.05$ | McWilliam & Lambert (1984) |
| HD152173 | Johnson | J | $2.82 \pm 0.05$ | Kenyon (1988) |
| HD152173 | Johnson | J | $2.82 \pm 0.05$ | Ducati (2002) |
| HD152173 | Johnson | H | $2.04 \pm 0.05$ | Kenyon (1988) |
| HD152173 | Johnson | H | $2.04 \pm 0.05$ | Ducati (2002) |
| HD152173 | 2200 | 361 | $122.90 \pm 11.30$ | Smith et al. (2004) |
| HD152173 | Johnson | K | $1.74 \pm 0.06$ | Neugebauer & Leighton (1969) |
| HD152173 | Johnson | K | $1.83 \pm 0.05$ | Ducati (2002) |
| HD152173 | 3500 | 898 | $61.70 \pm 5.70$ | Smith et al. (2004) |
| HD152173 | 4900 | 712 | $27.60 \pm 5.30$ | Smith et al. (2004) |
| HD152173 | 12000 | 6384 | $8.20 \pm 17.20$ | Smith et al. (2004) |
| HD152863 | Geneva | U | $8.05 \pm 0.08$ | Golay (1972) |
| HD152863 | Vilnius | U | $9.39 \pm 0.05$ | Jasevicius et al. (1990) |
| HD152863 | WBVR | W | $7.46 \pm 0.05$ | Kornilov et al. (1991) |
| HD152863 | Johnson | U | $7.62 \pm 0.05$ | Johnson & Knuckles (1957) |
| HD152863 | Johnson | U | $7.62 \pm 0.05$ | Johnson et al. (1966) |
| HD152863 | Johnson | U | $7.62 \pm 0.05$ | Nicolet (1978) |
| HD152863 | Johnson | U | $7.64 \pm 0.05$ | Mermilliod (1986) |
| HD152863 | Vilnius | P | $8.81 \pm 0.05$ | Jasevicius et al. (1990) |
| HD152863 | Geneva | B1 | $7.47 \pm 0.08$ | Golay (1972) |
| HD152863 | Vilnius | X | $7.94 \pm 0.05$ | Jasevicius et al. (1990) |
| HD152863 | Geneva | B | $6.25 \pm 0.08$ | Golay (1972) |
| HD152863 | WBVR | B | $7.03 \pm 0.05$ | Kornilov et al. (1991) |
| HD152863 | Johnson | B | $7.00 \pm 0.05$ | Johnson & Knuckles (1957) |
| HD152863 | Johnson | B | $7.00 \pm 0.05$ | Johnson et al. (1966) |
| HD152863 | Johnson | B | $7.00 \pm 0.05$ | Nicolet (1978) |
| HD152863 | Johnson | B | $7.01 \pm 0.05$ | Mermilliod (1986) |
| HD152863 | Johnson | B | $7.03 \pm 0.05$ | Miczaika (1954) |
| HD152863 | Geneva | B2 | $7.44 \pm 0.08$ | Golay (1972) |
| HD152863 | Vilnius | Y | $6.79 \pm 0.05$ | Jasevicius et al. (1990) |
| HD152863 | Vilnius | Z | $6.35 \pm 0.05$ | Jasevicius et al. (1990) |
| HD152863 | Geneva | V1 | $6.82 \pm 0.08$ | Golay (1972) |
| HD152863 | WBVR | V | $6.08 \pm 0.05$ | Kornilov et al. (1991) |

**Table 21** *(continued)*

| Star ID | System/Wvlen | Band/Bandpass | Value | Reference |
|---------|--------------|---------------|-------|-----------|
| HD152863 | Vilnius | V | $6.08 \pm 0.05$ | Jasevicius et al. (1990) |
| HD152863 | Geneva | V | $6.06 \pm 0.08$ | Golay (1972) |
| HD152863 | Johnson | V | $6.08 \pm 0.05$ | Johnson & Knuckles (1957) |
| HD152863 | Johnson | V | $6.08 \pm 0.05$ | Johnson et al. (1966) |
| HD152863 | Johnson | V | $6.08 \pm 0.05$ | Nicolet (1978) |
| HD152863 | Johnson | V | $6.08 \pm 0.05$ | Mermilliod (1986) |
| HD152863 | Johnson | V | $6.13 \pm 0.05$ | Miczaika (1954) |
| HD152863 | Geneva | G | $7.05 \pm 0.08$ | Golay (1972) |
| HD152863 | Vilnius | S | $5.38 \pm 0.05$ | Jasevicius et al. (1990) |
| HD152863 | WBVR | R | $5.39 \pm 0.05$ | Kornilov et al. (1991) |
| HD152863 | 2200 | 361 | $25.50 \pm 10.50$ | Smith et al. (2004) |
| HD152863 | 3500 | 898 | $14.00 \pm 9.20$ | Smith et al. (2004) |
| HD152863 | 4900 | 712 | $7.50 \pm 5.50$ | Smith et al. (2004) |
| HD152863 | 12000 | 6384 | $0.90 \pm 17.90$ | Smith et al. (2004) |
| HD153287 | KronComet | NH | $8.89 \pm 0.03$ | This work |
| HD153287 | KronComet | UVc | $8.64 \pm 0.02$ | This work |
| HD153287 | DDO | m35 | $9.09 \pm 0.05$ | McClure & Forrester (1981) |
| HD153287 | WBVR | W | $7.64 \pm 0.05$ | Kornilov et al. (1991) |
| HD153287 | Johnson | U | $7.70 \pm 0.30$ | This work |
| HD153287 | Johnson | U | $7.73 \pm 0.05$ | Guetter (1980) |
| HD153287 | Johnson | U | $7.79 \pm 0.01$ | Oja (1991) |
| HD153287 | DDO | m38 | $8.05 \pm 0.05$ | McClure & Forrester (1981) |
| HD153287 | KronComet | CN | $8.53 \pm 0.03$ | This work |
| HD153287 | DDO | m41 | $8.64 \pm 0.05$ | McClure & Forrester (1981) |
| HD153287 | DDO | m42 | $8.52 \pm 0.05$ | McClure & Forrester (1981) |
| HD153287 | KronComet | COp | $7.43 \pm 0.02$ | This work |
| HD153287 | WBVR | B | $7.23 \pm 0.05$ | Kornilov et al. (1991) |
| HD153287 | Johnson | B | $7.16 \pm 0.03$ | This work |
| HD153287 | Johnson | B | $7.18 \pm 0.05$ | Guetter (1980) |
| HD153287 | Johnson | B | $7.20 \pm 0.01$ | Oja (1991) |
| HD153287 | KronComet | Bc | $7.04 \pm 0.01$ | This work |
| HD153287 | DDO | m45 | $7.77 \pm 0.05$ | McClure & Forrester (1981) |
| HD153287 | DDO | m48 | $6.62 \pm 0.05$ | McClure & Forrester (1981) |
| HD153287 | KronComet | C2 | $6.43 \pm 0.01$ | This work |
| HD153287 | KronComet | Gc | $6.32 \pm 0.01$ | This work |
| HD153287 | WBVR | V | $6.29 \pm 0.05$ | Kornilov et al. (1991) |
| HD153287 | Johnson | V | $6.28 \pm 0.05$ | Guetter (1980) |
| HD153287 | Johnson | V | $6.29 \pm 0.01$ | Oja (1991) |
| HD153287 | Johnson | V | $6.34 \pm 0.01$ | This work |
| HD153287 | WBVR | R | $5.60 \pm 0.05$ | Kornilov et al. (1991) |
| HD153287 | KronComet | Rc | $5.31 \pm 0.01$ | This work |

**Table 21** *(continued)*

| Star ID | System/Wvlen | Band/Bandpass | Value | Reference |
|---------|-------------|---------------|-------|-----------|
| HD153698 | KronComet | NH | $12.59 \pm 0.23$ | This work |
| HD153698 | KronComet | UVc | $12.05 \pm 0.06$ | This work |
| HD153698 | WBVR | W | $10.70 \pm 0.05$ | Kornilov et al. (1991) |
| HD153698 | Johnson | U | $10.36 \pm 0.31$ | This work |
| HD153698 | Johnson | U | $10.76 \pm 0.05$ | Fernie (1983) |
| HD153698 | KronComet | CN | $10.86 \pm 0.03$ | This work |
| HD153698 | KronComet | COp | $9.38 \pm 0.02$ | This work |
| HD153698 | WBVR | B | $8.90 \pm 0.05$ | Kornilov et al. (1991) |
| HD153698 | Johnson | B | $8.73 \pm 0.03$ | This work |
| HD153698 | Johnson | B | $9.09 \pm 0.05$ | Fernie (1983) |
| HD153698 | KronComet | Bc | $8.66 \pm 0.01$ | This work |
| HD153698 | KronComet | C2 | $7.49 \pm 0.01$ | This work |
| HD153698 | KronComet | Gc | $7.33 \pm 0.01$ | This work |
| HD153698 | WBVR | V | $7.23 \pm 0.05$ | Kornilov et al. (1991) |
| HD153698 | Johnson | V | $7.31 \pm 0.01$ | This work |
| HD153698 | Johnson | V | $7.54 \pm 0.05$ | Fernie (1983) |
| HD153698 | WBVR | R | $5.53 \pm 0.05$ | Kornilov et al. (1991) |
| HD153698 | KronComet | Rc | $5.78 \pm 0.01$ | This work |
| HD153698 | 1250 | 310 | $86.50 \pm 8.00$ | Smith et al. (2004) |
| HD153698 | 2200 | 361 | $87.30 \pm 8.10$ | Smith et al. (2004) |
| HD153698 | Johnson | K | $2.08 \pm 0.05$ | Neugebauer & Leighton (1969) |
| HD153698 | 3500 | 898 | $44.20 \pm 6.00$ | Smith et al. (2004) |
| HD153698 | 4900 | 712 | $18.80 \pm 5.30$ | Smith et al. (2004) |
| HD153698 | 12000 | 6384 | $5.10 \pm 16.10$ | Smith et al. (2004) |
| HD153834 | KronComet | UVc | $9.34 \pm 0.02$ | This work |
| HD153834 | DDO | m35 | $9.77 \pm 0.05$ | McClure & Forrester (1981) |
| HD153834 | WBVR | W | $8.28 \pm 0.05$ | Kornilov et al. (1991) |
| HD153834 | Johnson | U | $8.15 \pm 0.34$ | This work |
| HD153834 | DDO | m38 | $8.58 \pm 0.05$ | McClure & Forrester (1981) |
| HD153834 | KronComet | CN | $9.12 \pm 0.02$ | This work |
| HD153834 | DDO | m41 | $8.95 \pm 0.05$ | McClure & Forrester (1981) |
| HD153834 | Oja | m41 | $8.50 \pm 0.05$ | Häggkvist & Oja (1970) |
| HD153834 | DDO | m42 | $8.55 \pm 0.05$ | McClure & Forrester (1981) |
| HD153834 | Oja | m42 | $8.10 \pm 0.05$ | Häggkvist & Oja (1970) |
| HD153834 | KronComet | COp | $7.40 \pm 0.02$ | This work |
| HD153834 | WBVR | B | $7.05 \pm 0.05$ | Kornilov et al. (1991) |
| HD153834 | Johnson | B | $6.90 \pm 0.02$ | This work |
| HD153834 | Johnson | B | $6.98 \pm 0.05$ | Haggkvist & Oja (1970) |
| HD153834 | KronComet | Bc | $6.84 \pm 0.01$ | This work |
| HD153834 | DDO | m45 | $7.51 \pm 0.05$ | McClure & Forrester (1981) |
| HD153834 | Oja | m45 | $6.70 \pm 0.05$ | Häggkvist & Oja (1970) |

**Table 21** *(continued)*

| Star ID | System/Wvlen | Band/Bandpass | Value | Reference |
|---------|--------------|---------------|-------|-----------|
| HD153834 | DDO | m48 | $6.16 \pm 0.05$ | McClure & Forrester (1981) |
| HD153834 | KronComet | C2 | $5.95 \pm 0.01$ | This work |
| HD153834 | KronComet | Gc | $5.81 \pm 0.00$ | This work |
| HD153834 | WBVR | V | $5.67 \pm 0.05$ | Kornilov et al. (1991) |
| HD153834 | Johnson | V | $5.65 \pm 0.05$ | Haggkvist & Oja (1970) |
| HD153834 | Johnson | V | $5.72 \pm 0.01$ | This work |
| HD153834 | WBVR | R | $4.77 \pm 0.05$ | Kornilov et al. (1991) |
| HD153834 | KronComet | Rc | $4.53 \pm 0.01$ | This work |
| HD153834 | 1250 | 310 | $67.80 \pm 20.40$ | Smith et al. (2004) |
| HD153834 | 2200 | 361 | $55.30 \pm 11.70$ | Smith et al. (2004) |
| HD153834 | Johnson | K | $2.80 \pm 0.06$ | Neugebauer & Leighton (1969) |
| HD153834 | 3500 | 898 | $26.00 \pm 9.00$ | Smith et al. (2004) |
| HD153834 | 4900 | 712 | $13.90 \pm 5.20$ | Smith et al. (2004) |
| HD153834 | 12000 | 6384 | $-0.40 \pm 17.00$ | Smith et al. (2004) |
| HD154301 | KronComet | COp | $8.51 \pm 0.02$ | This work |
| HD154301 | WBVR | B | $7.94 \pm 0.05$ | Kornilov et al. (1991) |
| HD154301 | Johnson | B | $7.79 \pm 0.02$ | This work |
| HD154301 | Johnson | B | $7.87 \pm 0.05$ | Haggkvist & Oja (1970) |
| HD154301 | Johnson | B | $7.87 \pm 0.05$ | Nicolet (1978) |
| HD154301 | KronComet | Bc | $7.74 \pm 0.01$ | This work |
| HD154301 | KronComet | C2 | $6.85 \pm 0.01$ | This work |
| HD154301 | KronComet | Gc | $6.57 \pm 0.01$ | This work |
| HD154301 | WBVR | V | $6.39 \pm 0.05$ | Kornilov et al. (1991) |
| HD154301 | Johnson | V | $6.35 \pm 0.05$ | Haggkvist & Oja (1970) |
| HD154301 | Johnson | V | $6.35 \pm 0.05$ | Nicolet (1978) |
| HD154301 | Johnson | V | $6.46 \pm 0.01$ | This work |
| HD154301 | WBVR | R | $5.18 \pm 0.05$ | Kornilov et al. (1991) |
| HD154301 | KronComet | Rc | $5.02 \pm 0.01$ | This work |
| HD154301 | 1250 | 310 | $65.30 \pm 8.80$ | Smith et al. (2004) |
| HD154301 | 2200 | 361 | $58.50 \pm 7.50$ | Smith et al. (2004) |
| HD154301 | Johnson | K | $2.58 \pm 0.05$ | Neugebauer & Leighton (1969) |
| HD154301 | 3500 | 898 | $27.20 \pm 6.10$ | Smith et al. (2004) |
| HD154301 | 4900 | 712 | $11.80 \pm 5.90$ | Smith et al. (2004) |
| HD154301 | 12000 | 6384 | $5.30 \pm 17.40$ | Smith et al. (2004) |
| HD155816 | WBVR | B | $8.23 \pm 0.05$ | Kornilov et al. (1991) |
| HD155816 | Johnson | B | $8.10 \pm 0.01$ | This work |
| HD155816 | KronComet | Bc | $8.05 \pm 0.01$ | This work |
| HD155816 | KronComet | C2 | $7.08 \pm 0.01$ | This work |
| HD155816 | KronComet | Gc | $6.79 \pm 0.01$ | This work |
| HD155816 | WBVR | V | $6.58 \pm 0.05$ | Kornilov et al. (1991) |
| HD155816 | Johnson | V | $6.65 \pm 0.01$ | This work |

**Table 21** *(continued)*

| Star ID | System/Wvlen | Band/Bandpass | Value | Reference |
|---------|--------------|---------------|-------|-----------|
| HD155816 | WBVR | R | $5.30 \pm 0.05$ | Kornilov et al. (1991) |
| HD155816 | KronComet | Rc | $5.18 \pm 0.01$ | This work |
| HD155816 | 1250 | 310 | $59.90 \pm 6.40$ | Smith et al. (2004) |
| HD155816 | 2200 | 361 | $55.90 \pm 6.50$ | Smith et al. (2004) |
| HD155816 | Johnson | K | $2.74 \pm 0.08$ | Neugebauer & Leighton (1969) |
| HD155816 | 3500 | 898 | $32.90 \pm 12.70$ | Smith et al. (2004) |
| HD155816 | 4900 | 712 | $14.40 \pm 8.80$ | Smith et al. (2004) |
| HD155816 | 12000 | 6384 | $6.10 \pm 16.20$ | Smith et al. (2004) |
| HD156966 | KronComet | COp | $9.04 \pm 0.05$ | This work |
| HD156966 | WBVR | B | $8.51 \pm 0.05$ | Kornilov et al. (1991) |
| HD156966 | Johnson | B | $8.27 \pm 0.09$ | This work |
| HD156966 | Johnson | B | $8.35 \pm 0.05$ | Roman (1955) |
| HD156966 | Johnson | B | $8.50 \pm 0.05$ | Mermilliod (1986) |
| HD156966 | KronComet | Bc | $8.21 \pm 0.05$ | This work |
| HD156966 | KronComet | C2 | $7.18 \pm 0.03$ | This work |
| HD156966 | KronComet | Gc | $6.92 \pm 0.03$ | This work |
| HD156966 | WBVR | V | $6.81 \pm 0.05$ | Kornilov et al. (1991) |
| HD156966 | Johnson | V | $6.70 \pm 0.05$ | Roman (1955) |
| HD156966 | Johnson | V | $6.84 \pm 0.05$ | Mermilliod (1986) |
| HD156966 | Johnson | V | $6.86 \pm 0.04$ | This work |
| HD156966 | WBVR | R | $5.37 \pm 0.05$ | Kornilov et al. (1991) |
| HD156966 | KronComet | Rc | $5.39 \pm 0.02$ | This work |
| HD156966 | 1250 | 310 | $62.00 \pm 4.70$ | Smith et al. (2004) |
| HD156966 | 2200 | 361 | $63.90 \pm 11.60$ | Smith et al. (2004) |
| HD156966 | Johnson | K | $2.38 \pm 0.05$ | Neugebauer & Leighton (1969) |
| HD156966 | 3500 | 898 | $37.70 \pm 18.00$ | Smith et al. (2004) |
| HD156966 | 4900 | 712 | $19.30 \pm 12.80$ | Smith et al. (2004) |
| HD156966 | 12000 | 6384 | $9.90 \pm 18.20$ | Smith et al. (2004) |
| HD157617 | KronComet | NH | $9.41 \pm 0.13$ | This work |
| HD157617 | KronComet | UVc | $9.13 \pm 0.11$ | This work |
| HD157617 | Vilnius | U | $10.17 \pm 0.05$ | Bartkevicius et al. (1973) |
| HD157617 | DDO | m35 | $9.69 \pm 0.05$ | McClure & Forrester (1981) |
| HD157617 | WBVR | W | $8.21 \pm 0.05$ | Kornilov et al. (1991) |
| HD157617 | Johnson | U | $7.67 \pm 0.03$ | This work |
| HD157617 | Johnson | U | $8.28 \pm 0.05$ | Cousins (1963c) |
| HD157617 | Johnson | U | $8.29 \pm 0.05$ | Johnson et al. (1966) |
| HD157617 | Vilnius | P | $9.44 \pm 0.05$ | Bartkevicius et al. (1973) |
| HD157617 | DDO | m38 | $8.46 \pm 0.05$ | McClure & Forrester (1981) |
| HD157617 | KronComet | CN | $8.93 \pm 0.07$ | This work |
| HD157617 | Vilnius | X | $8.28 \pm 0.05$ | Bartkevicius et al. (1973) |
| HD157617 | DDO | m41 | $8.85 \pm 0.05$ | McClure & Forrester (1981) |

**Table 21** *(continued)*

| Star ID | System/Wvlen | Band/Bandpass | Value | Reference |
|---------|--------------|---------------|-------|-----------|
| HD157617 | Oja | m41 | $8.39 \pm 0.05$ | Häggkvist & Oja (1970) |
| HD157617 | DDO | m42 | $8.49 \pm 0.05$ | McClure & Forrest (1981) |
| HD157617 | Oja | m42 | $8.01 \pm 0.05$ | Häggkvist & Oja (1970) |
| HD157617 | KronComet | COp | $7.28 \pm 0.05$ | This work |
| HD157617 | WBVR | B | $7.06 \pm 0.05$ | Kornilov et al. (1991) |
| HD157617 | Johnson | B | $6.90 \pm 0.09$ | This work |
| HD157617 | Johnson | B | $7.01 \pm 0.05$ | Cousins (1963c) |
| HD157617 | Johnson | B | $7.02 \pm 0.05$ | Johnson et al. (1966) |
| HD157617 | KronComet | Bc | $6.82 \pm 0.04$ | This work |
| HD157617 | DDO | m45 | $7.55 \pm 0.05$ | McClure & Forrest (1981) |
| HD157617 | Oja | m45 | $6.71 \pm 0.05$ | Häggkvist & Oja (1970) |
| HD157617 | Vilnius | Y | $6.72 \pm 0.05$ | Bartkevicius et al. (1973) |
| HD157617 | DDO | m48 | $6.26 \pm 0.05$ | McClure & Forrest (1981) |
| HD157617 | KronComet | C2 | $5.96 \pm 0.03$ | This work |
| HD157617 | Vilnius | Z | $6.14 \pm 0.05$ | Bartkevicius et al. (1973) |
| HD157617 | KronComet | Gc | $5.87 \pm 0.03$ | This work |
| HD157617 | WBVR | V | $5.76 \pm 0.05$ | Kornilov et al. (1991) |
| HD157617 | Vilnius | V | $5.77 \pm 0.05$ | Bartkevicius et al. (1973) |
| HD157617 | Johnson | V | $5.76 \pm 0.05$ | Cousins (1963c) |
| HD157617 | Johnson | V | $5.77 \pm 0.05$ | Johnson et al. (1966) |
| HD157617 | Johnson | V | $5.82 \pm 0.05$ | This work |
| HD157617 | Vilnius | S | $4.91 \pm 0.05$ | Bartkevicius et al. (1973) |
| HD157617 | WBVR | R | $4.87 \pm 0.05$ | Kornilov et al. (1991) |
| HD157617 | KronComet | Rc | $4.60 \pm 0.02$ | This work |
| HD157617 | 1250 | 310 | $73.20 \pm 21.00$ | Smith et al. (2004) |
| HD157617 | 2200 | 361 | $70.20 \pm 20.50$ | Smith et al. (2004) |
| HD157617 | Johnson | K | $2.92 \pm 0.08$ | Neugebauer & Leighton (1969) |
| HD157617 | 3500 | 898 | $36.80 \pm 10.70$ | Smith et al. (2004) |
| HD157617 | 4900 | 712 | $16.80 \pm 6.80$ | Smith et al. (2004) |
| HD157617 | 12000 | 6384 | $5.10 \pm 19.30$ | Smith et al. (2004) |
| HD163547 | KronComet | NH | $9.62 \pm 0.01$ | This work |
| HD163547 | KronComet | UVc | $9.30 \pm 0.01$ | This work |
| HD163547 | Vilnius | U | $10.07 \pm 0.05$ | Zdanavicius et al. (1972) |
| HD163547 | WBVR | W | $8.09 \pm 0.05$ | Kornilov et al. (1991) |
| HD163547 | Vilnius | P | $9.38 \pm 0.05$ | Zdanavicius et al. (1972) |
| HD163547 | KronComet | CN | $8.98 \pm 0.02$ | This work |
| HD163547 | Vilnius | X | $8.17 \pm 0.05$ | Zdanavicius et al. (1972) |
| HD163547 | Oja | m41 | $8.26 \pm 0.05$ | Häggkvist & Oja (1970) |
| HD163547 | Oja | m42 | $7.96 \pm 0.05$ | Häggkvist & Oja (1970) |
| HD163547 | KronComet | COp | $7.31 \pm 0.01$ | This work |
| HD163547 | WBVR | B | $6.92 \pm 0.05$ | Kornilov et al. (1991) |

**Table 21** *(continued)*

| Star ID | System/Wvlen | Band/Bandpass | Value | Reference |
|---------|--------------|---------------|-------|-----------|
| HD163547 | Johnson | B | $6.81 \pm 0.01$ | This work |
| HD163547 | Johnson | B | $6.83 \pm 0.05$ | Haggkvist & Oja (1970) |
| HD163547 | KronComet | Bc | $6.69 \pm 0.01$ | This work |
| HD163547 | Oja | m45 | $6.54 \pm 0.05$ | Häggkvist & Oja (1970) |
| HD163547 | Vilnius | Y | $6.53 \pm 0.05$ | Zdanavicius et al. (1972) |
| HD163547 | KronComet | C2 | $5.92 \pm 0.01$ | This work |
| HD163547 | Vilnius | Z | $6.02 \pm 0.05$ | Zdanavicius et al. (1972) |
| HD163547 | KronComet | Gc | $5.75 \pm 0.01$ | This work |
| HD163547 | WBVR | V | $5.62 \pm 0.05$ | Kornilov et al. (1991) |
| HD163547 | Vilnius | V | $5.62 \pm 0.05$ | Zdanavicius et al. (1972) |
| HD163547 | Johnson | V | $5.59 \pm 0.05$ | Haggkvist & Oja (1970) |
| HD163547 | Johnson | V | $5.69 \pm 0.01$ | This work |
| HD163547 | Vilnius | S | $4.78 \pm 0.05$ | Zdanavicius et al. (1972) |
| HD163547 | WBVR | R | $4.74 \pm 0.05$ | Kornilov et al. (1991) |
| HD163547 | KronComet | Rc | $4.47 \pm 0.01$ | This work |
| HD163547 | 1250 | 310 | $51.80 \pm 5.00$ | Smith et al. (2004) |
| HD163547 | 2200 | 361 | $41.00 \pm 6.60$ | Smith et al. (2004) |
| HD163547 | Johnson | K | $2.77 \pm 0.06$ | Neugebauer & Leighton (1969) |
| HD163547 | 3500 | 898 | $17.20 \pm 8.40$ | Smith et al. (2004) |
| HD163547 | 4900 | 712 | $8.40 \pm 5.40$ | Smith et al. (2004) |
| HD163547 | 12000 | 6384 | $-2.40 \pm 18.00$ | Smith et al. (2004) |
| HD163947 | KronComet | COp | $10.27 \pm 0.02$ | This work |
| HD163947 | Johnson | B | $9.86 \pm 0.02$ | This work |
| HD163947 | KronComet | Bc | $9.88 \pm 0.01$ | This work |
| HD163947 | KronComet | C2 | $8.52 \pm 0.01$ | This work |
| HD163947 | KronComet | Gc | $8.60 \pm 0.01$ | This work |
| HD163947 | Johnson | V | $8.54 \pm 0.01$ | This work |
| HD163947 | KronComet | Rc | $6.93 \pm 0.01$ | This work |
| HD163947 | 1250 | 310 | $66.80 \pm 7.70$ | Smith et al. (2004) |
| HD163947 | 2200 | 361 | $69.20 \pm 7.20$ | Smith et al. (2004) |
| HD163947 | Johnson | K | $2.32 \pm 0.07$ | Neugebauer & Leighton (1969) |
| HD163947 | 3500 | 898 | $33.50 \pm 7.50$ | Smith et al. (2004) |
| HD163947 | 4900 | 712 | $14.30 \pm 5.80$ | Smith et al. (2004) |
| HD163947 | 12000 | 6384 | $3.20 \pm 15.80$ | Smith et al. (2004) |
| HD163993 | 13c | m33 | $5.35 \pm 0.05$ | Johnson & Mitchell (1995) |
| HD163993 | Geneva | U | $5.77 \pm 0.08$ | Golay (1972) |
| HD163993 | 13c | m35 | $5.19 \pm 0.05$ | Johnson & Mitchell (1995) |
| HD163993 | DDO | m35 | $6.60 \pm 0.05$ | McClure & Forrester (1981) |
| HD163993 | WBVR | W | $5.16 \pm 0.05$ | Kornilov et al. (1991) |
| HD163993 | Johnson | U | $5.29 \pm 0.05$ | Mermilliod (1986) |
| HD163993 | Johnson | U | $5.31 \pm 0.05$ | Jennens & Helfer (1975) |

**Table 21** *(continued)*

| Star ID | System/Wvlen | Band/Bandpass | Value | Reference |
|---------|--------------|---------------|-------|-----------|
| HD163993 | Johnson | U | $5.33 \pm 0.05$ | Argue (1963) |
| HD163993 | Johnson | U | $5.33 \pm 0.05$ | Argue (1966) |
| HD163993 | Johnson | U | $5.34 \pm 0.05$ | Johnson et al. (1966) |
| HD163993 | Johnson | U | $5.34 \pm 0.05$ | Ducati (2002) |
| HD163993 | 13c | m37 | $5.27 \pm 0.05$ | Johnson & Mitchell (1995) |
| HD163993 | DDO | m38 | $5.56 \pm 0.05$ | McClure & Forrester (1981) |
| HD163993 | 13c | m40 | $5.15 \pm 0.05$ | Johnson & Mitchell (1995) |
| HD163993 | Geneva | B1 | $5.15 \pm 0.08$ | Golay (1972) |
| HD163993 | DDO | m41 | $6.19 \pm 0.05$ | McClure & Forrester (1981) |
| HD163993 | Oja | m41 | $5.79 \pm 0.05$ | Häggkvist & Oja (1970) |
| HD163993 | DDO | m42 | $5.97 \pm 0.05$ | McClure & Forrester (1981) |
| HD163993 | Oja | m42 | $5.57 \pm 0.05$ | Häggkvist & Oja (1970) |
| HD163993 | Geneva | B | $3.89 \pm 0.08$ | Golay (1972) |
| HD163993 | WBVR | B | $4.66 \pm 0.05$ | Kornilov et al. (1991) |
| HD163993 | Johnson | B | $4.61 \pm 0.05$ | Häggkvist & Oja (1966) |
| HD163993 | Johnson | B | $4.63 \pm 0.05$ | Argue (1966) |
| HD163993 | Johnson | B | $4.63 \pm 0.05$ | Jennens & Helfer (1975) |
| HD163993 | Johnson | B | $4.63 \pm 0.05$ | Mermilliod (1986) |
| HD163993 | Johnson | B | $4.64 \pm 0.05$ | Johnson et al. (1966) |
| HD163993 | Johnson | B | $4.64 \pm 0.05$ | Ducati (2002) |
| HD163993 | Johnson | B | $4.65 \pm 0.05$ | Argue (1963) |
| HD163993 | Geneva | B2 | $5.10 \pm 0.08$ | Golay (1972) |
| HD163993 | 13c | m45 | $4.38 \pm 0.05$ | Johnson & Mitchell (1995) |
| HD163993 | DDO | m45 | $5.18 \pm 0.05$ | McClure & Forrester (1981) |
| HD163993 | Oja | m45 | $4.38 \pm 0.05$ | Häggkvist & Oja (1970) |
| HD163993 | DDO | m48 | $4.03 \pm 0.05$ | McClure & Forrester (1981) |
| HD163993 | 13c | m52 | $3.94 \pm 0.05$ | Johnson & Mitchell (1995) |
| HD163993 | Geneva | V1 | $4.47 \pm 0.08$ | Golay (1972) |
| HD163993 | WBVR | V | $3.71 \pm 0.05$ | Kornilov et al. (1991) |
| HD163993 | Geneva | V | $3.69 \pm 0.08$ | Golay (1972) |
| HD163993 | Johnson | V | $3.68 \pm 0.05$ | Häggkvist & Oja (1966) |
| HD163993 | Johnson | V | $3.68 \pm 0.05$ | Argue (1966) |
| HD163993 | Johnson | V | $3.70 \pm 0.05$ | Johnson et al. (1966) |
| HD163993 | Johnson | V | $3.70 \pm 0.05$ | Jennens & Helfer (1975) |
| HD163993 | Johnson | V | $3.70 \pm 0.05$ | Mermilliod (1986) |
| HD163993 | Johnson | V | $3.70 \pm 0.05$ | Ducati (2002) |
| HD163993 | Johnson | V | $3.71 \pm 0.05$ | Argue (1963) |
| HD163993 | 13c | m58 | $3.50 \pm 0.05$ | Johnson & Mitchell (1995) |
| HD163993 | Geneva | G | $4.70 \pm 0.08$ | Golay (1972) |
| HD163993 | 13c | m63 | $3.22 \pm 0.05$ | Johnson & Mitchell (1995) |
| HD163993 | WBVR | R | $3.03 \pm 0.05$ | Kornilov et al. (1991) |

**Table 21** *(continued)*

| Star ID | System/Wvlen | Band/Bandpass | Value | Reference |
|---------|--------------|---------------|-------|-----------|
| HD163993 | 13c | m72 | $2.99 \pm 0.05$ | Johnson & Mitchell (1995) |
| HD163993 | 13c | m80 | $2.78 \pm 0.05$ | Johnson & Mitchell (1995) |
| HD163993 | 13c | m86 | $2.69 \pm 0.05$ | Johnson & Mitchell (1995) |
| HD163993 | 13c | m99 | $2.56 \pm 0.05$ | Johnson & Mitchell (1995) |
| HD163993 | 13c | m110 | $2.44 \pm 0.05$ | Johnson & Mitchell (1995) |
| HD163993 | Johnson | J | $2.14 \pm 0.05$ | Selby et al. (1988) |
| HD163993 | Johnson | J | $2.14 \pm 0.05$ | Blackwell et al. (1990) |
| HD163993 | Johnson | J | $2.15 \pm 0.05$ | Ducati (2002) |
| HD163993 | Johnson | J | $2.19 \pm 0.05$ | Johnson et al. (1966) |
| HD163993 | Johnson | J | $2.19 \pm 0.05$ | Shenavrin et al. (2011) |
| HD163993 | Johnson | H | $1.73 \pm 0.05$ | Shenavrin et al. (2011) |
| HD163993 | 2200 | 361 | $143.40 \pm 4.40$ | Smith et al. (2004) |
| HD163993 | Johnson | K | $1.58 \pm 0.04$ | Neugebauer & Leighton (1969) |
| HD163993 | Johnson | K | $1.58 \pm 0.05$ | Ducati (2002) |
| HD163993 | Johnson | K | $1.62 \pm 0.05$ | Johnson et al. (1966) |
| HD163993 | Johnson | K | $1.62 \pm 0.05$ | Shenavrin et al. (2011) |
| HD163993 | 3500 | 898 | $70.80 \pm 11.90$ | Smith et al. (2004) |
| HD163993 | 4900 | 712 | $43.60 \pm 14.00$ | Smith et al. (2004) |
| HD163993 | 12000 | 6384 | $12.50 \pm 20.50$ | Smith et al. (2004) |
| HD164064 | DDO | m35 | $10.76 \pm 0.05$ | McClure & Forrester (1981) |
| HD164064 | WBVR | W | $9.30 \pm 0.05$ | Kornilov et al. (1991) |
| HD164064 | Johnson | U | $9.32 \pm 0.05$ | Mermilliod (1986) |
| HD164064 | Johnson | U | $9.33 \pm 0.05$ | Cousins (1964a) |
| HD164064 | DDO | m38 | $9.41 \pm 0.05$ | McClure & Forrester (1981) |
| HD164064 | DDO | m41 | $9.48 \pm 0.05$ | McClure & Forrester (1981) |
| HD164064 | Oja | m41 | $8.98 \pm 0.05$ | Häggkvist & Oja (1970) |
| HD164064 | DDO | m42 | $9.21 \pm 0.05$ | McClure & Forrester (1981) |
| HD164064 | Oja | m42 | $8.67 \pm 0.05$ | Häggkvist & Oja (1970) |
| HD164064 | WBVR | B | $7.47 \pm 0.05$ | Kornilov et al. (1991) |
| HD164064 | Johnson | B | $7.42 \pm 0.05$ | Cousins (1964a) |
| HD164064 | Johnson | B | $7.42 \pm 0.05$ | Mermilliod (1986) |
| HD164064 | DDO | m45 | $7.88 \pm 0.05$ | McClure & Forrester (1981) |
| HD164064 | Oja | m45 | $7.03 \pm 0.05$ | Häggkvist & Oja (1970) |
| HD164064 | DDO | m48 | $6.50 \pm 0.05$ | McClure & Forrester (1981) |
| HD164064 | WBVR | V | $5.84 \pm 0.05$ | Kornilov et al. (1991) |
| HD164064 | Johnson | V | $5.86 \pm 0.05$ | Cousins (1964a) |
| HD164064 | Johnson | V | $5.86 \pm 0.05$ | Mermilliod (1986) |
| HD164064 | WBVR | R | $4.63 \pm 0.05$ | Kornilov et al. (1991) |
| HD164064 | 2200 | 361 | $103.20 \pm 10.80$ | Smith et al. (2004) |
| HD164064 | Johnson | K | $2.06 \pm 0.04$ | Neugebauer & Leighton (1969) |
| HD164064 | 3500 | 898 | $43.80 \pm 11.20$ | Smith et al. (2004) |

**Table 21** *(continued)*

| Star ID | System/Wvlen | Band/Bandpass | Value | Reference |
|---------|--------------|---------------|-------|-----------|
| HD164064 | 4900 | 712 | $21.90 \pm 6.70$ | Smith et al. (2004) |
| HD164064 | 12000 | 6384 | $-19.00 \pm 23.00$ | Smith et al. (2004) |
| HD165760 | 13c | m33 | $6.36 \pm 0.05$ | Johnson & Mitchell (1995) |
| HD165760 | 13c | m35 | $6.19 \pm 0.05$ | Johnson & Mitchell (1995) |
| HD165760 | DDO | m35 | $7.66 \pm 0.05$ | Dean (1981) |
| HD165760 | DDO | m35 | $7.69 \pm 0.05$ | McClure & Forrester (1981) |
| HD165760 | Stromgren | u | $7.60 \pm 0.08$ | Olsen (1993) |
| HD165760 | Stromgren | u | $7.60 \pm 0.08$ | Hauck & Mermilliod (1998) |
| HD165760 | WBVR | W | $6.20 \pm 0.05$ | Kornilov et al. (1991) |
| HD165760 | Johnson | U | $6.33 \pm 0.05$ | Johnson et al. (1966) |
| HD165760 | Johnson | U | $6.34 \pm 0.05$ | Argue (1963) |
| HD165760 | Johnson | U | $6.34 \pm 0.05$ | Mermilliod (1986) |
| HD165760 | Johnson | U | $6.35 \pm 0.05$ | Cousins (1963c) |
| HD165760 | Johnson | U | $6.35 \pm 0.05$ | Gutierrez-Moreno & et al. (1966) |
| HD165760 | Johnson | U | $6.35 \pm 0.05$ | Jennens & Helfer (1975) |
| HD165760 | 13c | m37 | $6.27 \pm 0.05$ | Johnson & Mitchell (1995) |
| HD165760 | DDO | m38 | $6.62 \pm 0.05$ | Dean (1981) |
| HD165760 | DDO | m38 | $6.64 \pm 0.05$ | McClure & Forrester (1981) |
| HD165760 | 13c | m40 | $6.13 \pm 0.05$ | Johnson & Mitchell (1995) |
| HD165760 | DDO | m41 | $7.17 \pm 0.05$ | Dean (1981) |
| HD165760 | DDO | m41 | $7.19 \pm 0.05$ | McClure & Forrester (1981) |
| HD165760 | Oja | m41 | $6.72 \pm 0.05$ | Häggkvist & Oja (1970) |
| HD165760 | Stromgren | v | $6.21 \pm 0.08$ | Olsen (1993) |
| HD165760 | Stromgren | v | $6.21 \pm 0.08$ | Hauck & Mermilliod (1998) |
| HD165760 | DDO | m42 | $6.97 \pm 0.05$ | McClure & Forrester (1981) |
| HD165760 | DDO | m42 | $6.97 \pm 0.05$ | Dean (1981) |
| HD165760 | Oja | m42 | $6.51 \pm 0.05$ | Häggkvist & Oja (1970) |
| HD165760 | WBVR | B | $5.63 \pm 0.05$ | Kornilov et al. (1991) |
| HD165760 | Johnson | B | $5.59 \pm 0.05$ | Cousins (1963c) |
| HD165760 | Johnson | B | $5.59 \pm 0.05$ | Häggkvist & Oja (1966) |
| HD165760 | Johnson | B | $5.60 \pm 0.05$ | Johnson et al. (1966) |
| HD165760 | Johnson | B | $5.60 \pm 0.05$ | Gutierrez-Moreno & et al. (1966) |
| HD165760 | Johnson | B | $5.60 \pm 0.05$ | Jennens & Helfer (1975) |
| HD165760 | Johnson | B | $5.60 \pm 0.05$ | Mermilliod (1986) |
| HD165760 | Johnson | B | $5.61 \pm 0.05$ | Argue (1963) |
| HD165760 | Johnson | B | $5.61 \pm 0.05$ | Moffett & Barnes (1979) |
| HD165760 | 13c | m45 | $5.33 \pm 0.05$ | Johnson & Mitchell (1995) |
| HD165760 | DDO | m45 | $6.16 \pm 0.05$ | McClure & Forrester (1981) |
| HD165760 | DDO | m45 | $6.16 \pm 0.05$ | Dean (1981) |
| HD165760 | Oja | m45 | $5.33 \pm 0.05$ | Häggkvist & Oja (1970) |
| HD165760 | Stromgren | b | $5.24 \pm 0.08$ | Olsen (1993) |

**Table 21** *(continued)*

| Star ID | System/Wvlen | Band/Bandpass | Value | Reference |
|---------|--------------|---------------|-------|-----------|
| HD165760 | Stromgren | b | $5.24 \pm 0.08$ | Hauck & Mermilliod (1998) |
| HD165760 | DDO | m48 | $4.99 \pm 0.05$ | McClure & Forrester (1981) |
| HD165760 | DDO | m48 | $4.99 \pm 0.05$ | Dean (1981) |
| HD165760 | 13c | m52 | $4.86 \pm 0.05$ | Johnson & Mitchell (1995) |
| HD165760 | WBVR | V | $4.65 \pm 0.05$ | Kornilov et al. (1991) |
| HD165760 | Stromgren | y | $4.65 \pm 0.08$ | Olsen (1993) |
| HD165760 | Stromgren | y | $4.65 \pm 0.08$ | Hauck & Mermilliod (1998) |
| HD165760 | Johnson | V | $4.63 \pm 0.05$ | Johnson et al. (1966) |
| HD165760 | Johnson | V | $4.64 \pm 0.05$ | Argue (1963) |
| HD165760 | Johnson | V | $4.64 \pm 0.05$ | Cousins (1963c) |
| HD165760 | Johnson | V | $4.64 \pm 0.05$ | Häggkvist & Oja (1966) |
| HD165760 | Johnson | V | $4.64 \pm 0.05$ | Gutierrez-Moreno & et al. (1966) |
| HD165760 | Johnson | V | $4.64 \pm 0.05$ | Jennens & Helfer (1975) |
| HD165760 | Johnson | V | $4.65 \pm 0.05$ | Moffett & Barnes (1979) |
| HD165760 | Johnson | V | $4.65 \pm 0.05$ | Mermilliod (1986) |
| HD165760 | 13c | m58 | $4.43 \pm 0.05$ | Johnson & Mitchell (1995) |
| HD165760 | 13c | m63 | $4.16 \pm 0.05$ | Johnson & Mitchell (1995) |
| HD165760 | WBVR | R | $3.95 \pm 0.05$ | Kornilov et al. (1991) |
| HD165760 | 13c | m72 | $3.96 \pm 0.05$ | Johnson & Mitchell (1995) |
| HD165760 | 13c | m80 | $3.75 \pm 0.05$ | Johnson & Mitchell (1995) |
| HD165760 | 13c | m86 | $3.65 \pm 0.05$ | Johnson & Mitchell (1995) |
| HD165760 | 13c | m99 | $3.52 \pm 0.05$ | Johnson & Mitchell (1995) |
| HD165760 | 13c | m110 | $3.35 \pm 0.05$ | Johnson & Mitchell (1995) |
| HD165760 | 1250 | 310 | $113.50 \pm 18.00$ | Smith et al. (2004) |
| HD165760 | Johnson | J | $2.98 \pm 0.05$ | Alonso et al. (1998) |
| HD165760 | Johnson | H | $2.57 \pm 0.05$ | Alonso et al. (1998) |
| HD165760 | 2200 | 361 | $94.00 \pm 26.50$ | Smith et al. (2004) |
| HD165760 | Johnson | K | $2.43 \pm 0.05$ | Neugebauer & Leighton (1969) |
| HD165760 | 3500 | 898 | $44.40 \pm 15.80$ | Smith et al. (2004) |
| HD165760 | 4900 | 712 | $24.60 \pm 8.10$ | Smith et al. (2004) |
| HD165760 | 12000 | 6384 | $8.40 \pm 19.40$ | Smith et al. (2004) |
| HD166013 | KronComet | COp | $9.51 \pm 0.01$ | This work |
| HD166013 | WBVR | B | $8.97 \pm 0.05$ | Kornilov et al. (1991) |
| HD166013 | Johnson | B | $8.85 \pm 0.01$ | This work |
| HD166013 | KronComet | Bc | $8.79 \pm 0.01$ | This work |
| HD166013 | KronComet | C2 | $7.63 \pm 0.01$ | This work |
| HD166013 | KronComet | Gc | $7.51 \pm 0.01$ | This work |
| HD166013 | WBVR | V | $7.35 \pm 0.05$ | Kornilov et al. (1991) |
| HD166013 | Johnson | V | $7.47 \pm 0.01$ | This work |
| HD166013 | WBVR | R | $5.65 \pm 0.05$ | Kornilov et al. (1991) |
| HD166013 | KronComet | Rc | $6.00 \pm 0.01$ | This work |

**Table 21** *(continued)*

| Star ID | System/Wvlen | Band/Bandpass | Value | Reference |
|---|---|---|---|---|
| HD166013 | 1250 | 310 | $67.20 \pm 4.40$ | Smith et al. (2004) |
| HD166013 | 2200 | 361 | $70.10 \pm 5.70$ | Smith et al. (2004) |
| HD166013 | Johnson | K | $2.14 \pm 0.05$ | Neugebauer & Leighton (1969) |
| HD166013 | 3500 | 898 | $39.90 \pm 9.40$ | Smith et al. (2004) |
| HD166013 | 4900 | 712 | $15.50 \pm 8.60$ | Smith et al. (2004) |
| HD166013 | 12000 | 6384 | $11.90 \pm 21.60$ | Smith et al. (2004) |
| HD168775 | 13c | m33 | $6.80 \pm 0.05$ | Johnson & Mitchell (1995) |
| HD168775 | Vilnius | U | $8.52 \pm 0.05$ | Kazlauskas et al. (2005) |
| HD168775 | 13c | m35 | $6.59 \pm 0.05$ | Johnson & Mitchell (1995) |
| HD168775 | DDO | m35 | $8.01 \pm 0.05$ | McClure & Forrester (1981) |
| HD168775 | WBVR | W | $6.57 \pm 0.05$ | Kornilov et al. (1991) |
| HD168775 | Johnson | U | $6.66 \pm 0.05$ | Mermilliod (1986) |
| HD168775 | Johnson | U | $6.69 \pm 0.05$ | Johnson et al. (1966) |
| HD168775 | Johnson | U | $6.70 \pm 0.05$ | Johnson et al. (1966) |
| HD168775 | Johnson | U | $6.70 \pm 0.05$ | Ducati (2002) |
| HD168775 | Johnson | U | $6.71 \pm 0.05$ | Argue (1963) |
| HD168775 | 13c | m37 | $6.63 \pm 0.05$ | Johnson & Mitchell (1995) |
| HD168775 | Vilnius | P | $7.89 \pm 0.05$ | Kazlauskas et al. (2005) |
| HD168775 | DDO | m38 | $6.86 \pm 0.05$ | McClure & Forrester (1981) |
| HD168775 | 13c | m40 | $6.27 \pm 0.05$ | Johnson & Mitchell (1995) |
| HD168775 | Vilnius | X | $6.71 \pm 0.05$ | Kazlauskas et al. (2005) |
| HD168775 | DDO | m41 | $7.30 \pm 0.05$ | McClure & Forrester (1981) |
| HD168775 | Oja | m41 | $6.87 \pm 0.05$ | Häggkvist & Oja (1970) |
| HD168775 | DDO | m42 | $6.96 \pm 0.05$ | McClure & Forrester (1981) |
| HD168775 | Oja | m42 | $6.52 \pm 0.05$ | Häggkvist & Oja (1970) |
| HD168775 | WBVR | B | $5.53 \pm 0.05$ | Kornilov et al. (1991) |
| HD168775 | Johnson | B | $5.47 \pm 0.05$ | Häggkvist & Oja (1966) |
| HD168775 | Johnson | B | $5.49 \pm 0.05$ | Mermilliod (1986) |
| HD168775 | Johnson | B | $5.50 \pm 0.05$ | Argue (1963) |
| HD168775 | Johnson | B | $5.51 \pm 0.05$ | Johnson et al. (1966) |
| HD168775 | Johnson | B | $5.51 \pm 0.05$ | Ducati (2002) |
| HD168775 | 13c | m45 | $5.15 \pm 0.05$ | Johnson & Mitchell (1995) |
| HD168775 | DDO | m45 | $5.99 \pm 0.05$ | McClure & Forrester (1981) |
| HD168775 | Oja | m45 | $5.17 \pm 0.05$ | Häggkvist & Oja (1970) |
| HD168775 | Vilnius | Y | $5.18 \pm 0.05$ | Kazlauskas et al. (2005) |
| HD168775 | DDO | m48 | $4.74 \pm 0.05$ | McClure & Forrester (1981) |
| HD168775 | Vilnius | Z | $4.68 \pm 0.05$ | Kazlauskas et al. (2005) |
| HD168775 | 13c | m52 | $4.63 \pm 0.05$ | Johnson & Mitchell (1995) |
| HD168775 | WBVR | V | $4.33 \pm 0.05$ | Kornilov et al. (1991) |
| HD168775 | Vilnius | V | $4.32 \pm 0.05$ | Kazlauskas et al. (2005) |
| HD168775 | Johnson | V | $4.31 \pm 0.05$ | Häggkvist & Oja (1966) |

**Table 21** *(continued)*

| Star ID | System/Wvlen | Band/Bandpass | Value | Reference |
|---------|--------------|---------------|-------|-----------|
| HD168775 | Johnson | V | $4.32 \pm 0.05$ | Argue (1963) |
| HD168775 | Johnson | V | $4.34 \pm 0.05$ | Johnson et al. (1966) |
| HD168775 | Johnson | V | $4.34 \pm 0.05$ | Ducati (2002) |
| HD168775 | Johnson | V | $4.35 \pm 0.05$ | Mermilliod (1986) |
| HD168775 | 13c | m58 | $4.08 \pm 0.05$ | Johnson & Mitchell (1995) |
| HD168775 | 13c | m63 | $3.75 \pm 0.05$ | Johnson & Mitchell (1995) |
| HD168775 | Vilnius | S | $3.53 \pm 0.05$ | Kazlauskas et al. (2005) |
| HD168775 | WBVR | R | $3.51 \pm 0.05$ | Kornilov et al. (1991) |
| HD168775 | 13c | m72 | $3.48 \pm 0.05$ | Johnson & Mitchell (1995) |
| HD168775 | 13c | m80 | $3.23 \pm 0.05$ | Johnson & Mitchell (1995) |
| HD168775 | 13c | m86 | $3.11 \pm 0.05$ | Johnson & Mitchell (1995) |
| HD168775 | 13c | m99 | $2.93 \pm 0.05$ | Johnson & Mitchell (1995) |
| HD168775 | 13c | m110 | $2.77 \pm 0.05$ | Johnson & Mitchell (1995) |
| HD168775 | 1250 | 310 | $167.30 \pm 12.70$ | Smith et al. (2004) |
| HD168775 | Johnson | J | $2.49 \pm 0.05$ | Johnson et al. (1966) |
| HD168775 | Johnson | J | $2.49 \pm 0.05$ | Ducati (2002) |
| HD168775 | 2200 | 361 | $116.90 \pm 9.00$ | Smith et al. (2004) |
| HD168775 | Johnson | K | $1.74 \pm 0.05$ | Johnson et al. (1966) |
| HD168775 | Johnson | K | $1.74 \pm 0.05$ | Ducati (2002) |
| HD168775 | Johnson | K | $1.80 \pm 0.05$ | Neugebauer & Leighton (1969) |
| HD168775 | 3500 | 898 | $56.40 \pm 6.20$ | Smith et al. (2004) |
| HD168775 | 4900 | 712 | $26.40 \pm 4.50$ | Smith et al. (2004) |
| HD168775 | 12000 | 6384 | $5.50 \pm 16.10$ | Smith et al. (2004) |
| HD169191 | Vilnius | U | $9.65 \pm 0.05$ | Zdanavicius et al. (1969) |
| HD169191 | Vilnius | U | $9.68 \pm 0.05$ | Forbes et al. (1993) |
| HD169191 | DDO | m35 | $9.15 \pm 0.05$ | McClure & Forrester (1981) |
| HD169191 | WBVR | W | $7.69 \pm 0.05$ | Kornilov et al. (1991) |
| HD169191 | Johnson | U | $7.81 \pm 0.05$ | Mermilliod (1986) |
| HD169191 | Johnson | U | $7.84 \pm 0.05$ | Argue (1963) |
| HD169191 | Vilnius | P | $9.03 \pm 0.05$ | Forbes et al. (1993) |
| HD169191 | Vilnius | P | $9.04 \pm 0.05$ | Zdanavicius et al. (1969) |
| HD169191 | DDO | m38 | $8.00 \pm 0.05$ | McClure & Forrester (1981) |
| HD169191 | Vilnius | X | $7.78 \pm 0.05$ | Zdanavicius et al. (1969) |
| HD169191 | Vilnius | X | $7.79 \pm 0.05$ | Forbes et al. (1993) |
| HD169191 | DDO | m41 | $8.30 \pm 0.05$ | McClure & Forrester (1981) |
| HD169191 | Oja | m41 | $7.81 \pm 0.05$ | Häggkvist & Oja (1970) |
| HD169191 | DDO | m42 | $8.05 \pm 0.05$ | McClure & Forrester (1981) |
| HD169191 | Oja | m42 | $7.57 \pm 0.05$ | Häggkvist & Oja (1970) |
| HD169191 | WBVR | B | $6.55 \pm 0.05$ | Kornilov et al. (1991) |
| HD169191 | Johnson | B | $6.51 \pm 0.05$ | Mermilliod (1986) |
| HD169191 | Johnson | B | $6.52 \pm 0.05$ | Argue (1963) |

**Table 21** *(continued)*

| Star ID | System/Wvlen | Band/Bandpass | Value | Reference |
|---|---|---|---|---|
| HD169191 | DDO | m45 | $7.00 \pm 0.05$ | McClure & Forrester (1981) |
| HD169191 | Oja | m45 | $6.16 \pm 0.05$ | Häggkvist & Oja (1970) |
| HD169191 | Vilnius | Y | $6.16 \pm 0.05$ | Zdanavicius et al. (1969) |
| HD169191 | Vilnius | Y | $6.18 \pm 0.05$ | Forbes et al. (1993) |
| HD169191 | DDO | m48 | $5.73 \pm 0.05$ | McClure & Forrester (1981) |
| HD169191 | Vilnius | Z | $5.63 \pm 0.05$ | Zdanavicius et al. (1969) |
| HD169191 | Vilnius | Z | $5.64 \pm 0.05$ | Forbes et al. (1993) |
| HD169191 | WBVR | V | $5.26 \pm 0.05$ | Kornilov et al. (1991) |
| HD169191 | Vilnius | V | $5.24 \pm 0.05$ | Zdanavicius et al. (1969) |
| HD169191 | Vilnius | V | $5.25 \pm 0.05$ | Forbes et al. (1993) |
| HD169191 | Johnson | V | $5.24 \pm 0.05$ | Mermilliod (1986) |
| HD169191 | Johnson | V | $5.25 \pm 0.05$ | Argue (1963) |
| HD169191 | Vilnius | S | $4.35 \pm 0.05$ | Zdanavicius et al. (1969) |
| HD169191 | Vilnius | S | $4.35 \pm 0.05$ | Forbes et al. (1993) |
| HD169191 | WBVR | R | $4.34 \pm 0.05$ | Kornilov et al. (1991) |
| HD169191 | 1250 | 310 | $111.70 \pm 10.20$ | Smith et al. (2004) |
| HD169191 | 2200 | 361 | $89.90 \pm 15.90$ | Smith et al. (2004) |
| HD169191 | Johnson | K | $2.29 \pm 0.07$ | Neugebauer & Leighton (1969) |
| HD169191 | 3500 | 898 | $43.40 \pm 10.50$ | Smith et al. (2004) |
| HD169191 | 4900 | 712 | $20.20 \pm 5.90$ | Smith et al. (2004) |
| HD169191 | 12000 | 6384 | $-2.60 \pm 17.90$ | Smith et al. (2004) |
| HD169305 | Oja | m41 | $8.20 \pm 0.05$ | Häggkvist & Oja (1970) |
| HD169305 | Oja | m42 | $8.00 \pm 0.05$ | Häggkvist & Oja (1970) |
| HD169305 | KronComet | COp | $7.32 \pm 0.01$ | This work |
| HD169305 | WBVR | B | $6.75 \pm 0.05$ | Kornilov et al. (1991) |
| HD169305 | Johnson | B | $6.46 \pm 0.02$ | This work |
| HD169305 | Johnson | B | $6.58 \pm 0.05$ | Mermilliod (1986) |
| HD169305 | Johnson | B | $6.71 \pm 0.05$ | Häggkvist & Oja (1966) |
| HD169305 | Johnson | B | $6.71 \pm 0.05$ | Ducati (2002) |
| HD169305 | KronComet | Bc | $6.46 \pm 0.02$ | This work |
| HD169305 | Oja | m45 | $6.31 \pm 0.05$ | Häggkvist & Oja (1970) |
| HD169305 | Vilnius | Y | $6.26 \pm 0.05$ | Jasevicius et al. (1990) |
| HD169305 | KronComet | C2 | $5.39 \pm 0.01$ | This work |
| HD169305 | Vilnius | Z | $5.59 \pm 0.05$ | Jasevicius et al. (1990) |
| HD169305 | KronComet | Gc | $5.13 \pm 0.01$ | This work |
| HD169305 | WBVR | V | $5.05 \pm 0.05$ | Kornilov et al. (1991) |
| HD169305 | Vilnius | V | $5.05 \pm 0.05$ | Jasevicius et al. (1990) |
| HD169305 | Johnson | V | $4.98 \pm 0.05$ | Mermilliod (1986) |
| HD169305 | Johnson | V | $5.03 \pm 0.01$ | This work |
| HD169305 | Johnson | V | $5.05 \pm 0.05$ | Häggkvist & Oja (1966) |
| HD169305 | Johnson | V | $5.05 \pm 0.05$ | Ducati (2002) |

**Table 21** *(continued)*

| Star ID | System/Wvlen | Band/Bandpass | Value | Reference |
|---------|--------------|---------------|-------|-----------|
| HD169305 | Vilnius | S | $3.88 \pm 0.05$ | Jasevicius et al. (1990) |
| HD169305 | WBVR | R | $3.62 \pm 0.05$ | Kornilov et al. (1991) |
| HD169305 | KronComet | Rc | $3.56 \pm 0.01$ | This work |
| HD169305 | 1250 | 310 | $344.40 \pm 15.50$ | Smith et al. (2004) |
| HD169305 | Johnson | J | $1.69 \pm 0.05$ | McWilliam & Lambert (1984) |
| HD169305 | Johnson | J | $1.69 \pm 0.05$ | Ducati (2002) |
| HD169305 | 2200 | 361 | $347.20 \pm 15.50$ | Smith et al. (2004) |
| HD169305 | Johnson | K | $0.57 \pm 0.05$ | Ducati (2002) |
| HD169305 | Johnson | K | $0.62 \pm 0.04$ | Neugebauer & Leighton (1969) |
| HD169305 | 3500 | 898 | $172.70 \pm 9.20$ | Smith et al. (2004) |
| HD169305 | 4900 | 712 | $75.10 \pm 6.90$ | Smith et al. (2004) |
| HD169305 | 12000 | 6384 | $16.10 \pm 15.80$ | Smith et al. (2004) |
| HD171779 | Geneva | U | $7.73 \pm 0.08$ | Golay (1972) |
| HD171779 | WBVR | W | $7.27 \pm 0.05$ | Kornilov et al. (1991) |
| HD171779 | Johnson | U | $7.36 \pm 0.05$ | Mermilliod (1986) |
| HD171779 | Johnson | U | $7.39 \pm 0.05$ | Eggen (1963) |
| HD171779 | Johnson | U | $7.41 \pm 0.05$ | Nicolet (1978) |
| HD171779 | Johnson | U | $7.44 \pm 0.05$ | Argue (1963) |
| HD171779 | Geneva | B1 | $6.97 \pm 0.08$ | Golay (1972) |
| HD171779 | Oja | m41 | $7.74 \pm 0.05$ | Häggkvist & Oja (1970) |
| HD171779 | Oja | m42 | $7.47 \pm 0.05$ | Häggkvist & Oja (1970) |
| HD171779 | Geneva | B | $5.66 \pm 0.08$ | Golay (1972) |
| HD171779 | WBVR | B | $6.50 \pm 0.05$ | Kornilov et al. (1991) |
| HD171779 | Johnson | B | $6.38 \pm 0.05$ | Häggkvist (1966) |
| HD171779 | Johnson | B | $6.45 \pm 0.05$ | Nicolet (1978) |
| HD171779 | Johnson | B | $6.47 \pm 0.05$ | Mermilliod (1986) |
| HD171779 | Johnson | B | $6.48 \pm 0.05$ | Eggen (1963) |
| HD171779 | Johnson | B | $6.48 \pm 0.05$ | Argue (1963) |
| HD171779 | Geneva | B2 | $6.80 \pm 0.08$ | Golay (1972) |
| HD171779 | Oja | m45 | $6.18 \pm 0.05$ | Häggkvist & Oja (1970) |
| HD171779 | Geneva | V1 | $6.10 \pm 0.08$ | Golay (1972) |
| HD171779 | WBVR | V | $5.38 \pm 0.05$ | Kornilov et al. (1991) |
| HD171779 | Geneva | V | $5.32 \pm 0.08$ | Golay (1972) |
| HD171779 | Johnson | V | $5.33 \pm 0.05$ | Häggkvist (1966) |
| HD171779 | Johnson | V | $5.36 \pm 0.05$ | Eggen (1963) |
| HD171779 | Johnson | V | $5.36 \pm 0.05$ | Nicolet (1978) |
| HD171779 | Johnson | V | $5.37 \pm 0.05$ | Argue (1963) |
| HD171779 | Johnson | V | $5.40 \pm 0.05$ | Mermilliod (1986) |
| HD171779 | Geneva | G | $6.29 \pm 0.08$ | Golay (1972) |
| HD171779 | WBVR | R | $4.61 \pm 0.05$ | Kornilov et al. (1991) |
| HD171779 | 1250 | 310 | $57.40 \pm 6.50$ | Smith et al. (2004) |

**Table 21** *(continued)*

| Star ID | System/Wvlen | Band/Bandpass | Value | Reference |
|---------|--------------|---------------|-------|-----------|
| HD171779 | 2200 | 361 | $41.00 \pm 4.80$ | Smith et al. (2004) |
| HD171779 | Johnson | K | $2.78 \pm 0.10$ | Neugebauer & Leighton (1969) |
| HD171779 | 3500 | 898 | $21.00 \pm 14.10$ | Smith et al. (2004) |
| HD171779 | 4900 | 712 | $9.60 \pm 7.50$ | Smith et al. (2004) |
| HD171779 | 12000 | 6384 | $2.90 \pm 15.50$ | Smith et al. (2004) |
| HD173213 | 1250 | 310 | $72.70 \pm 3.50$ | Smith et al. (2004) |
| HD173213 | 2200 | 361 | $90.00 \pm 9.90$ | Smith et al. (2004) |
| HD173213 | Johnson | K | $2.00 \pm 0.05$ | Neugebauer & Leighton (1969) |
| HD173213 | 3500 | 898 | $48.70 \pm 6.70$ | Smith et al. (2004) |
| HD173213 | 4900 | 712 | $19.30 \pm 5.30$ | Smith et al. (2004) |
| HD173213 | 12000 | 6384 | $12.60 \pm 17.60$ | Smith et al. (2004) |
| HD173215 | WBVR | B | $8.35 \pm 0.05$ | Kornilov et al. (1991) |
| HD173215 | WBVR | V | $6.60 \pm 0.05$ | Kornilov et al. (1991) |
| HD173215 | WBVR | R | $5.17 \pm 0.05$ | Kornilov et al. (1991) |
| HD173215 | Johnson | K | $2.24 \pm 0.06$ | Neugebauer & Leighton (1969) |
| HD173780 | 13c | m33 | $7.36 \pm 0.05$ | Johnson & Mitchell (1995) |
| HD173780 | Geneva | U | $7.79 \pm 0.08$ | Golay (1972) |
| HD173780 | Vilnius | U | $9.10 \pm 0.05$ | Kazlauskas et al. (2005) |
| HD173780 | 13c | m35 | $7.11 \pm 0.05$ | Johnson & Mitchell (1995) |
| HD173780 | DDO | m35 | $8.56 \pm 0.05$ | McClure & Forrester (1981) |
| HD173780 | WBVR | W | $7.14 \pm 0.05$ | Kornilov et al. (1991) |
| HD173780 | Johnson | U | $7.23 \pm 0.05$ | Mermilliod (1986) |
| HD173780 | Johnson | U | $7.25 \pm 0.05$ | Argue (1963) |
| HD173780 | Johnson | U | $7.26 \pm 0.05$ | Johnson et al. (1966) |
| HD173780 | Johnson | U | $7.26 \pm 0.05$ | Ducati (2002) |
| HD173780 | 13c | m37 | $7.19 \pm 0.05$ | Johnson & Mitchell (1995) |
| HD173780 | Vilnius | P | $8.46 \pm 0.05$ | Kazlauskas et al. (2005) |
| HD173780 | DDO | m38 | $7.42 \pm 0.05$ | McClure & Forrester (1981) |
| HD173780 | 13c | m40 | $6.80 \pm 0.05$ | Johnson & Mitchell (1995) |
| HD173780 | Geneva | B1 | $6.80 \pm 0.08$ | Golay (1972) |
| HD173780 | Vilnius | X | $7.26 \pm 0.05$ | Kazlauskas et al. (2005) |
| HD173780 | DDO | m41 | $7.77 \pm 0.05$ | McClure & Forrester (1981) |
| HD173780 | Oja | m41 | $7.36 \pm 0.05$ | Häggkvist & Oja (1970) |
| HD173780 | DDO | m42 | $7.52 \pm 0.05$ | McClure & Forrester (1981) |
| HD173780 | Oja | m42 | $7.09 \pm 0.05$ | Häggkvist & Oja (1970) |
| HD173780 | Geneva | B | $5.36 \pm 0.08$ | Golay (1972) |
| HD173780 | WBVR | B | $6.06 \pm 0.05$ | Kornilov et al. (1991) |
| HD173780 | Johnson | B | $6.01 \pm 0.05$ | Argue (1963) |
| HD173780 | Johnson | B | $6.01 \pm 0.05$ | Häggkvist & Oja (1966) |
| HD173780 | Johnson | B | $6.01 \pm 0.05$ | Mermilliod (1986) |
| HD173780 | Johnson | B | $6.04 \pm 0.05$ | Johnson et al. (1966) |

**Table 21** *(continued)*

| Star ID | System/Wvlen | Band/Bandpass | Value | Reference |
|---------|--------------|---------------|-------|-----------|
| HD173780 | Johnson | B | $6.04 \pm 0.05$ | Neckel (1974) |
| HD173780 | Johnson | B | $6.04 \pm 0.05$ | Moffett & Barnes (1979) |
| HD173780 | Johnson | B | $6.04 \pm 0.05$ | Ducati (2002) |
| HD173780 | Geneva | B2 | $6.44 \pm 0.08$ | Golay (1972) |
| HD173780 | 13c | m45 | $5.67 \pm 0.05$ | Johnson & Mitchell (1995) |
| HD173780 | DDO | m45 | $6.49 \pm 0.05$ | McClure & Forrester (1981) |
| HD173780 | Oja | m45 | $5.72 \pm 0.05$ | Häggkvist & Oja (1970) |
| HD173780 | Vilnius | Y | $5.71 \pm 0.05$ | Kazlauskas et al. (2005) |
| HD173780 | DDO | m48 | $5.25 \pm 0.05$ | McClure & Forrester (1981) |
| HD173780 | Vilnius | Z | $5.22 \pm 0.05$ | Kazlauskas et al. (2005) |
| HD173780 | 13c | m52 | $5.15 \pm 0.05$ | Johnson & Mitchell (1995) |
| HD173780 | Geneva | V1 | $5.60 \pm 0.08$ | Golay (1972) |
| HD173780 | WBVR | V | $4.84 \pm 0.05$ | Kornilov et al. (1991) |
| HD173780 | Vilnius | V | $4.83 \pm 0.05$ | Kazlauskas et al. (2005) |
| HD173780 | Geneva | V | $4.81 \pm 0.08$ | Golay (1972) |
| HD173780 | Johnson | V | $4.81 \pm 0.05$ | Argue (1963) |
| HD173780 | Johnson | V | $4.81 \pm 0.05$ | Häggkvist & Oja (1966) |
| HD173780 | Johnson | V | $4.82 \pm 0.05$ | Mermilliod (1986) |
| HD173780 | Johnson | V | $4.83 \pm 0.05$ | Neckel (1974) |
| HD173780 | Johnson | V | $4.83 \pm 0.05$ | Moffett & Barnes (1979) |
| HD173780 | Johnson | V | $4.84 \pm 0.05$ | Johnson et al. (1966) |
| HD173780 | Johnson | V | $4.84 \pm 0.05$ | Ducati (2002) |
| HD173780 | 13c | m58 | $4.56 \pm 0.05$ | Johnson & Mitchell (1995) |
| HD173780 | Geneva | G | $5.73 \pm 0.08$ | Golay (1972) |
| HD173780 | 13c | m63 | $4.22 \pm 0.05$ | Johnson & Mitchell (1995) |
| HD173780 | Vilnius | S | $4.00 \pm 0.05$ | Kazlauskas et al. (2005) |
| HD173780 | WBVR | R | $3.97 \pm 0.05$ | Kornilov et al. (1991) |
| HD173780 | 13c | m72 | $3.89 \pm 0.05$ | Johnson & Mitchell (1995) |
| HD173780 | 13c | m80 | $3.63 \pm 0.05$ | Johnson & Mitchell (1995) |
| HD173780 | 13c | m86 | $3.52 \pm 0.05$ | Johnson & Mitchell (1995) |
| HD173780 | 13c | m99 | $3.31 \pm 0.05$ | Johnson & Mitchell (1995) |
| HD173780 | 13c | m110 | $3.15 \pm 0.05$ | Johnson & Mitchell (1995) |
| HD173780 | 1250 | 310 | $106.70 \pm 9.30$ | Smith et al. (2004) |
| HD173780 | Johnson | H | $2.23 \pm 0.05$ | Voelcker (1975) |
| HD173780 | Johnson | H | $2.23 \pm 0.05$ | Ducati (2002) |
| HD173780 | 2200 | 361 | $74.20 \pm 8.50$ | Smith et al. (2004) |
| HD173780 | Johnson | K | $2.01 \pm 0.05$ | Neugebauer & Leighton (1969) |
| HD173780 | Johnson | K | $2.01 \pm 0.05$ | Ducati (2002) |
| HD173780 | Johnson | L | $1.86 \pm 0.05$ | Ducati (2002) |
| HD173780 | 3500 | 898 | $34.50 \pm 6.00$ | Smith et al. (2004) |
| HD173780 | 4900 | 712 | $15.70 \pm 5.20$ | Smith et al. (2004) |

**Table 21** *continued on next page*

**Table 21** *(continued)*

| Star ID | System/Wvlen | Band/Bandpass | Value | Reference |
|---------|--------------|---------------|-------|-----------|
| HD173780 | 12000 | 6384 | $1.00 \pm 16.60$ | Smith et al. (2004) |
| HD173954 | WBVR | W | $9.41 \pm 0.05$ | Kornilov et al. (1991) |
| HD173954 | Johnson | U | $9.49 \pm 0.05$ | Cousins (1965) |
| HD173954 | WBVR | B | $7.74 \pm 0.05$ | Kornilov et al. (1991) |
| HD173954 | Johnson | B | $7.71 \pm 0.05$ | Cousins (1965) |
| HD173954 | WBVR | V | $6.20 \pm 0.05$ | Kornilov et al. (1991) |
| HD173954 | Johnson | V | $6.20 \pm 0.05$ | Cousins (1965) |
| HD173954 | WBVR | R | $5.07 \pm 0.05$ | Kornilov et al. (1991) |
| HD173954 | Johnson | K | $2.60 \pm 0.06$ | Neugebauer & Leighton (1969) |
| HD176411 | 13c | m33 | $6.25 \pm 0.05$ | Johnson & Mitchell (1995) |
| HD176411 | 13c | m35 | $6.04 \pm 0.05$ | Johnson & Mitchell (1995) |
| HD176411 | WBVR | W | $5.98 \pm 0.05$ | Kornilov et al. (1991) |
| HD176411 | Johnson | U | $6.14 \pm 0.05$ | Argue (1963) |
| HD176411 | Johnson | U | $6.14 \pm 0.05$ | Johnson et al. (1966) |
| HD176411 | Johnson | U | $6.14 \pm 0.05$ | Ducati (2002) |
| HD176411 | 13c | m37 | $6.09 \pm 0.05$ | Johnson & Mitchell (1995) |
| HD176411 | 13c | m40 | $5.77 \pm 0.05$ | Johnson & Mitchell (1995) |
| HD176411 | Oja | m41 | $6.41 \pm 0.05$ | Häggkvist & Oja (1970) |
| HD176411 | Oja | m42 | $6.08 \pm 0.05$ | Häggkvist & Oja (1970) |
| HD176411 | WBVR | B | $5.12 \pm 0.05$ | Kornilov et al. (1991) |
| HD176411 | Johnson | B | $5.10 \pm 0.05$ | Johnson et al. (1966) |
| HD176411 | Johnson | B | $5.10 \pm 0.05$ | Ducati (2002) |
| HD176411 | Johnson | B | $5.11 \pm 0.05$ | Argue (1963) |
| HD176411 | Johnson | B | $5.12 \pm 0.05$ | Häggkvist & Oja (1966) |
| HD176411 | 13c | m45 | $4.79 \pm 0.05$ | Johnson & Mitchell (1995) |
| HD176411 | Oja | m45 | $4.80 \pm 0.05$ | Häggkvist & Oja (1970) |
| HD176411 | 13c | m52 | $4.31 \pm 0.05$ | Johnson & Mitchell (1995) |
| HD176411 | WBVR | V | $4.02 \pm 0.05$ | Kornilov et al. (1991) |
| HD176411 | Johnson | V | $4.02 \pm 0.05$ | Häggkvist & Oja (1966) |
| HD176411 | Johnson | V | $4.02 \pm 0.05$ | Johnson et al. (1966) |
| HD176411 | Johnson | V | $4.02 \pm 0.05$ | Ducati (2002) |
| HD176411 | Johnson | V | $4.03 \pm 0.05$ | Argue (1963) |
| HD176411 | 13c | m58 | $3.80 \pm 0.05$ | Johnson & Mitchell (1995) |
| HD176411 | 13c | m63 | $3.48 \pm 0.05$ | Johnson & Mitchell (1995) |
| HD176411 | WBVR | R | $3.26 \pm 0.05$ | Kornilov et al. (1991) |
| HD176411 | 13c | m72 | $3.26 \pm 0.05$ | Johnson & Mitchell (1995) |
| HD176411 | 13c | m80 | $3.05 \pm 0.05$ | Johnson & Mitchell (1995) |
| HD176411 | 13c | m86 | $2.93 \pm 0.05$ | Johnson & Mitchell (1995) |
| HD176411 | 13c | m99 | $2.77 \pm 0.05$ | Johnson & Mitchell (1995) |
| HD176411 | 13c | m110 | $2.59 \pm 0.05$ | Johnson & Mitchell (1995) |
| HD176411 | 1250 | 310 | $197.40 \pm 16.20$ | Smith et al. (2004) |

**Table 21** *(continued)*

| Star ID | System/Wvlen | Band/Bandpass | Value | Reference |
|---------|--------------|---------------|-------|-----------|
| HD176411 | Johnson | J | $2.25 \pm 0.05$ | Selby et al. (1988) |
| HD176411 | Johnson | J | $2.25 \pm 0.05$ | Blackwell et al. (1990) |
| HD176411 | Johnson | J | $2.33 \pm 0.05$ | Johnson et al. (1966) |
| HD176411 | Johnson | J | $2.33 \pm 0.05$ | Shenavrin et al. (2011) |
| HD176411 | Johnson | J | $2.63 \pm 0.05$ | Ducati (2002) |
| HD176411 | Johnson | H | $1.82 \pm 0.05$ | Shenavrin et al. (2011) |
| HD176411 | 2200 | 361 | $143.90 \pm 16.30$ | Smith et al. (2004) |
| HD176411 | Johnson | K | $1.50 \pm 0.06$ | Neugebauer & Leighton (1969) |
| HD176411 | Johnson | K | $1.69 \pm 0.05$ | Johnson et al. (1966) |
| HD176411 | Johnson | K | $1.69 \pm 0.05$ | Shenavrin et al. (2011) |
| HD176411 | Johnson | L | $1.55 \pm 0.05$ | Johnson et al. (1966) |
| HD176411 | 3500 | 898 | $72.20 \pm 14.10$ | Smith et al. (2004) |
| HD176411 | 4900 | 712 | $37.60 \pm 6.70$ | Smith et al. (2004) |
| HD176411 | Johnson | M | $3.68 \pm 0.05$ | Ducati (2002) |
| HD176411 | 12000 | 6384 | $26.80 \pm 22.30$ | Smith et al. (2004) |
| HD176670 | Vilnius | U | $9.92 \pm 0.05$ | Jasevicius et al. (1990) |
| HD176670 | DDO | m35 | $9.45 \pm 0.05$ | McClure & Forrester (1981) |
| HD176670 | WBVR | W | $7.98 \pm 0.05$ | Kornilov et al. (1991) |
| HD176670 | Johnson | U | $8.00 \pm 0.05$ | Jennens & Helfer (1975) |
| HD176670 | Johnson | U | $8.03 \pm 0.05$ | Mermilliod (1986) |
| HD176670 | Johnson | U | $8.07 \pm 0.05$ | Argue (1963) |
| HD176670 | Johnson | U | $8.11 \pm 0.05$ | Fernie (1983) |
| HD176670 | Vilnius | P | $9.20 \pm 0.05$ | Jasevicius et al. (1990) |
| HD176670 | DDO | m38 | $8.21 \pm 0.05$ | McClure & Forrester (1981) |
| HD176670 | Vilnius | X | $7.90 \pm 0.05$ | Jasevicius et al. (1990) |
| HD176670 | DDO | m41 | $8.44 \pm 0.05$ | McClure & Forrester (1981) |
| HD176670 | Oja | m41 | $7.98 \pm 0.05$ | Häggkvist & Oja (1970) |
| HD176670 | DDO | m42 | $8.07 \pm 0.05$ | McClure & Forrester (1981) |
| HD176670 | Oja | m42 | $7.61 \pm 0.05$ | Häggkvist & Oja (1970) |
| HD176670 | WBVR | B | $6.46 \pm 0.05$ | Kornilov et al. (1991) |
| HD176670 | Johnson | B | $6.38 \pm 0.05$ | Mermilliod (1986) |
| HD176670 | Johnson | B | $6.39 \pm 0.05$ | Jennens & Helfer (1975) |
| HD176670 | Johnson | B | $6.40 \pm 0.05$ | Argue (1963) |
| HD176670 | Johnson | B | $6.40 \pm 0.05$ | Häggkvist & Oja (1966) |
| HD176670 | Johnson | B | $6.43 \pm 0.05$ | Fernie (1983) |
| HD176670 | DDO | m45 | $6.92 \pm 0.05$ | McClure & Forrester (1981) |
| HD176670 | Oja | m45 | $6.09 \pm 0.05$ | Häggkvist & Oja (1970) |
| HD176670 | Vilnius | Y | $6.00 \pm 0.05$ | Jasevicius et al. (1990) |
| HD176670 | DDO | m48 | $5.51 \pm 0.05$ | McClure & Forrester (1981) |
| HD176670 | Vilnius | Z | $5.39 \pm 0.05$ | Jasevicius et al. (1990) |
| HD176670 | WBVR | V | $4.96 \pm 0.05$ | Kornilov et al. (1991) |

**Table 21** *(continued)*

| Star ID | System/Wvlen | Band/Bandpass | Value | Reference |
|---------|--------------|---------------|-------|-----------|
| HD176670 | Vilnius | V | $4.93 \pm 0.05$ | Jasevicius et al. (1990) |
| HD176670 | Johnson | V | $4.92 \pm 0.05$ | Mermilliod (1986) |
| HD176670 | Johnson | V | $4.93 \pm 0.05$ | Argue (1963) |
| HD176670 | Johnson | V | $4.93 \pm 0.05$ | Häggkvist & Oja (1966) |
| HD176670 | Johnson | V | $4.94 \pm 0.05$ | Jennens & Helfer (1975) |
| HD176670 | Johnson | V | $4.96 \pm 0.05$ | Fernie (1983) |
| HD176670 | Vilnius | S | $4.00 \pm 0.05$ | Jasevicius et al. (1990) |
| HD176670 | WBVR | R | $3.94 \pm 0.05$ | Kornilov et al. (1991) |
| HD176670 | Johnson | K | $1.69 \pm 0.05$ | Neugebauer & Leighton (1969) |
| HD176670 | 3500 | 898 | $62.30 \pm 12.60$ | Smith et al. (2004) |
| HD176670 | 4900 | 712 | $25.60 \pm 5.40$ | Smith et al. (2004) |
| HD176670 | 12000 | 6384 | $-4.90 \pm 16.80$ | Smith et al. (2004) |
| HD176844 | WBVR | B | $8.18 \pm 0.05$ | Kornilov et al. (1991) |
| HD176844 | Johnson | B | $8.18 \pm 0.01$ | Oja (1991) |
| HD176844 | WBVR | V | $6.58 \pm 0.05$ | Kornilov et al. (1991) |
| HD176844 | Johnson | V | $6.58 \pm 0.01$ | Oja (1991) |
| HD176844 | Johnson | V | $6.65 \pm 0.05$ | Ducati (2002) |
| HD176844 | WBVR | R | $4.78 \pm 0.05$ | Kornilov et al. (1991) |
| HD176844 | 1250 | 310 | $178.10 \pm 13.10$ | Smith et al. (2004) |
| HD176844 | Johnson | J | $2.27 \pm 0.05$ | McWilliam & Lambert (1984) |
| HD176844 | Johnson | J | $2.27 \pm 0.05$ | Ducati (2002) |
| HD176844 | 2200 | 361 | $195.10 \pm 9.30$ | Smith et al. (2004) |
| HD176844 | Johnson | K | $1.10 \pm 0.05$ | Ducati (2002) |
| HD176844 | Johnson | K | $1.13 \pm 0.03$ | Neugebauer & Leighton (1969) |
| HD176844 | 3500 | 898 | $101.80 \pm 20.40$ | Smith et al. (2004) |
| HD176844 | 4900 | 712 | $41.00 \pm 5.20$ | Smith et al. (2004) |
| HD176844 | 12000 | 6384 | $12.00 \pm 17.10$ | Smith et al. (2004) |
| HD176981 | Vilnius | X | $9.49 \pm 0.05$ | Zdanavicius et al. (1972) |
| HD176981 | WBVR | B | $7.96 \pm 0.05$ | Kornilov et al. (1991) |
| HD176981 | Johnson | B | $7.90 \pm 0.05$ | Argue (1966) |
| HD176981 | Johnson | B | $7.92 \pm 0.05$ | Johnson et al. (1966) |
| HD176981 | Johnson | B | $7.97 \pm 0.05$ | Moreno (1971) |
| HD176981 | Vilnius | Y | $7.50 \pm 0.05$ | Zdanavicius et al. (1972) |
| HD176981 | Vilnius | Z | $6.81 \pm 0.05$ | Zdanavicius et al. (1972) |
| HD176981 | WBVR | V | $6.29 \pm 0.05$ | Kornilov et al. (1991) |
| HD176981 | Vilnius | V | $6.30 \pm 0.05$ | Zdanavicius et al. (1972) |
| HD176981 | Johnson | V | $6.29 \pm 0.05$ | Argue (1966) |
| HD176981 | Johnson | V | $6.30 \pm 0.05$ | Johnson et al. (1966) |
| HD176981 | Johnson | V | $6.30 \pm 0.05$ | Moreno (1971) |
| HD176981 | Alexander | m608 | $4.06 \pm 0.05$ | Alexander et al. (1983) |
| HD176981 | Vilnius | S | $5.21 \pm 0.05$ | Zdanavicius et al. (1972) |

**Table 21** *(continued)*

| Star ID | System/Wvlen | Band/Bandpass | Value | Reference |
|---------|--------------|---------------|-------|-----------|
| HD176981 | Alexander | m683 | $4.52 \pm 0.05$ | Alexander et al. (1983) |
| HD176981 | WBVR | R | $5.07 \pm 0.05$ | Kornilov et al. (1991) |
| HD176981 | Alexander | m710 | $4.66 \pm 0.05$ | Alexander et al. (1983) |
| HD176981 | Alexander | m746 | $4.96 \pm 0.05$ | Alexander et al. (1983) |
| HD176981 | Johnson | K | $2.75 \pm 0.17$ | Neugebauer & Leighton (1969) |
| HD178690 | WBVR | B | $8.86 \pm 0.05$ | Kornilov et al. (1991) |
| HD178690 | WBVR | V | $6.73 \pm 0.05$ | Kornilov et al. (1991) |
| HD178690 | WBVR | R | $5.22 \pm 0.05$ | Kornilov et al. (1991) |
| HD178690 | 1250 | 310 | $77.60 \pm 8.30$ | Smith et al. (2004) |
| HD178690 | 2200 | 361 | $71.90 \pm 9.10$ | Smith et al. (2004) |
| HD178690 | Johnson | K | $2.01 \pm 0.06$ | Neugebauer & Leighton (1969) |
| HD178690 | 3500 | 898 | $39.70 \pm 9.60$ | Smith et al. (2004) |
| HD178690 | 4900 | 712 | $15.90 \pm 6.50$ | Smith et al. (2004) |
| HD178690 | 12000 | 6384 | $19.30 \pm 20.20$ | Smith et al. (2004) |
| HD180450 | Oja | m41 | $9.05 \pm 0.05$ | Häggkvist & Oja (1970) |
| HD180450 | Oja | m42 | $8.88 \pm 0.05$ | Häggkvist & Oja (1970) |
| HD180450 | Oja | m45 | $7.22 \pm 0.05$ | Häggkvist & Oja (1970) |
| HD180450 | Vilnius | Y | $7.16 \pm 0.05$ | Janulis (1986) |
| HD180450 | Vilnius | Z | $6.49 \pm 0.05$ | Janulis (1986) |
| HD180450 | WBVR | V | $5.90 \pm 0.05$ | Kornilov et al. (1991) |
| HD180450 | Vilnius | V | $5.93 \pm 0.05$ | Janulis (1986) |
| HD180450 | Johnson | V | $5.85 \pm 0.05$ | Haggkvist & Oja (1970) |
| HD180450 | Johnson | V | $5.85 \pm 0.05$ | Ducati (2002) |
| HD180450 | Johnson | V | $5.87 \pm 0.05$ | Mermilliod (1986) |
| HD180450 | Vilnius | S | $4.76 \pm 0.05$ | Janulis (1986) |
| HD180450 | WBVR | R | $4.49 \pm 0.05$ | Kornilov et al. (1991) |
| HD180450 | Johnson | J | $2.57 \pm 0.05$ | McWilliam & Lambert (1984) |
| HD180450 | Johnson | J | $2.57 \pm 0.05$ | Ducati (2002) |
| HD180450 | Johnson | K | $1.42 \pm 0.04$ | Neugebauer & Leighton (1969) |
| HD180450 | Johnson | K | $1.44 \pm 0.05$ | Ducati (2002) |
| HD180450 | 3500 | 898 | $75.30 \pm 11.20$ | Smith et al. (2004) |
| HD180450 | 4900 | 712 | $29.60 \pm 4.80$ | Smith et al. (2004) |
| HD180450 | 12000 | 6384 | $4.70 \pm 17.40$ | Smith et al. (2004) |
| HD184293 | Vilnius | U | $9.94 \pm 0.05$ | Zdanavicius et al. (1972) |
| HD184293 | DDO | m35 | $9.46 \pm 0.05$ | McClure & Forrester (1981) |
| HD184293 | WBVR | W | $8.00 \pm 0.05$ | Kornilov et al. (1991) |
| HD184293 | Vilnius | P | $9.28 \pm 0.05$ | Zdanavicius et al. (1972) |
| HD184293 | DDO | m38 | $8.31 \pm 0.05$ | McClure & Forrester (1981) |
| HD184293 | Vilnius | X | $8.09 \pm 0.05$ | Zdanavicius et al. (1972) |
| HD184293 | DDO | m41 | $8.62 \pm 0.05$ | McClure & Forrester (1981) |
| HD184293 | Oja | m41 | $8.16 \pm 0.05$ | Häggkvist & Oja (1970) |

**Table 21** *(continued)*

| Star ID | System/Wvlen | Band/Bandpass | Value | Reference |
|---------|--------------|---------------|-------|-----------|
| HD184293 | DDO | m42 | $8.38 \pm 0.05$ | McClure & Forrester (1981) |
| HD184293 | Oja | m42 | $7.91 \pm 0.05$ | Häggkvist & Oja (1970) |
| HD184293 | WBVR | B | $6.86 \pm 0.05$ | Kornilov et al. (1991) |
| HD184293 | Johnson | B | $6.78 \pm 0.05$ | Haggkvist & Oja (1970) |
| HD184293 | DDO | m45 | $7.31 \pm 0.05$ | McClure & Forrester (1981) |
| HD184293 | Oja | m45 | $6.47 \pm 0.05$ | Häggkvist & Oja (1970) |
| HD184293 | Vilnius | Y | $6.47 \pm 0.05$ | Zdanavicius et al. (1972) |
| HD184293 | DDO | m48 | $6.02 \pm 0.05$ | McClure & Forrester (1981) |
| HD184293 | Vilnius | Z | $5.95 \pm 0.05$ | Zdanavicius et al. (1972) |
| HD184293 | WBVR | V | $5.56 \pm 0.05$ | Kornilov et al. (1991) |
| HD184293 | Vilnius | V | $5.54 \pm 0.05$ | Zdanavicius et al. (1972) |
| HD184293 | Johnson | V | $5.53 \pm 0.05$ | Haggkvist & Oja (1970) |
| HD184293 | Vilnius | S | $4.69 \pm 0.05$ | Zdanavicius et al. (1972) |
| HD184293 | WBVR | R | $4.64 \pm 0.05$ | Kornilov et al. (1991) |
| HD184293 | 1250 | 310 | $89.70 \pm 22.00$ | Smith et al. (2004) |
| HD184293 | 2200 | 361 | $85.00 \pm 18.10$ | Smith et al. (2004) |
| HD184293 | Johnson | K | $2.59 \pm 0.06$ | Neugebauer & Leighton (1969) |
| HD184293 | 3500 | 898 | $40.30 \pm 27.90$ | Smith et al. (2004) |
| HD184293 | 4900 | 712 | $20.50 \pm 7.10$ | Smith et al. (2004) |
| HD184293 | 12000 | 6384 | $4.10 \pm 17.30$ | Smith et al. (2004) |
| HD185958 | 13c | m33 | $6.39 \pm 0.05$ | Johnson & Mitchell (1995) |
| HD185958 | Geneva | U | $6.87 \pm 0.08$ | Golay (1972) |
| HD185958 | Vilnius | U | $8.12 \pm 0.05$ | Jasevicius et al. (1990) |
| HD185958 | 13c | m35 | $6.21 \pm 0.05$ | Johnson & Mitchell (1995) |
| HD185958 | DDO | m35 | $7.65 \pm 0.05$ | McClure & Forrester (1981) |
| HD185958 | Johnson | U | $6.31 \pm 0.05$ | Johnson et al. (1966) |
| HD185958 | Johnson | U | $6.31 \pm 0.05$ | Cuffey (1973) |
| HD185958 | Johnson | U | $6.33 \pm 0.05$ | Argue (1963) |
| HD185958 | Johnson | U | $6.33 \pm 0.05$ | Ducati (2002) |
| HD185958 | Johnson | U | $6.35 \pm 0.05$ | Fernie (1983) |
| HD185958 | Geneva | B1 | $6.07 \pm 0.08$ | Golay (1972) |
| HD185958 | Oja | m41 | $6.64 \pm 0.05$ | Häggkvist & Oja (1970) |
| HD185958 | DDO | m42 | $6.80 \pm 0.05$ | McClure & Forrester (1981) |
| HD185958 | Oja | m42 | $6.35 \pm 0.05$ | Häggkvist & Oja (1970) |
| HD185958 | Geneva | B | $4.74 \pm 0.08$ | Golay (1972) |
| HD185958 | WBVR | B | $5.45 \pm 0.05$ | Kornilov et al. (1991) |
| HD185958 | Johnson | B | $5.40 \pm 0.05$ | Häggkvist & Oja (1966) |
| HD185958 | Johnson | B | $5.41 \pm 0.05$ | Cuffey (1973) |
| HD185958 | Johnson | B | $5.42 \pm 0.05$ | Johnson et al. (1966) |
| HD185958 | Johnson | B | $5.43 \pm 0.05$ | Fernie (1983) |
| HD185958 | Johnson | B | $5.43 \pm 0.05$ | Ducati (2002) |

**Table 21** *(continued)*

| Star ID | System/Wvlen | Band/Bandpass | Value | Reference |
|---------|--------------|---------------|-------|-----------|
| HD185958 | Johnson | B | $5.44 \pm 0.05$ | Argue (1963) |
| HD185958 | Geneva | B2 | $5.89 \pm 0.08$ | Golay (1972) |
| HD185958 | 13c | m45 | $5.14 \pm 0.05$ | Johnson & Mitchell (1995) |
| HD185958 | DDO | m45 | $5.96 \pm 0.05$ | McClure & Forrester (1981) |
| HD185958 | Oja | m45 | $5.14 \pm 0.05$ | Häggkvist & Oja (1970) |
| HD185958 | Vilnius | Y | $5.17 \pm 0.05$ | Jasevicius et al. (1990) |
| HD185958 | DDO | m48 | $4.74 \pm 0.05$ | McClure & Forrester (1981) |
| HD185958 | Vilnius | Z | $4.67 \pm 0.05$ | Jasevicius et al. (1990) |
| HD185958 | 13c | m52 | $4.64 \pm 0.05$ | Johnson & Mitchell (1995) |
| HD185958 | Geneva | V1 | $5.18 \pm 0.08$ | Golay (1972) |
| HD185958 | WBVR | V | $4.39 \pm 0.05$ | Kornilov et al. (1991) |
| HD185958 | Vilnius | V | $4.37 \pm 0.05$ | Jasevicius et al. (1990) |
| HD185958 | Geneva | V | $4.38 \pm 0.08$ | Golay (1972) |
| HD185958 | Johnson | V | $4.36 \pm 0.05$ | Häggkvist & Oja (1966) |
| HD185958 | Johnson | V | $4.36 \pm 0.05$ | Cuffey (1973) |
| HD185958 | Johnson | V | $4.37 \pm 0.05$ | Johnson et al. (1966) |
| HD185958 | Johnson | V | $4.38 \pm 0.05$ | Fernie (1983) |
| HD185958 | Johnson | V | $4.38 \pm 0.05$ | Ducati (2002) |
| HD185958 | Johnson | V | $4.39 \pm 0.05$ | Argue (1963) |
| HD185958 | 13c | m58 | $4.17 \pm 0.05$ | Johnson & Mitchell (1995) |
| HD185958 | Geneva | G | $5.38 \pm 0.08$ | Golay (1972) |
| HD185958 | 13c | m63 | $3.86 \pm 0.05$ | Johnson & Mitchell (1995) |
| HD185958 | Vilnius | S | $3.63 \pm 0.05$ | Jasevicius et al. (1990) |
| HD185958 | WBVR | R | $3.67 \pm 0.05$ | Kornilov et al. (1991) |
| HD185958 | 13c | m72 | $3.65 \pm 0.05$ | Johnson & Mitchell (1995) |
| HD185958 | 13c | m80 | $3.43 \pm 0.05$ | Johnson & Mitchell (1995) |
| HD185958 | 13c | m86 | $3.33 \pm 0.05$ | Johnson & Mitchell (1995) |
| HD185958 | 13c | m99 | $3.17 \pm 0.05$ | Johnson & Mitchell (1995) |
| HD185958 | 13c | m110 | $3.01 \pm 0.05$ | Johnson & Mitchell (1995) |
| HD185958 | Johnson | J | $2.74 \pm 0.05$ | Johnson et al. (1966) |
| HD185958 | Johnson | J | $2.74 \pm 0.05$ | Glass (1975) |
| HD185958 | Johnson | J | $2.74 \pm 0.05$ | Kodaira & Lenzen (1983) |
| HD185958 | Johnson | J | $2.74 \pm 0.05$ | Leitherer & Wolf (1984) |
| HD185958 | Johnson | J | $2.74 \pm 0.05$ | Ducati (2002) |
| HD185958 | Johnson | J | $2.74 \pm 0.05$ | Shenavrin et al. (2011) |
| HD185958 | Johnson | H | $2.20 \pm 0.05$ | Shenavrin et al. (2011) |
| HD185958 | Johnson | H | $2.21 \pm 0.05$ | Kodaira & Lenzen (1983) |
| HD185958 | Johnson | H | $2.21 \pm 0.05$ | Leitherer & Wolf (1984) |
| HD185958 | Johnson | H | $2.21 \pm 0.05$ | Ducati (2002) |
| HD185958 | Johnson | K | $2.09 \pm 0.06$ | Neugebauer & Leighton (1969) |
| HD185958 | Johnson | K | $2.13 \pm 0.05$ | Johnson et al. (1966) |

**Table 21** *(continued)*

| Star ID | System/Wvlen | Band/Bandpass | Value | Reference |
|---------|--------------|---------------|-------|-----------|
| HD185958 | Johnson | K | $2.13 \pm 0.05$ | Glass (1975) |
| HD185958 | Johnson | K | $2.13 \pm 0.05$ | Ducati (2002) |
| HD185958 | Johnson | K | $2.13 \pm 0.05$ | Shenavrin et al. (2011) |
| HD185958 | Johnson | L | $2.00 \pm 0.05$ | Johnson et al. (1966) |
| HD185958 | Johnson | L | $2.00 \pm 0.05$ | Glass (1975) |
| HD185958 | Johnson | L | $2.00 \pm 0.05$ | Ducati (2002) |
| HD185958 | Johnson | M | $1.89 \pm 0.05$ | Ducati (2002) |
| HD186675 | Geneva | U | $7.04 \pm 0.08$ | Golay (1972) |
| HD186675 | Vilnius | U | $8.24 \pm 0.05$ | Jasevicius et al. (1990) |
| HD186675 | DDO | m35 | $7.85 \pm 0.05$ | McClure & Forrester (1981) |
| HD186675 | WBVR | W | $6.38 \pm 0.05$ | Kornilov et al. (1991) |
| HD186675 | Johnson | U | $6.55 \pm 0.05$ | Argue (1963) |
| HD186675 | Johnson | U | $6.56 \pm 0.05$ | Mermilliod (1986) |
| HD186675 | Vilnius | P | $7.67 \pm 0.05$ | Jasevicius et al. (1990) |
| HD186675 | DDO | m38 | $6.81 \pm 0.05$ | McClure & Forrester (1981) |
| HD186675 | Geneva | B1 | $6.37 \pm 0.08$ | Golay (1972) |
| HD186675 | Vilnius | X | $6.77 \pm 0.05$ | Jasevicius et al. (1990) |
| HD186675 | DDO | m41 | $7.37 \pm 0.05$ | McClure & Forrester (1981) |
| HD186675 | Oja | m41 | $6.93 \pm 0.05$ | Häggkvist & Oja (1970) |
| HD186675 | DDO | m42 | $7.18 \pm 0.05$ | McClure & Forrester (1981) |
| HD186675 | Oja | m42 | $6.75 \pm 0.05$ | Häggkvist & Oja (1970) |
| HD186675 | Geneva | B | $5.11 \pm 0.08$ | Golay (1972) |
| HD186675 | WBVR | B | $5.86 \pm 0.05$ | Kornilov et al. (1991) |
| HD186675 | Johnson | B | $5.79 \pm 0.05$ | Häggkvist & Oja (1966) |
| HD186675 | Johnson | B | $5.86 \pm 0.05$ | Argue (1963) |
| HD186675 | Johnson | B | $5.87 \pm 0.05$ | Mermilliod (1986) |
| HD186675 | Geneva | B2 | $6.28 \pm 0.08$ | Golay (1972) |
| HD186675 | DDO | m45 | $6.39 \pm 0.05$ | McClure & Forrester (1981) |
| HD186675 | Oja | m45 | $5.56 \pm 0.05$ | Häggkvist & Oja (1970) |
| HD186675 | Vilnius | Y | $5.56 \pm 0.05$ | Jasevicius et al. (1990) |
| HD186675 | DDO | m48 | $5.22 \pm 0.05$ | McClure & Forrester (1981) |
| HD186675 | Vilnius | Z | $5.11 \pm 0.05$ | Jasevicius et al. (1990) |
| HD186675 | Geneva | V1 | $5.65 \pm 0.08$ | Golay (1972) |
| HD186675 | WBVR | V | $4.90 \pm 0.05$ | Kornilov et al. (1991) |
| HD186675 | Vilnius | V | $4.81 \pm 0.05$ | Jasevicius et al. (1990) |
| HD186675 | Geneva | V | $4.89 \pm 0.08$ | Golay (1972) |
| HD186675 | Johnson | V | $4.86 \pm 0.05$ | Häggkvist & Oja (1966) |
| HD186675 | Johnson | V | $4.90 \pm 0.05$ | Argue (1963) |
| HD186675 | Johnson | V | $4.92 \pm 0.05$ | Mermilliod (1986) |
| HD186675 | Geneva | G | $5.88 \pm 0.08$ | Golay (1972) |
| HD186675 | Vilnius | S | $4.11 \pm 0.05$ | Jasevicius et al. (1990) |

| Star ID | System/Wvlen | Band/Bandpass | Value | Reference |
|---------|--------------|---------------|-------|-----------|
| HD186675 | WBVR | R | $4.21 \pm 0.05$ | Kornilov et al. (1991) |
| HD186675 | Johnson | K | $2.73 \pm 0.05$ | Neugebauer & Leighton (1969) |
| HD186776 | Geneva | B1 | $8.96 \pm 0.08$ | Golay (1972) |
| HD186776 | DDO | m42 | $9.69 \pm 0.05$ | McClure & Forrester (1981) |
| HD186776 | Geneva | B | $7.40 \pm 0.08$ | Golay (1972) |
| HD186776 | WBVR | B | $8.08 \pm 0.05$ | Kornilov et al. (1991) |
| HD186776 | Johnson | B | $7.75 \pm 0.05$ | Mermilliod (1986) |
| HD186776 | Johnson | B | $7.95 \pm 0.05$ | Haggkvist & Oja (1970) |
| HD186776 | Johnson | B | $7.98 \pm 0.05$ | Nicolet (1978) |
| HD186776 | Johnson | B | $7.99 \pm 0.05$ | Carney (1983) |
| HD186776 | Johnson | B | $7.99 \pm 0.05$ | Ducati (2002) |
| HD186776 | Johnson | B | $8.00 \pm 0.05$ | Smak (1964) |
| HD186776 | Geneva | B2 | $8.40 \pm 0.08$ | Golay (1972) |
| HD186776 | DDO | m45 | $8.48 \pm 0.05$ | McClure & Forrester (1981) |
| HD186776 | Vilnius | Y | $7.59 \pm 0.05$ | Zdanavicius et al. (1969) |
| HD186776 | Vilnius | Y | $7.61 \pm 0.05$ | Sleivyte (1985) |
| HD186776 | DDO | m48 | $7.11 \pm 0.05$ | McClure & Forrester (1981) |
| HD186776 | Vilnius | Z | $6.79 \pm 0.05$ | Zdanavicius et al. (1969) |
| HD186776 | Vilnius | Z | $6.84 \pm 0.05$ | Sleivyte (1985) |
| HD186776 | Geneva | V1 | $7.15 \pm 0.08$ | Golay (1972) |
| HD186776 | WBVR | V | $6.42 \pm 0.05$ | Kornilov et al. (1991) |
| HD186776 | Vilnius | V | $6.30 \pm 0.05$ | Zdanavicius et al. (1969) |
| HD186776 | Vilnius | V | $6.33 \pm 0.05$ | Sleivyte (1985) |
| HD186776 | Geneva | V | $6.35 \pm 0.08$ | Golay (1972) |
| HD186776 | Johnson | V | $6.15 \pm 0.05$ | Mermilliod (1986) |
| HD186776 | Johnson | V | $6.32 \pm 0.05$ | Haggkvist & Oja (1970) |
| HD186776 | Johnson | V | $6.34 \pm 0.05$ | Nicolet (1978) |
| HD186776 | Johnson | V | $6.35 \pm 0.05$ | Smak (1964) |
| HD186776 | Johnson | V | $6.37 \pm 0.05$ | Carney (1983) |
| HD186776 | Johnson | V | $6.37 \pm 0.05$ | Ducati (2002) |
| HD186776 | Geneva | G | $7.27 \pm 0.08$ | Golay (1972) |
| HD186776 | Vilnius | S | $5.06 \pm 0.05$ | Zdanavicius et al. (1969) |
| HD186776 | Vilnius | S | $5.13 \pm 0.05$ | Sleivyte (1985) |
| HD186776 | WBVR | R | $4.80 \pm 0.05$ | Kornilov et al. (1991) |
| HD186776 | Johnson | J | $2.61 \pm 0.05$ | Ducati (2002) |
| HD186776 | Johnson | J | $2.64 \pm 0.05$ | Carney (1983) |
| HD186776 | Johnson | J | $2.69 \pm 0.05$ | Arribas & Martinez Roger (1987) |
| HD186776 | Johnson | J | $2.69 \pm 0.05$ | Kerschbaum & Hron (1994) |
| HD186776 | Johnson | H | $1.75 \pm 0.05$ | Carney (1983) |
| HD186776 | Johnson | H | $1.75 \pm 0.05$ | Arribas & Martinez Roger (1987) |
| HD186776 | Johnson | H | $1.75 \pm 0.05$ | Ducati (2002) |

**Table 21** *(continued)*

| Star ID | System/Wvlen | Band/Bandpass | Value | Reference |
|---------|--------------|---------------|-------|-----------|
| HD186776 | Johnson | H | $1.78 \pm 0.05$ | Kerschbaum & Hron (1994) |
| HD186776 | 2200 | 361 | $154.90 \pm 11.40$ | Smith et al. (2004) |
| HD186776 | Johnson | K | $1.48 \pm 0.05$ | Ducati (2002) |
| HD186776 | Johnson | K | $1.49 \pm 0.03$ | Neugebauer & Leighton (1969) |
| HD186776 | Johnson | L | $1.37 \pm 0.05$ | Ducati (2002) |
| HD186776 | 3500 | 898 | $73.10 \pm 12.00$ | Smith et al. (2004) |
| HD186776 | 4900 | 712 | $33.80 \pm 6.00$ | Smith et al. (2004) |
| HD186776 | 12000 | 6384 | $2.20 \pm 22.70$ | Smith et al. (2004) |
| HD187849 | WBVR | W | $8.87 \pm 0.05$ | Kornilov et al. (1991) |
| HD187849 | Johnson | U | $8.91 \pm 0.05$ | Ducati (2002) |
| HD187849 | Johnson | U | $9.03 \pm 0.05$ | Mermilliod (1986) |
| HD187849 | Oja | m41 | $8.32 \pm 0.05$ | Häggkvist & Oja (1970) |
| HD187849 | Oja | m42 | $8.14 \pm 0.05$ | Häggkvist & Oja (1970) |
| HD187849 | WBVR | B | $6.90 \pm 0.05$ | Kornilov et al. (1991) |
| HD187849 | Johnson | B | $6.76 \pm 0.05$ | Haggkvist & Oja (1970) |
| HD187849 | Johnson | B | $6.81 \pm 0.05$ | Ducati (2002) |
| HD187849 | Johnson | B | $6.82 \pm 0.05$ | Neckel (1974) |
| HD187849 | Johnson | B | $6.93 \pm 0.05$ | Mermilliod (1986) |
| HD187849 | Oja | m45 | $6.45 \pm 0.05$ | Häggkvist & Oja (1970) |
| HD187849 | WBVR | V | $5.17 \pm 0.05$ | Kornilov et al. (1991) |
| HD187849 | Johnson | V | $5.10 \pm 0.05$ | Haggkvist & Oja (1970) |
| HD187849 | Johnson | V | $5.12 \pm 0.05$ | Neckel (1974) |
| HD187849 | Johnson | V | $5.12 \pm 0.05$ | Ducati (2002) |
| HD187849 | Johnson | V | $5.26 \pm 0.05$ | Mermilliod (1986) |
| HD187849 | WBVR | R | $3.73 \pm 0.05$ | Kornilov et al. (1991) |
| HD187849 | Johnson | J | $1.78 \pm 0.05$ | McWilliam & Lambert (1984) |
| HD187849 | Johnson | J | $1.78 \pm 0.05$ | Ducati (2002) |
| HD187849 | 2200 | 361 | $318.20 \pm 17.10$ | Smith et al. (2004) |
| HD187849 | Johnson | K | $0.65 \pm 0.05$ | Ducati (2002) |
| HD187849 | Johnson | K | $0.71 \pm 0.05$ | Neugebauer & Leighton (1969) |
| HD187849 | 3500 | 898 | $168.80 \pm 12.80$ | Smith et al. (2004) |
| HD187849 | 4900 | 712 | $77.60 \pm 7.10$ | Smith et al. (2004) |
| HD187849 | 12000 | 6384 | $25.50 \pm 23.30$ | Smith et al. (2004) |
| HD188310 | 13c | m33 | $6.72 \pm 0.05$ | Johnson & Mitchell (1995) |
| HD188310 | Vilnius | U | $8.45 \pm 0.05$ | Kazlauskas et al. (2005) |
| HD188310 | 13c | m35 | $6.49 \pm 0.05$ | Johnson & Mitchell (1995) |
| HD188310 | DDO | m35 | $7.98 \pm 0.05$ | McClure & Forrester (1981) |
| HD188310 | WBVR | W | $6.52 \pm 0.05$ | Kornilov et al. (1991) |
| HD188310 | Johnson | U | $6.62 \pm 0.05$ | Johnson et al. (1966) |
| HD188310 | Johnson | U | $6.63 \pm 0.05$ | Mermilliod (1986) |
| HD188310 | Johnson | U | $6.68 \pm 0.05$ | Cousins (1963a) |

**Table 21** *(continued)*

| Star ID | System/Wvlen | Band/Bandpass | Value | Reference |
|---------|--------------|---------------|-------|-----------|
| HD188310 | Johnson | U | $6.69 \pm 0.05$ | Argue (1966) |
| HD188310 | 13c | m37 | $6.58 \pm 0.05$ | Johnson & Mitchell (1995) |
| HD188310 | Vilnius | P | $7.85 \pm 0.05$ | Kazlauskas et al. (2005) |
| HD188310 | DDO | m38 | $6.90 \pm 0.05$ | McClure & Forrester (1981) |
| HD188310 | 13c | m40 | $6.34 \pm 0.05$ | Johnson & Mitchell (1995) |
| HD188310 | Vilnius | X | $6.82 \pm 0.05$ | Kazlauskas et al. (2005) |
| HD188310 | DDO | m41 | $7.38 \pm 0.05$ | McClure & Forrester (1981) |
| HD188310 | Oja | m41 | $6.92 \pm 0.05$ | Häggkvist & Oja (1970) |
| HD188310 | DDO | m42 | $7.19 \pm 0.05$ | McClure & Forrester (1981) |
| HD188310 | Oja | m42 | $6.74 \pm 0.05$ | Häggkvist & Oja (1970) |
| HD188310 | WBVR | B | $5.79 \pm 0.05$ | Kornilov et al. (1991) |
| HD188310 | Johnson | B | $5.72 \pm 0.05$ | Häggkvist & Oja (1966) |
| HD188310 | Johnson | B | $5.73 \pm 0.05$ | Johnson et al. (1966) |
| HD188310 | Johnson | B | $5.76 \pm 0.05$ | Cousins (1963a) |
| HD188310 | Johnson | B | $5.76 \pm 0.05$ | Mermilliod (1986) |
| HD188310 | Johnson | B | $5.78 \pm 0.05$ | Argue (1966) |
| HD188310 | 13c | m45 | $5.44 \pm 0.05$ | Johnson & Mitchell (1995) |
| HD188310 | DDO | m45 | $6.28 \pm 0.05$ | McClure & Forrester (1981) |
| HD188310 | Oja | m45 | $5.45 \pm 0.05$ | Häggkvist & Oja (1970) |
| HD188310 | Vilnius | Y | $5.51 \pm 0.05$ | Kazlauskas et al. (2005) |
| HD188310 | DDO | m48 | $5.08 \pm 0.05$ | McClure & Forrester (1981) |
| HD188310 | Vilnius | Z | $5.05 \pm 0.05$ | Kazlauskas et al. (2005) |
| HD188310 | 13c | m52 | $4.98 \pm 0.05$ | Johnson & Mitchell (1995) |
| HD188310 | WBVR | V | $4.72 \pm 0.05$ | Kornilov et al. (1991) |
| HD188310 | Vilnius | V | $4.72 \pm 0.05$ | Kazlauskas et al. (2005) |
| HD188310 | Johnson | V | $4.67 \pm 0.05$ | Häggkvist & Oja (1966) |
| HD188310 | Johnson | V | $4.68 \pm 0.05$ | Johnson et al. (1966) |
| HD188310 | Johnson | V | $4.72 \pm 0.05$ | Cousins (1963a) |
| HD188310 | Johnson | V | $4.72 \pm 0.05$ | Mermilliod (1986) |
| HD188310 | Johnson | V | $4.73 \pm 0.05$ | Argue (1966) |
| HD188310 | 13c | m58 | $4.47 \pm 0.05$ | Johnson & Mitchell (1995) |
| HD188310 | 13c | m63 | $4.16 \pm 0.05$ | Johnson & Mitchell (1995) |
| HD188310 | Vilnius | S | $3.96 \pm 0.05$ | Kazlauskas et al. (2005) |
| HD188310 | WBVR | R | $3.95 \pm 0.05$ | Kornilov et al. (1991) |
| HD188310 | 13c | m72 | $3.93 \pm 0.05$ | Johnson & Mitchell (1995) |
| HD188310 | 13c | m80 | $3.71 \pm 0.05$ | Johnson & Mitchell (1995) |
| HD188310 | 13c | m86 | $3.60 \pm 0.05$ | Johnson & Mitchell (1995) |
| HD188310 | 13c | m99 | $3.43 \pm 0.05$ | Johnson & Mitchell (1995) |
| HD188310 | 13c | m110 | $3.24 \pm 0.05$ | Johnson & Mitchell (1995) |
| HD188310 | Johnson | K | $2.37 \pm 0.09$ | Neugebauer & Leighton (1969) |
| HD189695 | Vilnius | U | $11.26 \pm 0.05$ | Bartkevicius et al. (1973) |

**Table 21** *(continued)*

| Star ID | System/Wvlen | Band/Bandpass | Value | Reference |
|---------|--------------|---------------|-------|-----------|
| HD189695 | WBVR | W | $9.31 \pm 0.05$ | Kornilov et al. (1991) |
| HD189695 | Johnson | U | $9.31 \pm 0.05$ | Cousins (1964b) |
| HD189695 | Johnson | U | $9.32 \pm 0.05$ | Johnson et al. (1966) |
| HD189695 | Johnson | U | $9.32 \pm 0.05$ | Ducati (2002) |
| HD189695 | Vilnius | P | $10.42 \pm 0.05$ | Bartkevicius et al. (1973) |
| HD189695 | Vilnius | X | $9.03 \pm 0.05$ | Bartkevicius et al. (1973) |
| HD189695 | Oja | m41 | $8.96 \pm 0.05$ | Häggkvist & Oja (1970) |
| HD189695 | Oja | m42 | $8.70 \pm 0.05$ | Häggkvist & Oja (1970) |
| HD189695 | WBVR | B | $7.49 \pm 0.05$ | Kornilov et al. (1991) |
| HD189695 | Johnson | B | $7.42 \pm 0.05$ | Cousins (1964b) |
| HD189695 | Johnson | B | $7.43 \pm 0.05$ | Johnson et al. (1966) |
| HD189695 | Johnson | B | $7.43 \pm 0.05$ | Ducati (2002) |
| HD189695 | Oja | m45 | $7.09 \pm 0.05$ | Häggkvist & Oja (1970) |
| HD189695 | Vilnius | Y | $7.01 \pm 0.05$ | Bartkevicius et al. (1973) |
| HD189695 | Vilnius | Z | $6.44 \pm 0.05$ | Bartkevicius et al. (1973) |
| HD189695 | WBVR | V | $5.91 \pm 0.05$ | Kornilov et al. (1991) |
| HD189695 | Vilnius | V | $5.91 \pm 0.05$ | Bartkevicius et al. (1973) |
| HD189695 | Johnson | V | $5.90 \pm 0.05$ | Cousins (1964b) |
| HD189695 | Johnson | V | $5.91 \pm 0.05$ | Johnson et al. (1966) |
| HD189695 | Johnson | V | $5.91 \pm 0.05$ | Ducati (2002) |
| HD189695 | Vilnius | S | $4.86 \pm 0.05$ | Bartkevicius et al. (1973) |
| HD189695 | WBVR | R | $4.72 \pm 0.05$ | Kornilov et al. (1991) |
| HD189695 | 1250 | 310 | $106.90 \pm 9.70$ | Smith et al. (2004) |
| HD189695 | Johnson | J | $3.11 \pm 0.05$ | McWilliam & Lambert (1984) |
| HD189695 | Johnson | J | $3.11 \pm 0.05$ | Ducati (2002) |
| HD189695 | 2200 | 361 | $110.60 \pm 14.90$ | Smith et al. (2004) |
| HD189695 | Johnson | K | $2.16 \pm 0.05$ | Ducati (2002) |
| HD189695 | Johnson | K | $2.19 \pm 0.07$ | Neugebauer & Leighton (1969) |
| HD189695 | 3500 | 898 | $46.20 \pm 9.50$ | Smith et al. (2004) |
| HD189695 | 4900 | 712 | $22.80 \pm 6.50$ | Smith et al. (2004) |
| HD189695 | 12000 | 6384 | $3.80 \pm 19.10$ | Smith et al. (2004) |
| HD189849 | 13c | m33 | $5.13 \pm 0.05$ | Johnson & Mitchell (1995) |
| HD189849 | Vilnius | U | $6.93 \pm 0.05$ | Zdanavicius et al. (1969) |
| HD189849 | Vilnius | U | $6.93 \pm 0.05$ | Straizys & Meistas (1989) |
| HD189849 | Vilnius | U | $6.95 \pm 0.05$ | Kazlauskas et al. (2005) |
| HD189849 | 13c | m35 | $5.05 \pm 0.05$ | Johnson & Mitchell (1995) |
| HD189849 | Stromgren | u | $6.42 \pm 0.08$ | Crawford et al. (1966) |
| HD189849 | Stromgren | u | $6.42 \pm 0.08$ | Cameron (1966) |
| HD189849 | Stromgren | u | $6.42 \pm 0.08$ | Warren (1973) |
| HD189849 | Stromgren | u | $6.42 \pm 0.08$ | Hauck & Mermilliod (1998) |
| HD189849 | WBVR | W | $5.04 \pm 0.05$ | Kornilov et al. (1991) |

**Table 21** *(continued)*

| Star ID | System/Wvlen | Band/Bandpass | Value | Reference |
|---------|--------------|---------------|-------|-----------|
| HD189849 | Johnson | U | $4.95 \pm 0.05$ | Harmanec et al. (1980) |
| HD189849 | Johnson | U | $4.99 \pm 0.05$ | Johnson et al. (1966) |
| HD189849 | Johnson | U | $4.99 \pm 0.05$ | Mermilliod (1986) |
| HD189849 | 13c | m37 | $4.96 \pm 0.05$ | Johnson & Mitchell (1995) |
| HD189849 | Vilnius | P | $6.20 \pm 0.05$ | Zdanavicius et al. (1969) |
| HD189849 | Vilnius | P | $6.20 \pm 0.05$ | Straizys & Meistas (1989) |
| HD189849 | Vilnius | P | $6.24 \pm 0.05$ | Kazlauskas et al. (2005) |
| HD189849 | 13c | m40 | $4.94 \pm 0.05$ | Johnson & Mitchell (1995) |
| HD189849 | Vilnius | X | $5.38 \pm 0.05$ | Zdanavicius et al. (1969) |
| HD189849 | Vilnius | X | $5.38 \pm 0.05$ | Straizys & Meistas (1989) |
| HD189849 | Vilnius | X | $5.38 \pm 0.05$ | Kazlauskas et al. (2005) |
| HD189849 | Stromgren | v | $5.06 \pm 0.08$ | Crawford et al. (1966) |
| HD189849 | Stromgren | v | $5.06 \pm 0.08$ | Cameron (1966) |
| HD189849 | Stromgren | v | $5.06 \pm 0.08$ | Hauck & Mermilliod (1998) |
| HD189849 | Stromgren | v | $5.07 \pm 0.08$ | Warren (1973) |
| HD189849 | WBVR | B | $4.85 \pm 0.05$ | Kornilov et al. (1991) |
| HD189849 | Johnson | B | $4.80 \pm 0.05$ | Häggkvist & Oja (1966) |
| HD189849 | Johnson | B | $4.80 \pm 0.05$ | Harmanec et al. (1980) |
| HD189849 | Johnson | B | $4.83 \pm 0.05$ | Johnson et al. (1966) |
| HD189849 | Johnson | B | $4.84 \pm 0.05$ | Mermilliod (1986) |
| HD189849 | 13c | m45 | $4.81 \pm 0.05$ | Johnson & Mitchell (1995) |
| HD189849 | Vilnius | Y | $4.91 \pm 0.05$ | Kazlauskas et al. (2005) |
| HD189849 | Vilnius | Y | $4.92 \pm 0.05$ | Zdanavicius et al. (1969) |
| HD189849 | Vilnius | Y | $4.92 \pm 0.05$ | Straizys & Meistas (1989) |
| HD189849 | Stromgren | b | $4.76 \pm 0.08$ | Crawford et al. (1966) |
| HD189849 | Stromgren | b | $4.76 \pm 0.08$ | Cameron (1966) |
| HD189849 | Stromgren | b | $4.76 \pm 0.08$ | Hauck & Mermilliod (1998) |
| HD189849 | Stromgren | b | $4.77 \pm 0.08$ | Warren (1973) |
| HD189849 | Vilnius | Z | $4.75 \pm 0.05$ | Kazlauskas et al. (2005) |
| HD189849 | Vilnius | Z | $4.76 \pm 0.05$ | Zdanavicius et al. (1969) |
| HD189849 | Vilnius | Z | $4.76 \pm 0.05$ | Straizys & Meistas (1989) |
| HD189849 | 13c | m52 | $4.74 \pm 0.05$ | Johnson & Mitchell (1995) |
| HD189849 | WBVR | V | $4.67 \pm 0.05$ | Kornilov et al. (1991) |
| HD189849 | Vilnius | V | $4.64 \pm 0.05$ | Kazlauskas et al. (2005) |
| HD189849 | Vilnius | V | $4.65 \pm 0.05$ | Zdanavicius et al. (1969) |
| HD189849 | Vilnius | V | $4.65 \pm 0.05$ | Straizys & Meistas (1989) |
| HD189849 | Stromgren | y | $4.66 \pm 0.08$ | Crawford et al. (1966) |
| HD189849 | Stromgren | y | $4.66 \pm 0.08$ | Cameron (1966) |
| HD189849 | Stromgren | y | $4.66 \pm 0.08$ | Warren (1973) |
| HD189849 | Stromgren | y | $4.66 \pm 0.08$ | Hauck & Mermilliod (1998) |
| HD189849 | Johnson | V | $4.62 \pm 0.05$ | Häggkvist & Oja (1966) |

**Table 21** *continued on next page*

**Table 21** *(continued)*

| Star ID | System/Wvlen | Band/Bandpass | Value | Reference |
|---------|--------------|---------------|-------|-----------|
| HD189849 | Johnson | V | $4.62 \pm 0.05$ | Harmanec et al. (1980) |
| HD189849 | Johnson | V | $4.64 \pm 0.05$ | Mermilliod (1986) |
| HD189849 | Johnson | V | $4.65 \pm 0.05$ | Johnson et al. (1966) |
| HD189849 | 13c | m58 | $4.63 \pm 0.05$ | Johnson & Mitchell (1995) |
| HD189849 | 13c | m63 | $4.56 \pm 0.05$ | Johnson & Mitchell (1995) |
| HD189849 | Vilnius | S | $4.38 \pm 0.05$ | Zdanavicius et al. (1969) |
| HD189849 | Vilnius | S | $4.38 \pm 0.05$ | Straizys & Meistas (1989) |
| HD189849 | Vilnius | S | $4.40 \pm 0.05$ | Kazlauskas et al. (2005) |
| HD189849 | WBVR | R | $4.53 \pm 0.05$ | Kornilov et al. (1991) |
| HD189849 | 13c | m72 | $4.53 \pm 0.05$ | Johnson & Mitchell (1995) |
| HD189849 | 13c | m80 | $4.49 \pm 0.05$ | Johnson & Mitchell (1995) |
| HD189849 | 13c | m86 | $4.46 \pm 0.05$ | Johnson & Mitchell (1995) |
| HD189849 | 13c | m99 | $4.42 \pm 0.05$ | Johnson & Mitchell (1995) |
| HD189849 | 13c | m110 | $4.39 \pm 0.05$ | Johnson & Mitchell (1995) |
| HD189849 | Johnson | J | $4.27 \pm 0.05$ | Alonso et al. (1998) |
| HD189849 | Johnson | H | $4.20 \pm 0.05$ | Alonso et al. (1998) |
| HD191178 | WBVR | B | $8.25 \pm 0.05$ | Kornilov et al. (1991) |
| HD191178 | Johnson | B | $8.14 \pm 0.01$ | Oja (1991) |
| HD191178 | Johnson | B | $8.33 \pm 0.05$ | Mermilliod (1986) |
| HD191178 | Johnson | B | $8.33 \pm 0.05$ | Ducati (2002) |
| HD191178 | Vilnius | Y | $7.72 \pm 0.05$ | Zdanavicius et al. (1972) |
| HD191178 | Vilnius | Z | $6.91 \pm 0.05$ | Zdanavicius et al. (1972) |
| HD191178 | WBVR | V | $6.47 \pm 0.05$ | Kornilov et al. (1991) |
| HD191178 | Vilnius | V | $6.38 \pm 0.05$ | Zdanavicius et al. (1972) |
| HD191178 | Johnson | V | $6.40 \pm 0.01$ | Oja (1991) |
| HD191178 | Johnson | V | $6.42 \pm 0.05$ | Mermilliod (1986) |
| HD191178 | Johnson | V | $6.42 \pm 0.05$ | Ducati (2002) |
| HD191178 | Vilnius | S | $5.16 \pm 0.05$ | Zdanavicius et al. (1972) |
| HD191178 | WBVR | R | $4.83 \pm 0.05$ | Kornilov et al. (1991) |
| HD191178 | Johnson | J | $2.55 \pm 0.05$ | McWilliam & Lambert (1984) |
| HD191178 | Johnson | J | $2.55 \pm 0.05$ | Ducati (2002) |
| HD191178 | Johnson | K | $1.35 \pm 0.05$ | Ducati (2002) |
| HD191178 | Johnson | K | $1.46 \pm 0.04$ | Neugebauer & Leighton (1969) |
| HD193347 | KronComet | COp | $9.00 \pm 0.01$ | This work |
| HD193347 | WBVR | B | $8.31 \pm 0.05$ | Kornilov et al. (1991) |
| HD193347 | Johnson | B | $8.12 \pm 0.02$ | This work |
| HD193347 | KronComet | Bc | $8.08 \pm 0.01$ | This work |
| HD193347 | Vilnius | Y | $7.79 \pm 0.05$ | Zdanavicius & Cerniene (1985) |
| HD193347 | Vilnius | Y | $7.91 \pm 0.05$ | Straizys & Meistas (1989) |
| HD193347 | KronComet | C2 | $7.03 \pm 0.01$ | This work |
| HD193347 | Vilnius | Z | $7.14 \pm 0.05$ | Zdanavicius & Cerniene (1985) |

**Table 21** *(continued)*

| Star ID | System/Wvlen | Band/Bandpass | Value | Reference |
|---------|--------------|---------------|-------|-----------|
| HD193347 | Vilnius | Z | $7.26 \pm 0.05$ | Straizys & Meistas (1989) |
| HD193347 | KronComet | Gc | $6.77 \pm 0.01$ | This work |
| HD193347 | WBVR | V | $6.62 \pm 0.05$ | Kornilov et al. (1991) |
| HD193347 | Vilnius | V | $6.59 \pm 0.05$ | Zdanavicius & Cerniene (1985) |
| HD193347 | Vilnius | V | $6.70 \pm 0.05$ | Straizys & Meistas (1989) |
| HD193347 | Johnson | V | $6.64 \pm 0.01$ | This work |
| HD193347 | Vilnius | S | $5.44 \pm 0.05$ | Zdanavicius & Cerniene (1985) |
| HD193347 | Vilnius | S | $5.55 \pm 0.05$ | Straizys & Meistas (1989) |
| HD193347 | WBVR | R | $5.23 \pm 0.05$ | Kornilov et al. (1991) |
| HD193347 | KronComet | Rc | $5.21 \pm 0.01$ | This work |
| HD193347 | 1250 | 310 | $65.00 \pm 12.30$ | Smith et al. (2004) |
| HD193347 | 2200 | 361 | $47.20 \pm 6.90$ | Smith et al. (2004) |
| HD193347 | Johnson | K | $2.28 \pm 0.10$ | Neugebauer & Leighton (1969) |
| HD193347 | 3500 | 898 | $29.90 \pm 18.50$ | Smith et al. (2004) |
| HD193347 | 4900 | 712 | $10.70 \pm 7.30$ | Smith et al. (2004) |
| HD193347 | 12000 | 6384 | $-20.40 \pm 24.20$ | Smith et al. (2004) |
| HD193579 | Oja | m41 | $8.90 \pm 0.05$ | Häggkvist & Oja (1970) |
| HD193579 | Oja | m42 | $8.64 \pm 0.05$ | Häggkvist & Oja (1970) |
| HD193579 | KronComet | COp | $7.83 \pm 0.09$ | This work |
| HD193579 | WBVR | B | $7.37 \pm 0.05$ | Kornilov et al. (1991) |
| HD193579 | Johnson | B | $7.14 \pm 0.05$ | This work |
| HD193579 | Johnson | B | $7.29 \pm 0.05$ | Haggkvist & Oja (1970) |
| HD193579 | KronComet | Bc | $7.07 \pm 0.02$ | This work |
| HD193579 | Oja | m45 | $6.97 \pm 0.05$ | Häggkvist & Oja (1970) |
| HD193579 | Vilnius | Y | $6.89 \pm 0.05$ | Zdanavicius et al. (1972) |
| HD193579 | KronComet | C2 | $6.26 \pm 0.01$ | This work |
| HD193579 | Vilnius | Z | $6.36 \pm 0.05$ | Zdanavicius et al. (1972) |
| HD193579 | KronComet | Gc | $5.95 \pm 0.02$ | This work |
| HD193579 | WBVR | V | $5.82 \pm 0.05$ | Kornilov et al. (1991) |
| HD193579 | Vilnius | V | $5.81 \pm 0.05$ | Zdanavicius et al. (1972) |
| HD193579 | Johnson | V | $5.80 \pm 0.05$ | Haggkvist & Oja (1970) |
| HD193579 | Johnson | V | $5.88 \pm 0.03$ | This work |
| HD193579 | Vilnius | S | $4.77 \pm 0.05$ | Zdanavicius et al. (1972) |
| HD193579 | WBVR | R | $4.66 \pm 0.05$ | Kornilov et al. (1991) |
| HD193579 | KronComet | Rc | $4.50 \pm 0.01$ | This work |
| HD193579 | Johnson | K | $2.20 \pm 0.05$ | Neugebauer & Leighton (1969) |
| HD194097 | Vilnius | X | $8.86 \pm 0.05$ | Bartkevicius et al. (1973) |
| HD194097 | KronComet | COp | $8.00 \pm 0.01$ | This work |
| HD194097 | WBVR | B | $7.50 \pm 0.05$ | Kornilov et al. (1991) |
| HD194097 | Johnson | B | $7.35 \pm 0.02$ | This work |
| HD194097 | Johnson | B | $7.44 \pm 0.05$ | Haggkvist & Oja (1970) |

**Table 21** *(continued)*

| Star ID | System/Wvlen | Band/Bandpass | Value | Reference |
|---------|--------------|---------------|-------|-----------|
| HD194097 | KronComet | Bc | $7.26 \pm 0.01$ | This work |
| HD194097 | Vilnius | Y | $7.06 \pm 0.05$ | Bartkevicius et al. (1973) |
| HD194097 | KronComet | C2 | $6.43 \pm 0.01$ | This work |
| HD194097 | Vilnius | Z | $6.53 \pm 0.05$ | Bartkevicius et al. (1973) |
| HD194097 | KronComet | Gc | $6.22 \pm 0.01$ | This work |
| HD194097 | WBVR | V | $6.09 \pm 0.05$ | Kornilov et al. (1991) |
| HD194097 | Vilnius | V | $6.08 \pm 0.05$ | Bartkevicius et al. (1973) |
| HD194097 | Johnson | V | $6.08 \pm 0.01$ | This work |
| HD194097 | Johnson | V | $6.09 \pm 0.05$ | Haggkvist & Oja (1970) |
| HD194097 | Vilnius | S | $5.15 \pm 0.05$ | Bartkevicius et al. (1973) |
| HD194097 | WBVR | R | $5.09 \pm 0.05$ | Kornilov et al. (1991) |
| HD194097 | KronComet | Rc | $4.84 \pm 0.01$ | This work |
| HD194097 | Johnson | K | $2.76 \pm 0.08$ | Neugebauer & Leighton (1969) |
| HD194193 | KronComet | NH | $10.77 \pm 0.04$ | This work |
| HD194193 | KronComet | UVc | $10.37 \pm 0.04$ | This work |
| HD194193 | Vilnius | U | $11.41 \pm 0.05$ | Straizys et al. (1993) |
| HD194193 | Vilnius | U | $11.49 \pm 0.05$ | Straizys et al. (1989a) |
| HD194193 | WBVR | W | $9.39 \pm 0.05$ | Kornilov et al. (1991) |
| HD194193 | Johnson | U | $8.90 \pm 0.02$ | This work |
| HD194193 | Johnson | U | $9.51 \pm 0.05$ | Ducati (2002) |
| HD194193 | Johnson | U | $9.53 \pm 0.05$ | Mermilliod (1986) |
| HD194193 | Vilnius | P | $10.57 \pm 0.05$ | Straizys et al. (1993) |
| HD194193 | Vilnius | P | $10.63 \pm 0.05$ | Straizys et al. (1989a) |
| HD194193 | KronComet | CN | $9.75 \pm 0.11$ | This work |
| HD194193 | Vilnius | X | $9.20 \pm 0.05$ | Straizys et al. (1993) |
| HD194193 | Vilnius | X | $9.25 \pm 0.05$ | Straizys et al. (1989a) |
| HD194193 | Oja | m41 | $9.13 \pm 0.05$ | Häggkvist & Oja (1970) |
| HD194193 | Oja | m42 | $8.90 \pm 0.05$ | Häggkvist & Oja (1970) |
| HD194193 | KronComet | COp | $8.08 \pm 0.09$ | This work |
| HD194193 | WBVR | B | $7.59 \pm 0.05$ | Kornilov et al. (1991) |
| HD194193 | Johnson | B | $7.35 \pm 0.05$ | This work |
| HD194193 | Johnson | B | $7.53 \pm 0.05$ | Haggkvist & Oja (1970) |
| HD194193 | Johnson | B | $7.53 \pm 0.05$ | Ducati (2002) |
| HD194193 | Johnson | B | $7.55 \pm 0.05$ | Mermilliod (1986) |
| HD194193 | KronComet | Bc | $7.34 \pm 0.02$ | This work |
| HD194193 | Oja | m45 | $7.20 \pm 0.05$ | Häggkvist & Oja (1970) |
| HD194193 | Vilnius | Y | $7.10 \pm 0.05$ | Straizys et al. (1993) |
| HD194193 | Vilnius | Y | $7.12 \pm 0.05$ | Straizys et al. (1989a) |
| HD194193 | KronComet | C2 | $6.36 \pm 0.01$ | This work |
| HD194193 | Vilnius | Z | $6.48 \pm 0.05$ | Straizys et al. (1993) |
| HD194193 | Vilnius | Z | $6.50 \pm 0.05$ | Straizys et al. (1989a) |

**Table 21** (continued)

| Star ID | System/Wvlen | Band/Bandpass | Value | Reference |
|---------|-------------|---------------|-------|-----------|
| HD194193 | KronComet | Gc | $6.09 \pm 0.01$ | This work |
| HD194193 | WBVR | V | $5.94 \pm 0.05$ | Kornilov et al. (1991) |
| HD194193 | Vilnius | V | $5.94 \pm 0.05$ | Straizys et al. (1993) |
| HD194193 | Vilnius | V | $5.95 \pm 0.05$ | Straizys et al. (1989a) |
| HD194193 | Johnson | V | $5.92 \pm 0.05$ | Mermilliod (1986) |
| HD194193 | Johnson | V | $5.93 \pm 0.05$ | Haggkvist & Oja (1970) |
| HD194193 | Johnson | V | $5.93 \pm 0.05$ | Ducati (2002) |
| HD194193 | Johnson | V | $5.99 \pm 0.03$ | This work |
| HD194193 | Vilnius | S | $4.83 \pm 0.05$ | Straizys et al. (1993) |
| HD194193 | Vilnius | S | $4.84 \pm 0.05$ | Straizys et al. (1989a) |
| HD194193 | WBVR | R | $4.66 \pm 0.05$ | Kornilov et al. (1991) |
| HD194193 | KronComet | Rc | $4.54 \pm 0.01$ | This work |
| HD194193 | Johnson | J | $2.95 \pm 0.05$ | McWilliam & Lambert (1984) |
| HD194193 | Johnson | J | $2.95 \pm 0.05$ | Ducati (2002) |
| HD194193 | Johnson | J | $3.12 \pm 0.05$ | Bergner et al. (1995) |
| HD194193 | Johnson | H | $2.00 \pm 0.05$ | Ito et al. (1995) |
| HD194193 | Johnson | H | $2.28 \pm 0.05$ | Bergner et al. (1995) |
| HD194193 | Johnson | K | $1.93 \pm 0.05$ | Ducati (2002) |
| HD194193 | Johnson | K | $1.93 \pm 0.06$ | Neugebauer & Leighton (1969) |
| HD194317 | Oja | m41 | $7.21 \pm 0.05$ | Häggkvist & Oja (1970) |
| HD194317 | DDO | m42 | $7.40 \pm 0.05$ | McClure & Forrester (1981) |
| HD194317 | Oja | m42 | $6.94 \pm 0.05$ | Häggkvist & Oja (1970) |
| HD194317 | WBVR | B | $5.82 \pm 0.05$ | Kornilov et al. (1991) |
| HD194317 | Johnson | B | $5.71 \pm 0.05$ | Roman (1955) |
| HD194317 | Johnson | B | $5.75 \pm 0.05$ | Häggkvist & Oja (1966) |
| HD194317 | Johnson | B | $5.77 \pm 0.05$ | Johnson et al. (1966) |
| HD194317 | Johnson | B | $5.77 \pm 0.05$ | Ducati (2002) |
| HD194317 | Johnson | B | $5.78 \pm 0.05$ | Argue (1963) |
| HD194317 | 13c | m45 | $5.38 \pm 0.05$ | Johnson & Mitchell (1995) |
| HD194317 | DDO | m45 | $6.24 \pm 0.05$ | McClure & Forrester (1981) |
| HD194317 | Oja | m45 | $5.42 \pm 0.05$ | Häggkvist & Oja (1970) |
| HD194317 | DDO | m48 | $4.93 \pm 0.05$ | McClure & Forrester (1981) |
| HD194317 | 13c | m52 | $4.79 \pm 0.05$ | Johnson & Mitchell (1995) |
| HD194317 | WBVR | V | $4.44 \pm 0.05$ | Kornilov et al. (1991) |
| HD194317 | Johnson | V | $4.40 \pm 0.05$ | Roman (1955) |
| HD194317 | Johnson | V | $4.42 \pm 0.05$ | Häggkvist & Oja (1966) |
| HD194317 | Johnson | V | $4.43 \pm 0.05$ | Argue (1963) |
| HD194317 | Johnson | V | $4.44 \pm 0.05$ | Johnson et al. (1966) |
| HD194317 | Johnson | V | $4.44 \pm 0.05$ | Ducati (2002) |
| HD194317 | 13c | m58 | $4.10 \pm 0.05$ | Johnson & Mitchell (1995) |
| HD194317 | 13c | m63 | $3.72 \pm 0.05$ | Johnson & Mitchell (1995) |

**Table 21** *(continued)*

| Star ID | System/Wvlen | Band/Bandpass | Value | Reference |
|---------|--------------|---------------|-------|-----------|
| HD194317 | WBVR | R | $3.45 \pm 0.05$ | Kornilov et al. (1991) |
| HD194317 | 13c | m72 | $3.37 \pm 0.05$ | Johnson & Mitchell (1995) |
| HD194317 | 13c | m80 | $3.07 \pm 0.05$ | Johnson & Mitchell (1995) |
| HD194317 | 13c | m86 | $2.94 \pm 0.05$ | Johnson & Mitchell (1995) |
| HD194317 | 13c | m99 | $2.69 \pm 0.05$ | Johnson & Mitchell (1995) |
| HD194317 | 13c | m110 | $2.50 \pm 0.05$ | Johnson & Mitchell (1995) |
| HD194317 | Johnson | J | $2.15 \pm 0.05$ | McWilliam & Lambert (1984) |
| HD194317 | Johnson | J | $2.22 \pm 0.05$ | Ducati (2002) |
| HD194317 | Johnson | J | $2.29 \pm 0.05$ | Johnson et al. (1966) |
| HD194317 | Johnson | K | $1.38 \pm 0.05$ | Ducati (2002) |
| HD194317 | Johnson | K | $1.38 \pm 0.06$ | Neugebauer & Leighton (1969) |
| HD194317 | Johnson | K | $1.44 \pm 0.05$ | Johnson et al. (1966) |
| HD194317 | Johnson | L | $1.32 \pm 0.05$ | Johnson et al. (1966) |
| HD194317 | Johnson | L | $1.32 \pm 0.05$ | Ducati (2002) |
| HD194526 | DDO | m35 | $11.23 \pm 0.05$ | McClure & Forrester (1981) |
| HD194526 | WBVR | W | $9.82 \pm 0.05$ | Kornilov et al. (1991) |
| HD194526 | Johnson | U | $9.79 \pm 0.05$ | Cousins (1964b) |
| HD194526 | Johnson | U | $9.80 \pm 0.05$ | Johnson et al. (1966) |
| HD194526 | Johnson | U | $9.80 \pm 0.05$ | Argue (1966) |
| HD194526 | Johnson | U | $9.83 \pm 0.05$ | Mermilliod (1986) |
| HD194526 | DDO | m38 | $9.91 \pm 0.05$ | McClure & Forrester (1981) |
| HD194526 | DDO | m41 | $9.92 \pm 0.05$ | McClure & Forrester (1981) |
| HD194526 | DDO | m42 | $9.70 \pm 0.05$ | McClure & Forrester (1981) |
| HD194526 | WBVR | B | $7.96 \pm 0.05$ | Kornilov et al. (1991) |
| HD194526 | Johnson | B | $7.87 \pm 0.05$ | Argue (1966) |
| HD194526 | Johnson | B | $7.88 \pm 0.05$ | Cousins (1964b) |
| HD194526 | Johnson | B | $7.89 \pm 0.05$ | Johnson et al. (1966) |
| HD194526 | Johnson | B | $7.89 \pm 0.05$ | Mermilliod (1986) |
| HD194526 | DDO | m45 | $8.36 \pm 0.05$ | McClure & Forrester (1981) |
| HD194526 | DDO | m48 | $6.96 \pm 0.05$ | McClure & Forrester (1981) |
| HD194526 | WBVR | V | $6.33 \pm 0.05$ | Kornilov et al. (1991) |
| HD194526 | Johnson | V | $6.31 \pm 0.05$ | Argue (1966) |
| HD194526 | Johnson | V | $6.32 \pm 0.05$ | Cousins (1964b) |
| HD194526 | Johnson | V | $6.32 \pm 0.05$ | Mermilliod (1986) |
| HD194526 | Johnson | V | $6.33 \pm 0.05$ | Johnson et al. (1966) |
| HD194526 | WBVR | R | $5.11 \pm 0.05$ | Kornilov et al. (1991) |
| HD194526 | 1250 | 310 | $60.70 \pm 15.60$ | Smith et al. (2004) |
| HD194526 | 2200 | 361 | $59.10 \pm 11.10$ | Smith et al. (2004) |
| HD194526 | Johnson | K | $2.46 \pm 0.06$ | Neugebauer & Leighton (1969) |
| HD194526 | 3500 | 898 | $25.30 \pm 10.30$ | Smith et al. (2004) |
| HD194526 | 4900 | 712 | $9.80 \pm 5.50$ | Smith et al. (2004) |

**Table 21** *(continued)*

| Star ID | System/Wvlen | Band/Bandpass | Value | Reference |
|---------|--------------|---------------|-------|-----------|
| HD194526 | 12000 | 6384 | $-7.50 \pm 21.20$ | Smith et al. (2004) |
| HD196036 | KronComet | NH | $12.93 \pm 0.14$ | This work |
| HD196036 | KronComet | UVc | $12.21 \pm 0.04$ | This work |
| HD196036 | Johnson | U | $10.50 \pm 0.03$ | This work |
| HD196036 | KronComet | CN | $11.05 \pm 0.04$ | This work |
| HD196036 | KronComet | COp | $9.63 \pm 0.01$ | This work |
| HD196036 | Johnson | B | $8.99 \pm 0.02$ | This work |
| HD196036 | KronComet | Bc | $8.96 \pm 0.01$ | This work |
| HD196036 | KronComet | C2 | $7.63 \pm 0.01$ | This work |
| HD196036 | KronComet | Gc | $7.57 \pm 0.01$ | This work |
| HD196036 | Johnson | V | $7.47 \pm 0.01$ | This work |
| HD196036 | KronComet | Rc | $5.95 \pm 0.01$ | This work |
| HD196036 | Johnson | K | $1.98 \pm 0.05$ | Neugebauer & Leighton (1969) |
| HD196360 | Vilnius | U | $10.08 \pm 0.05$ | Straizys et al. (1993) |
| HD196360 | WBVR | W | $8.16 \pm 0.05$ | Kornilov et al. (1991) |
| HD196360 | Johnson | U | $8.26 \pm 0.02$ | This work |
| HD196360 | Johnson | U | $8.32 \pm 0.05$ | Appenzeller (1966) |
| HD196360 | Vilnius | P | $9.49 \pm 0.05$ | Straizys et al. (1993) |
| HD196360 | KronComet | CN | $9.15 \pm 0.02$ | This work |
| HD196360 | Vilnius | X | $8.58 \pm 0.05$ | Straizys et al. (1993) |
| HD196360 | Oja | m41 | $8.69 \pm 0.05$ | Häggkvist & Oja (1970) |
| HD196360 | Oja | m42 | $8.52 \pm 0.05$ | Häggkvist & Oja (1970) |
| HD196360 | KronComet | COp | $7.85 \pm 0.01$ | This work |
| HD196360 | WBVR | B | $7.62 \pm 0.05$ | Kornilov et al. (1991) |
| HD196360 | Johnson | B | $7.54 \pm 0.02$ | This work |
| HD196360 | Johnson | B | $7.58 \pm 0.05$ | Appenzeller (1966) |
| HD196360 | KronComet | Bc | $7.42 \pm 0.01$ | This work |
| HD196360 | Oja | m45 | $7.33 \pm 0.05$ | Häggkvist & Oja (1970) |
| HD196360 | Vilnius | Y | $7.38 \pm 0.05$ | Straizys et al. (1993) |
| HD196360 | KronComet | C2 | $6.77 \pm 0.01$ | This work |
| HD196360 | Vilnius | Z | $6.93 \pm 0.05$ | Straizys et al. (1993) |
| HD196360 | KronComet | Gc | $6.69 \pm 0.01$ | This work |
| HD196360 | WBVR | V | $6.65 \pm 0.05$ | Kornilov et al. (1991) |
| HD196360 | Vilnius | V | $6.64 \pm 0.05$ | Straizys et al. (1993) |
| HD196360 | Johnson | V | $6.64 \pm 0.01$ | This work |
| HD196360 | Johnson | V | $6.65 \pm 0.05$ | Appenzeller (1966) |
| HD196360 | Vilnius | S | $5.93 \pm 0.05$ | Straizys et al. (1993) |
| HD196360 | WBVR | R | $5.93 \pm 0.05$ | Kornilov et al. (1991) |
| HD196360 | KronComet | Rc | $5.67 \pm 0.01$ | This work |
| HD197912 | 13c | m33 | $6.19 \pm 0.05$ | Johnson & Mitchell (1995) |
| HD197912 | Geneva | U | $6.64 \pm 0.08$ | Golay (1972) |

**Table 21** *(continued)*

| Star ID | System/Wvlen | Band/Bandpass | Value | Reference |
|---------|--------------|---------------|-------|-----------|
| HD197912 | 13c | m35 | $5.98 \pm 0.05$ | Johnson & Mitchell (1995) |
| HD197912 | Johnson | U | $6.13 \pm 0.05$ | Johnson (1964) |
| HD197912 | Johnson | U | $6.14 \pm 0.05$ | Argue (1966) |
| HD197912 | Johnson | U | $6.15 \pm 0.05$ | Argue (1963) |
| HD197912 | Johnson | U | $6.15 \pm 0.05$ | Mermilliod (1986) |
| HD197912 | Johnson | U | $6.16 \pm 0.05$ | Nicolet (1978) |
| HD197912 | Johnson | U | $6.16 \pm 0.05$ | Ducati (2002) |
| HD197912 | Johnson | U | $6.17 \pm 0.05$ | Johnson et al. (1966) |
| HD197912 | Geneva | B1 | $5.88 \pm 0.08$ | Golay (1972) |
| HD197912 | Oja | m41 | $6.42 \pm 0.05$ | Häggkvist & Oja (1970) |
| HD197912 | Oja | m42 | $6.22 \pm 0.05$ | Häggkvist & Oja (1970) |
| HD197912 | Geneva | B | $4.56 \pm 0.08$ | Golay (1972) |
| HD197912 | WBVR | B | $5.28 \pm 0.05$ | Kornilov et al. (1991) |
| HD197912 | Johnson | B | $5.24 \pm 0.05$ | Argue (1966) |
| HD197912 | Johnson | B | $5.25 \pm 0.05$ | Häggkvist & Oja (1966) |
| HD197912 | Johnson | B | $5.26 \pm 0.05$ | Argue (1963) |
| HD197912 | Johnson | B | $5.27 \pm 0.05$ | Johnson (1964) |
| HD197912 | Johnson | B | $5.27 \pm 0.05$ | Nicolet (1978) |
| HD197912 | Johnson | B | $5.27 \pm 0.05$ | Mermilliod (1986) |
| HD197912 | Johnson | B | $5.29 \pm 0.05$ | Johnson et al. (1966) |
| HD197912 | Johnson | B | $5.29 \pm 0.05$ | Ducati (2002) |
| HD197912 | Geneva | B2 | $5.71 \pm 0.08$ | Golay (1972) |
| HD197912 | 13c | m45 | $4.92 \pm 0.05$ | Johnson & Mitchell (1995) |
| HD197912 | Oja | m45 | $4.96 \pm 0.05$ | Häggkvist & Oja (1970) |
| HD197912 | 13c | m52 | $4.45 \pm 0.05$ | Johnson & Mitchell (1995) |
| HD197912 | Geneva | V1 | $5.01 \pm 0.08$ | Golay (1972) |
| HD197912 | WBVR | V | $4.22 \pm 0.05$ | Kornilov et al. (1991) |
| HD197912 | Geneva | V | $4.22 \pm 0.08$ | Golay (1972) |
| HD197912 | Johnson | V | $4.20 \pm 0.05$ | Argue (1963) |
| HD197912 | Johnson | V | $4.20 \pm 0.05$ | Häggkvist & Oja (1966) |
| HD197912 | Johnson | V | $4.20 \pm 0.05$ | Argue (1966) |
| HD197912 | Johnson | V | $4.22 \pm 0.05$ | Johnson (1964) |
| HD197912 | Johnson | V | $4.22 \pm 0.05$ | Nicolet (1978) |
| HD197912 | Johnson | V | $4.22 \pm 0.05$ | Mermilliod (1986) |
| HD197912 | Johnson | V | $4.23 \pm 0.05$ | Johnson et al. (1966) |
| HD197912 | Johnson | V | $4.23 \pm 0.05$ | Ducati (2002) |
| HD197912 | 13c | m58 | $3.97 \pm 0.05$ | Johnson & Mitchell (1995) |
| HD197912 | Geneva | G | $5.19 \pm 0.08$ | Golay (1972) |
| HD197912 | 13c | m63 | $3.66 \pm 0.05$ | Johnson & Mitchell (1995) |
| HD197912 | WBVR | R | $3.46 \pm 0.05$ | Kornilov et al. (1991) |
| HD197912 | 13c | m72 | $3.39 \pm 0.05$ | Johnson & Mitchell (1995) |

**Table 21** *(continued)*

| Star ID | System/Wvlen | Band/Bandpass | Value | Reference |
|---------|--------------|---------------|-------|-----------|
| HD197912 | 13c | m80 | $3.17 \pm 0.05$ | Johnson & Mitchell (1995) |
| HD197912 | 13c | m86 | $3.07 \pm 0.05$ | Johnson & Mitchell (1995) |
| HD197912 | 13c | m99 | $2.89 \pm 0.05$ | Johnson & Mitchell (1995) |
| HD197912 | 13c | m110 | $2.74 \pm 0.05$ | Johnson & Mitchell (1995) |
| HD197912 | Johnson | J | $2.56 \pm 0.05$ | Johnson et al. (1966) |
| HD197912 | Johnson | J | $2.56 \pm 0.05$ | Ducati (2002) |
| HD197912 | 2200 | 361 | $125.80 \pm 27.70$ | Smith et al. (2004) |
| HD197912 | Johnson | K | $1.78 \pm 0.04$ | Neugebauer & Leighton (1969) |
| HD197912 | Johnson | K | $1.87 \pm 0.05$ | Johnson et al. (1966) |
| HD197912 | Johnson | K | $1.87 \pm 0.05$ | Ducati (2002) |
| HD197912 | Johnson | L | $1.67 \pm 0.05$ | Johnson et al. (1966) |
| HD197912 | Johnson | L | $1.67 \pm 0.05$ | Ducati (2002) |
| HD197912 | 3500 | 898 | $50.50 \pm 12.80$ | Smith et al. (2004) |
| HD197912 | 4900 | 712 | $25.00 \pm 6.30$ | Smith et al. (2004) |
| HD197912 | 12000 | 6384 | $15.70 \pm 21.80$ | Smith et al. (2004) |
| HD198237 | KronComet | NH | $11.68 \pm 0.32$ | This work |
| HD198237 | KronComet | UVc | $10.98 \pm 0.05$ | This work |
| HD198237 | Vilnius | U | $11.92 \pm 0.05$ | Straizys et al. (1989b) |
| HD198237 | WBVR | W | $9.94 \pm 0.05$ | Kornilov et al. (1991) |
| HD198237 | Johnson | U | $10.03 \pm 0.05$ | Mermilliod (1986) |
| HD198237 | Johnson | U | $9.98 \pm 0.05$ | Guetter & Hewitt (1984) |
| HD198237 | Vilnius | P | $11.05 \pm 0.05$ | Straizys et al. (1989b) |
| HD198237 | KronComet | CN | $10.39 \pm 0.07$ | This work |
| HD198237 | Vilnius | X | $9.64 \pm 0.05$ | Straizys et al. (1989b) |
| HD198237 | KronComet | COp | $8.72 \pm 0.04$ | This work |
| HD198237 | WBVR | B | $8.08 \pm 0.05$ | Kornilov et al. (1991) |
| HD198237 | Johnson | B | $8.01 \pm 0.05$ | Neckel (1974) |
| HD198237 | Johnson | B | $8.01 \pm 0.05$ | Guetter & Hewitt (1984) |
| HD198237 | Johnson | B | $8.01 \pm 0.05$ | Mermilliod (1986) |
| HD198237 | Johnson | B | $8.26 \pm 0.37$ | This work |
| HD198237 | KronComet | Bc | $7.84 \pm 0.05$ | This work |
| HD198237 | Vilnius | Y | $7.56 \pm 0.05$ | Straizys et al. (1989b) |
| HD198237 | KronComet | C2 | $6.87 \pm 0.07$ | This work |
| HD198237 | Vilnius | Z | $6.94 \pm 0.05$ | Straizys et al. (1989b) |
| HD198237 | KronComet | Gc | $6.82 \pm 0.31$ | This work |
| HD198237 | WBVR | V | $6.44 \pm 0.05$ | Kornilov et al. (1991) |
| HD198237 | Vilnius | V | $6.40 \pm 0.05$ | Straizys et al. (1989b) |
| HD198237 | Johnson | V | $6.40 \pm 0.05$ | Neckel (1974) |
| HD198237 | Johnson | V | $6.40 \pm 0.05$ | Mermilliod (1986) |
| HD198237 | Johnson | V | $6.41 \pm 0.05$ | Guetter & Hewitt (1984) |
| HD198237 | Johnson | V | $6.71 \pm 0.34$ | This work |

**Table 21** *(continued)*

| Star ID | System/Wvlen | Band/Bandpass | Value | Reference |
|---------|--------------|---------------|-------|-----------|
| HD198237 | Vilnius | S | $5.29 \pm 0.05$ | Straizys et al. (1989b) |
| HD198237 | WBVR | R | $5.11 \pm 0.05$ | Kornilov et al. (1991) |
| HD198237 | Johnson | K | $2.32 \pm 0.06$ | Neugebauer & Leighton (1969) |
| HD199101 | Vilnius | U | $10.84 \pm 0.05$ | Zdanavicius et al. (1972) |
| HD199101 | WBVR | W | $8.82 \pm 0.05$ | Kornilov et al. (1991) |
| HD199101 | Johnson | U | $8.86 \pm 0.05$ | Guetter & Hewitt (1984) |
| HD199101 | Johnson | U | $8.86 \pm 0.05$ | Ducati (2002) |
| HD199101 | Vilnius | P | $10.04 \pm 0.05$ | Zdanavicius et al. (1972) |
| HD199101 | Vilnius | X | $8.65 \pm 0.05$ | Zdanavicius et al. (1972) |
| HD199101 | Oja | m41 | $8.52 \pm 0.05$ | Häggkvist & Oja (1970) |
| HD199101 | Oja | m42 | $8.29 \pm 0.05$ | Häggkvist & Oja (1970) |
| HD199101 | WBVR | B | $7.07 \pm 0.05$ | Kornilov et al. (1991) |
| HD199101 | Johnson | B | $6.99 \pm 0.05$ | Haggkvist & Oja (1970) |
| HD199101 | Johnson | B | $6.99 \pm 0.05$ | Guetter & Hewitt (1984) |
| HD199101 | Johnson | B | $6.99 \pm 0.05$ | Ducati (2002) |
| HD199101 | Oja | m45 | $6.64 \pm 0.05$ | Häggkvist & Oja (1970) |
| HD199101 | Vilnius | Y | $6.59 \pm 0.05$ | Zdanavicius et al. (1972) |
| HD199101 | Vilnius | Z | $6.03 \pm 0.05$ | Zdanavicius et al. (1972) |
| HD199101 | WBVR | V | $5.49 \pm 0.05$ | Kornilov et al. (1991) |
| HD199101 | Vilnius | V | $5.49 \pm 0.05$ | Zdanavicius et al. (1972) |
| HD199101 | Johnson | V | $5.46 \pm 0.05$ | Guetter & Hewitt (1984) |
| HD199101 | Johnson | V | $5.47 \pm 0.05$ | Haggkvist & Oja (1970) |
| HD199101 | Johnson | V | $5.47 \pm 0.05$ | Ducati (2002) |
| HD199101 | Vilnius | S | $4.44 \pm 0.05$ | Zdanavicius et al. (1972) |
| HD199101 | WBVR | R | $4.32 \pm 0.05$ | Kornilov et al. (1991) |
| HD199101 | Johnson | J | $2.72 \pm 0.05$ | McWilliam & Lambert (1984) |
| HD199101 | Johnson | J | $2.72 \pm 0.05$ | Alonso et al. (1998) |
| HD199101 | Johnson | J | $2.72 \pm 0.05$ | Ducati (2002) |
| HD199101 | Johnson | H | $1.97 \pm 0.05$ | Alonso et al. (1998) |
| HD199101 | Johnson | K | $1.74 \pm 0.05$ | Ducati (2002) |
| HD199101 | Johnson | K | $1.82 \pm 0.06$ | Neugebauer & Leighton (1969) |
| HD199101 | 3500 | 898 | $77.70 \pm 22.30$ | Smith et al. (2004) |
| HD199101 | 12000 | 6384 | $0.00 \pm 19.60$ | Smith et al. (2004) |
| HD199697 | Vilnius | U | $10.31 \pm 0.05$ | Zdanavicius et al. (1972) |
| HD199697 | DDO | m35 | $9.78 \pm 0.05$ | McClure & Forrester (1981) |
| HD199697 | WBVR | W | $8.34 \pm 0.05$ | Kornilov et al. (1991) |
| HD199697 | Vilnius | P | $9.59 \pm 0.05$ | Zdanavicius et al. (1972) |
| HD199697 | DDO | m38 | $8.54 \pm 0.05$ | McClure & Forrester (1981) |
| HD199697 | Vilnius | X | $8.24 \pm 0.05$ | Zdanavicius et al. (1972) |
| HD199697 | DDO | m41 | $8.67 \pm 0.05$ | McClure & Forrester (1981) |
| HD199697 | Oja | m41 | $8.20 \pm 0.05$ | Häggkvist & Oja (1970) |

Table 21 *(continued)*

| Star ID | System/Wvlen | Band/Bandpass | Value | Reference |
|---------|--------------|---------------|-------|-----------|
| HD199697 | DDO | m42 | $8.40 \pm 0.05$ | McClure & Forrester (1981) |
| HD199697 | Oja | m42 | $7.92 \pm 0.05$ | Häggkvist & Oja (1970) |
| HD199697 | WBVR | B | $6.77 \pm 0.05$ | Kornilov et al. (1991) |
| HD199697 | Johnson | B | $6.71 \pm 0.05$ | Haggkvist & Oja (1970) |
| HD199697 | DDO | m45 | $7.18 \pm 0.05$ | McClure & Forrester (1981) |
| HD199697 | Oja | m45 | $6.37 \pm 0.05$ | Häggkvist & Oja (1970) |
| HD199697 | Vilnius | Y | $6.34 \pm 0.05$ | Zdanavicius et al. (1972) |
| HD199697 | DDO | m48 | $5.84 \pm 0.05$ | McClure & Forrester (1981) |
| HD199697 | Vilnius | Z | $5.80 \pm 0.05$ | Zdanavicius et al. (1972) |
| HD199697 | WBVR | V | $5.31 \pm 0.05$ | Kornilov et al. (1991) |
| HD199697 | Vilnius | V | $5.32 \pm 0.05$ | Zdanavicius et al. (1972) |
| HD199697 | Johnson | V | $5.31 \pm 0.05$ | Haggkvist & Oja (1970) |
| HD199697 | Vilnius | S | $4.36 \pm 0.05$ | Zdanavicius et al. (1972) |
| HD199697 | WBVR | R | $4.27 \pm 0.05$ | Kornilov et al. (1991) |
| HD199697 | Johnson | K | $2.06 \pm 0.05$ | Neugebauer & Leighton (1969) |
| HD199697 | 3500 | 898 | $66.20 \pm 10.90$ | Smith et al. (2004) |
| HD199697 | 4900 | 712 | $28.10 \pm 5.10$ | Smith et al. (2004) |
| HD199697 | 12000 | 6384 | $6.50 \pm 18.70$ | Smith et al. (2004) |
| HD199799 | KronComet | COp | $9.44 \pm 0.03$ | This work |
| HD199799 | WBVR | B | $9.30 \pm 0.05$ | Kornilov et al. (1991) |
| HD199799 | Johnson | B | $9.00 \pm 0.05$ | Landolt (1975) |
| HD199799 | Johnson | B | $9.02 \pm 0.02$ | This work |
| HD199799 | KronComet | Bc | $8.91 \pm 0.07$ | This work |
| HD199799 | KronComet | C2 | $7.42 \pm 0.07$ | This work |
| HD199799 | KronComet | Gc | $7.36 \pm 0.04$ | This work |
| HD199799 | WBVR | V | $7.55 \pm 0.05$ | Kornilov et al. (1991) |
| HD199799 | Johnson | V | $7.33 \pm 0.05$ | Landolt (1975) |
| HD199799 | Johnson | V | $7.41 \pm 0.03$ | This work |
| HD199799 | WBVR | R | $5.26 \pm 0.05$ | Kornilov et al. (1991) |
| HD199799 | KronComet | Rc | $5.55 \pm 0.03$ | This work |
| HD199799 | Johnson | K | $1.23 \pm 0.06$ | Neugebauer & Leighton (1969) |
| HD202109 | 13c | m33 | $4.96 \pm 0.05$ | Johnson & Mitchell (1995) |
| HD202109 | Geneva | U | $5.44 \pm 0.08$ | Golay (1972) |
| HD202109 | Vilnius | U | $6.73 \pm 0.05$ | Zdanavicius et al. (1969) |
| HD202109 | 13c | m35 | $4.82 \pm 0.05$ | Johnson & Mitchell (1995) |
| HD202109 | DDO | m35 | $6.24 \pm 0.05$ | McClure & Forrester (1981) |
| HD202109 | Stromgren | u | $6.15 \pm 0.08$ | Grønbech et al. (1976) |
| HD202109 | Stromgren | u | $6.15 \pm 0.08$ | Pilachowski (1978) |
| HD202109 | Stromgren | u | $6.15 \pm 0.08$ | Olsen (1983) |
| HD202109 | Stromgren | u | $6.17 \pm 0.08$ | Crawford & Barnes (1970) |
| HD202109 | Stromgren | u | $6.17 \pm 0.08$ | Hauck & Mermilliod (1998) |

**Table 21** *(continued)*

| Star ID | System/Wvlen | Band/Bandpass | Value | Reference |
|---------|-------------|---------------|-------|-----------|
| HD202109 | Stromgren | u | $6.18 \pm 0.08$ | Olsen (1993) |
| HD202109 | Stromgren | u | $6.19 \pm 0.08$ | Gray & Olsen (1991) |
| HD202109 | WBVR | W | $4.79 \pm 0.05$ | Kornilov et al. (1991) |
| HD202109 | Johnson | U | $4.93 \pm 0.05$ | Mermilliod (1986) |
| HD202109 | Johnson | U | $4.95 \pm 0.05$ | Kraft & Hiltner (1961) |
| HD202109 | Johnson | U | $4.95 \pm 0.05$ | Johnson (1964) |
| HD202109 | Johnson | U | $4.95 \pm 0.05$ | Johnson (1965b) |
| HD202109 | Johnson | U | $4.95 \pm 0.05$ | Johnson et al. (1966) |
| HD202109 | Johnson | U | $4.96 \pm 0.05$ | Argue (1966) |
| HD202109 | Johnson | U | $4.96 \pm 0.05$ | Ducati (2002) |
| HD202109 | Johnson | U | $5.01 \pm 0.05$ | Fernie (1983) |
| HD202109 | 13c | m37 | $4.89 \pm 0.05$ | Johnson & Mitchell (1995) |
| HD202109 | Vilnius | P | $6.15 \pm 0.05$ | Zdanavicius et al. (1969) |
| HD202109 | DDO | m38 | $5.21 \pm 0.05$ | McClure & Forrester (1981) |
| HD202109 | DDO | m38 | $5.21 \pm 0.05$ | Lu et al. (1983) |
| HD202109 | 13c | m40 | $4.78 \pm 0.05$ | Johnson & Mitchell (1995) |
| HD202109 | Geneva | B1 | $4.79 \pm 0.08$ | Golay (1972) |
| HD202109 | Vilnius | X | $5.23 \pm 0.05$ | Zdanavicius et al. (1969) |
| HD202109 | DDO | m41 | $5.86 \pm 0.05$ | McClure & Forrester (1981) |
| HD202109 | DDO | m41 | $5.86 \pm 0.05$ | Lu et al. (1983) |
| HD202109 | Oja | m41 | $5.44 \pm 0.05$ | Häggkvist & Oja (1970) |
| HD202109 | Stromgren | v | $4.83 \pm 0.08$ | Pilachowski (1978) |
| HD202109 | Stromgren | v | $4.84 \pm 0.08$ | Crawford & Barnes (1970) |
| HD202109 | Stromgren | v | $4.84 \pm 0.08$ | Grønbech et al. (1976) |
| HD202109 | Stromgren | v | $4.84 \pm 0.08$ | Olsen (1993) |
| HD202109 | Stromgren | v | $4.84 \pm 0.08$ | Hauck & Mermilliod (1998) |
| HD202109 | Stromgren | v | $4.85 \pm 0.08$ | Olsen (1983) |
| HD202109 | Stromgren | v | $4.86 \pm 0.08$ | Gray & Olsen (1991) |
| HD202109 | DDO | m42 | $5.57 \pm 0.05$ | McClure & Forrester (1981) |
| HD202109 | DDO | m42 | $5.57 \pm 0.05$ | Lu et al. (1983) |
| HD202109 | Oja | m42 | $5.13 \pm 0.05$ | Häggkvist & Oja (1970) |
| HD202109 | Geneva | B | $3.49 \pm 0.08$ | Golay (1972) |
| HD202109 | WBVR | B | $4.22 \pm 0.05$ | Kornilov et al. (1991) |
| HD202109 | Johnson | B | $4.17 \pm 0.05$ | Häggkvist & Oja (1966) |
| HD202109 | Johnson | B | $4.19 \pm 0.05$ | Johnson (1964) |
| HD202109 | Johnson | B | $4.19 \pm 0.05$ | Johnson (1965b) |
| HD202109 | Johnson | B | $4.19 \pm 0.05$ | Johnson et al. (1966) |
| HD202109 | Johnson | B | $4.20 \pm 0.05$ | Kraft & Hiltner (1961) |
| HD202109 | Johnson | B | $4.20 \pm 0.05$ | Argue (1966) |
| HD202109 | Johnson | B | $4.20 \pm 0.05$ | Ducati (2002) |
| HD202109 | Johnson | B | $4.21 \pm 0.05$ | Moffett & Barnes (1979) |

**Table 21** *(continued)*

| Star ID | System/Wvlen | Band/Bandpass | Value | Reference |
|---------|--------------|---------------|-------|-----------|
| HD202109 | Johnson | B | $4.21 \pm 0.05$ | Mermilliod (1986) |
| HD202109 | Johnson | B | $4.24 \pm 0.05$ | Fernie (1983) |
| HD202109 | Geneva | B2 | $4.66 \pm 0.08$ | Golay (1972) |
| HD202109 | 13c | m45 | $3.93 \pm 0.05$ | Johnson & Mitchell (1995) |
| HD202109 | DDO | m45 | $4.73 \pm 0.05$ | McClure & Forrester (1981) |
| HD202109 | DDO | m45 | $4.73 \pm 0.05$ | Lu et al. (1983) |
| HD202109 | Oja | m45 | $3.92 \pm 0.05$ | Häggkvist & Oja (1970) |
| HD202109 | Vilnius | Y | $3.96 \pm 0.05$ | Zdanavicius et al. (1969) |
| HD202109 | Stromgren | b | $3.79 \pm 0.08$ | Olsen (1983) |
| HD202109 | Stromgren | b | $3.80 \pm 0.08$ | Crawford & Barnes (1970) |
| HD202109 | Stromgren | b | $3.80 \pm 0.08$ | Grønbech et al. (1976) |
| HD202109 | Stromgren | b | $3.80 \pm 0.08$ | Pilachowski (1978) |
| HD202109 | Stromgren | b | $3.80 \pm 0.08$ | Gray & Olsen (1991) |
| HD202109 | Stromgren | b | $3.80 \pm 0.08$ | Olsen (1993) |
| HD202109 | Stromgren | b | $3.80 \pm 0.08$ | Hauck & Mermilliod (1998) |
| HD202109 | DDO | m48 | $3.54 \pm 0.05$ | McClure & Forrester (1981) |
| HD202109 | DDO | m48 | $3.54 \pm 0.05$ | Lu et al. (1983) |
| HD202109 | Vilnius | Z | $3.49 \pm 0.05$ | Zdanavicius et al. (1969) |
| HD202109 | 13c | m52 | $3.45 \pm 0.05$ | Johnson & Mitchell (1995) |
| HD202109 | Geneva | V1 | $3.99 \pm 0.08$ | Golay (1972) |
| HD202109 | WBVR | V | $3.21 \pm 0.05$ | Kornilov et al. (1991) |
| HD202109 | Vilnius | V | $3.20 \pm 0.05$ | Zdanavicius et al. (1969) |
| HD202109 | Stromgren | y | $3.21 \pm 0.08$ | Crawford & Barnes (1970) |
| HD202109 | Stromgren | y | $3.21 \pm 0.08$ | Grønbech et al. (1976) |
| HD202109 | Stromgren | y | $3.21 \pm 0.08$ | Pilachowski (1978) |
| HD202109 | Stromgren | y | $3.21 \pm 0.08$ | Olsen (1983) |
| HD202109 | Stromgren | y | $3.21 \pm 0.08$ | Gray & Olsen (1991) |
| HD202109 | Stromgren | y | $3.21 \pm 0.08$ | Olsen (1993) |
| HD202109 | Stromgren | y | $3.21 \pm 0.08$ | Hauck & Mermilliod (1998) |
| HD202109 | Geneva | V | $3.22 \pm 0.08$ | Golay (1972) |
| HD202109 | Johnson | V | $3.18 \pm 0.05$ | Häggkvist & Oja (1966) |
| HD202109 | Johnson | V | $3.19 \pm 0.05$ | Johnson (1964) |
| HD202109 | Johnson | V | $3.19 \pm 0.05$ | Argue (1966) |
| HD202109 | Johnson | V | $3.20 \pm 0.05$ | Kraft & Hiltner (1961) |
| HD202109 | Johnson | V | $3.20 \pm 0.05$ | Johnson (1965b) |
| HD202109 | Johnson | V | $3.20 \pm 0.05$ | Johnson et al. (1966) |
| HD202109 | Johnson | V | $3.21 \pm 0.05$ | Ducati (2002) |
| HD202109 | Johnson | V | $3.22 \pm 0.05$ | Moffett & Barnes (1979) |
| HD202109 | Johnson | V | $3.23 \pm 0.05$ | Mermilliod (1986) |
| HD202109 | Johnson | V | $3.26 \pm 0.05$ | Fernie (1983) |
| HD202109 | 13c | m58 | $3.01 \pm 0.05$ | Johnson & Mitchell (1995) |

**Table 21** *(continued)*

| Star ID | System/Wvlen | Band/Bandpass | Value | Reference |
|---------|--------------|---------------|-------|-----------|
| HD202109 | Geneva | G | $4.22 \pm 0.08$ | Golay (1972) |
| HD202109 | 13c | m63 | $2.74 \pm 0.05$ | Johnson & Mitchell (1995) |
| HD202109 | Vilnius | S | $2.53 \pm 0.05$ | Zdanavicius et al. (1969) |
| HD202109 | WBVR | R | $2.54 \pm 0.05$ | Kornilov et al. (1991) |
| HD202109 | 13c | m72 | $2.53 \pm 0.05$ | Johnson & Mitchell (1995) |
| HD202109 | 13c | m80 | $2.32 \pm 0.05$ | Johnson & Mitchell (1995) |
| HD202109 | 13c | m86 | $2.22 \pm 0.05$ | Johnson & Mitchell (1995) |
| HD202109 | 13c | m99 | $2.08 \pm 0.05$ | Johnson & Mitchell (1995) |
| HD202109 | 13c | m110 | $1.94 \pm 0.05$ | Johnson & Mitchell (1995) |
| HD202109 | 1250 | 310 | $374.10 \pm 16.60$ | Smith et al. (2004) |
| HD202109 | Johnson | J | $1.52 \pm 0.05$ | Noguchi et al. (1981) |
| HD202109 | Johnson | J | $1.60 \pm 0.05$ | Blackwell et al. (1979) |
| HD202109 | Johnson | J | $1.60 \pm 0.05$ | Ducati (2002) |
| HD202109 | Johnson | J | $1.65 \pm 0.05$ | Johnson et al. (1966) |
| HD202109 | Johnson | J | $1.65 \pm 0.05$ | Shenavrin et al. (2011) |
| HD202109 | Johnson | J | $1.66 \pm 0.05$ | Johnson (1965c) |
| HD202109 | Johnson | H | $1.15 \pm 0.05$ | Noguchi et al. (1981) |
| HD202109 | Johnson | H | $1.15 \pm 0.05$ | Ducati (2002) |
| HD202109 | Johnson | H | $1.16 \pm 0.05$ | Blackwell et al. (1979) |
| HD202109 | Johnson | H | $1.20 \pm 0.05$ | Shenavrin et al. (2011) |
| HD202109 | 2200 | 361 | $247.10 \pm 18.10$ | Smith et al. (2004) |
| HD202109 | Johnson | K | $1.07 \pm 0.03$ | Neugebauer & Leighton (1969) |
| HD202109 | Johnson | K | $1.08 \pm 0.05$ | Ducati (2002) |
| HD202109 | Johnson | K | $1.09 \pm 0.05$ | Johnson et al. (1966) |
| HD202109 | Johnson | K | $1.09 \pm 0.05$ | Shenavrin et al. (2011) |
| HD202109 | Johnson | L | $0.98 \pm 0.05$ | Johnson et al. (1966) |
| HD202109 | Johnson | L | $0.98 \pm 0.05$ | Ducati (2002) |
| HD202109 | 3500 | 898 | $117.20 \pm 9.00$ | Smith et al. (2004) |
| HD202109 | 4900 | 712 | $57.50 \pm 6.70$ | Smith et al. (2004) |
| HD202109 | Johnson | M | $1.15 \pm 0.05$ | Ducati (2002) |
| HD202109 | 12000 | 6384 | $19.50 \pm 16.70$ | Smith et al. (2004) |
| HD205435 | Geneva | B1 | $5.34 \pm 0.08$ | Golay (1972) |
| HD205435 | DDO | m41 | $6.36 \pm 0.05$ | McClure & Forrester (1981) |
| HD205435 | Oja | m41 | $5.91 \pm 0.05$ | Häggkvist & Oja (1970) |
| HD205435 | DDO | m42 | $6.23 \pm 0.05$ | McClure & Forrester (1981) |
| HD205435 | Oja | m42 | $5.80 \pm 0.05$ | Häggkvist & Oja (1970) |
| HD205435 | Geneva | B | $4.13 \pm 0.08$ | Golay (1972) |
| HD205435 | WBVR | B | $4.90 \pm 0.05$ | Kornilov et al. (1991) |
| HD205435 | Johnson | B | $4.88 \pm 0.05$ | Häggkvist & Oja (1966) |
| HD205435 | Johnson | B | $4.89 \pm 0.05$ | Jennens & Helfer (1975) |
| HD205435 | Johnson | B | $4.89 \pm 0.05$ | Mermilliod (1986) |

Table 21 *(continued)*

| Star ID | System/Wvlen | Band/Bandpass | Value | Reference |
|---------|--------------|---------------|-------|-----------|
| HD205435 | Johnson | B | $4.90 \pm 0.05$ | Johnson & Knuckles (1957) |
| HD205435 | Johnson | B | $4.90 \pm 0.05$ | Argue (1966) |
| HD205435 | Johnson | B | $4.91 \pm 0.05$ | Johnson et al. (1966) |
| HD205435 | Johnson | B | $4.91 \pm 0.05$ | Ducati (2002) |
| HD205435 | Johnson | B | $4.94 \pm 0.05$ | Johnson (1964) |
| HD205435 | Johnson | B | $4.97 \pm 0.05$ | Miczaika (1954) |
| HD205435 | Geneva | B2 | $5.34 \pm 0.08$ | Golay (1972) |
| HD205435 | 13c | m45 | $4.62 \pm 0.05$ | Johnson & Mitchell (1995) |
| HD205435 | DDO | m45 | $5.45 \pm 0.05$ | McClure & Forrester (1981) |
| HD205435 | Oja | m45 | $4.65 \pm 0.05$ | Häggkvist & Oja (1970) |
| HD205435 | Vilnius | Y | $4.71 \pm 0.05$ | Kazlauskas et al. (2005) |
| HD205435 | DDO | m48 | $4.31 \pm 0.05$ | McClure & Forrester (1981) |
| HD205435 | Vilnius | Z | $4.28 \pm 0.05$ | Kazlauskas et al. (2005) |
| HD205435 | 13c | m52 | $4.20 \pm 0.05$ | Johnson & Mitchell (1995) |
| HD205435 | Geneva | V1 | $4.76 \pm 0.08$ | Golay (1972) |
| HD205435 | WBVR | V | $4.00 \pm 0.05$ | Kornilov et al. (1991) |
| HD205435 | Vilnius | V | $4.01 \pm 0.05$ | Kazlauskas et al. (2005) |
| HD205435 | Geneva | V | $3.99 \pm 0.08$ | Golay (1972) |
| HD205435 | Johnson | V | $3.98 \pm 0.05$ | Häggkvist & Oja (1966) |
| HD205435 | Johnson | V | $4.00 \pm 0.05$ | Argue (1966) |
| HD205435 | Johnson | V | $4.00 \pm 0.05$ | Mermilliod (1986) |
| HD205435 | Johnson | V | $4.01 \pm 0.05$ | Jennens & Helfer (1975) |
| HD205435 | Johnson | V | $4.02 \pm 0.05$ | Johnson & Knuckles (1957) |
| HD205435 | Johnson | V | $4.02 \pm 0.05$ | Johnson et al. (1966) |
| HD205435 | Johnson | V | $4.02 \pm 0.05$ | Ducati (2002) |
| HD205435 | Johnson | V | $4.04 \pm 0.05$ | Johnson (1964) |
| HD205435 | Johnson | V | $4.09 \pm 0.05$ | Miczaika (1954) |
| HD205435 | 13c | m58 | $3.78 \pm 0.05$ | Johnson & Mitchell (1995) |
| HD205435 | Geneva | G | $5.00 \pm 0.08$ | Golay (1972) |
| HD205435 | 13c | m63 | $3.51 \pm 0.05$ | Johnson & Mitchell (1995) |
| HD205435 | Vilnius | S | $3.32 \pm 0.05$ | Kazlauskas et al. (2005) |
| HD205435 | WBVR | R | $3.33 \pm 0.05$ | Kornilov et al. (1991) |
| HD205435 | 13c | m72 | $3.20 \pm 0.05$ | Johnson & Mitchell (1995) |
| HD205435 | 13c | m80 | $2.99 \pm 0.05$ | Johnson & Mitchell (1995) |
| HD205435 | 13c | m86 | $2.91 \pm 0.05$ | Johnson & Mitchell (1995) |
| HD205435 | 13c | m99 | $2.78 \pm 0.05$ | Johnson & Mitchell (1995) |
| HD205435 | 13c | m110 | $2.60 \pm 0.05$ | Johnson & Mitchell (1995) |
| HD205435 | Johnson | J | $2.40 \pm 0.05$ | Noguchi et al. (1981) |
| HD205435 | Johnson | J | $2.47 \pm 0.05$ | Ducati (2002) |
| HD205435 | Johnson | J | $2.55 \pm 0.05$ | Johnson et al. (1966) |
| HD205435 | Johnson | J | $2.55 \pm 0.05$ | Shenavrin et al. (2011) |

**Table 21** *(continued)*

| Star ID | System/Wvlen | Band/Bandpass | Value | Reference |
|---------|--------------|---------------|-------|-----------|
| HD205435 | Johnson | H | $2.06 \pm 0.05$ | Noguchi et al. (1981) |
| HD205435 | Johnson | H | $2.06 \pm 0.05$ | Ducati (2002) |
| HD205435 | Johnson | H | $2.09 \pm 0.05$ | Shenavrin et al. (2011) |
| HD205435 | Johnson | K | $1.87 \pm 0.06$ | Neugebauer & Leighton (1969) |
| HD205435 | Johnson | K | $1.96 \pm 0.05$ | Ducati (2002) |
| HD205435 | Johnson | K | $1.97 \pm 0.05$ | Johnson et al. (1966) |
| HD205435 | Johnson | K | $1.97 \pm 0.05$ | Shenavrin et al. (2011) |
| HD205435 | Johnson | L | $1.88 \pm 0.05$ | Johnson et al. (1966) |
| HD205435 | Johnson | L | $1.88 \pm 0.05$ | Ducati (2002) |
| HD205435 | 3500 | 898 | $234.40 \pm 265.80$ | Smith et al. (2004) |
| HD205435 | 4900 | 712 | $42.50 \pm 12.00$ | Smith et al. (2004) |
| HD205435 | 12000 | 6384 | $82.00 \pm 81.60$ | Smith et al. (2004) |
| HD205733 | KronComet | COp | $10.83 \pm 0.08$ | This work |
| HD205733 | Johnson | B | $10.32 \pm 0.09$ | This work |
| HD205733 | KronComet | Bc | $10.20 \pm 0.09$ | This work |
| HD205733 | KronComet | C2 | $8.59 \pm 0.08$ | This work |
| HD205733 | KronComet | Gc | $8.56 \pm 0.09$ | This work |
| HD205733 | Johnson | V | $8.48 \pm 0.10$ | This work |
| HD205733 | KronComet | Rc | $6.75 \pm 0.10$ | This work |
| HD205733 | 1250 | 310 | $96.80 \pm 18.60$ | Smith et al. (2004) |
| HD205733 | 2200 | 361 | $125.70 \pm 40.10$ | Smith et al. (2004) |
| HD205733 | Johnson | K | $1.93 \pm 0.05$ | Neugebauer & Leighton (1969) |
| HD205733 | 3500 | 898 | $78.70 \pm 24.50$ | Smith et al. (2004) |
| HD205733 | 4900 | 712 | $35.80 \pm 19.50$ | Smith et al. (2004) |
| HD205733 | 12000 | 6384 | $29.20 \pm 22.00$ | Smith et al. (2004) |
| HD206445 | Vilnius | U | $10.66 \pm 0.05$ | Zdanavicius et al. (1972) |
| HD206445 | DDO | m35 | $10.19 \pm 0.05$ | McClure & Forrester (1981) |
| HD206445 | WBVR | W | $8.70 \pm 0.05$ | Kornilov et al. (1991) |
| HD206445 | Johnson | U | $8.73 \pm 0.05$ | Rybka (1969) |
| HD206445 | Johnson | U | $8.77 \pm 0.05$ | Cousins (1964b) |
| HD206445 | Johnson | U | $8.78 \pm 0.05$ | Johnson et al. (1966) |
| HD206445 | Vilnius | P | $9.97 \pm 0.05$ | Zdanavicius et al. (1972) |
| HD206445 | DDO | m38 | $8.94 \pm 0.05$ | McClure & Forrester (1981) |
| HD206445 | Vilnius | X | $8.66 \pm 0.05$ | Zdanavicius et al. (1972) |
| HD206445 | DDO | m41 | $9.10 \pm 0.05$ | McClure & Forrester (1981) |
| HD206445 | Oja | m41 | $8.62 \pm 0.05$ | Häggkvist & Oja (1970) |
| HD206445 | DDO | m42 | $8.80 \pm 0.05$ | McClure & Forrester (1981) |
| HD206445 | Oja | m42 | $8.26 \pm 0.05$ | Häggkvist & Oja (1970) |
| HD206445 | WBVR | B | $7.17 \pm 0.05$ | Kornilov et al. (1991) |
| HD206445 | Johnson | B | $7.08 \pm 0.05$ | Rybka (1969) |
| HD206445 | Johnson | B | $7.10 \pm 0.05$ | Cousins (1964b) |

**Table 21** (continued)

| Star ID | System/Wvlen | Band/Bandpass | Value | Reference |
|---------|--------------|---------------|-------|-----------|
| HD206445 | Johnson | B | $7.11 \pm 0.05$ | Johnson et al. (1966) |
| HD206445 | DDO | m45 | $7.61 \pm 0.05$ | McClure & Forrester (1981) |
| HD206445 | Oja | m45 | $6.75 \pm 0.05$ | Häggkvist & Oja (1970) |
| HD206445 | Vilnius | Y | $6.74 \pm 0.05$ | Zdanavicius et al. (1972) |
| HD206445 | DDO | m48 | $6.23 \pm 0.05$ | McClure & Forrester (1981) |
| HD206445 | Vilnius | Z | $6.15 \pm 0.05$ | Zdanavicius et al. (1972) |
| HD206445 | WBVR | V | $5.67 \pm 0.05$ | Kornilov et al. (1991) |
| HD206445 | Vilnius | V | $5.67 \pm 0.05$ | Zdanavicius et al. (1972) |
| HD206445 | Johnson | V | $5.64 \pm 0.05$ | Rybka (1969) |
| HD206445 | Johnson | V | $5.66 \pm 0.05$ | Cousins (1964b) |
| HD206445 | Johnson | V | $5.67 \pm 0.05$ | Johnson et al. (1966) |
| HD206445 | Vilnius | S | $4.71 \pm 0.05$ | Zdanavicius et al. (1972) |
| HD206445 | WBVR | R | $4.62 \pm 0.05$ | Kornilov et al. (1991) |
| HD206445 | 1250 | 310 | $81.30 \pm 6.00$ | Smith et al. (2004) |
| HD206445 | Johnson | J | $3.15 \pm 0.05$ | Alonso et al. (1998) |
| HD206445 | Johnson | H | $2.47 \pm 0.05$ | Alonso et al. (1998) |
| HD206445 | 2200 | 361 | $68.70 \pm 6.50$ | Smith et al. (2004) |
| HD206445 | Johnson | K | $2.31 \pm 0.07$ | Neugebauer & Leighton (1969) |
| HD206445 | 3500 | 898 | $32.40 \pm 6.00$ | Smith et al. (2004) |
| HD206445 | 4900 | 712 | $13.50 \pm 4.70$ | Smith et al. (2004) |
| HD206445 | 12000 | 6384 | $-0.50 \pm 22.10$ | Smith et al. (2004) |
| HD206749 | Oja | m41 | $8.73 \pm 0.05$ | Häggkvist & Oja (1970) |
| HD206749 | Oja | m42 | $8.48 \pm 0.05$ | Häggkvist & Oja (1970) |
| HD206749 | WBVR | B | $7.17 \pm 0.05$ | Kornilov et al. (1991) |
| HD206749 | Johnson | B | $7.08 \pm 0.05$ | Mermilliod (1986) |
| HD206749 | Johnson | B | $7.09 \pm 0.05$ | Haggkvist & Oja (1970) |
| HD206749 | Oja | m45 | $6.77 \pm 0.05$ | Häggkvist & Oja (1970) |
| HD206749 | WBVR | V | $5.51 \pm 0.05$ | Kornilov et al. (1991) |
| HD206749 | Johnson | V | $5.48 \pm 0.05$ | Mermilliod (1986) |
| HD206749 | Johnson | V | $5.49 \pm 0.05$ | Haggkvist & Oja (1970) |
| HD206749 | WBVR | R | $4.14 \pm 0.05$ | Kornilov et al. (1991) |
| HD206749 | 1250 | 310 | $192.80 \pm 10.30$ | Smith et al. (2004) |
| HD206749 | 2200 | 361 | $178.60 \pm 7.10$ | Smith et al. (2004) |
| HD206749 | Johnson | K | $1.31 \pm 0.05$ | Neugebauer & Leighton (1969) |
| HD206749 | 3500 | 898 | $89.10 \pm 5.90$ | Smith et al. (2004) |
| HD206749 | 4900 | 712 | $38.20 \pm 5.40$ | Smith et al. (2004) |
| HD206749 | 12000 | 6384 | $10.40 \pm 17.80$ | Smith et al. (2004) |
| HD207001 | KronComet | COp | $10.03 \pm 0.06$ | This work |
| HD207001 | Johnson | B | $9.30 \pm 0.02$ | This work |
| HD207001 | KronComet | Bc | $9.16 \pm 0.02$ | This work |
| HD207001 | KronComet | C2 | $8.02 \pm 0.02$ | This work |

**Table 21** *(continued)*

| Star ID | System/Wvlen | Band/Bandpass | Value | Reference |
|---------|--------------|---------------|-------|-----------|
| HD207001 | KronComet | Gc | $7.75 \pm 0.02$ | This work |
| HD207001 | Johnson | V | $7.61 \pm 0.04$ | This work |
| HD207001 | KronComet | Rc | $6.03 \pm 0.01$ | This work |
| HD207001 | Johnson | K | $2.92 \pm 0.10$ | Neugebauer & Leighton (1969) |
| HD207328 | KronComet | C2 | $7.81 \pm 0.01$ | This work |
| HD207328 | KronComet | Gc | $7.62 \pm 0.01$ | This work |
| HD207328 | Johnson | V | $7.36 \pm 0.05$ | Garrison & Kormendy (1976) |
| HD207328 | Johnson | V | $7.42 \pm 0.05$ | Neckel (1974) |
| HD207328 | Johnson | V | $7.42 \pm 0.05$ | Nicolet (1978) |
| HD207328 | Johnson | V | $7.50 \pm 0.04$ | This work |
| HD207328 | KronComet | Rc | $5.70 \pm 0.02$ | This work |
| HD207328 | 1250 | 310 | $124.70 \pm 16.20$ | Smith et al. (2004) |
| HD207328 | 2200 | 361 | $147.20 \pm 15.40$ | Smith et al. (2004) |
| HD207328 | Johnson | K | $1.79 \pm 0.05$ | Neugebauer & Leighton (1969) |
| HD207328 | 3500 | 898 | $89.00 \pm 10.40$ | Smith et al. (2004) |
| HD207328 | 4900 | 712 | $45.70 \pm 8.30$ | Smith et al. (2004) |
| HD207328 | 12000 | 6384 | $166.50 \pm 38.60$ | Smith et al. (2004) |
| HD209857 | KronComet | UVc | $10.89 \pm 0.04$ | This work |
| HD209857 | Johnson | U | $9.11 \pm 0.05$ | This work |
| HD209857 | Johnson | U | $9.48 \pm 0.05$ | Mermilliod (1986) |
| HD209857 | Johnson | U | $9.49 \pm 0.05$ | Johnson et al. (1966) |
| HD209857 | Johnson | U | $9.49 \pm 0.05$ | Mendoza (1967) |
| HD209857 | Johnson | U | $9.50 \pm 0.05$ | Fernie (1969) |
| HD209857 | Johnson | U | $9.50 \pm 0.05$ | Ducati (2002) |
| HD209857 | KronComet | CN | $9.90 \pm 0.02$ | This work |
| HD209857 | KronComet | COp | $8.39 \pm 0.06$ | This work |
| HD209857 | Johnson | B | $7.72 \pm 0.05$ | Johnson et al. (1966) |
| HD209857 | Johnson | B | $7.72 \pm 0.05$ | Mendoza (1967) |
| HD209857 | Johnson | B | $7.74 \pm 0.05$ | Ducati (2002) |
| HD209857 | Johnson | B | $7.75 \pm 0.05$ | Mermilliod (1986) |
| HD209857 | Johnson | B | $7.76 \pm 0.05$ | Fernie (1969) |
| HD209857 | Johnson | B | $7.82 \pm 0.05$ | Haggkvist & Oja (1970) |
| HD209857 | Johnson | B | $7.84 \pm 0.06$ | This work |
| HD209857 | KronComet | Bc | $7.70 \pm 0.03$ | This work |
| HD209857 | KronComet | C2 | $6.47 \pm 0.03$ | This work |
| HD209857 | KronComet | Gc | $6.38 \pm 0.03$ | This work |
| HD209857 | Johnson | V | $6.12 \pm 0.05$ | Johnson et al. (1966) |
| HD209857 | Johnson | V | $6.12 \pm 0.05$ | Mendoza (1967) |
| HD209857 | Johnson | V | $6.13 \pm 0.05$ | Fernie (1969) |
| HD209857 | Johnson | V | $6.13 \pm 0.05$ | Ducati (2002) |
| HD209857 | Johnson | V | $6.15 \pm 0.05$ | Mermilliod (1986) |

**Table 21** *(continued)*

| Star ID | System/Wvlen | Band/Bandpass | Value | Reference |
|---------|-------------|---------------|-------|-----------|
| HD209857 | Johnson | V | $6.18 \pm 0.05$ | Haggkvist & Oja (1970) |
| HD209857 | Johnson | V | $6.33 \pm 0.04$ | This work |
| HD209857 | KronComet | Rc | $4.83 \pm 0.03$ | This work |
| HD209857 | 1250 | 310 | $261.60 \pm 19.30$ | Smith et al. (2004) |
| HD209857 | Johnson | J | $2.03 \pm 0.05$ | Kerschbaum (1995) |
| HD209857 | Johnson | J | $2.11 \pm 0.05$ | Johnson et al. (1966) |
| HD209857 | Johnson | J | $2.11 \pm 0.05$ | Mendoza (1967) |
| HD209857 | Johnson | J | $2.11 \pm 0.05$ | Voelcker (1975) |
| HD209857 | Johnson | J | $2.11 \pm 0.05$ | Ducati (2002) |
| HD209857 | Johnson | H | $1.22 \pm 0.05$ | Kerschbaum (1995) |
| HD209857 | Johnson | H | $1.39 \pm 0.05$ | Voelcker (1975) |
| HD209857 | Johnson | H | $1.39 \pm 0.05$ | Ducati (2002) |
| HD209857 | 2200 | 361 | $280.80 \pm 18.80$ | Smith et al. (2004) |
| HD209857 | Johnson | K | $0.92 \pm 0.05$ | Neugebauer & Leighton (1969) |
| HD209857 | Johnson | K | $0.98 \pm 0.05$ | Johnson et al. (1966) |
| HD209857 | Johnson | K | $0.98 \pm 0.05$ | Mendoza (1967) |
| HD209857 | Johnson | K | $0.98 \pm 0.05$ | Ducati (2002) |
| HD209857 | Johnson | L | $0.77 \pm 0.05$ | Johnson et al. (1966) |
| HD209857 | Johnson | L | $0.77 \pm 0.05$ | Mendoza (1967) |
| HD209857 | Johnson | L | $0.77 \pm 0.05$ | Ducati (2002) |
| HD209857 | 3500 | 898 | $136.70 \pm 13.70$ | Smith et al. (2004) |
| HD209857 | 4900 | 712 | $59.20 \pm 6.10$ | Smith et al. (2004) |
| HD209857 | 12000 | 6384 | $9.50 \pm 17.00$ | Smith et al. (2004) |
| HD210514 | KronComet | COp | $9.42 \pm 0.02$ | This work |
| HD210514 | WBVR | B | $8.90 \pm 0.05$ | Kornilov et al. (1991) |
| HD210514 | Johnson | B | $8.88 \pm 0.09$ | This work |
| HD210514 | KronComet | Bc | $8.73 \pm 0.05$ | This work |
| HD210514 | KronComet | C2 | $7.48 \pm 0.04$ | This work |
| HD210514 | KronComet | Gc | $7.38 \pm 0.03$ | This work |
| HD210514 | WBVR | V | $7.23 \pm 0.05$ | Kornilov et al. (1991) |
| HD210514 | Johnson | V | $7.33 \pm 0.05$ | This work |
| HD210514 | WBVR | R | $5.44 \pm 0.05$ | Kornilov et al. (1991) |
| HD210514 | KronComet | Rc | $5.81 \pm 0.03$ | This work |
| HD210514 | 1250 | 310 | $100.10 \pm 5.60$ | Smith et al. (2004) |
| HD210514 | 2200 | 361 | $104.10 \pm 6.00$ | Smith et al. (2004) |
| HD210514 | Johnson | K | $1.90 \pm 0.06$ | Neugebauer & Leighton (1969) |
| HD210514 | 3500 | 898 | $53.60 \pm 8.60$ | Smith et al. (2004) |
| HD210514 | 4900 | 712 | $22.10 \pm 7.20$ | Smith et al. (2004) |
| HD210514 | 12000 | 6384 | $3.60 \pm 17.90$ | Smith et al. (2004) |
| HD211800 | WBVR | B | $8.63 \pm 0.05$ | Kornilov et al. (1991) |
| HD211800 | Johnson | B | $8.56 \pm 0.03$ | This work |

**Table 21** *(continued)*

| Star ID | System/Wvlen | Band/Bandpass | Value | Reference |
|---------|--------------|---------------|-------|-----------|
| HD211800 | Johnson | B | $8.57 \pm 0.01$ | Oja (1987) |
| HD211800 | KronComet | Bc | $8.41 \pm 0.02$ | This work |
| HD211800 | KronComet | C2 | $7.39 \pm 0.02$ | This work |
| HD211800 | KronComet | Gc | $7.07 \pm 0.02$ | This work |
| HD211800 | WBVR | V | $6.92 \pm 0.05$ | Kornilov et al. (1991) |
| HD211800 | Johnson | V | $6.91 \pm 0.01$ | Oja (1987) |
| HD211800 | Johnson | V | $6.97 \pm 0.03$ | This work |
| HD211800 | WBVR | R | $5.56 \pm 0.05$ | Kornilov et al. (1991) |
| HD211800 | KronComet | Rc | $5.48 \pm 0.02$ | This work |
| HD211800 | 2200 | 361 | $47.00 \pm 6.10$ | Smith et al. (2004) |
| HD211800 | Johnson | K | $2.75 \pm 0.10$ | Neugebauer & Leighton (1969) |
| HD211800 | 3500 | 898 | $23.40 \pm 4.60$ | Smith et al. (2004) |
| HD211800 | 4900 | 712 | $11.20 \pm 4.80$ | Smith et al. (2004) |
| HD211800 | 12000 | 6384 | $1.20 \pm 18.60$ | Smith et al. (2004) |
| HD212470 | KronComet | COp | $9.03 \pm 0.04$ | This work |
| HD212470 | Johnson | B | $8.71 \pm 0.04$ | This work |
| HD212470 | KronComet | Bc | $8.60 \pm 0.03$ | This work |
| HD212470 | KronComet | C2 | $7.14 \pm 0.05$ | This work |
| HD212470 | KronComet | Gc | $7.18 \pm 0.01$ | This work |
| HD212470 | Johnson | V | $7.11 \pm 0.02$ | This work |
| HD212470 | KronComet | Rc | $5.39 \pm 0.01$ | This work |
| HD212470 | 1250 | 310 | $218.80 \pm 13.10$ | Smith et al. (2004) |
| HD212470 | 2200 | 361 | $264.90 \pm 13.00$ | Smith et al. (2004) |
| HD212470 | Johnson | K | $0.92 \pm 0.04$ | Neugebauer & Leighton (1969) |
| HD212470 | 3500 | 898 | $139.40 \pm 10.70$ | Smith et al. (2004) |
| HD212470 | 4900 | 712 | $68.90 \pm 7.10$ | Smith et al. (2004) |
| HD212470 | 12000 | 6384 | $28.70 \pm 20.50$ | Smith et al. (2004) |
| HD212496 | 13c | m33 | $6.24 \pm 0.05$ | Johnson & Mitchell (1995) |
| HD212496 | Geneva | U | $6.74 \pm 0.08$ | Golay (1972) |
| HD212496 | Vilnius | U | $8.02 \pm 0.05$ | Kazlauskas et al. (2005) |
| HD212496 | 13c | m35 | $6.01 \pm 0.05$ | Johnson & Mitchell (1995) |
| HD212496 | DDO | m35 | $7.54 \pm 0.06$ | McClure & Forrester (1981) |
| HD212496 | Johnson | U | $6.21 \pm 0.05$ | Argue (1966) |
| HD212496 | Johnson | U | $6.21 \pm 0.05$ | Moreno (1971) |
| HD212496 | Johnson | U | $6.22 \pm 0.05$ | Nicolet (1978) |
| HD212496 | Johnson | U | $6.23 \pm 0.05$ | Johnson et al. (1966) |
| HD212496 | Johnson | U | $6.24 \pm 0.05$ | Jennens & Helfer (1975) |
| HD212496 | Johnson | U | $6.24 \pm 0.05$ | Mermilliod (1986) |
| HD212496 | Johnson | U | $6.25 \pm 0.05$ | Johnson (1964) |
| HD212496 | Johnson | U | $6.26 \pm 0.05$ | Ducati (2002) |
| HD212496 | Geneva | B1 | $6.03 \pm 0.08$ | Golay (1972) |

**Table 21** *(continued)*

| Star ID | System/Wvlen | Band/Bandpass | Value | Reference |
|---------|--------------|---------------|-------|-----------|
| HD212496 | Oja | m41 | 6.54 ± 0.05 | Häggkvist & Oja (1970) |
| HD212496 | DDO | m42 | 6.85 ± 0.05 | McClure & Forrester (1981) |
| HD212496 | Oja | m42 | 6.42 ± 0.05 | Häggkvist & Oja (1970) |
| HD212496 | Geneva | B | 4.74 ± 0.08 | Golay (1972) |
| HD212496 | WBVR | B | 5.47 ± 0.05 | Kornilov et al. (1991) |
| HD212496 | Johnson | B | 5.42 ± 0.05 | Argue (1966) |
| HD212496 | Johnson | B | 5.43 ± 0.05 | Häggkvist & Oja (1966) |
| HD212496 | Johnson | B | 5.44 ± 0.05 | Moreno (1971) |
| HD212496 | Johnson | B | 5.45 ± 0.05 | Jennens & Helfer (1975) |
| HD212496 | Johnson | B | 5.45 ± 0.05 | Nicolet (1978) |
| HD212496 | Johnson | B | 5.46 ± 0.05 | Johnson (1964) |
| HD212496 | Johnson | B | 5.46 ± 0.05 | Johnson et al. (1966) |
| HD212496 | Johnson | B | 5.46 ± 0.05 | Ducati (2002) |
| HD212496 | Johnson | B | 5.48 ± 0.05 | Mermilliod (1986) |
| HD212496 | Geneva | B2 | 5.90 ± 0.08 | Golay (1972) |
| HD212496 | 13c | m45 | 5.12 ± 0.05 | Johnson & Mitchell (1995) |
| HD212496 | DDO | m45 | 5.98 ± 0.05 | McClure & Forrester (1981) |
| HD212496 | Oja | m45 | 5.18 ± 0.05 | Häggkvist & Oja (1970) |
| HD212496 | Vilnius | Y | 5.22 ± 0.05 | Kazlauskas et al. (2005) |
| HD212496 | DDO | m48 | 4.80 ± 0.05 | McClure & Forrester (1981) |
| HD212496 | Vilnius | Z | 4.75 ± 0.05 | Kazlauskas et al. (2005) |
| HD212496 | 13c | m52 | 4.66 ± 0.05 | Johnson & Mitchell (1995) |
| HD212496 | Geneva | V1 | 5.21 ± 0.08 | Golay (1972) |
| HD212496 | WBVR | V | 4.44 ± 0.05 | Kornilov et al. (1991) |
| HD212496 | Vilnius | V | 4.44 ± 0.05 | Kazlauskas et al. (2005) |
| HD212496 | Geneva | V | 4.42 ± 0.08 | Golay (1972) |
| HD212496 | Johnson | V | 4.40 ± 0.05 | Argue (1966) |
| HD212496 | Johnson | V | 4.41 ± 0.05 | Häggkvist & Oja (1966) |
| HD212496 | Johnson | V | 4.43 ± 0.05 | Moreno (1971) |
| HD212496 | Johnson | V | 4.43 ± 0.05 | Nicolet (1978) |
| HD212496 | Johnson | V | 4.44 ± 0.05 | Johnson (1964) |
| HD212496 | Johnson | V | 4.44 ± 0.05 | Johnson et al. (1966) |
| HD212496 | Johnson | V | 4.44 ± 0.05 | Jennens & Helfer (1975) |
| HD212496 | Johnson | V | 4.44 ± 0.05 | Ducati (2002) |
| HD212496 | Johnson | V | 4.46 ± 0.05 | Mermilliod (1986) |
| HD212496 | 13c | m58 | 4.18 ± 0.05 | Johnson & Mitchell (1995) |
| HD212496 | Geneva | G | 5.40 ± 0.08 | Golay (1972) |
| HD212496 | 13c | m63 | 3.88 ± 0.05 | Johnson & Mitchell (1995) |
| HD212496 | Vilnius | S | 3.68 ± 0.05 | Kazlauskas et al. (2005) |
| HD212496 | WBVR | R | 3.68 ± 0.05 | Kornilov et al. (1991) |
| HD212496 | 13c | m72 | 3.60 ± 0.05 | Johnson & Mitchell (1995) |

**Table 21** *(continued)*

| Star ID | System/Wvlen | Band/Bandpass | Value | Reference |
|---------|-------------|---------------|-------|-----------|
| HD212496 | 13c | m80 | $3.36 \pm 0.05$ | Johnson & Mitchell (1995) |
| HD212496 | 13c | m86 | $3.26 \pm 0.05$ | Johnson & Mitchell (1995) |
| HD212496 | 13c | m99 | $3.09 \pm 0.05$ | Johnson & Mitchell (1995) |
| HD212496 | 13c | m110 | $2.91 \pm 0.05$ | Johnson & Mitchell (1995) |
| HD212496 | Johnson | J | $2.69 \pm 0.05$ | Johnson et al. (1966) |
| HD212496 | Johnson | J | $2.69 \pm 0.05$ | Voelcker (1975) |
| HD212496 | Johnson | J | $2.69 \pm 0.05$ | Bergeat & Lunel (1980) |
| HD212496 | Johnson | J | $2.69 \pm 0.05$ | Shenavrin et al. (2011) |
| HD212496 | Johnson | J | $2.70 \pm 0.05$ | Ducati (2002) |
| HD212496 | Johnson | H | $2.17 \pm 0.05$ | Bergeat & Lunel (1980) |
| HD212496 | Johnson | H | $2.19 \pm 0.05$ | Shenavrin et al. (2011) |
| HD212496 | Johnson | H | $2.21 \pm 0.05$ | Voelcker (1975) |
| HD212496 | Johnson | H | $2.21 \pm 0.05$ | Ducati (2002) |
| HD212496 | Johnson | K | $1.98 \pm 0.05$ | Neugebauer & Leighton (1969) |
| HD212496 | Johnson | K | $2.07 \pm 0.05$ | Johnson et al. (1966) |
| HD212496 | Johnson | K | $2.07 \pm 0.05$ | Shenavrin et al. (2011) |
| HD212496 | Johnson | K | $2.10 \pm 0.05$ | Ducati (2002) |
| HD212496 | Johnson | L | $1.95 \pm 0.05$ | Ducati (2002) |
| HD214868 | 13c | m33 | $7.36 \pm 0.05$ | Johnson & Mitchell (1995) |
| HD214868 | Vilnius | U | $9.08 \pm 0.05$ | Kazlauskas et al. (2005) |
| HD214868 | 13c | m35 | $7.02 \pm 0.05$ | Johnson & Mitchell (1995) |
| HD214868 | DDO | m35 | $8.56 \pm 0.05$ | McClure & Forrester (1981) |
| HD214868 | Johnson | U | $7.15 \pm 0.05$ | Johnson et al. (1966) |
| HD214868 | Johnson | U | $7.15 \pm 0.05$ | Ducati (2002) |
| HD214868 | Johnson | U | $7.19 \pm 0.05$ | Mermilliod (1986) |
| HD214868 | Johnson | U | $7.21 \pm 0.05$ | Jennens & Helfer (1975) |
| HD214868 | Oja | m41 | $7.20 \pm 0.05$ | Häggkvist & Oja (1970) |
| HD214868 | DDO | m42 | $7.38 \pm 0.05$ | McClure & Forrester (1981) |
| HD214868 | Oja | m42 | $6.94 \pm 0.05$ | Häggkvist & Oja (1970) |
| HD214868 | WBVR | B | $5.85 \pm 0.05$ | Kornilov et al. (1991) |
| HD214868 | Johnson | B | $5.79 \pm 0.05$ | Häggkvist & Oja (1966) |
| HD214868 | Johnson | B | $5.79 \pm 0.05$ | Johnson et al. (1966) |
| HD214868 | Johnson | B | $5.79 \pm 0.05$ | Ducati (2002) |
| HD214868 | Johnson | B | $5.81 \pm 0.05$ | Mermilliod (1986) |
| HD214868 | Johnson | B | $5.82 \pm 0.05$ | Jennens & Helfer (1975) |
| HD214868 | 13c | m45 | $5.39 \pm 0.05$ | Johnson & Mitchell (1995) |
| HD214868 | DDO | m45 | $6.30 \pm 0.05$ | McClure & Forrester (1981) |
| HD214868 | Oja | m45 | $5.50 \pm 0.05$ | Häggkvist & Oja (1970) |
| HD214868 | Vilnius | Y | $5.47 \pm 0.05$ | Kazlauskas et al. (2005) |
| HD214868 | DDO | m48 | $4.99 \pm 0.05$ | McClure & Forrester (1981) |
| HD214868 | Vilnius | Z | $4.92 \pm 0.05$ | Kazlauskas et al. (2005) |

Table 21 (continued)

| Star ID | System/Wvlen | Band/Bandpass | Value | Reference |
|---------|--------------|---------------|-------|-----------|
| HD214868 | 13c | m52 | $4.82 \pm 0.05$ | Johnson & Mitchell (1995) |
| HD214868 | WBVR | V | $4.51 \pm 0.05$ | Kornilov et al. (1991) |
| HD214868 | Vilnius | V | $4.51 \pm 0.05$ | Kazlauskas et al. (2005) |
| HD214868 | Johnson | V | $4.46 \pm 0.05$ | Johnson et al. (1966) |
| HD214868 | Johnson | V | $4.46 \pm 0.05$ | Ducati (2002) |
| HD214868 | Johnson | V | $4.50 \pm 0.05$ | Häggkvist & Oja (1966) |
| HD214868 | Johnson | V | $4.50 \pm 0.05$ | Jennens & Helfer (1975) |
| HD214868 | Johnson | V | $4.50 \pm 0.05$ | Mermilliod (1986) |
| HD214868 | 13c | m58 | $4.17 \pm 0.05$ | Johnson & Mitchell (1995) |
| HD214868 | 13c | m63 | $3.81 \pm 0.05$ | Johnson & Mitchell (1995) |
| HD214868 | Vilnius | S | $3.60 \pm 0.05$ | Kazlauskas et al. (2005) |
| HD214868 | WBVR | R | $3.55 \pm 0.05$ | Kornilov et al. (1991) |
| HD214868 | 13c | m72 | $3.45 \pm 0.05$ | Johnson & Mitchell (1995) |
| HD214868 | 13c | m80 | $3.18 \pm 0.05$ | Johnson & Mitchell (1995) |
| HD214868 | 13c | m86 | $3.03 \pm 0.05$ | Johnson & Mitchell (1995) |
| HD214868 | 13c | m99 | $2.80 \pm 0.05$ | Johnson & Mitchell (1995) |
| HD214868 | 13c | m110 | $2.60 \pm 0.05$ | Johnson & Mitchell (1995) |
| HD214868 | 1250 | 310 | $200.70 \pm 8.90$ | Smith et al. (2004) |
| HD214868 | Johnson | J | $2.25 \pm 0.05$ | Selby et al. (1988) |
| HD214868 | Johnson | J | $2.25 \pm 0.05$ | Blackwell et al. (1990) |
| HD214868 | Johnson | J | $2.26 \pm 0.05$ | Alonso et al. (1998) |
| HD214868 | Johnson | J | $2.28 \pm 0.05$ | Ducati (2002) |
| HD214868 | Johnson | J | $2.36 \pm 0.05$ | Johnson et al. (1966) |
| HD214868 | Johnson | J | $2.36 \pm 0.05$ | Shenavrin et al. (2011) |
| HD214868 | Johnson | H | $1.61 \pm 0.05$ | Alonso et al. (1998) |
| HD214868 | Johnson | H | $1.68 \pm 0.05$ | Shenavrin et al. (2011) |
| HD214868 | 2200 | 361 | $164.10 \pm 12.40$ | Smith et al. (2004) |
| HD214868 | Johnson | K | $1.41 \pm 0.07$ | Neugebauer & Leighton (1969) |
| HD214868 | Johnson | K | $1.46 \pm 0.05$ | Ducati (2002) |
| HD214868 | Johnson | K | $1.51 \pm 0.05$ | Johnson et al. (1966) |
| HD214868 | Johnson | K | $1.51 \pm 0.05$ | Shenavrin et al. (2011) |
| HD214868 | 3500 | 898 | $77.40 \pm 8.60$ | Smith et al. (2004) |
| HD214868 | 4900 | 712 | $36.10 \pm 5.30$ | Smith et al. (2004) |
| HD214868 | 12000 | 6384 | $-5.30 \pm 18.30$ | Smith et al. (2004) |
| HD215182 | 13c | m33 | $4.36 \pm 0.05$ | Johnson & Mitchell (1995) |
| HD215182 | 13c | m35 | $4.18 \pm 0.05$ | Johnson & Mitchell (1995) |
| HD215182 | WBVR | W | $4.23 \pm 0.05$ | Kornilov et al. (1991) |
| HD215182 | Johnson | U | $4.31 \pm 0.05$ | Johnson & Morgan (1953b) |
| HD215182 | Johnson | U | $4.34 \pm 0.05$ | Mermilliod (1986) |
| HD215182 | Johnson | U | $4.35 \pm 0.05$ | Johnson (1964) |
| HD215182 | Johnson | U | $4.36 \pm 0.05$ | Argue (1966) |

**Table 21** (continued)

| Star ID | System/Wvlen | Band/Bandpass | Value | Reference |
|---------|--------------|---------------|-------|-----------|
| HD215182 | Johnson | U | $4.38 \pm 0.05$ | Johnson et al. (1966) |
| HD215182 | Johnson | U | $4.38 \pm 0.05$ | Ducati (2002) |
| HD215182 | 13c | m37 | $4.24 \pm 0.05$ | Johnson & Mitchell (1995) |
| HD215182 | 13c | m40 | $4.20 \pm 0.05$ | Johnson & Mitchell (1995) |
| HD215182 | Oja | m41 | $4.79 \pm 0.05$ | Häggkvist & Oja (1970) |
| HD215182 | Oja | m42 | $4.65 \pm 0.05$ | Häggkvist & Oja (1970) |
| HD215182 | WBVR | B | $3.81 \pm 0.05$ | Kornilov et al. (1991) |
| HD215182 | Johnson | B | $3.76 \pm 0.05$ | Oja (1963) |
| HD215182 | Johnson | B | $3.76 \pm 0.05$ | Ljunggren & Oja (1965) |
| HD215182 | Johnson | B | $3.79 \pm 0.05$ | Argue (1966) |
| HD215182 | Johnson | B | $3.80 \pm 0.05$ | Johnson & Morgan (1953b) |
| HD215182 | Johnson | B | $3.80 \pm 0.05$ | Johnson (1964) |
| HD215182 | Johnson | B | $3.80 \pm 0.05$ | Mermilliod (1986) |
| HD215182 | Johnson | B | $3.81 \pm 0.05$ | Johnson et al. (1966) |
| HD215182 | Johnson | B | $3.81 \pm 0.05$ | Ducati (2002) |
| HD215182 | 13c | m45 | $3.53 \pm 0.05$ | Johnson & Mitchell (1995) |
| HD215182 | Oja | m45 | $3.57 \pm 0.05$ | Häggkvist & Oja (1970) |
| HD215182 | 13c | m52 | $3.13 \pm 0.05$ | Johnson & Mitchell (1995) |
| HD215182 | WBVR | V | $2.94 \pm 0.05$ | Kornilov et al. (1991) |
| HD215182 | Johnson | V | $2.89 \pm 0.05$ | Oja (1963) |
| HD215182 | Johnson | V | $2.92 \pm 0.05$ | Argue (1966) |
| HD215182 | Johnson | V | $2.93 \pm 0.05$ | Ljunggren & Oja (1965) |
| HD215182 | Johnson | V | $2.95 \pm 0.05$ | Johnson (1964) |
| HD215182 | Johnson | V | $2.95 \pm 0.05$ | Johnson et al. (1966) |
| HD215182 | Johnson | V | $2.95 \pm 0.05$ | Mermilliod (1986) |
| HD215182 | Johnson | V | $2.95 \pm 0.05$ | Ducati (2002) |
| HD215182 | Johnson | V | $2.96 \pm 0.05$ | Johnson & Morgan (1953b) |
| HD215182 | 13c | m58 | $2.73 \pm 0.05$ | Johnson & Mitchell (1995) |
| HD215182 | 13c | m63 | $2.48 \pm 0.05$ | Johnson & Mitchell (1995) |
| HD215182 | WBVR | R | $2.30 \pm 0.05$ | Kornilov et al. (1991) |
| HD215182 | 13c | m72 | $2.26 \pm 0.05$ | Johnson & Mitchell (1995) |
| HD215182 | 13c | m80 | $2.04 \pm 0.05$ | Johnson & Mitchell (1995) |
| HD215182 | 13c | m86 | $1.95 \pm 0.05$ | Johnson & Mitchell (1995) |
| HD215182 | 13c | m99 | $1.81 \pm 0.05$ | Johnson & Mitchell (1995) |
| HD215182 | 13c | m110 | $1.66 \pm 0.05$ | Johnson & Mitchell (1995) |
| HD215182 | 1250 | 310 | $423.20 \pm 18.10$ | Smith et al. (2004) |
| HD215182 | Johnson | J | $1.41 \pm 0.05$ | Selby et al. (1988) |
| HD215182 | Johnson | J | $1.41 \pm 0.05$ | Blackwell et al. (1990) |
| HD215182 | Johnson | J | $1.42 \pm 0.05$ | Ducati (2002) |
| HD215182 | Johnson | J | $1.46 \pm 0.05$ | Johnson et al. (1966) |
| HD215182 | 2200 | 361 | $278.40 \pm 10.30$ | Smith et al. (2004) |

**Table 21** *(continued)*

| Star ID | System/Wvlen | Band/Bandpass | Value | Reference |
|---------|-------------|---------------|-------|-----------|
| HD215182 | Johnson | K | $0.82 \pm 0.05$ | Neugebauer & Leighton (1969) |
| HD215182 | Johnson | K | $0.89 \pm 0.05$ | Ducati (2002) |
| HD215182 | Johnson | K | $0.93 \pm 0.05$ | Johnson et al. (1966) |
| HD215182 | 3500 | 898 | $130.70 \pm 11.10$ | Smith et al. (2004) |
| HD215182 | Johnson | N | $0.50 \pm 0.05$ | Ducati (2002) |
| HD215182 | 4900 | 712 | $63.60 \pm 6.30$ | Smith et al. (2004) |
| HD215182 | 12000 | 6384 | $12.50 \pm 22.50$ | Smith et al. (2004) |
| HD215547 | KronComet | COp | $10.01 \pm 0.05$ | This work |
| HD215547 | Johnson | B | $9.52 \pm 0.01$ | Oja (1986) |
| HD215547 | Johnson | B | $9.52 \pm 0.05$ | Oja (1986) |
| HD215547 | Johnson | B | $9.58 \pm 0.08$ | This work |
| HD215547 | KronComet | Bc | $9.41 \pm 0.05$ | This work |
| HD215547 | KronComet | C2 | $8.13 \pm 0.06$ | This work |
| HD215547 | KronComet | Gc | $8.10 \pm 0.06$ | This work |
| HD215547 | Johnson | V | $7.92 \pm 0.01$ | Oja (1986) |
| HD215547 | Johnson | V | $7.92 \pm 0.05$ | Oja (1986) |
| HD215547 | Johnson | V | $8.04 \pm 0.06$ | This work |
| HD215547 | KronComet | Rc | $6.47 \pm 0.09$ | This work |
| HD215547 | 1250 | 310 | $86.50 \pm 10.80$ | Smith et al. (2004) |
| HD215547 | 2200 | 361 | $92.60 \pm 13.30$ | Smith et al. (2004) |
| HD215547 | Johnson | K | $2.13 \pm 0.05$ | Neugebauer & Leighton (1969) |
| HD215547 | 3500 | 898 | $51.80 \pm 10.10$ | Smith et al. (2004) |
| HD215547 | 4900 | 712 | $20.20 \pm 6.00$ | Smith et al. (2004) |
| HD215547 | 12000 | 6384 | $8.40 \pm 16.50$ | Smith et al. (2004) |
| HD216131 | 13c | m33 | $5.12 \pm 0.05$ | Johnson & Mitchell (1995) |
| HD216131 | Geneva | U | $5.63 \pm 0.08$ | Golay (1972) |
| HD216131 | Vilnius | U | $6.83 \pm 0.05$ | Straizys et al. (1989a) |
| HD216131 | Vilnius | U | $6.89 \pm 0.05$ | Kazlauskas et al. (2005) |
| HD216131 | 13c | m35 | $4.93 \pm 0.05$ | Johnson & Mitchell (1995) |
| HD216131 | DDO | m35 | $6.42 \pm 0.05$ | McClure & Forrester (1981) |
| HD216131 | WBVR | W | $4.96 \pm 0.05$ | Kornilov et al. (1991) |
| HD216131 | Johnson | U | $5.04 \pm 0.05$ | Mermilliod (1986) |
| HD216131 | Johnson | U | $5.10 \pm 0.05$ | Johnson et al. (1966) |
| HD216131 | Johnson | U | $5.10 \pm 0.05$ | Ducati (2002) |
| HD216131 | Johnson | U | $5.12 \pm 0.05$ | Jennens & Helfer (1975) |
| HD216131 | 13c | m37 | $5.03 \pm 0.05$ | Johnson & Mitchell (1995) |
| HD216131 | Vilnius | P | $6.27 \pm 0.05$ | Straizys et al. (1989a) |
| HD216131 | Vilnius | P | $6.33 \pm 0.05$ | Kazlauskas et al. (2005) |
| HD216131 | DDO | m38 | $5.40 \pm 0.05$ | McClure & Forrester (1981) |
| HD216131 | 13c | m40 | $4.94 \pm 0.05$ | Johnson & Mitchell (1995) |
| HD216131 | Geneva | B1 | $4.99 \pm 0.08$ | Golay (1972) |

**Table 21** *(continued)*

| Star ID | System/Wvlen | Band/Bandpass | Value | Reference |
|---------|--------------|---------------|-------|-----------|
| HD216131 | Vilnius | X | $5.39 \pm 0.05$ | Straizys et al. (1989a) |
| HD216131 | Vilnius | X | $5.43 \pm 0.05$ | Kazlauskas et al. (2005) |
| HD216131 | DDO | m41 | $5.96 \pm 0.05$ | McClure & Forrester (1981) |
| HD216131 | Oja | m41 | $5.51 \pm 0.05$ | Häggkvist & Oja (1970) |
| HD216131 | DDO | m42 | $5.81 \pm 0.05$ | McClure & Forrester (1981) |
| HD216131 | Oja | m42 | $5.36 \pm 0.05$ | Häggkvist & Oja (1970) |
| HD216131 | Geneva | B | $3.74 \pm 0.08$ | Golay (1972) |
| HD216131 | WBVR | B | $4.46 \pm 0.05$ | Kornilov et al. (1991) |
| HD216131 | Johnson | B | $4.41 \pm 0.05$ | Bouigue et al. (1961) |
| HD216131 | Johnson | B | $4.42 \pm 0.05$ | Häggkvist & Oja (1966) |
| HD216131 | Johnson | B | $4.42 \pm 0.05$ | Johnson et al. (1966) |
| HD216131 | Johnson | B | $4.42 \pm 0.05$ | Mermilliod (1986) |
| HD216131 | Johnson | B | $4.42 \pm 0.05$ | Ducati (2002) |
| HD216131 | Johnson | B | $4.45 \pm 0.05$ | Jennens & Helfer (1975) |
| HD216131 | Geneva | B2 | $4.93 \pm 0.08$ | Golay (1972) |
| HD216131 | 13c | m45 | $4.17 \pm 0.05$ | Johnson & Mitchell (1995) |
| HD216131 | DDO | m45 | $5.00 \pm 0.05$ | McClure & Forrester (1981) |
| HD216131 | Oja | m45 | $4.18 \pm 0.05$ | Häggkvist & Oja (1970) |
| HD216131 | Vilnius | Y | $4.20 \pm 0.05$ | Straizys et al. (1989a) |
| HD216131 | Vilnius | Y | $4.26 \pm 0.05$ | Kazlauskas et al. (2005) |
| HD216131 | DDO | m48 | $3.84 \pm 0.05$ | McClure & Forrester (1981) |
| HD216131 | Vilnius | Z | $3.75 \pm 0.05$ | Straizys et al. (1989a) |
| HD216131 | Vilnius | Z | $3.81 \pm 0.05$ | Kazlauskas et al. (2005) |
| HD216131 | 13c | m52 | $3.73 \pm 0.05$ | Johnson & Mitchell (1995) |
| HD216131 | Geneva | V1 | $4.31 \pm 0.08$ | Golay (1972) |
| HD216131 | WBVR | V | $3.51 \pm 0.05$ | Kornilov et al. (1991) |
| HD216131 | Vilnius | V | $3.48 \pm 0.05$ | Straizys et al. (1989a) |
| HD216131 | Vilnius | V | $3.53 \pm 0.05$ | Kazlauskas et al. (2005) |
| HD216131 | Geneva | V | $3.53 \pm 0.08$ | Golay (1972) |
| HD216131 | Johnson | V | $3.48 \pm 0.05$ | Johnson et al. (1966) |
| HD216131 | Johnson | V | $3.48 \pm 0.05$ | Mermilliod (1986) |
| HD216131 | Johnson | V | $3.48 \pm 0.05$ | Ducati (2002) |
| HD216131 | Johnson | V | $3.49 \pm 0.05$ | Häggkvist & Oja (1966) |
| HD216131 | Johnson | V | $3.50 \pm 0.05$ | Bouigue et al. (1961) |
| HD216131 | Johnson | V | $3.51 \pm 0.05$ | Jennens & Helfer (1975) |
| HD216131 | 13c | m58 | $3.30 \pm 0.05$ | Johnson & Mitchell (1995) |
| HD216131 | Geneva | G | $4.54 \pm 0.08$ | Golay (1972) |
| HD216131 | 13c | m63 | $3.04 \pm 0.05$ | Johnson & Mitchell (1995) |
| HD216131 | Vilnius | S | $2.80 \pm 0.05$ | Straizys et al. (1989a) |
| HD216131 | Vilnius | S | $2.82 \pm 0.05$ | Kazlauskas et al. (2005) |
| HD216131 | WBVR | R | $2.83 \pm 0.05$ | Kornilov et al. (1991) |

**Table 21** *(continued)*

| Star ID | System/Wvlen | Band/Bandpass | Value | Reference |
|---------|--------------|---------------|-------|-----------|
| HD216131 | 13c | m72 | $2.75 \pm 0.05$ | Johnson & Mitchell (1995) |
| HD216131 | 13c | m80 | $2.55 \pm 0.05$ | Johnson & Mitchell (1995) |
| HD216131 | 13c | m86 | $2.46 \pm 0.05$ | Johnson & Mitchell (1995) |
| HD216131 | 13c | m99 | $2.33 \pm 0.05$ | Johnson & Mitchell (1995) |
| HD216131 | 13c | m110 | $2.20 \pm 0.05$ | Johnson & Mitchell (1995) |
| HD216131 | 1250 | 310 | $283.90 \pm 20.80$ | Smith et al. (2004) |
| HD216131 | Johnson | J | $1.90 \pm 0.05$ | Selby et al. (1988) |
| HD216131 | Johnson | J | $1.90 \pm 0.05$ | Blackwell et al. (1990) |
| HD216131 | Johnson | J | $1.93 \pm 0.05$ | Ducati (2002) |
| HD216131 | Johnson | J | $2.01 \pm 0.05$ | Johnson et al. (1966) |
| HD216131 | Johnson | J | $2.01 \pm 0.05$ | Shenavrin et al. (2011) |
| HD216131 | Johnson | H | $1.55 \pm 0.05$ | Shenavrin et al. (2011) |
| HD216131 | 2200 | 361 | $195.10 \pm 24.90$ | Smith et al. (2004) |
| HD216131 | Johnson | K | $1.35 \pm 0.05$ | Neugebauer & Leighton (1969) |
| HD216131 | Johnson | K | $1.37 \pm 0.05$ | Ducati (2002) |
| HD216131 | Johnson | K | $1.43 \pm 0.05$ | Johnson et al. (1966) |
| HD216131 | Johnson | K | $1.43 \pm 0.05$ | Shenavrin et al. (2011) |
| HD216131 | 3500 | 898 | $95.20 \pm 14.60$ | Smith et al. (2004) |
| HD216131 | 4900 | 712 | $49.70 \pm 8.10$ | Smith et al. (2004) |
| HD216131 | 12000 | 6384 | $18.70 \pm 19.50$ | Smith et al. (2004) |
| HD216930 | KronComet | UVc | $12.48 \pm 0.09$ | This work |
| HD216930 | KronComet | CN | $11.59 \pm 0.06$ | This work |
| HD216930 | KronComet | COp | $10.06 \pm 0.05$ | This work |
| HD216930 | Johnson | B | $9.47 \pm 0.04$ | This work |
| HD216930 | KronComet | Bc | $9.29 \pm 0.03$ | This work |
| HD216930 | KronComet | C2 | $8.09 \pm 0.02$ | This work |
| HD216930 | KronComet | Gc | $7.94 \pm 0.02$ | This work |
| HD216930 | Johnson | V | $7.90 \pm 0.04$ | This work |
| HD216930 | KronComet | Rc | $6.37 \pm 0.01$ | This work |
| HD216930 | 1250 | 310 | $44.60 \pm 7.10$ | Smith et al. (2004) |
| HD216930 | 2200 | 361 | $48.40 \pm 6.80$ | Smith et al. (2004) |
| HD216930 | Johnson | K | $2.64 \pm 0.06$ | Neugebauer & Leighton (1969) |
| HD216930 | 3500 | 898 | $24.50 \pm 4.20$ | Smith et al. (2004) |
| HD216930 | 4900 | 712 | $9.30 \pm 5.30$ | Smith et al. (2004) |
| HD216930 | 12000 | 6384 | $-1.30 \pm 20.90$ | Smith et al. (2004) |
| HD218452 | WBVR | W | $8.37 \pm 0.05$ | Kornilov et al. (1991) |
| HD218452 | Johnson | U | $8.41 \pm 0.05$ | Jennens & Helfer (1975) |
| HD218452 | Johnson | U | $8.46 \pm 0.05$ | Johnson & Knuckles (1957) |
| HD218452 | Johnson | U | $8.46 \pm 0.05$ | Johnson et al. (1966) |
| HD218452 | Oja | m41 | $8.29 \pm 0.05$ | Häggkvist & Oja (1970) |
| HD218452 | Oja | m42 | $7.93 \pm 0.05$ | Häggkvist & Oja (1970) |

**Table 21** *continued on next page*

**Table 21** *(continued)*

| Star ID | System/Wvlen | Band/Bandpass | Value | Reference |
|---------|--------------|---------------|-------|-----------|
| HD218452 | WBVR | B | $6.77 \pm 0.05$ | Kornilov et al. (1991) |
| HD218452 | Johnson | B | $6.71 \pm 0.05$ | Miczaika (1954) |
| HD218452 | Johnson | B | $6.72 \pm 0.05$ | Jennens & Helfer (1975) |
| HD218452 | Johnson | B | $6.74 \pm 0.05$ | Johnson & Knuckles (1957) |
| HD218452 | Johnson | B | $6.74 \pm 0.05$ | Johnson et al. (1966) |
| HD218452 | Oja | m45 | $6.36 \pm 0.05$ | Häggkvist & Oja (1970) |
| HD218452 | WBVR | V | $5.31 \pm 0.05$ | Kornilov et al. (1991) |
| HD218452 | Johnson | V | $5.31 \pm 0.05$ | Jennens & Helfer (1975) |
| HD218452 | Johnson | V | $5.32 \pm 0.05$ | Miczaika (1954) |
| HD218452 | Johnson | V | $5.33 \pm 0.05$ | Johnson & Knuckles (1957) |
| HD218452 | Johnson | V | $5.33 \pm 0.05$ | Johnson et al. (1966) |
| HD218452 | WBVR | R | $4.27 \pm 0.05$ | Kornilov et al. (1991) |
| HD218452 | 1250 | 310 | $104.10 \pm 9.70$ | Smith et al. (2004) |
| HD218452 | 2200 | 361 | $92.80 \pm 12.20$ | Smith et al. (2004) |
| HD218452 | Johnson | K | $1.97 \pm 0.07$ | Neugebauer & Leighton (1969) |
| HD218452 | 3500 | 898 | $38.10 \pm 14.50$ | Smith et al. (2004) |
| HD218452 | 4900 | 712 | $17.60 \pm 5.70$ | Smith et al. (2004) |
| HD218452 | 12000 | 6384 | $-5.50 \pm 16.90$ | Smith et al. (2004) |
| HD219945 | Vilnius | U | $9.06 \pm 0.05$ | Kazlauskas et al. (2005) |
| HD219945 | DDO | m35 | $8.58 \pm 0.05$ | McClure & Forrester (1981) |
| HD219945 | WBVR | W | $7.13 \pm 0.05$ | Kornilov et al. (1991) |
| HD219945 | Johnson | U | $7.25 \pm 0.05$ | Mermilliod (1986) |
| HD219945 | Johnson | U | $7.29 \pm 0.05$ | Argue (1966) |
| HD219945 | Vilnius | P | $8.47 \pm 0.05$ | Kazlauskas et al. (2005) |
| HD219945 | DDO | m38 | $7.53 \pm 0.05$ | McClure & Forrester (1981) |
| HD219945 | Vilnius | X | $7.50 \pm 0.05$ | Kazlauskas et al. (2005) |
| HD219945 | DDO | m41 | $8.04 \pm 0.05$ | McClure & Forrester (1981) |
| HD219945 | Oja | m41 | $7.59 \pm 0.05$ | Häggkvist & Oja (1970) |
| HD219945 | DDO | m42 | $7.86 \pm 0.05$ | McClure & Forrester (1981) |
| HD219945 | Oja | m42 | $7.42 \pm 0.05$ | Häggkvist & Oja (1970) |
| HD219945 | WBVR | B | $6.49 \pm 0.05$ | Kornilov et al. (1991) |
| HD219945 | Johnson | B | $6.46 \pm 0.05$ | Mermilliod (1986) |
| HD219945 | Johnson | B | $6.47 \pm 0.05$ | Argue (1966) |
| HD219945 | DDO | m45 | $7.00 \pm 0.05$ | McClure & Forrester (1981) |
| HD219945 | Oja | m45 | $6.17 \pm 0.05$ | Häggkvist & Oja (1970) |
| HD219945 | Vilnius | Y | $6.24 \pm 0.05$ | Kazlauskas et al. (2005) |
| HD219945 | DDO | m48 | $5.81 \pm 0.05$ | McClure & Forrester (1981) |
| HD219945 | Vilnius | Z | $5.77 \pm 0.05$ | Kazlauskas et al. (2005) |
| HD219945 | WBVR | V | $5.46 \pm 0.05$ | Kornilov et al. (1991) |
| HD219945 | Vilnius | V | $5.46 \pm 0.05$ | Kazlauskas et al. (2005) |
| HD219945 | Johnson | V | $5.42 \pm 0.05$ | Mermilliod (1986) |

**Table 21** *(continued)*

| Star ID | System/Wvlen | Band/Bandpass | Value | Reference |
|---------|--------------|---------------|-------|-----------|
| HD219945 | Johnson | V | $5.44 \pm 0.05$ | Argue (1966) |
| HD219945 | Vilnius | S | $4.71 \pm 0.05$ | Kazlauskas et al. (2005) |
| HD219945 | WBVR | R | $4.71 \pm 0.05$ | Kornilov et al. (1991) |
| HD219945 | 3500 | 898 | $170.10 \pm 43.80$ | Smith et al. (2004) |
| HD219945 | 4900 | 712 | $96.80 \pm 38.20$ | Smith et al. (2004) |
| HD219945 | 12000 | 6384 | $42.50 \pm 23.30$ | Smith et al. (2004) |
| HD220211 | KronComet | COp | $9.75 \pm 0.02$ | This work |
| HD220211 | WBVR | B | $9.01 \pm 0.05$ | Kornilov et al. (1991) |
| HD220211 | Johnson | B | $9.01 \pm 0.03$ | This work |
| HD220211 | KronComet | Bc | $8.86 \pm 0.03$ | This work |
| HD220211 | KronComet | C2 | $7.66 \pm 0.03$ | This work |
| HD220211 | KronComet | Gc | $7.40 \pm 0.02$ | This work |
| HD220211 | WBVR | V | $7.18 \pm 0.05$ | Kornilov et al. (1991) |
| HD220211 | Johnson | V | $7.31 \pm 0.04$ | This work |
| HD220211 | WBVR | R | $5.58 \pm 0.05$ | Kornilov et al. (1991) |
| HD220211 | KronComet | Rc | $5.70 \pm 0.03$ | This work |
| HD220211 | 1250 | 310 | $78.10 \pm 7.10$ | Smith et al. (2004) |
| HD220211 | 2200 | 361 | $76.30 \pm 5.50$ | Smith et al. (2004) |
| HD220211 | Johnson | K | $2.30 \pm 0.07$ | Neugebauer & Leighton (1969) |
| HD220211 | 3500 | 898 | $38.20 \pm 4.30$ | Smith et al. (2004) |
| HD220211 | 4900 | 712 | $16.80 \pm 5.00$ | Smith et al. (2004) |
| HD221115 | 13c | m33 | $6.23 \pm 0.05$ | Johnson & Mitchell (1995) |
| HD221115 | Geneva | U | $6.67 \pm 0.08$ | Golay (1972) |
| HD221115 | Vilnius | U | $7.98 \pm 0.05$ | Straizys et al. (1989a) |
| HD221115 | Vilnius | U | $7.98 \pm 0.05$ | Kazlauskas et al. (2005) |
| HD221115 | 13c | m35 | $6.06 \pm 0.05$ | Johnson & Mitchell (1995) |
| HD221115 | DDO | m35 | $7.49 \pm 0.05$ | McClure & Forrester (1981) |
| HD221115 | Johnson | U | $6.18 \pm 0.05$ | Argue (1966) |
| HD221115 | Johnson | U | $6.24 \pm 0.05$ | Johnson et al. (1966) |
| HD221115 | Johnson | U | $6.24 \pm 0.05$ | Ducati (2002) |
| HD221115 | Geneva | B1 | $6.01 \pm 0.08$ | Golay (1972) |
| HD221115 | Oja | m41 | $6.60 \pm 0.05$ | Häggkvist & Oja (1970) |
| HD221115 | DDO | m42 | $6.83 \pm 0.05$ | McClure & Forrester (1981) |
| HD221115 | Oja | m42 | $6.41 \pm 0.05$ | Häggkvist & Oja (1970) |
| HD221115 | Geneva | B | $4.75 \pm 0.08$ | Golay (1972) |
| HD221115 | WBVR | B | $5.50 \pm 0.05$ | Kornilov et al. (1991) |
| HD221115 | Johnson | B | $5.45 \pm 0.05$ | Häggkvist & Oja (1966) |
| HD221115 | Johnson | B | $5.47 \pm 0.05$ | Argue (1966) |
| HD221115 | Johnson | B | $5.50 \pm 0.05$ | Johnson et al. (1966) |
| HD221115 | Johnson | B | $5.50 \pm 0.05$ | Ducati (2002) |
| HD221115 | Geneva | B2 | $5.92 \pm 0.08$ | Golay (1972) |

**Table 21** (continued)

| Star ID | System/Wvlen | Band/Bandpass | Value | Reference |
|---------|--------------|---------------|-------|-----------|
| HD221115 | 13c | m45 | $5.21 \pm 0.05$ | Johnson & Mitchell (1995) |
| HD221115 | DDO | m45 | $6.02 \pm 0.05$ | McClure & Forrester (1981) |
| HD221115 | Oja | m45 | $5.22 \pm 0.05$ | Häggkvist & Oja (1970) |
| HD221115 | Vilnius | Y | $5.28 \pm 0.05$ | Straizys et al. (1989a) |
| HD221115 | Vilnius | Y | $5.29 \pm 0.05$ | Kazlauskas et al. (2005) |
| HD221115 | DDO | m48 | $4.87 \pm 0.05$ | McClure & Forrester (1981) |
| HD221115 | Vilnius | Z | $4.84 \pm 0.05$ | Straizys et al. (1989a) |
| HD221115 | Vilnius | Z | $4.84 \pm 0.05$ | Kazlauskas et al. (2005) |
| HD221115 | 13c | m52 | $4.77 \pm 0.05$ | Johnson & Mitchell (1995) |
| HD221115 | Geneva | V1 | $5.31 \pm 0.08$ | Golay (1972) |
| HD221115 | WBVR | V | $4.55 \pm 0.05$ | Kornilov et al. (1991) |
| HD221115 | Vilnius | V | $4.55 \pm 0.05$ | Straizys et al. (1989a) |
| HD221115 | Vilnius | V | $4.55 \pm 0.05$ | Kazlauskas et al. (2005) |
| HD221115 | Geneva | V | $4.54 \pm 0.08$ | Golay (1972) |
| HD221115 | Johnson | V | $4.52 \pm 0.05$ | Häggkvist & Oja (1966) |
| HD221115 | Johnson | V | $4.53 \pm 0.05$ | Argue (1966) |
| HD221115 | Johnson | V | $4.56 \pm 0.05$ | Johnson et al. (1966) |
| HD221115 | Johnson | V | $4.56 \pm 0.05$ | Ducati (2002) |
| HD221115 | 13c | m58 | $4.31 \pm 0.05$ | Johnson & Mitchell (1995) |
| HD221115 | Geneva | G | $5.53 \pm 0.08$ | Golay (1972) |
| HD221115 | 13c | m63 | $4.04 \pm 0.05$ | Johnson & Mitchell (1995) |
| HD221115 | Vilnius | S | $3.85 \pm 0.05$ | Straizys et al. (1989a) |
| HD221115 | Vilnius | S | $3.85 \pm 0.05$ | Kazlauskas et al. (2005) |
| HD221115 | WBVR | R | $3.87 \pm 0.05$ | Kornilov et al. (1991) |
| HD221115 | 13c | m72 | $3.82 \pm 0.05$ | Johnson & Mitchell (1995) |
| HD221115 | 13c | m80 | $3.60 \pm 0.05$ | Johnson & Mitchell (1995) |
| HD221115 | 13c | m86 | $3.51 \pm 0.05$ | Johnson & Mitchell (1995) |
| HD221115 | 13c | m99 | $3.37 \pm 0.05$ | Johnson & Mitchell (1995) |
| HD221115 | 13c | m110 | $3.20 \pm 0.05$ | Johnson & Mitchell (1995) |
| HD221115 | 1250 | 310 | $101.00 \pm 5.70$ | Smith et al. (2004) |
| HD221115 | Johnson | J | $2.94 \pm 0.05$ | Selby et al. (1988) |
| HD221115 | Johnson | J | $2.94 \pm 0.05$ | Blackwell et al. (1990) |
| HD221115 | Johnson | J | $2.95 \pm 0.05$ | Alonso et al. (1998) |
| HD221115 | Johnson | J | $2.96 \pm 0.05$ | Ducati (2002) |
| HD221115 | Johnson | J | $2.98 \pm 0.01$ | Laney et al. (2012) |
| HD221115 | Johnson | J | $3.00 \pm 0.05$ | Johnson et al. (1966) |
| HD221115 | Johnson | H | $2.51 \pm 0.05$ | Alonso et al. (1998) |
| HD221115 | Johnson | H | $2.55 \pm 0.01$ | Laney et al. (2012) |
| HD221115 | 2200 | 361 | $64.30 \pm 5.30$ | Smith et al. (2004) |
| HD221115 | Johnson | K | $2.37 \pm 0.07$ | Neugebauer & Leighton (1969) |
| HD221115 | Johnson | K | $2.41 \pm 0.05$ | Ducati (2002) |

**Table 21** *(continued)*

| Star ID | System/Wvlen | Band/Bandpass | Value | Reference |
|---------|--------------|---------------|-------|-----------|
| HD221115 | Johnson | K | $2.43 \pm 0.01$ | Laney et al. (2012) |
| HD221115 | Johnson | K | $2.44 \pm 0.05$ | Johnson et al. (1966) |
| HD221115 | 3500 | 898 | $29.60 \pm 5.00$ | Smith et al. (2004) |
| HD221115 | 4900 | 712 | $14.50 \pm 5.00$ | Smith et al. (2004) |
| HD221115 | 12000 | 6384 | $2.30 \pm 21.30$ | Smith et al. (2004) |
| HD221345 | Vilnius | U | $8.86 \pm 0.05$ | Zdanavicius et al. (1969) |
| HD221345 | Vilnius | U | $8.92 \pm 0.05$ | Straizys et al. (1989a) |
| HD221345 | Vilnius | U | $8.92 \pm 0.05$ | Kazlauskas et al. (2005) |
| HD221345 | DDO | m35 | $8.45 \pm 0.05$ | McClure & Forrester (1981) |
| HD221345 | WBVR | W | $6.98 \pm 0.05$ | Kornilov et al. (1991) |
| HD221345 | Johnson | U | $7.10 \pm 0.05$ | Roman (1955) |
| HD221345 | Johnson | U | $7.10 \pm 0.05$ | Johnson et al. (1966) |
| HD221345 | Johnson | U | $7.12 \pm 0.05$ | Argue (1966) |
| HD221345 | Johnson | U | $7.13 \pm 0.05$ | Jennens & Helfer (1975) |
| HD221345 | Vilnius | P | $8.28 \pm 0.05$ | Zdanavicius et al. (1969) |
| HD221345 | Vilnius | P | $8.31 \pm 0.05$ | Kazlauskas et al. (2005) |
| HD221345 | Vilnius | P | $8.33 \pm 0.05$ | Straizys et al. (1989a) |
| HD221345 | DDO | m38 | $7.37 \pm 0.05$ | McClure & Forrester (1981) |
| HD221345 | Vilnius | X | $7.26 \pm 0.05$ | Zdanavicius et al. (1969) |
| HD221345 | Vilnius | X | $7.31 \pm 0.05$ | Straizys et al. (1989a) |
| HD221345 | Vilnius | X | $7.31 \pm 0.05$ | Kazlauskas et al. (2005) |
| HD221345 | DDO | m41 | $7.83 \pm 0.05$ | McClure & Forrester (1981) |
| HD221345 | Oja | m41 | $7.39 \pm 0.05$ | Häggkvist & Oja (1970) |
| HD221345 | DDO | m42 | $7.68 \pm 0.05$ | McClure & Forrester (1981) |
| HD221345 | Oja | m42 | $7.23 \pm 0.05$ | Häggkvist & Oja (1970) |
| HD221345 | WBVR | B | $6.29 \pm 0.05$ | Kornilov et al. (1991) |
| HD221345 | Johnson | B | $6.24 \pm 0.05$ | Roman (1955) |
| HD221345 | Johnson | B | $6.24 \pm 0.05$ | Johnson et al. (1966) |
| HD221345 | Johnson | B | $6.24 \pm 0.05$ | Argue (1966) |
| HD221345 | Johnson | B | $6.27 \pm 0.05$ | Jennens & Helfer (1975) |
| HD221345 | DDO | m45 | $6.79 \pm 0.05$ | McClure & Forrester (1981) |
| HD221345 | Oja | m45 | $5.98 \pm 0.05$ | Häggkvist & Oja (1970) |
| HD221345 | Vilnius | Y | $5.98 \pm 0.05$ | Zdanavicius et al. (1969) |
| HD221345 | Vilnius | Y | $6.01 \pm 0.05$ | Straizys et al. (1989a) |
| HD221345 | Vilnius | Y | $6.02 \pm 0.05$ | Kazlauskas et al. (2005) |
| HD221345 | DDO | m48 | $5.60 \pm 0.05$ | McClure & Forrester (1981) |
| HD221345 | Vilnius | Z | $5.55 \pm 0.05$ | Zdanavicius et al. (1969) |
| HD221345 | Vilnius | Z | $5.55 \pm 0.05$ | Straizys et al. (1989a) |
| HD221345 | Vilnius | Z | $5.56 \pm 0.05$ | Kazlauskas et al. (2005) |
| HD221345 | WBVR | V | $5.24 \pm 0.05$ | Kornilov et al. (1991) |
| HD221345 | Vilnius | V | $5.22 \pm 0.05$ | Zdanavicius et al. (1969) |

**Table 21** *(continued)*

| Star ID | System/Wvlen | Band/Bandpass | Value | Reference |
|---------|--------------|---------------|-------|-----------|
| HD221345 | Vilnius | V | $5.22 \pm 0.05$ | Straizys et al. (1989a) |
| HD221345 | Vilnius | V | $5.24 \pm 0.05$ | Kazlauskas et al. (2005) |
| HD221345 | Johnson | V | $5.21 \pm 0.05$ | Argue (1966) |
| HD221345 | Johnson | V | $5.22 \pm 0.05$ | Roman (1955) |
| HD221345 | Johnson | V | $5.22 \pm 0.05$ | Johnson et al. (1966) |
| HD221345 | Johnson | V | $5.24 \pm 0.05$ | Jennens & Helfer (1975) |
| HD221345 | Vilnius | S | $4.43 \pm 0.05$ | Zdanavicius et al. (1969) |
| HD221345 | Vilnius | S | $4.47 \pm 0.05$ | Straizys et al. (1989a) |
| HD221345 | Vilnius | S | $4.47 \pm 0.05$ | Kazlauskas et al. (2005) |
| HD221345 | WBVR | R | $4.46 \pm 0.05$ | Kornilov et al. (1991) |
| HD221345 | Johnson | J | $3.38 \pm 0.05$ | Alonso et al. (1998) |
| HD221345 | Johnson | H | $2.88 \pm 0.05$ | Alonso et al. (1998) |
| HD221345 | 2200 | 361 | $50.10 \pm 5.70$ | Smith et al. (2004) |
| HD221345 | Johnson | K | $2.73 \pm 0.09$ | Neugebauer & Leighton (1969) |
| HD221345 | 3500 | 898 | $23.00 \pm 9.20$ | Smith et al. (2004) |
| HD221345 | 4900 | 712 | $10.80 \pm 4.70$ | Smith et al. (2004) |
| HD221345 | 12000 | 6384 | $-1.20 \pm 18.50$ | Smith et al. (2004) |
| HD221673 | WBVR | W | $7.88 \pm 0.05$ | Kornilov et al. (1991) |
| HD221673 | Johnson | U | $7.98 \pm 0.05$ | Nicolet (1978) |
| HD221673 | Johnson | U | $7.99 \pm 0.05$ | Argue (1966) |
| HD221673 | Johnson | U | $8.02 \pm 0.05$ | Eggen (1965) |
| HD221673 | Johnson | U | $8.02 \pm 0.05$ | Mermilliod (1986) |
| HD221673 | Oja | m41 | $7.85 \pm 0.05$ | Häggkvist & Oja (1970) |
| HD221673 | Oja | m42 | $7.55 \pm 0.05$ | Häggkvist & Oja (1970) |
| HD221673 | WBVR | B | $6.40 \pm 0.05$ | Kornilov et al. (1991) |
| HD221673 | Johnson | B | $6.33 \pm 0.05$ | Häggkvist & Oja (1966) |
| HD221673 | Johnson | B | $6.35 \pm 0.05$ | Häggkvist (1966) |
| HD221673 | Johnson | B | $6.36 \pm 0.05$ | Nicolet (1978) |
| HD221673 | Johnson | B | $6.37 \pm 0.05$ | Eggen (1965) |
| HD221673 | Johnson | B | $6.37 \pm 0.05$ | Argue (1966) |
| HD221673 | Johnson | B | $6.37 \pm 0.05$ | Mermilliod (1986) |
| HD221673 | Oja | m45 | $6.01 \pm 0.05$ | Häggkvist & Oja (1970) |
| HD221673 | WBVR | V | $4.98 \pm 0.05$ | Kornilov et al. (1991) |
| HD221673 | Johnson | V | $4.95 \pm 0.05$ | Eggen (1965) |
| HD221673 | Johnson | V | $4.95 \pm 0.05$ | Mermilliod (1986) |
| HD221673 | Johnson | V | $4.96 \pm 0.05$ | Häggkvist & Oja (1966) |
| HD221673 | Johnson | V | $4.98 \pm 0.05$ | Häggkvist (1966) |
| HD221673 | Johnson | V | $4.98 \pm 0.05$ | Argue (1966) |
| HD221673 | Johnson | V | $4.98 \pm 0.05$ | Nicolet (1978) |
| HD221673 | WBVR | R | $3.97 \pm 0.05$ | Kornilov et al. (1991) |
| HD221673 | 1250 | 310 | $165.50 \pm 26.30$ | Smith et al. (2004) |

**Table 21** *(continued)*

| Star ID | System/Wvlen | Band/Bandpass | Value | Reference |
|---------|-----------|-------------|-------------------|-------------------------|
| HD221673 | 2200 | 361 | $133.90 \pm 20.40$ | Smith et al. (2004) |
| HD221673 | Johnson | K | $1.76 \pm 0.07$ | Neugebauer & Leighton (1969) |
| HD221673 | 3500 | 898 | $72.20 \pm 12.60$ | Smith et al. (2004) |
| HD221673 | 4900 | 712 | $32.70 \pm 6.90$ | Smith et al. (2004) |
| HD221673 | 12000 | 6384 | $6.20 \pm 18.90$ | Smith et al. (2004) |
| HD221905 | KronComet | COp | $8.97 \pm 0.03$ | This work |
| HD221905 | WBVR | B | $8.26 \pm 0.05$ | Kornilov et al. (1991) |
| HD221905 | Johnson | B | $8.15 \pm 0.05$ | Guetter & Hewitt (1984) |
| HD221905 | Johnson | B | $8.18 \pm 0.05$ | Oja (1983) |
| HD221905 | Johnson | B | $8.19 \pm 0.01$ | Oja (1991) |
| HD221905 | Johnson | B | $8.24 \pm 0.02$ | This work |
| HD221905 | KronComet | Bc | $8.08 \pm 0.03$ | This work |
| HD221905 | KronComet | C2 | $6.94 \pm 0.02$ | This work |
| HD221905 | KronComet | Gc | $6.66 \pm 0.02$ | This work |
| HD221905 | WBVR | V | $6.47 \pm 0.05$ | Kornilov et al. (1991) |
| HD221905 | Johnson | V | $6.45 \pm 0.05$ | Oja (1983) |
| HD221905 | Johnson | V | $6.45 \pm 0.05$ | Guetter & Hewitt (1984) |
| HD221905 | Johnson | V | $6.47 \pm 0.01$ | Oja (1991) |
| HD221905 | Johnson | V | $6.56 \pm 0.03$ | This work |
| HD221905 | WBVR | R | $4.98 \pm 0.05$ | Kornilov et al. (1991) |
| HD221905 | KronComet | Rc | $5.02 \pm 0.02$ | This work |
| HD221905 | 1250 | 310 | $118.10 \pm 11.00$ | Smith et al. (2004) |
| HD221905 | 2200 | 361 | $117.80 \pm 9.80$ | Smith et al. (2004) |
| HD221905 | Johnson | K | $1.86 \pm 0.06$ | Neugebauer & Leighton (1969) |
| HD221905 | 3500 | 898 | $59.20 \pm 7.10$ | Smith et al. (2004) |
| HD221905 | 4900 | 712 | $26.00 \pm 4.70$ | Smith et al. (2004) |
| HD221905 | 12000 | 6384 | $6.90 \pm 18.10$ | Smith et al. (2004) |
| HD222842 | 13c | m33 | $6.51 \pm 0.05$ | Johnson & Mitchell (1995) |
| HD222842 | Geneva | U | $7.00 \pm 0.08$ | Golay (1972) |
| HD222842 | 13c | m35 | $6.33 \pm 0.05$ | Johnson & Mitchell (1995) |
| HD222842 | WBVR | W | $6.38 \pm 0.05$ | Kornilov et al. (1991) |
| HD222842 | Johnson | U | $6.50 \pm 0.05$ | Argue (1966) |
| HD222842 | Johnson | U | $6.51 \pm 0.05$ | Nicolet (1978) |
| HD222842 | Johnson | U | $6.52 \pm 0.05$ | Johnson et al. (1966) |
| HD222842 | 13c | m37 | $6.43 \pm 0.05$ | Johnson & Mitchell (1995) |
| HD222842 | 13c | m40 | $6.34 \pm 0.05$ | Johnson & Mitchell (1995) |
| HD222842 | Geneva | B1 | $6.38 \pm 0.08$ | Golay (1972) |
| HD222842 | Oja | m41 | $6.95 \pm 0.05$ | Häggkvist & Oja (1970) |
| HD222842 | Oja | m42 | $6.80 \pm 0.05$ | Häggkvist & Oja (1970) |
| HD222842 | Geneva | B | $5.13 \pm 0.08$ | Golay (1972) |
| HD222842 | WBVR | B | $5.89 \pm 0.05$ | Kornilov et al. (1991) |

**Table 21** *continued on next page*

**Table 21** (continued)

| Star ID | System/Wvlen | Band/Bandpass | Value | Reference |
|---------|--------------|---------------|-------|-----------|
| HD222842 | Johnson | B | $5.85 \pm 0.05$ | Häggkvist & Oja (1966) |
| HD222842 | Johnson | B | $5.85 \pm 0.05$ | Argue (1966) |
| HD222842 | Johnson | B | $5.86 \pm 0.05$ | Moffett & Barnes (1979) |
| HD222842 | Johnson | B | $5.88 \pm 0.05$ | Nicolet (1978) |
| HD222842 | Johnson | B | $5.89 \pm 0.05$ | Johnson et al. (1966) |
| HD222842 | Geneva | B2 | $6.32 \pm 0.08$ | Golay (1972) |
| HD222842 | 13c | m45 | $5.57 \pm 0.05$ | Johnson & Mitchell (1995) |
| HD222842 | Oja | m45 | $5.61 \pm 0.05$ | Häggkvist & Oja (1970) |
| HD222842 | 13c | m52 | $5.13 \pm 0.05$ | Johnson & Mitchell (1995) |
| HD222842 | Geneva | V1 | $5.69 \pm 0.08$ | Golay (1972) |
| HD222842 | WBVR | V | $4.94 \pm 0.05$ | Kornilov et al. (1991) |
| HD222842 | Geneva | V | $4.92 \pm 0.08$ | Golay (1972) |
| HD222842 | Johnson | V | $4.92 \pm 0.05$ | Häggkvist & Oja (1966) |
| HD222842 | Johnson | V | $4.92 \pm 0.05$ | Argue (1966) |
| HD222842 | Johnson | V | $4.93 \pm 0.05$ | Johnson et al. (1966) |
| HD222842 | Johnson | V | $4.93 \pm 0.05$ | Nicolet (1978) |
| HD222842 | Johnson | V | $4.94 \pm 0.05$ | Moffett & Barnes (1979) |
| HD222842 | 13c | m58 | $4.69 \pm 0.05$ | Johnson & Mitchell (1995) |
| HD222842 | Geneva | G | $5.91 \pm 0.08$ | Golay (1972) |
| HD222842 | 13c | m63 | $4.42 \pm 0.05$ | Johnson & Mitchell (1995) |
| HD222842 | WBVR | R | $4.24 \pm 0.05$ | Kornilov et al. (1991) |
| HD222842 | 13c | m72 | $4.17 \pm 0.05$ | Johnson & Mitchell (1995) |
| HD222842 | 13c | m80 | $3.95 \pm 0.05$ | Johnson & Mitchell (1995) |
| HD222842 | 13c | m86 | $3.85 \pm 0.05$ | Johnson & Mitchell (1995) |
| HD222842 | 13c | m99 | $3.71 \pm 0.05$ | Johnson & Mitchell (1995) |
| HD222842 | 13c | m110 | $3.53 \pm 0.05$ | Johnson & Mitchell (1995) |
| HD222842 | 1250 | 310 | $83.00 \pm 8.50$ | Smith et al. (2004) |
| HD222842 | 2200 | 361 | $55.60 \pm 10.00$ | Smith et al. (2004) |
| HD222842 | Johnson | K | $2.69 \pm 0.10$ | Neugebauer & Leighton (1969) |
| HD222842 | 3500 | 898 | $24.40 \pm 9.50$ | Smith et al. (2004) |
| HD222842 | 4900 | 712 | $12.50 \pm 5.50$ | Smith et al. (2004) |
| HD222842 | 12000 | 6384 | $3.90 \pm 20.00$ | Smith et al. (2004) |
| HD224303 | KronComet | COp | $8.58 \pm 0.02$ | This work |
| HD224303 | WBVR | B | $7.85 \pm 0.05$ | Kornilov et al. (1991) |
| HD224303 | Johnson | B | $7.75 \pm 0.05$ | Haggkvist & Oja (1970) |
| HD224303 | Johnson | B | $7.78 \pm 0.05$ | Mermilliod (1986) |
| HD224303 | Johnson | B | $7.85 \pm 0.02$ | This work |
| HD224303 | KronComet | Bc | $7.67 \pm 0.02$ | This work |
| HD224303 | KronComet | C2 | $6.66 \pm 0.01$ | This work |
| HD224303 | KronComet | Gc | $6.38 \pm 0.01$ | This work |
| HD224303 | WBVR | V | $6.19 \pm 0.05$ | Kornilov et al. (1991) |

**Table 21** *(continued)*

| Star ID | System/Wvlen | Band/Bandpass | Value | Reference |
|---------|--------------|---------------|-------|-----------|
| HD224303 | Johnson | V | $6.15 \pm 0.05$ | Haggkvist & Oja (1970) |
| HD224303 | Johnson | V | $6.17 \pm 0.05$ | Mermilliod (1986) |
| HD224303 | Johnson | V | $6.26 \pm 0.02$ | This work |
| HD224303 | WBVR | R | $4.82 \pm 0.05$ | Kornilov et al. (1991) |
| HD224303 | KronComet | Rc | $4.82 \pm 0.01$ | This work |
| HD224303 | 1250 | 310 | $106.60 \pm 6.00$ | Smith et al. (2004) |
| HD224303 | 2200 | 361 | $102.30 \pm 6.60$ | Smith et al. (2004) |
| HD224303 | Johnson | K | $1.96 \pm 0.05$ | Neugebauer & Leighton (1969) |
| HD224303 | 3500 | 898 | $50.60 \pm 6.40$ | Smith et al. (2004) |
| HD224303 | 4900 | 712 | $22.30 \pm 4.70$ | Smith et al. (2004) |
| HD224303 | 12000 | 6384 | $-0.70 \pm 19.40$ | Smith et al. (2004) |
| HIP8682 | KronComet | C2 | $9.87 \pm 0.01$ | This work |
| HIP8682 | KronComet | Gc | $10.13 \pm 0.02$ | This work |
| HIP8682 | Johnson | V | $9.97 \pm 0.05$ | This work |
| HIP8682 | KronComet | Rc | $7.99 \pm 0.02$ | This work |
| HIP8682 | Johnson | K | $2.12 \pm 0.06$ | Neugebauer & Leighton (1969) |
| HIP8682 | 3500 | 898 | $72.20 \pm 11.60$ | Smith et al. (2004) |
| HIP8682 | 4900 | 712 | $31.20 \pm 5.50$ | Smith et al. (2004) |
| HIP8682 | 12000 | 6384 | $8.80 \pm 18.60$ | Smith et al. (2004) |
| HIP35915 | KronComet | CN | $12.51 \pm 0.05$ | This work |
| HIP35915 | KronComet | COp | $11.50 \pm 0.03$ | This work |
| HIP35915 | Johnson | B | $11.36 \pm 0.01$ | This work |
| HIP35915 | KronComet | Bc | $11.38 \pm 0.03$ | This work |
| HIP35915 | KronComet | C2 | $9.87 \pm 0.01$ | This work |
| HIP35915 | KronComet | Gc | $10.14 \pm 0.02$ | This work |
| HIP35915 | Johnson | V | $10.20 \pm 0.05$ | Ducati (2002) |
| HIP35915 | Johnson | V | $9.96 \pm 0.01$ | This work |
| HIP35915 | KronComet | Rc | $8.10 \pm 0.01$ | This work |
| HIP35915 | 1250 | 310 | $31.70 \pm 17.40$ | Smith et al. (2004) |
| HIP35915 | 2200 | 361 | $41.50 \pm 12.50$ | Smith et al. (2004) |
| HIP35915 | Johnson | K | $2.84 \pm 0.10$ | Neugebauer & Leighton (1969) |
| HIP35915 | Johnson | K | $2.87 \pm 0.05$ | Ducati (2002) |
| HIP35915 | Johnson | L | $2.61 \pm 0.05$ | Ducati (2002) |
| HIP35915 | 3500 | 898 | $22.00 \pm 13.50$ | Smith et al. (2004) |
| HIP35915 | Johnson | N | $1.97 \pm 0.05$ | Ducati (2002) |
| HIP35915 | 4900 | 712 | $8.20 \pm 7.00$ | Smith et al. (2004) |
| HIP35915 | Johnson | M | $2.72 \pm 0.05$ | Ducati (2002) |
| HIP68357 | KronComet | COp | $10.48 \pm 0.05$ | This work |
| HIP68357 | Johnson | B | $10.02 \pm 0.23$ | This work |
| HIP68357 | KronComet | Bc | $10.62 \pm 0.03$ | This work |
| HIP68357 | KronComet | C2 | $8.92 \pm 0.02$ | This work |

**Table 21** *continued on next page*

**Table 21** *(continued)*

| Star ID | System/Wvlen | Band/Bandpass | Value | Reference |
|---------|--------------|---------------|-------|-----------|
| HIP68357 | KronComet | Gc | $9.30 \pm 0.01$ | This work |
| HIP68357 | Johnson | V | $9.11 \pm 0.04$ | This work |
| HIP68357 | Johnson | V | $9.46 \pm 0.05$ | Oja (1985a) |
| HIP68357 | KronComet | Rc | $7.24 \pm 0.01$ | This work |
| HIP68357 | 1250 | 310 | $132.30 \pm 7.20$ | Smith et al. (2004) |
| HIP68357 | 2200 | 361 | $157.30 \pm 5.50$ | Smith et al. (2004) |
| HIP68357 | Johnson | K | $1.45 \pm 0.05$ | Neugebauer & Leighton (1969) |
| HIP68357 | 3500 | 898 | $86.90 \pm 7.60$ | Smith et al. (2004) |
| HIP68357 | 4900 | 712 | $44.40 \pm 4.60$ | Smith et al. (2004) |
| HIP68357 | 12000 | 6384 | $24.40 \pm 15.30$ | Smith et al. (2004) |
| HIP86677 | Johnson | B | $8.48 \pm 0.10$ | This work |
| HIP86677 | Johnson | V | $6.60 \pm 0.10$ | This work |
| HIP86677 | 1250 | 310 | $148.60 \pm 6.70$ | Smith et al. (2004) |
| HIP86677 | 2200 | 361 | $162.20 \pm 8.00$ | Smith et al. (2004) |
| HIP86677 | 3500 | 898 | $84.20 \pm 11.40$ | Smith et al. (2004) |
| HIP86677 | 4900 | 712 | $35.40 \pm 6.60$ | Smith et al. (2004) |
| HIP86677 | 12000 | 6384 | $16.40 \pm 20.10$ | Smith et al. (2004) |
| HIP95024 | KronComet | NH | $13.14 \pm 0.18$ | This work |
| HIP95024 | KronComet | UVc | $12.88 \pm 0.05$ | This work |
| HIP95024 | Johnson | U | $11.84 \pm 0.04$ | This work |
| HIP95024 | KronComet | CN | $12.27 \pm 0.06$ | This work |
| HIP95024 | KronComet | COp | $11.93 \pm 0.01$ | This work |
| HIP95024 | Johnson | B | $11.84 \pm 0.01$ | This work |
| HIP95024 | KronComet | Bc | $11.75 \pm 0.01$ | This work |
| HIP95024 | KronComet | C2 | $11.36 \pm 0.01$ | This work |
| HIP95024 | KronComet | Gc | $11.31 \pm 0.01$ | This work |
| HIP95024 | Johnson | V | $11.31 \pm 0.01$ | This work |
| HIP95024 | KronComet | Rc | $10.62 \pm 0.01$ | This work |
| HIP113390 | KronComet | COp | $11.37 \pm 0.10$ | This work |
| HIP113390 | Johnson | B | $11.32 \pm 0.10$ | This work |
| HIP113390 | KronComet | Bc | $11.42 \pm 0.08$ | This work |
| HIP113390 | KronComet | C2 | $9.63 \pm 0.06$ | This work |
| HIP113390 | KronComet | Gc | $9.89 \pm 0.08$ | This work |
| HIP113390 | Johnson | V | $9.70 \pm 0.10$ | This work |
| HIP113390 | KronComet | Rc | $7.62 \pm 0.05$ | This work |
| HIP113390 | 1250 | 310 | $62.70 \pm 14.10$ | Smith et al. (2004) |
| HIP113390 | Johnson | J | $3.68 \pm 0.05$ | Kerschbaum & Hron (1994) |
| HIP113390 | Johnson | H | $2.67 \pm 0.05$ | Kerschbaum & Hron (1994) |
| HIP113390 | 2200 | 361 | $82.60 \pm 9.10$ | Smith et al. (2004) |
| HIP113390 | Johnson | K | $2.05 \pm 0.08$ | Neugebauer & Leighton (1969) |
| HIP113390 | 3500 | 898 | $44.90 \pm 19.20$ | Smith et al. (2004) |

Table 21 (continued)

| Star ID | System/Wvlen | Band/Bandpass | Value | Reference |
|---------|--------------|---------------|-------|-----------|
| HIP113390 | 4900 | 712 | $22.50 \pm 5.40$ | Smith et al. (2004) |
| HIP113390 | 12000 | 6384 | $19.20 \pm 18.70$ | Smith et al. (2004) |
| IRC+30095 | KronComet | COp | $12.31 \pm 0.13$ | This work |
| IRC+30095 | Johnson | B | $11.96 \pm 0.05$ | This work |
| IRC+30095 | KronComet | Bc | $12.08 \pm 0.12$ | This work |
| IRC+30095 | KronComet | C2 | $10.13 \pm 0.04$ | This work |
| IRC+30095 | KronComet | Gc | $10.21 \pm 0.06$ | This work |
| IRC+30095 | Johnson | V | $10.15 \pm 0.04$ | This work |
| IRC+30095 | KronComet | Rc | $7.77 \pm 0.04$ | This work |
| IRC+30095 | 2200 | 361 | $106.70 \pm 26.40$ | Smith et al. (2004) |
| IRC+30095 | Johnson | K | $1.92 \pm 0.05$ | Neugebauer & Leighton (1969) |
| IRC+30095 | 3500 | 898 | $72.40 \pm 29.90$ | Smith et al. (2004) |
| IRC+30095 | 4900 | 712 | $19.20 \pm 4.30$ | Smith et al. (2004) |
| IRC+30095 | 12000 | 6384 | $0.60 \pm 26.30$ | Smith et al. (2004) |
| IRC+40448 | KronComet | UVc | $16.27 \pm 0.01$ | This work |
| IRC+40448 | Johnson | U | $15.93 \pm 0.06$ | This work |
| IRC+40448 | KronComet | CN | $16.76 \pm 0.06$ | This work |
| IRC+40448 | KronComet | COp | $16.68 \pm 0.27$ | This work |
| IRC+40448 | Johnson | B | $16.97 \pm 0.01$ | This work |
| IRC+40448 | KronComet | Bc | $15.83 \pm 0.01$ | This work |
| IRC+40448 | KronComet | Gc | $15.59 \pm 0.15$ | This work |
| IRC+40448 | KronComet | Rc | $12.03 \pm 0.02$ | This work |
| IRC+40448 | Johnson | J | $4.57 \pm 0.05$ | Johnson et al. (1965c) |
| IRC+40448 | Johnson | J | $4.57 \pm 0.05$ | Johnson et al. (1965d) |
| IRC+40448 | Johnson | J | $4.57 \pm 0.05$ | Johnson (1967) |
| IRC+40448 | Johnson | J | $4.64 \pm 0.05$ | Johnson et al. (1965a) |
| IRC+40448 | Johnson | J | $4.64 \pm 0.05$ | Johnson et al. (1965b) |
| IRC+40448 | Johnson | J | $4.78 \pm 0.05$ | Low et al. (1970) |
| IRC+40448 | Johnson | J | $4.78 \pm 0.05$ | Voelcker (1975) |
| IRC+40448 | Johnson | J | $4.89 \pm 0.05$ | Wisniewski et al. (1967) |
| IRC+40448 | Johnson | J | $4.90 \pm 0.05$ | Hyland et al. (1969) |
| IRC+40448 | Johnson | J | $4.90 \pm 0.05$ | Hyland et al. (1972) |
| IRC+40448 | Johnson | J | $4.91 \pm 0.05$ | Wisniewski et al. (1967) |
| IRC+40448 | Johnson | H | $2.40 \pm 0.05$ | Voelcker (1975) |
| IRC+40448 | Johnson | H | $2.50 \pm 0.05$ | Hyland et al. (1972) |
| IRC+40448 | Johnson | H | $2.55 \pm 0.05$ | Hyland et al. (1969) |
| IRC+40448 | Johnson | K | $0.38 \pm 0.05$ | Johnson et al. (1965a) |
| IRC+40448 | Johnson | K | $0.38 \pm 0.05$ | Johnson et al. (1965c) |
| IRC+40448 | Johnson | K | $0.44 \pm 0.05$ | Wisniewski et al. (1967) |
| IRC+40448 | Johnson | K | $0.49 \pm 0.05$ | Wisniewski et al. (1967) |
| IRC+40448 | Johnson | K | $0.62 \pm 0.03$ | Neugebauer & Leighton (1969) |

**Table 21** *(continued)*

| Star ID | System/Wvlen | Band/Bandpass | Value | Reference |
|---|---|---|---|---|
| IRC+40448 | Johnson | L | $-1.65 \pm 0.05$ | Wisniewski et al. (1967) |
| IRC+40448 | Johnson | L | $-1.85 \pm 0.05$ | Wisniewski et al. (1967) |
| IRC+40448 | Johnson | L | $-1.91 \pm 0.05$ | Johnson et al. (1965c) |
| IRC+40448 | Johnson | L | $-1.92 \pm 0.05$ | Johnson et al. (1965a) |
| IRC+40448 | 3500 | 898 | $931.20 \pm 91.30$ | Smith et al. (2004) |
| IRC+40448 | Johnson | N | $-5.39 \pm 0.05$ | Johnson et al. (1965a) |
| IRC+40448 | Johnson | N | $-5.39 \pm 0.05$ | Johnson et al. (1965c) |
| IRC+40448 | Johnson | N | $-5.39 \pm 0.05$ | Johnson (1967) |
| IRC+40448 | Johnson | M | $-3.47 \pm 0.05$ | Johnson et al. (1965a) |
| IRC+40448 | Johnson | M | $-3.47 \pm 0.05$ | Johnson et al. (1965c) |
| IRC+40448 | Johnson | M | $-3.47 \pm 0.05$ | Johnson (1967) |
| IRC+40448 | 12000 | 6384 | $3,882.70 \pm 280.80$ | Smith et al. (2004) |
| IRC+40533 | KronComet | NH | $13.34 \pm 0.07$ | This work |
| IRC+40533 | KronComet | UVc | $13.34 \pm 0.04$ | This work |
| IRC+40533 | KronComet | CN | $12.15 \pm 0.10$ | This work |
| IRC+40533 | KronComet | COp | $11.40 \pm 0.04$ | This work |
| IRC+40533 | Johnson | B | $11.37 \pm 0.02$ | This work |
| IRC+40533 | KronComet | Bc | $11.47 \pm 0.12$ | This work |
| IRC+40533 | KronComet | C2 | $9.97 \pm 0.07$ | This work |
| IRC+40533 | KronComet | Gc | $10.20 \pm 0.02$ | This work |
| IRC+40533 | Johnson | V | $9.99 \pm 0.03$ | This work |
| IRC+40533 | KronComet | Rc | $7.88 \pm 0.05$ | This work |
| IRC+40533 | 1250 | 310 | $119.00 \pm 20.70$ | Smith et al. (2004) |
| IRC+40533 | Johnson | K | $1.43 \pm 0.03$ | Neugebauer & Leighton (1969) |
| IRC+40533 | 3500 | 898 | $86.30 \pm 24.50$ | Smith et al. (2004) |
| IRC+40533 | 4900 | 712 | $43.20 \pm 11.30$ | Smith et al. (2004) |
| IRC+40533 | 12000 | 6384 | $1.20 \pm 19.20$ | Smith et al. (2004) |
| IRC+60005 | Johnson | V | $8.30 \pm 0.05$ | Ducati (2002) |
| IRC+60005 | 1250 | 310 | $99.00 \pm 10.40$ | Smith et al. (2004) |
| IRC+60005 | Johnson | H | $2.20 \pm 0.05$ | Swings & Allen (1972) |
| IRC+60005 | Johnson | H | $2.20 \pm 0.05$ | Ducati (2002) |
| IRC+60005 | 2200 | 361 | $128.10 \pm 10.40$ | Smith et al. (2004) |
| IRC+60005 | Johnson | K | $1.81 \pm 0.05$ | Swings & Allen (1972) |
| IRC+60005 | Johnson | K | $1.81 \pm 0.05$ | Ducati (2002) |
| IRC+60005 | Johnson | K | $1.85 \pm 0.05$ | Neugebauer & Leighton (1969) |
| IRC+60005 | Johnson | L | $1.27 \pm 0.05$ | Swings & Allen (1972) |
| IRC+60005 | Johnson | L | $1.27 \pm 0.05$ | Ducati (2002) |
| IRC+60005 | 3500 | 898 | $86.50 \pm 7.60$ | Smith et al. (2004) |
| IRC+60005 | 4900 | 712 | $48.10 \pm 5.50$ | Smith et al. (2004) |
| IRC+60005 | 12000 | 6384 | $-22.30 \pm 24.70$ | Smith et al. (2004) |